%% file: Thesis.tex
\documentclass[11pt,oneside,openany,headings=optiontoheadandtoc,numbers=noenddot]{scrbook}
\usepackage[utf8]{inputenc}
\pdfoutput=1

\input{preamble/mypreamble}

\input{preamble/definitions}

\begin{document}

\frontmatter
\input{chapters/frontmatter}
\pdfbookmark{\contentsname}{toc}\tableofcontents

\mainmatter
\thispagestyle{empty}
\addtocounter{page}{-1}
\input{chapters/introduction}
\input{chapters/qft}

\input{chapters/grassmann}
\input{chapters/kinematicspaces}
\input{chapters/amplitudes}

\input{chapters/positivegeom}

\input{chapters/dualspace}
\input{chapters/conclusion}

\appendix
\input{appendices/alggeom}
\input{appendices/schubert}

\input{appendices/black-white}
\input{appendices/scatt-eq}
\input{appendices/chambers}

\bibliographystyle{nb}
\bibliography{bibliography}

\end{document}

%% file: preamble/mypreamble.tex
\makeatletter
\DeclareOldFontCommand{\rm}{\normalfont\rmfamily}{\mathrm}
\DeclareOldFontCommand{\sf}{\normalfont\sffamily}{\mathsf}
\DeclareOldFontCommand{\tt}{\normalfont\ttfamily}{\mathtt}
\DeclareOldFontCommand{\bf}{\normalfont\bfseries}{\mathbf}
\DeclareOldFontCommand{\it}{\normalfont\itshape}{\mathit}
\DeclareOldFontCommand{\sl}{\normalfont\slshape}{\@nomath\sl}
\DeclareOldFontCommand{\sc}{\normalfont\scshape}{\@nomath\sc}
\makeatother

\usepackage[a4paper,width=150mm,top=25mm,bottom=25mm]{geometry}

\usepackage[dvipsnames]{xcolor}
\usepackage{hyperref}
\hypersetup{
	pdftoolbar=false,				% show Acrobat’s toolbar?
	pdfmenubar=false,				% show Acrobat’s menu?
	pdffitwindow=false,				% window fit to page when opened
	pdfstartview={FitH},			% fits the width of the page to the window
	pdftitle={Positive Geometries for Scattering Amplitudes in N=4 SYM and ABJM},							% title
	pdfauthor={Jonah Stalknecht},	% author
	pdfsubject={PhD Thesis},		% subject of the document
	pdfcreator={Jonah Stalknecht},	% creator of the document
	pdfproducer={Jonah Stalknecht},	% producer of the document
	pdfkeywords={Quantum Field Theory}{Scattering Amplitudes} {Positive Geometry} {Amplituhedron} {Theoretical Physics},			% list of keywords
	pdfnewwindow=true,				% links in new window
	colorlinks=true,				% false: boxed links; true: colored links
	linkcolor=Mahogany,				% color of internal links (change box color with linkbordercolor)
	citecolor=MidnightBlue,			% color of links to bibliography
	filecolor=magenta,				% color of file links
	urlcolor=MidnightBlue,          % color of external links
	linktocpage=true,				% make page numbers linked in ToC, rather than chapter name
	pdfdisplaydoctitle=true			% display document title instead of file name in title bar
}

\usepackage{booktabs}

\usepackage{nameref}

\usepackage{cite}

\usepackage{graphicx}
\graphicspath{ {images/} }

\usepackage{abstract}

\usepackage{amsfonts,amsmath,amsthm,amssymb,amsbsy}
\usepackage{mathrsfs}  
\usepackage{physics}
\usepackage{tikz}
\usepackage{dsfont}
\usepackage{xcolor}
\usepackage{mathtools}
\usepackage{bbold}

\usepackage{fancyhdr}
\usepackage{array}
\usepackage{arydshln}

\usepackage{bm}

\usepackage{slashed}

\usepackage{xspace}

\usepackage{wasysym}

\usepackage{subcaption}

\usepackage{mathtools}

\usepackage{nameref}

\usepackage{titling}

\newtheorem{theorem}{Theorem}

\newtheoremstyle{colon}
{}
{}
{}
{}
{\bfseries}
{:}
{ }
{}
\theoremstyle{colon}

%% file: preamble/definitions.tex
\DeclareMathOperator{\sgn}{sgn}
\DeclareMathOperator{\conv}{Conv}
\DeclareMathOperator{\diag}{diag}
\DeclareMathOperator{\vol}{vol}
\DeclareMathOperator{\PT}{PT}

\newcommand{\Grob}{Gr\"o{}bner\xspace}
\newcommand{\Pluck}{Pl\"ucker\xspace}
\newcommand{\nf}{$\smash{\mathcal{N}=4}$ SYM\xspace}
\newcommand{\MHVbar}{$\overline{\text{MHV}}$\xspace}
\newcommand{\ABHY}{\text{ABHY}}

\renewcommand{\tr}[1]{\mathop{\mathrm{Tr}} \left(#1\right)}
\newcommand{\brac}[1]{\left( #1 \right)}

\newcommand{\overbar}[1]{{\mkern 1.5mu\overline{\mkern-1.5mu#1\mkern-1.5mu}\mkern 1.5mu}}
\newcommand{\<}{\langle}
\renewcommand{\>}{\rangle}

\newcommand{\mcl}[1]{\mathcal{#1}}

\newcommand{\ls}{\ell^\star}
\newcommand{\lstil}{\tilde{\ell}^\star}

\newcommand{\eps}{\varepsilon}

\newcommand{\unit}{\mathds{1}}
\newcommand{\nul}{{\mathbb{0}}}
\newcommand{\Rbb}{\mathds{R}}
\newcommand{\Pbb}{\mathds{P}}
\newcommand{\Cbb}{\mathds{C}}
\newcommand{\Kbb}{\mathds{K}}
\newcommand{\Zbb}{\mathds{Z}}
\newcommand{\Qbb}{\mathds{Q}}
\newcommand{\Dbb}{\mathds{D}}
\newcommand{\Ical}{\mathcal{I}}
\newcommand{\Vcal}{\mathcal{V}}
\newcommand{\Acal}{\mathcal{A}}
\newcommand{\Lcal}{\mathcal{L}}
\newcommand{\Ncal}{\mathcal{N}}
\newcommand{\Mcal}{\mathcal{M}}
\newcommand{\Zcal}{\mathcal{Z}}
\newcommand{\Tcal}{\mathcal{T}}
\newcommand{\Ccal}{\mathcal{C}}
\newcommand{\Ycal}{\mathcal{Y}}
\newcommand{\Ocal}{\mathcal{O}}
\newcommand{\Gcal}{\mathcal{G}}
\newcommand{\Jcal}{\mathcal{J}}
\newcommand{\Kcal}{\mathcal{K}}
\newcommand{\Bcal}{\mathcal{B}}
\newcommand{\Xcal}{\mathcal{X}}
\newcommand{\Qcal}{\mathcal{Q}}
\newcommand{\Wcal}{\mathcal{W}}
\newcommand{\Ecal}{\mathcal{E}}

\newcommand{\Mfrak}{\mathfrak{M}}
\newcommand{\cfrak}{\mathfrak{c}}
\newcommand{\Cfrak}{\mathfrak{C}}
\newcommand{\Ascr}{\mathscr{A}}

\newcommand{\alphadot}{{\dot{\alpha}}}
\newcommand{\betadot}{{\dot{\beta}}}

\newcommand{\adot}{{\dot{a}}}
\newcommand{\bdot}{{\dot{b}}}

\let\originalexists\exists
\renewcommand{\exists}{\originalexists\,}
\let\originalforall\forall
\renewcommand{\forall}{\originalforall\,}

\let\emptyset\varnothing

\newcommand{\precdot}{\prec\mathrel{\mkern-5mu}\mathrel{\cdot}}
\newcommand{\succdot}{\mathrel{\reflectbox{$\precdot$}}}

\makeatletter
\renewcommand{\@chapapp}{}% Not necessary...
\newenvironment{chapquote}[2][2em]
{\setlength{\@tempdima}{#1}%
	\def\chapquote@author{#2}%
	\parshape 1 \@tempdima \dimexpr\textwidth-2\@tempdima\relax%
	\itshape}
{\par\normalfont\hfill--\ \chapquote@author\hspace*{\@tempdima}\par\bigskip}
\makeatother

\DeclareFontFamily{U}{wncy}{}
\DeclareFontShape{U}{wncy}{m}{n}{<->wncyr10}{}
\DeclareSymbolFont{mcy}{U}{wncy}{m}{n}
\DeclareMathSymbol{\Sh}{\mathord}{mcy}{"58} 

\makeatletter
\newcommand{\thickhline}{%
	\noalign {\ifnum 0=`}\fi \hrule height 1pt
	\futurelet \reserved@a \@xhline
}
\newcolumntype{"}{@{\hskip\tabcolsep\vrule width 1pt\hskip\tabcolsep}}
\makeatother

%% file: chapters/frontmatter.tex
% Title page

\begin{titlepage}\pdfbookmark[0]{Front page}{label:frontpage}%
	
	\begin{center}
		\vspace*{1cm}
		
		\Huge
		\vspace{0.2cm}

		\begin{tabular}{@{}p{\textwidth}@{}}
			\toprule[2pt]
			\midrule
			\begin{center}
				\textbf{Positive Geometries for Scattering Amplitudes in $\mathbf{\Ncal=4}$ SYM \\and ABJM}\vspace{-0.22cm}
			\end{center}
			\\
			\midrule
			\toprule[2pt]
		\end{tabular}
		
		\LARGE

		\vfill
		
		\textbf{Jonah Stalknecht}
		
		\vfill
		
		\Large
		
		\textit{Submitted to the University of Hertfordshire in partial fulfilment \\of the 
		requirements of the degree of Doctor of Philosophy}

		\vspace{2cm}
		
		\Large
		University of Hertfordshire\\
		June 2024
		
	\end{center}
\end{titlepage}

% Empty page after title
\newpage\null\thispagestyle{empty}\addtocounter{page}{-2}\newpage

% Abstract
\thispagestyle{plain}
\begin{center}
	\Large
	\textbf{Positive Geometries for Scattering Amplitudes\\ in \nf and ABJM}
	
	\vspace{0.4cm}
	\textbf{Jonah Stalknecht}
	
	\vspace{0.9cm}
	\textbf{Abstract}
\end{center}
\noindent
This thesis investigates geometric descriptions of scattering amplitudes, with a specific focus on scattering amplitudes in \nf and ABJM theory. The recent development of the field of \emph{positive geometries} provides us with a suitable framework for this endeavour. In particular, we will give a detailed account of the \emph{amplituhedron} and the \emph{momentum amplituhedron}, which describe amplitudes in \nf, and the \emph{ABJM momentum amplituhedron} for ABJM theory. Alongside these geometries, we will also discuss the \emph{ABHY associahedron}, which encapsulates tree-level scattering amplitudes in bi-adjoint scalar theory. We provide a detailed introduction to these positive geometries, which includes a comprehensive discussion of their structure. For the momentum amplituhedron, ABJM momentum amplituhedron, and ABHY associahedron we give a full stratification of their boundaries, which equivalently elucidates the singularity structure of the tree-level scattering amplitudes. Notably, we show that the ABJM momentum amplituhedron has an Euler characteristic equal to one. Furthermore, we explore the interconnections between these, and other, positive geometries. These connections are in part obtained via \emph{push forwards through the scattering equations}. We develop techniques to calculate these push forwards which circumvents the necessity to solve the scattering equations explicitly. Beyond tree-level, we illustrate how positive geometries can be used to describe loop integrands in planar \nf and ABJM. A new framework is established to investigate these loop geometries in the space of \emph{dual momenta}. The construction relies solely on lightcones and their intersections, and the framework simultaneously encompasses the loop level structure of the amplituhedron, momentum amplituhedron, and the ABJM momentum amplituhedron. This further leads to compact general formulae for all one-loop integrands in \nf and ABJM.

% Declaration
\newpage
\chapter*{Author's Declaration}
I hereby declare that I am the sole author of this thesis. To the best of my knowledge this thesis contains no material previously published by any other person except where due acknowledgement has been made. This thesis contains no material which has been accepted as part of the requirements
of any other academic degree or non-degree program. The contents presented in this thesis were obtained during my doctoral studies at the University of Hertfordshire, under supervision of Tomasz \L{}ukowski. The results of this thesis are in part based on the articles \cite{Lukowski:2021fkf, Lukowski:2022fwz, Lukowski:2023nnf, Ferro:2023qdp}. Other works produced during my doctoral studies include \cite{Lukowski:2021amu,Ferro:2024vwn}.

% Dedication
\newpage\thispagestyle{plain}
\topskip0pt
\vspace*{\fill}
\begin{flushright}
	\large{\textit{Dedicated to my mother}}
\end{flushright}
\vspace*{\fill}

% Acknowledgements
\newpage
\chapter*{Acknowledgements}
First and foremost, I would like to thank my supervisor Tomasz \L{}ukowski. I am incredibly grateful for his time, excellent advice, and the many enlightening discussions. His guidance has been invaluable for my development as a researcher and I feel lucky to have been his student. I would also like to thank the other members of my supervisory team, Elias Brinks and Vidas Regelskis, and my external assessor James Drummond. I would further like to thank Livia Ferro, Ross Glew, Tomasz \L{}ukowski, and Robert Moerman for collaboration on various research projects over the past few years. Without them, much of the research outlined in this thesis would not have been possible, and I am thankful for all the notes, discussions, and feedback that we shared. 

I am grateful for the excellent Mathematical and Theoretical Physics community at the University of Hertfordshire, which has provided me with a stimulating environment and many interesting seminars. In particular I would like to thank the staff members Luigi Alfonsi, Leron Borsten, Livia Ferro, Tomasz \L{}ukowski, Yann Peresse, Charles Strickland-Constable, and Charles Young, the postdocs Ross Glew, Hyungrok Kim, and Fridrich Valach, and my fellow PhD students Martin Christensen, Tommaso Franzini, Simon Jonsson, Julian Kupka, Robert Moerman, and Alessandro Palazio for many intriguing discussions over lunch at the Forum, or in evenings at the pub, which has greatly enhanced my PhD experience. 

In the part of my PhD which took place after the Covid lockdown, I am grateful for the opportunity to attended numerous conferences, workshops, and PhD schools around the world. I have met countless fellow researchers who have shared some of their wisdom with me. There are many people to thank here, but to keep the list short, I would like to say a special thanks to Christoph Bartsch, Gabrielle Dian, and Carolina Figueiredo, with whom I have had many enriching conversations.

I am further thankful to the many attendants of the Geometry and Scattering Amplitudes Journal Club, which I co-organised with Henrik Munch and Joseph Fluegemann. Our semi-regular meetings have given me a valuable glimpse at the various facets of scattering amplitudes research. I would like to thank David Damgaard and Robert Moerman for founding this journal club, and I thank Alessandro Palazio for continuing this wonderful initiative after me.

Last but not least, a big thanks must go to my father, stepmother, and my partner Stasy. Their continual support has been nothing short of amazing, and without them I would never have made it this far. 

%% file: chapters/introduction.tex
\chapter{Introduction}\label{sec:INT}

\begin{chapquote}{Sir Michael Atiyah%, \textit{Geometry vs Algebra. An excerpt from Mathematics in the 20th century}
	}
	``Algebra is the offer made by the devil to the mathematician. The
	devil says: I will give you this powerful machine, it will answer any question
	you like. All you need to do is give me your soul: give up geometry and
	you will have this marvellous machine.''
\end{chapquote}
\noindent
Ancient Greeks believed that the fundamental structure of the universe was geometric. The perfect `Platonic solids' were to explain both the smallest and the largest aspects of nature. In his dialogue \emph{Timaeus} \cite{Zeyl2008-ZEYPT, Horan:2021} (c.a. 360 B.C.), Plato associated geometric objects to the classical elements, the fundamental constituents of nature: the \emph{cube} represents earth, the \emph{octahedron} air, the \emph{icosahedron} water, and the \emph{tetrahedron} represents fire. The fifth and last Platonic solid, the \emph{dodecahedron}, was associated to the `arrangements of the constellations on the whole heaven'. Nearly two and a half millennia later, we find that our fundamental description of nature is still rooted in geometric ideas: both general relativity and gauge theory are built on the framework of differential geometry. However, these formulations would not even remotely be interpreted as `geometry' by the ancients. Interestingly, recent developments in theoretical physics relate fundamental aspects of nature to a geometric picture which is much closer to the Platonic idea. Both the fundamental constituents of matter (interactions of elementary particles) and the arrangements of large scale structures (cosmological correlators) can be described by \emph{positive geometries}. These positive geometries are in some sense much closer to the classical idea of geometry, the main difference being that they typically live in more than three dimensions, and often have `curvy' aspects. 

In this thesis we will study these positive geometries, and we will see how they can be used to give a geometric interpretation of the interactions of fundamental particles. We will discuss these fundamental particles from the framework of \emph{quantum field theory (QFT)}. Quantum field theory originates as the merger of quantum mechanics and Einstein's theory of special relativity, both of which stem from the early 20\textsuperscript{th} century. One of the major triumphs of QFT is the development of the \emph{standard model} of particle physics, which is, by some measures, the most successful scientific theory of all time. The standard model classifies all known elementary particles, and it describes how they behave and interact within the framework of QFT. The standard model has been experimentally verified to an extraordinary degree: it predicted the existence of the Higgs boson, accurately predicts the value of the electron $g-2$ to at least ten significant figures, and has been instrumental for the understanding of high energy physics at particle accelerators such as the LHC at CERN. The link between the mathematical framework of QFT and the quantities we observe in experiments are \emph{scattering amplitudes}. The modulus squared of the scattering amplitude is directly proportional to the cross sections and decay rates we observe in experiments, and the mathematical expression for these scattering amplitudes can be calculated directly from QFT. It is precisely these scattering amplitudes for which we will find a geometric formulation. At the moment, the standard model is still out of reach of such geometric formulations, and we will instead focus on certain other quantum field theories. In particular, in this thesis we will focus on the scattering amplitudes of \emph{$\Ncal=4$ supersymmetric Yang-Mills theory (\nf)}, and \emph{$\Ncal=6$ supersymmetric matter Chern-Simons theory (ABJM) theory}. These theories should be regarded as toy models, as we do not expect them to describe any real world phenomena. However, as far as toy models go, \nf is remarkably useful for making predictions for particle accelerators, as it accurately describes the leading order scattering of the gluons in the standard model.

\section{What are Scattering Amplitudes?}

\begin{figure}
	\centering
	\includegraphics[width=0.5\textwidth]{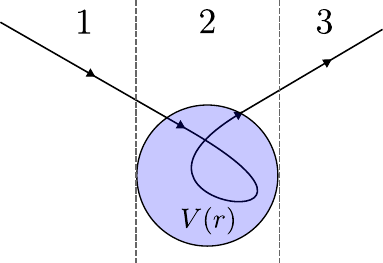}
	\caption{This figure depicts the scattering of a classical particle in a potential $V(r)$ divided into three parts.}
	\label{fig:scattering-classical}
\end{figure}

Let us start with a light-hearted introduction to scattering amplitudes, interweaved with some historical context. As a warm up, we follow \cite{taylor2012scattering} and start by considering the following classical picture of a scattering process, which is depicted in figure \ref{fig:scattering-classical}:
\begin{enumerate}
	\item A particle comes in from afar,
	\item scatters off some stationary potential,
	\item and flies away into the distance.
\end{enumerate}
The statements ``comes in from afar'' and ``into the distance'' refers to the particle starting and ending outside of the range of the interacting potential, and the particle (asymptotically) continues in a straight line, which are said to be `asymptotic states'. The interesting part of the scattering happens in step 2, the region of interaction. If we use this set-up to describe the scattering of an electron off some stationary atom, then the region of interaction is roughly the size of a few atomic diameters and is in practice unobservable. Rather, an experimentalist will perform measurements on the asymptotic states, from which we can build a model to describe the process in step 2\footnote{This simple set-up captures the essence of many classical nuclear experiments. For example, the famous Rutherford gold foil experiments (performed by Geiger and Marsden), which demonstrated that atoms are mostly empty with a positively charged nucleus, essentially follows this set-up.}. In this classical case, given sufficient initial data, we can calculate the precise trajectory of the particle in regions 2 and 3, however this will no longer be the case when we move into the realm of quantum field theory. In this thesis, we will take the perspective that whatever happens in step 2 is a `black box': we can't observe what happens there anyway! This is a slight shift of paradigm: we no longer concern ourselves with `what happens in a scattering process', but we only ask the question how the initial and the final asymptotic states are related. 

In the quantum version of this simple picture, the asymptotic states are state vectors $\ket{\psi_{\text{in}}}$ and $\ket{\psi_{\text{out}}}$ in some Hilbert space. The interesting part of the interaction is encoded in the \emph{S-matrix}, which relates these two states:
\begin{align}
	\ket{\psi_{\text{out}}}= S \ket{\psi_{\text{in}}}\,.
\end{align}
The probability of the scattering of $\ket{\psi_{\text{in}}}$ into $\ket{\psi_{\text{out}}}$ is given by $|\bra{\psi_{\text{out}}}S\ket{\psi_{\text{in}}}|^2$, and hence it is appropriate to interpret the element of the S-matrix $\bra{\psi_{\text{out}}}S\ket{\psi_{\text{in}}}$ as the probability amplitude of this process. If the S-matrix is known explicitly, then we have `solved' the scattering problem, as it gives us access to any probability amplitudes we desire. To make sure that probabilities always add up to one, we require $\<\psi_{\text{in}}|\psi_{\text{in}}\>=\<\psi_{\text{out}}|\psi_{\text{out}}\>=1$, which is ensured by the \emph{unitarity} of the S-matrix: $S^\dagger S=\unit$.

When we move on to quantum field theory, our asymptotic non-interacting states are elements of \emph{Fock space}, instead of merely a Hilbert space. The elements of Fock space are `free fields'. A consequence of this is that the number of particles is no longer necessarily preserved in a scattering process. It is customary to split up the S-matrix into a trivial part $\unit$, which reflects the part where no scattering happens at all, and a transition matrix, which encodes the non-trivial scattering information, as $S=\unit+iT$. Let us consider the scattering of $m$ asymptotic states into $n-m$ asymptotic states. We can represent the elements of Fock space by momentum eigenstates $|p_{1},\ldots,p_{m}\>_{\text{in}}$ and $|p_{m+1},\ldots,p_{n}\>_{\text{out}}$. The matrix elements
\begin{align}
	\<p_{m+1},\ldots,p_{n}|iT|p_{1},\ldots,p_{m}\> = \delta^4\big(\sum_{i=1}^m p^\mu_{i}-\sum_{j=m+1}^{n} p^\mu_{j}\big) A(\{p_{1},\ldots,p_{m}\}\to \{p_{m+1},\ldots,p_{n}\})\,,
\end{align}
are what we call the \emph{scattering amplitudes}. Assuming \emph{crossing symmetry}, we can flip the direction of all incoming momenta and interpret them as outgoing instead. The scattering amplitude $A(p_1,\ldots,p_n)$ is what we refer to as \emph{the $n$-particle (or $n$-point) amplitude}. Different scattering processes can be obtained from $A(p_1,\ldots,p_n)$ by analytic continuation. Using this `all-out' convention, momentum conservation simply reads
\begin{align}
	\sum_{i=1}^n p_i^\mu=0\,.
\end{align}
In addition to the momenta of the external particles, a scattering amplitude depends on the `particle type', which includes a specification of appropriate quantum numbers for the asymptotic states, such as spin or colour/flavour group. When there is little room for confusion, we often suppress the explicit dependence, and denote the scattering amplitude as $A_n$.

The transition from quantum mechanics to QFT introduces some new structure to the problem. In addition to the \emph{unitarity} of the S-matrix, which ensures that the time evolution of a system gives positive probabilities which sum to one, there are two more fundamental principles which coalesce in QFT. \emph{Causality} is a direct consequence of Einstein's theory of relativity, which says that no information can transfer faster than the speed of light. As a consequence, any spacetime event can only have an influence on another event which is inside its lightcone, \emph{i.e.} the two events are time-like separated. The principle of \emph{locality} stems from classical field theory, and postulates that for any spacetime event to exert an influence on another event, something (\emph{e.g.} a particle or wave) must mediate between these two events. Since this `something' cannot travel faster than light, any event can only directly influence events in its immediate surroundings. These three cornerstone principles place important constraints on what scattering amplitudes in QFT can look like. 

The traditional approach to describe and define quantum field theories is in terms of their action. We write down a Lagrangian density $\Lcal$ consisting of a free term and an interaction term, and we retrieve the scattering amplitudes from renormalised Green's functions through the Lehman-Symanzik-Zimmermann reduction formula. Each interaction term comes equipped with a coupling constant, which we denote $g$. Assuming $g$ is small, it is customary to do a \emph{perturbative expansion} of the scattering amplitude as
\begin{align}
	A_n= \sum_{L=0}^\infty g^{2L} A_n^{(L)}\,.
\end{align}
The $L$\textsuperscript{th} term in this expansion is known as the \emph{$L$-loop amplitude}, with $L=0$ being referred to as the \emph{tree amplitude}. There is a completely algorithmic way to calculate the $L$-loop amplitude due to Feynman. From the Lagrangian we read off the Feynman rules, and to find $A_n^{(L)}$ we simply sum over all $L$-loop \emph{Feynman diagrams}. For example, for a scalar theory with $g\phi^3$ interaction term, we can depict the scattering amplitudes as
\begin{align}
	A_4^{\text{tree}}&=\vcenter{\hbox{\includegraphics[width=35mm]{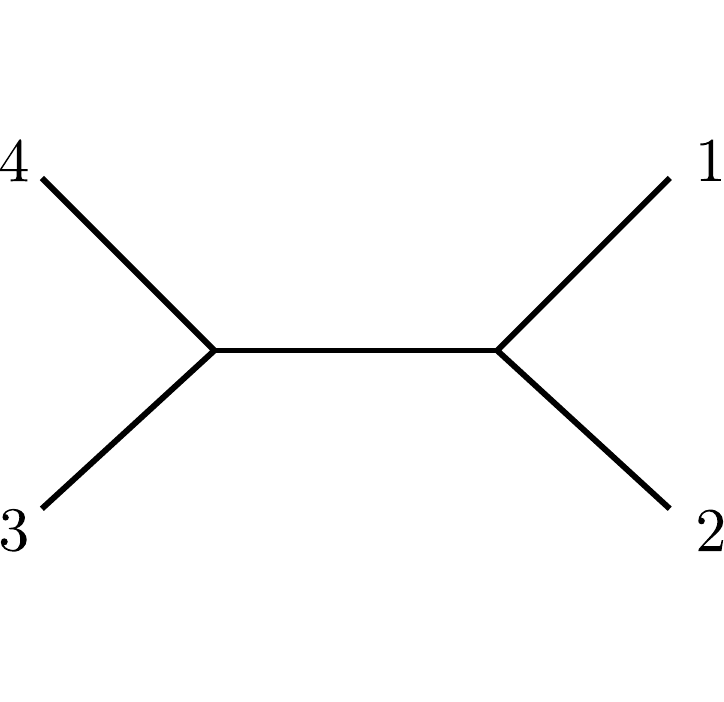}}}+\vcenter{\hbox{\includegraphics[width=35mm]{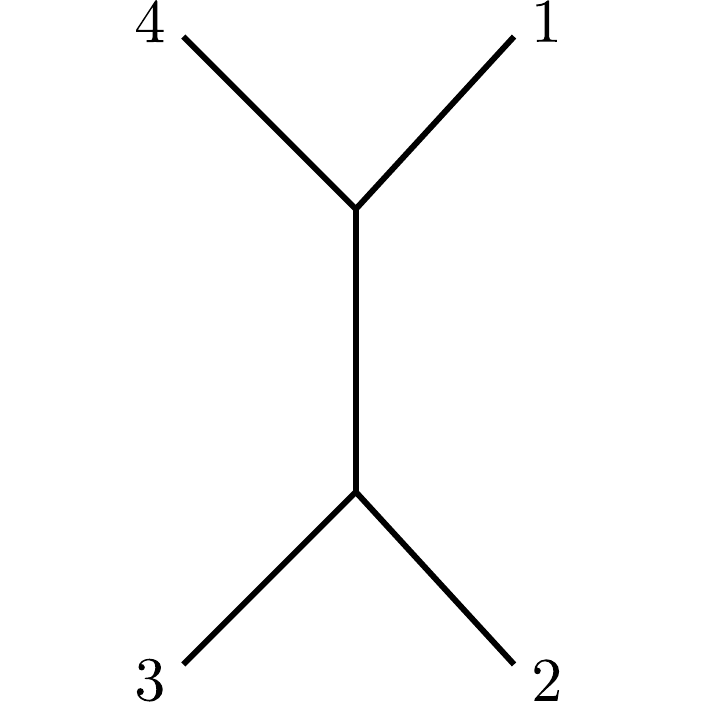}}}+\vcenter{\hbox{\includegraphics[width=35mm]{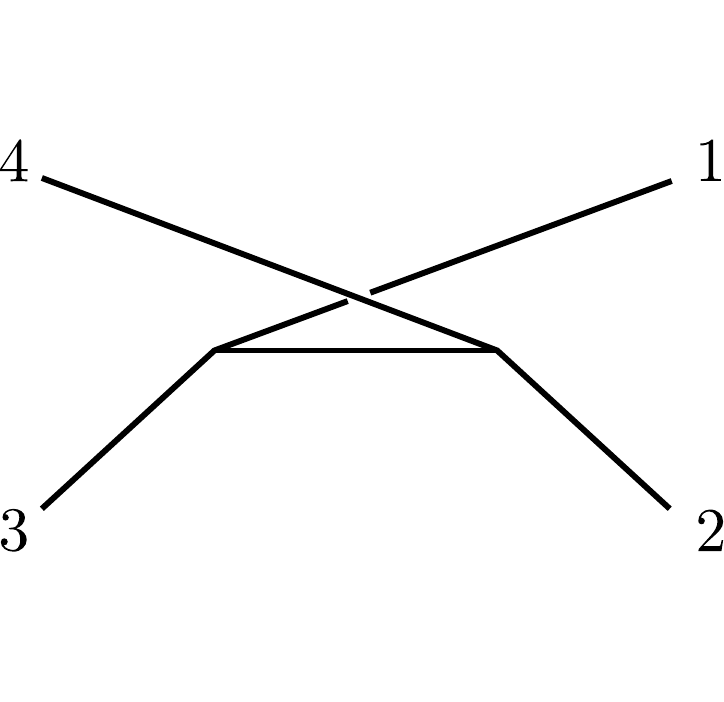}}}\\
	A_4^{(1)}&=\vcenter{\hbox{\includegraphics[width=30mm]{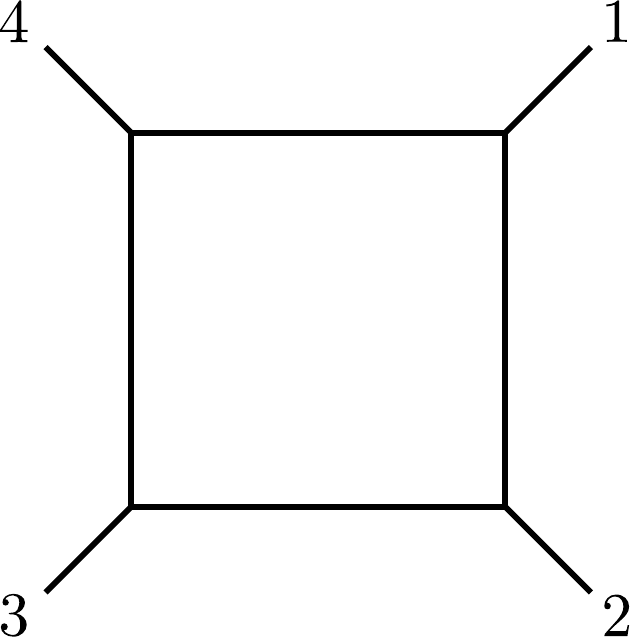}}}+\ldots
\end{align}
The locality of QFT is manifest in these Feynman diagrams: the internal edges of the diagrams are interpreted as (virtual) particles which mediate the interaction. Unitarity is ensured by the fact that we are summing over \emph{all} possible spacetime processes. Causality is slightly less trivial to see, and is related to the $i\epsilon$ prescription of the propagators.

The Feynman diagrams and Feynman rules give an algorithm to calculate scattering amplitudes in a way which manifests unitarity, causality, and locality, and can be used to calculate any scattering amplitude we desire. However, there is a down side to all this. For instance, the Lagrangian description of quantum field theory is riddled with redundancy. This is famously the case for gauge redundancy, to describe particles with spin in a Lorentz invariant way, we necessarily introduce non-physical degrees of freedom which need to be gauge fixed. But there is also an infinite-dimensional redundancy for scalar particles coming from field redefinitions. Since the path integral treats the fields as integration parameters, we can simply redefine our fields $\phi\to f(\phi)$ (assuming $f'(0)=1$), and the result of the path integral will be unchanged. To illustrate, we borrow an example from \cite{Cheung:2017pzi} and consider the Lagrangian
\begin{align}
	\Lcal = \frac{1}{2}\partial_\mu\phi\partial^\mu\phi + g_1 \phi \partial_\mu\phi\partial^\mu\phi + \frac{1}{2!}g_2\phi^2\partial_\mu\phi\partial^\mu\phi + \frac{1}{3!}g_3\phi^3 \partial_\mu\phi\partial^\mu\phi+\ldots
\end{align}
We can now use Feynman diagrams to start calculating the scattering amplitudes corresponding to this QFT. Remarkably, after adding together all the Feynman diagrams, we find that the scattering amplitudes vanish order by order in perturbation theory. The reason for this is that this Lagrangian is just a field redefinition of a free Lagrangian, and hence it is obvious that all scattering amplitudes must vanish. However, this is far from clear by just staring at the Lagrangian, and the individual Feynman diagrams give finite answers, it is only in the \emph{sum} that the miraculous cancellation takes place. 

The traditional approach to scattering amplitudes obscures many of the symmetries which the amplitudes enjoy, and these redundancies force us to consider an often enormous sum of Feynman diagrams which dramatically simplify at the end. This is a pattern which also persists for theories of more physical interest. Let us consider tree-level gluon amplitudes in quantum chromodynamics (QCD). The number of Feynman diagrams we have to calculate are listed in table \ref{tab:gluon-amplitude}.
\begin{table}
	\centering
	\begin{tabular}{l|lllllll}
	$n$ & 4 & 5 & 6 & 7 & 8 & 9 & 10 \\\hline
	\# Diagrams & 4 & 25 & 220 & 2485 & 34300 & 559405 & 10525900
	\end{tabular}
	\caption{The number of Feynman diagrams needed to calculate tree-level gluon amplitudes \cite{Mangano:1990by}.}
	\label{tab:gluon-amplitude}
\end{table}
These scattering amplitudes were of genuine importance for experiments when they were first calculated by Parke and Taylor, who studied the case where particles $1$ and $2$ have negative helicity, and the remaining particles have a positive helicity. Even for the relatively simple case of $n=6$, after expanding many pages of Lorentz invariant contractions of momentum and polarisation vectors, the final result was messy and spanned around eight pages \cite{Parke:1985ax}. However, at the end of their paper they made the remark that they hoped to \textit{``obtain a simple analytic form for the answer, making our result not only an experimentalist's, but also a theorist's delight''}, a pursuit in which they succeeded by making use of \emph{spinor-helicity variables}. The $n$-particle result reads \cite{Parke:1986gb}
\begin{align}
	A_n(1^-2^-3^+\cdots n^+)=\frac{\<12\>^4}{\<12\>\<23\>\cdots\<n1\>}\,.
\end{align}
The definition of these variables will follow in section \ref{sec:KIN_spin-hel}. 

The complexity of the Feynman diagram expansion and the miraculous simplicity of the final answer motivates us to look for an alternative method which might allow us to arrive at this scattering amplitude. Rather than introducing the auxiliary concepts of `unitary evolution in spacetime through the exchange of off-shell virtual particles', which manifests the unitarity and locality of the Feynman diagram method, we again ask the question \emph{what are scattering amplitudes?} It is clear that scattering amplitudes are just functions (or distributions, if we include the momentum conserving delta functions) depending on Lorentz invariant combinations of the on-shell external momenta. Our perspective will be that we don't try to fill in the `black box' where the scattering happens, but we rather ask how the momenta of the outgoing particles are related to the momenta of the incoming particles. Rather than interpreting unitarity, causality, and locality as dictating what happens during the scattering process, we instead interpret them as adding \emph{constraints} on the type of functions we can encounter. This is morally equivalent to what was attempted by the `analytic S-matrix bootstrap' programme from the sixties \cite{eden2002analytic}, which was originally notoriously unsuccessful. However, a modern approach over the past two decades has fared significantly better. Among the differences is the modern focus on \emph{on-shell methods}, and, instead of trying to uncover non-perturbative properties of scattering amplitudes, the modern methods are comfortable treating things one loop order at a time. We will be going one step further, however. Instead of trying to `bootstrap' scattering amplitudes, we will want to describe the scattering amplitudes \emph{as a whole}, \emph{directly} in the kinematic space. Given that scattering amplitudes are just functions of the kinematics, we might wonder what type of questions we can ask in the space of kinematic variables whose answer is the scattering amplitude. We are only given a list of momentum vectors as input, which rather restricts the type of questions we can ask, they essentially have to be \emph{combinatoric} or \emph{geometric} in nature.

This motivates our search for a geometric description of scattering amplitudes. To motivate this even further, we review some of the modern on-shell methods for scattering amplitudes in a partially historical way, and we will see that a geometric picture naturally emerges. Many of these methods will be explained in more detail throughout this thesis. As a first remark, we note the general trend that theories with more symmetries will have simpler scattering amplitudes. This is exemplified by \nf in four dimensions, a theory which enjoys \emph{superconformal}, \emph{dual superconformal} \cite{Drummond:2008vq}, and an infinite-dimensional \emph{Yangian} symmetry \cite{Drummond:2009fd}. \nf is sometimes claimed to be the simplest quantum field theory \cite{Arkani-Hamed:2008owk}. Many of the modern advances in scattering amplitudes were first formulated for this maximally simple theory, and, more often than not, some slight variation of these methods subsequently proved themselves useful for a more general class of theories. As a second remark, we point out the importance of using the right kinematic variables. The marvellous simplicity of the Parke-Taylor formula is in part due to the use of spinor-helicity variables, which manifest the masslessness of the gluons. If we use a description which trivialise or linearise the symmetries of the theory, then the scattering amplitudes are typically simpler and reveal new structures.

In one way or another, many of the modern techniques can be traced back to Witten's formulation of scattering amplitudes in \nf from \emph{twistor string theory} \cite{Witten:2003nn}. The idea is that scattering amplitudes in \nf only have support on certain holomorphic curves in \emph{twistor space}. Twistor variables have been around since the sixties when they were introduced by Penrose \cite{Penrose:1967wn}, one of the benefits of these variables for scattering amplitudes is that they linearise conformal symmetry. Shortly after Witten's seminal paper, Roiban, Spradlin, and Volovich gave a formulation of the twistor string result as an integral over moduli space which completely localises on the support of the \emph{Witten-RSV equations} \cite{Roiban:2004vt, Roiban:2004yf}. This was followed by a surge of new methods to calculate scattering amplitudes in \nf and beyond. Notably, it lead to on-shell recursion relations such as the BCFW recursion \cite{Britto:2004ap, Britto:2005fq}. The idea is to use the fact that locality dictates what type of poles appear in tree-level scattering amplitudes, and then use Cauchy's theorem to write scattering amplitudes as a sum over products of lower-point amplitudes. The recursion is not unique, and this method can be used to find many distinct expressions for the amplitude. The fact that all these expressions must be equal, leads to some highly non-trivial identities between the terms in these expansions. These identities emerge from the global residue theorem, and they are sometimes referred to as \emph{homological identities}.

The BCFW recursion for \nf can be expressed very naturally in terms of \emph{momentum twistors}. These are a new type of twistor variable introduced by Hodges \cite{Hodges:2009hk}, which linearise the dual conformal symmetry instead. Expressions for scattering amplitudes in \nf are often simplest when written in terms of momentum twistors, including the terms in the BCFW expansion. Hodges noticed that the terms appearing for the so-called `NMHV' amplitude (corresponding to the case where three gluons have negative helicity in QCD) look remarkably similar to the volume of a simplex in \emph{projective geometry}. Taking this one step further, the NMHV amplitude can then be interpreted as the object obtained when these simplices are glued together into a larger polytope, and the homological identities arise by slicing up this polytope in different ways! Hodges' polytopes gave a fully geometric interpretation of NMHV scattering amplitudes in \nf, and stands as a major victory in our quest to find amplitudes from geometry, although this method doesn't generalise beyond NMHV.

In a parallel line of investigations, scattering amplitudes in \nf were investigated in terms of the geometric space of $k$-planes in an $n$-dimensional vector space, known as the \emph{Grassmannian} $G(k,n)$. It was observed that the Witten-RSV equations can be given a natural geometric interpretation in the Grassmannian \cite{Arkani-Hamed:2009kmp} (see also \cite{Bullimore:2009cb}), and that scattering amplitudes can be obtained as integrals over the Grassmannian \cite{Arkani-Hamed:2009ljj}. These Grassmannian formulations are simplest when paired with momentum twistors, in which case any Grassmannian integral is manifestly Yangian invariant \cite{Arkani-Hamed:2009nll}. Furthermore, the terms appearing in the BCFW expansion can be recursed down to sums over \emph{on-shell diagrams} consisting of many three-particle amplitudes glued together \cite{Arkani-Hamed:2012zlh}. These type of diagrams, known to mathematicians as \emph{plabic graphs}, were being used around the same time by Postnikov to study a specific subset of the Grassmannian: the \emph{positive Grassmannian} $G_+(k,n)$ \cite{Postnikov:2006kva}. This ultimately lead to a formulation of scattering amplitudes in \nf from the positive Grassmannian \cite{Arkani-Hamed:2012zlh}. 

These ideas culminated with the introduction of the \emph{amplituhedron} by Arkani-Hamed and Trnka in 2013 \cite{Arkani-Hamed:2013jha}, which is the archetypical example and a prototype for all \emph{positive geometries}. The amplituhedron is a geometric object which fully encodes the tree-level scattering amplitudes and planar integrands of \nf, even beyond NMHV level, and it can further be extended to include loop integrands as well. The original definition is remarkably simple: it is the image of a positive linear map between two Grassmannian spaces. Whereas Hodges' polytopes encode the scattering amplitudes as their volume, the amplituhedron instead encodes it in its \emph{canonical form}. This canonical form is a logarithmic differential form with poles exactly at the boundaries of the amplituhedron. Instead of defining the amplituhedron in some auxiliary Grassmannian space, it was later realised that there exists a topological description \emph{directly in momentum twistor space} \cite{Arkani-Hamed:2017vfh}. The canonical form of the amplituhedron can then \emph{literally} be interpreted as the scattering amplitude. Note that this geometric construction of scattering amplitudes is manifestly different from traditional descriptions. The definition of the amplituhedron is agnostic about the Lagrangian and the various redundancies it induces, there is no reference to processes happening in spacetime, and nowhere in this construction did we need to introduce locality or unitarity as input. Rather, we find the scattering amplitude as the canonical form to some easily defined geometric object in kinematic space, from which properties such as locality and unitarity somehow \emph{emerge}.

The amplituhedron gives a conceptually satisfying new way of finding scattering amplitudes, and the following emergence of the field of positive geometries has been an active area of research for physicists and mathematicians alike. The next step was to try to generalise the amplituhedron to different theories. This is easier said than done, as the amplituhedron construction manifestly depends on momentum twistors, which are only well-defined for massless planar theories in four dimensions, and are most naturally used for theories which enjoy a dual conformal symmetry. Fortunately, there exists another positive geometry which encodes scattering amplitudes in \nf, this time in the much less constrained spinor-helicity space: the \emph{momentum amplituhedron} \cite{Damgaard:2019ztj}. The fact that there are two distinct ways to describe amplitudes in \nf stems from the \emph{scattering amplitude -- Wilson loop duality} \cite{Alday:2008yw}, a property which has been an important topic of study in \nf \cite{Alday:2007hr, Drummond:2007aua, Brandhuber:2007yx, Alday:2007he, Drummond:2007cf, Drummond:2007bm, Alday:2008yw, Anastasiou:2009kna, Mason:2010yk, Caron-Huot:2010ryg, Adamo:2011pv}. This duality has its origins in \emph{T-duality} \cite{Berkovits:2008ic}. In fact, it might be more appropriate to interpret the amplituhedron as describing Wilson loops, whereas the momentum amplituhedron truly captures the scattering amplitudes. Another benefit of this is that the boundaries of the momentum amplituhedron are in one-to-one correspondence to the singularities of the corresponding scattering amplitude.

Another success story is the construction of the \emph{Arkani-Hamed--Bai--He--Yuan (ABHY) associahedron} \cite{Arkani-Hamed:2017mur}. This is a positive geometry in the space of planar Mandelstam invariants, and it describes tree-level amplitudes in \emph{bi-adjoint $\phi^3$ theory}. This is a scalar theory which became notable because of its relation to the \emph{CHY formalism} \cite{Cachazo:2013iaa, Cachazo:2013gna, Cachazo:2013hca, Cachazo:2013iea, Cachazo:2014nsa, Cachazo:2014xea}. We recall that the Witten-RSV formula allows us to calculate scattering amplitudes in \nf by doing an integral over moduli space which completely localises on the support of some rational equations. This turns out to be a far more general concept, and the CHY formalism allows us to calculate scattering amplitudes in a wealth of different theories in general dimensions as an integral over moduli space. This integral completely localises on the support of the rational \emph{scattering equations}. Bi-adjoint $\phi^3$ theory arises because it has a particularly natural description in the CHY formalism. We see that both the ABHY associahedron and the momentum amplituhedron describe theories which, in addition to a geometric description, also have a CHY formulation. In fact, it was argued in \cite{Arkani-Hamed:2017mur, He:2018okq, He:2021llb} that it is possible to find the canonical form of these positive geometries by calculating the \emph{push forward through the scattering equations}. 

Since there are a multitude of theories with a CHY description, it is natural to ask if we can find positive geometries for other theories by calculating push forwards through the scattering equations. This was proven successful for \emph{Aharony-Bergman-Jafferis-Maldacena (ABJM) theory} \cite{Aharony:2008ug}, and it lead to the construction of the \emph{ABJM momentum amplituhedron} \cite{He:2021llb, Huang:2021jlh}. ABJM is another theory which enjoys a remarkable amount of symmetry, including superconformal, dual superconformal, and Yangian symmetry \cite{Bargheer:2010hn, Huang:2010qy, Gang:2010gy} (see \cite{Henn:2010ps, Bianchi:2011dg, Wiegandt:2011uu, Chen:2011vv} for comments on the amplitudes/Wilson loop duality in ABJM and its connection to \nf). Following the success of \nf, many of the various formulations for scattering amplitudes have eventually also found a purpose for ABJM theory. At first sight, the two theories could hardly be more different: ABJM theory is a supersymmetric version of matter Chern-Simons in three dimensions, whereas \nf is a supersymmetric version of Yang-Mills in four dimensions. However, it turns out, that the scattering amplitudes in these two theories allow for very similar constructions, including in terms of positive geometries. In addition to the ABJM momentum amplituhedron, also an ABJM version of the amplituhedron has recently been introduced \cite{He:2023rou}.

In this thesis we will investigate positive geometries and how they relate to scattering amplitudes, with a particular focus on amplitudes in \nf and ABJM. We will study the various positive geometries, their structure, and interconnections.

\section{Outline}

This thesis will be largely self contained, and all the necessary background information will be introduced along the way. This includes an introduction to modern techniques in scattering amplitudes, and a fairly detailed overview of the various positive geometries. However, this thesis is not meant as an introductory or pedagogical text, as the motivation and context of a subject might not be clear when introduced, only to be called upon much later. Instead, we introduce concepts in a hierarchical way: we start from the most stand-alone and easily definable subjects, and we work our way up to more complicated and interconnected topics. The hope is that this thesis can serve as a useful reference for future researchers who already have some basic familiarity with the subject.

\begin{itemize}
	\item In chapter \ref{sec:QFT}, we will briefly review some basic concepts of quantum field theory. We will first review the symmetries in QFT, which we already hinted will play an important role in the simplicity of scattering amplitudes. After this, we will give a short introduction to the main theories of interest: \nf, ABJM theory, and bi-adjoint $\phi^3$ theory. 
	
	\item Chapter \ref{sec:GRASS} will be rather mathematical in nature. We will review the constructions of projective geometry and the Grassmannian. Since we will ultimately be interested in describing positive geometries in these spaces, we will put some emphasis on projective polytopes and their volume. On the Grassmannian side, we will spend the majority of our efforts on studying the positive Grassmannian, and the related positive orthogonal Grassmannian, as they will prove to be useful for descriptions of scattering amplitudes in \nf and ABJM, respectively.
	
	\item After we have the tools from projective and Grassmannian geometry under our belt, we move on in chapter \ref{sec:KIN} to introduce the various kinematic spaces which we shall call upon later. As we mentioned, using the correct kinematic variables can make your life maximally easy, and the various positive geometries we encounter naturally live in some kinematic space. This is where we introduce spinor-helicity variables, which hold some natural connections to the Grassmannian. In this chapter we further introduce twistors, momentum twistors, and embedding space, which rely on an understanding of projective space. The discussion of {dual momenta} in section \ref{sec:KIN_dual} will prove of importance, as this dual space is where we later define positive geometries for loop integrands.
	
	\item In chapter \ref{sec:AMP} we turn to a review of modern scattering amplitudes methods. This includes an introduction to BCFW recursion, the CHY formalism, twistor strings, on-shell diagrams, and Grassmannian integrals. Many of these discussions will be fairly streamlined due to the previous introductions of the various kinematic variables and the Grassmannian. Of particular interest is how these methods can be used for \nf, and we will introduce the analogous expressions for ABJM.
	
	\item Then, in chapter \ref{sec:POS}, we finally arrive at the main topic of interest: positive geometries. After giving a basic definition and reviewing some important properties such as triangulations and push forwards, we will give an introduction to the various positive geometries of interest. We start with the ABHY associahedron, as it is arguably the simplest one to define and describe. After this warm-up, we move on to study the amplituhedron. We subsequently study the momentum amplituhedron and ABJM momentum amplituhedron, whose definition and discussion closely mimic the section on the amplituhedron. However, a benefit which these momentum amplituhedra have with respect to the amplituhedron is in their boundary structure, which captures the singularity structure of scattering amplitudes. In particular, we give a detailed account of the boundaries of the ABJM momentum amplituhedron, based on the results from \cite{Lukowski:2021fkf}. In the last section of this chapter we turn to the idea that we can find canonical forms of positive geometries by calculating the push forward through the scattering equations. In particular, we provide algorithms to calculate these push forwards based on tools from algebraic geometry \cite{Lukowski:2022fwz}. Many of the technical details for this section will be delegated to appendix \ref{sec:APP_alg-geom}.
	
	\item Chapter \ref{sec:DUAL} will be dedicated to the study of positive geometries in dual space. After a review of the notion of \emph{chambers}, we give a detailed overview of lightcone (or null-cone) geometries in dual space, which completely capture the loop integrands of both ABJM and \nf. We will see how to characterise these lightcone geometries in terms of their vertices, which will allow us to find a general formula for their canonical form. This ultimately leads to a general formula for the one-loop integrand for both ABJM and \nf. This chapter is largely based on \cite{Lukowski:2023nnf} and \cite{Ferro:2023qdp}.
	
	\item We conclude this thesis with some closing remarks and a brief summary in chapter \ref{sec:CONC}. We further give an outlook on some open problems which naturally follow from the topics discussed in this thesis.
\end{itemize}

We will give a short summary at the end of each chapter which highlights the most important concepts. The intent is that this will make it more clear where the respective concepts belong in the broader context of this thesis.

\section{Conventions}

We will work in the mostly plus signature of Minkowski space with $\eta=\diag(-1,1,\ldots,1)$. In later chapters we will encounter $\Rbb^{2,2}$, in which case we will work with the metric $\eta=\diag(1,1,-1,-1)$. 

We frequently use multiindex notation throughout this thesis. We define $[n]$ to be the set of natural numbers (starting from 1) up to $n$: $[n]=\{1,2,\ldots,n\}$. Furthermore, for some given set $S$ and integer $k$, we let $\binom{S}{k}$ be the set of all $k$ element subsets of $S$, which we will usually encounter in the case where $S=[n]$. Another important point about these sets: when they appear as indices, we will always assume the elements of a set to be ordered. That is, if $I=\{i_1,i_2,\ldots,i_k\}\in\binom{[n]}{k}$ with $i_1<i_2<\cdots<i_k$, then $X_I=X_{i_1,i_2,\ldots,i_k}$, whatever $X$ may be. This is important to keep in mind when encountering determinants, such as \Pluck variables and other minors of matrices, as an inconsistent ordering of the indices can yield crucial minus sign errors.

The character $\eta$ is perhaps overused in this thesis, as it is used for the flat Minkowski metric, anti-commuting Grassmann variables in both three and four dimensional superspace, and an $n\times n$ `metric' on the space of particle indices which we will use to define the positive orthogonal Grassmannian. These are all standard conventions, and rather than inventing new notation we will keep it as is. The appropriate interpretation of $\eta$ should be clear from context, and to minimise any chance at confusion we regularly re-emphasise the definition of $\eta$ at hand.

Lastly, a few points regarding our notation for scattering amplitudes. General scattering amplitudes will be denoted by $A$, and we often use a subscript to indicate the number of particles: $A_n$ is a general $n$-particle amplitude. Scattering amplitudes are always assumed to be dressed with a momentum conserving delta function, although we will rarely write down the explicit $\delta^D(\sum p)$ in practice. Occasionally we will abuse terminology and refer to scattering amplitudes or loop integrands as \emph{(rational) functions}, in which case it is understood that we mean the function which is multiplying the delta function. For scalar theories such as bi-adjoint $\phi^3$ and $\tr{\phi^3}$, we will use $m$ to denote their scattering amplitudes. In the literature, superamplitudes are often denoted by the calligraphic $\Acal$ or $\Mcal$, however we reserve these symbols for the amplituhedron and momentum amplituhedron, respectively. We will simply use the symbol $A$ for superamplitudes as well. When discussing four-dimensional (super) Yang-Mills, we use $k$ to denote the number of negative helicity gluons in an amplitude, and $K$ to denote the N\textsuperscript{$K$}MHV sector. They are related via $K=k-2$. When discussing objects which are related through `T-duality', we will usually indicate the dual with a hat: $\hat{C}$ is T-dual to $C$, $\hat{\sigma}$ is T-dual to $\sigma$, and so on. Notably, we will also use this notation for superamplitudes: the T-dual of $A$ is $\hat{A}$, which is nothing but $A$ divided by an MHV amplitude. We hope that this will not cause any confusion with a similar notation which is used to indicate BCFW shifted amplitudes as $\hat{A}$, which are not the T-dual of anything.

%% file: chapters/qft.tex
\chapter[head={Aspects of QFT},tocentry={Aspects of Quantum Field Theory}]{Aspects of Quantum Field Theory}\label{sec:QFT}

In this chapter we will give a a very brief review of certain aspects of quantum field theory. The field of QFT is extraordinarily rich and remains to this day one of the most well-researched areas of physics. It is beyond the scope of this thesis to introduce or review QFT, and it is assumed that the reader is familiar with the basics. Instead, for the sake of completeness, we will use this chapter to highlight a few select topics with relevance for scattering amplitudes.

\section[head={Symmetries in QFT},tocentry={Symmetries in Quantum Field Theory}]{Symmetries in Quantum Field Theory}

Symmetries have been an important aspect of quantum field theory since its conception, giving rise to various conservation laws and Wigner's classification of elementary particles, among other important uses. In this thesis our interest in symmetries comes from the restrictions they place on scattering amplitudes. The invariance under some symmetry transformation gives non-trivial information on what sort of structures the amplitudes can exhibit. Furthermore, these symmetries often dictate what the appropriate variables are to study a given problem. In addition to the discrete symmetries, such as \emph{cyclic} or \emph{dihedral} invariance which ordered amplitudes enjoy, there are a few spacetime symmetries which will make a recurring appearance. We will give a brief introduction to the spacetime symmetries of interest.

\subsection{Poincar\'e Invariance}

The Poincar\'e group is the isometry group of Minkowski space, \emph{i.e.} it consists of transformation which leaves the distance $(x-y)^2$ between two points $x,y$ in Minkowski space unchanged. Included in the Poincar\'e group are the Lorentz group (consisting of boosts and rotations) and translations. Physical observables, such as scattering amplitudes, need to be invariant under Poincar\'e transformations.

Scattering amplitudes depend on a set of external momenta $p_1^\mu,\ldots, p_n^\mu$. More precisely, as a consequence of Poincar\'e invariance, the scattering amplitude can only depend on Lorentz invariant combinations of these momenta. Lorentz transformations act on momentum vectors as
\begin{align}
	p^\mu\to\Lambda^\mu_\nu p^\nu\,,
\end{align}
where 
\begin{align}
	\Lambda^T\cdot\eta\cdot\Lambda=\eta\,.
\end{align}
Contracting the Lorentz indices of two vectors $p^\mu q_\mu\coloneqq \eta_{\mu\nu}p^\mu q^\nu$ gives a manifestly Lorentz invariant object. For scattering amplitudes in the `all-out' convention, the remaining translation invariance is ensured by requiring \emph{momentum conservation} $p_1^\mu+\ldots+p_n^\mu=0$.

\subsection{Conformal Symmetry}

We can extend the Poincar\'e group to a larger group of spacetime symmetries. Conformal transformations which map $x\to x'$ are required to leave the metric $g_{\mu\nu}(x)$ unchanged up to a scale:
\begin{align}
	g_{\mu\nu}(x)\to g'_{\mu\nu}(x')= \Omega(x)g_{\mu\nu}(x)\,.
\end{align}
These transformations are called \emph{conformal} because they preserve the angle between the crossing of two arbitrary curves. In addition to the expected Poincar\'e transformations (corresponding to $\Omega(x)=1$), the conformal group contains a \emph{dilation}
\begin{align}
	{x'}^\mu = \lambda x^\mu\,,
\end{align}
and \emph{special conformal transformations}
\begin{align}
	{x'}^\mu=\frac{x^\mu-b^\mu x^2}{1-2b\cdot x +b^2x^2}\,.
\end{align}
If we define the \emph{inversion} operator
\begin{align}
	\Ical(x^\mu)=\frac{x^\mu}{x^2}\,,
\end{align}
then we can equivalently define special conformal transformations as an inversion, followed by a translation, followed by an inversion. A consequence of this is that that the inversion operator generates the entire conformal symmetry algebra from the Poincar\'e algebra. 

\subsection{Supersymmetry}

Supersymmetry is an interesting symmetry which relates bosonic degrees of freedom and fermionic degrees of freedom, and has been a rich topic of study in QFT, string theory, and quantum mechanics (there exist many introductory texts, such as \cite{Martin:1997ns}). In quantum field theory, fermions are matter particles and bosons are force carriers, and supersymmetry unifies these two otherwise distinct notions. Lagrangian descriptions of supersymmetric quantum field theories are notoriously unwieldy and cumbersome to work with, which makes certain calculations particularly difficult. It is therefore somewhat surprising that the results of these calculations are typically much simpler than analogous calculations for their non supersymmetric counterparts. In fact, supersymmetric theories often admit exact solutions for problems which are practically impossible to solve in traditional QFT. For this reason, supersymmetry forms an important playground for theoretical physicists to explore otherwise inaccessible realms of QFT. This emergent simplicity is precisely because supersymmetry puts very strong constraints on the type of interactions a theory can have. Supersymmetric theories are among the most maximally constrained and simple theories we can have, while still having a rich structure and being physically interesting. 

Many topics covered in this thesis will benefit greatly from the additional simplicity that supersymmetry provides. As such, supersymmetry often pervades the discussion at hand, however it will mainly take the role of an auxiliary component rather than a focal point of the discussion. Hence, for our purposes, a rudimentary introduction to some core concepts of supersymmetry will suffice.

The Noether charges associated with supersymmetry transformations are the anti-commuting \emph{supercharges} $Q_\alpha$ and $\tilde{Q}_\beta$, which satisfy
\begin{align}
	\{Q_\alpha,Q_\beta\} = \{ \tilde{Q}_\alpha,\tilde{Q}_\beta \}=0\,,
\end{align}
and which transform as left-handed and right-handed Weyl spinors, respectively. These supercharges extend the Poincar\'e algebra to a \emph{super Poincar\'e algebra}, which has both bosonic and fermionic generators. A central point of supersymmetry is that the supercharges satisfy the anti-commutation relation
\begin{align}\label{eq:INY_susy-anti-comm}
	\{Q_\alpha,\tilde{Q}_\beta\} = 2 P^\mu (\Gamma_\mu)_{\alpha\beta}\,,
\end{align}
where the gamma matrices $\Gamma_\mu$ generate the Clifford algebra. It is possible (and often desirable) for theories to have more than one supersymmetry. For extended supersymmetry we will have multiple supercharges, typically denoted by $\Ncal$,
\begin{align}
	Q_\alpha^I\,, \tilde{Q}_\beta^I\,,\quad I=1,\ldots,\Ncal\,.
\end{align}
The anti-commutation relations \eqref{eq:INY_susy-anti-comm} are then generalised to
\begin{align}\label{eq:INY_extended-susy-anti-comm}
	\{Q_\alpha^I,\tilde{Q}_\beta^J\} = 2P^\mu (\Gamma_\mu)_{\alpha\beta} \delta^{IJ}\,.
\end{align}
Extended supersymmetry allows for the more general anti-commutation relations between the supercharges: $\{ Q_\alpha^I, Q_\beta^J \} = Z^{IJ}$, where $Z^{IJ} = -Z^{JI}$ is the \emph{central charge}. In this thesis we will exclusively deal with theories without a central charge, and hence we will assume $Z=0$ from now on. There is an additional global symmetry which allows us to rotate the supercharges into each other, known as \emph{R-symmetry}. The full R-symmetry is $SU(\Ncal)$ in four spacetime dimensions and $SO(\Ncal)$ in three spacetime dimensions, which may or may not be fully realised for a given model.

The supercharges relate bosonic and fermionic states, and hence supersymmetry imposes constrains on the possible particle content of a theory. Starting from a state in our theory we can find the closure under action by the supercharges to find the \emph{supermultiplet} it belongs to. In addition, \emph{CPT symmetry} implies the existence of a CPT dual supermultiplet, which may or may not be distinct from the original supermultiplet. 

The large particle content, intricate transformation rules and unwieldy Lagrangian descriptions make supersymmetry initially seem rather complicated. We can reduce the difficulty and manifest supersymmetry by combining the bosonic and fermionic states in a supermultiplet into a single \emph{superfield}, and extending spacetime to \emph{superspace} by introducing anti-commuting Grassmann degrees of freedom. In the superspace formalism we extend our traditional spacetime manifold to a \emph{supermanifold} which has both commuting and anti-commuting dimensions. We will encounter these ideas in more detail in section \ref{sec:INT_nf}.

\section{Theories of Interest}

There are a few specific quantum field theories that make a recurring appearance in this thesis. In particular \nf, ABJM, and $\tr{\phi^3}$ theory will be of great interest. These theories are of interest to us because their scattering amplitudes admit multiple interesting descriptions (in particular, their scattering amplitudes can be described using \emph{positive geometries}, \emph{c.f.} section \ref{sec:POS}), which we will investigate in detail later in this thesis. Before delving into their scattering amplitudes, we will take this moment to define these theories and provide some basic information regarding their structure.

\subsection{\texorpdfstring{\nf}{N=4 SYM}}\label{sec:INT_nf}

Four dimensional maximally supersymmetric ($\Ncal=4$) $SU(N)$ Yang-Mills theory, denoted `\nf' throughout this thesis, is an important playground for modern techniques in scattering amplitudes. It was originally identified as an interesting theory due to its holographic duality to type IIB string theory compactified on $AdS_5\times S^5$ \cite{Maldacena:1997re}. Although this theory should for all intents and purposes be considered a toy model, it is worth noting that, as far as toy models go, this theory is remarkably accurate at describing theories of physical interests. Notably, gluon tree-level amplitudes in \nf are equivalent to tree-level gluon amplitudes in non-supersymmetric Yang-Mills theory.

\nf is a supersymmetric extension of Yang-Mills with 4 supercharges $Q^A, A=1,\ldots,4$. In this theory, the $SU(4)_R$ R-symmetry constrains all matter fields, and the only parameters left unfixed are the gauge group and the coupling constant. The theory is CPT self-conjugate and its spectrum consists of a single supermultiplet consisting of 16 particles given in table \ref{tab:nf-particles}.
\begin{table}
	\centering
	\begin{tabular}{|l|lllll|}
		\hline
		Multiplicity & 1 & 4 & 6 & 4 & 1 \\
		Helicity & +1 & +1/2 & 0 & -1/2 & -1  \\
		Symbol & $g^+$ & $\psi^A$ & $\phi^{AB}$ & $\bar\psi^{A}$ & $g^-$ \\
		Name & gluon & gluino & scalar & gluino & gluon \\\hline
	\end{tabular}
	\caption{The particle content of \nf.}
	\label{tab:nf-particles}
\end{table} 
All these particles transform into one another under $SU(4)_R$ supersymmetry transformations, which allows us to combine the states into a single chiral on-shell \emph{superfield}
\begin{equation}\label{eq:INT_supermultiplet-nf}
	\Phi = g^+ + \eta_A \psi^A - \frac{1}{2!} \epsilon_{ABCD} \eta^A \eta^B \phi^{CD} - \frac{1}{3!}\epsilon_{ABCD} \eta^A \eta^B \eta^C \bar\psi^{D} + \frac{1}{4!} \epsilon_{ABCD} \eta^A \eta^B \eta^C \eta^D g^-,
\end{equation}
where the $\eta$s are anti-commuting Grassmann variables, and $\phi^{AB}=-\phi^{BA}$.

The \nf action can be written compactly as
\begin{align}
	S= \int \dd^4x \text{Tr}\,{\Big (}&-\frac{1}{4}F^{\mu\nu}F_{\mu\nu} - \frac{1}{2}(D_\mu\phi^{AB})^2 -\frac{1}{2}[\phi_{AB},\phi_{CD}]^2+i\bar\psi_A\slashed{D}\psi^A \notag\\&-\frac{i}{2}\bar\psi^A[\phi_{AB},\bar\psi^B] -\frac{i}{2}\psi^A[\phi_{AB},\psi^B]{\Big )}\,,
\end{align}
where the trace indicates a summation over the adjoint indices of the gauge group, and $D_\mu= \partial_\mu - ig[A_\mu,\bullet]$ is the covariant derivative.

This action famously enjoys a tremendous number of classical symmetries. Alongside the obvious Poincar\'e symmetry, $SU(N)$ gauge symmetry and (global) supersymmetry, at the origin of moduli space (where the vacuum expectation value of the scalar potential vanish, $\<\phi^{AB}\>=0$) all states are massless, and the theory develops a further \emph{conformal symmetry}. This conformal symmetry further combines with supersymmetry into a \emph{superconformal symmetry}. Furthermore, for a fixed order of the external momenta, we can define the \emph{dual momenta} $x_i^\mu$ which satisfy $x^\mu_{i+1}-x^\mu_i=p_i^\mu$ (we will treat these dual momenta in more detail in section \ref{sec:KIN_dual}). When taking the \emph{planar limit} (see section \ref{sec:AMP_large-N}), we have such a manifest ordering, and \nf develops a further \emph{dual superconformal symmetry} in this space of dual momenta \cite{Drummond:2008vq}. On top of all of these symmetries, the superconformal and dual superconformal symmetries close into an infinite dimensional \emph{Yangian symmetry} \cite{Drummond:2009fd}. 

The presence of this infinite dimensional Yangian symmetry is often seen as the hallmark for \emph{integrability}, and as such it is expected that the full S-matrix can be solved exactly. In contrast, many of the remarkable modern developments surrounding the computation of scattering amplitudes in \nf rarely exploit this integrability, and many of these techniques can (and have) been used for non-integrable theories as well. All these symmetries impose many constraints on the structure of \nf, while still remaining non-trivial and interesting. 

Lastly, we point out that \emph{parity symmetry} relates the positive helicity particles in \nf to their negative helicity counterparts, and vice versa. The superfield \eqref{eq:INT_supermultiplet-nf} is a \emph{chiral} superfield, which manifestly breaks this symmetry by giving a different Grassmann weight to $g^+$ and $g^-$. As an alternative to chiral superspace, we can use the \emph{non-chiral superspace formalism} \cite{Huang:2011um}. The two formalisms are effectively related by a half-Fourier transform (see also the Penrose transform introduced in section \ref{sec:KIN_twistors})
\begin{align}
	\Phi' &=\left.\int \dd \eta^3\dd\eta^4 e^{\eta^3\tilde\eta_{\dot3} + \eta^4\tilde\eta_{\dot4}} \Phi\right|_{\eta\leftrightarrow\tilde\eta}\\
	&= \eta^2 g^+ +\tilde\eta^2 g^- + \phi + \eta^\alpha\tilde\eta^{\alphadot}\phi'_{\alpha\alphadot} + \eta^2\tilde\eta^2\phi'' +\eta^\alpha \psi_\alpha + \eta^2\tilde\eta^\alphadot\psi_\alphadot + \tilde\eta^\alphadot\bar\psi_\alphadot+\tilde\eta^2\eta^\alpha \bar\psi_\alpha\,,
\end{align}
where $\eta^2=\eta^\alpha\eta_\alpha/2\,,\tilde\eta^2=\tilde\eta^\alphadot\tilde\eta_\alphadot/2$, and we have split up the six $\phi^{AB}$ fields into a $\phi$, four $\phi'_{\alpha\beta}$, and a $\phi''$ field. We have swapped $\eta$ and $\tilde\eta$ in $\Phi'$ to be consistent with the notation of \cite{He:2018okq}. In this representation the R-symmetry is $SU(2)\times SU(2)$.

\subsection{ABJM Theory}\label{sec:QFT_ABJM}

The second theory of interest is the three-dimensional theory of Chern-Simons matter with $\Ncal=6$ supersymmetry, often called \emph{Aharony-Bergman-Jafferis-Maldacena (ABJM) theory} \cite{Aharony:2008ug, Hosomichi:2008jb} (see \emph{e.g.} \cite{Elvang:2013cua} for an amplitudes oriented introduction). We are motivated to find theories with a maximal amount of symmetry. In three dimensions, the only theories which can have both supersymmetry and conformal invariance need to be supersymmetric extensions of Chern-Simons theory, due to the dimensionless coupling (see also \cite{Gustavsson:2007vu, Bagger:2007jr} for a discussion on $\Ncal=8$ supersymmetric matter-Chern-Simons theory, known as \emph{BLG theory}). Although physically distinct, ABJM theory has many intriguing parallels to \nf. ABJM theory is holographically dual to M-theory compactified on $AdS_4\times S^7$, and, as we will see in section \ref{sec:AMP_ABJM}, their amplitudes allow for constructions very similar to \nf. The theory has $SO(6)=SU(4)$ R-symmetry, and the spectrum consists of four complex scalars $X_A$, four complex fermions $\psi^{Aa}$, together with their complex conjugates $\bar{X}^{\mathrm{A}}, \bar{\psi}_{\mathrm{A}a}$, all of which transform in the anti-fundamental of $SU(4)$, where $\mathrm{A}=1,2,3,4$. We combine these fields into a bosonic and fermionic on-shell superfield
\begin{align}\label{eq:INT_ABJM-fields}
	\Phi^{\mathcal{N}=6} & = X_4 + \eta_A \psi^A - \frac{1}{2} \epsilon^{ABC}\eta_A\eta_B X_C-\eta_1\eta_2\eta_3\psi^4\,\\
	\bar{\Psi}^{\Ncal=6} & = \bar{\psi}_4+\eta_A \bar{X}^A - \frac{1}{2} \epsilon^{ABC} \eta_A \eta_B \bar{\psi}_C-\eta_1\eta_2\eta_3\bar{X}^4\,,
\end{align} 
where the $\eta$s again denote anti-commuting Grassmann variables. We have split the R-symmetry indices $\mathrm{A}\to (A,4)$, with $A=1,2,3$, this superspace formalism thus only manifests and $SU(3)_R$ subgroup of $SU(4)_R$. The two superfields are conjugate to each other under R-symmetry.

ABJM theory contains two gauge fields $A^a_b$, $\hat{A}^\adot_\bdot$ with gauge group $U(N)\times U(N)$. The Lagrangian can be written as \cite{Benna:2008zy, Bandres:2008ry}
\begin{align}
	\Lcal = \frac{k}{2\pi} \Big[ &\frac{1}{2}\epsilon^{\mu\nu\rho}\left(A_\mu\partial_\nu A_\rho + \frac{2i}{3} A_\mu A_\nu A_\rho - \hat{A}_\mu\partial_\nu \hat{A}_\rho - \frac{2i}{3} \hat{A}_\mu \hat{A}_\nu\hat{A}_\rho\right)\\
	&-(D^\mu X_\mathrm{A})^\dagger D_\mu X_\mathrm{A} + i \bar{\psi}_\mathrm{A} \slashed{D} \psi^\mathrm{A} + \Lcal_4 + \Lcal_6 \Big]\,,
\end{align}
where the covariant derivatives are defined as
\begin{align}
	D^\mu X_\mathrm{A} &\coloneqq \partial_\mu X_\mathrm{A} + i \hat{A}_\mu X_\mathrm{A} - i X_\mathrm{A} A_\mu\,,\\
	(D^\mu X_\mathrm{A})^\dagger & \coloneqq \partial_\mu \bar{X}^\mathrm{A} + i A_\mu \bar{X}^\mathrm{A} - i \bar{X}^\mathrm{A} \hat{A}_\mu\,,
\end{align}
and similarly for $\psi$ and $\bar{\psi}$. $\Lcal_4$ and $\Lcal_6$ are defined as
\begin{alignat}{2}
	\Lcal_4 &= i \text{Tr}\Big( && \bar{X}^\mathrm{B} X_\mathrm{B} \bar{\psi}_\mathrm{A} \psi^\mathrm{A} - X_\mathrm{B} \bar{X}^\mathrm{B} \psi^\mathrm{A} \bar{\psi}_\mathrm{A} + 2 X_\mathrm{A} \bar{X}^\mathrm{B} \psi^\mathrm{A}\bar{\psi}_\mathrm{B} -2 \bar{X}^\mathrm{A} X_\mathrm{B} \bar\psi_\mathrm{A} \psi^\mathrm{B} \notag \\ 
	& && -\epsilon_{\mathrm{A}\mathrm{B}\mathrm{C}\mathrm{D}} \bar{X}^\mathrm{A} \psi^\mathrm{B} \bar{X}^\mathrm{C} \psi^\mathrm{D} + \epsilon^{\mathrm{A}\mathrm{B}\mathrm{C}\mathrm{D}} X_\mathrm{A} \bar\psi_\mathrm{B} X_\mathrm{C} \bar\psi_D \Big)\,, \\
	\Lcal_6 &= \frac{1}{3} \text{Tr} \Big( && X_\mathrm{A} \bar{X}^\mathrm{A} X_\mathrm{B} \bar{X}^\mathrm{B} X_\mathrm{C} \bar{X}^\mathrm{C} + \bar{X}^\mathrm{A} X_\mathrm{A} \bar{X}^\mathrm{B} X_\mathrm{B} \bar{X}^\mathrm{C} X_\mathrm{C} + 4 \bar{X}^\mathrm{A} X_\mathrm{B} \bar{X}^\mathrm{C} X_\mathrm{A} \bar{X}^\mathrm{B} X_\mathrm{C} \notag \\ 
	& && -6 X_\mathrm{A} \bar{X}^\mathrm{B} X_\mathrm{B} \bar{X}^\mathrm{A} X_\mathrm{C} \bar{X}^\mathrm{C} \Big)\,.
\end{alignat}
Fortunately, we will not have to work with Lagrangians in the remainder of this thesis, and we record this Lagrangian here only for the sake of completeness.

ABJM theory has an $OSp(6|4)$ superconformal symmetry, which consists of an $SO(6)$ R-symmetry and an $Sp(4)$ conformal symmetry. In the planar limit, the theory further also has an $OSp(6|4)$ dual superconformal symmetry \cite{Huang:2010qy}, which close into an infinite dimensional Yangian symmetry \cite{Bargheer:2010hn}.

In later chapters we will put some emphasis on a specific supersymmetric reduction of ABJM theory. We now only need two anti-commuting variables $\eta^A$, $A=1,2$ and define the four superfields
\begin{subequations}\label{eq:QFT_susy-reduced-superfield}
	\begin{alignat}{2}
		\Phi^{\Ncal=4} &= \Phi^{\Ncal=6}|_{\eta^3\to0} &&= X_4 + \eta_I\psi^I +\eta_1\eta_2 X_3\,,\\
		\bar{\Phi}^{\Ncal=4} &=  \int \dd\eta^3 \bar{\Psi}^{\Ncal=6} && = \bar{X}^3 + \eta_I \bar{\psi}^I - \eta_1\eta_2 \bar{X}^4\,,\\
		\Psi^{\Ncal=4} &=  \int \dd\eta^3 \Phi^{\Ncal=6} && = \psi^3 + \eta_I X^I - \eta_1\eta_2 \psi^4\,,\\
		\bar{\Psi}^{\Ncal=4} &= \bar{\Psi}^{\Ncal=6}|_{\eta_3\to0} && = \bar{\psi}_4 + \eta_I \bar{X}^I + \eta_1\eta_2 \bar{\psi}_3\,.
	\end{alignat}
\end{subequations}

\subsection{\texorpdfstring{Bi-Adjoint $\phi^3$}{Bi-Adjoint Scalar}}\label{sec:INT_phi3}

When taking an introductory course on quantum field theory, among the first theories one usually encounters are scalar theories with a $\phi^3$ interaction term. Contrary to the previous theories we discussed, $\phi^3$ theory is very bare with a minimal number of additional symmetries or constraints. This lack of structure is precisely what gives it such a simple Lagrangian and diagrammatic description.

We will be interested in a slight modification of this simple cubic scalar theory. We enrich the structure of the theory by having our scalar field carry two adjoint indices of $U(N)\times U(\tilde{N})$. This new theory goes by the name \emph{bi-adjoint scalar theory}, or \emph{bi-adjoint $\phi^3$}. It was first introduced in \cite{Cachazo:2013iea} from considerations from the CHY formalism, a topic which we will return to in section \ref{sec:AMP_CHY}. The Lagrangian of this theory has an interaction term of the form
\begin{align}
	f_{abc}\tilde{f}_{a'b'c'}\phi^{aa'}\phi^{bb'}\phi^{cc'}\,,
\end{align}
where $f^{abc}=\tr{[T^a,T^b]T^c}$ and $\tilde{f}^{a'b'c'}=\text{Tr}({[\tilde{T}^{a'},\tilde{T}^{b'}]\tilde{T}^{c'}})$ are the structure constants of $U(N)$ and $U(\tilde{N})$, respectively. 

Contrary to the proclaimed intent of this section, let us have a quick look at the scattering amplitudes in this theory. To calculate these amplitudes, we will not make use of any of the elegant machinery we will encounter later in this thesis, and shall instead take the classical Feynman diagram approach. We will restrict ourselves to tree-level scattering amplitudes only, and we will assume the fields to be massless, although all formulas can easily be generalised to the massive case as well.

Given some Feynman diagram, the Feynman rules tell us to include a factor $f_{abc}\tilde{f}_{a'b'c'}$ at the vertex where particles $\phi^{aa'},\phi^{bb'},\phi^{cc'}$ meet. The only other relevant part of the Feynman diagram are the propagators, the rule for which is to include a factor $1/P^2$ for each internal line with momentum $P^\mu$. When expanding out the sum over all Feynman diagrams, it is possible to decompose the contraction of the adjoint indices in a double colour decomposition:
\begin{align}
	A_n = \sum_{\alpha\in S_n/Z_n}\sum_{\beta\in S_n/Z_n} \tr{T^{a_{\alpha(1)}}\cdots T^{a_{\alpha({n})}}}\tr{\tilde{T}^{b_{\beta(1)}}\cdots \tilde{T}^{b_{\beta({n})}}} m_n(\alpha|\beta)\,,
\end{align}
where we introduce the \emph{double colour ordered amplitudes} $m_n(\alpha|\beta)$. The Feynman diagrams which contribute to such a double colour ordered amplitude are precisely those which are \emph{mutually compatible} with the orderings $\alpha$ and $\beta$, which means that the Feynman diagram is a planar graph if we order the external particles on a disk according to both $\alpha$ and $\beta$. This also show that we only need to concern ourselves with the \emph{relative ordering} between $\alpha$ and $\beta$. We can therefore take $\alpha=\{1,2,\ldots,n\}\equiv\unit$ without loss of generality.

We will be interested in partial amplitudes where $\alpha=\beta$. In this case, the amplitudes $m_n\equiv m_n(\alpha|\alpha)$ are precisely those of so-called $\tr{\phi^3}$ theory, and we shall refer to the theory as such. All planar Feynman diagrams contribute to this amplitude. Since we have stripped off all colour factors, the only relevant information in the Feynman diagram is the set of propagators. Since the diagrams are planar, we can only encounter sums of consecutive momenta in the propagators, which motivates the definition of the \emph{planar Mandelstam variables} $X_{ij}=(p_{i}+p_{i+1}+\ldots+p_{j-1})^2$.

\begin{figure}
	\centering
	\includegraphics[scale=0.4]{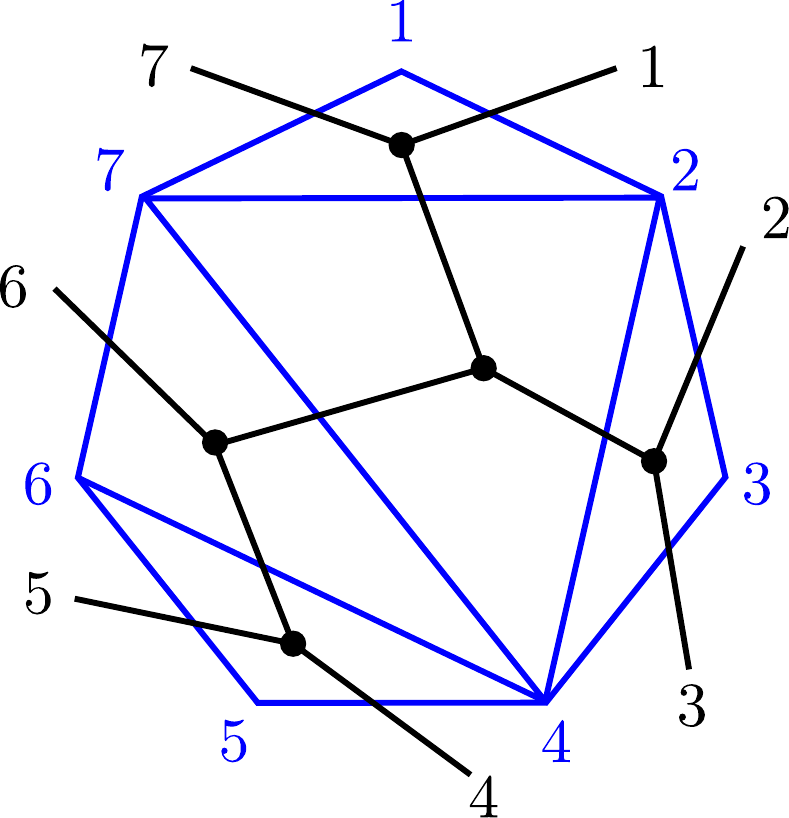}
	\caption{We see that the dual of an $n$-particle planar Feynman diagram is a triangulation of an $n$-gon.}
	\label{fig:feynman-triangulation}
\end{figure}

The \emph{dual} of a Feynman diagram contributing to $m_n$ is a triangulation of an $n$-gon, as is illustrated in figure \ref{fig:feynman-triangulation}.
The chords of such a triangulation correspond to a propagator in the Feynman diagram. Specifically, the chord $(i,j)$ corresponds to the propagator $1/X_{ij}$. If we let $\Tcal=\{(i_1,j_i),\ldots,(i_{n-3},j_{n-3})\}$ denote the set of chords in some triangulation of an $n$-gon, then the dual Feynman diagram contributes $(X_{i_1,j_1}\cdots X_{i_{n-3},j_{n-3}})^{-1}$. Hence, we can write the full tree-level $\tr{\phi^3}$ amplitude as
\begin{align}
	m_n = \sum_{\Tcal}\prod_{(i,j)\in\Tcal}\frac{1}{X_{ij}}\,,
\end{align}
where the sum is over all triangulations $\Tcal$ of the $n$-gon. The number of triangulations of an $n$-gon, and hence the number of Feynman diagrams, is given by $C_{n-2}$, where $C_p$ is the $p$\textsuperscript{th} \emph{Catalan number}
\begin{align}
	C_p=\frac{1}{p+1}\binom{2p}{p}\,.
\end{align} 

\section{Summary}

In this chapter we have introduced some essential elements from QFT. We have introduced \emph{Poincar\'e symmetry}, \emph{conformal symmetry}, and \emph{supersymmetry}, which place important constraints on the structure of scattering amplitudes. The symmetries of a theory will additionally guide us to consider the appropriate set of kinematic variables. After this, we introduced the main theories of interest for this thesis: \emph{\nf}, \emph{ABJM}, and \emph{bi-adjoint $\phi^3$ theory}. These theories will make a recurring appearance in this thesis. We will be particularly interested in a positive geometric description of the scattering amplitudes for these theories in section \ref{sec:POS}.

%% file: chapters/grassmann.tex
\chapter{Grassmannian Geometry}\label{sec:GRASS}

We continue with an investigation of \emph{projective geometry} and \emph{Grassmannian geometry}. Although physically motivated, we will embrace the mathematical nature of these subjects and proceed in this chapter with little to no reference to physics, although we keep a physicists level of rigour. Throughout this thesis, we will encounter several descriptions of scattering amplitudes which are rooted in projective and Grassmannian geometry. For example, as we will see in chapter \ref{sec:KIN}, (momentum) twistors are naturally understood as elements of projective space $\Pbb^3$, and momentum conservation in terms of spinor-helicity variables can be linearised using the Grassmannian. Furthermore, in chapter \ref{sec:AMP} we will see that we can find scattering amplitudes by studying certain Grassmannian integrals. And last but not least, in chapter \ref{sec:POS} we will study the amplituhedron, which can be understood as a generalisation of projective polytopes into the Grassmannian. 

The aim of this chapter is to develop the tools and terminology which we will need in future chapters. This includes a detailed study of projective polytopes, the positive (orthogonal) Grassmannian, and the combinatorics of positroid and orthitroid cells. If the reader is familiar with these notions, then the majority of this thesis should be understandable, even upon omission of this chapter. However, for a deeper understanding of the geometric formulations of scattering amplitudes which we will encounter in chapter \ref{sec:POS}, it is essential to develop an intuition for Grassmannian geometry.

\section{Projective Geometry}\label{sec:GRASS_proj}

Given a field $K$, we define the \emph{projective space} $K\Pbb^n$ as a set of equivalence classes in $K^{n+1}\setminus\{0\}$ where $x \sim y$ if $y= \lambda x$ for some $\lambda\in K$. An interpretation which we will call on repeatedly is that $K\Pbb^n$ is the space of rays passing through the origin of $K^{n+1}$. In the cases we are interested in, we will exclusively encounter $K=\Rbb$ or $K=\Cbb$. In cases where there is no room for confusion, or where the distinction is irrelevant, we will often simply write $\Pbb^n$ rather than $\Rbb\Pbb^n$ or $\Cbb\Pbb^n$, the underlying field of this projective space should be clear from context. 

We can denote elements of projective space $K\Pbb^n$ by their \emph{homogeneous coordinates} $x=(X^1,\ldots,X^{n+1})$, which can be understood as defining a vector in $K^{n+1}$ whose span defines the element of projective space. By definition, this $x$ is equivalent to any rescaling of $x$, which means that homogeneous coordinates have a $GL(1)$ redundancy. In the patch where $X^1\neq 0$, we can use this redundancy to `gauge fix' the first entry to be one. The variables $Y^1,\ldots, Y^{n}$, where $Y^i=X^{i+1}/X^1$, are now free and unconstrained, which shows that this patch is isomorphic to $K^n$. The complement to this patch, \emph{i.e.} the subvariety where $X^1=0$, admits homogeneous coordinates $(0,X^2,\ldots, X^{n+1})$. Since we haven't fixed the $GL(1)$ redundancy, there is still a rescaling freedom, and hence this subvariety is isomorphic to $K\Pbb^{n-1}$. This argument shows that $K\Pbb^n \simeq K^n \bigcup K\Pbb^{n-1}$. As an example, let us consider the case where $K=\Rbb$. First, we take $n=1$, in which case the above arguments shows that $\Rbb\Pbb^1\simeq \Rbb\bigcup \Rbb\Pbb^0$. Since $\Rbb\Pbb^0$ is just a point, we can think of the \emph{real projective line} $\Rbb\Pbb^1$ as an extension of $\Rbb$ by `a point at infinity'. Continuing to $n=2$, the \emph{real projective plane} $\Rbb\Pbb^2$ is isomorphic to $\Rbb^2\bigcup \Rbb\Pbb^1$, which we interpret as an extension of the Euclidean plane by a (projective) line at infinity. This continues for higher $n$. Statements such as `two (distinct) lines intersect in a point' are almost true in the Euclidean plane, since we have to account for the possibility that two lines are parallel. In the projective plan, however, this statement is exact, and two `parallel' lines intersect at infinity. In higher dimensions, the analogous statement is that two hyperplanes intersect in a codimension-2 plane in projective space.

The automorphism group of $\Pbb^n$ is $SL(n+1)$ which sends $X^I\to L^I_J X^J$ for some $L\in SL(n+1)$. We could consider $GL(n+1)$ transformations instead, however due to the projective nature we can effectively always rescale the determinant to one. One example, which will make an occurrence later, is the automorphism group of the Riemann sphere $\Cbb\Pbb^1$. We can parametrise a patch of the Riemann sphere by $z\in \Cbb$, corresponding to the point $(1,z)$ in $\Cbb\Pbb^1$. The automorphism group acts on this parametrisation as 
\begin{align}
	\begin{pmatrix}	1\\z \end{pmatrix} \to \begin{pmatrix}
	A & B \\ C & D
\end{pmatrix} \begin{pmatrix} 1\\z \end{pmatrix} = \begin{pmatrix} A+Bz\\C+Dz \end{pmatrix}\,.
\end{align}
After rescaling the first component to one, we see that this transformation sends
\begin{align}
	z\to \frac{C+Dz}{A+Bz},\quad AD-BC=1\,.
\end{align}
These are the well-known \emph{M\"obius transformations}. We note that this transformation is invariant when replacing $(A,B,C,D)\to(-A,-B,-C,-D)$, which means that the automorphism group is technically $PSL(2,\Cbb)$, rather than $SL(2,\Cbb)$, however this subtlety is often ignored.

\subsection{Points, Lines, Planes, etc.}

Projective space is non-metric, meaning that there is no notion of distance between points. Instead, we concern ourselves with questions regarding the configurations of points, lines, planes, etc, and their incidence (\emph{e.g.} whether three points lie on a line). A collection of $n+1$ points $\smash{\{x_i\}_{i=1}^{n+1}}$ in $\Pbb^n$ with homogeneous coordinates $X_i^A$ are linearly dependent if the determinant made up from their homogeneous coordinates vanishes. If we define $\smash{\epsilon_{A_1\cdots A_{n+1}}}$ as the fully anti-symmetry Levi-Civita tensor, then this determinant can be written as $\smash{\epsilon_{A_1\cdots A_{n+1}}X_1^{A_1}\cdots X_{n+1}^{A_{n+1}}}$. We will often denote this determinant as $\<X_1\cdots X_{n+1}\>$, or, when there is no room for confusion, as $\<1\cdots n+1\>$.

To build some intuition, let us focus on $\Rbb\Pbb^2$. We recall that this is essentially $\Rbb^2$, as long as we take care not to make any lines parallel, and hence this is easy to visualise. We use the homogeneous coordinates $X=(x,y,z)$. A line in $\Rbb\Pbb^2$ can be defined as the subvariety where $Ax+By+Cz=0$ for some coefficients $A,B,C$. We can record this in the vector $\Lcal_I=(A,B,C)$, such that the line is defined by $(\Lcal\cdot X) = \sum_{I=1}^3\Lcal_IX^I =0$. It is appropriate to give $\Lcal$ a downstairs index, since under an $SL(3)$ transformation $X^I\to L^I_J X^J$, the line transforms as $\Lcal_I\to (L^{-1})_I^J\Lcal_J$. Furthermore, it is clear that some rescaling $\Lcal \to \lambda\Lcal$ still defines the same line, hence we can interpret $\Lcal$ as a point in some \emph{dual projective space}. We can equivalently define a line by any two points $X_1^I$ and $X_2^I$ on the line. A point $X$ lies on this line when it is linearly dependent on $X_1$ and $X_2$, which thus means that $\epsilon_{IJK}X_1^IX_2^JX^K=0$. Hence, the point in dual projective space corresponding to the line $(X_1X_2)$ is given by $\Lcal_I=\epsilon_{IJK}X_1^JX_2^K$. Furthermore, we know that two lines $\Lcal_1$ and $\Lcal_2$ intersect in a point, which is given by $X^I=\epsilon^{IJK}\Lcal_{1\,J}\Lcal_{2\,K}$. We note the similarity between the statements
\begin{alignat}{4}
	&\text{\textit{``Two points define a line''}}\quad &&\leftrightarrow\quad &&\Lcal_I=\epsilon_{IJK}X_1^JX_2^K\\
	& \text{\textit{``Two lines define a point''}} &&\leftrightarrow && X^I=\epsilon^{IJK}\Lcal_{1\,J}\Lcal_{2\,K}\,.
\end{alignat} 
This is an example of \emph{projective duality}, which essentially states that we can take any true statement in $\Rbb\Pbb^2$ and interchange the words line and point to find another true statement. From our current discussion this is completely trivial, as we can simply interpret the statement in dual projective space instead. Said equivalently, we simply raise/lower all indices to find the dual statement.

Next, we move on to $\Pbb^3$. A point $X^I$ has one upstairs index, and a 2-plane $Y_I$ has one downstairs index and corresponds to a point in dual projective space. We can define a line by giving two points $A^I$ and $B^I$. Any $GL(2)$ transformation acting on the $2\times 4$ matrix $\begin{pmatrix} A^I & B^I \end{pmatrix} $ will give two new points on this line. It is therefore natural to interpret the line as having two anti-symmetrised indices: $(AB)^{[I,J]}=A^IB^J-A^JB^I$, as this combination is projectively invariant under a $GL(2)$ transformation. Two lines $(AB)^{[I,J]}$ and $(CD)^{[I,J]}$ intersect precisely when $\<ABCD\>=0$, which is a statement which will be important when discussing twistor geometry later on. These considerations trivially generalise to higher-dimensional projective spaces. 

\subsection{Projective Polytopes}\label{sec:GRASS_proj_poly}

\begin{figure}
	\centering
	\includegraphics[scale=0.5]{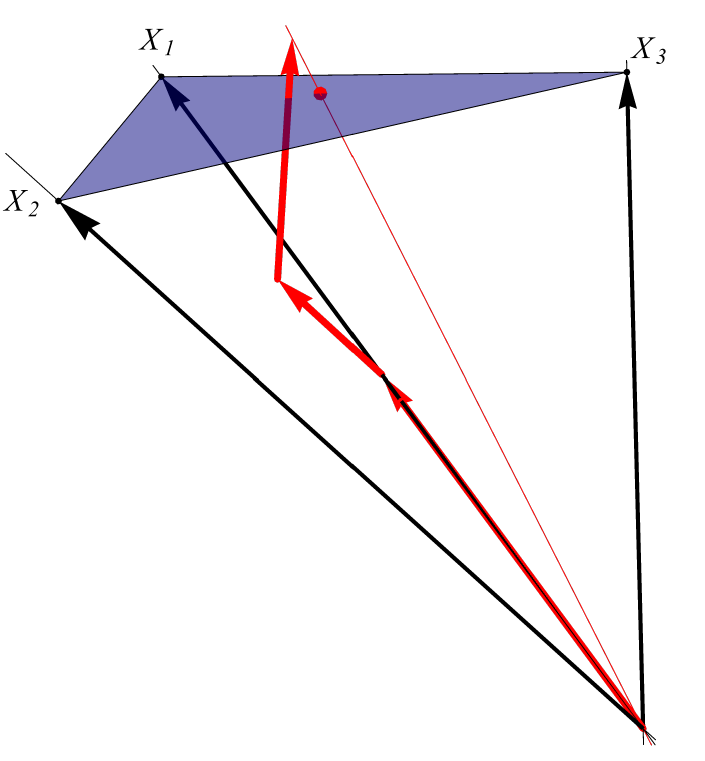}
	\caption{Any positive linear combinations of three points in $\Rbb\Pbb^2$ is in the convex hull of these points.}
	\label{fig:conv-hull}
\end{figure}

Consider two points $X_1=(1,x_1),X_2=(1,x_2)$ in $\Rbb\Pbb^1$. If we consider $c_1,c_2\in \Rbb^+$, then $c_1 X_1+c_2 X_2$, once rescaled to fix its first component to one, will lie somewhere between $x_1$ and $x_2$. Explicitly, we get a weighted sum of $x_1$ and $x_2$:
\begin{align}
	c_1 X_1 + c_2 X_2 = \begin{pmatrix}
		c_1 +c_2\\ c_1 x_1 + c_2 x_2
	\end{pmatrix} \sim \begin{pmatrix}
	1\\ \frac{c_1 x_1+c_2x_2}{c_1+c_2}
\end{pmatrix}\,.
\end{align}
We thus see that we can define the \emph{convex hull} of points $X_1$ and $X_2$ (in $\Rbb^2$ this would be the cone spanned by $X_1$ and $X_2$, or in $\Rbb^1$ this would be the line-segment between $x_1$ and $x_2$) as their positive span\footnote{Note that we are now talking about a slightly restricted notion of projective geometry. In the standard notion we could rescale $X_i\to - X_i$, so the restriction for the $c_i$ to be positive would be meaningless. Instead, we now only identify $X_i \sim \lambda X_i$ for \emph{positive} $\lambda$. We are thus no longer identifying rays through the origin, but half-rays.}:
\begin{align}
	\conv(X_1,X_2) = \{c_1 X_1+c_2 X_2 \colon c_1,c_2\in\Rbb^+\}\,.
\end{align}
This generalises to higher dimensions. Given $n$ points $x_1,\ldots,x_n$ in $\Rbb\Pbb^k$, we find their convex hull as
\begin{align}
	\conv(X_1,\ldots,X_n)=\{\sum_{i=1}^n c_i X_i\colon c_i > 0\}\,.
\end{align}
This is illustrated in figure \ref{fig:conv-hull}. 

Said equivalently, given $n$ points in $\Rbb\Pbb^k$ with homogeneous coordinates $(X_{i}^1,\ldots,X_{i}^{k+1})$, then we define a $(k+1)\times n$ matrix
\begin{align}
	X = \begin{pmatrix}
		X_{1}^1 & X_{2}^1 & \cdots & X_{n}^1\\
		\vdots & \vdots & \ddots & \vdots \\
		X_{1}^{k+1} & X_{2}^{k+1} & \cdots & X_{n}^{k+1}
	\end{pmatrix}\,,
\end{align} 
which is defined up to a torus action $(\Rbb^+)^n$ (a positive rescaling of the $n$ columns), then the convex hull can be defined as the image of the linear map
\begin{align}
	\Phi\colon (\Rbb^{+})^n &\to \Rbb\Pbb^k\\
	C &\mapsto C\cdot X^T\,,
\end{align}
where $C = \begin{pmatrix}c_1 & \cdots & c_n\end{pmatrix} \in (\Rbb^{+})^n$. Due to the projective nature of the image, the matrix $C$ should be considered up to $GL(1)$. We will see in section \ref{sec:GRASS_pos} that the domain $(\Rbb^{+})^n/GL(1)$ is the \emph{positive Grassmannian} $G_+(1,n-1)$. We note that this definition of the convex hull is not always projectively well-defined, as we have to ensure that the sum $\sum_i c_i X_i,\,c_i>0$ cannot be zero, since the zero-vector does not represent a point in projective space. This can be ensured by requiring that all the ordered maximal minors of the matrix $X$ are positive. We have thus shown that the projective polytope defined as the convex hull of $n$ points in $\Rbb\Pbb^k$ can be found as the image of the positive Grassmannian $G_+(1,n-1)$ under a positive linear map $\Phi$. This is an important observation which will later motivate us to consider the \emph{amplituhedron} (\emph{c.f.} section \ref{sec:POS_amplituhedron}).

We have seen how we can define convex polytopes through their vertices, but they can equivalently be defined by their facets. A collection of $k$ vertices $X_{a_1},\ldots,X_{a_k}$ define a facet of the polytope if all other $X$s are on the same side of the hyperplane $W_I=\epsilon_{IJ_1\cdots J_k}X_{a_1}^{J_1}\cdots X_{a_k}^{J_k}$. That is, $(W\cdot X_b)=\<X_{a_1}\cdots X_{a_k} X_b\>$ has the same sign for all $b$. Given a collection of points in dual projective space $W_1,\ldots, W_r$, where $r$ denotes the number of facets, we can then equivalently define a projective polytope as the set of all $Y\in \Rbb\Pbb^k$ such that $(Y\cdot W_i)>0$ for all $i$. In the dual projective space we can define a \emph{dual polytope} as the convex hull of the points $W_1,\ldots,W_r$, which can equivalently be `cut out' by the inequalities $(\tilde{Y}\cdot X_i)>0$, where $\tilde{Y}$ represents an arbitrary point in dual projective space. Due to the projective duality it is clear that vertices of the polytope correspond to facets of the dual polytope, edges of the polytope correspond to codimension-2 boundaries of the dual polytope, and so on. In particular, it is clear that the facets of the polytope correspond to the vertices of the dual polytope. 

\subsection{Volumes}\label{sec:GRASS_projective-volume}

We take this moment to make a remark about the volume of simplices. The notion of a volume does not depend on a metric, however it is not projectively well-defined. Rather, it is an \emph{affine} notion. We can talk about affine notions from the point of view of projective geometry by fixing some hyperplane `at infinity'. That is, if we fix a hyperplane $Y_I$, then the subgroup of the automorphism group of $\Pbb^n$ which leave this hyperplane fixed are precisely the affine transformations, and it will allow us to define a notion of volume.

Let us consider an $n$-simplex $\Delta_n$ in $\Rbb^n$ with vertices $x_1,\ldots,x_{n+1}$. We can embed this in projective space by defining points with homogeneous coordinates $X_i=(1,x_i)$. The oriented volume of this simplex is then given by 
\begin{align}\label{eq:GRASS_dual-simplex-volume-fixed}
	\vol(\Delta_n)=\frac{\<X_1\cdots X_{n+1}\>}{n!}=\frac{1}{n!}\begin{vmatrix}
		1 & 1 & \cdots & 1\\ x_1 & x_2 & \cdots & x_{n+1}
	\end{vmatrix}\,.
\end{align}
We note that this formula is crucially dependent on the way we retrieve our points $x_i$ in $\Rbb^n$ from the points $X_i$ in $\Rbb\Pbb^n$, which we do by intersecting the rays in $\Rbb^{n+1}$ with the `plane at infinity' given by $X^1=1$. We can write this more invariantly by considering a general hyperplane at infinity $Y_I$, which allows us to write the volume as
\begin{align}
	\vol(\Delta_n)=\frac{1}{n!}\frac{\<X_1\cdots X_{n+1}\>}{(Y\cdot X_1)\cdots (Y\cdot X_{n+1})}\,.
\end{align}
We note that this representation is invariant under rescaling $X_i\to\lambda X_i$. 

Rather than defining the simplex by its vertices, we can equivalently have defined it by its facets $Z_{1I},\ldots, Z_{n+1I}$. Clearly the facets are defined by $n$ of the vertices, which allows us to write $Z_{1I}=\epsilon_{IJ_2\cdots J_{n+1}}X_2^{J_2}\cdots X_{n+1}^{J_{n+1}}$, and so on. In terms of these $Z_i$ we can write the volume as
\begin{align}\label{eq:GRASS_dual-simplex-volume}
	\vol(\Delta_n) = \frac{1}{n!}\frac{\<Z_1\cdots Z_{n+1}\>^{n}}{\<YZ_1Z_2\cdots Z_n\>\cdots \<YZ_2Z_3\cdots Z_{n+1}\>}\,.
\end{align}
For future reference we explicitly record the volume of a simplex in $\Pbb^4$ with facets given by the points $Z_a, Z_b, Z_c, Z_d, Z_e$ in dual projective space:
\begin{align}\label{eq:GRASS_dual-simplex-volume-4}
	\vol(\Delta_4) = \frac{\<abcde\>^4}{\<Yabcd\>\<Yabce\>\<Yabde\>\<Yacde\>\<Ybcde\>}\,.
\end{align}
\paragraph{Projecting Through $Y$.}
We recall that we went from a simplex in $\Rbb\Pbb^n$ to a simplex in $\Rbb^n$ by intersecting with the hyperplane $Y_I$. In dual projective space, the hyperplane at infinity corresponds to a point `at the origin', and the dual notion of intersecting with the hyperplane is now \emph{projecting through $Y$}. Informally, we are looking at our dual simplex `in the direction of $Y$'. Mathematically, we can define such a projection by considering the equivalence class of a point $W$ in dual projective space defined as
\begin{align}
	[W]=\{W+\alpha Y\colon \alpha\in\Rbb\}\,.
\end{align}
For example, in equation \eqref{eq:GRASS_dual-simplex-volume-fixed} we intersected with the hyperplane defined by $Y=(1,0,\ldots,0)$. In dual projective space, the projection through this $Y$ tells us to equate 
\begin{align}
	[W]=\{(\alpha,W^2,\ldots,W^{n+1})\colon\alpha\in\Rbb\} \,,
\end{align}
which means that we can associate the $n$-vector $(W^2,\ldots,W^{n+1})\in\Rbb^n$ to the equivalence class $[W]$. We will encounter the idea of `projecting through $Y$' again in section \ref{sec:POS_amplituhedron}, where we return to it when discussing the amplituhedron in momentum twistor space.

\section{The Grassmannian}\label{sec:GRASS_grass}

Given a vector space $V$, the Grassmannian $G(k,V)$ is defined as the set of $k$-dimensional linear subspaces of $V$. In the cases where $K=\Rbb^n$ or $\Cbb^n$ we talk about the real or complex Grassmannian, respectively. With abuse of notation, we will denote both the real and the complex Grassmannian simply as $G(k,n)$. Phrased differently, the Grassmannian $G(k,n)$ is the space of all $k$-planes passing through the origin of an $n$-dimensional vector space. From this phrasing, it should be clear that the Grassmannian can be considered a generalisation of projective space $\Pbb^{n-1}$, which is defined as the space of \emph{lines} passing through the origin of an $n$-dimensional vector space. This shows the simple isomorphism $G(1,n)\simeq \Pbb^{n-1}$.

Any element of the Grassmannian $G(k,n)$ can be specified by a set of $k$ $n$-vectors that span the plane, which we combine into a $k\times n$ matrix $C$. Clearly, any linear combination of these vectors still spans the same $k$-plane, so the matrix $C$ is defined up to a $GL(k)$ transformation. Hence we can equivalently define the Grassmannian as the set of equivalence classes of $k\times n$ matrices, where $C\sim G\cdot C$ for some $G\in GL(k)$. This shows that the dimensions of $G(k,n)$ is equal to $k\times n - k^2= k(n-k)$. The maximal ($k\times k$) minors of this matrix $C$ are known as the \emph{\Pluck} coordinates $p_I(C)$, where $I\in\binom{[n]}{k}$ is some multiindex of cardinality $k$. In this thesis we will usually follow the physics literature and denote \Pluck coordinates in bracket notation: $p_{i_1,\cdots, i_k}(C) \equiv (i_1\cdots i_k)$. If we want to explicitly specify to which matrix a \Pluck variable refers, we will include it in the subscript of the bracket as $(i_1\cdots i_k)_C$. 

It is important to note that the \Pluck variables are not independent; they are related through the \emph{\Pluck relations}. Instead of considering the matrix $C$ as consisting of $k$ $n$-vectors, we can equivalently consider it as a collection of $n$ $k$-vectors $\bm{c}_1,\ldots,\bm{c}_n$. Since any $k$-vector can be expanded in an arbitrary basis of $k$ linearly independent $k$-vectors, we find the following relations between these $n$ vectors (known as \emph{Cramer's rule} in linear algebra)
\begin{align}\label{eq:GRASS_cramer}
	&\sum_{l=1}^{k+1} (-1)^{l+1} \bm{c}_{i_l} (i_1\cdots \hat{i}_l\cdots i_{k+1}) \notag\\&
	= \bm{c}_{i_1}(i_2 i_3\cdots i_{k+1}) - \bm{c}_{i_2}(i_1 i_3\cdots i_{k+1})+ \cdots + (-1)^k \bm{c}_{i_{k+1}}(i_1 i_2\cdots i_k) = \bm{0}\,,
\end{align}
where the hat indicates omission. We can `contract' this identity by taking an arbitrary set $\bm{c}_{j_1},\ldots,\bm{c}_{j_{k-1}}$ of column vectors of $C$ and taking the determinant with the rewriting of the zero vector in \eqref{eq:GRASS_cramer}. This gives us the \emph{(Grassmann-)\Pluck relations}
\begin{align}
	&\sum_{l=1}^{k+1} (-1)^{l+1} (j_1\cdots j_{k-1}i_l) (i_1\cdots \hat{i}_l\cdots i_{k+1})=\notag\\ & (j_1\cdots j_{k-1}i_1)(i_2 i_3\cdots i_{k+1}) + \cdots + (-1)^k (j_1\cdots j_{k-1}i_{k+1})(i_1 i_2\cdots i_k) = 0\,.
\end{align}
Especially for the case when $k=2$, both Cramer's rule and the \Pluck relation are often referred to as the \emph{Schouten identity} in the physics literature.

Since $p_I(G\cdot C) = \det G \;p_I(C)$, the \Pluck coordinates encode the $SL(k)$ invariant information of $C$, the full $GL(k)$ information is thus encoded in the ratios of \Pluck variables. From this, it is clear that the $\binom{n}{k}$ \Pluck variables are only defined up to a rescaling, and thus provide a map from the Grassmannian into $\smash{\Pbb^{\binom{n}{k}-1}}$. We can then interpret the full Grassmannian $G(k,n)$ as $k(n-k)$-dimensional subvariety of $\smash{\Pbb^{\binom{n}{k}-1}}$ generated by the \Pluck relations. This is known as \emph{\Pluck embedding} of the Grassmannian.

In the patch of the Grassmannian $G(k,n)$ where the first Pl\"ucker $p_{12\cdots k}(C)\neq0$, we can use the $GL(k)$ redundancy to `gauge fix' the first $k\times k$ minor of $C$ to the identity matrix:
\begin{align}\label{eq:GRASS_cmat_gauge_fixed}
	C= \begin{pmatrix}
		\unit_{k\times k} | c
	\end{pmatrix}\,,
\end{align} 
where $c_{k\times (n-k)}$ is some $k\times(n-k)$ matrix which now uniquely specifies all elements of the Grassmannian in this patch.

Every $k$-plane represented by a matrix $C\in G(k,n)$ has an associated \emph{orthogonal complement} $C^\perp\in G(n-k,n)$ such that
\begin{align}
	C\cdot (C^\perp)^T = \nul_{k\times (n-k)}\,.
\end{align}
This provides us with a natural isomorphism between $G(k,n)$ and $G(n-k,n)$. If we take $C$ in the form \eqref{eq:GRASS_cmat_gauge_fixed}, it is easy to see that the orthogonal complement is given by
\begin{align}
	C^\perp = \begin{pmatrix}
		-c^T | \unit_{(n-k)\times (n-k)}
	\end{pmatrix}\,.
\end{align}
The \Pluck variables of $C$ and $C^\perp$ are related by
\begin{align}
	p_I(C) = \pm p_{[n]\setminus I}(C^\perp)\,.
\end{align}
The Grassmannian has deep ties to the mathematical field of combinatorics. This can be seen most clearly through the study of \emph{matroids}, which were introduced in the 1930's to study and generalise the notion of dependence relations among a set of vectors. We follow the definitions and notation of \cite{Williams:2021zph}. Given some finite set $E$, we define a matroid as the pair $(E,M)$, where $M$ is a nonempty set of subsets of $E$ (called bases), which satisfy that for all distinct bases $B_1$, $B_2$ such that $b_1\in B_1\setminus B_2$, then there must exist $b_2\in B_2\setminus B_1$ such that $(B_1\setminus\{b_1\})\cup b_2\in M$. Any element of the Grassmannian $G(k,n)$ gives rise to a matroid $([n],M)$ if we define
\begin{align}
	M\coloneqq \left\{I\in\binom{[n]}{k}\colon\quad p_I(C)\neq 0\right\}\,.
\end{align}
All matroids which can arise from some matrix $C$ in this way are called \emph{realisable}.

It is possible to subdivide the Grassmannian into \emph{matroid strata}. If we take some subset $M\subseteq \binom{[n]}{k}$, then we define the \emph{matroid strata} $S_M$ of $G(k,n)$ as
\begin{align}
	S_M\coloneqq \{C\in G(k,n)\colon \quad
	p_I(C)\neq 0 \text{ for } I\in M,\; p_I(C) = 0 \text{ for } I \not\in M\}\,.
\end{align}
The study of the matroid stratification of the Grassmannian is notoriously hard. In fact, it was proven in \cite{Mnev1988} that a matroid stratum can have a topology which is as bad as \emph{any} algebraic variety. This is a result known as \emph{Mn\"ev's universality theorem}. Luckily for us, our main interest is in a specific subset of the Grassmannian which admits a far more well-behaved stratification. This is the \emph{positive Grassmannian}, which we will introduce in section \ref{sec:GRASS_pos}.

\section{The Geometry of Planes}\label{sec:GRASS_binary}

The definition of the Grassmannian as outlined above inherently deals with planes in some $n$-dimensional vector space. We will take this section to study how these planes intersect, combine, and complement each other. Many of these concepts will be encountered later, and this discussion will streamline the treatment of the (momentum) amplituhedron in chapter \ref{sec:POS}. The discussion presented in this section is largely technical in nature. The most important things to take away from this section are the definitions of the binary operations and their explicit formulae in terms of matrices.

\subsection{Binary Operations}

Take $C_1\in G(k_1,n)$ and $C_2\in G(k_2,n)$. We define the following binary operations:
\begin{itemize}
	\item The \emph{union} $C_1 \cup C_2$ is the span of the union of all vectors in $C_1$ and $C_2$. For generic $C_1,\, C_2$, if $k_1+k_2<n$ then $C_1 \cup C_2\in G(k_1+k_2,n)$, and $C_1 \cup C_2=\Rbb^n$ if $k_1+k_2\ge n$. In matrix form this can be represented as \begin{align}\arraycolsep=1.4pt\def\arraystretch{1.2}
		C_1 \cup C_2 = \left(\begin{array}{c c} C_1 \\ \hdashline[2pt/2pt]\ C_2 \end{array} \right).
	\end{align}
	\item The \emph{intersection} $C_1 \cap C_2$ is the maximal common subplane of $C_1$ and $C_2$. For generic $C_1,\, C_2$, $C_1 \cap C_2=\{0\}$ if $k_1+k_2<n$, and $C_1 \cap C_2\in G(k_1+k_2-n,n)$ if $k_1+k_2>n$
	\item The \emph{complement} $C_1\setminus C_2$ is the maximal subplane of $C_1$ that is orthogonal to $C_2$. For generic $C_1, \, C_2$, assuming $k_1>k_2$, then $C_1\setminus C_2\in G(k_1-k_2,n)$.
\end{itemize}
In what follows, we will assume $C_1,\,C_2$ generic. 

The interpretation of the matrix multiplication $C_1\cdot C_2^T\in G(k_1,C_2)$ is that of the projection of $C_1$ into $C_2$. The resulting $k_1\times k_2$ matrix represents a $k_1$-plane in the $k_2$-dimensional vector space $C_2$. If we use a $GL(k_2)$ transformation to make the row-vectors of $C_2$ orthogonal, then we can embed this $k_1$ plane into $\Rbb^n$ as $C_1\cdot C_2^T\cdot C_2$. The resulting element of $G(k_1,n)$ can be understood geometrically as first extending $C_1$ in all directions orthogonal to $C_2$ and seeing where the resulting $(n+k_1-k_2)$-plane intersects $C_2$:
\begin{align}
	C_1\cdot C_2^T\cdot C_2 = \big(C_1\cup C_2^\perp\big)\cap C_2 \in G(k_1,n)\,.
\end{align}

\begin{figure}
	\centering
	\includegraphics[scale=0.6]{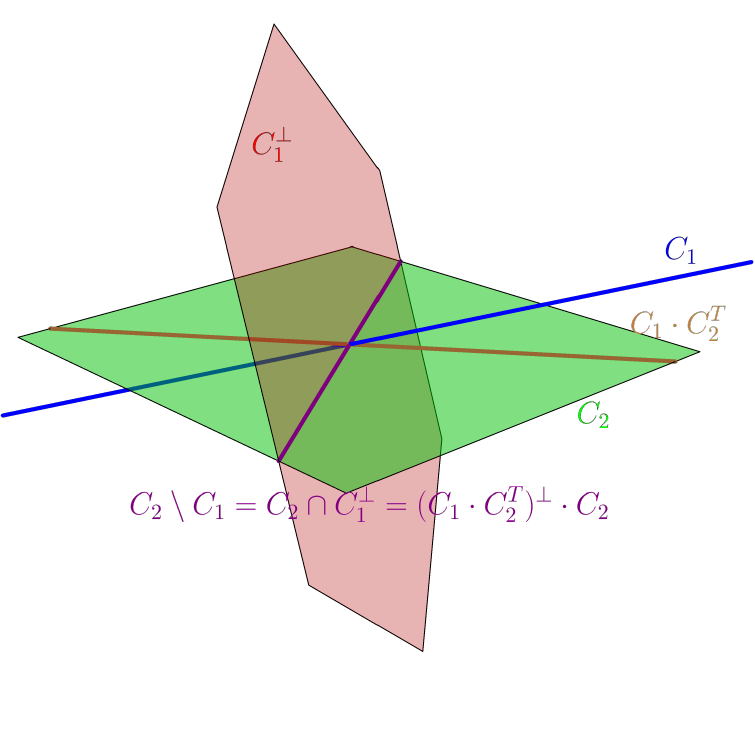}
	\caption{This example shows some $C_1\in G(1,3)$ in blue, $C_2\in G(2,3)$ in green, $C_1^\perp\in G(2,3)$ in red, $C_1\cdot C_2^T\in G(1,C_2)$ in brown, and $C_2\setminus C_1\in G(1,3)$ in purple. We see that $C_2\setminus C_1$ can be found both as the intersection $C_2\cap C_1^\perp$, as well as the orthogonal complement of $C_1\cdot C_2^T$ in $C_2$ embedded in three-dimensional space.}
	\label{fig:grass-intersect}
\end{figure}

Starting from $C_1\cdot C_2^T\in G(k_1,k_2)$ we can look at the orthogonal complement of $C_1$ in $C_2$, $\big(C_1\cdot C_2^T\big)^\perp\in G(k_2-k_1,C_2)$, which we can subsequently embed in $\Rbb^n$ as $\big(C_1\cdot C_2^T\big)^\perp\cdot C_2\in G(k_2-k_1,n)$. The resulting $(k_2-k_1)$-plane is contained inside $C_2$ and is orthogonal to $C_1$. This is exactly the complement $C_2\setminus C_1$:
\begin{align}
	C_2\setminus C_1 = \big(C_1\cdot C_2^T\big)^\perp\cdot C_2\in G(k_2-k_1,n)\,.
\end{align}
Interestingly, the above formula holds even when we don't take the columns of $C_2$ to be orthogonal. Using the identities
\begin{align}
	C_1\cap C_2 = C_2\setminus C_1^\perp =C_1\setminus C_2^\perp,\quad C_1 \cup C_2 = \big(C_1^\perp \cap C_2^\perp\big)^\perp\,,
\end{align}
we arrive at the following equations for the binary operations introduced above:
\begin{subequations}\label{eq:binary-matrices}
	\begin{align}
		C_1\setminus C_2 &= \big(C_2\cdot C_1^T\big)^\perp\cdot C_1 \in G(k_1-k_2,n)\,,\\
		C_1 \cap C_2 &= \big(C_2^\perp\cdot C_1^T\big)^\perp\cdot C_1 = \big(C_1^\perp\cdot C_2^T\big)^\perp\cdot C_2\in G(k_1+k_2-n,n)\,,\\
		C_1 \cup C_2 &= \big((C_1\cdot (C_2^\perp)^T)^\perp\cdot C_2^\perp\big)^\perp = \big((C_2\cdot (C_1^\perp)^T)^\perp\cdot C_1^\perp\big)^\perp\in G(k_1+k_2,n)\,. 
	\end{align}
\end{subequations}
We can use this to define an orthogonality-invariant version of the projection of $C_1$ onto $C_2$:
\begin{align}
	\mcl{P}(C_1\to C_2) = \big(C_1\cup C_2^\perp\big)\cap C_2 =\big((C_1\cdot C_2^T)^\perp\cdot C_2\cdot C_2^T\big)^\perp\cdot C_2\in G(k_1,n)\,,
\end{align}
which reduces to the simple $C_1\cdot C_2^T\cdot C_2$ when the columns of $C_2$ are orthogonal. To illustrate some of the definitions in this section, we have depicted a simple three-dimensional example in figure \ref{fig:grass-intersect}.

We note here several identities between the binary operations, starting with the ones we have already introduced earlier:
\begin{subequations}\label{eq:binary-relations}
	\begin{align}
		C_1\cap C_2 &= C_2\setminus C_1^\perp =C_1\setminus C_2^\perp\,,\\C_1 \cup C_2 &= \big(C_1^\perp \cap C_2^\perp\big)^\perp\,,\\
		\mcl{P}(C_1\to C_2) &= \big(C_1\cup C_2^\perp\big)\cap C_2 = C_2\setminus (C_2\setminus C_1)\,,\\
		C_1\setminus(C_2\cup C_3)&=\big(C_1 \setminus C_2\big)\setminus C_3 = \big(C_1 \setminus C_3\big)\setminus C_2\,.
	\end{align}
\end{subequations}
We include these identities for the sake of future reference.

\subsection{Volume Forms}

To each $k$-plane in $C\in G(k,n)$ we can associate a volume form of degree $k$. If $C$ has elements $c_{\alpha i}$, $\alpha=1,\ldots,k,\,i=1,\ldots,n$ this volume form is
\begin{align}
	\Omega(C)=\bigwedge_{\alpha=1}^k \sum_{i=1}^n c_{\alpha i}\dd x_i=\sum_{I \in \binom{[n]}{k}} p_I(C) \dd \bm{x}_I\,,
\end{align}
where we define $\dd \bm{x}_{\{i_1,\ldots,i_k\}}=\dd x_{i_1}\wedge\cdots\wedge\dd x_{i_k}$ (assuming $i_1<\ldots<i_k$). It is generally true that 
\begin{align}
	\Omega(C^\perp)=\star \Omega(C),
\end{align}
where $\star$ indicates the Hodge dual which acts on $\dd \bm{x}_I$ as
\begin{align}
	\star \;\dd x_{i_1}\wedge\cdots\wedge\dd x_{i_k} = \frac{1}{(n-k)!} \epsilon_{i_1\ldots i_k j_1 \ldots j_{n-k}} \dd x_{j_1}\wedge\cdots\wedge \dd x_{j_{n-k}}\,,
\end{align}
which we can also write in multiindex notation as
\begin{align}
	\star \;\dd\bm{x}_I = \text{sgn}(I\cup \bar{I})\dd\bm{x}_{\bar{I}}\,.
\end{align}
For generic $C_1\in G(k_1,n),\, C_2\in G(k_2,n)$ with $k_1+k_2<n$ we have the relation
\begin{align}
	\Omega(C_1 \cup C_2) = \Omega(C_1)\wedge \Omega(C_2)\,.
\end{align}
Together with the relations between binary operations, this can be used to derive relations between the Pl\"ucker variables. For instance, we find
\begin{subequations}\label{eq:GRASS_vol_form_plucker_relations}
	\begin{align}
		p_I(C_1\setminus C_2) &= \sum_{J \in \binom{[n]}{k_2}} p_{J}(C_2) p_{I \cup J} (C_1),\qquad &&I \in \binom{[n]}{k_1-k_2}\,,\\
		%	p_I(C_1 \cup C_2) &= \sum_{J\in\binom{I}{k_1}}\text{sgn}(J\cup \big(I\setminus J)\big) p_I(C_1) p_{I\setminus J}(C_2)&&I \in \binom{[n]}{k_1+k_2}\,,\\
		p_I(C_1 \cup C_2) &= \text{sgn}(I\cup\bar{I}) p_{\bar{I}}(C_1^\perp \setminus C_2)\,,&& I\in\binom{[n]}{k_1+k_2}\,.
	\end{align}
\end{subequations}
A particular example which will become useful later is the case where $Z\in G(K+4,n)$, $\hat{C}\in G(K,n)$, and we consider $z=Z\setminus \hat{C}\in G(4,n)$. The \Pluck variables of $z$ can then be written as
\begin{align}
	p_{i_1i_2i_3i_4}(z) = \sum_{1\leq j_1<\cdots<j_K\leq n} p_{j_1\cdots j_K}(\hat{C}) p_{i_1i_2i_3i_4j_1\cdots j_K}(Z)\,,
\end{align}
or in physicist's notation
\begin{align}\label{eq:GRASS_mom-twistor-bracket}
	\<i_1i_2i_3i_4\>_z=\sum_{1\leq j_1<\cdots<j_K\leq n} (j_1\cdots j_K)_{\hat{C}} \<i_1i_2i_3i_4j_1\cdots j_K\>_Z\,.
\end{align}

\section{The Positive Grassmannian}\label{sec:GRASS_pos}

We define the \emph{totally nonnegative Grassmannian} $G_{\geq 0}(k,n)$ as the region of the Grassmannian where all \Pluck variables are nonnegative\footnote{Recall that our convention is for the multiindex $I$ to be ordered: for $I=\{i_1,\ldots,i_k\},\, i_1<\cdots<i_k$, then $p_I(C)=p_{i_1\cdots i_k}(C)$.}:
\begin{align}
	G_{\geq 0}(k,n)\coloneqq\{C\in G(k,n)\colon p_I(C)\geq 0 \quad \forall I \in \binom{[n]}{k}\}\,.
\end{align}
We similarly define the \emph{positive Grassmannian} $G_+(k,n)$ to be the part of Grassmannian where all \Pluck variables are strictly positive. As is customary in the physics literature, we will abuse notation and refer to both the totally nonnegative as the positive Grassmannian simply as the `positive Grassmannian', denoted $G_+(k,n)$. The matroid strata of the positive Grassmannian are called \emph{positroid cells}:
\begin{align}
	S_M^+\coloneqq \{C\in G_+(k,n)\colon \quad
	p_I(C)> 0 \text{ for } I\in M,\; p_I(C) = 0 \text{ for } I \not\in M\}\,.
\end{align}
The positroid stratification of the positive Grassmannian is much simpler than the matroid stratification of the Grassmannian, and it was studied in detail in \cite{Postnikov:2006kva}. In this paper, it was proven that each positroid cell is homeomorphic to an open ball, and that the decomposition of $G_+(k,n)$ into the union of positroids is a CW-complex. It was subsequently proven in \cite{Galashin:2017onl} that $G_+(k,n)$ is homeomorphic to a closed ball.

Our definition of the positive Grassmannian is manifestly rooted in a notion of ordering on the columns of the $C$ matrix. We note, however, that this notion naturally leads to a \emph{cyclic} structure. Explicitly, if we have $C\in G_+(k,n)$ with ordered columns $C=\begin{pmatrix}
	\bm{c}_1 & \bm{c}_2 & \cdots & \bm{c}_n
\end{pmatrix}$, then the shift
\begin{align}
	\bm{c}_1\to \bm{c}_2, \bm{c}_2\to \bm{c}_3,\ldots, \bm{c}_n\to (-1)^{k-1} \bm{c}_1\,,
\end{align}
yields another positive matrix. The presence of the $(-1)^k$ when cycling $\bm{c}_n \to \bm{c}_1$ is crucial. This slight modification of a cyclic shift is known as \emph{twisted cyclic symmetry}.

The top-dimensional positroid cell, called the \emph{top-cell}, is the positive Grassmannian as a whole. Its codimension-1 boundaries are given by positroid cell which have a single cyclic \Pluck variable set to zero, $p_{i,i+1,\ldots,i+k-1}(C)=0$. From this point on, we can find lower-dimensional positroid cells by continually setting additional \Pluck variables to zero. Given some positroid cell $S$, the set of all positroid cells which can be obtained from $S$ by setting additional \Pluck variables to zero is called the \emph{positroid stratification} of $S$. The set of all positroid cells which have $S$ in their positroid stratification is called the \emph{inverse positroid stratification} of $S$. It should be noted that, due to the \Pluck relations and positivity, there are restrictions to which \Pluck variables can be set to zero to find the codimension-1 boundaries of some positroid cell $S$. Setting some \Pluck variables to zero might necessarily set some other \Pluck variables to zero, which could mean that these \Pluck variables can only be set to zero for some higher codimension positroid cells. It us usually not practical to study positroid cells purely by keeping track of the vanishing \Pluck variables. Instead, there are many \emph{combinatorial} labels which completely characterise the positroid cell, which are usually more practical for finding their dimension, their positroid stratification, or a \emph{positive parametrisation} of the cell. This was originally studied in detail by Postnikov in \cite{Postnikov:2006kva}, and we will use the remainder of this section to introduce some of these ideas.

\subsection{Combinatorics of the Positive Grassmannian}\label{sec:GRAS_positroid-combinatorics}

Positroid cells are known to be in bijection to several combinatorial objects. We will define some of them here, and refer to \cite{Postnikov:2006kva} for proofs of the bijections. There are several more combinatorial objects we could add to this list, such as Grassmann necklaces and Le diagrams. However, we will not encounter them in the remainder of this thesis, and for this reason we omit them from our discussion.

\subsubsection{Decorated Permutations}

Firstly, a particularly convenient way to characterise positroid cells is through \emph{decorated permutations}. A decorated permutation is an injective map $\sigma\colon [n]\to [2n]$ which satisfies $i\leq \sigma(i)\leq i+n$, which we typically label as $\sigma=\{\sigma(1),\ldots,\sigma(n)\}$. To relate this to a regular permutation we simple reduce $\{\sigma(1),\ldots,\sigma(n)\} \mod n$. This process is always invertible, except for the fixed points where $\sigma(i)=i$, or $\sigma(i)=i+n$, which are sometimes referred to as \emph{black} and \emph{white} fixed points, respectively (or \emph{loops} and \emph{coloops}, respectively). Hence, these decorated permutations can equivalently be interpreted as regular permutations on $[n]$ where the fixed points can be `decorated' either black or white. To relate these decorated permutations to an element $C\in G_+(k,n)$, we let $\bm{c}_a$ denote the $a$\textsuperscript{th} column vector of $C$, and for each $a\in[n]$ we let $\sigma(a)$ be the \emph{first} number such that $\bm{c}_a\in\text{span}\{\bm{c}_{a+1},\bm{c}_{a+2},\ldots,\bm{c}_{\sigma(a)}\}$. The indices on the column vectors are defined cyclically ($\bm{c}_{a+n}\equiv \bm{c}_a$), and we define $\sigma(a)=a$ in the case where $\bm{c}_a=\bm{0}$. This procedure allows us to associate a decorated permutation to each element of the positive Grassmannian.

We define an \emph{anti-exceedance} of a decorated permutation $\sigma$ as a number $a$ which is mapped `beyond' $n$: $\sigma(a)>n$. The number $k$ of anti-exceedances of a decorated permutation is called the \emph{helicity} of the decorated permutation, and is equivalently given by the mean
\begin{align}
	k=\frac{1}{n}\sum_{i=1}^n(\sigma(i)-i)\,.
\end{align}
A decorated permutation on $[n]$ with helicity $k$ is said to be of \emph{type} $(k,n)$. It was shown in \cite{Postnikov:2006kva} that decorated permutations of type $(k,n)$ are in bijection to the positroid cells of $G_+(k,n)$. As an example, the decorated permutation
\begin{align}
	\sigma = \{8, 2, 11, 9, 12, 6, 13, 15\}\,,
\end{align}
is of type $(5,8)$. 

In practice, these decorated permutations provide a convenient way to uniquely and concisely specify a positroid cell. For this reason, we often use the permutation to index a positroid cell. That is, we use the notation $S_\sigma$ to denote the positroid cell corresponding to the decorated permutation $\sigma$, and we use $C_\sigma$ to denote the matrix representative of an element of $S_\sigma$. 

\begin{figure}
	\centering
	\includegraphics[scale=0.3]{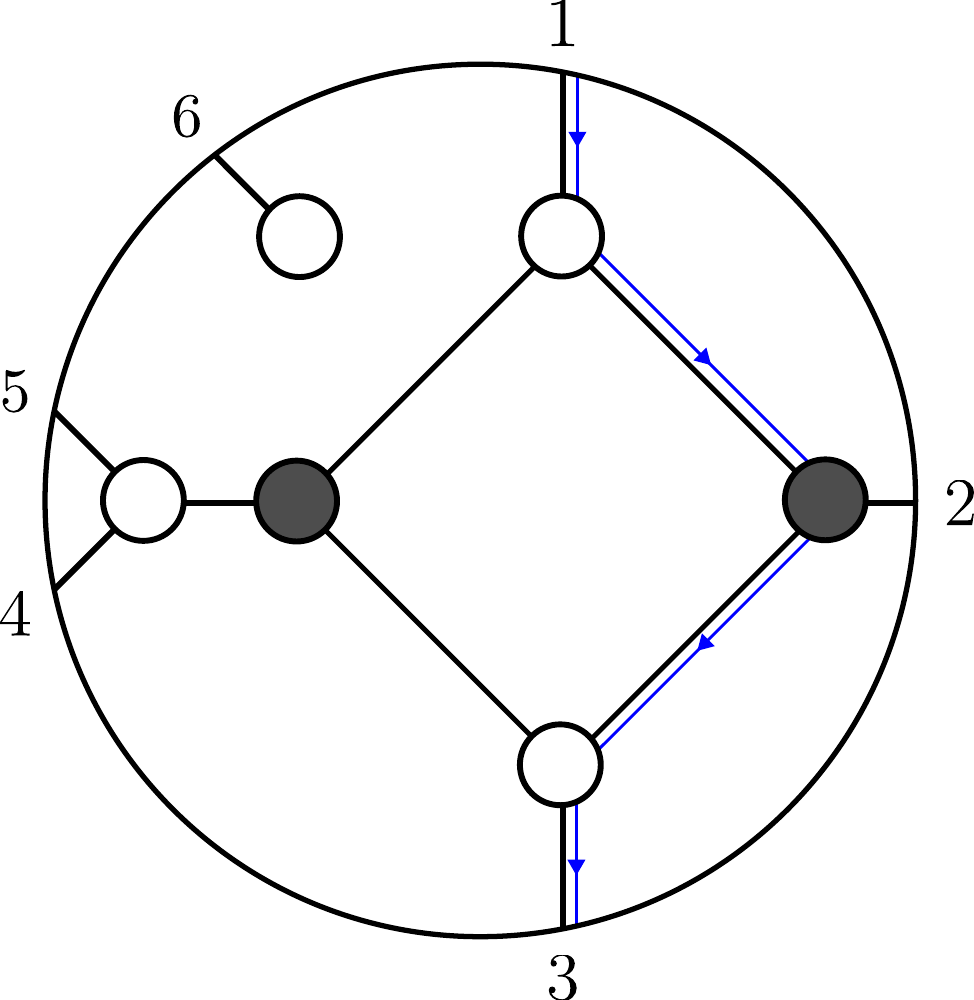}
	\caption{An example of a plabic graph in $G_+(3,6)$. The blue path shows that the rules of the road tell us that $\sigma(1)=3$. The full decorated permutation associated to this plabic graph is $\{3,4,7,5,8,12\}$.}
	\label{fig:plabic_graph}
\end{figure}

\subsubsection{Plabic Graphs}

Another way to label positroid cells is through \emph{plabic graphs} (\emph{pla}nar \emph{bic}oloured graphs). These are planar graphs with $n$ external legs whose internal vertices can be coloured either black or white. The external legs are labelled $1,\ldots,n$ clockwise. Plabic graphs are often embedded in a closed disk. We have depicted an example of a plabic graph in figure \ref{fig:plabic_graph}.

From a plabic graph we can read off the decorated permutation associated to the corresponding positroid cells by following `the rules of the road':
\begin{enumerate}
	\item Start at external edge $i$, and follow it into the graph,
	\item When reaching a white vertex, take the first exit clockwise (\emph{i.e.} turn left), and when reaching a black vertex, take the first exit counter-clockwise (\emph{i.e.} turn right),
	\item We continue the previous step until we exit the graph via external edge $j$. This then means that $\sigma(i)=j$.
\end{enumerate}
In addition, the special white and black fixed points of a decorated permutation correspond to white and black \emph{lollipops}, respectively, which are isolated subgraphs consisting of an external edge connected to a single internal vertex. An example route dictated by the rules of the road has been depicted in the plabic graph \ref{fig:plabic_graph}.

\begin{figure}
	\centering
	\includegraphics[width=0.8\textwidth]{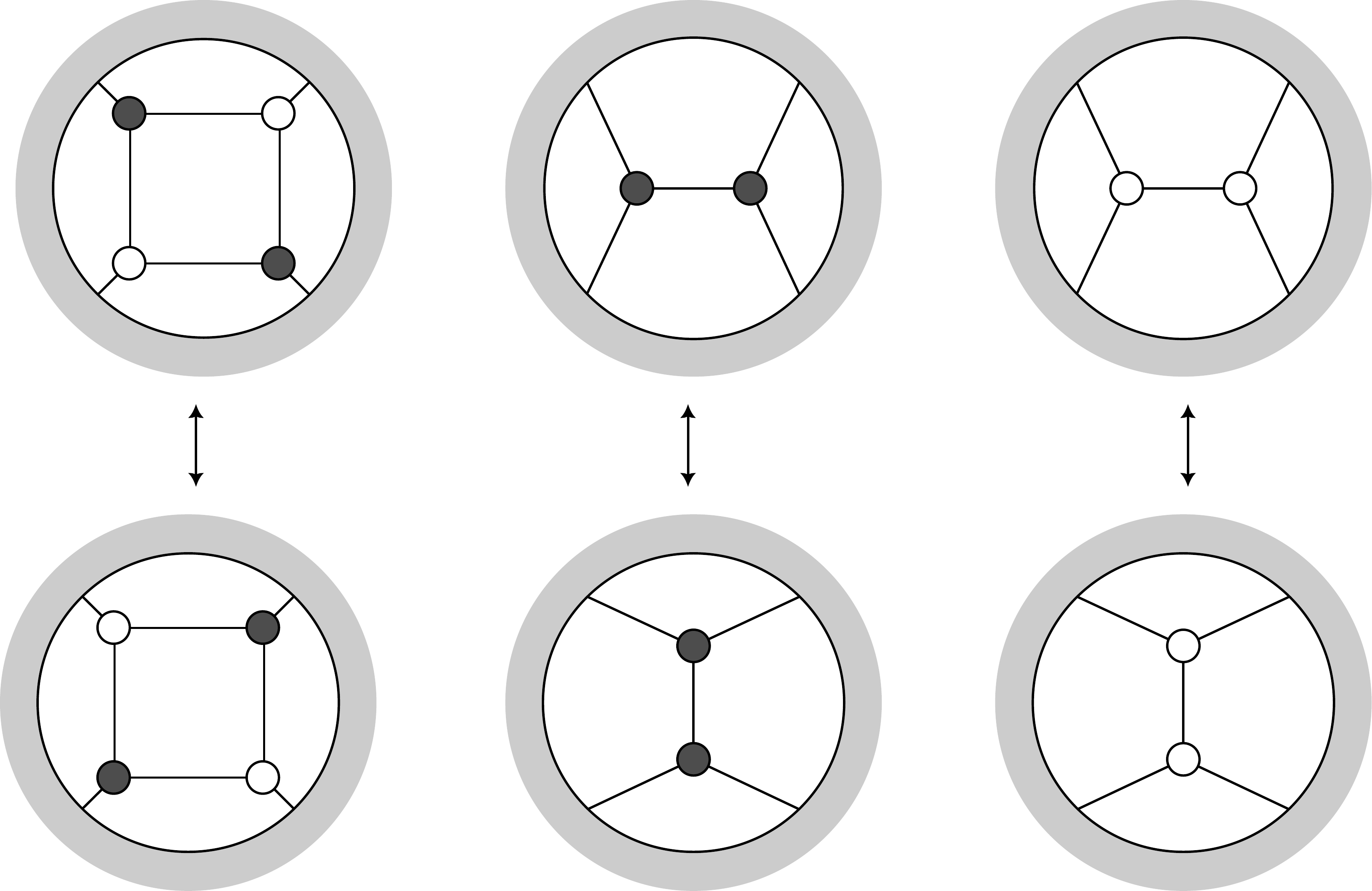}
	\caption{The moves one can perform on a plabic graph to leave the associated decorated permutation invariant consist of the square move (left), and the merge-expand move (centre and right). The grey outline is used to indicate that these diagrams can be part of some larger plabic graph}
	\label{fig:plabic_moves}
\end{figure}

There are several `moves' one can perform to a plabic graph. These are the so-called \emph{square move} and \emph{merge-expand move}, depicted in figure \ref{fig:plabic_moves}.
It is easy to see that these moves leave the associated decorated permutation invariant. Hence, two plabic graphs which are related by a sequence of these moves label the same positroid cell. To each plabic graph we associate an integer number called the \emph{helicity}, which is equal to $n_B + 2n_W - n_E$, where $n_B$ is the number of black vertices, $n_W$ the number of white vertices, and $n_E$ the number of internal edges. A plabic graph with $n$ external edges and a helicity of $k$ is said to be of \emph{type} $(k,n)$.
Equivalence classes of plabic graphs of type $(k,n)$ under the moves defined above are in bijection to positroid cells of $G_+(k,n)$\footnote{Technically, we should restrict ourselves to equivalence classes of \emph{reduced} plabic graphs. That is, we only consider plabic graphs which do not have any isolated connected components and have no internal vertices of degree 2. We should further remove any `bubbles', which can occur when a white and a black vertex share two edges between them, in which case we remove one of the edges.}.

\subsubsection{Grassmannian Graphs}

\begin{figure}
	\centering
	\includegraphics[scale=0.4]{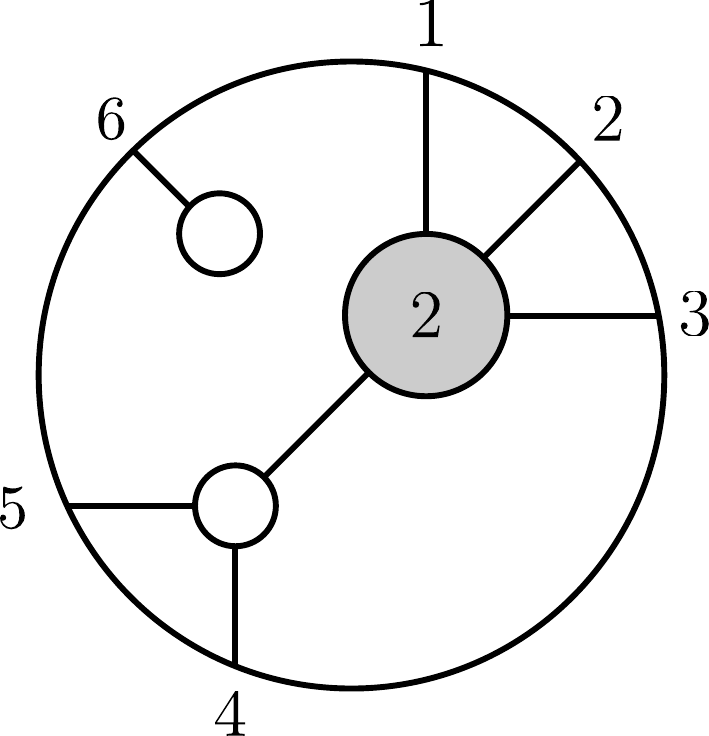}
	\caption{An example of a Grassmannian graph. The numbers inside the vertices indicate their helicity, with a white vertex having helicity 1. The associated positroid cell is equivalent to the one in figure \ref{fig:plabic_graph}.}
	\label{fig:grassmann-graph}
\end{figure}

Grassmannian graphs \cite{Postnikov:2018jfq} are similar in nature to plabic graphs. They are also planar graphs embedded in a disk with some decoration on the internal vertices. The decoration we allow on a vertex $v$ is now any integer number between one and $\deg(v)-1$, called the \emph{helicity}. These graphs can similarly be related to decorated permutations by following a similar set of rules for the road: the difference is that at a vertex with helicity $h$, we now take the $h$\textsuperscript{th} exit counter-clockwise. The total helicity of the Grassmannian graph is defined to be sum of all helicities of internal vertices, minus the number of internal edges. As an example, we depict a Grassmannian graph and its associated permutation in figure \ref{fig:grassmann-graph}.

We note that a Grassmannian graphs where all vertices have helicity $1$ or $n-1$ have the same rules of the road as plabic graphs with white and black vertices, respectively. For this reason, we often refer to vertices with helicity $1$ as white, and $n-1$ as black, and colour them correspondingly in a Grassmannian graph.

\subsection{Positive Parametrisation}

We have seen that positroid cells are in bijection to several combinatorial objects. We should not forget, however, that first and foremost positroid cells are subsets of the Grassmannian. Elements of a positroid cell are thus $k$-planes in $n$ dimensions, which can be represented by $k\times n$ matrices. 

It is possible to find a \emph{positive parametrisation} of a $k\times n$ matrix for any positroid cell, such that the matrix is inside the positroid if and only if all parameters are positive numbers. If we let $\sigma$ be the decorated permutation of some positroid cell $S_\sigma$, and $\alpha_1,\ldots,\alpha_d$ the parameters, then we denote the positive parametrisation as $C_\sigma(\alpha_1,\ldots,\alpha_d)\equiv C_\sigma(\bm{\alpha})$. We are particularly interested in the case where the number of positive parameters is equal to the dimension of the positroid cell. In this case, the positive parametrisation provides a bijection between the (interior of) the positroid cell and $(\Rbb^+)^d$. As an example, the top cell of $G_+(2,4)$ (which has a decorated permutation $\{3,4,5,6\}$) has a positive parametrisation
\begin{align}
	C_{\{3,4,5,6\}}(\bm{\alpha})=\begin{pmatrix}
		1 & \alpha_1 & 0 & -\alpha_2 \\ 0 & \alpha_3 & 1 & \alpha_4
	\end{pmatrix}\,.
\end{align}
It is easy to check that all maximal minors of this matrix are positive (and hence is an element of the top cell of $G_+(2,4)$) if and only if $\alpha_1>0,\alpha_2>0,\alpha_3>0,\alpha_4>0$.

It should be noted that a given positive parametrisations gives points in the interior of a positroid cell, but not necessarily on its boundary component. That is, we only parametrise some chart of the positroid cell. This is analogous to homogeneous coordinates of projective space where we have fixed one of the coordinates to 1. In the example above, we see that a positive parametrisation of the codimension-1 boundaries of $\{3,4,5,6\}$ can be obtained by setting one of the parameters to zero. However, in the current parametrisation we can never obtain any positroid cell which has $(13)=0$. To see these boundaries, we need to perform an appropriate $GL(2)$ transformation to this matrix.

There are algorithms which give explicit realisations of a positive parametrisations of positroid cells, starting from any decorated permutation, plabic graph, or Grassmannian graph of choice \cite{Postnikov:2006kva, Postnikov:2018jfq, Arkani-Hamed:2012zlh}. This has been implemented in the \texttt{Mathematica} package \texttt{positroids} \cite{Bourjaily:2012gy}, which can be used to efficiently find explicit positive parametrisations of a positroid cell of choice.

\section[The Orthogonal Grassmannian]{The (Positive) Orthogonal Grassmannian}\label{sec:GRASS_orth}

In addition to the positive Grassmannian, there is another subset of the Grassmannian that will be of importance. The definition of the \emph{orthogonal Grassmannian} requires a non-degenerate symmetric bilinear form $\<\cdot,\cdot\>\colon \Cbb^n\times \Cbb^n\to\Cbb$. The orthogonal Grassmannian $OG(k,n)$ is defined to be the set of $k$-dimensional subspaces of $\Cbb^n$ which are isotropic (orthogonal) with respect to this bilinear form:
\begin{align}\label{eq:GRASS_orth-grass-def}
	OG(k,n)\coloneqq \{C\in G(k,n)\colon \<C,C\>=\nul_{k\times k}\}\,,
\end{align}
where $\<C,C\>$ is a $k\times k$ matrix whose entries are $[\<C,C\>]_{a,b}=\<C_a,C_b\>$ for row-vectors $C_a$ and $C_b$. Since, by definition of the bilinear form, $\<C,C\>$ is a symmetric matrix, this orthogonality constraint imposes $k(k+1)/2$ constraints, and hence
\begin{align}\label{eq:GRASS_OG-dim-gen}
	\dim OG(k,n) = k(n-k)- k(k+1)/2 = \frac{k(2n-3k-1)}{2}\,.
\end{align}
If $\Cbb^n$ has basis vectors $\{\bm{e}_i\}_{i=1}^n$, then we record the bilinear form in a symmetric $n\times n$ matrix $\eta$
\begin{align}
	\eta_{ij}\coloneqq \<\bm{e}_i,\bm{e}_j\>\,,
\end{align}
such that
\begin{align}
	\<C,D\>=C\cdot \eta\cdot D^T\,.
\end{align}
We can always choose a basis of $\Cbb$ such that $\eta$ is diagonal with entries $\pm 1$. Our main interest will be in the special case when $n=2k$, to which we will restrict ourselves from now on.

It is known that $OG(k,2k)$ is isomorphic to the coset $O(2k)/U(k)$. This shows that $OG(k,2k)$ has two branches, since $O(2k)$ splits up into $SO_+(2k)$ and $SO_-(2k)$. These branches can be understood explicitly by inspecting the consequences of orthogonality on the \Pluck variables. Orthogonality implies
\begin{align}
	\<C,C\> = \nul \implies 0 = \sum_{i=1}^{2k} \eta_{ii} C_{ai} C_{bi} = \sum_{i\in I} \eta_{ii} C_{ai} C_{bi} + \sum_{j\in \bar{I}} \eta_{jj} C_{aj} C_{bj}\,,
\end{align}
for some $I\in\binom{[2k]}{k}$, $\bar{I}=[2k]\setminus I$. Taking the determinant of this equation implies
\begin{align}
	|\eta|_{I}^I \;p_I(C)^2 = (-1)^{k} |\eta|_{\bar{I}}^{\bar{I}}\; p_{\bar{I}}(C)^2\,,
\end{align}
where $|\eta|_{I}^I$ denotes the minor of $\eta$ with rows and columns indexed by $I$. Since we have assumed $\eta$ to be diagonal with entries $\pm1$, the minor $|\eta|_{I}^I$ is just $-1$ to the power of the number of minus signs in $\eta$ with indices in $I$, and hence $|\eta|_{I}^I/|\eta|_{\bar{I}}^{\bar{I}} = |\eta|_{I}^I \;|\eta|_{\bar{I}}^{\bar{I}} $ just counts the total number of minus signs and equals $\det\eta$. Taking the square root of the equation above, we find
\begin{align}
	p_I(C) = \pm \sqrt{(-1)^{k} \det\eta} \; p_{\bar{I}}(C)\,.
\end{align}
The relative sign distinguishes the two branches.

When considering the \emph{positive} part of the orthogonal Grassmannian, we need the \Pluck variables to be real valued, which highlights the importance of taking $\eta$ to have exactly $k$ minuses, such that
\begin{align}
	p_I(C) = \pm p_{\bar{I}}(C)\,.
\end{align}
We will follow the conventions of \cite{Huang:2014xza} and take
\begin{align}
	\eta = \diag (1,-1,1,\ldots,-1)\,.
\end{align}
We will occasionally abuse terminology and use the term `orthogonal Grassmannian' to only refer to the \emph{positive branch}, where $p_I(C)=p_{\bar{I}}(C)$, which we denote $OG(k)$:
\begin{align}
	OG(k)\coloneqq \{C \in G(k,2k)\colon p_I(C)=p_{[n]\setminus I}(C)\quad\forall I \in \binom{[2k]}{k}\}\,.
\end{align}
It follows from \eqref{eq:GRASS_OG-dim-gen} that
\begin{align}\label{eq:GRASS_OG-dim}
	\dim OG(k) = \frac{k(k-1)}{2}\,.
\end{align}
We are now ready to define the \emph{totally nonnegative orthogonal Grassmannian} as
\begin{align}
	OG_{\geq 0}(k)\coloneqq \{C\in OG(k)\colon p_I(C)\geq 0\quad \forall I \in \binom{[2k]}{k}\}\,,
\end{align}
and similarly for the \emph{positive orthogonal Grassmannian}. As was the case for the standard positive Grassmannian, we will abuse terminology and refer to both the nonnegative as the positive orthogonal Grassmannian as the `positive orthogonal Grassmannian', denoted $OG_+(k)$. We could equivalently have defined $OG_+(k)$ as the subset of $G_+(k,2k)$ that satisfies $\<C,C\>=0$.

\paragraph{Examples.}
Let us look at a few examples. First, in the simple case for $k=1$ we start from some $1\times 2$ matrix $C$. Using $GL(1)$ to fix the first entry to 1, $C=\begin{pmatrix}	1 & c \end{pmatrix}$, orthogonality takes the form $1-c^2=0$, which implies $c=\pm 1$. This gives us two options for $C$, corresponding to the positive and the negative branch:
\begin{align}
	C=\begin{pmatrix} 1 & 1 \end{pmatrix},\quad \text{ or } C=\begin{pmatrix} 1 & -1 \end{pmatrix}\,.
\end{align}
Since all the `minors' of the first matrix are positive, it is an element (in fact, the only element) of $OG_+(1)$. There are no free parameters, hence $OG(1)$ is zero-dimensional, in agreement with \eqref{eq:GRASS_OG-dim}. 

The first non-trivial example is for $k=2$. Elements from $OG(2)$ are $2\times 4$ matrices. We use $GL(2)$ redundancy to fix the minor $(13)$ to the identity matrix:
\begin{align}
	C = \begin{pmatrix}
		1 & \alpha & 0 &\beta\\ 0 & \gamma & 1 & \delta
	\end{pmatrix}\,.
\end{align}
Orthogonality now reads
\begin{align}
	c\cdot c^T = \unit,\quad c= \begin{pmatrix}
		\alpha & \beta \\ \gamma & \delta
	\end{pmatrix}\,,
\end{align}
which is equivalent to the constraint that $c\in O(2)$. The positive branch,
\begin{align}
	C=\begin{pmatrix}
		1 & \sin\theta & 0 & -\cos\theta\\ 0 & \cos\theta & 1 & \sin\theta
	\end{pmatrix}\,,
\end{align}
is part of $OG_+(2)$ for values of $\theta$ where $\sin\theta$ and $\cos\theta$ are positive. 

\subsection{Orthitroid Cells and Combinatorics}\label{sec:GRASS_orthitroid}

Many of the properties of the positive Grassmannian described in section \ref{sec:GRASS_pos} can be applied to $OG_+(k)$ as well, albeit with some slight modifications. The analogue of positroid cells are \emph{orthitroid cells}:
\begin{align}
	O_M^+\coloneqq \{C\in OG_+(k)\colon \quad
	p_I(C)> 0 \text{ for } I\in M,\; p_I(C) = 0 \text{ for } I \not\in M\}\,.
\end{align}
These orthitroid cells are in bijection to several combinatorial objects which are similar to those introduced in section \ref{sec:GRAS_positroid-combinatorics} for positroid cells \cite{Kim:2014hva}. 

The analogue of decorated permutations are a subset of permutations on $[2k]$ which can be written as the product of $k$ disjoint transpositions. For example, $\{\{1,3\},\{2,4\}\}$ labels the permutation $\sigma(1)=3, \sigma(2)=4, \sigma(3)=1, \sigma(4)=2$, and it corresponds to the top-cell $OG_+(2)$.

\begin{figure}[t]
	\centering
	\includegraphics[width=40mm]{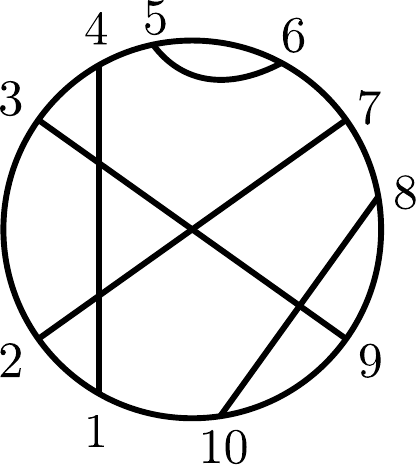}
	\caption{An example of a crossing diagram. The permutation associated to this diagram is $\{\{1,4\}, \{2,7\}, \{3,9\}, \{5,6\}, \{8,10\}\}$.}
	\label{fig:crossing-diagram}
\end{figure}

\begin{figure}[t]
	\centering
	\begin{align*}
		\vcenter{\hbox{\includegraphics[width=40mm]{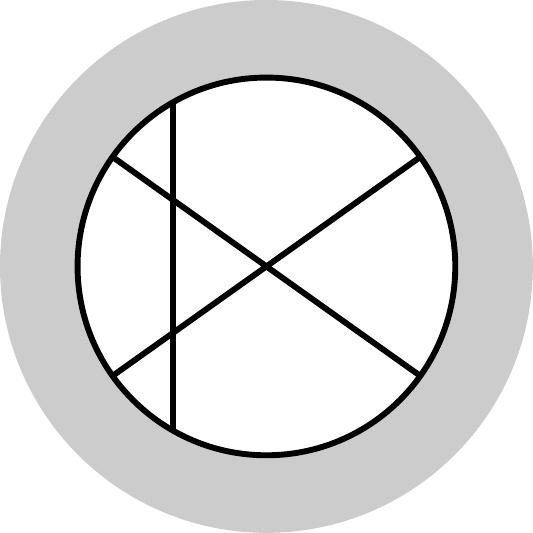}}}\quad\text{\Huge{$\longleftrightarrow$}} \quad \vcenter{\hbox{\includegraphics[width=40mm]{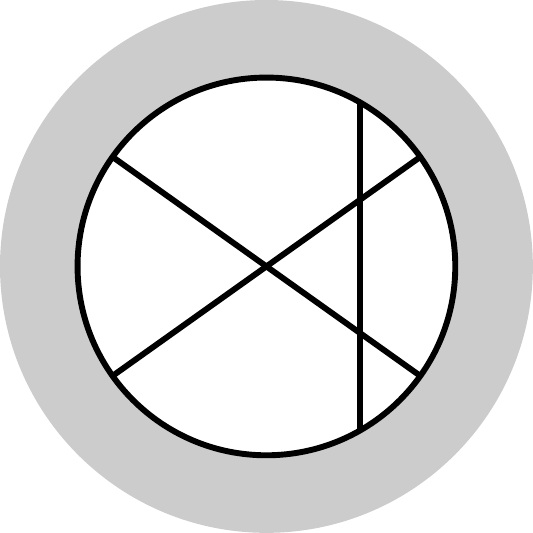}}}
	\end{align*}
	\caption{The so-called Yang-Baxter move, which leaves the permutation associated to crossing diagram invariant. The grey disk is used to indicate that these diagrams can be part of some larger diagram.}
	\label{fig:Yang-baxter}
\end{figure}

The plabic graphs of the positive orthogonal Grassmannian are \emph{crossing diagrams}, which are planar graphs with $2k$ external edges labelled $1,\ldots,2k$ with internal vertices of degree four. Again, these diagrams are typically embedded in a disk. We can read off the associated permutation by defining a new set of rules of the road:
\begin{enumerate}
	\item Start at external edge $i$, and follow it into the graph,
	\item When reaching a vertex, take the second exit (\emph{i.e.} go straight),
	\item We continue the previous step until we exit the graph via external edge $j$. This then means that $\sigma(i)=j$.
\end{enumerate}
As an example, we depict a crossing diagram in \ref{fig:crossing-diagram}.
The square move of plabic graphs also has an analogue, which is called the \emph{Yang-Baxter move}, depicted in figure \ref{fig:Yang-baxter}. This move leaves the permutation of a crossing diagram invariant.

An orthogonal version of Grassmannian graphs is given by \emph{OG graphs}. These are similar to crossing diagrams, except that we allow internal vertices with any even degree. The rules of the road for OG graphs are again similar, where you still go `straight' at every internal vertex. That is, for a vertex with degree $2m$ we take the $m$\textsuperscript{th} exit. We see that OG graphs can equivalently be interpreted as Grassmannian graphs of type $(k,2k)$ which only have internal vertices of degree $2m$ and associated helicity $m$.

\section{Summary}

We have introduced the notions of \emph{projective geometry} and \emph{Grassmannian geometry}. Projective geometry will be important in the following sections because of their importance in \emph{twistor} and \emph{momentum twistor} geometry. We additionally focussed on \emph{projective polytopes} and their volumes, which can serve as important examples of positive geometries. Additionally, we defined projective polytopes as the image of a positive linear map from the positive Grassmannian $G_+(1,n)$. A natural generalisation of this will lead to the definition of the \emph{amplituhedron} in section \ref{sec:POS_amplituhedron}, and similar definitions will be encountered for the \emph{momentum amplituhedron} and \emph{ABJM momentum amplituhedron}.

Additionally, we have introduced the Grassmannian, and we have studied various important subspaces of the Grassmannian, namely the \emph{positive Grassmannian} and the \emph{positive orthogonal Grassmannian}. The Grassmannian will naturally appear when using \emph{spinor-helicity variables} (which we will introduce in the next chapter), where the Lorentz invariant information is encoded in two elements of $G(2,n)$. The Grassmannian will additionally be used to linearise momentum conservation in these variables. The positive Grassmannian and positive orthogonal Grassmannian will play an important role in the study of scattering amplitudes in \nf and ABJM, respectively. We will see in chapter \ref{sec:AMP} that their amplitudes and other \emph{on-shell functions} are naturally associated to certain \emph{positroid} and \emph{orthitroid} cells. As already noted, the definition of the (momentum) amplituhedra is also crucially reliant on these positive Grassmannians. Additionally, we will see that we can define the amplituhedron and momentum amplituhedron directly in momentum twistor space and spinor-helicity space, respectively, which is most naturally done by using the binary operations of planes defined in section \ref{sec:GRASS_binary}. In practice, we can define these positive geometries by using only the complement $C_1\setminus C_2$ of $C_1\in G_+(k_1,n)$ and $C_2\in G_+(k_2,n)$. However, it will be useful to keep the other binary operations in mind as well, as it allows for easy manipulation of some expressions we encounter.

%% file: chapters/kinematicspaces.tex
\chapter{Kinematic Spaces}\label{sec:KIN}

In this chapter we will give an overview of the various kinematic variables which we will encounter throughout the thesis. After stripping of a colour group dependent factor (see section \ref{sec:AMP_large-N}), we will typically think of scattering amplitudes as complex-valued functions (or, if you include a momentum conserving delta function, a distribution) which depend on an ordered set of momenta of the scattering particles. However, momentum vectors are very redundant and are hardly ever the appropriate choice of variables when discussing scattering amplitudes. This is already clear from Lorentz invariance: if we expand some scattering amplitude in the components of the momentum vectors, then any Lorentz invariance will be completely obscured. Instead, it is better to express our scattering amplitudes in terms of variables which manifest Lorentz invariance, such as Mandelstam variables $s_{ij}=(p_i+p_j)^2$. Dependent on the problem at hand, a different set of kinematic variables might be even more well-suited to describe the scattering amplitude. The most famous example of this is the \emph{Parke-Taylor} formula, which we already encountered in the introduction. The remarkable simplicity of the $n$-particle gluon amplitude
\begin{align}
	A_n(1^+2^+3^-\cdots n^-)=\frac{\<12\>^4}{\<12\>\<23\>\cdots\<n1\>}\,,
\end{align}
is in part due to the use of \emph{spinor-helicity variables}, which we will define in section \ref{sec:KIN_spin-hel}. 

Using the `correct' choice of kinematic variables can make your life maximally easy when describing or deriving scattering amplitudes. In addition, a remarkable amount of progress can be made by simply reinterpreting old results in different kinematic spaces, which might expose hitherto unknown properties. One of the central themes in this thesis is the juxtaposition of positive geometries in various kinematic spaces, and many of the results rely on translating between these different kinematic spaces. As such, we will use this chapter to review the different kinematic variables of interest, and record formulas which will help us go from one type of variable to another.

\section{Mandelstam Invariants}\label{sec:KIN_mand}

In addition to the dot product $p_i\cdot p_j = \eta_{\mu\nu}p_i^\mu p_j^\nu$, the most common form of Lorentz invariants are the \emph{Mandelstam invariants}:
\begin{align}
	s_{i j\cdots k} = (p_i+p_j+\ldots+p_k)^2\,.
\end{align}
From this definition it is clear that the Mandelstam invariants are completely invariant under permutations of the indices. They are of particular importance for scattering amplitudes because they appear in the propagators of Feynman diagrams: an edge with momentum $p_i+p_j+\ldots+p_k$ and mass $m$ will contribute $1/(s_{ij\cdots k}+m^2)$ to the Feynman diagram.

When considering massless particles, $p^\mu p_\mu=0$, we have 
\begin{align}
	s_{ij}=(p_i+p_j)^2=2p_i\cdot p_j\,.
\end{align}
The Mandelstam invariants are not independent, for example for massless particles
\begin{align}
	s_{123}=(p_1+p_2+p_3)^2=(p_1+p_2)^2+(p_1+p_3)^2+(p_2+p_3)^2 = s_{12} + s_{13} + s_{23}\,.
\end{align}
This identity generalises, for some $A\subset[n]$ we have
\begin{align}
	s_A = \sum_{B\in\binom{A}{2}}s_B\,.
\end{align}
Furthermore, momentum conservation implies that
\begin{align}
	\sum_{i=1}^n s_{ij} = 0,\quad \forall j\in[n]\,.
\end{align}
At this stage, we thus have $\binom{n}{2}-n=n(n-3)/2$ independent Mandelstam invariants. In $D$ spacetime dimensions with $n\leq D+1$, this is forms a basis for the space of Mandelstam invariants, which we denote $\Kbb_n$. A useful basis for $\Kbb_n$ is given by the \emph{planar Mandelstam invariants}
\begin{align}\label{eq:KIN_SYM_planar-mand}
	X_{ij}=(p_i+p_{i+1}+\ldots+p_{j-1})^2\,,
\end{align}
which, for massless particles, satisfy $X_{ii+1}=X_{1n}=0$. We already encountered these variables when discussing $\tr{\phi^3}$ theory in section \ref{sec:INT_phi3}.

In general there are further relations among the Mandelstam variants. This comes from the basic observation that in $D$ dimensions, at most $D$ vectors can be linearly independent. This is commonly encoded in the so-called \emph{Gram determinant relations}. The Gram matrix $\Gcal\equiv\Gcal(p_1,p_2,\ldots,p_n)$ is defined to have entries $[\Gcal]_{ij}=2p_i\cdot p_j = s_{ij}$, the Gram determinant conditions then state that any minor of this matrix of size greater than $D\times D$ will vanish. A basis for these relations is given by the minors
\begin{align}\label{eq:KIN_gram}
	\begin{vmatrix}
		0 & s_{12} & s_{13}&\cdots & s_{1D} & s_{1i}\\
		s_{12} & 0 & s_{23}&\cdots & s_{2D} & s_{2i}\\
		s_{13} & s_{23} & 0 &\cdots & s_{3D} & s_{3i}\\
		\vdots &\vdots & \vdots & \ddots &\vdots & \vdots\\
		s_{1D} & s_{2D} & s_{3D} &\cdots & 0 & s_{Di}\\
		s_{1j} & s_{2j} & s_{3j} & \cdots & s_{Dj} & s_{ij}
	\end{vmatrix}=0\,.
\end{align}
We denote the space $\Kbb_n$ modulo Gram determinant relations as $\Kbb_n^D$. We find that
\begin{align}
	\dim\Kbb_n^D =\frac{n(n-3)}{2} - \frac{(n-D)(n-D-1)}{2} = (D-1)n - \frac{D(D+1)}{2}\,.
\end{align}
A natural choice of a basis for this space is given by the planar Mandelstam variables $X_{ij}$ with $i<D, i<j$.

\section{Spinor-Helicity}

For most of this thesis, we will be considering the scattering of massless particles in three or four dimensions. In these cases, it is often useful to make use of \emph{spinor-helicity variables}, which trivialises the masslessness constraint, as well as the Gram determinant relations. This section will serve as a short introduction to the spinor-helicity formalism in three and four dimensions, we refer the interested reader to \cite{Elvang:2013cua} for more details. Extensions to five \cite{Chiodaroli:2022ssi,Pokraka:2024fao} and six dimensions \cite{Cheung:2009dc} are also known, but these will not be relevant for the context of this thesis.

\subsection{Spinor-Helicity in Four Dimensions}\label{sec:KIN_spin-hel}

We start by writing any four-momentum $\smash{p^\mu\in\Rbb^{1,3}}$ in a Hermitian $2\times 2$ matrix as
\begin{align}\label{eq:KIN_SH_p-to-mat}
	p^\mu \to p^{\alpha \dot\alpha} =\begin{pmatrix} p^0 + p^3 && p^1-i p^2\\p^1+i p^2 && p^0 - p^3\end{pmatrix}.
\end{align}
It follows immediately that $\det p^{\alpha\dot\alpha} = -p^\mu p_\mu = m^2$. If we consider the $SL(2,\Cbb)$ transformation $p\to p' = S p S^\dagger$, with $S\in SL(2,\Cbb)$, then clearly, $p'$ is still a Hermitian matrix satisfying $\det p'=\det p$, hence this map is a Lorentz transformation\footnote{In fact, it can be shown that every Lorentz transformation can be written in this way.}.
In the case where we are considering massless particles, $p^\mu p_\mu=0$, the matrix $p^{\alpha\dot\alpha}$ is a $2\times 2$ matrix with zero determinant. Any $2\times 2$ matrix with zero determinant can be written as an outer product of two 2-vectors:
\begin{align}
	p^{\alpha\dot\alpha}=\lambda^\alpha \tilde\lambda^{\dot\alpha}\,.
\end{align} 
Since $p^{\alpha\dot\alpha}$ is Hermitian, this implies that 
\begin{align}
	\tilde\lambda^{\dot\alpha}=\pm (\lambda^\alpha)^*\,.
\end{align}
The Lorentz group acts on these variables as
\begin{align}
	\lambda^\alpha\to S^\alpha_{\beta}\lambda^\beta,\quad\tilde\lambda^{\dot\alpha}\to S_{\dot\beta}^{\dagger\,\dot\alpha}\tilde\lambda^{\dot\beta}\,.
\end{align}
We will often consider complexified momenta, in which case the matrix $p^{\alpha\alphadot}$ is no longer Hermitian, and there are no relations between $\lambda$ and $\tilde\lambda$. The Lorentz group acts on $\lambda$ and $\tilde\lambda$ as two separate copies of $SL(2,\Cbb)$. There is a $GL(1)$ subgroup of the Lorentz group which leaves the momentum $p^{\alpha\alphadot}$ invariant: $\lambda \to t \lambda, \tilde\lambda\to t^{-1}\tilde\lambda$. This is known as the \emph{little group}. 

Consider a scattering process of $n$ massless particles. We take all particles outgoing, such that momentum conservation reads
\begin{align}
	\sum_{i=1}^n p_i^{\alpha\alphadot}=\sum_{i=1}^n \lambda_i^\alpha\tilde\lambda_i^{\dot\alpha}=0\,.
\end{align}
We can organise our kinematic data into two $2\times n$ matrices
\begin{align}
	\lambda = \begin{pmatrix}
		\lambda_1^1 & \lambda_2^1 &\cdots &\lambda_n^1\\
		\lambda_1^2 & \lambda_2^2 &\cdots &\lambda_n^2
	\end{pmatrix},\quad
	\tilde\lambda = \begin{pmatrix}
		\tilde\lambda_1^{\dot{1}} & \tilde\lambda_2^{\dot{1}} &\cdots &\tilde\lambda_n^{\dot{1}}\\
		\tilde\lambda_1^{\dot{2}} & \tilde\lambda_2^{\dot{2}} &\cdots &\tilde\lambda_n^{\dot{2}}
	\end{pmatrix}\,.
\end{align}
Instead of interpreting the matrix $\lambda$ as a collection of $n$ 2-vectors $\lambda_1,\ldots,\lambda_n$, we can equivalently interpret it as 2 $n$-vectors $\lambda^1,\lambda^2$. The advantage of this is that Lorentz transformations act on $\lambda$ as an $SL(2)$ transformation, which we then interpret as some linear combination of these two $n$-vectors. We thus see that the Lorentz invariant information is encoded in the plane spanned by these vectors. This implies that we should consider $\lambda$ and $\tilde\lambda$ to be elements of the Grassmannian $G(2,n)$. Momentum conservation then reads
\begin{align}\label{eq:KIN_spin-hel-mom-cons}
	\lambda\cdot\tilde\lambda^T=\nul_{2\times 2}\,,
\end{align}
which is equivalent to the statement that the two $2$-planes $\lambda$ and $\tilde\lambda$ are orthogonal to each other.

The Pl\"ucker variables of $\lambda$ and $\tilde\lambda$ are Lorentz invariant objects and are often written as `angle' and `square' brackets:
\begin{align}
	\<ij\> &\coloneqq \eps_{\alpha\beta}\lambda_i^\alpha\lambda_j^\beta=p_{ij}(\lambda)\,,\\
	[ij] &\coloneqq \eps_{\dot\alpha\dot\beta}\tilde\lambda_i^{\dot\alpha}\tilde\lambda_j^{\dot\beta}=p_{ij}(\tilde\lambda)\,.
\end{align}
They are related to the Mandelstam invariants via
\begin{align}
	s_{ij} = (p_i+p_j)^2 = \<ij\>[ij]\,.
\end{align}
Since these angle and square brackets are Pl\"ucker variables, they also satisfy the Pl\"ucker relations, in this case often referred to as the \emph{Schouten identity}:
\begin{alignat}{7}
	&\<ij\>&&\<kl\>&&+\<ik\>&&\<lj\>&&+\<il\>&&\<jk\>&&=0\\
	&[ij]&&[kl]&&+[ik]&&[lj]&&+[il]&&[jk]&&=0\,.
\end{alignat}
The set of brackets of the form $\<i i+1\>$ and $\<1 i\>$ form a basis for all angle brackets. We can express any angle bracket in terms of this basis by repeatedly applying the Schouten identity. Explicitly, we find
\begin{align}
	\<a b\> = \<1 a\> \<1 b\> \sum_{i=a}^{b-1} \frac{\<i i+1\>}{\<1 i\>\<1 i+1\>}\,.
\end{align}
We note that when we consider momenta in $\Rbb^{2,2}$ instead of $\Rbb^{1,3}$, we can keep the matrix $p^{\alpha\alphadot}$ real for uncomplexified momenta. In practice, we can just Wick rotate $p^2\to i p^2$, and we see that the matrix $p^{\alpha\dot\alpha}$ in \eqref{eq:KIN_SH_p-to-mat} is completely real. The Lorentz group now acts on $\lambda,\tilde\lambda$ as two independent copies of $SL(2,\Rbb)$.

We further note that momentum conservation from equation \eqref{eq:KIN_spin-hel-mom-cons} can be linearised by introducing an auxiliary $k$-plane in $n$-dimensions: $C\in G(k,n)$. If we require that
\begin{align}\label{eq:KIN-lambda-C}
	\lambda\subseteq C\,,\quad \tilde\lambda\subseteq C^\perp\,,
\end{align}
then momentum conservation is trivially satisfied. As constraints this can be written as
\begin{align}
	\lambda\cdot (C^\perp)^T=\nul_{2\times(n-k)}\,,\quad \tilde\lambda\cdot C^T = \nul_{2\times k}\,.
\end{align}
Equivalently, the fact that $\lambda\subseteq C$ means that there must exist a $2\times k$ matrix $\rho$ such that $\lambda = \rho\cdot C$. This is a first glimpse at the importance of Grassmannian geometry in scattering amplitudes, a topic which we will return to in section \ref{sec:AMP_nf}.

\subsection{Spinor-Helicity in Three Dimensions}

We can recycle the four-dimensional spinor-helicity formalism to three-dimensional momenta. We can reduce $\Rbb^{1,3}$ (or $\Rbb^{2,2}$) to $\Rbb^{1,2}$ by setting $p^2\to 0$ in equation \eqref{eq:KIN_SH_p-to-mat}. Relabelling $p^3\to p^2$ for consistent labelling, we represent the momentum vectors in matrix form as
\begin{align}
	p^\mu \to p^{\alpha \beta} =\begin{pmatrix} p^0 + p^2 && p^1\\p^1 && p^0 - p^2 \end{pmatrix},
\end{align} 
where we still have the relation $\det p^{\alpha\beta}=p^\mu p_\mu =m^2$. For massless particles, the matrix $p^{\alpha\beta}$ is now a \emph{symmetric} $2\times 2$ matrix of rank 1, which means we can write $p^{\alpha\beta}=\lambda^\alpha\lambda^\beta$, for the \emph{same} $\lambda$. Again, the Lorentz group acts on these spinor-helicity variables as $SL(2)$, however the little group is only $\Zbb_{2}$, as $\lambda\to-\lambda$ keeps $p^\mu$ invariant. We can again collect the kinematic data of an $n$-particle scattering process as a $2\times n$ matrix $\lambda$, where the Lorentz invariant information is encoded in the plane spanned by the row-vectors of $\lambda$. Momentum conservation with all particles outgoing then reads
\begin{align}
	\sum_{i=1}^n \lambda_i^\alpha\lambda_i^\beta =0\implies \lambda\cdot\lambda^T =\nul\,.
\end{align}
This has the interpretation that the 2-plane $\lambda$ is orthogonal to itself, hence we consider it to be a part of the orthogonal Grassmannian $\lambda\in OG(2,n)$. There are no real solutions to planes being orthogonal to themselves, thus this requires complex momenta. If we want to keep the momenta real instead, we can consider the particles to be alternatingly incoming and outgoing, assuming an even number of external particles. The 2-plane $\lambda$ is then an element of the orthogonal Grassmannian $OG(2,n)$ with respect to the $n\times n$ metric $\eta=\text{diag}(1,-1,1,\cdots,-1)$, \emph{i.e.}
\begin{align}
	\lambda\cdot\eta\cdot\lambda^T = \nul\,.
\end{align}

\section{Twistors}\label{sec:KIN_twistors}

Twistor theory was introduced by Penrose in 1967 \cite{Penrose:1967wn}. It was originally used to describe flat four-dimensional spacetime, and it was hoped that could lead to a possible formulation of quantum gravity \cite{PENROSE1973241}. It is a rich topic with many applications in physics and mathematics, but it has remained on the sidelines of QFT research until its adoption by scattering amplitudes in the 2000s \cite{Hodges:2015kla}. One of the main reasons for its recent success stems from the fact that twistor variables make conformal symmetry in Minkowski space manifest. 

Let us start by considering two points in Minkowski space which are light-like separated: $(x^{(1)}-x^{(2)})^2=0$. From our discussion on spinor-helicity variables, we know that the matrix $(x^{(1)}-x^{(2)})_{\alpha\alphadot}$ has rank 1, which means that there must exist $\tilde\lambda^\alphadot$ and $\tilde\mu_\alpha$ such that $\tilde\lambda^\alphadot= (x^{(1)})^{\alpha\alphadot}\tilde\mu_\alpha=(x^{(2)})^{\alpha\alphadot}\tilde\mu_\alpha$. Hence, the \emph{twistor}
\begin{align}
	W^A\coloneqq \begin{pmatrix}
		\tilde\lambda^\alphadot\\\tilde\mu^\alpha
	\end{pmatrix}\,,
\end{align}
defines a light-ray in Minkowski space through the \emph{incidence relations}
\begin{align}
	x_{\alpha\alphadot} \tilde\lambda^\alphadot=\tilde\mu_\alpha\,.
\end{align}
That is, for a given $\tilde\lambda$ and $\tilde\mu$, the set of all $x$ which satisfy the incidence relations all lie on a light-ray, and vice versa. Since $W^I$ and $t W^I$ define the same light-ray in spacetime, we can interpret the twistors projectively as elements of $\Pbb^3$. 

These twistor variables linearise conformal transformations. To see this explicitly, assume we are given $n$ twistors $W_i^A$ which we combine into a $4\times n$ matrix $W$. A general conformal transformation acts on this matrix as $SL(4)$, which means that the conformally invariant information is encoded in the plane spanned by the four $n$-vectors which make up $W$. That is, we should interpret $W$ as an element of $G(4,n)$, and the \Pluck variables of $W$ are conformal invariant quantities\footnote{Note that we have not taken the projective nature of twistors into account. There is still a redundant $\Cbb^{*(n-1)}$ torus action, which really reduces $W$ to an element of $\text{Conf}_n(\Pbb^3)$, \emph{i.e.} the configuration space of $n$ points in $\Pbb^3$. Instead of considering $\text{Conf}_n(\Pbb^3)$, we will go back and forth between thinking of twistors as being points in a projective space, or as defining an element of the Grassmannian $G(4,n)$.}.

We can go from spinor-helicity space to twistor space through a \emph{Penrose transform}. To motivate this, let us consider the massless wave equation and its Fourier transform
\begin{align}
	\Box \phi(x) = 0 \implies p^2 \tilde\phi=0\,,
\end{align}
which is solved by
\begin{align}
	\phi(x)=\int \dd^4 p \delta(p^2)e^{i p\cdot x} \tilde\phi(p)\,.
\end{align}
The delta function constraint $p^2=0$ can be solved by using spinor-helicity variables:
\begin{align}
	\phi(x)=\int\frac{\dd^2\lambda\dd^2\tilde\lambda}{\vol[GL(1)]} e^{i\lambda^\alpha\tilde\lambda^\alphadot x_{\alpha\alphadot}}\tilde\phi(\lambda,\tilde\lambda) = \int\frac{\dd^2\tilde\lambda}{\vol[GL(1)]}\tilde\phi(\tilde\lambda^\alphadot,x_{\alpha\alphadot}\lambda^\alpha)\,.
\end{align} 
We recognise $\tilde\mu_\alphadot=x_{\alpha\alphadot}\lambda^\alpha$ in the argument of $\tilde\phi$ through the incidence relations. Hence, after a `half Fourier transform' we end up with twistor variables.

The generators of the conformal group do not have a uniform number of derivatives in terms of spinor-helicity variables, as the translation generator $P^{\alpha\alphadot}$ acts multiplicative, and the special conformal transformation generator $K^{\alpha\alphadot}$ has two derivates. This is not the case for twistors. We can translate between the generators as
\begin{align}
	\lambda_i^\alpha \leftrightarrow \frac{\partial}{\partial \tilde\mu_i^\alpha}\,,\quad \frac{\partial}{\partial \lambda_i^\alpha}\leftrightarrow -i\tilde\mu_{i\alpha}\,.
\end{align} 
In terms of twistor variables, the generators of the conformal group become the nice and uniform
\begin{align}
	G^A_B = \sum_{i=1}^n \left( W_i^A\partial_{W_i^B}-\frac{1}{4}\delta^A_B W_i^C\partial_{W_i^C}\right)\,.
\end{align}

\section{Dual Momenta}\label{sec:KIN_dual}

Consider an ordered set of $n$ momenta $p_1^\mu,\ldots,p_n^\mu$. Momentum conservation implies that the sum of these vectors adds to zero, which we can interpret geometrically as saying that the vectors form a closed polygon. Inspired by this observation, we introduce the \emph{dual variables}, or \emph{dual momenta}, $x_i^\mu$ that satisfy 
\begin{align}\label{eq:KIN_DUAL_dual-def}
	x_{i+1}^\mu-x_i^\mu=p_i^\mu\,,
\end{align}
which represent the corners of the polygon (we identify $x_{n+1}\equiv x_1$). From the definition it follows immediately that some translation $x_i^\mu\to x_i^\mu+a^\mu$ leaves the momenta invariant. The consequence of this is that momentum conservation is manifest in these dual variables, for any set of $n$ points in \emph{dual space} (i.e. the space of dual variables), the momenta defined as in \eqref{eq:KIN_DUAL_dual-def} automatically satisfy momentum conservation. We further define
\begin{align}
	x_{ij}\coloneqq x_i-x_j=p_i+p_{i+1}+\ldots+p_{j-1}\,.
\end{align}
We note that the planar Mandelstam variables from \eqref{eq:KIN_SYM_planar-mand}
\begin{align}
	X_{ij}=(p_i+p_{i+1}+\ldots+p_{j-1})^2 = (x_i-x_j)^2 = x_{ij}^2\,,
\end{align}
can thus be interpreted as the square distance between points $x_i$ and $x_j$ in dual space.

We emphasise that the momentum vectors need to have a well-defined ordering to be able to define the dual-space polygon. In practice, as we will see in section \ref{sec:AMP_large-N}, we can generally define such an ordering from the \emph{colour ordering} of tree-level amplitudes in theories with a non-abelian colour group. For loop amplitudes we instead have to consider the so-called \emph{planar limit}, which is the leading order contribution in the large-$N$ expansion.

We will often consider the scattering of massless particles, in which case the condition $p_i^2=0$ implies that $(x_{i+1}-x_i)^2=0$. This means that the points $x_i$ and $x_{i+1}$ are light-like separated, and the momenta are thus encoded in a \emph{null-polygon} in dual space. This emphasises the importance of light-cone geometry in dual space (or mass-shell geometry in the case of massive particles), an idea we will repeatedly return to in this thesis. Of particular importance is the observation that $D$ points in dual space $x_1^\mu,\ldots,x_D^\mu$ generically define two (possibly complex) special points in dual space: $q^\pm_{12\cdots D}$, which are specified by the intersection of the $D$ lightcones centred at points $x_i$. An explicit formula for $q^\pm_{12\cdots D}$ is given in appendix \ref{sec:APP_schubert}. In addition to this, the geometry of lightcones can give us non-trivial information about the possible momenta involved in a scattering process. For example, in $D=4$ spacetime dimensions, the fact that amplitudes can be split up into different helicity sectors is a direct consequence from the fact that $4D$ null-polygons can have quantitatively different structures, which we can classify by an integer number: the helicity.

We have shown how we can relate a null-polygon in dual space to an ordered set of $n$ massless momenta. These dual variables manifest momentum conservation as translational invariance. To go in the opposite direction, and find such a null-polygon from given momentum vectors, we have to break this translation invariance. This is typically done by fixing $x_1$ to the origin, in which case we can invert equation \eqref{eq:KIN_DUAL_dual-def} to give
\begin{align}\label{eq:KIN_mom-to-dual}
	x_i = \sum_{j=1}^{i-1} p_j\,.
\end{align}

\section{Embedding Space}\label{sec:KIN_embedding}

In the embedding space formalism, first introduced by Dirac in 1936 \cite{Dirac:1936fq}, we embed our $D$-dimensional spacetime into a `projective null-cone' in $D+2$-dimensional spacetime in such a way that the conformal transformations are linearised in this embedding space.

We consider points $X$ in $\Rbb^{d_1+1,d_2+1}$ which lie on the null-cone 
\begin{align}
	\Ncal\coloneqq \{X\in \Rbb^{d_1+1,d_2+1}\colon X^2=0\}\,,
\end{align}
and we further treat the variables in this space projectively, \emph{i.e.} we identify
\begin{align}
	X\sim \lambda X\,.
\end{align}
We will rotate an $\Rbb^{1,1}$ subspace into light-cone coordinates, such that $X$ has components $X^A$, where $A=\{+,-,\mu\}$. In these coordinates the metric becomes $\eta_{+-}=\eta_{-+}=\frac{1}{2}$, $\eta^{+-}=\eta^{-+}=2$. Then
\begin{align}
	X^2=X^A X^B\eta_{AB} = X^+ X^- + X_\mu X^\mu\,,
\end{align}
such that the null condition $X^2=0$ implies that
\begin{align}
	X^- = -\frac{X_\mu X^\mu}{X^+}\,.
\end{align}
We note that
\begin{align}
	(X_i-X_j)^2&=-2X_i\cdot X_j = -2\left[ \frac{X_i^+ X_j^-}{2}+\frac{X_j^+ X_i^-}{2}+X_{i\mu}X_j^\mu \right]\\
	&=X_i^+ X_j^+ (x_i-x_j)_\mu (x_i-x_j)^\mu\,,
\end{align}
where we have defined (for $X^+\neq 0$)
\begin{align}
	x^\mu = \frac{X^\mu}{X^+}\,.
\end{align}
Using projectivity to fix $X^+=1$, we find
\begin{align}
	X^-&=-x^2\,,\\ 
	(x_i-x_j)^2&=-2 X_i\cdot X_j\,,\label{eq:KIN_EMB_dist-sqr}\\
	X^A&=\begin{pmatrix} X^+\\X^-\\X^\mu \end{pmatrix} = \begin{pmatrix}1\\-x^2\\x^\mu \end{pmatrix}\,,
\end{align} 
where $x^2=x_\mu x^\mu$ is now with respect to the induced metric $\eta_{\mu\nu}$ on $\Rbb^{d_1,d_2}$.

The natural isometries of embedding space $X^A\to G^A_B X^B$ are elements of the group $G^A_B\in O(d_1+1,d_2+1)$, and hence satisfy $\eta_{CD} G^C_A G^D_B = \eta_{AB}$. From $X^2=0$ we find the identity
\begin{align}
	\dd X^+ X^- + X^+ \dd X^- + 2 X_\mu \dd X^\mu = 0\,,
\end{align}
which we can use to derive
\begin{align}
	\eta_{AB}\dd X^A \dd X^B = (X^+)^2 \eta_{\mu\nu}\dd x^\mu\dd x^\nu\,.
\end{align}
This shows that any transformation $X^A\to X'^A = G^A_B X^B$ which leaves $\eta_{AB}\dd X^A \dd X^B$ invariant, will induce the transformation
\begin{align}
	\eta_{\mu\nu}\dd x^\mu\dd x^\nu \to \left( \frac{X^+}{X'^+} \right)^2 \eta_{\mu\nu}\dd x^\mu\dd x^\nu\,.
\end{align}
We see that the isometries in embedding space induce conformal transformations in the $x^\mu$ variables. Using the embedding space formalism, we see that conformal transformations are linearised, and, from \eqref{eq:KIN_EMB_dist-sqr} also the distance squared between two points is linearised. 

\section{Momentum Twistors}\label{sec:KIN_mom-twistor}

Momentum twistors are similar objects to the twistors introduced in section \ref{sec:KIN_twistors}, except that our starting point is dual space, rather than Minkowski space. We start by assuming that we have $n$ points $x_i^\mu$ in dual space, which encodes the scattering data of $n$ ordered massless particles. Since $p_i=(x_{i+1}-x_i)=\lambda_i\tilde\lambda_i$, we have
\begin{align}
	0=\lambda_{i\alpha} p_i^{\alpha\alphadot} = \lambda_{i\alpha} (x_{i+1}-x_i)^{\alpha\alphadot}\implies \lambda_{i\alpha}x_{i}^{\alpha\alphadot}=\lambda_{i\alpha}x_{i+1}^{\alpha\alphadot}\,,
\end{align}
which motivates the definition of $\mu_i^\alphadot$ through the \emph{incidence relations}
\begin{align}
	\mu_i^\alphadot \coloneqq \lambda_{i\alpha}x_{i}^{\alpha\alphadot}=\lambda_{i\alpha}x_{i+1}^{\alpha\alphadot}\,.
\end{align}
We then define the \emph{momentum twistors} as
\begin{align}
	z_i^A = \begin{pmatrix}
		\lambda_i^\alpha\\\mu_i^\alphadot
	\end{pmatrix}\,.
\end{align}
We note that $\mu_i$ has the same little group weight as $\lambda_i$, and hence under a little group transformation $z_i \to t_i z_i$. We see that rescaling the momentum twistors leaves the corresponding momentum vectors invariant, and we therefore consider momentum twistors projectively. We see that two null-separated points $x_i$ and $x_{i+1}$ in dual space define a point $z_i$ in momentum twistor space $\Pbb^3$.

\begin{figure}
	\centering
	\includegraphics[scale=0.45]{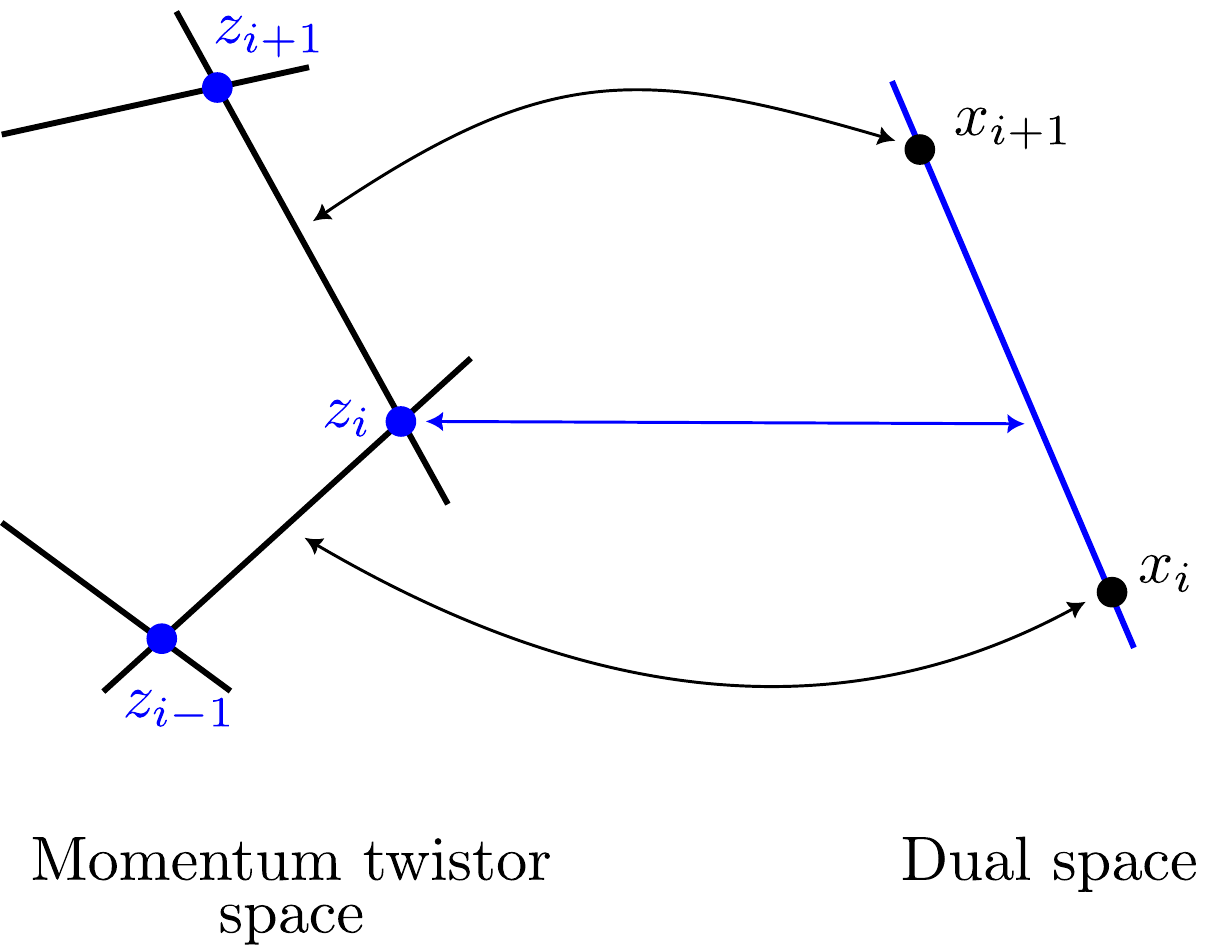}
	\caption{This diagram schematically shows the correspondence between points in momentum twistor space and light rays in dual space (blue), and the correspondence between lines in momentum twistor space and points in dual space (black). Using the incidence relations we can go between a point $z_i$ in momentum twistor space and the light-ray $(x_{i}x_{i+1})$ in dual space. We go between the line $(z_{i-1} z_i)$ in momentum twistor space and the point $x_i$ in dual space through equation \eqref{eq:KIN_MT_xi-from-mom-twistor}. The fact that the lines $(z_{i-1}z_i)$ and $(z_i z_{i+1})$ intersect (they intersect in the point $z_i$) implies that the points $x_i$ and $x_{i+1}$ must be light-like separated.}
	\label{fig:mom-twistor-dual-space}
\end{figure}

We can invert the incidence relations and map a line in momentum twistor space to a point in dual space. We can rewrite the combination $\lambda_i^\alpha \mu_{i-1}^\alphadot - \lambda_{i-1}^\alpha \mu_{i}^\alphadot$ as
\begin{align}
	\lambda_i^\alpha \mu_{i-1}^\alphadot - \lambda_{i-1}^\alpha \mu_{i}^\alphadot = \epsilon_{\beta\gamma}( \lambda_i^\alpha \lambda_{i-1}^\beta - \lambda_{i-1}^\alpha \lambda_{i}^\beta) x_{i}^{\gamma\alphadot}\,.
\end{align}
By inspection, we find that 
\begin{align}
	 \lambda_i^\alpha \lambda_{i-1}^\beta - \lambda_{i-1}^\alpha \lambda_{i}^\beta = \< i i-1\> \epsilon^{\alpha\beta}\,,
\end{align}
which means that we can solve for $x_i$ as
\begin{align}\label{eq:KIN_MT_xi-from-mom-twistor}
	x_i^{\alpha\alphadot} = \frac{\lambda_{i-1}^\alpha \mu_i^\alphadot - \lambda_i^\alpha \mu_{i-1}^\alphadot}{\<i\,i-1\>}\,.
\end{align}
Furthermore, any point on the line in momentum twistor space passing through the points points $z_i, z_{i+1}$ can be written as a $GL(2)$ transformation of the $2\times 4$ matrix $\begin{pmatrix}	z_i z_{i+1} \end{pmatrix}$. It is clear that \eqref{eq:KIN_MT_xi-from-mom-twistor} is invariant under such a transformation, and this formula has the interpretation of a map from a line in momentum twistor space to a point in dual space. Hence, we have shown that a \emph{light-ray} in dual space corresponds to a \emph{point} in momentum twistor space, and a \emph{line} in momentum twistor space corresponds to a \emph{point} in dual space. This is summarised schematically in figure \ref{fig:mom-twistor-dual-space}.

We can use equation \eqref{eq:KIN_MT_xi-from-mom-twistor} to go one step further and solve for $\tilde\lambda_i$ from the momentum twistors:
\begin{subequations}
\begin{align}
	\lambda_i^\alpha\tilde\lambda_i^\alphadot& = x_{i+1}^{\alpha\alphadot} - x_i^{\alpha\alphadot} = \frac{\lambda_{i}^\alpha \mu_{i+1}^\alphadot - \lambda_{i+1}^\alpha \mu_{i}^\alphadot}{\<i i+1\>} - \frac{\lambda_{i-1}^\alpha \mu_i^\alphadot - \lambda_i^\alpha \mu_{i-1}^\alphadot}{\<i-1 i\>}\\
	&=\frac{\<i i+1\> \lambda_i \mu_{i-1} - (\<i-1 i\> \lambda_{i+1} - \<i i+1\> \lambda_{i-1})\mu_i + \<i-1 i\> \lambda_i\mu_{i+1}}{\<i-1 i\> \<i i+1\>}\\
	&= \lambda_i \frac{\<i i+1\> \mu_{i-1} + \< i-1 i+1\> \mu_i + \<i-1 i\> \mu_{i+1}}{\<i-1 i\>\<i i+1\>}\,,
\end{align} 
\end{subequations}
where we use the Schouten identity in the last step. Hence, if we introduce the $n\times n$ matrix $Q$ with elements
\begin{align}\label{eq:KIN_Q-mat-def}
	Q_{ij} = \frac{\<i i+1\> \delta_{i-1\, j} + \< i-1 i+1\> \delta_{i\,j} + \<i-1 i\> \delta_{i+1 \,j}}{\<i-1 i\>\<i i+1\>}\,,
\end{align}
then we can map from momentum twistor space to spinor-helicity space by 
\begin{align}\label{eq:KIN_lambda-tilde=muQ}
	\tilde\lambda = \mu\cdot Q\,,
\end{align}
where $\tilde\lambda$ and $\mu$ are now considered as $2\times n$ matrices. We note that $\ker Q = \lambda$, and hence $Q$ has rank $n-2$. 

We recall from section \ref{sec:KIN_twistors} that twistor variables linearise conformal transformations in Minkowski space. In a similar line, momentum twistors linearise \emph{dual conformal transformations}, \emph{i.e.} conformal transformations in dual space. That is, if we have a $4\times n$ matrix $z$ which contains $n$ momentum twistors $z_i$, then a dual conformal transformation acts on this matrix as an $SL(4)$ transformation. The dual conformal invariant information is therefore encoded in the four-plane spanned by the columns of the $4\times n $ matrix $z$, which allows us to interpret $z \in G(4,n)$. Any expression which can be written solely in terms of the \Pluck variables of $z$, often denoted $\<ijkl\>\coloneqq p_{ijkl}(z)=\epsilon_{ABCD}z_i^Az_j^Bz_k^Cz_l^D$, is manifestly dual conformal invariant.

Assume we are given $z\in G(4,n)$ and want to extract the corresponding momentum vectors or spinor helicity variables. The $SL(4)$ transformations can mix $\lambda$ and $\mu$, and to isolate $\lambda$ it is therefore necessary to break the dual conformal invariance. This is usually done by introducing an `infinity twistor' $I_\infty$. We will take $I_\infty=\begin{pmatrix} \unit_{2\times 2} & \nul_{2\times 2} \end{pmatrix}$, and define $\lambda = I_\infty \cdot z$, which for this choice of infinity twistor just isolates the first two rows.

We can use equation \eqref{eq:KIN_MT_xi-from-mom-twistor} to find a point in dual space for any line in momentum twistor space, not just for adjacent momentum twistors. For example, we denote by $\ls_{ij}$ the point in dual space defined by two arbitrary momentum twistors $z_i$ and $z_j$, which is given by
\begin{align}\label{eq:KIN_ls-pure}
	{\ls_{ij}}^{\alpha\alphadot}=\frac{\lambda_j^\alpha\mu_i^\alphadot-\lambda_i^\alpha\mu_j^\alphadot}{\<ij\>}\,.
\end{align}
When two lines in momentum twistor space intersect, the corresponding points in dual space are null-separated. Since the line $(z_i z_j)$ intersects the lines $(z_{i-1} z_i)$,$(z_{i} z_{i+1})$, $(z_{j-1} z_j)$ and $(z_{j} z_{j+1})$, we see that the point $\ls_{ij}$ in dual space is null-separated from $x_{i},x_{i+1},x_j,x_{j+1}$. We know that there are two such points in dual space, this one corresponds to $q^+_{ii+1jj+1}$. The other solution is denoted $\lstil_{ij}$, and in momentum twistor space is $(i-1 i i+1)\cap (j-1 j j+1)$. Here we use the notation $(i-1 i i+1)\cap (j-1 j j+1)$ to denote the line in momentum twistor space which defined as the intersection of the plane spanned by $z_{i-1},z_i,z_{i+1}$ and the plane spanned by $z_{j-1},z_j,z_{j+1}$. Explicitly, we can write the intersection of lines and planes in momentum twistor space as
\begin{align}
	(ab)\bigcap (cde) &= z_a \<bcde\> - z_b \<acde\>\,,\\
	(abc)\bigcap (def) &= (z_a z_b)\<cdef\> + (z_c z_a)\<bedf\> + (z_b z_c) \<adef\>\,.
\end{align}
Using \eqref{eq:KIN_mom-to-dual} to find points in dual space from some given set of momentum vectors, we can write \eqref{eq:KIN_ls-pure}
\begin{align}\label{eq:KIN_ls-def}
	\ls_{ij} = \frac{1}{\<ij\>}\left( \sum_{l=1}^{j-1}\<lj\>\lambda_i\tilde\lambda_l - \sum_{l=1}^{i-1}\<li\>\lambda_j\tilde\lambda_l \right)\,,
\end{align}
and its parity conjugate (\emph{e.g.} $q^-_{ii+1jj+1}$) as
\begin{align}\label{eq:KIN_lstil-def}
	\lstil_{ij} = \frac{1}{[ij]} \left( \sum_{l=1}^{i-1}[li]\lambda_l\tilde\lambda_i - \sum_{l=1}^{j-1}[lj]\lambda_k\tilde\lambda_i \right)\,.
\end{align}

Given four points $z_A,z_B,z_C,z_D$ in momentum twistor space, we construct the Lorentz invariant $\<ABCD\>$. We can interpret this as a Lorentz invariant related to the lines $(AB)$ and $(CD)$, and hence in dual space it should be a Lorentz invariant constructed from $\ls_{AB}$ and $\ls_{CD}$, and hence should be proportional to $(\ls_{AB}-\ls_{CD})^2$. This quantity does not have the correct little group weight, and we need to introduce the angle brackets $\<AB\>$ and $\<CD\>$ to remedy this. We find
\begin{align}
	\<ABCD\> = \<AB\>\<CD\> (\ls_{AB}-\ls_{CD})^2,\\
\end{align}
which allows us to derive other useful identities such as
\begin{align}
	\frac{\<ij\>\<kl\>}{\<ik\>\<lj\>} = \frac{(\ls_{ik}-\ls_{jl})^2}{(\ls_{ij}-\ls_{kl})^2}\,.
\end{align}

\section{Summary}

In this chapter we have introduced the various kinematic spaces which are of importance for the study of scattering amplitudes. We have placed some emphasis on which symmetries and constraints are manifest or linearised by using these variables. To summarise,
\begin{itemize}
	\item \emph{Mandelstam invariants} manifest Lorentz invariance.
	\item \emph{Spinor-helicity} variables trivialise masslessness, and can be used for theories in three or four dimensions. The Lorentz invariant information is encoded in elements of $G(2,n)$.
	\item \emph{Twistors} linearise conformal invariance, and can be used in four dimensions. The conformal invariant information is encoded in an element of $G(4,n)$.
	\item \emph{Dual momenta} manifest momentum conservation. They can be used in any number of dimensions, but it requires a notion of ordering on the external particles.
	\item \emph{Embedding space} linearises conformal transformations, and it can be used in arbitrary dimensions. When using the embedding formalism for dual space, rather than spacetime, it linearises dual conformal transformations instead.
	\item \emph{Momentum twistors} linearise dual conformal transformations and trivialise momentum conservation. They can be used in four dimensions, and require a notion of ordering. The dual conformal invariant information is encoded in an element of $G(4,n)$. Momentum twistors are completely unconstrained, and any set of $n$ momentum twistors defines for us the kinematic data of $n$ ordered massless momentum vectors which satisfy momentum conservation.
\end{itemize}
Notably absent in this discussion is the notion of \emph{supersymmetry} and \emph{supermomentum conservation}. Many of the kinematic variables will have a suitable supersymmetric extension as well, however the precise nature of this is dependent on the physical theory at hand. To keep the discussion in this chapter general and applicable to generic theories, we postpone any supersymmetric extensions (such as \emph{supertwistors} and \emph{supermomentum twistors}) to chapter \ref{sec:AMP}. 

Our interest in the study of positive geometries will be aided by a good understanding of the various kinematic spaces. These kinematic spaces will play the role of an ambient space for the positive geometries of interest. The ABHY associahedron lives in the space of planar Mandelstam variables, the amplituhedron lives in momentum twistor space, the momentum amplituhedron in four-dimensional spinor-helicity space, and the ABJM momentum amplituhedron lives in three-dimensional spinor-helicity space. Additionally, in chapter \ref{sec:DUAL} we will give a general framework for positive geometries for loop integrands in dual space. We will regularly want to relate these positive geometries, which relies on our ability to translate between the various kinematic variables. 

%% file: chapters/amplitudes.tex
\chapter{Scattering Amplitudes}\label{sec:AMP}

We are now ready to delve into the modern techniques for scattering amplitudes. Having introduced Grassmannian geometry and the various kinematic spaces and their interrelations in the previous chapters should significantly streamline some of the topics discussed in this chapter. The topic of scattering amplitudes is immensely broad and this thesis is not intended to give a comprehensive literary overview. There are many books \cite{Elvang:2013cua, Henn:2014yza} and review articles/lecture notes \cite{Badger:2023eqz,Cheung:2017pzi, Taylor:2017sph,Dixon:2013uaa} which are far more well-suited for this purpose. The aim of this chapter is rather to give a concise and largely self-contained introduction of the aspects of scattering amplitudes which are of importance to the topic of positive geometries which we will encounter in the coming chapters.

\section[head={Basic Properties},tocentry={Basic Properties of Scattering Amplitudes}]{Basic Properties of Scattering Amplitudes}

We will start by reviewing some of the basic properties of scattering amplitudes. Our starting point is QFT, which places some fundamental constraints on the type of functions that can appear as scattering amplitudes. These are most commonly summarised as \emph{unitarity}, \emph{locality}, and \emph{causality}. Unitarity is a consequence of the quantum nature, and is often interpreted as being equivalent to `the sum of probabilities must add up to one'. In terms of scattering, this is reflected in the unitarity of the $S$-matrix: $S^\dagger S=\unit$. Locality tells us that interactions must happen through the exchange of some (virtual) particle. The consequence for scattering amplitudes is \emph{factorisation}, which can be understood most easily by looking at tree-level Feynman diagrams. The internal edges of Feynman diagrams represent off-shell momenta and contribute a factor $1/(P^2-m^2)$. When $P^2-m^2$ approaches zero, the contribution from diagrams that include this particular pole will dominate, and the internal line can be interpreted as going on-shell. This means that if we take the \emph{residue} on this pole, we get contributions from Feynman diagrams calculating a `left' and a `right' amplitude, connected through an on-shell state. Unitarity requires us to sum over all the intermediate on-shell states. This is often represented diagrammatically as:
\begin{align}
	\mathop{\Res}_{P^2-m^2\to 0} A_n =\quad \vcenter{\hbox{\includegraphics[width=45mm]{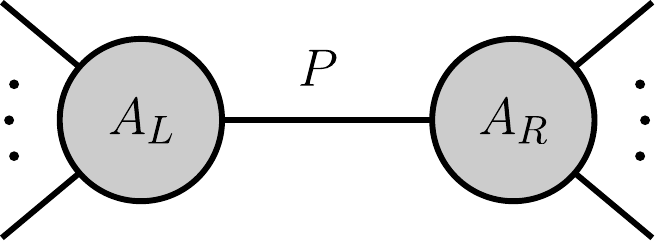}}}\,.
\end{align}
Causality is related to the traditional $i\epsilon$ prescription, however the way in which causality is reflected in higher loop amplitudes is currently not well understood. One of the consequences of causality is that scattering amplitudes have to satisfy the \emph{Steinman relations}, which places constraints on which consecutive branch cuts of an amplitude need to vanish. A not fully understood extension of the Steinman relations \cite{Caron-Huot:2019bsq, Caron-Huot:2019vjl, Caron-Huot:2020bkp} has recently been important in calculations of the \emph{symbol} of scattering amplitudes \cite{Goncharov:2010jf,Papathanasiou:2022lan}, a mathematical tool which allows one to efficiently record the structure of certain higher loop amplitudes.

\subsection{\texorpdfstring{Large-$N$ Limit and Colour Ordering}{Large-N Limit and Colour Ordering}}\label{sec:AMP_large-N}

To introduce the large-$N$ limit and colour ordering we will focus on $SU(N)$ Yang-Mills theory, however the ideas presented here can be applied more broadly to any theory with a $U(N)$ or $SU(N)$ colour/flavour group. In Yang-Mills theory, a three-points interaction is proportional to the structure constant $f^{abc}=\tr{[T^a,T^b]T^c}$, and the propagator is proportional to
\begin{align}
	\<A_{\mu\,b}^{a}A_{\nu\,a}^b\>\propto \delta^a_d \delta^b_c-\frac{1}{N} \delta^a_b\delta^c_d\,.
\end{align}
The last term can be ignored when considering pure gluon amplitudes due to the so-called \emph{$U(1)$ decoupling identity}. In the present work, we will often ignore this term for the reason that it is suppressed by a factor of $N$, and it thus gives negligible contributions in the \emph{large-$N$ limit}.

\begin{figure}
	\centering
	\includegraphics[width=0.6\textwidth]{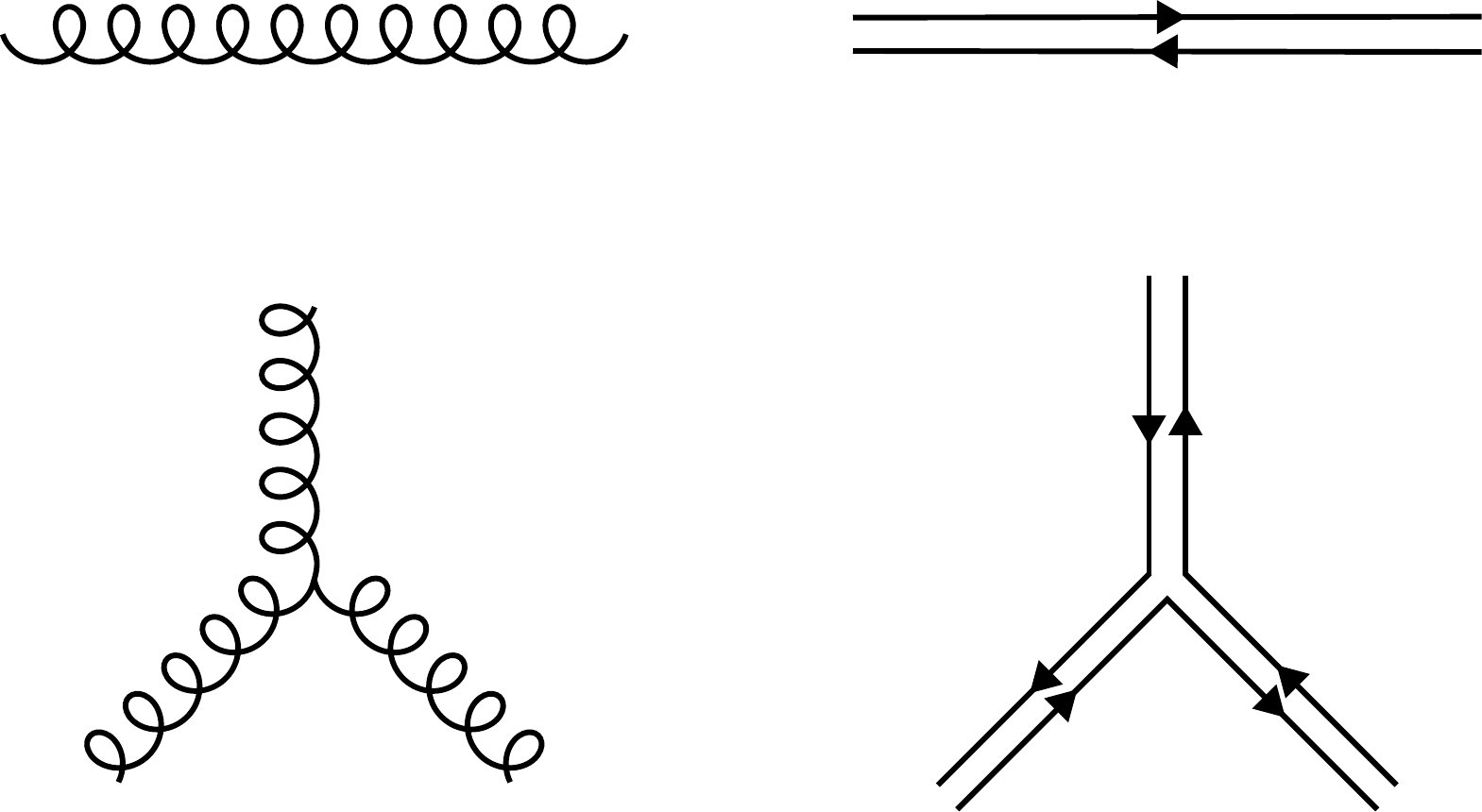}
	\caption{The double line notation for the gluon propagator (top), and the tree-point vertex (bottom).}
	\label{fig:double-line}
\end{figure}

In \cite{tHooft:1973alw}, 't Hooft introduced the \emph{double line notation}, where we replace the gluon lines in a Feynman diagram by a double line which keeps track of `where the colour indices flow'. This is illustrated in figure \ref{fig:double-line}.
Any closed loop in one of these \emph{fat graphs} contributes a factor of $N$, a propagator a factor $1/N$, and a three-points vertex a factor $N$. This suggests a rearranging of the scattering amplitude in terms of powers of $N$. It is clear that this is a `topological' expansion of the scattering amplitude, where each order in the expansion consists of Feynman diagrams with a given topology. It turns out that the leading order is given by planar Feynman diagrams, this is why the large-$N$ limit is often referred to as the \emph{planar limit}. Scattering amplitudes simplify drastically when considering the planar limit, as was already observed by `t Hooft in \cite{tHooft:1974pnl}. This added simplicity has been an important factor to many modern scattering amplitude techniques.

When expanding a Feynman diagram with the above rules for colour factors, it is clear that we obtain a large number of traces of the generators $T^a$, which are related through Fierz identities. The different topological sectors in the large-$N$ expansion correspond to different trace structures. For tree-level amplitudes, we can expand the amplitude into pieces with a single trace as
\begin{align}
	A_n^{\text{tree}} = \sum_{\text{perms. }\sigma} A_n[1\sigma(2\cdots n)] \tr{T^{a_1}T^{a_{\sigma(2)}}\cdots T^{a_{\sigma(n)}}}\,,
\end{align}
where $A_n[1\sigma(2\cdots n)]$ is the \emph{colour-ordered}, or \emph{partial} amplitude. This essentially tells us that all tree-level Feynman diagrams are planar for a certain ordering of the external particles. For higher loops this is not the case, and there are diagrams which have products of multiple traces in their colour structure. However, these diagrams are necessarily non-planar, and in the large-$N$ limit they can be ignored. Hence, when focusing on the leading term in the $1/N$ expansion, we always have a well-defined notion of particle ordering. In the remainder of this thesis we will exclusively deal with colour-ordered amplitudes in the large-$N$ limit. The relation to the full `colour-dressed' amplitudes is always implied, and we shall abuse terminology by referring to these partial amplitudes simply as `amplitudes'.

Lastly, we mention that these colour ordered amplitudes satisfy certain non-trivial relations. These are known as the \emph{Kleiss-Kuijf relations} \cite{Kleiss:1988ne, DelDuca:1999rs}, and can be summarised as 
\begin{align}
	A_n^{\text{tree}}=(-1)^{n_\beta}\sum_{\sigma\in\alpha \Sh \beta^T} A_n^{\text{tree}}(1,\sigma,n)\,,
\end{align}
where $n_\beta$ denotes the number of elements in the ordered set $\beta$, and $\beta^T$ is $\beta$ in reverse order. The \emph{shuffle product} $\alpha \Sh \beta^T$ denotes the set of all orderings on $\alpha\cup\beta$ which preserves the individual ordering of $\alpha$ and $\beta^T$. The Kleiss-Kuijf relations include the $U(1)$ decoupling relations mentioned above:
\begin{align}
	A_n(12\cdots n) + A_n(213\cdots n) + \ldots + A_n(23\cdots 1n)=0\,.
\end{align}
The Kleiss-Kuijf relations reduce the basis of partial amplitudes to $(n-2)!$. Yang-Mills, as well as a large set of other theories, actually satisfies a stronger set of relations, known as the \emph{BCJ relations} \cite{Bern:2008qj}, which further reduces the number of independent partial amplitudes to $(n-3)!$. An introduction to the BCJ relations and the related \emph{colour-kinematics duality} are beyond the scope of this thesis.

\subsection{Loop Integrands}

When considering loop amplitudes, the most difficult part is often the integration over loop momenta. The loop \emph{integrand}, the expression that we need to integrate to get the full amplitude, is more similar to tree-level amplitudes, being rational functions of Lorentz invariants with simple poles as internal propagators go on-shell. We denote the integrand for the $L$-loop amplitude as $A_n^{(L)}$. Since we only consider integrands in this thesis, there is no room for confusion with the integrated answer.

For some diagram which contributes to $A_n^{(L)}$, there is generally no way to uniquely assign a loop momenta to each loop, which makes it difficult to construct a full integrand. However, we are in luck when we consider amplitudes in the planar limit, where this is not an issue. Using the dual momenta from section \ref{sec:KIN_dual}, we assign a \emph{zone variable} $y_i$ to each loop, which renders the loop integrand a well-defined notion. As an example, to the scalar box integral we attribute the integrand
\begin{align}
	I^\square(p_1,p_2,p_3,p_4)=\vcenter{\hbox{\includegraphics[width=38mm]{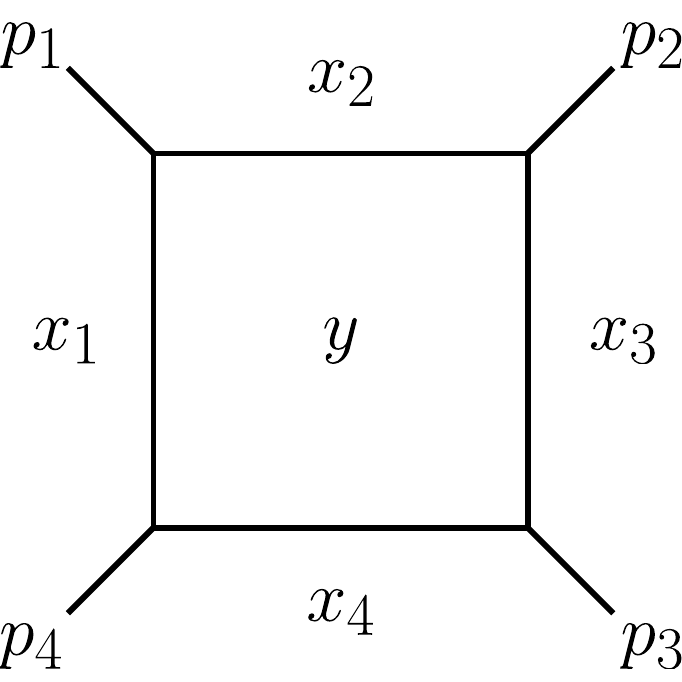}}} = \frac{\dd^Dy}{(y-x_1)^2(y-x_2)^2(y-x_3)^2(y-x_4)^2}\,,
\end{align}
which we can write in the standard momentum space as
\begin{align}
	I^\square(p_1,p_2,p_3,p_4)=\frac{\dd^D\ell}{\ell^2(\ell+p_1)^2(\ell+p_1+p_2)^2(\ell+p_1+p_2+p_3)^2}\,.
\end{align}
It is often useful to decompose the full integrand in a basis of scalar `master integrands'
\begin{align}\label{eq:AMP_loop-basis-expansion}
	A^{(L)} = \sum_\alpha c_\alpha I^{(L)}_\alpha\,,
\end{align}
where $c_\alpha$ are some (external kinematic dependent) coefficients, and the $I^{(L)}_\alpha$ are some basis of $L$-loop scalar integrands. For example, it can be shown \cite{vanNeerven:1983vr, Bern:1992em, Bern:1993kr, tHooft:1978jhc} that one-loop integrands in $D$ dimensions admit a basis of $m$-gon scalar integrands, $m=2,\ldots,D$. The coefficients $c_\alpha$ are traditionally determined by using the \emph{generalised unitarity method}. The idea is similar to factorisation of tree-level amplitudes: if we localise the loop momenta such that $(\ell-P_1)^2=(\ell-P_2)^2=0$, then only Feynman diagrams with these internal lines will contribute. Taking the residue, we find that the integrand factorises into the product of two integrands with a lower number of particles and loops. Setting a propagator which includes a loop momentum on-shell is referred to as a \emph{(unitarity) cut}. In general, we can consider an $N$-line cut ($2\leq N\leq D$), which will isolate some product of simpler amplitudes. The case where $N=D$ is referred to as a \emph{maximal cut}. Requiring that the expansion \eqref{eq:AMP_loop-basis-expansion} correctly reproduces these cuts imposes some (linear) constraints on the coefficients, which can subsequently be determined by basic linear algebra. 

The right choice of basis can significantly simplify these linear dependencies. If for some theory we have a given set of cuts which are sufficient to uniquely determine all coefficients, then we can `linearise' with respect to these cuts. Each cut then only gets a contribution from a single integrand, and no linear algebra needs to be done at all, the coefficients are uniquely determined by the residue at the corresponding cut. A basis which satisfies these properties is said to be \emph{prescriptive} \cite{Bourjaily:2017wjl}.

\section{Recursion Relations}\label{sec:AMP_BCFW}

Recursion relations are a powerful tool to construct tree-level amplitudes from lower-point amplitudes, by utilising the knowledge of the singularity structure of the amplitudes. The idea is to impose a one-parameter deformation of the momenta as
\begin{align}
	p_i^\mu\to \hat{p}_i^\mu(z) = p_i^\mu+ z q_i^\mu\,,
\end{align}
where $q_i^\mu$ are $n$ complex-valued vectors which satisfy
\begin{itemize}
	\item $\displaystyle \sum_{i=1}^n q_i^\mu = 0\,,$
	\item $\displaystyle q_i\cdot q_j=0\quad\forall i,j=1,\ldots,n\,,$
	\item $\displaystyle p_i\cdot q_i=0\quad\forall i=1,\ldots,n\,.$
\end{itemize}
These conditions imply that the shifted momenta satisfy momentum conservation and masslessness:
\begin{align}
	\sum_{i=1}^n \hat{p}_i^\mu(z)=0\,,\quad \hat{p}_i^2=0\,.
\end{align}
Furthermore, any Mandelstam variables in these shifted momenta are at most \emph{linear} in $z$: if we define
\begin{align}
	P_I=\sum_{i\in I}p_i\,,\quad \hat{P}_I(z)=\sum_{i\in I}\hat{p}_i(z)\,,\quad Q_I=\sum_{i\in I}q_i\,,\quad \text{ for some } I \subseteq [n]\,,
\end{align}
then
\begin{align}
	\hat{P}_I^2 =-\frac{s_I}{z_I}(z-z_I)\,,\quad \text{ where } z_I\coloneqq -\frac{s_I}{2 P_I\cdot Q_I}\,.
\end{align}
We use the notation $s_I=P_I^2$ for Mandelstam variables, as introduced in section \ref{sec:KIN_mand}. 

We define the shifted amplitude $\hat{A}_n(z)$ to be the amplitude $A_n$ of some theory evaluated on the shifted momenta. It is then clear that $A_n=\hat{A}_n(0)$. We can now use Cauchy's theorem to find the scattering amplitude:
\begin{align}
	A_n=\hat{A}_n(0)=\mathop{\Res}_{z=0}\frac{\hat{A}_n(z)}{z}=-\sum_{z_I} \mathop{\Res}_{z=z_I} \frac{\hat{A}_n(z)}{z} + B_\infty\,,
\end{align}
where the sum is over all simple poles of $\hat{A}_n(z)$, and $B_\infty$ is a potential boundary term coming from a residue at $z=\infty$. The simple poles of $\hat{A}_n(z)$ can occur only when some shifted propagator $1/\hat{P}_I^2$ goes on shell. In this limit, the shifted amplitude \emph{factorises} into two subamplitudes with a lower number of particles. Assuming that $B_\infty=0$, this then allows us to write the amplitude $A_n$ as 
\begin{align}
	A_n=\sum_I \frac{\hat{A}_L(z_I) \hat{A}_R(z_I)}{P_I^2}= \sum_I\quad \vcenter{\hbox{\includegraphics[width=45mm]{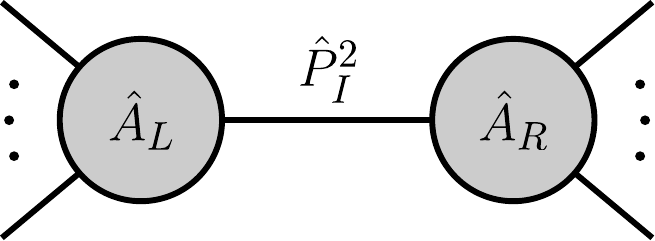}}}\,,
\end{align}
where the sum is over all factorisation channels where $\hat{P}_I^2$ is dependent on $z$. For all theories considered in this thesis, it is safe to assume that $B_\infty$ indeed vanishes.%\footnote{For \nf and ABJM the absence of a `pole at infinity' is guaranteed by the dual conformal invariance of the theories. For $\tr{\phi^3}$ this comes from the hidden `projective invariance', which we will encounter in section \ref{sec:POS_ABHY}.}.

\subsection{BCFW Recursion in Four Dimensions}

A particularly simple incarnation of these on-shell recursion relations is the so-called \emph{Britto-Cachazo-Feng-Witten (BCFW) recursion} \cite{Britto:2004ap, Britto:2005fq}. Specialising to four spacetime dimensions and using spinor-helicity variables, the BCFW shift can be summarised as
\begin{align}
	\tilde\lambda_1&\to\tilde\lambda_1-\alpha \tilde\lambda_n\,,\\
	\lambda_n &\to \lambda_n+\alpha\lambda_1\,,
\end{align}
which send
\begin{align}\label{eq:AMP_BCFW-shift-mom-vec}
	p_1\to\;&\hat{p}_1(\alpha)=p_1-\alpha \lambda_1\tilde\lambda_n\,,\\
	p_n\to\;&\hat{p}_n(\alpha)=p_n+\alpha \lambda_1\tilde\lambda_n\,.
\end{align}
We then only have to consider factorisation channels where particles $1$ and $n$ are on opposite sides of the factorisation. Of course, the choice to single out legs $1$ and $n$ in this manner is purely a choice, and we can equivalently choose any other pair of legs. A different choice will typically have different terms in the recursive expansion, however when summed together all these representations will give the same scattering amplitude.

For planar theories we can also consider a BCFW shift on momentum twistors instead. Since momentum twistors are unconstrained, any shift will leave momentum conservation and masslessness manifest. We consider the shift
\begin{align}
	z_n\to\hat{z}_n = z_n + \beta z_{1}\,,
\end{align}
for some complex parameter $\beta$. Geometrically, we can interpret this shift as moving $\hat{z}_n$ along the line spanned by the points $z_n$ and $z_1$. 

We further note that the BCFW shift in equation \eqref{eq:AMP_BCFW-shift-mom-vec} has a natural geometric interpretation in dual space as well, which is explained further in appendix \ref{sec:APP_WB-BCFW}.

\subsection{BCFW Recursion in Three Dimensions}

In three dimensions, something special happens. Since the deformation vectors satisfy $q^2=0$, we must be able to write it as $q^{\alpha\beta}=\lambda_q^\alpha\lambda_q^\beta$. Since $\lambda_q$ is a two-vector, we can always write $\lambda_q=\alpha \lambda_i+\beta\lambda_j$ (assuming $\<ij\>\neq 0$). This then means that the requirements $q\cdot p_i=q\cdot p_j=0$ only has the trivial solution $q=0$. To find on-shell recursion relations for three-dimensional theories, we give up on the requirement that $\hat{p}_i(z)$ is linear in $z$, and instead define the `BCFW' deformation \cite{Gang:2010gy}:
\begin{align}
	\begin{pmatrix}
		\hat\lambda_i\\ \hat\lambda_j
	\end{pmatrix} = R(z) \begin{pmatrix}
	\lambda_i\\\lambda_j
\end{pmatrix}\,,
\end{align}
where $R(z)$ is a $2\times 2$ matrix. Momentum conservation requires $R(z)$ to be an orthogonal matrix, which we parametrise as
\begin{align}
	R(z)=\begin{pmatrix}
		\frac{z+1/z}{2} & -\frac{z-1/z}{2i} \\ \frac{z-1/z}{2i} & \frac{z+1/z}{2}
	\end{pmatrix}\,.
\end{align}
We recover unshifted momenta for $z=1$, and hence 
\begin{align}
	A_n=\frac{1}{2\pi i}\oint_{z=1} \frac{\hat{A}_n(z)}{z-1}\,.
\end{align}
We again assume that there is no contribution from a pole at infinity, and proceed as before.

\section{The CHY Formalism}\label{sec:AMP_CHY}

In this section we review the \emph{CHY formalism}, named after Cachazo, He, and Yuang \cite{Cachazo:2013iaa, Cachazo:2013gna, Cachazo:2013hca, Cachazo:2013iea, Cachazo:2014nsa, Cachazo:2014xea}, which allows us to write tree-level scattering amplitudes in a wealth of different theories as an integral of some theory dependent integrand over the \emph{moduli space} $\Mfrak_{0,n}$ on the support of the \emph{scattering equations}.

A point in $\mathfrak{M}_{0,n}$ is specified by a set of $n$ holomorphic variables $(z_1,\ldots,z_n)\in \Cbb\Pbb^1$ which are defined up to the automorphism group $PSL(2,\Cbb)$ (the $P$ is often omitted) of the Riemann sphere. As we saw in section \ref{sec:GRASS_proj}, this group consists of \emph{M\"obius transformations}
\begin{align}\label{eq:AMP_Mobius}
	z\mapsto \psi(z)=\frac{A z+B}{C z+D},\quad A,B,C,D \in \Cbb, \text{where }A D - B C =1.
\end{align}

\subsection{The Scattering Equations}
The scattering equations are a set of equations which relate the moduli space to the space of kinematic invariants for massless $n$-particle scattering. To motivate the scattering equations, we follow \cite{Cachazo:2013gna} and search for a meromorphic 1-form $\omega^\mu$ on $\Cbb\Pbb^1$ that has simple poles at all $z_a$ and satisfies
\begin{align}\label{eq:AMP_1form-SE-def}
	\mathop{\Res}\limits_{z=z_a}\omega^\mu=k_a^\mu.
\end{align}
There is a unique solution given by
\begin{align}
	\omega^\mu = \sum_{a=1}^n \frac{k_a^\mu}{z-z_a}\dd z = \sum_{a=1}^n \frac{P^\mu(z)}{\prod_{b=1}^n (z-z_b)}\dd z,
\end{align}
where we define
\begin{align}
	P^\mu(z)\coloneqq \sum_{a=1}^n k_a^\mu \prod_{b\neq a} (z-z_b).
\end{align}
We note that $P^2(z_a)=0$ for all $a\in[n]$. The scattering equations can be obtained by requiring that $P^2(z)$ vanishes for all $z$. This means that $P^\mu(z)$ defines a map from $\Cbb\Pbb^1$ into the space of null-vectors in (complexified) momentum space. 

The condition $P^2(z)=0$ implies $P(z)\cdot\partial P (z)=0$. Evaluating this on $z_a$ yields
\begin{align}
	P(z_a)\cdot \partial P(z_a)=\left(\prod_{b \neq a} (z_a-z_b)^2\right)\left( \sum_{b \neq a} \frac{k_a\cdot k_b}{z_a-z_b}\right).
\end{align}
Since this has to vanish for general nonsingular configurations, this implies the \emph{scattering equations}
\begin{equation}\label{eq:AMP_Scatt-Eq-CHY}
	E_a\coloneqq \sum_{b\neq a} \frac{s_{ab}}{z_a-z_b}=0,\quad a\in[n].
\end{equation}
The equations $E_a$ are implicitly functions of $\bm{z}=\{z_a\}_{a=1}^n$ and $\bm{s}=\{s_{ij}\}_{1\leq i<j\leq n}$, which is sometimes made explicit by writing $E_a(\bm{z})$, or $E_a(\bm{z};\bm{s})$. Of these $n$ equations only $n-3$ are independent. An explicit set of dependencies is given by
\begin{align}\label{eq:AMP_CHY-Dependencies}
	\sum_{a=1}^n E_a = \sum_{a=1}^n z_a E_a =\sum_{a=1}^n z_a^2 E_a =0.
\end{align}
The scattering equations provide $n-3$ equations for $n-3$ variables, which, for generic kinematics, admits $(n-3)!$ solutions.

\subsubsection{Polynomial Form of the Scattering Equations}
It will be useful to introduce the \emph{polynomial form} of the scattering equations, first introduced by Dolan and Goddard in \cite{Dolan:2014ega}. We start by defining
\begin{align}
	g_m=\sum_{a=1}^n z_a^{m+1} E_a\,.
\end{align}
From equation \eqref{eq:AMP_CHY-Dependencies}, $g_{-1},g_0$ and $g_1$ vanish identically. Since 
\begin{align}
	\frac{z_a^{m+1}-z_b^{m+1}}{z_a-z_b}=\sum_{r=0}^m z_a^r z_b^{m-r}\,,
\end{align}
the equations $g_m$ are polynomial in the $z_i$ variables. We find
\begin{align}
	g_m=\sum_{b\neq a} s_{ab}\sum_{r=0}^m z_a^r z_b^{m-r}\,.
\end{align}
If we define the $n\times n$-matrix $Z$ to have components $Z_{ab}=z_a^{b+1}$, then the above definition is equivalent to $\bm{g}=Z \bm{E}$. The determinant of $Z$ is the Vandermonde determinant,
\begin{align}
	\det Z =\prod_{1\le a<b \le n} (z_a-z_b)\,,
\end{align}
which is nonvanishing for generic configurations. Hence, the equations $g_m=0,\,m=2,\ldots,n-2$ are equivalent to the scattering equations. We have already succeeded in casting the scattering equations into an equivalent polynomial form. In \cite{Dolan:2014ega}, it is shown that these equations can be further simplified into the \emph{polynomial scattering equations}
\begin{align}\label{eq:AMP_scatt-eq-poly}
	f_m \coloneqq \sum_{A \in \binom{[n]}{m}} z_A s_A, \quad m =2,\ldots,n-2\,,
\end{align}
where $z_A = \prod_{i\in A}z_i$. 

\subsection{Gauge Fixing the CHY Integral}

In the CHY formalism, we ultimately want to write tree-level scattering amplitudes by integrating some integrand over $\Mfrak_{0,n}$ on the support of the scattering equations. However, an integral of the form
\begin{align}\label{eq:AMP_CHY-int-naive}
	\int_{\mathfrak{M}_{0,n}}I(\bm{z}) \bigwedge_{a=1}^n\dd z_a \prod_{a=1}^n\delta(E_a)\,,
\end{align}
is obviously not well-defined due to the $SL(2)$ `gauge' freedom on $\Mfrak_{0,n}$ and the relations \eqref{eq:AMP_CHY-Dependencies} relating the scattering equations, which reduces both the number of integration variables and the number of delta functions to $n-3$. To make this work, we need to `gauge' fix this integral in an $SL(2)$ invariant way.

In complete generality, we can use this M\"obius redundancy to fix $z_i \to \sigma_i, \, z_j\to \sigma_j$, and $z_k \to \sigma_k$. Explicitly, the M\"obius transformation which accomplishes this is given by
\begin{subequations}\label{eq:AMP_Mobius-gauge-fix-gen}
	\begin{alignat}{2}
		A&=\sigma_i\sigma_j(z_i-z_j)&&/\Delta+\text{cyclic}\,,\\
		B&=-\sigma_i\sigma_j z_k(z_i-z_j)&&/\Delta+\text{cyclic}\,,\\
		C&=-\sigma_i(z_j-z_k)&&/\Delta+\text{cyclic}\,,\\
		D&=\sigma_i z_i(z_j-z_k)&&/\Delta+\text{cyclic}\,,
	\end{alignat}
\end{subequations}
where $\Delta\coloneqq\sqrt{(z_i-z_j)(z_j-z_k)(z_k-z_i)(\sigma_i-\sigma_j)(\sigma_j-\sigma_k)(\sigma_k-\sigma_i)}$, and `$+ \text{ cyclic}$' refers to the sum over cyclic shifts in $(i,j,k)$. This M\"obius transformation sends an arbitrary point $z\in\Cbb\Pbb^1$ to
\begin{align}\label{eq:gauge-fix-general}
	\psi(z)=-\frac{\sigma_i\sigma_j (z_i-z_j)(z_k-z)+\text{cyclic}}{\sigma_i (z_j-z_k)(z_i-z)+ \text{cyclic}}\,.
\end{align}
For this choice of gauge fixing, we have to remove $\dd z_i\wedge\dd z_j\wedge\dd z_k$ from our integration measure in an $SL(2,\mathbb{C})$ invariant way. To this extent, note that under a generic $z\to \frac{Az + B}{Cz+D}$, we have $\dd z \to \frac{AD-BC}{(Cz+D)^2}\dd z$. In our case, the form $\dd z_i\wedge\dd z_j\wedge\dd z_k$ picks up a factor
\begin{align}
	\frac{1}{[(C z_i-D)(C z_j-D)(C z_k-D)]^2} = \frac{(z_i-z_j)(z_j-z_k)(z_k-z_i)}{(\sigma_i-\sigma_j)(\sigma_j-\sigma_k)(\sigma_k-\sigma_i)}\,,
\end{align}
and hence the measure
\begin{align}
	\dd \mu \coloneqq \frac{\dd z_i\wedge\dd z_j\wedge\dd z_k}{(z_i-z_j)(z_j-z_k)(z_k-z_i)}\,,
\end{align}
is M\"obius invariant. The appropriate top-form on $\mathfrak{M}_{0,n}$ is thus given by $\bigwedge_{a=1}^n \dd z_a/\dd\mu$. 

Next, we want remove three of the delta functions in $\prod_{a=1}^n\delta(E_a)$, say $E_p, E_q,$ and $E_r$. We note that under \eqref{eq:AMP_Mobius} the delta functions transform as $\delta(E_a)\to (Cz_a+D)^{-2}\delta(E_a)$, and that $z_{p}-z_{q}$ transforms as $z_{p}-z_{q}\to(z_{p}-z_{q})[(Cz_p+D)(Cz_q+D)]^{-1}$. Hence, we find that the combination $\delta(E_p)\delta(E_q)\delta(E_r)/[(z_p-z_q)(z_q-z_r)(z_r-z_p)]$ is $SL(2)$ invariant. This motivates us to define
\begin{align}
	\prod_{a=1}^n\delta(E_a)\to {\prod_a}' \delta(E_a)\coloneqq (z_p-z_q)(z_q-z_r)(z_r-z_p)\prod_{a\neq p,q,r} \delta(E_a).
\end{align}

\subsection{Scattering Amplitudes from CHY}

Now that we know how to properly gauge fix all the redundancies in \eqref{eq:AMP_CHY-int-naive}, we can consider the well-defined
\begin{align}\label{eq:CHY-integral}
	A_n=\int_{\Mfrak_{0,n}} I(\bm{z})\dd \mu_n^{\text{CHY}}=\int_{\Mfrak_{0,n}} I(\bm{z}) z_{ij}z_{jk}z_{ki}z_{pq}z_{qr}z_{rp}\bigwedge_{a\neq i,j,k}\dd z_a \prod_{a\neq p,q,r} \delta(E_a)\,,
\end{align}
where the second equality defines the `CHY measure' $\dd \mu_n^{\text{CHY}}$, and $z_{ab}\coloneqq z_a-z_b$. This integral completely localizes on the support of the scattering equations 
\begin{align}
	A_n=\sum_{\text{sols.}\,\bm{\xi}_i} \frac{I(\bm{\xi}_i)}{J(\bm{\xi}_i)}\,,
\end{align}
where 
\begin{align}\label{eq:CHY-integral-GF}
	J(\bm{z})=\left|\frac{\partial\bm{E}}{\partial\bm{z}}\right|/z_{ij}z_{jk}z_{ki}z_{pq}z_{qr}z_{rp}\,,
\end{align}
with ${\partial\bm{E}}/{\partial\bm{z}}$ the $(n-3)\times(n-3)$-dimensional Jacobian matrix with entries $\partial E_a/\partial z_b$ where the indices run over $a\in[n]\setminus\{p,q,r\},\,b\in[n]\setminus\{i,j,k\}$. 

The only distinguishing feature of scattering amplitudes in various theories are the different integrands $I(\bm{z})$. A list of `CHY constructable' theories and their associated integrands can be found in \cite{Cachazo:2016njl}. It is worth noting that the integrands typically split up into two `half integrands', which are the natural building blocks for the different integrands. The simplest of the half integrands is the so-called \emph{Parke-Taylor factor}
\begin{align}
	I^{\PT}(\alpha)\coloneqq\frac{1}{(z_{\alpha(1)}-z_{\alpha(2)})(z_{\alpha(2)}-z_{\alpha(3)})\cdots (z_{\alpha(n)}-z_{\alpha(1)})}\,,
\end{align} 
which roughly corresponds to the addition of a colour structure to the resulting amplitude. We recall that the only structure in bi-adjoint $\phi^3$ theory are the two independent colour structures. The CHY integrand for the double colour ordered amplitudes consist of two Parke-Taylor factors, and we have
\begin{align}\label{eq:AMP_CHY-bas}
	m_n(\alpha|\beta) = \int_{\mathfrak{M}_{0,n}} I^{\PT}(\alpha) I^{\PT}(\beta)\dd\mu_n^{\text{CHY}}\,.
\end{align}

\subsection{Scattering Equations in Four Dimensions}\label{sec:AMP_scatt-eq-4D}

The CHY formalism and the scattering equations presented above should be interpreted in `generic' spacetime dimensions. When restricting to a specific number of $D$ dimensions, the Mandelstam variables appearing in \eqref{eq:AMP_Scatt-Eq-CHY} can no longer be considered as independent due to the Gram determinant conditions \eqref{eq:KIN_gram}. When considering four-dimensional spacetime, the Gram conditions can be trivialised by using spinor-helicity variables. 

Reformulating the scattering equations in spinor-helicity space allows one to split up the $n$-particle scattering equations into $n-3$ disjoint sectors labelled by $k=2,\ldots,n-2$. The $k$\textsuperscript{th} sector admits $E_{n-3,k-2}$ solutions \cite{Cachazo:2013iaa}, where $E_{n,k}$ are the \emph{Eulerian numbers} which count the number of permutations of $[n]$ with exactly $k$ ascents. Since summing over all such sets simply recovers the full set of permutations of $[n]$, we find that
\begin{align}
	\sum_{k=2}^{n-3} E_{n-3,k-2}=(n-3)!\,,
\end{align}
and we recover the full set of solutions to the scattering equations when summing over all helicity sectors. For the cases where $k=2,n-2$ the scattering equations only admit a single solution. These `MHV' and `\MHVbar' solutions takes a particularly simple form, and can be found in appendix \ref{sec:scatt-eq-4D}.

There are various different ways in which the four-dimensional scattering equations can be represented, several of which will be encountered in this thesis. To summarise, a few particularly useful equivalent ways to write the scattering equations are
\begin{subequations}
\begin{alignat}{2}
	&\lambda_a^\alpha = t_a \sum_{m=0}^{k-1} \rho_m^\alpha z_a^m,\quad &&0=\sum_{a=1}^n t_a\tilde\lambda_a^\alphadot z_a^m,\,\\
	& C^\perp(\bm{z},\bm{t})\cdot \lambda^T = \nul_{(n-k)\times 2},\quad && C(\bm{z},\bm{t})\cdot \tilde\lambda^T=\nul_{k\times 2}\,,\\
	&\lambda_j^\alpha = \sum_{i=1}^k \frac{\lambda_i^\alpha}{(ij)}, && \tilde\lambda_i^\alphadot = \sum_{j=k+1}^n \frac{\tilde\lambda_j^\alphadot}{(ji)}\,.
\end{alignat}
\end{subequations}
The first equations are called the \emph{Witten-RSV equations} \cite{Witten:2003nn, Roiban:2004vt, Roiban:2004yf}, and they depend on the additional auxiliary variables $t_a$ and $\rho_m^\alpha$.
In the second equations, which we shall refer to as the \emph{Grassmannian scattering equations} \cite{Arkani-Hamed:2009kmp}, we define $C(\bm{z},\bm{t})$ as the $k\times n$ matrix with entries $t_a z_a^{m-1},\,a=1,\ldots,n,\,m=1,\ldots,k$, and $\lambda$, $\tilde\lambda$ are interpreted as $2\times n$ matrices. In the last equation the indices run over $j=k+1,\ldots,n$, $i=1,\ldots,k$, and we define $(ab)=(z_a-z_b)/s_as_b$ for some auxiliary variables $s_a$. These last equations are referred to as the \emph{ambitwistor scattering equations} \cite{Geyer:2014fka}, as they were first discovered in the framework of \emph{ambitwistor string theory} \cite{Mason:2013sva, Geyer:2022cey}. The way to interpret the presence of these auxiliary variables is as follows: we have more equations in more unknowns, but the solutions for the $z$s in terms of the kinematic variables $\lambda,\tilde\lambda$ are equivalent for the different formulae. When joining the solutions of the $z$s over all helicity sectors, we recover the full set of solutions to the scattering equations \eqref{eq:AMP_Scatt-Eq-CHY}. In appendix \ref{sec:scatt-eq-4D} we show how these different formulations of the scattering equations are related. 

\section[head={Amplitudes in $\Ncal=4$ SYM},tocentry={Scattering Amplitudes in $\Ncal=4$ SYM}]{\texorpdfstring{Amplitudes in $\Ncal=4$ SYM}{Scattering Amplitudes in N=4 SYM}}\label{sec:AMP_nf}

As we have seen in section \ref{sec:INT_nf}, planar \nf is a theory with a tremendous amount if symmetry, which allows for many particularly simple descriptions of its scattering amplitudes. The superconformal, dual superconformal, and Yangian symmetries which hold at tree-level for \nf, also hold for loop integrands in the planar limit. These simple descriptions have fuelled many of the recent advances in scattering amplitudes, and these techniques have often later found their way to more general applications for different theories. This trend also persists in this thesis, and the results for planar \nf will form a guiding example for many of the discussion to follow. This section will be used to summarise some of the most important techniques which have been developed for \nf. But first, let us note a few general properties of gluon amplitudes in non-supersymmetric non-planar Yang-Mills theory in four dimensions, which are also applicable for \nf. 

Apart from their momentum and colour charge, the only meaningful property of gluons in four spacetime dimensions is their \emph{helicity}. It is customary (and useful) to organise (supersymmetric) Yang-Mills scattering amplitudes in terms of the total helicity of the outgoing particles. We let $k$ denote the number of negative helicity gluons, assuming all particles outgoing. At tree-level, all scattering amplitudes with $k=0,1,n-1,n$ vanish for $n>3$, a property that holds to all orders in perturbation theory for \nf. For this reason, the $k=2$ case is often known as the \emph{maximally helicity violating (MHV)} sector. The $k=3$ sector is known as \emph{next-to-MHV (NMHV)}, and so on. In general, the N\textsuperscript{$K$}MHV sector describes amplitudes with $k=K+2$ negative helicity gluons. Swapping all positive and negative helicities is known as \emph{parity conjugation} and swaps $k \leftrightarrow n-k$. The corresponding `anti-MHV' amplitude is often denoted \MHVbar, sometimes called `googly'. The exception to this classification is for $n=3$, in which case the $k=1,2$ amplitudes are non-zero for complexified momenta.

We further note that scattering amplitudes in four-dimensional Yang-Mills are \emph{covariant} under a little group transformation $\lambda_i\to t_i\lambda_i$, $\tilde\lambda_i\to t_i^{-1}\tilde\lambda_i$. That is, if we consider an amplitude where particle $i$ has helicity $h_i$, then under such a little group transformation the amplitude transforms as
\begin{align}
	A(1^{h_1}2^{h_2}\cdots n^{h_n})\to \prod_{i=1}^n t_i^{-2h_i} A(1^{h_1}2^{h_2}\cdots n^{h_n})\,.
\end{align}
This can be argued from the fact that polarisation vectors are written in terms of spinor-helicity variables as
\begin{align}
	e_+^{\alpha\alphadot} = \frac{\xi^\alpha\tilde\lambda^\alphadot}{\<\lambda\xi\>}\,,\quad e_-^{\alpha\alphadot} = \frac{\lambda^\alpha\tilde\xi^\alphadot}{[\tilde\lambda\tilde\xi]}\,,
\end{align}
for arbitrary generic reference spinors $\xi$ and $\tilde\xi$. Since these scale as $e_h\to t^{-2h} e_h\,,h=\pm1$, the full amplitude must pick up such a factor for each particle.

\subsection{Superamplitudes}\label{sec:AMP_superamplitudes}

Using the on-shell superspace introduced in the introduction \ref{sec:INT_nf}, we can define scattering amplitudes of the superfields $A(\Phi_1,\ldots,\Phi_n)$, called \emph{superamplitudes}. We can isolate a `component' scattering amplitude of certain particles in the supermultiplet by considering the part multiplying a specific combination of $\eta$s. Referring back to \eqref{eq:INT_supermultiplet-nf}, we can select a specific state for the $i$\textsuperscript{th} particle with the operators listed in table \ref{tab:AMP_superstate}.
\begin{table}
	\centering
\begin{tabular}{l|c|c|c|c|c}
	state & $g^+$ & $\psi^A$ & $\phi^{AB}$ & $\bar\psi^{A}$ & $g^-$ \\\hline
	operator & 1 & $\partial_i^A$ & $\partial_i^A\partial_i^B$ & $\partial_i^B\partial_i^C\partial_i^D$ & $\partial_i^1\partial_i^2\partial_i^3\partial_i^4$
\end{tabular} 
\caption{Operators that select specific states of the supermultiplet. We use $\partial_i^A = \partial/\partial\eta_i^A$.}
\label{tab:AMP_superstate}
\end{table}
For example, the gluon amplitude with negative helicity particles at indices $i_1,\ldots,i_l$, and positive helicities at the remaining particles, can be obtained as
\begin{align}
	\partial_{i_1}^4\cdots\partial_{i_l}^4 A(\Phi_1,\ldots,\Phi_n)\,,
\end{align}
where $\partial_i^4=\partial_i^1\partial_i^2\partial_i^3\partial_i^4$. In the chiral on-shell superspace formalism the supercharges take the form
\begin{align}
	Q^A = \sum_{i=1}^n \tilde\lambda_i\frac{\partial}{\partial \eta_{iA}}\,,\quad \tilde{Q}_A = \sum_{i=1}^n \lambda_i \eta_{iA}\,.
\end{align}
Explicitly, 
\begin{align}
	\delta^8(\tilde{Q})= \frac{1}{2^4}\prod_{A=1}^4\prod_{i,j=1}^n\<ij\>\eta_{iA}\eta_{jA}\,.
\end{align}
When expanding the superamplitude in the Grassmann variables, $SU(4)_R$ symmetry requires each monomial to have a Grassmann degree $4(K+2)$, which precisely corresponds to the N\textsuperscript{$K$}MHV sector. The full superamplitude can then be organised as
\begin{equation}
	A_n = A_{n,2}+ A_{n,3}+\ldots+A_{n,n-2},
\end{equation}
where $A_{n,k}$ has Grassmann degree $4(K+2)$. In practice, we will often factor out a `supermomentum conserving' delta function $\delta^8(\tilde{Q})=\delta^8(\lambda\cdot \eta^T)$ from the superamplitudes, in which case the remaining function has Grassmann degree $4K=4(k-2)$. This delta function manifests $\tilde{Q}$ conservation, to also manifest $Q$ conservation we have to make sure that the superamplitude is invariant under translation of $\eta$ by $\eta_a^I\to \eta_a^I+\<\xi^I\lambda_a\>$. Under such a shift $\delta^8(\tilde{Q})$ is invariant, since $\delta^8(\lambda_a\eta_a^I+\xi_\alpha^I\sum_{a=1}^n\lambda_a^\alpha\tilde\lambda_a^\alphadot)=\delta^8(\tilde{Q})$, where we use that we are localised on the support of momentum conservation $\sum_{a=1}^n\lambda_a^\alpha\tilde\lambda_a^\alphadot=0$.

In non-chiral superspace, we instead have the supercharges
\begin{align}
	\tilde{q}^{\alphadot\alpha}=\sum_{i=1}^n \tilde\lambda_i^\alphadot\eta_i^\alpha\,,\quad q^{\alpha\alphadot} = \sum_{i=1}^n \lambda_i^\alpha\tilde\eta_i^\alphadot\,. 
\end{align}
The superamplitudes $A_{n,k}$ are of Grassmann degree $(2(n-k),2k)$ in $(\eta,\tilde\eta)$. However, as argued in \cite{He:2018okq}, $\delta^4(q)\delta^4(\tilde{q})$ vanishes on the support of momentum conservation, hence it is necessary to strip off a $\delta^4(q)$ or $\delta^4(\tilde{q})$ from the superamplitude (both choices are equivalent up to a potential overall sign), and the total superamplitude has Grassmann degree $2n-4$.

The three-particle amplitudes are exceptions to the constraint that $2\leq k \leq n-2$. For complexified momenta, there are two non-trivial 3-point amplitudes corresponding to helicities $k=1,2$. In chiral superspace they are explicitly given by
\begin{align}
	A_{3,2} &= \frac{\delta^8(\tilde{Q})}{\<12\>\<23\>\<31\>}\,,\\
	A_{3,1} &= \frac{\delta^{4}([12]\eta^A_3+[23]\eta^A_1+[31]\eta^A_2)}{[12][23][31]}\,.
\end{align}
The $Q$ conservation of $A_{3,1}$ is ensured by the Schouten identity, and no momentum conservation is necessary. A general MHV superamplitude has Grassmann degree 8, which is completely fixed by $\delta^8(\tilde{Q})$. At tree-level, the MHV superamplitudes are
\begin{align}
	A_{n,2} = \frac{\delta^{8}(\tilde{Q})}{\<12\>\<23\>\cdots\<n1\>}\,.
\end{align} 
To streamline future discussions, we note that we can uplift twistors $W_a^A$ to \emph{supertwistors}
\begin{align}
	\Wcal_a^{\text{A}}\coloneqq \begin{pmatrix}
		W_a^A\\ \eta_a^I
	\end{pmatrix} = \begin{pmatrix}
	\tilde\lambda_a^\alphadot\\ \tilde\mu^\alpha\\ \eta_a^I
\end{pmatrix}\,,
\end{align}
which are to be understood as points in $\Cbb\Pbb^{3|4}$, and superconformal transformations act on these variables as $SL(4|4)$. We can also enrich the space of dual momenta by Grassmann variables $\theta$ to manifest supermomentum conservation. Explicitly, we define 
\begin{align}
	\theta_{a,I}^\alpha -\theta_{a+1,I}^\alpha = \lambda_a^\alpha \eta_{a I}\,.
\end{align}
To go to momentum twistor space, we define $\chi$ through a Grassmannian analogue of the incidence relations as
\begin{align}
	\chi_a^I = \theta_a^{\alpha I}\lambda_{a\alpha} = \theta_{a+1}^{\alpha I}\lambda_{a \alpha}\,.
\end{align}
Then, we define the \emph{supermomentum twistors}
\begin{align}
	\Zcal_a^{\text{A}}\coloneqq\begin{pmatrix}
		z_a^A\\ \chi_a^I
	\end{pmatrix} = \begin{pmatrix}
	\lambda_a^\alpha\\ \mu_a^\alphadot\\ \chi_a^I
\end{pmatrix}\,.
\end{align}
Much like the standard momentum twistors, these supermomentum twistors are unconstrained variables. We can retrieve the usual supervariables by analogy to equation \eqref{eq:KIN_lambda-tilde=muQ} as
\begin{align}
	\eta = \chi\cdot Q\,.
\end{align}
Similar to how twistors and momentum twistors linearise conformal and dual conformal transformations, respectively, these supertwistors and supermomentum twistors linearise superconformal and dual superconformal transformations.

\subsection{Singularity Structure}

We will take a brief moment to discuss some of the singularities that are present in \nf scattering amplitudes. As already discussed at several points before, tree-level scattering amplitudes, including those in \nf, have \emph{factorisation channels} when a propagator goes on shell. In the colour ordered case, this corresponds to a planar Mandelstam variables $X_{ij}$ going to zero. At this pole, the scattering amplitude factorises into the product of two smaller amplitudes. In the case where one of the factorised amplitudes is a three-points amplitude, we notice that $X_{ii+2}=s_{ii+1}=\<ii+1\>[ii+1]\to0$ has the two solutions $\<ii+1\>\to0$ and $[ii+1]\to0$. These two scenarios occur when $\lambda_i\propto \lambda_{i+1}$ and $\tilde\lambda_i\propto\tilde\lambda_{i+1}$, respectively, and they are known as \emph{collinear limits}. We can consider two adjacent collinear limits by sending, for example, $\<i-1i\>\to0\,,\<ii+1\>\to0$. If we want to keep $\<i-1i+1\>$ generic, then the only option for such a limit is to send $\lambda_i^\alpha\to0$. Alternatively, by taking $[i-1i]\to0\,,[ii+1]\to0$ we have $\tilde\lambda_i^\alphadot\to0$. These are known as \emph{soft limits}, and they correspond to the momentum becoming `soft': $p_i^\mu\to0$. These collinear and soft singularities have been an important topic of study since the '70s, see, for example, \cite{Dixon:2013uaa} or \cite{Henn:2014yza} and references therein.

For loop integrands we have an additional important singularity, coming from a loop propagator going on-shell. We will focus on one-loop integrands for the moment. Such a locus has the interpretation of an on-shell particle running in this edge of the loop, and the resulting residue at this singularity is therefore a tree-level amplitude in the \emph{forward limit}. To elaborate, using the same logic we used to motivate factorisation at tree-level, the result of such a residue can be interpreted as a tree-level amplitude with two extra legs with equal and opposite momentum. In general, for an $L$-loop $n$ particle integrand, this residue will yield an $(L-1)$-loop $n+2$ particle integrand where the two extra particles are in the forward limit. These operations might not be well-defined, as evaluating tree-level amplitudes on the forward limit can yield divergences. However, it was shown in \cite{Caron-Huot:2010fvq} that they exist at one-loop for any supersymmetric theory, and at all loop orders for \nf. We denote these residues pictorially as
\begin{align}
	\mathop{\Res}_{(y-x_1)^2=0}\quad \vcenter{\hbox{\includegraphics[width=25mm]{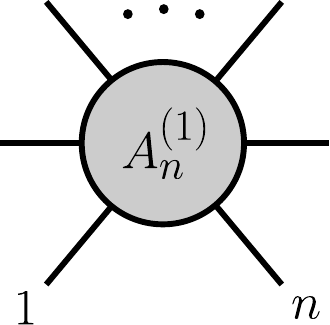}}}\quad=\quad\vcenter{\hbox{\includegraphics[width=25mm]{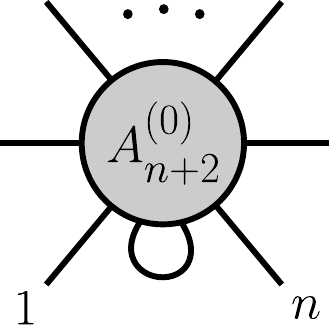}}}\,,
\end{align}
where we use dual momenta to define the cut. The loop at the bottom of the diagram on the right should be interpreted as an on-shell momentum loop which feeds into the tree-level amplitude as two external particles with equal and opposite momenta.

\subsection{Twistor Strings}

We recall from section \ref{sec:KIN_twistors} that we can go from spinor-helicity space to twistor space by doing a half Fourier transform. If we want to write a scattering amplitude $A(\lambda,\tilde\lambda)$ in twistor variables, we can therefore do the integral
\begin{align}
	\int\dd^2\lambda_1\cdots\dd^2\lambda_n\prod_{a=1}^n e^{i\lambda_a^\alpha \tilde\mu_{\alpha\alphadot}}A(\lambda,\tilde\lambda)\delta^4(\sum_a \lambda_a\tilde\lambda_a)\,.
\end{align}
To aid in the integration it is useful to write the momentum conserving delta function as
\begin{align}
	\delta^4(\sum_a\lambda_a\tilde\lambda_a)=\int\dd^4 xe^{-ix_{\alpha\alphadot}\sum_{a=1}^n \lambda_a^\alpha\tilde\lambda_a^\alphadot}\,.
\end{align}
If we now consider an amplitude which is only dependent on $\tilde\lambda$, such as the \MHVbar amplitude, then it can be factored out of the integral, and we are left with
\begin{align}
	A_n^{\text{\MHVbar}}(\tilde\lambda)\int \dd^4 x \prod_{a=1}^n \delta^2(\tilde\mu_{\alpha\alphadot}-x_{\alpha\alphadot}\tilde\lambda_a^\alphadot)\,.
\end{align}
These delta functions imply that every particle must satisfy $\tilde\mu_{\alpha\alphadot}=x_{\alpha\alphadot}\tilde\lambda_a^\alphadot$ for the same $x$. In twistor space this means that all $n$ twistors $W_a^I$ must lie on a line.

This observation is meant to serve as a motivation for the more general ideas relating scattering amplitudes to curves in twistor space. We note that our conventions for twistors differ from those in \cite{Witten:2003nn}, and the results we now present differ by parity conjugation from the discussion above. In \cite{Witten:2003nn} Witten conjectured that $L$-loop N\textsuperscript{$K$}MHV scattering amplitudes are supported on holomorphic curves of degree $K+L+1$, whose genus is bounded by $L$. The natural interpretation is that this curve is the world-sheet of a string, and the amplitudes arise from gluons coupling to the string \cite{Cachazo:2005ga}. One such proposal is Witten's twistor string theory \cite{Witten:2003nn}, whose tree-level amplitudes exactly give the scattering amplitudes of $\mathcal{N}=4$ SYM. Although the support of the amplitudes is naively on both connected and disconnected curves in twistor space and one should sum over all these contributions, it was shown by Roiban, Spradlin, and Volovich (RSV) \cite{Roiban:2004yf} that is sufficient to consider only connected curves.

We can parametrize two degree-$d$ curves in the moduli space $\Mfrak_{0,n}$ as $P_i^\alpha = \sum_{m=0}^d \rho_m^\alpha z_i^m$, where $\alpha=1,2$. The Witten-RSV formula for tree-level N\textsuperscript{$k-2$}MHV scattering amplitudes in $\mathcal{N}=4$ SYM is then given by
\begin{align}\label{eq:AMP_twistor-string}
	A_{n,k}=\int\dd \mu_{n,d}\prod_{i=1}^n\delta^2\big(\lambda_i^\alpha-t_i P_i^\alpha\big)\prod_{m=0}^d \delta^2\big(\sum_{i=1}^n t_i z_i^m\tilde\lambda_i^{\dot\alpha}\big)\delta^4\big(\sum_{i=1}^n t_i z_i^m \eta_{i A}\big),
\end{align}
where the integration measure is defined as
\begin{align}
	\dd \mu_{n,d} = \frac{\dd^{2d+2}\rho\; \dd^nz\;\dd^nt}{\vol[GL(2)]}\prod_{i=1}^n \frac{1}{t_i(z_i-z_{i+1})},
\end{align}
and $d=K+1$ labels the degree of the curve in twistor space. We note that the $2d+2+n+n-4$ integrations that we need to do are completely fixed by the $2n+2(d+1)$ delta functions in the integrand. After pulling out an overall momentum conserving delta function the remaining delta functions completely cancel all integrations. Hence, \eqref{eq:AMP_twistor-string} can be calculated by summing over the solutions to the $2n+2d+2$ polynomial equations
\begin{subequations}\label{eq:RSV-scatt-eq}
	\begin{alignat}{2}
		\lambda_i^\alpha &=\sum_{m=0}^d t_i z_i^m \rho_m^\alpha,\quad&&\text{for }i=1,\ldots,n\\
		0&=\sum_{i=1}^n t_iz_i^m \tilde\lambda_i^{\dot\alpha},&&\text{for } m=0,\ldots,d.
	\end{alignat}
\end{subequations}
These are the \emph{Witten-RSV equations}, and, as we noted in section \ref{sec:AMP_CHY}, they are equivalent to the scattering equations in four dimensions, as is derived in appendix \ref{sec:scatt-eq-4D}. This formula was a historical precursor to the CHY formalism.

We already noted that the four-dimensional scattering equations can be cast in a Grassmannian language:
\begin{align*}
	C^\perp(\bm{z},\bm{t})\cdot \lambda^T = \nul_{(n-k)\times 2},\quad C(\bm{z},\bm{t})\cdot \tilde\lambda^T=\nul_{k\times 2}\,,
\end{align*}
with $C_{ma}=t_az_a^{m-1}$, $a=1,\ldots,n$, $m=1,\ldots,k$. The $k\times n$ matrix $C$ is defined up to a $GL(k)$ transformation, and can thus be understood as an element of the Grassmannian $G(k,n)$. We further note that, as explained in section \ref{sec:KIN_spin-hel}, the $2\times n$ matrices $\lambda$ and $\tilde\lambda$ can be interpreted as elements of $G(2,n)$. The Grassmannian scattering equations are then given the following geometric interpretation: $C$ is a $k$-plane in $n$ dimensions which contains the 2-plane $\lambda$, and its orthogonal complement $C^\perp$ is an $(n-k)$-plane which contains the 2-plane $\tilde\lambda$. This then immediately implies that $\lambda$ is orthogonal to $\tilde\lambda$:
\begin{align*}
	\lambda\cdot\tilde\lambda^T=\nul_{2\times2}\,,
\end{align*}
which is equivalent to the statement of momentum conservation, as noted in equation \eqref{eq:KIN_spin-hel-mom-cons}. These Grassmannian equations linearise the otherwise quadratic constrain of momentum conservation.

Using 
\begin{align}
	\int \dd^{2\times k}\rho \;\delta^{2\times n}\big(\rho\cdot C - \lambda^T\big) = \delta^{2\times(n-k)}\big(C^\perp\cdot\lambda^T\big)\,,
\end{align}
we can rewrite \eqref{eq:AMP_twistor-string} as 
\begin{align}\label{eq:AMP_twistor-string-Grassmannian}
	A_{n,k} = \int \frac{\dd^n z\dd^n t}{\vol[GL(2)]}\frac{\delta^{2k}(C\cdot\tilde\lambda^T)\delta^{2(n-k)}(C^\perp\cdot\lambda^T)\delta^{4k}(C\cdot\eta^T)}{\prod_{i=1}^n t_i (z_i-z_{i+1})}\,.
\end{align}

\subsection{On-Shell Diagrams and Leading Singularities}

Scattering amplitudes are important examples of a more general class of functions known as \emph{on-shell functions}, which are physically relevant functions which are only dependent on the on-shell kinematical data of a scattering process. Other examples of on-shell functions include factorisation channels or lower dimensional residues of scattering amplitudes. On-shell functions were studied in detail in \cite{Arkani-Hamed:2012zlh}.

We can find on-shell functions by `gluing' together lower point amplitudes using `on-shell propagators'. This essentially just means that we consider the product of two or more lower-point amplitudes and integrate over the internal phase space of the on-shell particle which connects them. The rule for an on-shell propagator with momentum $p_I$ is
\begin{align}
	\int \frac{\dd^2\lambda_I\dd^2\tilde\lambda_I\dd^4\eta_I}{\vol[GL(1)]}\,.
\end{align}
We summarise these on-shell functions by their \emph{on-shell diagrams}, which are nothing but graphical representations of the amplitudes and on-shell propagators that make up the on-shell function. For example, the on-shell diagram associated to a factorisation channel is given by
\begin{align}
	 \vcenter{\hbox{\includegraphics[width=45mm]{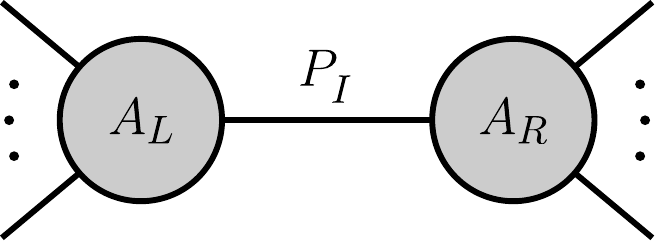}}} = \int \frac{\dd^2 \lambda_I\dd^2\tilde\lambda_I \dd^4\eta_I}{\vol[GL(1)]}A_L(\ldots,\lambda_I,\tilde\lambda_I,\eta_I)A_R(\lambda_I,-\tilde\lambda_I,-\eta_I,\ldots)\,.
\end{align}
We note that the integral is completely localised by the momentum conserving delta functions. There are 8 constraints coming from momentum conservation in $A_L$ and $A_R$. After doing the three integrations, we are left with five constraints, which impose overall momentum conservation, and the additional constraint that $(\sum_{i\in L} p_i)^2=0$. In general the number of excess delta functions is given by 
\begin{align}
	n_\delta = 4 n_V -3 n_I -4\,,
\end{align}
where $n_V$ is the number of vertices and $n_I$ the number of internal edges in the on-shell diagram, and the `$-4$' comes from the overall momentum conservation. In the example considered above, we have $n_\delta=1$, which signifies that there is \emph{one} additional constraint on the external kinematics: $(\sum_{i\in L} p_i)^2=0$. When $n_\delta<0$, a non-trivial integration needs to be done, and we have to specify a contour over which to integrate.

As fundamental building blocks of on-shell diagrams, it is natural to start with the two three-particle amplitudes $A_{3,1}$ and $A_{3,2}$, which we shall denote with white and black vertices, respectively.

\subsubsection{BCFW Bridges} 

We saw in section \ref{sec:AMP_BCFW} that we can recursively build up scattering amplitudes from lower-point amplitudes through the BCFW recursion. This has a natural interpretation in terms of on-shell diagrams. Explicitly, if we attach a \emph{BCFW bridge} to a factorisation channel by adding a white vertex to leg $1$ and a black vertex to leg $n$ connected through an internal edge, then this precisely imposes the BCFW kinematic shift! This means that the BCFW recursion gives us an expression for scattering amplitudes as sums of on-shell diagrams:
\begin{align}\label{eq:AMP_BCFW-on-shell}
	A_n = \sum_{L,R}\quad\vcenter{\hbox{\includegraphics[width=35mm]{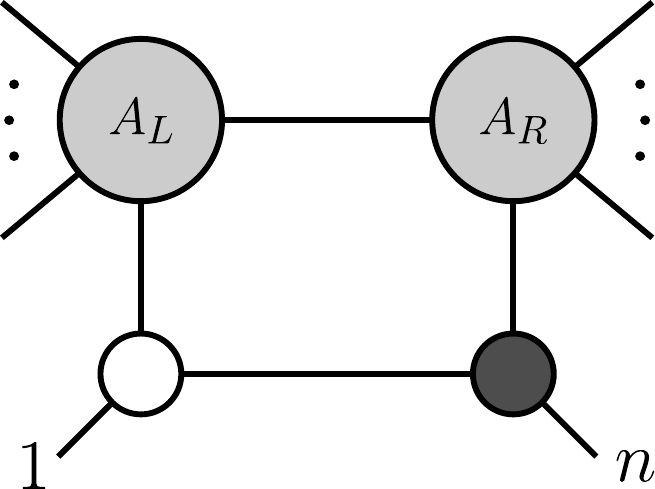}}}\,.
\end{align}
We note that, since we are considering superamplitudes, we also shift $\eta_1^A\to\hat{\eta}_1^A=\eta_1^A-\alpha \eta_n^A$. We can use this recursive relation to express all these on-shell diagrams in terms of fundamental three-particle amplitudes. For example, the four particle amplitude can be expressed as
\begin{align}\label{eq:AMP_n4-on-shell}
	A_{4,2} = \vcenter{\hbox{\includegraphics[width=35mm]{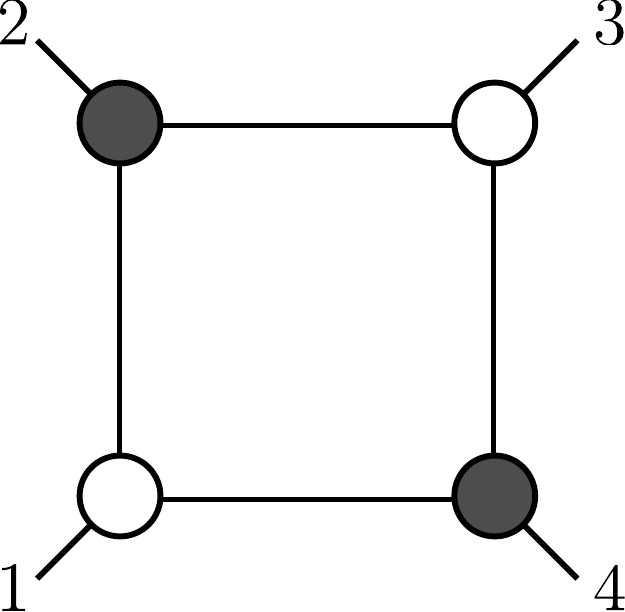}}}\,.
\end{align}
We note that these BCFW diagrams always have four vertices and four internal edges. This implies that $n_\delta=0$, meaning that the integrals are completely localised without imposing any additional constraints on the external kinematics.

We briefly mention that for planar \nf we can further extend the BCFW recursion to loop integrands as well \cite{Arkani-Hamed:2010zjl}. The additional poles coming from forward limits add an extra term in the expansion, which we can write schematically in terms of on-shell diagrams as
\begin{align}\label{eq:AMP_BCFW-on-shell-loop}
	A_n^{(\ell)} = \sum_{L,R}\quad\vcenter{\hbox{\includegraphics[width=40mm]{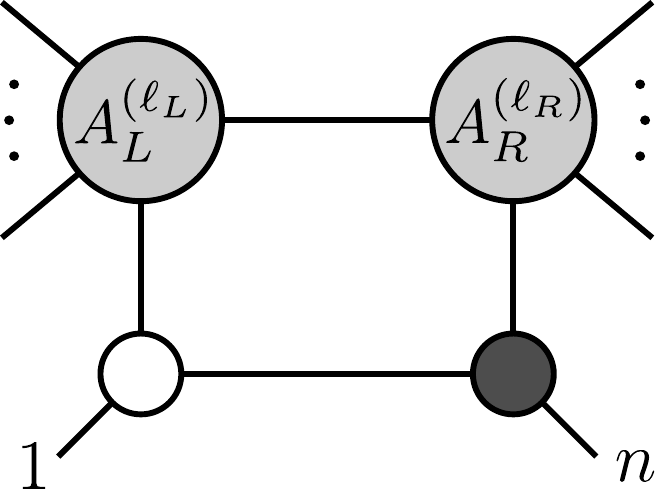}}} \quad+\quad \vcenter{\hbox{\includegraphics[width=35mm]{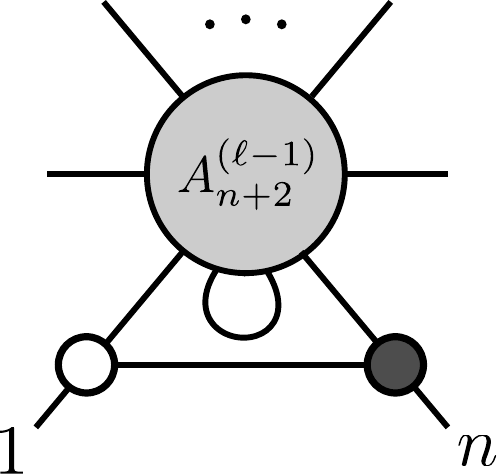}}}\,.
\end{align}

\subsubsection{Leading Singularities}

\begin{figure}
	\centering
	\includegraphics[scale=0.35]{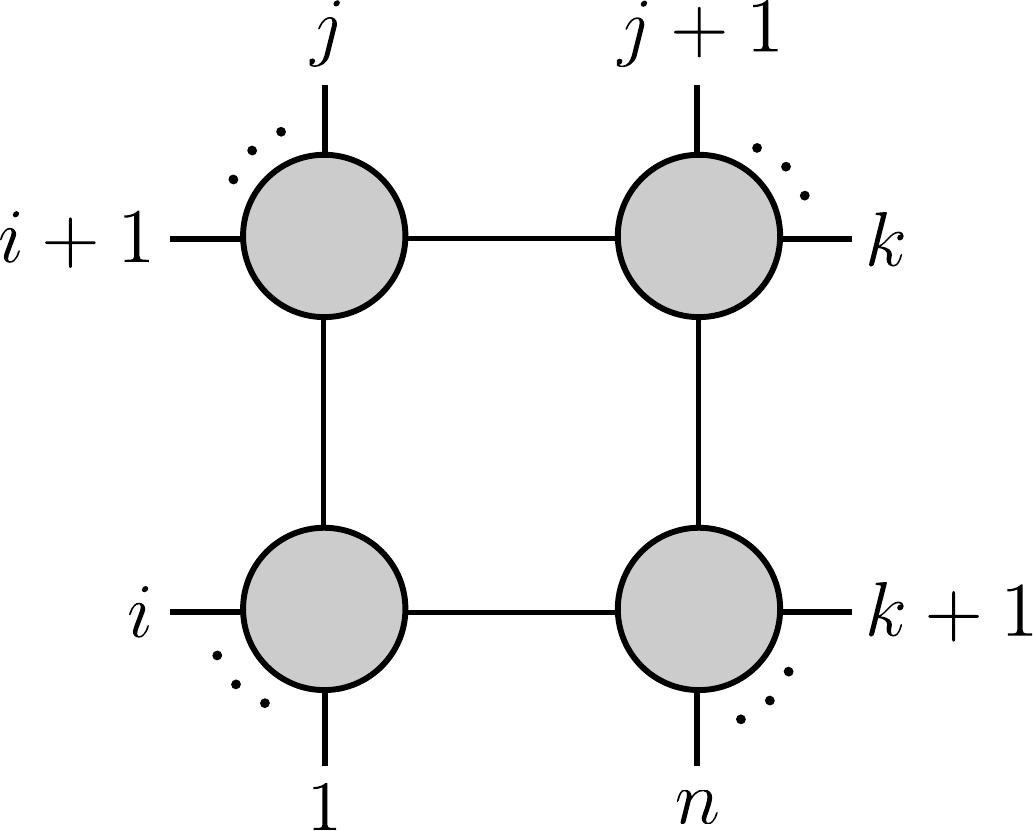}
	\caption{The on-shell diagram corresponding to the leading singularity with\\ $\ell^2=(\ell-x_{1,i+1})^2=(\ell-x_{i+1,j+1})^2=(\ell-x_{j+1,k+1})^2=0$.}
	\label{fig:on-shell-LS}
\end{figure}

Another important type of on-shell functions go by the name of \emph{leading singularities}. They arise by taking a loop integrand and considering the residue where the loop variables are maximally cut (\textit{i.e.} four cuts per loop variable). We will mainly be interested in leading singularities of one-loop integrands, in which case the leading singularities are given by residues where $\ell^2=(\ell-x_{1,i+1})^2=(\ell-x_{i+1,j+1})^2=(\ell-x_{j+1,k+1})^2=0$, where $x_{1,i+1}=p_1+p_2+\ldots+p_i$, $x_{i+1,j+1}=p_{i+1}+\ldots+p_j$, $x_{j+1,k+1}=p_{j+1}+\ldots+p_k$. The result of these residues yields the product of four tree-level amplitudes, summed over internal states. That is, it is an on-shell function with corresponding diagram given in figure \ref{fig:on-shell-LS}.

Since we are in four dimensions, we can choose a basis of bubble, triangle, and box integrands in equation \eqref{eq:AMP_loop-basis-expansion}. The only Feynman diagram which will survive such a residue is the box integrand. This allows us to write a large chunk of the one-loop integrand as a sum over the product of leading singularities and box integrands. In general, there can be additional contributions coming from the bubble and triangle integrands. However, loop integrands in planar \nf are dual conformal invariant, and box integrands are the only one-loop scalar integrands with this property, and hence this is an \emph{exact} expansion \cite{Britto:2004nc, Bern:1994zx}, meaning that the full one-loop integrand can be written as
\begin{align}\label{eq:AMP_1-loop-nf}
	A_n^{(1)}=\sum_{\text{maximal cuts}}\vcenter{\hbox{\includegraphics[width=30mm]{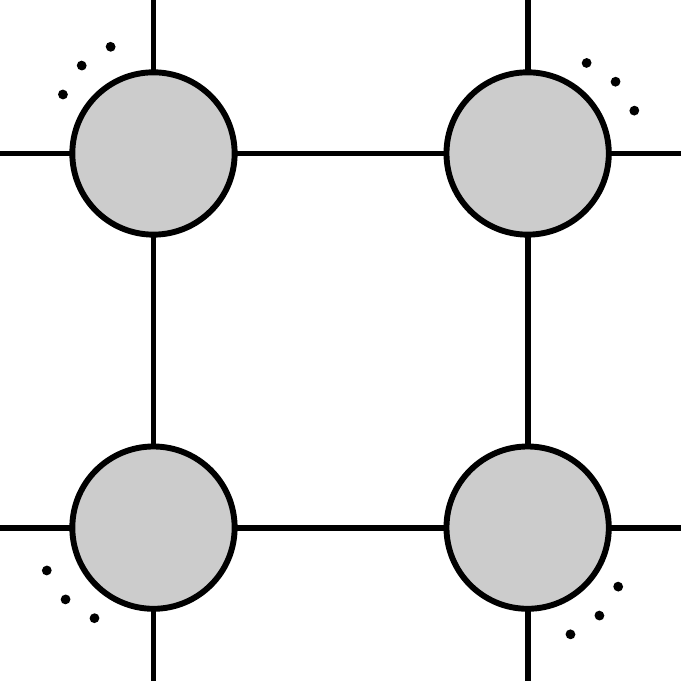}}}\times\vcenter{\hbox{\includegraphics[width=30mm]{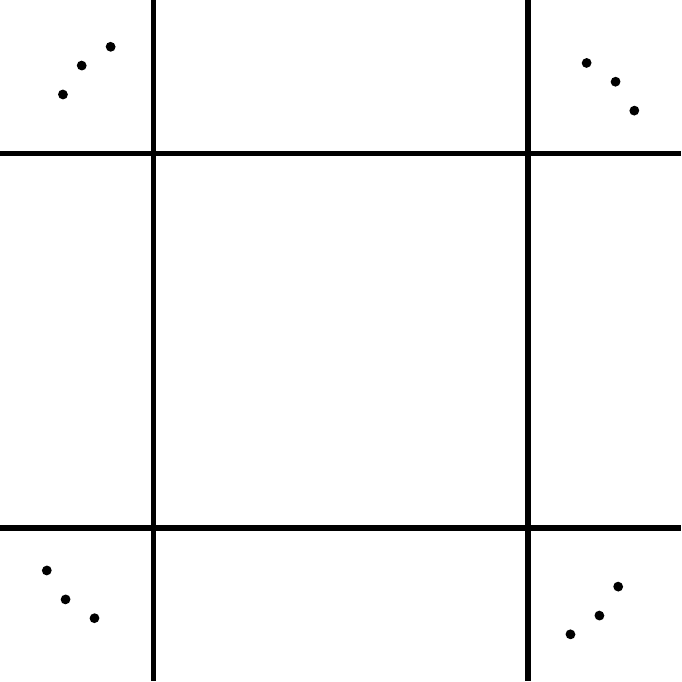}}}\,,
\end{align}
where the first term should be understood as an on-shell diagram, and the second term as a Feynman diagram. For example, for $n=4$ the maximal cut conditions $\ell^2=(\ell+p_1)^2=(\ell+p_1+p_2)^2=(\ell-p_4)^2=0$ has two solutions: $\ell=\ls_{13}$, and $\ell=\ls_{24}$, where $\ls_{ij}$ is defined in \eqref{eq:KIN_ls-def}. The corresponding leading singularities are given be the on-shell diagram in \eqref{eq:AMP_n4-on-shell} and the same diagram with black and white vertices exchanged. The leading singularities are equal, and, as argued around \eqref{eq:AMP_n4-on-shell}, this on-shell diagram equals the four-point tree-level amplitude. That is,
\begin{align}\label{eq:AMP_n4-dual-on-shell}
	\mathop{\Res}_{\ell=\ls_{13}}A_{4,2}^{(1)}=\mathop{\Res}_{\ell=\ls_{24}}A_{4,2}^{(1)}=A_{4,2}=\vcenter{\hbox{\includegraphics[width=35mm]{on-shell-n4.pdf}}}=\vcenter{\hbox{\includegraphics[width=35mm]{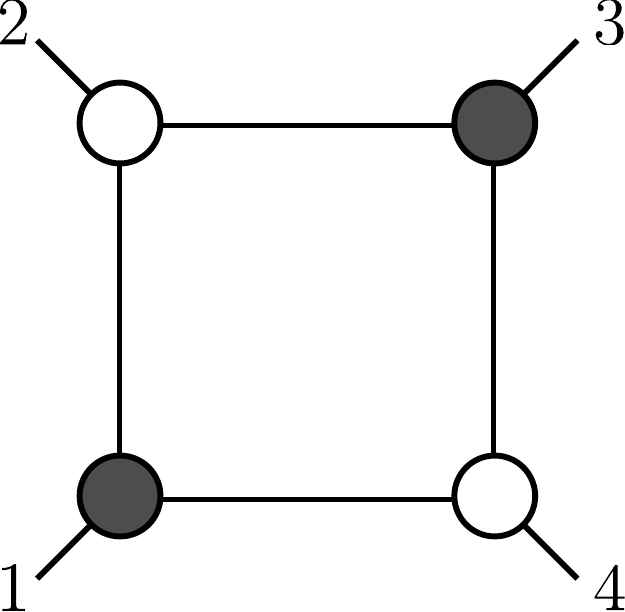}}}\,.
\end{align}
Just like the four-point case, the general maximal cut conditions (as in figure \ref{fig:on-shell-LS}) have two solutions. If we translate these conditions into dual space, then this is equivalent to finding the quadruple intersection of four lightcones. An explicit solution to this problem can be found in appendix \ref{sec:APP_schubert}.

\subsection{Scattering Amplitudes and the (Positive) Grassmannian}

In the previous section we saw that scattering amplitudes, their boundaries, and leading singularities can be expressed in terms of on-shell diagrams. Furthermore, through the BCFW recursion, all of these on-shell diagrams can be reduced to on-shell diagrams which only have three-point vertices. However, these on-shell diagrams are not unique representatives of their on-shell function.

The on-shell diagrams with trivalent vertices remind us of the plabic diagrams we encountered when studying the positive Grassmannian is section \ref{sec:GRASS_pos}, and the more general on-shell diagrams look like Grassmannian graphs. In fact, from \eqref{eq:AMP_n4-dual-on-shell} we see that even the \emph{square move} of figure \ref{fig:plabic_moves} is satisfied. It is not difficult to see that also to merge-expand moves are satisfied by on-shell diagrams. This means that we can really interpret these on-shell diagrams as corresponding to positroid cells! In particular, we can associate to each on-shell diagram a decorated permutation by simply reading off the rules of the road introduced in section \ref{sec:GRASS_pos}. If two on-shell diagrams have the same corresponding permutation (\textit{i.e.} the Grassmannian graphs represent the same positroid cell), then they have the same on-shell function. The correspondence between on-shell diagrams and the positive Grassmannian has been investigated in great detail in \cite{Arkani-Hamed:2012zlh}, and this work has been a cornerstone for many results presented in this thesis.

We see that on-shell functions are naturally associated to positroid cells. The results of \cite{Arkani-Hamed:2012zlh} go one step further, and give a way we can find this on-shell function as a \emph{Grassmannian integral} over the positroid cell. To make this precise, we recall from section \ref{sec:KIN_spin-hel} that kinematic data is naturally encoded in the Grassmannian by requiring
\begin{align}
	C^\perp\cdot\lambda^T=0\,,\quad C\cdot\tilde\lambda^T=0\,.
\end{align} 
Furthermore, supermomentum conservation $\delta(\lambda\cdot \eta^T)$ can be encoded in the Grassmannian by the additional constraints
\begin{align}
	C\cdot \eta^T=0\,.
\end{align}
If we let $C_\sigma(\bm{\alpha})$ be a positive parametrisation of some $d$-dimensional positroid cell $S_\sigma$ of $G_+(k,n)$, then the on-shell function associated to $\sigma$ can be obtained by the integral
\begin{align}\label{eq:AMP_grass-integral}
	f_\sigma = \int_{C_{\sigma}} \frac{\dd \alpha_1\wedge\cdots\wedge\dd\alpha_d}{\alpha_1\cdots\alpha_d} \delta^{(n-k)\times2}( C^\perp\cdot\lambda^T)\delta^{ k\times 2}(C\cdot\tilde\lambda^T) \delta^{k\times4}(C\cdot\eta^T)\,.
\end{align}
We can obtain all such differential forms by considering residues of the differential form associated to the top-cell $G_+(k,n)$. There is a nice parametrisation invariant way to write the `canonical form'\footnote{The wedge product of the $\dd\log\alpha_i$ appearing in equation \eqref{eq:AMP_grass-integral} can be interpreted as the `canonical form' of the positroid cell. We will make this statement more precise in chapter \ref{sec:POS}.} of the top-cell, which allows us to write the Grassmannian integral as
\begin{align}\label{eq:AMP_Lnk}
	\Lcal_{n,k} = \int \frac{\dd^{k\times n} C}{\vol[GL(k)]}\frac{\delta^{2\times(n-k)}(C^\perp\cdot\lambda^T)\delta^{2\times k}(C\cdot\tilde\lambda^T) \delta^{4\times k}( C\cdot\eta^T)}{(12\cdots k)(23\cdots k+1)\cdots (n1\cdots k-1)}\,.
\end{align}
It is customary to denote this Grassmannian integral by $\Lcal_{n,k}$, following \cite{Arkani-Hamed:2009ljj}. The canonical form of lower-dimensional positroid cells can be obtained from residues of the canonical form of the top cell, which is something that will be made more precise in chapter \ref{sec:POS} on positive geometries. This means that the full scattering amplitude (as well as other on-shell functions) can be obtained as an integral as in $\Lcal_{n,k}$ over a contour which encircles the positroid cells necessary. In the case of $A_{n,k}$, the BCFW recursion \eqref{eq:AMP_BCFW-on-shell} tells us which positroid cells to include.

As explained in section \ref{sec:KIN_twistors}, we can write these formula's in terms of twistor variables by doing a half Fourier transform on the $\lambda$ variables. An explicit calculation shows that
\begin{align}
	&\int \dd^{2\times n}\lambda e^{i \lambda\cdot\tilde\mu}\delta^{2\times (n-k)}(C^\perp\cdot\lambda^T) = \int \dd^{2\times n}\lambda\dd^{2\times k}\rho e^{i\lambda\cdot\tilde\mu} \delta^{2\times n}(\rho\cdot C-\lambda)\notag\\
	=&\int\dd^{2\times k }\rho e^{i (\rho\cdot C)\cdot\tilde\mu}=\delta^{2\times k}(C\cdot \tilde\mu^T)\,.
\end{align}
Hence, we find that the delta functions appearing in the Grassmannian integrals have taken a very concise form in terms of supertwistors:
\begin{align}
	\delta^{2\times(n-k)}(C^\perp\cdot\lambda^T)\delta^{2\times k}(C\cdot\tilde\lambda^T) \delta^{4\times k}( C\cdot\eta^T) \implies \delta^{4k|4k}(C\cdot\Wcal^T)\,.
\end{align}
In twistor space, we write the Grassmannian integral as
\begin{align}\label{eq:AMP_Grassmannian-integral-twistor}
	\Lcal_{n,k}=\int \frac{\dd^{k\times n}C}{\vol[GL(k)]}\frac{\delta^{4k|4k}(C\cdot\Wcal^T)}{(1\cdot k)\cdots (n\cdots k-1)}\,.
\end{align}
This delta function makes the $SL(4|4)$ superconformal invariance of the on-shell forms manifest. We note that a Fourier transform does not yield functions which depend on generic supertwistors, and only $2n-4$ of the (bosonic) delta functions can be used to localise the Grassmannian integral. The remaining delta functions impose constraints on the configurations of the supertwistors.

\subsubsection{Grassmannian Formulations in Momentum Twistor Space}

The fact that $\lambda\subseteq C$ means that it is possible to write the $C$ matrix as
\begin{align}\label{eq:AMP_C-with-lambda}
	C= \begin{pmatrix}
		\lambda_1^1 & \lambda_2^1 & \cdots & \lambda_n^1\\
		\lambda_1^2 & \lambda_2^2 & \cdots & \lambda_n^2\\
		c_{11} & c_{12} & \cdots & c_{1n}\\
		\vdots & \vdots & \ddots & \dots \\
		c_{K1} & c_{K2} & \cdots & c_{Kn} 
	\end{pmatrix} \equiv \left(
	\begin{array}{c}
	\lambda\\ \hdashline[2pt/2pt]
	c
	\end{array}\right)\,,
\end{align}
where $K=k-2$, and the new $K\times n$ matrix $c$ is defined up to a $GL(K)$ transformation, \emph{and} any translation by $\lambda$. After fixing this translation invariance, it is then possible to interpret $c$ as an element of $G(K,n)$. Following the arguments of \cite{Arkani-Hamed:2012zlh}, a natural choice would be to use translations by $\lambda$ to fix $c$ to be orthogonal to $\lambda$, which can be done by introducing an $n\times n$ matrix $Q$ which `projects' onto $\lambda^\perp$ (that is, $\ker(Q)=\lambda$). If we then define $\hat{C}= c\cdot Q$, then it immediately follows that $\hat{C}\cdot\lambda^T=\nul$. The constraints that $\tilde\lambda\subseteq \lambda^\perp$ and $\eta\subseteq \lambda^\perp$ can now be solved by introducing $\mu$ and $\chi$ which satisfy
\begin{align}
	\tilde\lambda= \mu\cdot Q\,,\quad \eta=\chi\cdot Q\,.
\end{align}
From this definition, we can start from any \emph{unconstrained} $\lambda$, $\mu$ and $\chi$ and end up with $\lambda,\tilde\lambda,\eta$ which automatically satisfy momentum and supermomentum conservation. 

There are infinitely many matrices which satisfy $\ker(Q)=\lambda$, and hence we have to specify which of these matrices we are considering. Remarkably, the simplest possible choice would be the matrix introduced in equation \eqref{eq:KIN_Q-mat-def}, which is a natural candidate to consider from the Schouten identity. Explicitly, this $Q$ matrix has the form
\begin{align}\label{eq:AMP_Q-mat}
	Q = \begin{pmatrix}
		-\frac{\<2n\>}{\<12\>\<1n\>} & \frac{1}{\<12\>} & 0 &\cdots & 0 & -\frac{1}{\<1n\>}\\
		\frac{1}{\<12\>} & - \frac{\<13\>}{\<12\>\<23\>} & \frac{1}{\<23\>}&\cdots & 0 & 0\\
		0 & \frac{1}{\<23\>} & -\frac{\<24\>}{\<23\>\<34\>} & \cdots & 0 & 0\\
		\vdots & \vdots & \vdots & \ddots & \vdots & \vdots \\
		0 & 0 & 0 & \cdots & -\frac{\<n-2 n\>}{\<n-2 n-1\> \<n-1 n\>} & \frac{1}{\<n-1 n\>} \\
		-\frac{1}{\<1n\>} & 0 & 0 & \cdots & \frac{1}{\<n-1n\>} & -\frac{\<1 n-1\>}{\<1n\>\<n-1n\>}
	\end{pmatrix}\,.
\end{align}
Making this choice, we see that our newly defined variables are exactly the same as those introduced when defining momentum twistors in section \ref{sec:KIN_mom-twistor}! That is, we define the momentum twistors $z$ and super momentum twistors $\Zcal$ as
\begin{align}
	z_a \equiv \begin{pmatrix}
		\lambda_a \\ \mu_a
	\end{pmatrix}\,, \quad \Zcal_a \equiv \begin{pmatrix}
	z_a \\ \chi_a
\end{pmatrix}\,,
\end{align}
such that the constraints $C\cdot \lambda^T=\nul$, $C\cdot \mu^T = \nul$, $C\cdot\eta^T=\nul$ can be neatly summarised as $\hat{C}\cdot\Zcal^T=\nul$.

There is a slight flaw in the logic presented above. Our choice of $Q$ matrix, although simple, is from some perspectives not a very natural one. Particularly, we motivated the $Q$ matrix as a `projection' matrix onto $\lambda^\perp$. However, this is not strictly correct, as $Q^2\neq Q$, and hence we cannot simply interpret $Q$ as a projection matrix. The consequence of this is that $\hat{C}$ is no longer a subspace of $C$. One can define projection matrices $P$ which satisfy $\ker(P)=\lambda$ and $P^2=P$\footnote{For example by doing a $GL(n-2)$ transformation on $\lambda^\perp$ which renders all column vectors mutually orthogonal, then $P=\lambda^\perp\cdot(\lambda^\perp)^T$ satisfies these properties.}, and the resulting matrix $c\cdot P$ is both orthogonal to $\lambda$ \emph{and} contained in $C$ (in the notation of section \ref{sec:GRASS_binary}, we have $c\cdot P=C\setminus\lambda$). From this perspective, it might be more natural to consider a matrix $Q$ of this type. Against these arguments, the $Q$ matrix of equation \eqref{eq:AMP_Q-mat} ends up being the correct choice. The magic comes from the fact that this specific choice of $Q$ maps consecutive chains of columns onto consecutive chains of columns:
\begin{align}
	\text{span}(\hat{\bm{c}}_a,\hat{\bm{c}}_{a+1},\ldots,\hat{\bm{c}}_b)\subseteq \text{span}(\bm{c}_{a-1},\bm{c}_a,\ldots,\bm{c}_b,\bm{c}_{b+1})\,.
\end{align}
This means that the consecutive minors of $C$ and $\hat{C}$ are directly proportional to each other:
\begin{align}\label{eq:AMP_C-Ccheck-minor-relation}
	(12\cdots k)_C=\<12\>\cdots\<k-1 k\> (23\cdots k-1)_{\hat{C}}\,.
\end{align}
Hence, up to some $\lambda$-dependent overall factor, $Q$ relates the top-form of $G_+(k,n)$ onto the top-form of $G_+(K,n)$. If we additionally assume that $\lambda$ satisfies $\<ii+1\>>0$, then we can interpret $Q$ as mapping positroid cells of $G_+(k,n)$ to positroid cells of $G_+(K,n)$: for $C\in G_+(k,n)$ with $\lambda\subseteq C$, then $\hat{C}=c\cdot Q =(C\setminus\lambda)\cdot Q \in G_+(K,n)$. The decorated permutations labelling these positroid cells are also related in a straightforward manner: $Q$ maps a positroid cell with decorated permutation $\sigma$ to a positroid cell with permutation $\hat{\sigma}$ satisfying
\begin{align}
	\hat{\sigma}(a)=\sigma(a-1)-1\,.
\end{align} 
This map between positroid cells is known as \emph{T-duality}.

We can use the above discussion to translate \eqref{eq:AMP_Lnk} into momentum twistor variables. As argued above, the delta functions are neatly summarised by a $\delta^{4K|4K}(\hat{C}\cdot \Zcal^T)$, and the minors from \eqref{eq:AMP_C-Ccheck-minor-relation} together with the Jacobian from the change of variables combine neatly into an MHV amplitude! That is, 
\begin{align}
	\Lcal_{n,k} = \frac{\delta^{2\times 4}(\lambda\cdot\tilde\eta^T)\delta^{2\times2}(\lambda\cdot\tilde\lambda^T)}{\<12\>\<23\>\cdots\<n1\>}\hat\Lcal_{n,k}=A_{n,2}\hat\Lcal_{n,k}\,,
\end{align}
where
\begin{align}
	\hat\Lcal_{n,k}=\int\frac{\dd^{K\times n}\hat{C}}{\vol[GL(K)]}\frac{\delta^{4K|4K}(\hat{C}\cdot \Zcal^T)}{(1\cdots K)(2\cdots K+1)\cdots (n\cdots K-1)}\,.
\end{align}
Just as the supertwistor form of the Grassmannian integral presented in \eqref{eq:AMP_Grassmannian-integral-twistor} manifests superconformal symmetry, this supermomentum twistor formulation manifests dual superconformal symmetry. In fact, these Grassmannian integrals are invariant under the full Yangian symmetry. The associated T-dual amplitude $\hat{A}_{n,K}=A_{n,k}/A_{n,2}$ is cyclically invariant and is also invariant (rather than covariant) under a little group transformation $\Zcal_a\to t_a \Zcal_a$, and can be understood as the expectation value of a Wilson loop rather than a scattering amplitude, which are T-dual in \nf \cite{Alday:2008yw, Berkovits:2008ic}.

More generally, given a $d$-dimensional positroid cell $C_\sigma$ of $G_+(K,n)$ with a positive parametrisation $C_\sigma(\bm\alpha)$, then it has an associated Yangian invariant on-shell function given by
\begin{align}
	\hat{f}_\sigma = \int \frac{\dd \alpha_1\wedge\cdots\wedge\dd\alpha_d}{\alpha_1\cdots\alpha_d} \delta^{4K|4K}(C\cdot\Zcal^T)\,.
\end{align}

\section[head={Scattering Amplitudes in ABJM},tocentry={Scattering Amplitudes in ABJM Theory}]{Scattering Amplitudes in ABJM Theory}\label{sec:AMP_ABJM}

\emph{Aharony-Bergman-Jafferis-Maldacena (ABJM) theory} is the nickname for $\Ncal=6$ supersymmetric Chern-Simons matter theory, which we encountered in section \ref{sec:QFT_ABJM}. There are surprisingly many connections between scattering amplitudes in ABJM and \nf that have been uncovered over the years. In particular, much like in the previous section, we will see that ABJM also has a twistor string description, an all-loop BCFW recursion, and a natural on-shell description which is deeply related to the Grassmannian (in this case the positive orthogonal Grassmannian). We will introduce these concepts in this section, making frequent comparisons to the analogous formulations of \nf. But first, let us deal with some of the basic properties of ABJM amplitudes.

We will be considering ABJM theory in the planar limit, in which case the theory has an $OSp(6|4)$ dual superconformal symmetry. For any massless theory in three dimensions, any three-particle amplitude $A_3$ must vanish, since all Lorentz invariants $s_{ij}=\<ij\>^2$ vanish for generic kinematics. Furthermore, since the Chern-Simons gauge field does not carry any physical degrees of freedom, any scattering process with a gauge field on an external leg must have a vanishing contribution. As a consequence, we only have even particle amplitudes in ABJM theory, and they consist purely of matter states. From here on, we assume that we are dealing with an even number $n=2k$ of particles.

The super charges take a similar form to those we encountered in the previous section:
\begin{align}
	\tilde{Q}_{\mathrm{A}}^\alpha = \sum_i \lambda_i^\alpha \eta_i^{\mathrm{A}}\,,\quad Q^{\alpha{\mathrm{A}}} = \sum_i\lambda_i^\alpha \frac{\partial}{\partial \eta_{i {\mathrm{A}}}}\,.
\end{align}
Colour decomposition for theories whose matter states are in the bi-fundamental is given in terms of products of Kronecker deltas \cite{Bargheer:2010hn}:
\begin{align}
	\sum_{\substack{\sigma\in S_k\\ \bar\sigma \in \bar{S}_{k-1}}} A_n[\bar{1},\sigma_1,\bar{\sigma}_1,\ldots,\bar{\sigma}_{k-1},\sigma_k] \delta^{\adot_{\sigma_k}}_{\adot_{\bar{1}}} \delta^{a_{\bar{\sigma}_1}}_{a_{\sigma_1}} \cdots \delta^{a_1}_{a_{\sigma_k}}\,,
\end{align}
where $S_k$ runs over the permutations of even sites, and $\bar{S}_{k-1}$ runs over the permutations of odd sites. This naturally provides us with a notion of colour ordered (super) amplitudes. Note that the colour ordered amplitudes have alternating external states $\overbar\Psi$ and $\Phi$. This implies that there are classes of amplitudes
\begin{align}
	A_n[\bar{1},2,\bar{3},\ldots,n]\,,\quad A_n[1,\bar{2},3,\ldots,\bar{n}]\,,
\end{align}
where the state $i$ denotes $\Phi_i$, and $\bar{j}$ denotes $\overline{\Psi}_j$. The little group in three dimension is $\Zbb_2$, and under a little group transformation fermionic fields pick up a minus sign. Hence, if the $i$\textsuperscript{th} state of the superamplitude is the fermionic field $\overline{\Psi}$ introduced in \eqref{eq:INT_ABJM-fields}, the amplitude must pick up an overall minus sign under the substitution $\lambda_i\to-\lambda_i, \eta_i\to-\eta_i$.

Scattering amplitudes in ABJM are cyclic symmetric by two sites (up to a sign), sometimes called $\Lambda$-parity:
\begin{align}
	A_{2k}[\bar{1},2,\ldots,2k] = (-1)^{k+1}A_{2k}[\bar{3}, 4,\ldots,2k,\bar{1},2]\,.
\end{align}
Since ABJM does not have a three-point amplitude, the most fundamental amplitude is the four-point amplitude. At tree-level it is given by \cite{Bargheer:2010hn}
\begin{align}
	A_4[\overline{\Psi}_1\Phi_2\overline{\Psi}_3\Phi_4]= \frac{\delta^{(6)}(\tilde{Q})}{\<14\>\<43\>}\,.
\end{align}
We can then build higher-point amplitudes using the three-dimensional BCFW recursion introduced in section \ref{sec:AMP_BCFW}. In general, ABJM superamplitudes have a Grassmann degree $3k$. Taking the residue at one of the poles does not factorise $A_4$ into two three-point amplitudes, since the three-point amplitudes don't exist. Instead, they are analogous to `soft limits'. Setting $\<14\>\to 0$ implies that $p_1+p_4=p_2+p_3=0$, and setting $\<34\>\to 0$ implies $p_1+p_2=p_3+p_4=0$. These type of special poles are exclusive to four-particle amplitudes \cite{Huang:2013owa}.

On-shell functions and on-shell diagrams are defined analogously as for \nf in section \ref{sec:AMP_nf} by taking the product of amplitudes and integrating over the internal phase space. The BCFW recursion we encountered in section \ref{sec:AMP_BCFW} can be written in terms of on-shell diagrams as
\begin{align}\label{eq:AMP_BCFW-on-shell-ABJM}
	A_n = \sum_{L,R}\quad\vcenter{\hbox{\includegraphics[width=40mm]{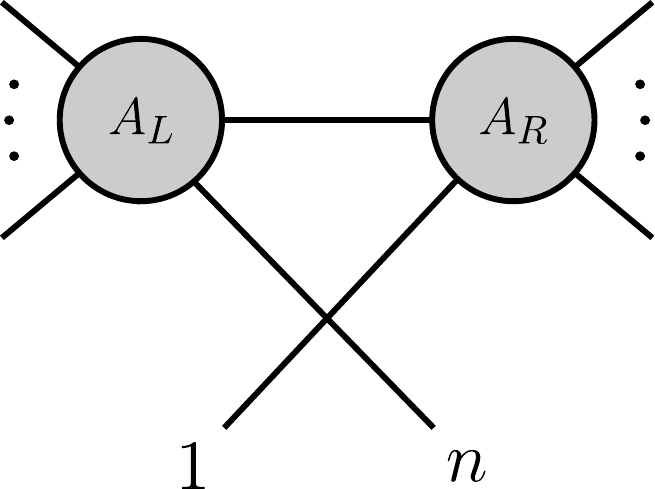}}}\,,
\end{align}
where the intersection point of the two edges on the bottom of this diagram should be interpreted as a four-point amplitude. Similar to equation \eqref{eq:AMP_BCFW-on-shell-loop}, we can extend the BCFW recursion to include loop integrand extension as \cite{Huang:2014xza}
\begin{align}
	A_n^{(\ell)}=\sum_{L,R}\quad\vcenter{\hbox{\includegraphics[width=40mm]{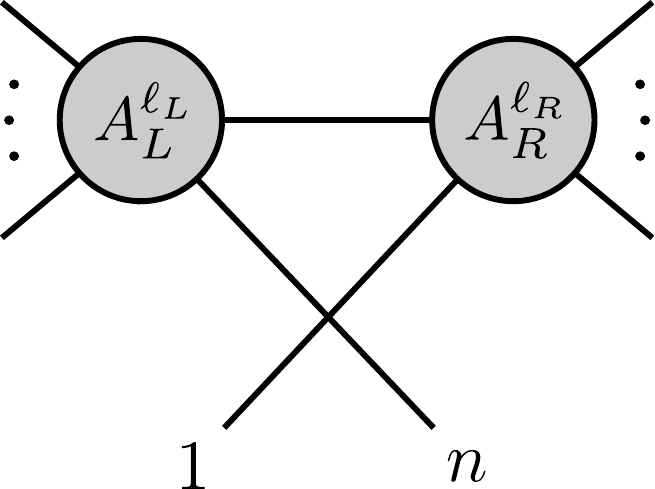}}}\quad+\vcenter{\hbox{\includegraphics[width=40mm]{
				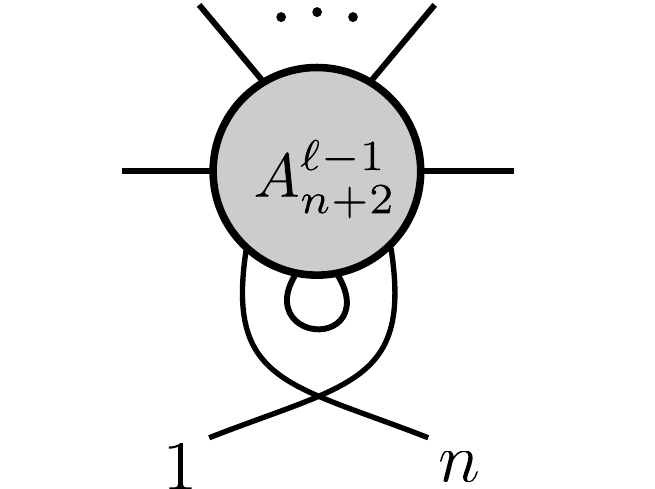}}}\,.
\end{align}
On-shell functions also represent the leading singularities of the one-loop integrands. The analogous statement to equation \eqref{eq:AMP_n4-dual-on-shell} is now a statement about the six-point amplitude:
\begin{align}\label{eq:AMP_ABJM-Yang-Baxter}
	\mathop{\Res}_{\ell=\ell^\star_{135}} A_6^{(1)} = \mathop{\Res}_{\ell=\ell^\star_{246}} A_6^{(1)} = A_6 = \vcenter{\hbox{\includegraphics[width=25mm]{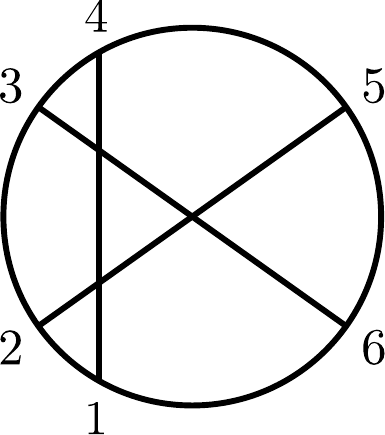}}} = \vcenter{\hbox{\includegraphics[width=25mm]{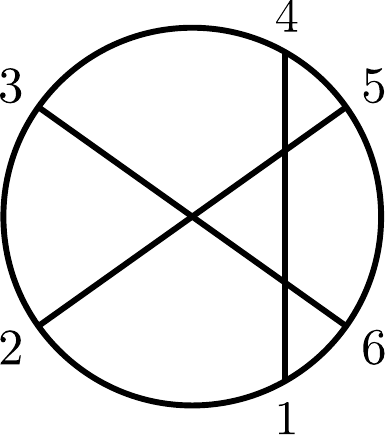}}}\,,
\end{align}
where $\ell^\star_{135}$ and $\ell^\star_{246}$ are solutions to the equations $\ell^2=(\ell-p_1+p_2)^2=(\ell-p_5+p_6)^2=0$, and $(\ell-p_1)^2=(\ell-p_1+p_2-p_3)^2=(\ell+p_6)^2=0$ (recall that we have an alternating convention for incoming and outgoing particles in 3D, see section \ref{sec:KIN_spin-hel}). In dual space, these cut conditions correspond to $(y-x_1)^2=(y-x_3)^2=(y-x_5)^2=0$ and $(y-x_2)^2=(y-x_4)^2=(y-x_6)^2=0$, respectively. This can again be given the interpretation as the intersection of three lightcones, an explicit formula for which is presented in appendix \ref{sec:APP_schubert}. The equivalence of the two on-shell diagrams in equation \eqref{eq:AMP_ABJM-Yang-Baxter} is known as the \emph{Yang-Baxter move} (which we already encountered in section \ref{sec:GRASS_orthitroid}), and it is a graphical description of the Yang-Baxter equation, which is of importance for integrable theories. 

Thanks to the Yang-Baxter move, we can find a correspondence between these on-shell diagrams and the crossing diagrams or OG graphs introduced for the orthogonal Grassmannian in section \ref{sec:GRASS_orthitroid}. Thus, each on-shell diagram labels some orthitroid cell, and two on-shell diagrams have the same on-shell function if their associated permutation (following the rules of the road defined in section \ref{sec:GRASS_orthitroid}) are equivalent. 

An analogue of the twistor string formula \eqref{eq:AMP_twistor-string-Grassmannian} was proposed by Huang and Lee in \cite{Huang:2012vt}. In terms of an integral over $G(2,n)$ it can be written as
\begin{align}\label{eq:AMP_ABJM-twistor-string}
	A_{2k}=\int\frac{\dd^{2\times 2k }C}{\vol[GL(2)]}\frac{J\Delta \delta^2(C\cdot \lambda^T)\delta^3(C\cdot\eta^T)}{(12)(23)\cdots (n1)}\,,
\end{align}	
where the numerator includes the explicit Jacobian
\begin{align}
	J=\frac{\prod_{1\leq i<j\leq n-1}(ij)}{\prod_{1\leq i<j\leq n}(2i-1,2j-1)}\,,
\end{align}
and $\Delta$ imposes $n-1$ additional delta functions
\begin{align}
	\Delta = \prod_{i=1}^{2k-1} \delta\left(\sum_{j=1}^n (-1)^{j+1} C_{1,j}^{n-i-1} C_{2,j}^{i-1}\right)\,.
\end{align}
On-shell functions corresponding to orthitroid cells can be found via Grassmannian integrals. The ABJM versions of Grassmannian integrals was proposed by Lee in \cite{Lee:2010du} and is given by
\begin{align}\label{eq:AMP_ABJM-Grass-integral}
	\Lcal_{2k}=\int\frac{\dd^{k\times 2k}C}{\vol[GL(k)]}\frac{\delta^{k(k+1)/2}(C\cdot C^T)\delta^{2k|3k}(C\cdot (\lambda | \eta))}{\prod_{i=1}^k (i\,i+1\cdots i+k)}\,.
\end{align}
We recall from section \ref{sec:GRASS_orth} that the orthogonal Grassmannian splits up into two branches. Furthermore, also the scattering amplitudes in ABJM decompose into topologically distinct branches. For example, for $n=4$ the orthogonality of $\lambda$ implies that $\<12\>^2=\<34\>^2$, which gives us two distinct configurations of the external kinematics: $\<12\>= \pm \<34\>$. The scattering amplitude needs to have support on both of these branches, because else it would become non-analytic \cite{Huang:2013owa}. In the Grassmannian integral \eqref{eq:AMP_ABJM-Grass-integral} the two branches of $OG(2,4)$ precisely correspond to the two branches of $A_4$, and hence we are required to sum over both branches. To generalise this observation to higher points amplitudes, we follow the arguments from \cite{Huang:2013owa}. We note that a three-dimensional null-momentum vector can be written as $p_i= E_i (1,\cos\theta_i,\sin\theta_i)$. Thus, projectively, we can label each momentum vector as a point on $S^1$. Each different configuration of $n=2k$ points on $S^1$ labels a distinct topological sector for the scattering amplitude. In the remainder of this thesis, we will mainly focus on a single branch of the amplitude: the \emph{positive branch}.

We will later want to interpret the superamplitudes $A_{2k}$ as differential forms by associating $\eta_a^\alpha\to\dd\lambda_a^\alpha$. To ensure that this has a chance of working, we need to go to a \emph{supersymmetry reduced} $\Ncal=4$ description of ABJM theory. We introduced the supersymmetry reduced superfields in equation \eqref{eq:QFT_susy-reduced-superfield}. We will put our emphasis on the partial superamplitudes
\begin{align}
	A_{2k}[\overbar{1},2,\ldots,\overbar{2k-1},{2k}] \coloneqq A_{2k}(\bar{\Phi}^{\mathcal{N}=4}_1, {\Phi}^{\mathcal{N}=4}_2,\ldots,\bar{\Phi}^{\mathcal{N}=4}_{2k-1},{\Phi}^{\mathcal{N}=4}_{2k} )\,.
\end{align}
It was shown in \cite{He:2021llb} that these amplitudes have a much simpler form of the twistor string equations \eqref{eq:AMP_ABJM-twistor-string}. We introduce $C(\bm{z})$ as the \emph{Veronese embedding} of $G(2,2k)/T$ into $OG(k,2k)$:
\begin{align}
	\begin{pmatrix}
		1 & 1 & \cdots &1\\ z_1 & z_2 & \cdots & z_n
	\end{pmatrix} \mapsto C(\bm{z})= \begin{pmatrix}
		t_1 & t_2 & \cdots & t_n\\
		t_1 z_1 & t_2 z_2 & \cdots & t_n z_n\\
		\vdots & \vdots & \ddots & \vdots \\
		t_1 z_1^{k-1} & t_2 z_2^{k-1}& \cdots & t_n z_n^{k-1}
	\end{pmatrix}\,,
\end{align}
with
\begin{align}\label{eq:AMP_C-mat-veronese}
	t_i = \sqrt{(-1)^i \frac{\prod_{j\neq n} (z_n-z_j)}{\prod_{j\neq i} (z_i-z_j)}}\,.
\end{align}
Here we use $G(2,2k)/T$ to denote the Grassmannian modulo the torus action, which allows us to rescale each column to have $1$ in its first entry, and the condition \eqref{eq:AMP_C-mat-veronese} ensures that $C(\bm{z})\cdot\eta\cdot C(\bm{z})^T=\nul$. Having defined the matrix $C(\bm{z})$, we find that the twistor string formula takes the form
\begin{align}\label{eq:AMP_ABJM-reduced-twistor-string}
	A_{2k} = \int\frac{\dd^{n}z}{\vol[SL(2)]}\frac{\delta(C(\bm{z})\cdot\lambda^T)\delta(C(\bm{z})\cdot\eta^T)}{(z_1-z_2)(z_2-z_3)\cdots(z_n-z_1)}\,.
\end{align}
We note in passing that $G_+(2,2k)/T$ is isomorphic to the \emph{positive moduli space} $\Mfrak_{0,2k}^+$, which will be important in section \ref{sec:POS_pf}.

\section{Summary}

In this chapter we have introduced some essential elements of the modern scattering amplitudes programme. We have seen some basic ideas such as colour ordering and unitarity methods for loop integrands, and we further defined BCFW recursion in both three and four dimensions. We have further seen a description of the CHY formalism and the scattering equations, with some focus on four dimensions. We have seen numerous different descriptions for superamplitudes in \nf, including a formulation in terms of twistor strings (which is related to the CHY formalism and the scattering equations), and in terms of a Grassmannian integrals for spinor-helicity variables and a T-dual version for momentum twistors. We further saw that we can define a more general class of on-shell functions which are naturally associated to positroid cells. Among these on-shell functions are terms in the BCFW expansion and leading singularities of loop integrands. We have also seen that essentially all these properties of scattering amplitudes in \nf have an analogue for ABJM theory, which also admits a twistor string and Grassmannian integral formulation, and whose on-shell functions are associated to orthitroid cells. We briefly encountered supersymmetry reduced ABJM theory, which has a more natural twistor-string formula.

The ideas and formulae outlined in this chapter have been important breakthroughs in the modern study of scattering amplitudes. For us, the value of these concepts comes from the fact that they coalesce into the various positive geometries which we will describe in the next chapters. The formulation of scattering amplitudes from the positive (orthogonal) Grassmannian will be important for the definition of amplituhedra, and the positroid cells corresponding to BCFW terms will be given the interpretation of triangulating these geometries. We will further see that some of the positive geometries we encounter can be obtained by taking the \emph{push forward through the scattering equations}. Additionally, the positroid cells corresponding to leading singularities will play an important role for the positive geometric description of loop integrands in dual space in chapter \ref{sec:DUAL}.

%% file: chapters/positivegeom.tex
\chapter{Positive Geometry}\label{sec:POS}

We are now finally equipped to investigate the main topic of this thesis: positive geometries and their relation to scattering amplitudes. Historically first considered in the amplituhedron, which describes scattering amplitudes in \nf in momentum twistor space, positive geometries have since been found in different kinematic spaces (such as the momentum amplituhedron, which describes \nf amplitudes in spinor-helicity space), and for different theories (such as the ABJM momentum amplituhedron and the ABHY associahedron, which describe scattering amplitudes in ABJM and bi-adjoint $\phi^3$ theory, respectively). We will define and study all of these positive geometries in this chapter. In addition, the field of positive geometries has been studied outside the context of scattering amplitudes as well: it has been investigated from the point of view of pure mathematics \cite{Arkani-Hamed:2017tmz, Karp:2016uax, Karp:2017ouj, Lukowski:2020dpn, Parisi:2021oql, Moerman:2021cjg, Even-Zohar:2021sec, Even-Zohar:2023del, Even-Zohar:2024nvw, Parisi:2024psm, Bao:2019bfe, Galashin:2018fri, Lam:2024gyg}, and on the physics side it has also found applications for correlators in \nf \cite{Eden:2017fow, Dian:2021idl, He:2024xed}, the conformal bootstrap \cite{Arkani-Hamed:2018ign}, cosmological correlators \cite{Arkani-Hamed:2017fdk, Arkani-Hamed:2018bjr, Benincasa:2022gtd, Benincasa:2019vqr, Benincasa:2020uph, Benincasa:2024leu}, and effective field theory \cite{Arkani-Hamed:2020blm, Chiang:2021ziz, Berman:2023jys}.

To motivate positive geometries, and in particular the amplituhedron, we will repeat some of the semi-historical overview from the introduction. With our current understanding of projective and Grassmannian geometry, kinematic spaces, and scattering amplitudes, we can give this overview a more explicit context. To start, we define the \emph{R-invariants}
\begin{align}\label{eq:POS_R_inv-def}
	[i,j,k,l,m] = \frac{(\eta_{iA}\<jklm\> + \text{ cyclic})^4}{\<ijkl\>\<jklm\>\<klmi\>\<lmij\>\<mijk\>}\,,
\end{align}
which are the Yangian invariant on-shell functions associated to the positroid cell of $G_+(1,n)$ whose only non-zero entries are indexed by $i$, $j$, $k$, $l$, and $m$. For example, in the notation of section \ref{sec:AMP_nf} with $n=6$, we have $[1,2,3,4,5]=\hat{f}_{\{2,3,4,5,7,6\}}$. We note the similarity to the volume of a simplex in $\Pbb^4$, which we encountered in equation \eqref{eq:GRASS_dual-simplex-volume-4}:
\begin{align}
	\vol(\Delta_4) = \frac{\<ijklm\>^4}{\<Yijkl\>\<Yjklm\>\<Yklmi\>\<Ylmij\>\<Ymijk\>}\,.
\end{align}
Even the numerator in equation \eqref{eq:POS_R_inv-def} looks like a $5\times 5$ determinant if we expend with respect to a `row of $\eta$s'. Furthermore, the BCFW recursion relations provide a way to write all \nf NMHV amplitudes in terms of these R-invariants as
\begin{align}
	\hat{A}_n^{\text{NMHV}}=\sum_{i<j} [1,i,i+1,j,j+1]\,.
\end{align}
We note that the explicit presence of `1' in all terms in this expansions comes from our choice of BCFW shift. We could equivalently have written the same amplitude with an arbitrary $a\in [n]$ in the first slot of the R-invariants. This shows that these R-invariants satisfy several non-trivial identities, for example for $n=6$:
\begin{align}\label{eq:POS_NMHV6-R-inv}
	\hat{A}^{\text{NMHV}}_6 &= [1,2,3,4,5]+[1,2,3,5,6]+[1,3,4,5,6] \\ &=[1,2,3,4,6]+[1,2,4,5,6]+[2,3,4,5,6]\,.
\end{align}
The identity
\begin{align}
	[1,2,3,4,5]-[1,2,3,4,6]+[1,2,3,5,6]-[1,2,4,5,6]+[1,3,4,5,6]-[2,3,4,5,6]=0\,,
\end{align}
is reminiscent of the boundary of a simplex in simplicial homology, for this reason these identities are known as \emph{homological identities}. These homological identities are very non-trivial from an algebraic point of view, however based on the observation that the R-invariants look like volumes of simplices, we are tempted to give them a geometric interpretation instead. This was achieved by Hodges in \cite{Hodges:2009hk}, where it was pointed out that all NMHV amplitudes can be interpreted as the volume of certain projective polytopes. The different formulas for the amplitude in equation \eqref{eq:POS_NMHV6-R-inv} can then be understood as emerging from different ways to \emph{triangulate} Hodges' polytope in terms of simplices.

Although Hodges' polytopes do not generalise beyond NMHV, the idea itself can be generalised to any amplitude in \nf. It is exactly the \emph{dual polytope} to Hodges' polytopes (see section \ref{sec:GRASS_proj}) which lend themselves to such a generalisation. These dual polytopes do not encode the scattering amplitude in their volume, but rather in their \emph{canonical form} (which we will introduce below). This is precisely what will lead to the definition of the amplituhedron, which was introduced by Arkani-Hamed and Trnka in \cite{Arkani-Hamed:2013jha}. 

To be more explicit, we recall from section \ref{sec:GRASS_proj_poly} that we can define a projective polytope as the image of $G_+(1,n)$ under a positive linear map. We further recall from section \ref{sec:AMP_nf} that N\textsuperscript{$K$}MHV amplitudes are associated with the positive Grassmannian $G_+(K,n)$. For NMHV amplitudes we see that $G_+(1,n)$ makes an appearance not only in these Grassmannian integrals, but also as the domain of a positive linear map. It is then natural to ask if this generalises, and if a positive linear map with as domain $G_+(K,n)$ yields N\textsuperscript{$K$}MHV amplitudes. This turns out to be the case, and we have just rediscovered the original definition of the amplituhedron! The image of positroid cells under this map give rise to a canonical form which reproduces the correct on-shell diagrams, which shows the validity of the resulting amplitude. We have ignored various details in this overview, and part of this chapter will be dedicated to understanding this construction more precisely.

\section[head={Positive Geometries},tocentry={Positive Geometries and Their Canonical Forms}]{Positive Geometries and Their Canonical Forms}\label{sec:POS_def}

The term \emph{positive geometries} encompasses a framework for studying certain geometric objects with an associated \emph{canonical form}. A formal definition of what a `positive geometry' is was given in \cite{Arkani-Hamed:2017tmz}, which will be reviewed in the following section. Physicists typically use the term in a less rigorous manner, and many of the objects we consider below have not been proven to be positive geometries following the definition. Furthermore, loop extensions of the amplituhedron have been proven \emph{not} to be positive geometries in this strict sense \cite{Dian:2022tpf}, and instead belong to the more general class of \emph{weighted positive geometries}. This is among several examples from recent years which seem to indicate that it will be necessary to move away from the traditional definition of positive geometries. Nevertheless, we will not concern ourselves with these intricacies, and we shall refer to the objects introduced in the following sections under the umbrella term `positive geometries'. 

\subsection{Definition}

\begin{figure}
	\centering
	\includegraphics[scale=0.4]{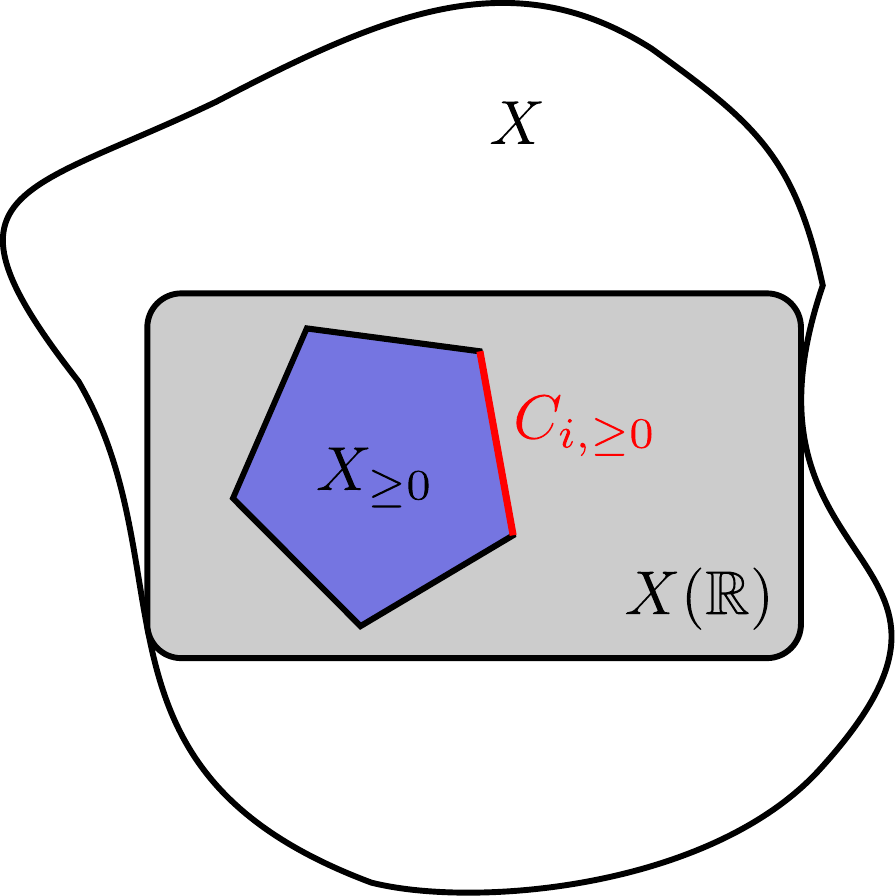}
	\caption{A schematic illustration of the sets $X, X(\Rbb),X_{\geq0}$ and $C_{i,\geq0}$.}
	\label{fig:pos-geom-def}
\end{figure}

We follow the definition from \cite{Arkani-Hamed:2017tmz}, see also the review articles \cite{Ferro:2020ygk, Herrmann:2022nkh, Williams:2021zph}. We start by considering a pair $(X,X_{\geq0})$, illustrated schematically in figure \ref{fig:pos-geom-def}, where 
\begin{itemize}
	\item $X$ is an algebraic variety of complex dimension $d$. This is an ambient `embedding' space, and will typically be a Grassmannian or projective space.
	\item $X_{\geq0}$ is a real, oriented, closed subset of $X(\Rbb)$, the real part of $X$. We require $X_{\geq0}$ to have boundaries of all codimensions.
\end{itemize}
To make the definition of a `boundary' explicit, we equip $X(\Rbb)$ with the standard topology, and let $X_{>0}$ be the interior of $X_{\geq0}$, which is a real, oriented, $d$-dimensional manifold. We can always recover $X_{\geq0}$ as the closure of $X_{>0}$. The \emph{boundary} of $X_{\geq0}$, denoted $\partial X_{\geq0}$, is defined as the set $X_{\geq 0}\setminus X_{>0}$. We further define $\partial X$ to be the \emph{Zariski closure} of $\partial X_{\geq0}$, which means that it is the largest subset of $X$ with the property that all homogeneous polynomials which vanish on $\partial X_{\geq0}$ also vanish on $\partial X$. $\partial X$ contains a number of irreducible components of codimension-1, which we label $C_1,\ldots,C_r$. We define $C_{i,> 0}$ to be the interior of $C_i\cap \partial X_{\geq0}$, and denote its closure as $C_{i,\geq0}$. We then define the \emph{boundary components}, or \emph{facets}, of $(X,X_{\geq0})$ as the collection $(C_i,C_{i,\geq0})$, $i=1,\ldots,r$. 

A $d$-dimensional positive geometry is defined to be the pair $X,X_{\geq 0}$ equipped by a unique (up to a constant) \emph{canonical form} $\Omega(X,X_{\geq0})$ which satisfies the following recursive properties:
\begin{itemize}
	\item $\Omega(X,X_{\geq0})$ is a meromorphic top-form with logarithmic singularities exactly at the boundaries of $X_{\geq 0}$.
	\item If $d=0$, $X_{\geq0}$ consists of a single point, and the canonical form is $\Omega(X,X_{\geq0})=\pm1$.
	\item for $d>0$, any boundary component $(C,C_{\geq0})$ of $(X,X_{\geq0})$ is itself a positive geometry of dimension $d-1$, with canonical form $\Omega(C,C_{\geq0})=\Res_{C} \Omega(X,X_{\geq0})$\,.
\end{itemize}
The residue operator which appears in the definition of a positive geometry is defined similarly to the well-known definition from complex analysis. Consider a $d$-form $\omega$ with a logarithmic singularity along the hypersurface $H$. We can locally parametrise $H$ as the zero set of some holomorphic coordinate $\alpha$. It is always possible to write $\omega$ such that
\begin{align}
	\omega = \omega' \wedge \dd\log\alpha + \eta\,,
\end{align}
where $\eta$ is a $d$-form that does not have a singularity at $\alpha=0$, and $\omega'$ is some $(d-1)-$form. The residue of $\omega$ along $H$ is then defined as the restriction of $\omega'$ to $H$:
\begin{align}
	\Res_{H}\omega = \omega'|_{H}\,.
\end{align}
If a differential form does not have a simple pole at $H$, then we define the residue to be zero. In practice, we often ignore the importance of the ambient space, and abuse terminology by referring to $X_{\geq0}$ as the `positive geometry'. This also extends to the canonical form, which we often denote as $\Omega(X_{\geq0})$.

\subsection{Examples}

Before we move on, it will be instructive to look at some simple examples of positive geometries.

\subsubsection{Projective Simplices}

\begin{figure}
	\centering
	\includegraphics[width=0.7\textwidth]{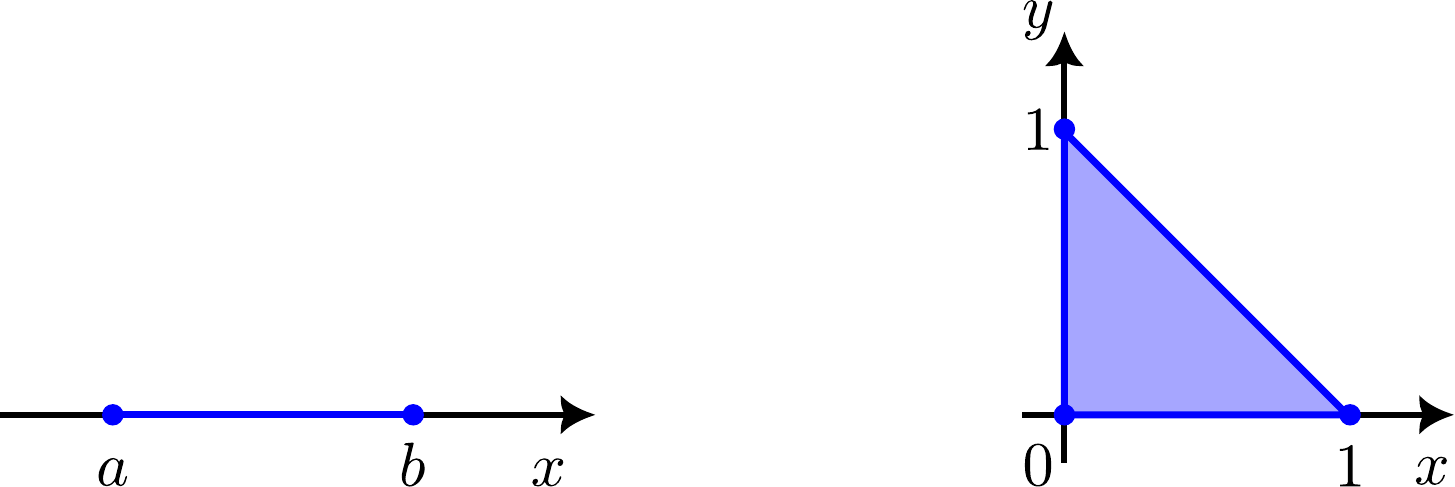}
	\caption{Examples of a one-dimensional positive geometry (left), and a two-dimensional positive geometry (right). We only depict the positive geometry $X_{\geq 0}$ inside $X(\Rbb)$, although it is understood that these geometries are embedded in some ambient space $X$.}
	\label{fig:pos-geom-example}
\end{figure}

Beyond zero-dimensional positive geometries (which are defined to have canonical form $\pm1$), the next simplest positive geometry is a line segment. To be completely explicit, we take our ambient space $X$ to be the projective line $\Pbb^1$ with real slice $\Pbb^1(\Rbb)\cong S^1$. We define our one-dimensional positive geometry to be the line segment $X_{\geq0} = [a,b]\coloneqq \{(1,x)\colon a\leq x\leq b\}\subseteq \Pbb^1(\Rbb)$, as depicted in figure \ref{fig:pos-geom-example}. This positive geometry has a canonical form
\begin{align}
	\Omega([a,b])=\dd\log\frac{x-a}{x-b} = \frac{b-a}{(x-a)(x-b)}\dd x\,.
\end{align} 
By inspection it is clear that this canonical form is meromorphic with logarithmic singularities at the points $x=a$ and $x=b$. Since we can write
\begin{align}
	\Omega([a,b]) = \dd \log (x-a) - \dd \log (x-b)\,,
\end{align}
it is clear that the residue at the boundaries $x=a$ and $x=b$ are $+1$ and $-1$, respectively, which means that the boundary components are zero-dimensional positive geometries. 

Next, moving to two-dimensional positive geometries, the simplest examples are triangles. The ambient space is $\Pbb^2$ and we use holomorphic coordinates $(1,x,y)$ on a patch of $\Pbb^2(\Rbb)$. We define our positive geometry $\Delta$ to be the convex hull of the points $(1,x,y)=(1,0,0),(1,1,0),(1,0,1)$. The boundary components lie on the hypersurfaces $x=0$, $y=0$, and $1-x-y=0$, as depicted in figure \ref{fig:pos-geom-example}. It has a canonical form
\begin{align}
	\Omega(\Delta) = \dd\log\frac{x}{1-x-y}\wedge\dd\log\frac{y}{1-x-y}=\frac{\dd x\wedge\dd y}{xy(1-x-y)}\,.
\end{align}
The residue at the boundary given by $y=0$ is 
\begin{align}
	\mathop{\Res}_{y=0}\Omega(\Delta) = \left.-\dd\log\frac{x}{1-x-y}\right|_{y=0}=\dd\log\frac{x}{1-x}\,,
\end{align}
which is precisely the canonical form of the line segment $[0,1]$, and hence this boundary is a one-dimensional positive geometry.

Next, we consider a general projective $m$-simplex $(\Pbb^m,\Delta_m)$ with ambient space $\Pbb^m$, and the positive region $\Delta_m$ is defined as the convex hull of $m+1$ points in $\Pbb^m(\Rbb)$ (see section \ref{sec:GRASS_proj_poly}) $X_1,\ldots,X_{m+1}$:
\begin{align}
	\Delta_m \coloneqq \{Y \in \Pbb^m(\Rbb)\colon Y = \sum_{i=1}^{m+1}C_i X_i,\quad C_i> 0\}\,.
\end{align}
The facets of $\Delta_m$ lie on hyperplanes where $\<Y X_{i_1}\cdots X_{i_m}\>$ vanishes. The canonical form is given by
\begin{align}\label{eq:POS_proj-simplex-form}
	\Omega(\Delta_m) = \bigwedge_{i=2}^{m+1}\dd\log\frac{\<Y X_1 X_2\cdots \hat{X}_i \cdots X_m X_{m+1} \>}{\<Y X_2 X_3 \cdots X_{m+1} \>}\,,
\end{align}
where the hat denotes omission. This form clearly has logarithmic singularities as $Y$ approaches any of the boundaries. From this representation of the canonical form it might look like the boundary $\<Y23\cdots m+1\>=0$ has some special status over the other facets. This is not the case, and we could have chosen any boundary in the numerator. This would leave the canonical form invariant up to a potential minus sign with which we do not concern ourselves. Expanding out \eqref{eq:POS_proj-simplex-form} we get
\begin{align}
	\Omega(\Delta_m) =\frac{\<X_1\cdots X_{m+1}\>}{\prod_{i=1}^{m+1}\<Y X_1\cdots \hat{X}_i\cdots X_{m+1}\>}\frac{\<Y\dd^m Y\>}{m!}\,,
\end{align}
where we use the projective measure
\begin{align}
	\<Y\dd^mY\>&\coloneqq \epsilon_{I_1\cdots I_{m+1}}Y^{I_1}\dd Y^{I_2}\wedge\cdots\wedge\dd Y^{I_{m+1}}\\
	&=m! \sum_{i=1}^{m+1} (-1)^i Y^i \dd Y^1\wedge\cdots\wedge\dd\hat{Y}^i\wedge\cdots\wedge\dd Y^{m+1}\,.
\end{align}
The numerator in equation \eqref{eq:POS_proj-simplex-form} ensures that the canonical form is invariant under $X_i\to t X_i$, which means that it is projectively well-defined. If our positive geometry lives in the patch where $Y^1\neq0$, we can use projectivity to fix $Y = (1,y^1,\ldots,y^m)$, and the projective measure becomes 
\begin{align}
	\frac{\<Y\dd^mY\>}{m!} = \dd y^1\wedge\cdots\wedge\dd y^m\,.
\end{align}
As was remarked in section \ref{sec:GRASS_proj_poly}, we can label facets by a \emph{dual vector} 
\begin{align}
	W_{i I_i}=\epsilon_{I_1\cdots I_{m+1}} Y_1^{I_1}\cdots \hat{Y}_i^{I_i}\cdots Y_{m+1}^{I_{m+1}}\,.	
\end{align}
The facets of $\Delta_m$ are then described by $Y\cdot W_i=0$, and the canonical form can be written as
\begin{align}
	\Omega(\Delta_m) = \bigwedge_{i=1}^m \dd\log\frac{Y\cdot W_i}{Y\cdot W_{m+1}} =\frac{\<W_1\cdots W_{m+1}\>}{\prod_{i=1}^{m+1}Y\cdot W_i}\frac{\<Y\dd^mY\>}{m!}\,.
\end{align}

\subsubsection{The Positive Grassmannian}

An important example of a positive geometry is the positive Grassmannian. The embedding space is the complex Grassmannian $G(k,n)$, and the boundaries are given by the codimension-one positroid cells, which have one of the cyclic minors going to zero. The canonical form of $G_+(k,n)$ is given by
\begin{align}
	\Omega(G_+(k,n))= \frac{\dd^{k\times n}C}{\vol[GL(k)]}\frac{1}{(1\cdots k)\cdots (n\cdots k-1)}\,.
\end{align}
If we have a positive parametrisation, such that $C(\bm{\alpha}) \in G_+(k,n)$ for $\alpha_i>0,\,i=1,\ldots,k(n-k)$, then the canonical form takes the form
\begin{align}
	\Omega(G_+(k,n))= \dd\log\alpha_1\wedge\dd\log\alpha_2\wedge\cdots\wedge\dd\log\alpha_{k(n-k)}\,.
\end{align}
This also extends to lower dimensional positroid cells. If a $d$-dimensional positroid cell $S_\sigma$ corresponding to some permutation $\sigma$ has a positive parametrisation $C_\sigma(\bm{\alpha})$ in terms of parameters $\alpha_i>0,\,i=1,\ldots,d$, then it has a canonical form
\begin{align}
	\Omega(S_\sigma)=\dd\log\alpha_1\wedge\cdots\wedge\dd\log\alpha_d\,.
\end{align}

\subsection{Triangulations}

A useful way to find the canonical form of more complicated positive geometries is by \emph{triangulating} it in terms of simpler pieces whose canonical form is known. Given a dissection of the positive geometry $\Acal$ in terms of finitely many pieces $\Acal_i$, then the canonical form of $\Acal$ can be obtained by summing the canonical form of all the $\Acal_i$.
Contrary to what the name might suggest, the pieces $\Acal_i$ of a `triangulation' do not need to be triangles or simplices. Since the canonical form of $\Acal$ can be obtained from any number of different triangulations, and no echo of the individual pieces is left in $\Omega(\Acal)$, we say that the canonical form is \emph{triangulation independent}.

To be more explicit, given a positive geometry $\Acal$, we say that a collection of positive geometries $\Acal_i$ in the same ambient space \emph{triangulates} $\Acal$ if 
\begin{itemize}
	\item $\Acal_i\subseteq\Acal$, and the orientations agree,
	\item $\Acal_i\cap \Acal_j=\emptyset$ for all $i\neq j$,
	\item $\Acal = \bigcup_{i} \Acal_i$.
\end{itemize}
If this is satisfied, then
\begin{align}
	\Omega(\Acal)=\sum_i \Omega(\Acal_i)\,.
\end{align}
The triangles\footnote{Again, not necessarily triangles.} $\Acal_i$ in a triangulation typically have boundaries which are not boundaries of $\Acal$, these are called \emph{spurious boundaries}. When adding the canonical forms of $\Acal_i$, the spurious poles must cancel out, as $\Omega(\Acal)$ cannot have a pole there.

We can also think about `subtracting' a positive geometry from another. This can be interpreted more formally by rearranging terms such that all signs on either side of the equals sign are positive, in which case it again becomes a statement of a triangulation of some larger geometry. This allows us to `triangulate' a positive geometry externally. An external triangulation of a positive geometry where none of the pieces have any spurious boundaries is said to be in the \emph{local form}. 

This allows us to construct the canonical form of general projective polytopes by triangulating it in terms of projective simplices, whose canonical forms we found in the previous section. We point out that the \emph{canonical function} (the function multiplying the top-form) of the canonical form of a simplex given in equation \eqref{eq:POS_proj-simplex-form} is \emph{precisely} the formula we found for the volume of its dual simplex in equation \eqref{eq:GRASS_dual-simplex-volume}, if we interpret $Y$ as specifying the point at the origin. By triangulating some polytope in terms of simplices, it is clear that this property will persist: the canonical function of a polytope is the volume of its dual polytope.

\begin{figure}
	\centering
	\includegraphics[width=0.75\textwidth]{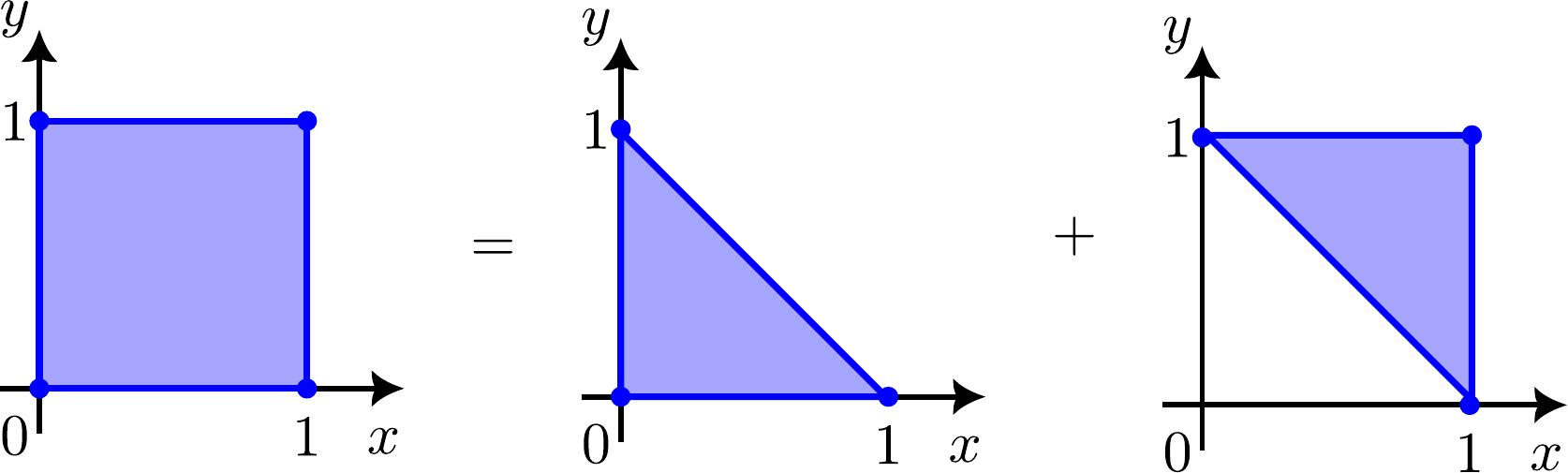}
	\caption{An example of a square being triangulated by two triangles.}
	\label{fig:triangulation-example}
\end{figure}

\paragraph{Examples.}
As an example, we can consider a projective square $\Acal$ in $\Pbb^2(\Rbb)$ with vertices $(1,0,0)$, $(1,1,0)$, $(1,0,1)$, and $(1,1,1)$. This square can be triangulated by two triangles $\Acal_1$ with vertices $(1,0,0)$, $(1,1,0)$, $(1,0,1)$, and $\Acal_2$ with vertices $(1,1,0)$, $(1,0,1)$, $(1,1,1)$, as depicted in figure \ref{fig:triangulation-example}. Their canonical forms are given by
\begin{align}
	\Omega(\Acal_1)&=\dd\log\frac{x}{1-x-y}\wedge\dd\log\frac{y}{1-x-y} = \frac{\dd x \wedge \dd y}{x y (1-x-y)}\,,\\
	\Omega(\Acal_2) &= \dd\log\frac{1-x}{1-x-y}\wedge\dd\log\frac{1-y}{1-x-y}=- \frac{\dd x\wedge \dd y}{(1-x)(1-y)(1-x-y)}\,.
\end{align}
The canonical form of the square is therefore given by
\begin{align}
	\Omega(\Acal) = \Omega(\Acal_1)+\Omega(\Acal_2) = \frac{\dd x\wedge \dd y}{x y (1-x)(1-y)}\,.
\end{align}
We note that the relative minus sign is important here, as this reflects that $\Acal_1$ and $\Acal_2$ have a compatible orientation. We see that both $\Omega(\Acal_1)$ and $\Omega(\Acal_2)$ have a spurious boundary at $1-x-y=0$, however this pole is no longer present in the sum. 

We take this moment to remark that the canonical form of $\Omega(\Acal)$ can equivalently be written as
\begin{align}
	\Omega(\Acal)=\dd\log\frac{x}{1-x}\wedge\dd\log\frac{y}{1-y}=\Omega([0,1])^2\,.
\end{align}
The square $\Acal$ is equal to the Cartesian product $[0,1]\times[0,1]$, and this is reflected in its canonical form. In general, if some positive geometry $X$ can be written as $X=Y\times Z$, then $\Omega(X)=\Omega(Y)\wedge\Omega(Z)$.

\subsection{Push Forwards}\label{sec:POS_push-forward}

We now come to another important method to find the canonical form of positive geometries, namely calculating the \emph{push forward} of an already known canonical form. This is a topic which will be of importance later, and we will give a fairly detailed overview. Rather than focussing solely on push forwards between positive geometries, we will discuss push forward between general differential forms instead. Before we do so, however, we will need to define the \emph{pull back}.

\subsubsection{Pull Back} 

Suppose we give $\Cbb^m$ coordinates $\bm{y}=(y_1,\ldots,y_m)$, $\Cbb^n$ coordinates $\bm{z}=(z_1,\ldots,z_n)$, and we have a map $\bm\phi\colon \Cbb^m\to\Cbb^n$ which maps $\bm{y}\mapsto \bm{\phi}(\bm{y}) = (\phi_1(\bm{y}),\ldots,\phi_n(\bm{y}))$. Now suppose we have some meromorphic $p$-form $\omega$ on the space $\Cbb^n$ ($p\leq n$), then we define the \emph{pull back} of $\omega$ though $\bm\phi$ as
\begin{align}
	\bm\phi^*\omega = \omega|_{\bm{z}=\bm{\phi}(\bm{y})}\,.
\end{align}
Without loss of generality, we take $\omega$ to be
\begin{align}\label{eq:POS_omega-defs}
	\omega = \sum_{I\in\binom{[n]}{p}}{\omega}_I(\bm{z})\dd\bm{z}^I\,,
\end{align}
where we use $\dd\bm{z}^I=\dd z_{i_1}\wedge\cdots\wedge\dd z_{i_p}$ for $I=\{i_1,\ldots,i_p\}$, and the ${\omega}_I={\omega}_{i_1,\ldots,i_p}$ are some rational functions of $z$. In this case we use the chain rule $\dd\phi_i=\sum_j \partial \phi_i/\partial z_j \dd z_j$ to write the pullback as
\begin{align}
	\bm\phi^*\omega = \sum_{I\in\binom{[n]}{p}}\sum_{J\in\binom{[m]}{p}}{\omega}_I(\bm{\phi}(\bm{y})) \left|\frac{\partial \bm\phi}{\partial \bm{y}}\right|^I_J \dd\bm{y}^J\,,
\end{align}
where $\partial\bm{\phi}/\partial\bm{y}$ is the $m\times n$ Jacobian matrix whose entries are the partial derivatives $\partial \phi_j/\partial y_i$, and $|M|^I_J$ denotes the minor of the matrix $M$ made from the rows $I$ and columns $J$.

\subsubsection{Push Forward}

\begin{figure}
	\centering
	\includegraphics[width=0.6\textwidth]{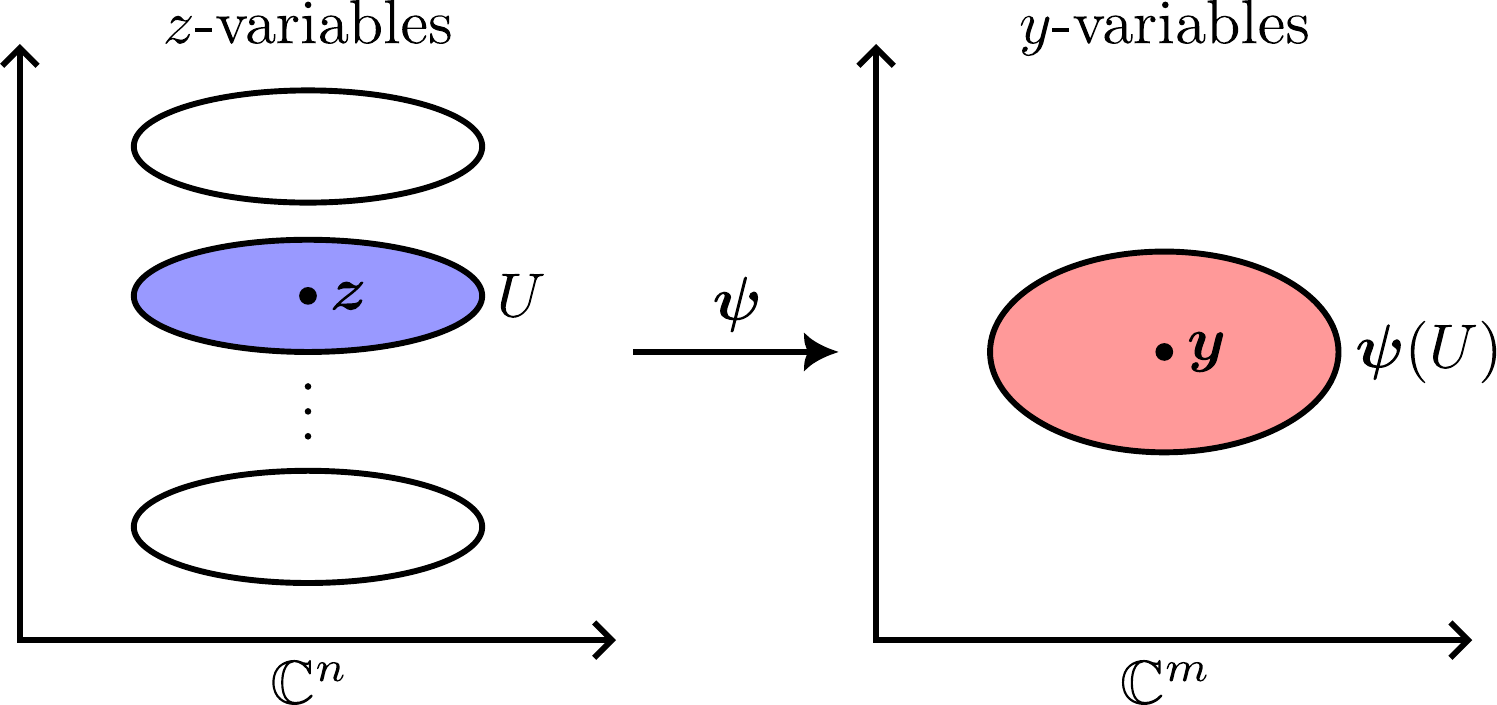}
	\caption{The map $\bm{\psi}\colon\Cbb^n\to\Cbb^m$ sends $\bm{z}\in\Cbb^n$ to $\bm{y}=\bm{\psi}(\bm{z})\in \Cbb^m$. A neighbourhood $U$ of $\bm{z}$ is sent to a neighbourhood $\psi(U)$ of $\bm\psi(\bm{z})$}
	\label{fig:psi}
\end{figure}
\begin{figure}
	\centering
	\includegraphics[width=0.6\textwidth]{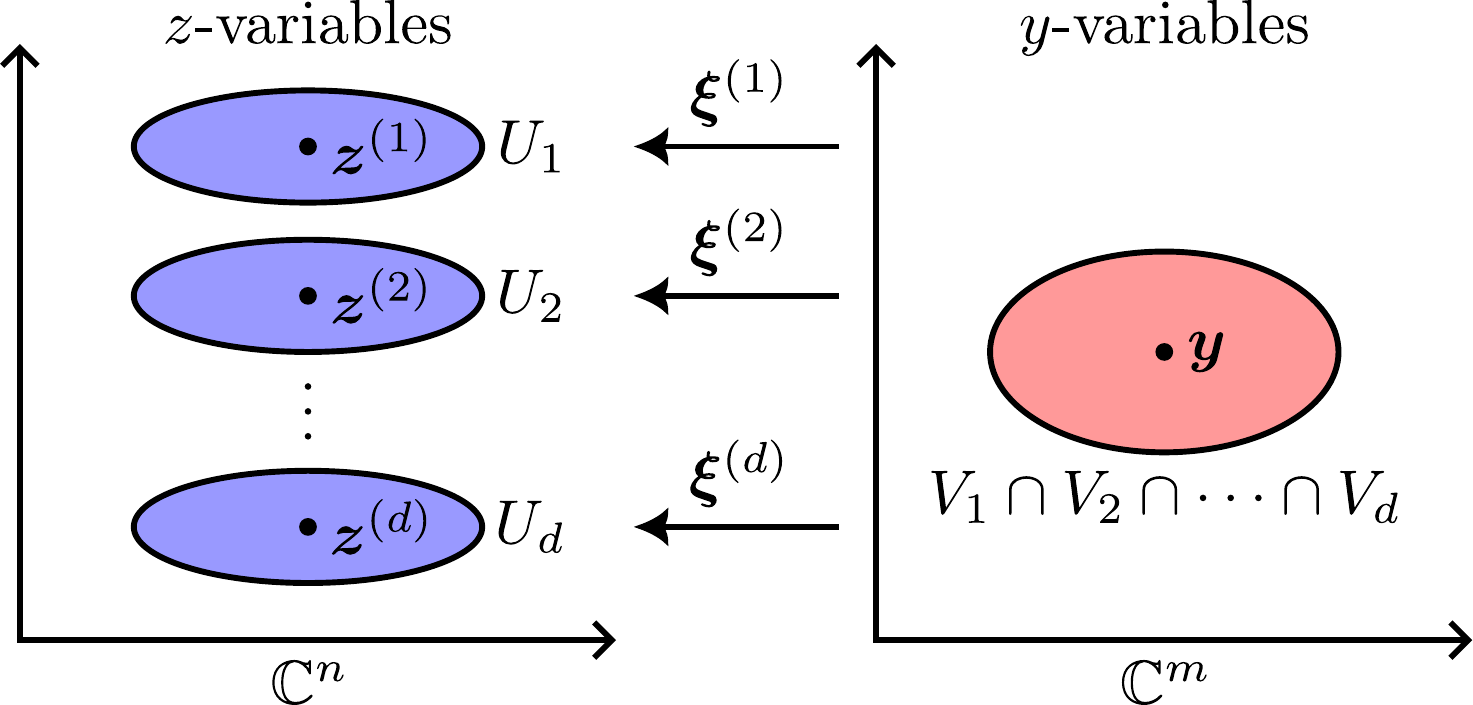}
	\caption{In some neighbourhood of the point $\bm{y}\in\Cbb^m$, the map $\bm{\psi}:\Cbb^n\to\Cbb^m$ has $d$ local inverse maps \smash{$\bm{\xi}^{(\alpha)}\coloneqq\bm{\psi}\big|_{U_\alpha}^{-1}:V_\alpha\to U_\alpha$} which send $\bm{y}$ to \smash{$\bm{z}^{(\alpha)}\coloneqq\bm{\xi}^{(\alpha)}(\bm{y})$}.}
	\label{fig:xi}
\end{figure}
\begin{figure}
	\centering
	\includegraphics[width=0.6\textwidth]{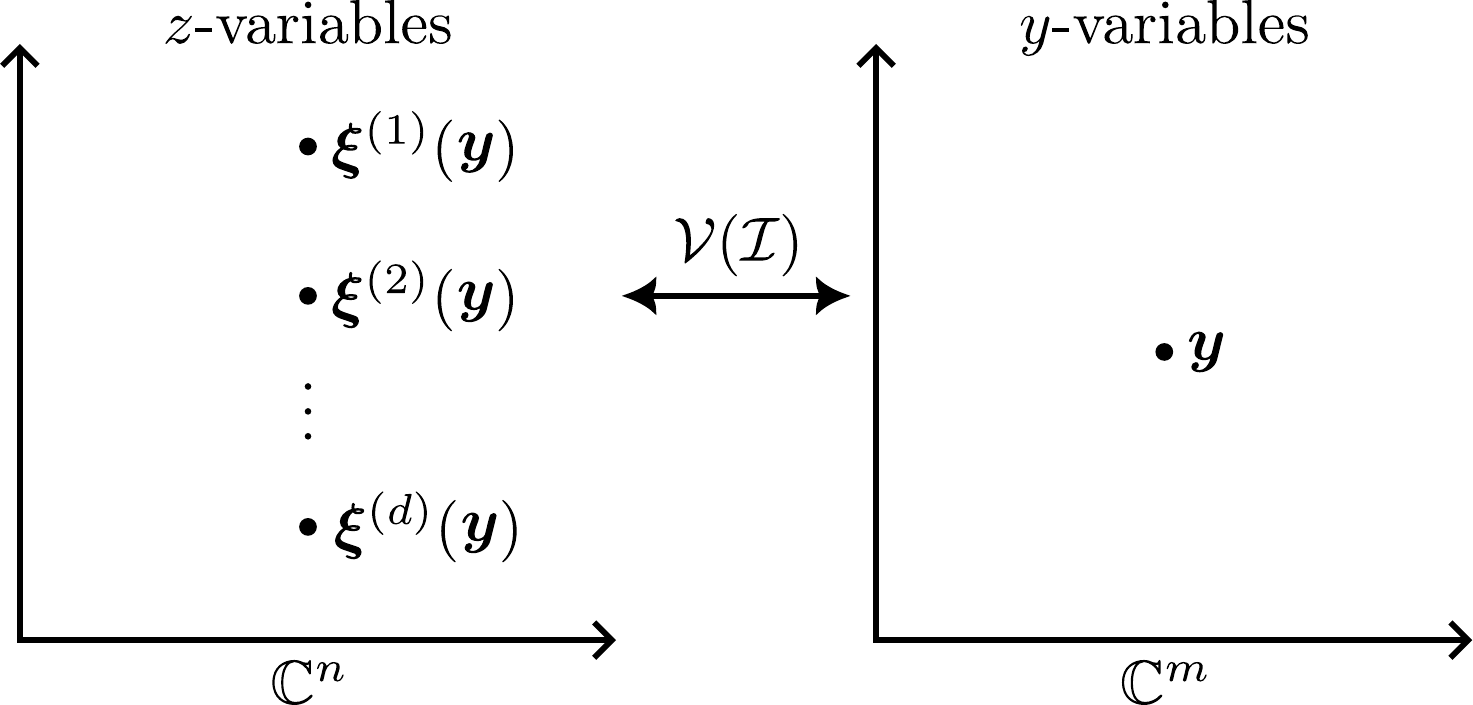}
	\caption{The variety $\Vcal(\Ical)=\{\bm{\xi}^{(\alpha)}\}_{\alpha=1}^d$ consists of $d$ maps $\bm{\xi}^{(\alpha)}\colon\Cbb^m\to\Cbb^n$.}
	\label{fig:variety}
\end{figure}

To define the push forward we consider a map $\bm\psi\colon \Cbb^n\to\Cbb^m$ which sends $\bm{z}\in\Cbb^n$ to $\bm{z}\mapsto \bm{\psi}(\bm{z})=(\psi_1(\bm{z}),\ldots,\psi_m(\bm{z}))$, depicted in figure \ref{fig:psi}. We assume that $\bm\psi$ is a meromorphic map of degree $d$, which means that a general point $\bm{y}\in\Cbb^m$ has $d$ points in its preimage:
\begin{align}
	\bm\psi^{-1}(\bm{y}) = \{\bm{z}^{(1)},\ldots,\bm{z}^{(d)}\}\,,
\end{align}
for some points $z^{(\alpha)}\in \Cbb^n$. We then define the \emph{local inverse maps} $\bm{\xi}^{(\alpha)}\coloneqq \bm{\psi}|_{U_\alpha}^{-1}\colon V_\alpha\to U_\alpha$, where $V_\alpha$ is an open neighbourhood of $\Cbb^m$ containing $\bm{y}$, and $U_\alpha$ an open neighbourhood of $\Cbb^n$ containing $\bm{z}^{(\alpha)}$. Figure \ref{fig:xi} schematically depicts these local inverse maps. We then define the \emph{push forward of $\omega$ through $\psi$} as the sum over pullbacks through $\bm{\xi}^{(\alpha)}$:
\begin{align}
	\bm{\psi}_*\omega\coloneqq \sum_{\alpha=1}^d \bm{\xi}^{(\alpha)*}\omega\,.
\end{align}
In particular, we will be interested in considering push forwards in cases where the map $\bm\psi$ is only defined implicitly through some set of polynomial equations. We consider the polynomials $f_1,\ldots,f_n$ in variables $\bm{z}$ with rational coefficients in $\bm{y}$ such that the ideal $\Ical=\<f_1,\ldots,f_n\>$ is zero dimensional (see appendix \ref{sec:APP_alg-geom-gen} for a brief review of some basic concepts in algebraic geometry). We then define the \emph{push forward through the ideal $\Ical$} as
\begin{align}
	\Ical_*\omega\coloneqq \sum_{\bm{\xi}\in\Vcal(\Ical)}\bm\xi^*\omega\,,
\end{align}
where the elements $\bm{\xi}$ of $\Vcal(\Ical)=\{\bm\xi^{(\alpha)}\}_{\alpha=1}^d$ should generically be considered maps from $\Cbb^m$ to $\Cbb^n$, as depicted in figure \ref{fig:variety}.

We continue working with $\omega$ as defined in equation \eqref{eq:POS_omega-defs}. Using the equation for the pullback derived above, we find
\begin{align}\label{eq:POS_push-forward-dxi}
	\Ical_*\omega=\sum_{J\in\binom{[m]}{p}}\left[ \sum_{\bm{\xi}\in\Vcal(\Ical)}\sum_{I\in\binom{[n]}{p}}\omega_I(\bm{\xi})\left| \frac{\partial \bm{\xi}}{\partial\bm{y}} \right|^I_J \right]\dd\bm{y}^J\,.
\end{align}
Since, by definition, $\bm{f}=\bm{0}$ when evaluated on a point $\bm{\xi}\in\Vcal(\Ical)$, we find
\begin{align}
	0=\frac{\dd f_i}{\dd y_j}= \frac{\partial f_i}{\partial y_j} + \sum_l\frac{\partial f_i}{\partial z_l}\frac{\partial \xi_l}{\partial y_j}\implies \frac{\partial \bm{\xi}}{\partial\bm{y}} = - \left.\left[ \frac{\partial \bm{f}}{\partial \bm{z}} \right]^{-1} \frac{\partial \bm{f}}{\partial \bm{y}}\right|_{\bm{z}=\bm{\xi}}\,,
\end{align}
and hence
\begin{align}
	\left| \frac{\partial \bm{\xi}}{\partial\bm{y}} \right|^I_J = (-1)^p \left|\left[\frac{\partial\bm{f}}{\partial\bm{z}}(\bm{\xi})\right]^{-1}\frac{\partial\bm{f}}{\partial\bm{y}}(\bm{\xi})\right|^I_J\,,
\end{align}
where all Jacobians are now evaluated at $\bm{z}=\bm{\xi}$. Putting everything together, we arrive at the following formula for a push forward of a general $p$-form $\omega$ through the ideal $\Ical$:
\begin{align}\label{eq:POS_pushforward-ideal}
	\Ical_*\omega = \sum_{J\in\binom{[m]}{p}}\big(\Ical_*\overline{\omega}_J\big)\dd \bm{y}^J\,,
\end{align}
where we have defined
\begin{align}\label{eq:POS_pushforward-special-fn}
	\overline\omega_J(\bm{z};\bm{y})\coloneqq (-1)^p \sum_{I\in\binom{[n]}{p}}\omega_I(\bm{z})\left| \left[\frac{\partial\bm{f}}{\partial\bm{z}}\right]^{-1}\frac{\partial\bm{f}}{\partial\bm{y}} \right|^I_J\,,
\end{align}
which are rational functions of $\bm{z}$ and $\bm{y}$ for each index set $J\in\binom{[m]}{p}$. We further point out that we defined the pushforward is for general $p$-forms, including rational functions (0-forms):
\begin{align}
	\Ical_*\overline\omega_J = \sum_{\bm{\xi}\in\Vcal(\Ical)}\overline\omega_J(\bm\xi)\,.
\end{align}
We have thus reduced the question of finding the pushforward of a general $p$-form $\omega$ to being able to find the pushforward of the rational functions $\overline\omega_J$. In section \ref{sec:POS_pf} we will discuss several ways this can be done explicitly.

\subsubsection{Push Forwards of Canonical Forms}

We consider two $d$-dimensional positive geometries $(X,X_{\geq0})$ and $(Y,Y_{\geq0})$. A \emph{morphism} $\Phi$ from $(X,X_{\geq0})$ to $(Y,Y_{\geq0})$ consists of a meromorphic map $\Phi\colon X \to Y$ which restricts to an orientation preserving diffeomorphism between the interiors $X_{>0}$ and $Y_{>0}$. We can then find the canonical form of $Y_{\geq0}$ from $\Omega(X_{\geq0})$ by taking the pushforward:
\begin{align}
	Y_{>0} \stackrel{\text{diffeo}}{=} \Phi(X_{>0})\quad\implies\quad	\Omega(Y_\geq0) = \Phi_* \Omega(X_{\geq0})\,.
\end{align}
If we use coordinates $\bm{x}=(x_1,\ldots,x_d)$ on $X$ and $\bm{y}=(y_1,\ldots,y_d)$ on $Y$, we can write the canonical forms as
\begin{align}
	\Omega(X_{\geq0}) = \underline\Omega(X_{\geq 0}) \dd^d x\,,\quad \Omega(Y_{\geq0}) = \underline\Omega(Y_{\geq 0}) \dd^d y\,,
\end{align}
where $\underline\Omega(X_{\geq0})$ and $\underline\Omega(Y_{\geq0})$ are known as the \emph{canonical (rational) functions}. The \emph{delta function expression} of the pushforward is then given by
\begin{align}\label{eq:POS_delta-push-forward}
	\underline\Omega(Y_{\geq0})(\bm{y}) = \int \dd^d x\, \underline\Omega(X_{\geq0})(\bm{x}) \delta^d(\bm{y}-\Phi(\bm{x}))\,.
\end{align}
Of course, we can again define the pushforward through an implicit map given by a zero-dimensional ideal $\Ical = \<f_1,\ldots, f_d\>$. Since the canonical forms are top-forms, this becomes
\begin{align}
	\Ical_*\Omega(X_{\geq0}) = \sum_{\bm{\xi}\in\Vcal(\Ical)} (-1)^d \underline\Omega(X_{\geq0})|_{\bm{x}=\bm{\xi} }\left.\frac{|\partial\bm{f}/\partial\bm{y}|}{|\partial\bm{f}/\partial\bm{x}|} \right|_{\bm{x}=\bm{\xi}} \dd^dy\,.
\end{align}
Hence, the canonical function of $Y_{\geq0}$ can be calculated as
\begin{align}
	\underline\Omega(Y_{\geq0}) = \Ical_*\left( (-1)^n \underline\Omega(X_{\geq0}) \frac{|\partial\bm{f}/\partial\bm{y}|}{|\partial\bm{f}/\partial\bm{x}|} \right)\,.
\end{align}

\paragraph{Examples.}

As a simple example, let us consider the morphism $\Phi$ between the line segments $[1,2]$ and $[1,4]$ given by $y=\Phi(x)=x^2$. As we have seen above, the canonical form of $[1,2]$ is given by
\begin{align}
	\Omega([1,2])=\dd\log\frac{x-1}{x-2}\,.
\end{align}
The map $\Phi$ has two local inverses given by $\xi^{(1)}=\sqrt{x}$, and $\xi^{(2)}=-\sqrt{x}$. Then, following the definition, the push forward of $\Omega([1,2])$ through this map is given by
\begin{align}
	\Phi_*\Omega([1,2]) &= \xi^{(1)*}\Omega([1,2])+\xi^{(2)*}\Omega([1,2])= \dd\log\frac{\sqrt{y}-1}{\sqrt{y}-2}+\dd\log\frac{-\sqrt{y}-1}{-\sqrt{y}-2}\\
	&= \dd\log \frac{y-1}{y-4}=\Omega([1,4])\,.
\end{align}
We see that we indeed end up with the canonical form of the segment $[1,4]$.

\begin{figure}
	\centering
	\includegraphics[width=0.9\textwidth]{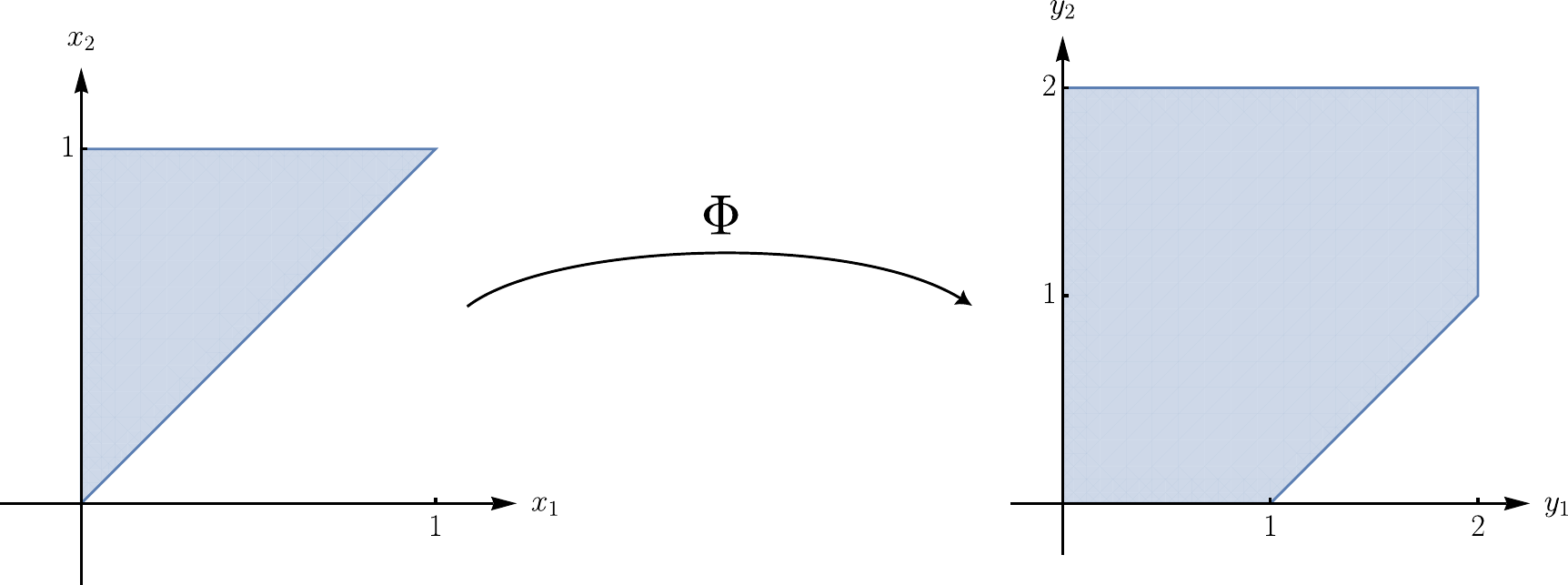}
	\caption{The map $\Phi$ maps the triangle $\Acal$ to the pentagon $\Bcal$.}
	\label{fig:ABHY-map-pentagon}
\end{figure}

As a less trivial example, which will turn out to be relevant later, we consider a map from a triangle $\Acal$ to a pentagon $\Bcal$, as depicted in figure \ref{fig:ABHY-map-pentagon}. The triangle $\Acal$ defined by $0\leq x_1\leq x_2\leq 1$, and the map
\begin{align}
	(y_1,y_2)=\Phi (x_1,x_2) = \left(\frac{x_1(1+x_2)}{x_2}, \frac{x_2-x_1+x_2(1-x_1)}{1-x_1}\right)\,,
\end{align}
provides a morphism to the pentagon $\Bcal$.
The canonical form of the triangle $\Acal$ is given by
\begin{align}
	\Omega(\Acal) = \frac{\dd x_1\wedge\dd x_2}{x_1(x_1-x_2)(1-x_2)}\,.
\end{align}
This map again has two local inverses, given explicitly by
\begin{align}
	\bm{\xi}^{\pm} = \Big(&\notag\frac{2-y_1+y_2+y_1y_2\pm \sqrt{-4 y_1 y_2^2 + (2+y_2-y_1(1-y_2))^2}}{2y_2},\\
	&\frac{2-y_1+y_2+y_1y_2\mp \sqrt{-4 y_1 y_2^2 + (2+y_2-y_1(1-y_2))^2}}{2(y_1-2)}\Big)\,.
\end{align}
Although the local inverses are somewhat unwieldy, the push forward takes the relatively simple form
\begin{align}\label{eq:POS_ABHY-pentagon-form-c=1}
	\Phi^*\Omega(\Acal) = \bm\xi^{+*}\Omega(\Acal)+\bm\xi^{-*}\Omega(\Acal)= \frac{4y_2+4-y_1 y_2 - 2 y_1}{y_1y_2(2-y_1)(2-y_2)(1+y_2-y_1)}\dd y_1\wedge\dd y_2\,,
\end{align}
which is the canonical form of the pentagon $\Omega(\Bcal)$. In this example we see very clearly that the map $\Phi$ is only a diffeomorphism between the \emph{interiors} of $\Acal$ and $\Bcal$, which does not extend to the boundaries. Most of the boundaries of $\Acal$ map appropriately to boundaries of $\Bcal$, except that the map $\Phi$ is ill-defined on the point $(x_1,x_2)=(1,0)$, and ignoring this point, we don't end up with all the boundaries of $\Bcal$\footnote{It is possible to recover all the boundaries of $\Bcal$ by considering a \emph{blow-up} of the point $(1,0)$. This essentially amounts to keeping track of how this vertex is approached by inserting a projective line at this point.}.

We notice that in both of the examples above the result of the push forward is some rational differential form, which is far from obvious given the square roots which appear in the local inverses. This is a general statement which will become clear after our discussion in section \ref{sec:POS_pf}.

\subsection{Simple Polytopes}\label{sec:POS_simple-polytopes}

We take this moment to review a particular way to find canonical forms for a special subset of projective polytopes. A $d$-dimensional polytope is \emph{simple} if all vertices are adjacent to exactly $d$ facets (or, equivalently, $d$ edges). These simple polytopes have a particularly simple formula for their canonical form, which we will call upon in future sections. For some vertex $v$ of our simple polytope $\Acal$, we can write the facets adjacent to this vertex as $Y\cdot W_a=0$, $a=1,\ldots,d$ for some dual vectors $W_a$ (see section \ref{sec:GRASS_proj_poly}). The canonical form can then be written as
\begin{align}\label{eq:POS_simple-polytope}
	\Omega(\Acal) = \sum_{v\in \Vcal(\Acal)} \text{sign}(v) \bigwedge_{a=1}^d \dd\log Y\cdot W_a\,,
\end{align}
where $\Vcal(\Acal)$ is the vertex set of $\Acal$. The $\sgn(v)$ denotes the orientation of the facets $W_1,\ldots,W_d$. We note that two vertices $v_1,v_2\in \Vcal(\Acal)$ are connected by an edge if they share all but one facets. This means that their contribution to \eqref{eq:POS_simple-polytope} only differs by a single term in the wedge product. If we keep the order of all other terms the same, only interchanging these two different facets, then $\sgn(v_1)=-\sgn(v_2)$. This rule allows us to fix all the relative signs in equation \eqref{eq:POS_simple-polytope}.

Following \cite{Arkani-Hamed:2017mur}, the validity of formula \eqref{eq:POS_simple-polytope} can be argued inductively. The base case with $d=0$ is trivial. Assuming that \eqref{eq:POS_simple-polytope} holds for simple polytopes with dimensions less than $d$, then it is sufficient to show that \eqref{eq:POS_simple-polytope} has the correct codimension-one poles and residues. Clearly it follows from the definition that all poles in \eqref{eq:POS_simple-polytope} correspond to facets of the geometry, and that all these poles are simple. Taking the residue at some facet $Y\cdot W=0$, only terms coming from vertices adjacent to this facet will survive. Hence we are left with a very similar equation to \eqref{eq:POS_simple-polytope}, except we are only summing over the vertices adjacent to this facet. By the induction hypothesis this is the correct canonical form for the facet, thus completing the proof. 

\paragraph{Example.}
As an example we consider the pentagon $\Bcal$ depicted on the right hand side of figure \ref{fig:ABHY-map-pentagon}, which is a simple polytope (all polygons are simple). There are five vertices with incident facets given by $(y_1=0,y_2=0),(y_1=0,2-y_2=0),(2-y_1=0,2-y_2=0),(2-y_1=0,1-y_1-y_2=0),(y_2=0,1-y_1-y_2=0)$. Filling this in into \eqref{eq:POS_simple-polytope} we get
\begin{align}
	\Omega(\Bcal)&=\dd\log y_1\wedge\dd\log y_2-\dd\log y_1\wedge \dd\log (2-y_2)+\dd\log (2-y_1)\wedge\dd\log(2-y_2) \notag\\
	&-\dd\log(2-y_1)\wedge\dd\log(1+y_2-y_1)+\dd\log y_2\wedge\dd\log (1-y_1-y_2)\\
	&=-\frac{4y_2+4-y_1 y_2 - 2 y_1}{y_1y_2(2-y_1)(2-y_2)(1+y_2-y_1)}\dd y_1\wedge\dd y_2\,,
\end{align}
which agrees with the form we found in equation \eqref{eq:POS_ABHY-pentagon-form-c=1} up to an overall sign.

\section{The ABHY Associahedron}\label{sec:POS_ABHY}

We now turn to a simple positive geometry which is relevant in physics. The \emph{ABHY associahedron}, also called the \emph{kinematic associahedron}, describes tree-level amplitudes in bi-adjoint $\phi^3$ theory. We have encountered the scattering amplitudes of this theory in section \ref{sec:INT_phi3}. We will mainly be interested in the case where both colour orderings are identical: $\alpha=\beta=\unit$, also known as $\tr{\phi^3}$ theory. We recall that these double partial amplitudes $m_n\equiv m_n[\unit,\unit]$ get contributions from planar tree trivalent Feynman diagrams, which are dual to triangulations of an $n$-gon. 

\begin{figure}
	\centering
	\begin{subfigure}[b]{0.45\textwidth}
		\centering
		\includegraphics[width=\textwidth]{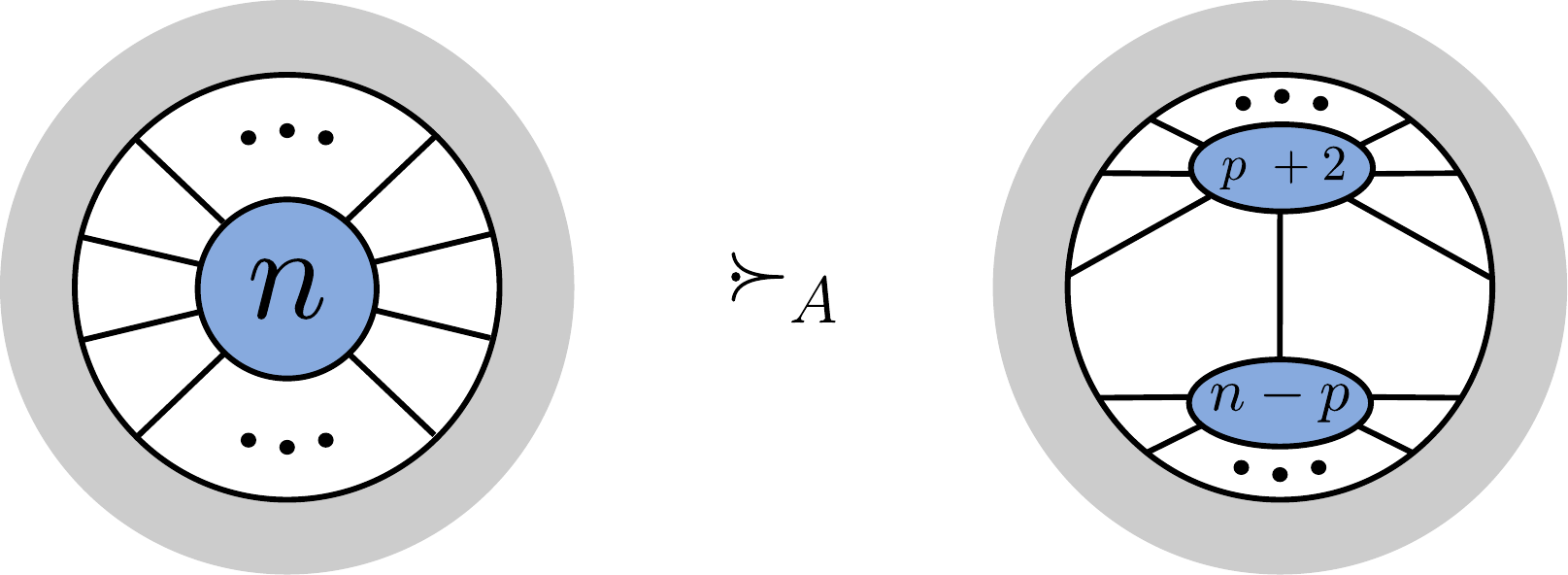}
		\caption{The covering relations for factorisation graphs. The number inside a vertex indicates its degree.}
		\label{fig:covering-relations-phi3}
	\end{subfigure}
	\hfill
	\begin{subfigure}[b]{0.45\textwidth}
		\centering
		\includegraphics[width=\textwidth]{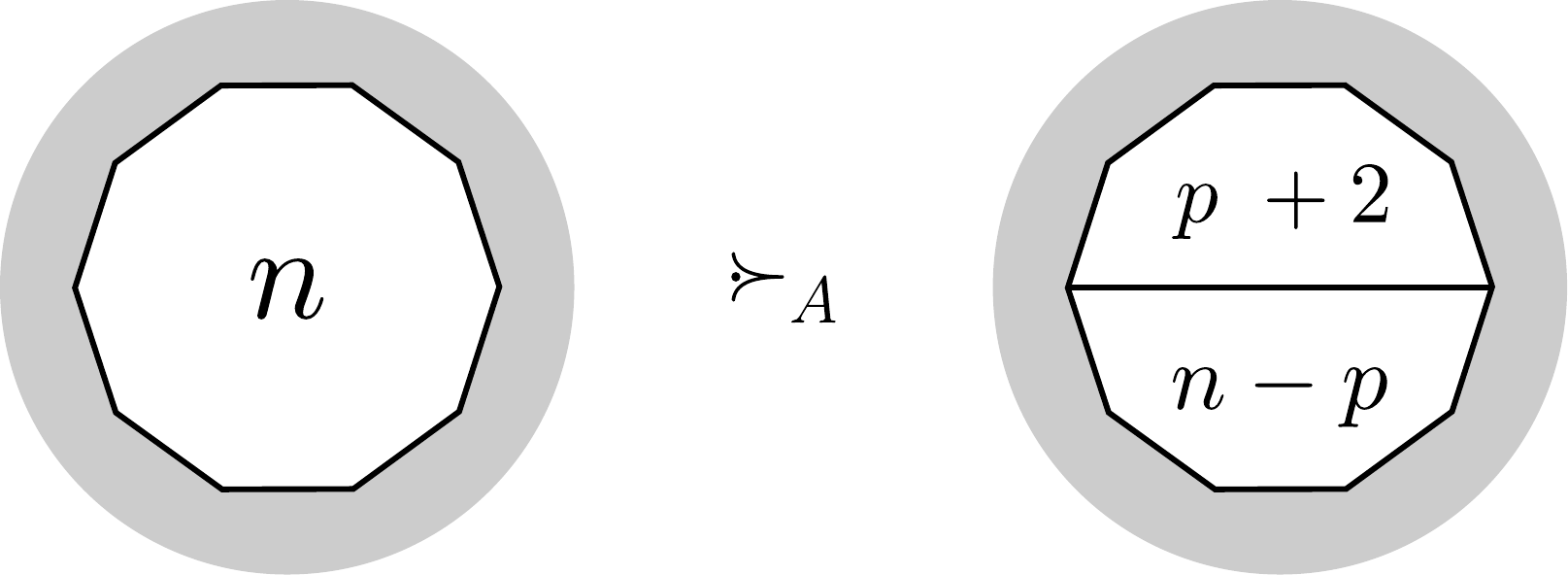}
		\caption{The covering relations for triangulations of $n$-gons. The number inside a polygon indicates how many edges it has.}
		\label{fig:covering-relations-associahedron}
	\end{subfigure}
	\caption{The covering relations for singularities in $\tr{\phi^3}$ theory. The grey outline is meant to indicate that these diagrams can be part of some larger diagram. We assume $n>3$, and $1\leq p \leq n-3$}
	\label{fig:covering-relations}
\end{figure}

To motivate the geometric description of these scattering amplitudes, let us first consider the `boundary stratification' of $m_n$, by which we mean the poset of singularities. In this exceedingly simple theory, the only singularities of the amplitude are factorisations. Starting from $m_n$, which we denote by a single $n$-point vertex, the codimension-1 boundaries consist of all ways to split the $n$-point amplitude into an $(n-p)$-point amplitude connected by an internal edge to a $(p+2)$-point amplitude, for some $1\leq p\leq n-3$. We can then repeat this process on the subsequent amplitudes, and so on. This process comes to a natural halt when the diagram consists solely of trivalent vertices, since the three-point amplitude does not have any singularities. We refer to these diagrams as \emph{factorisation diagrams}. The full boundary structure can be generated from the \emph{covering relations} depicted in figure \ref{fig:covering-relations-phi3}.
These covering relations extend transitively to a partial order on the set of factorisation diagrams, which summarises the boundary stratification of $m_n$. 

We note that we can equivalently record these factorisation graphs in their dual diagram, as depicted in figure \ref{fig:covering-relations-associahedron}. Starting from an empty $n$-gon, a factorisation on the pole $X_{ij}=0$ is then depicted by a chord connecting corners $i$ and $j$. In this way, we have an equivalent formulation of a boundary of $m_n$ in terms of non-overlapping chords on an $n$-gon. These diagrams are known to label exactly the boundaries of the \emph{associahedron}, or \emph{Stashef polytope} \cite{c507c782-a486-3e42-a801-97778df5e634, 58e41a3e-2db4-3e2f-8edb-8effbf4ad0ce}, an $(n-3)$-dimensional simple polytope. We thus see that the singularity structure of scattering amplitudes in $\tr{\phi^3}$ theory can be neatly summarised by the boundary structure of an associahedron. This is a non-trivial statement that is completely obscured when looking at the amplitude from the point of view of Feynman diagrams. 

\begin{figure}
	\centering
	\includegraphics[scale=0.3]{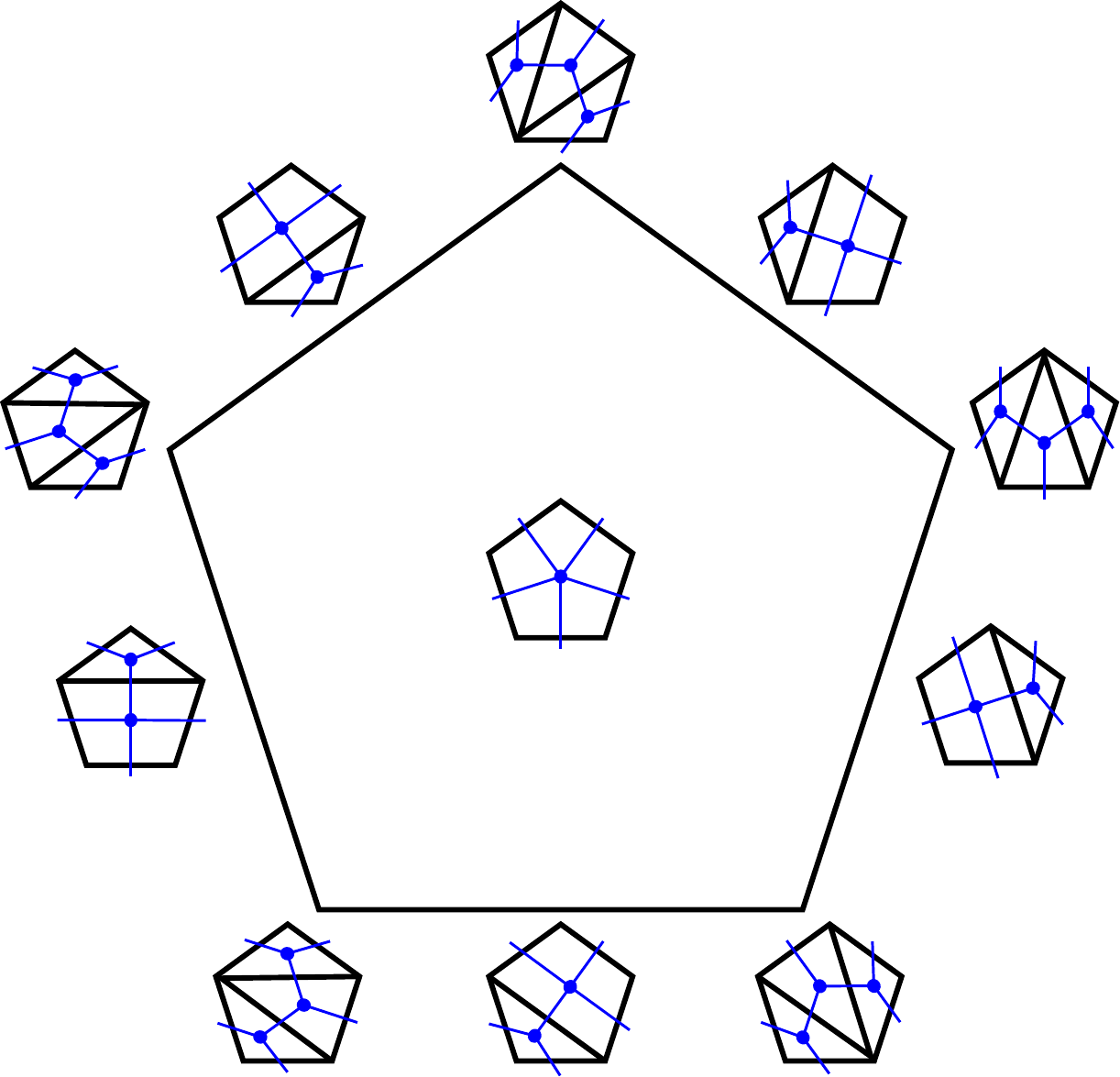}
	\caption{The $d$-dimensional boundaries of a 2-dimensional associahedron (a pentagon) can be labelled by a triangulation of a 5-gon (also a pentagon) by $2-d$ non-intersecting chords. The dual factorisation diagrams are depicted in blue.}
	\label{fig:pentagon-triangulation-boundaries}
\end{figure}
\begin{figure}
	\centering
	\includegraphics[scale=0.5]{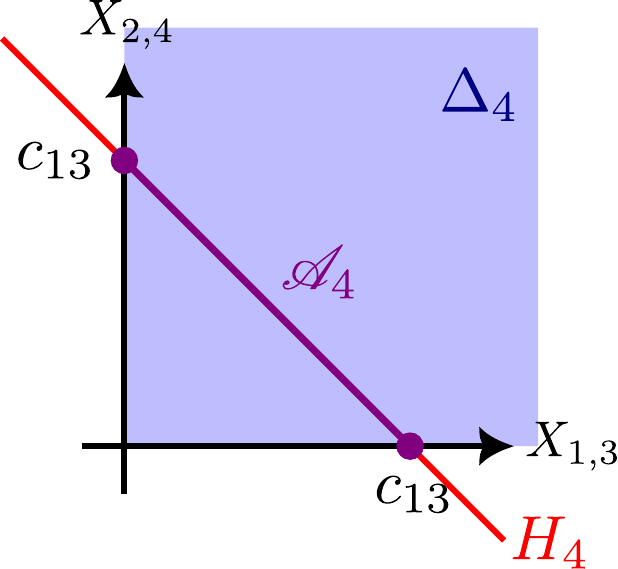}
	\caption{The associahedron $\Ascr_{4}$ (in purple) is defined as the intersection of the positive quadrant $\Delta_4$ (in blue) with the hypersurface $H_4$ (in red).}
	\label{fig:ABHY-4}
\end{figure}

Of course, just knowing that the boundaries of $m_n$ form a polytopal poset is not the same as having an actual positive geometric description of the amplitudes. This is achieved by a specific associahedron living in kinematic space: the \emph{kinematic associahedron}, or the \emph{ABHY associahedron}. The appropriate kinematic space $\Kbb_n$ consists of all planar Mandelstam variables $X_{ij}$ (see section \ref{sec:KIN_mand}). This (complex) space will form the ambient space of our positive geometry. We further consider the nonnegative region $\Delta_n$ of the real slice of the kinematic space $\Kbb_n(\Rbb)$ defined to be the region where all planar Mandelstam variables are nonnegative:
\begin{align}
	X_{ij}\geq 0\quad \forall i,j=1,\ldots,n\,.
\end{align}
Next, we define an affine hyperplane $H_n\subset \Kbb_n$ where all non-planar two-particle Mandelstam variables $s_{ij}$ are equal to some fixed constant $-c_{ij}$, where $c_{ij}>0$. We can write this as the region of $\Kbb_n$ where
\begin{align}\label{eq:POS_ABHY-surface-def}
	X_{ij}+X_{i+1\,j+1}-X_{i+1\,j}-X_{i\,j+1}=c_{ij},\quad\forall 1\leq i< j<n\,.
\end{align}
We then define the ABHY associahedron as the intersection
\begin{align}
	\Ascr_n = \Delta_n\cap H_n\,.
\end{align}
We illustrate this definition for the simple case where $n=4$ in figure \ref{fig:ABHY-4}. By counting the number of constraints, we find
\begin{align}
	\dim \Ascr_n= \dim \Kbb_n - \frac{(n-2)(n-3)}{2}=n-3\,.
\end{align}

\begin{figure}
	\centering
	\includegraphics[width=0.9\textwidth]{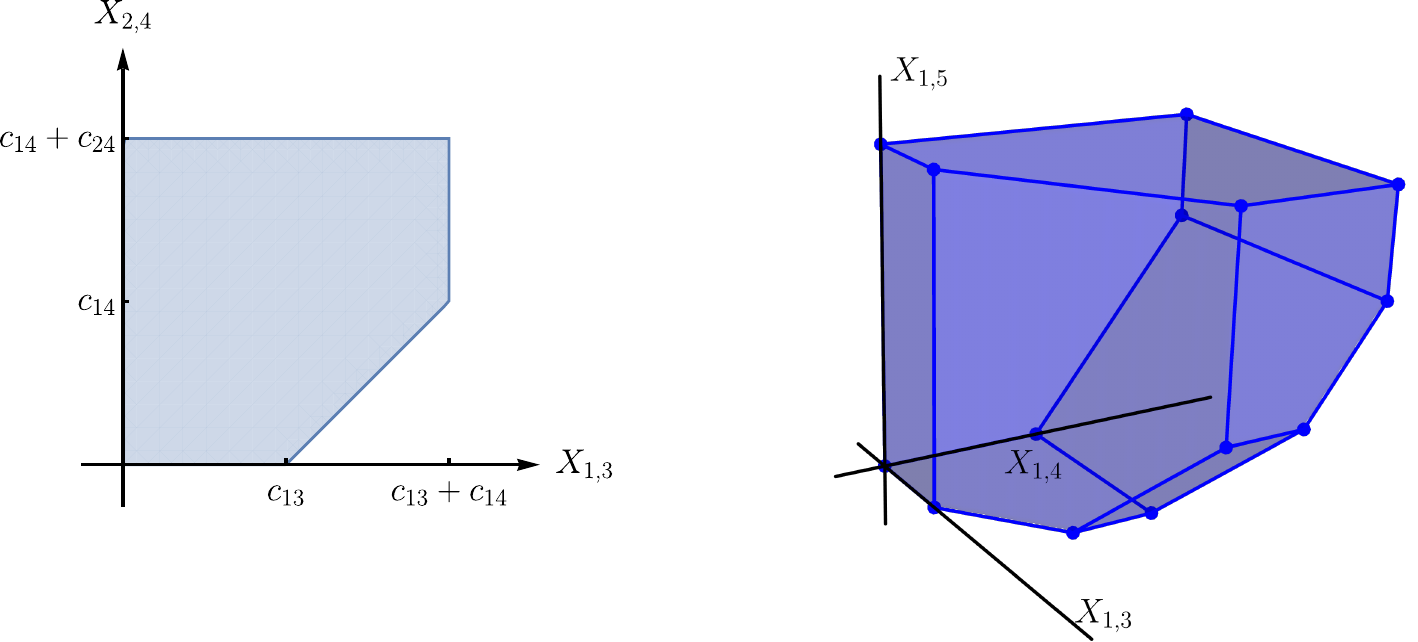}
	\caption{The ABHY associahedra $\Ascr_5$ and $\Ascr_6$.}
	\label{fig:ABHY-56}
\end{figure}

The claim that this is an associahedron can be understood as follows. First, from the definition of $\Delta_n$ it is clear that the boundaries of $\Ascr_n$ are given by some $X_{ij}=0$, which we label by adding a chord $(i,j)$ to the triangulation of our $n$-gon. Next, from \eqref{eq:POS_ABHY-surface-def} we find
\begin{align}
	X_{jk}+X_{il} = X_{ik} + X_{jl} - \sum_{\substack{i\leq a <j\\k \leq b < l}}c_{ab}\,.
\end{align} 
We assume that $(i,j,k,l)$ are cyclically ordered, which means that the chords $(i,k)$ and $(j,l)$ cross. We then see that we cannot set $X_{ik}$ and $X_{jl}$ to zero at the same time, as this would imply that $X_{jk}+X_{il}$ is negative, which contradicts the definition of $\Ascr_n$. Hence, the boundaries of $\Ascr_n$ are described by collections of \emph{non crossing} chords, \emph{i.e.} triangulations of an $n$-gon, which is exactly the boundary structure of the associahedron we encountered above. The ABHY associahedra for $n=5,6$ are depicted in figure \ref{fig:ABHY-56}. 

Since an associahedron is a simple polytope, we can use the results from section \ref{sec:POS_simple-polytopes} to find its canonical form as a sum over vertices. We recall that the vertices are in correspondence with full triangulations of an $n$-gon, and we are thus summing over all triangulations (or, from the dual perspective, all Feynman diagrams). We label a triangulation $T=\{(i_1,j_1),\ldots,(i_{n-3},j_{n-3})\}$ by the chords $(i,j)$ part if this triangulation. If we let $\Tcal_n$ denote the total set of all triangulations of an $n$-gon, then
\begin{align}
	\Omega(\Ascr_n) = \sum_{T\in \Tcal_n} \sgn(T)\bigwedge_{(i,j)\in T}\dd\log X_{ij}\,.
\end{align}
The relative signs can be determined following the procedure explained in section \ref{sec:POS_simple-polytopes}.

There is another way we can fix the $\sgn(T)$ which will be important later on. We can fix the signs by requiring that the final answer is \emph{projective invariant}: the canonical form $\omega_n^\ABHY$ should be invariant under the substitution $X_{ij}\to \Lambda(X) X_{ij}$, where we specify that $\Lambda$ is a function of $X$ to signify that it is not a constant with respect to the exterior derivative: $\dd \Lambda \neq 0$. This requirement essentially means that the canonical form can be written in a way such that it only depends on ratios $X_{ij}/X_{kl}$.

Using the rules for simple polytopes described above, we find
\begin{align}
	\omega_4^{\text{ABHY}} &= \dd\log X_{13}-\dd\log X_{24} = \dd\log\frac{X_{13}}{X_{24}}\,,\\
	\omega_5^{\text{ABHY}} &= \dd\log X_{14}\wedge\dd\log X_{13} + \dd\log X_{13}\wedge\dd\log X_{35} + \dd\log X_{35}\wedge\dd\log X_{25}\\\nonumber
	&+\dd\log X_{25}\wedge\dd\log X_{24} + \dd\log X_{24}\wedge\dd\log X_{14}\\\nonumber
	&=\dd\log\frac{X_{13}}{X_{24}}\wedge\dd\log\frac{X_{13}}{X_{14}}+\dd\log\frac{X_{13}}{X_{25}}\wedge\dd\log\frac{X_{35}}{X_{24}}\,.
\end{align}
The forms presented above should be understood as forms on the ambient space $\Kbb_n$. To extract the scattering amplitudes $m_n$ and to properly define the canonical forms, we need to pull these forms back onto the hypersurface $H_n$, where we can find $m_n$ as the canonical function (\emph{i.e.} the function multiplying the top-dimensional measure on $H_n$). We need to choose a basis on $H_n$, the most natural choice is to use a subset of $n-3$ planar Mandelstam variables. We are free to pick any set of $X_{ij}$ where the chords $(i,j)$ do not cross. We will choose $(X_{13},X_{14},\ldots,X_{1\,n-1})$ as our basis. Using \eqref{eq:POS_ABHY-surface-def} we find
\begin{align}
	X_{i>1,j} = X_{1j}-X_{1\,i+1}+C_{ij}\,,
\end{align}
where $C_{ij}=\sum_{r=1}^{i-1}\sum_{s=i+1}^{j-1} c_{rs}$. From this we find that
\begin{align}
	\dd X_{i>1,j} = \dd X_{1j}-\dd X_{1i+1}\,.
\end{align}
Substituting this into our canonical form we find the amplitude as
\begin{align}
	\Omega(\Ascr_n) = m_n \dd^{n-3}X\,,
\end{align}
where $\dd^{n-3}X=\dd X_{13}\wedge\cdots\wedge\dd X_{1\,n-1}$. For example for the $n=4,5$ cases:
\begin{align}
	\Omega(\Ascr_4) & = \brac{\frac{1}{X_{13}}+\frac{1}{X_{24}}}\dd X_{13}\,,\\
	\Omega(\Ascr_5) & = \brac{\frac{1}{X_{13}X_{14}}+\frac{1}{X_{13}X_{35}}+\frac{1}{X_{25}X_{35}}+\frac{1}{X_{24}X_{25}}+\frac{1}{X_{14}X_{24}}}\dd X_{13}\wedge\dd X_{14}\,.
\end{align}
Technically, since $\Ascr_n$ lives on the hypersurface $H_n$, the appropriate variables to use for its canonical form are the $X_{1i}$ and some number of $c_{ij}$'s. However, for the purpose of scattering amplitudes it is more convenient to write $\Omega(\Ascr_n)$ in terms of the $X_{ij}$'s instead, as it immediately gives the correct scattering amplitudes
\begin{align}
	m_4 &= \frac{1}{X_{13}}+\frac{1}{X_{24}}\,,\\
	m_5 &= \frac{1}{X_{13}X_{14}}+\frac{1}{X_{13}X_{35}}+\frac{1}{X_{25}X_{35}}+\frac{1}{X_{24}X_{25}}+\frac{1}{X_{14}X_{24}}\,.
\end{align}
Unlike the amplituhedron (which we will encounter in the next section), the ABHY associahedron only captures tree-level scattering amplitudes. An extension to one-loop integrands has been achieved by the \emph{halohedron} \cite{Salvatori:2018fjp, Salvatori:2018aha}, or by \emph{cluster polytopes} \cite{Arkani-Hamed:2019vag}. The positive geometric description of higher-loop amplitudes in $\tr{\phi^3}$ has proven to be a more difficult subject. However, the search for such a geometric description ultimately lead to an all-loop all-topology description of scattering amplitudes in $\tr{\phi^3}$ as a `counting problem' \cite{Arkani-Hamed:2023lbd}. This new combinatorial way of describing amplitudes is distinct from the positive geometries which we discuss in this thesis, although it follows a very similar set of guiding principles. It has recently been an active area of research, which lead to the discovery of new factorisations of scattering amplitudes and formulae for different theories \cite{Arkani-Hamed:2023mvg, Arkani-Hamed:2023swr, Arkani-Hamed:2023jry, Arkani-Hamed:2024nhp, Arkani-Hamed:2024vna, Arkani-Hamed:2024yvu, Cao:2024gln, Bartsch:2024amu, Li:2024qfp, Arkani-Hamed:2024fyd}.

\section{The Amplituhedron}\label{sec:POS_amplituhedron}

The amplituhedron is the historical seed for the field of positive geometries and has been studied extensively by both physicists \cite{Arkani-Hamed:2013kca, Bai:2014cna, Franco:2014csa, Arkani-Hamed:2014dca, Bai:2015qoa, Ferro:2015grk, Dennen:2016mdk, Ferro:2016zmx, Ferro:2016ptt, Arkani-Hamed:2017vfh, Ferro:2018vpf, Arkani-Hamed:2018rsk, Langer:2019iuo, YelleshpurSrikant:2019meu, Herrmann:2020qlt, Dian:2022tpf, Arkani-Hamed:2023epq, Kojima:2020tjf, Arkani-Hamed:2021iya, Brown:2023mqi} and mathematicians \cite{Lam:2014jda, Karp:2017ouj, Lukowski:2019kqi, Lukowski:2019sxw, Gurdogan:2020tip, Mohammadi:2020plf, Akhmedova:2023wcf, Arkani-Hamed:2017tmz, Karp:2016uax, Parisi:2021oql, Moerman:2021cjg, Even-Zohar:2023del, Even-Zohar:2024nvw, Parisi:2024psm, Even-Zohar:2021sec, Bao:2019bfe, Lam:2024gyg, Galashin:2018fri}. It was first introduced in 2013 by Arkani-Hamed and Trnka \cite{Arkani-Hamed:2013jha}, and it captures the tree amplitudes and $L$-loop integrands for planar \nf. The formulation of the amplituhedron partially rests upon the Grassmannian formulation of scattering amplitudes in momentum twistor variables, as were introduced toward the end of section \ref{sec:AMP_nf}. As such, the amplituhedron doesn't describe the full amplitude, but rather the amplitude normalised by an MHV `Parke-Taylor' amplitude, which we denote $\hat{A}_{n,K}$ in the N\textsuperscript{$K$}MHV sector. As we noted before, these quantities are more naturally associated to Wilson loops which are dual to scattering amplitudes in \nf, a property which stems from T-duality in string theory. Since its first conception, several distinct yet equivalent descriptions of the amplituhedron have been introduced, we will use them somewhat interchangeably, depending on which description is most natural for the case at hand. We will start our discussion with the original definition in an auxiliary Grassmannian space, and afterwards we will consider the topological description in momentum twistor space. 

As an initial motivation, we can interpret the amplituhedron as a generalisation of projective polytopes into the Grassmannian. We recall from section \ref{sec:GRASS_proj_poly} that we can find the convex hull of $n$ points $Z_1,\ldots,Z_n$ in $\Rbb\Pbb^m=G(1,m+1)$ as the image of the map
\begin{align}
	\Phi\colon G_+(1,n)&\to G(1,m+1)\\
	C &\mapsto C\cdot Z^T\,,
\end{align}
where the $1\times n$ matrix $C=\begin{pmatrix} c_1 & \cdots & c_n \end{pmatrix}$ parametrises the interior of $G_+(1,n)$, and $Z$ is a $(m+1)\times n$-matrix encoding the homogeneous coordinates of the points $Z_i$. We require that the matrix $Z$ is \emph{positive}, in the same sense as the positive Grassmannian:
\begin{align}
	\<Z_{i_1}\cdots Z_{i_{m+1}}\> > 0\,,\quad \forall i_1<\ldots<i_{m+1}\,.
\end{align} 
This positivity implies that any simplex made from $m+1$ of these points has the same orientation, and ensures that the map is projectively well-defined. If some of the maximal minors of $Z$ were negative, it would be possible for a point $Z_i$ to lie in the convex hull of the remaining points, in which case it would be possible to find a positive linear combination of $Z$s which equals zero, which would make the map ill-defined.

We generalise this map by upgrading the matrix $Z$ to a positive $(K+m)\times n$-matrix\footnote{The positivity of the matrix $Z$ is sufficient to ensure that the amplituhedron map is well-defined, however it is not a necessary condition \cite{Lam:2015uma, Karp:2016uax}. The images $\Phi_Z(G_+(K,n))$ for arbitrary $Z$ go by the name of \emph{Grasstopes}, and have recently been studied for the case $m=1$ in \cite{mandelshtam2023combinatorics}.},
\begin{align}
	Z = \begin{pmatrix}
		Z_{1}^1 & Z_{2}^1 & \cdots & Z_{n}^1\\
		\vdots & \vdots & \ddots & \vdots \\
		Z_{1}^{K+m} & Z_{2}^{K+m} & \cdots & Z_{n}^{K+m}
	\end{pmatrix}\,,
\end{align}
and taking $G_+(K,n)$ as the domain:
\begin{align}
	\Phi_Z\colon G_+(K,n)&\to G(K,K+m)\,,\\
	C&\mapsto Y = C\cdot Z^T\,.
\end{align}
The image of this map is the amplituhedron $\Acal_{n,K,m}$. The amplituhedron is full-dimensional on the $K m$-dimensional ambient space $G(K,K+m)$. The $SL(K)$ invariant measure on $G(K,K+m)$ is $\prod_{\alpha=1}^K \<Y_1\cdots Y_k\dd^m Y_\alpha\>$. It is conjectured that the qualitative properties of the amplituhedron do not depend on the specific form of the matrix $Z$. The amplituhedron is invariant under the \emph{twisted cyclic shift}
\begin{align}
	\bm{c}_1\to \bm{c}_2,\cdots, \bm{c}_n\to (-1)^{K-1}\bm{c}_1\,,\quad \bm{Z}_1\to \bm{Z}_2,\cdots, \bm{Z}_n\to (-1)^{K+m-1}\bm{Z}_1\,,
\end{align}
where $\bm{c}_i$ ($\bm{Z}_i$) denotes the $i$\textsuperscript{th} column-vector of $C$ ($Z$).

A natural set of coordinates to use are determinants of the form
\begin{align}
	\<Y Z_{i_1}\cdots Z_{i_m}\>=\epsilon_{I_1\cdots I_k J_1\cdots J_M} Y_1^{I_1} \cdots Y_k^{I_k} Z_{i_1}^{J_1}\cdots Z_{i_m}^{J_M}\,,
\end{align}
which are sometimes called \emph{twistor coordinates}. Expanding this by using $Y_{i}^I = C_{ia}Z_a^I$, we find (note the similarity to equation \eqref{eq:GRASS_mom-twistor-bracket})
\begin{align}
	\<Y Z_{i_1}\cdots Z_{i_m}\> = \sum_{a_1<\cdots<a_k} (a_1\cdots a_k)_C \<a_1\cdots a_k i_1\cdots i_m\>_Z\,.
\end{align}
From this it follows that for all $Y\in\Acal_{n,K,m}$ 
\begin{align}
	\<Y i_1 i_1+1\cdots i_{\frac{m}{2}} i_{\frac{m}{2}}+1\> >0\,,
\end{align}
for $m$ even, and
\begin{align}
	(-1)^K \<Y 1 i_1 i_1+1\cdots i_{\frac{m-1}{2}} i_{\frac{m-1}{2}}+1 \>>0\,,\quad \<Y i_1 i_1+1\cdots i_{\frac{m-1}{2}} i_{\frac{m-1}{2}}+1 n \>>0\,,
\end{align}
for $m$ odd. The codimension-1 boundaries of the amplituhedron are described by one of these brackets going to zero.

Most relevant for physics is the $m=4$ case, often referred to as \emph{the} amplituhedron. The canonical form of $\Acal_{n,K}\equiv\Acal_{n,K,4}$ encodes $\text{N}^K\text{MHV}$ amplitudes of \nf. We define the `bosonisation' of Grassmann parameters $\chi_{iA}$, $i=1,\ldots,n,\,A=1,\ldots,K+4$ to be $\sum_A \phi_\alpha^A \chi_{iA}$, where $\phi_\alpha^A,\,\alpha=1,\ldots,K$ are auxiliary Grassmann-odd parameters. If we collect the kinematic variables in a $(K+4)\times n$ matrix
\begin{align}
	Z = \begin{pmatrix}
		z_i^a \\ \phi_a^A \chi_{1A} \\ \vdots \\ \phi_1^A \chi_{k A}
	\end{pmatrix}\,,
\end{align}
which we interpret as the matrix $Z$ appearing in the definition of the amplituhedron. The canonical form of the amplituhedron
\begin{align}
	\Omega(\Acal_{n,K,4}) = \omega_{n,K}(Y;Z)\prod_{\alpha=1}^{K}\<Y_1\cdots Y_K\dd^4 Y_\alpha\>\,,
\end{align}
then encodes the \emph{bosonised superamplitude} in its canonical function. We can retrieve the standard superamplitude by localising $Y$ on 
\begin{align}
	Y_0=\left(
	\begin{array}{c}
		\nul_{4\times K} \\ \hdashline[2pt/2pt]
		\unit_{K\times K} \\
	\end{array}
	\right)\,,
\end{align}
such that
\begin{align}\label{eq:POS_projection-through-Y0}
	\<Y_0ijkl\>=\<z_iz_jz_kz_l\>\,,
\end{align}
and integrating out the auxiliary Grassmann variables:
\begin{align}\label{eq:POS_amplituhedron-to-amplitude}
	\hat{A}_{n,K} = \int \dd^4 \phi_1\cdots\dd^4\phi_k \omega_{n,K}(Y_0;Z)\,.
\end{align}
Starting from the $4K$-dimensional positroid cell $S_\sigma$ whose image is injective, the image $\Phi_Z(C_\sigma)$ is called a \emph{tile}, and triangulations in terms of tiles are called \emph{tilings}. If we have a positive parametrisation $C_\sigma(\bm\alpha)$ of a positroid cell $S_\sigma$ in terms of variables $\alpha_1,\ldots,\alpha_{4K}$, such that
\begin{align}
	\Omega(S_\sigma)=\bigwedge_{i=1}^{4K}\dd\log\alpha_i\,,
\end{align}
then the canonical form of the tile is given by the pushforward through $\Phi_Z$:
\begin{align}
	\Omega(\Phi_Z(S_\sigma)) = \Phi_{Z*}\Omega(S_\sigma)\,.
\end{align}
After integrating out the auxiliary Grassmann variables, we find that the superfunction obtained from the image of this positroid cell is exactly
\begin{align}
	\int \dd\log\alpha_1\wedge\cdots\wedge\dd\log\alpha_{4K} \delta^{4K|4K}(C\cdot \Zcal^T)\,,
\end{align}
where $\smash{\Zcal_a = \begin{pmatrix} z_a & \chi_a \end{pmatrix}}$ are supermomentum twistors. We recall from section \ref{sec:AMP_nf} that this is the result for the Yangian invariant on-shell diagram corresponding to this positroid cell. 

It was recently proven \cite{Even-Zohar:2021sec, Even-Zohar:2023del, Even-Zohar:2024nvw} that the amplituhedron is triangulated by the tiles corresponding to a BCFW representation of the scattering amplitude (known as a \emph{BCFW triangulation}, or \emph{BCFW tiling}). It thus follows that equation \eqref{eq:POS_amplituhedron-to-amplitude} correctly reproduces the $\text{N}^K\text{MHV}$ amplitudes of \nf.

\paragraph{Examples.}

There are a few special cases of amplituhedra. We have already seen examples of the case where $K=1$, in which case the amplituhedron is just the convex hull of the points $Z_a$ in projective space. The condition that $Z$ is a positive matrix turns these polytopes into so-called \emph{cyclic polytopes} \cite{STURMFELS1988275}. When the $Z$ matrix is square, \emph{i.e.} $m+K=n$, then the amplituhedron is isomorphic to the positive Grassmannian $G_+(K,n)$.

\begin{figure}
	\centering
	\includegraphics[width=\textwidth]{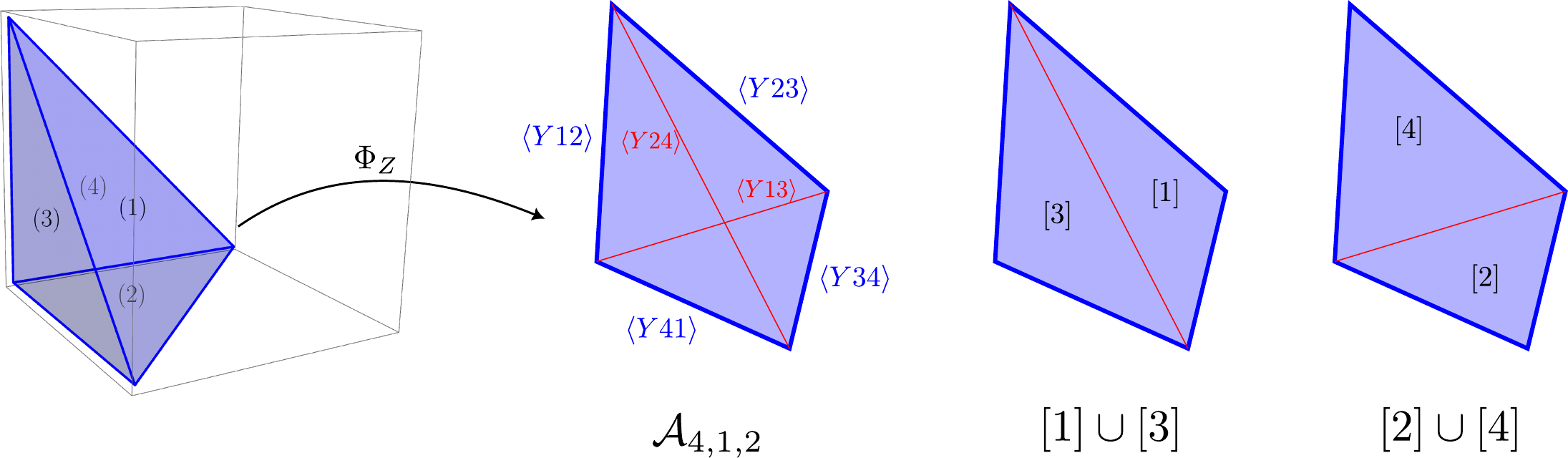}
	\caption{A schematic depiction of the map $\Phi_Z$ from $G_+(1,4)$ to the amplituhedron $\Acal_{4,1,2}$. We see that $\Acal_{4,1,2}$ can be triangulated by the positroid tiles $[1]\cup[3]$, or $[2]\cup[4]$.}
	\label{fig:m2-amplituhedron-example}
\end{figure}

Let us have a detailed look at a simple example. We consider the amplituhedron $\Acal_{4,1,2}$. We assume we have a positive matrix $Z\in M_+(3,n)$. The domain of the map $\Phi_Z$ is $G_+(1,4)$, and the amplituhedron is a 2-dimensional object in the space $G(1,3)\cong\Pbb^2$. The positive Grassmannian $G_+(1,4)$ is topologically a tetrahedron, whereas the amplituhedron $\Acal_{4,1,2}$ is a quadrilateral, we have depicted the map schematically in figure \ref{fig:m2-amplituhedron-example}. We choose the positive parametrisation $\smash{C=\begin{pmatrix}	\alpha_1&\alpha_2&\alpha_3&\alpha_4 \end{pmatrix}/GL(1)\in G_+(1,4)}$ for the matrix $C$. The amplituhedron is defined as the space
\begin{align}
	\Acal_{4,1,2} = \{Y^I= \alpha_1 Z_1^I+ \alpha_2 Z_2^I + \alpha_3 Z_3^I + \alpha_4 Z_4^I\;\colon \alpha_i>0\}\,.
\end{align}
We see that the twistor coordinates
\begin{alignat}{2}
	&\<Y12\> = \alpha_3\<123\>_Z + \alpha_4 \<124\>_Z\,,\qquad &&\<Y23\> = \alpha_1 \<123\>_Z + \alpha_4\<234\>_Z\,,\\
	&\<Y34\> = \alpha_1 \<134\>_Z + \alpha_2 \<234\>_Z \,,\qquad &&\<Y41\> = \alpha_2\<124\>_Z + \alpha_3 \<134\>_Z\,,
\end{alignat}
are all manifestly positive and are the facets of $\Acal_{4,1,2}$, whereas
\begin{align}\label{eq:POS_m2-amplituhedron-example-spurious-boundaries}
	\<Y13\> = -\alpha_2\<123\>_Z+\alpha_4\<134\>\,,\qquad \<Y24\> = \alpha_1\<124\>_Z - \alpha_3\<234\>_Z\,,
\end{align}
do not have a definite sign and are not facets.

The codimension-1 positroid cells are parametrised by one of the $\alpha_i$ going to zero. They are two-dimensional cells whose image is also two-dimensional, their images are \emph{positroid tiles}, and they are the analogue of BCFW tiles for the $m=4$ amplituhedron. We label these positroid cells by the location of the zero element of the $C$ matrix, \emph{e.g.} $(1)=\begin{pmatrix}	0&\alpha_2&\alpha_3&\alpha_4 \end{pmatrix}$, and we denote the associated tiles by $[1]=\Phi_Z\big((1)\big)$. From \eqref{eq:POS_m2-amplituhedron-example-spurious-boundaries} we see that $Y\in[1]$ satisfies $\<Y24\><0$, and $Y\in[3]$ satisfies $\<Y24\>>0$. These tiles are therefore on the opposite side of the spurious boundary $\<Y24\>=0$. The tiles $[1]$ and $[3]$ triangulate the amplituhedron, as is clear from figure \ref{fig:m2-amplituhedron-example}. Similarly, it is clear that the tiles $[2]$ and $[4]$ are on opposite sides of the spurious boundary $\<Y13\>=0$, and they also triangulate the amplituhedron.

We can use $GL(1)$ transformation to write
\begin{align}
	\begin{pmatrix}	0&\alpha_2&\alpha_3&\alpha_4 \end{pmatrix}\sim \begin{pmatrix}	0&\frac{\alpha_2}{\alpha_4}&\frac{\alpha_3}{\alpha_4}&1 \end{pmatrix}\,,
\end{align}
from which we see that this positroid cell has the canonical form
\begin{align}
	\Omega\big((1)\big) = \dd\log\frac{\alpha_2}{\alpha_4}\wedge\dd\log\frac{\alpha_3}{\alpha_4}\,.
\end{align}
We find that the map $\Phi_Z$ acting on this positroid cell only has one `local inverse', which we can express in terms of twistor coordinates as
\begin{align}
	\frac{\alpha_2}{\alpha_4} = \frac{\<Y34\>}{\<Y23\>}\,,\quad \frac{\alpha_3}{\alpha_4} = -\frac{\<Y24\>}{\<Y23\>}\,.
\end{align}
We then find the canonical form of the positroid tile $[1]$ as the pushforward
\begin{align}
	\Omega([1])=\Phi_{Z*}\left( \dd\log\frac{\alpha_2}{\alpha_4}\wedge\dd\log\frac{\alpha_3}{\alpha_4}\right) = \dd\log\frac{\<Y34\>}{\<Y23\>}\wedge\dd\log\frac{\<Y24\>}{\<Y23\>}\,.
\end{align}
One can easily find the canonical form of the remaining three positroid tiles using similar arguments. This allows us to find the canonical form of the full amplituhedron from its triangulation as
\begin{align}
	\Omega(\Acal_{4,1,2}) &= \Omega([1])+\Omega([3]) = \dd\log\frac{\<Y34\>}{\<Y23\>}\wedge\dd\log\frac{\<Y24\>}{\<Y23\>} + \dd\log\frac{\<Y12\>}{\<Y41\>}\wedge\dd\log\frac{\<Y24\>}{\<Y41\>}\\
	&=\frac{\<134\>\<234\>\<Y12\>+\<123\>\<124\>\<Y34\>}{\<Y12\>\<Y23\>\<Y34\>\<Y41\>}\frac{\<Y\dd^2Y\>}{2!}\,.
\end{align}
Although it may not look like it at first glance, the numerator is actually cyclically invariant. This is a consequence of the \Pluck relations on the $3\times 5$ matrix $\begin{pmatrix}
	Y & Z
\end{pmatrix}$. For example, we note that the \Pluck relations imply $\<234\>\<Y12\>=\<124\>\<Y23\>-\<123\>\<Y24\>$, and $\<124\>\<Y34\>=\<134\>\<Y24\>-\<234\>\<Y14\>$. Hence, we can equivalently write
\begin{align}
	\Omega(\Acal_{4,1,2})=\frac{\<124\>\<134\>\<Y23\>-\<123\>\<234\>\<Y14\>}{\<Y12\>\<Y23\>\<Y34\>\<Y41\>}\frac{\<Y\dd^2Y\>}{2!}\,.
\end{align}
We observe that in the preceding example we have not made any reference to the specific form of the matrix $Z$ (in fact, we haven't even defined a matrix $Z$), we only made use of its positivity. The results about its boundaries, positroid tilings, and canonical form are therefore independent of the specific form of the matrix $Z$.

\subsection{The Loop Amplituhedron}\label{sec:POS_loop-amp}

We have seen that tree-level scattering amplitudes in \nf are encoded in the canonical form of the amplituhedron $\Acal_{n,K,4}$. A natural question is whether we can extend this to loop amplitudes. This is, however, not possible in our current framework. The canonical form of a positive geometry only has simple poles. This is rather convenient for tree-level amplitudes, which indeed only have simple poles, but it forms an obstacle for loop amplitudes, which have a more intricate structure involving branch cuts. Fortunately for us, the loop \emph{integrand} is far more similar to tree amplitudes and also only has simple poles. The amplituhedron can indeed be extended such that its canonical form encodes loop integrands in planar \nf, sometimes called the \emph{loop amplituhedron}. 

We start by defining a loop extension of the Grassmannian $G(K,n;L)$, which we define as the space of $K$-planes $C\in G(K,n)$ together with $L$ two-planes $\{D^{(l)}\}_{l=1}^L$ that live in the orthogonal complement of $C$: $D^{(l)}\in G(2,n),\,D^{(l)}\subseteq C^\perp$. We represent a point $\Ccal\in G(K,n;L)$ by a $(K+2L)\times n$ matrix
\begin{align}
	\Ccal = \left(
	\begin{array}{c}
		D^{(1)}\\ \hdashline[2pt/2pt]\rule{0pt}{2.6ex}
		D^{(2)}\\ \hdashline[2pt/2pt]\rule{0pt}{2.6ex}
		\vdots\\ \hdashline[2pt/2pt]\rule{0pt}{2.6ex}
		D^{(L)}\\ \hdashline[2pt/2pt]\rule{0pt}{2.6ex}
		C
	\end{array}
	\right)\,.
\end{align}
We further extend the notion of positivity to the \emph{positive loop Grassmannian} $G_+(K,n;L)$, where we require the maximal minors of some number of $D$ matrices stacked on top of $C$ to be positive. That is, all matrices of the form
\begin{align}\label{eq:POS_loop-positivity}
	\left(
	\begin{array}{c}
		C
	\end{array}
	\right)\quad \left(
	\begin{array}{c}
		D^{(l_1)}\\ \hdashline[2pt/2pt]\rule{0pt}{2.6ex}
		C
	\end{array}
	\right)\quad \left(
	\begin{array}{c}
		D^{(l_1)}\\ \hdashline[2pt/2pt]\rule{0pt}{2.6ex}
		D^{(l_2)}\\ \hdashline[2pt/2pt]\rule{0pt}{2.6ex}
		C
	\end{array}
	\right)\quad\cdots\quad \left(
	\begin{array}{c}
		D^{(l_1)}\\ \hdashline[2pt/2pt]\rule{0pt}{2.6ex}
		\vdots\\ \hdashline[2pt/2pt]\rule{0pt}{2.6ex}
		D^{(l_L)}\\ \hdashline[2pt/2pt]\rule{0pt}{2.6ex}
		C
	\end{array}
	\right)\,,
\end{align}
are positive.

The loop amplituhedron is then defined similarly to the tree amplituhedron as the image of the map 
\begin{align}
	\Phi_Z^{(L)}\colon G_+(K,n;L)&\to G(K,K+4;L)\\
	\Ccal &\mapsto \Ycal = \Ccal\cdot Z^T\,.
\end{align} 
The image $\Ycal$ is naturally decomposed into the tree-level $Y=C\cdot Z^T$ together with $L$ 2-planes $\Lcal^{(l)}=D^{(l)}\cdot Z^T$ living in the four-dimensional orthogonal complement of $Y$:
\begin{align}
	\Ycal = \left(
	\begin{array}{c}
		\Lcal^{(1)}\\ \hdashline[2pt/2pt]\rule{0pt}{2.6ex}
		\Lcal^{(2)}\\ \hdashline[2pt/2pt]\rule{0pt}{2.6ex}
		\vdots\\ \hdashline[2pt/2pt]\rule{0pt}{2.6ex}
		\Lcal^{(L)}\\ \hdashline[2pt/2pt]\rule{0pt}{2.6ex}
		Y
	\end{array}
	\right)\,.
\end{align}

\subsection{Topological Description}

It was argued in \cite{Arkani-Hamed:2017vfh} that the amplituhedron admits a topological description purely in terms of sign-flips. A point $Y$ is inside the amplituhedron $\Acal_{n,K,m}$ if 
\begin{subequations}\label{eq:POS_boundary-def-amp-Y}
\begin{alignat}{2}
	&\<Y i_1\, i_1+1\,\cdots i_{\frac{m}{2}} \,i_{\frac{m}{2}}+1\>>0,\quad &&m\text{ even}\,,\\
	&\begin{cases}
		(-1)^k\<Y1\,i_1\, i_1+1\,\cdots i_{\frac{m-1}{2}} \,i_{\frac{m-1}{2}}+1\> >0\\
		\<Y\,i_1\, i_1+1\,\cdots i_{\frac{m-1}{2}} \,i_{\frac{m-1}{2}}+1\, 1\,n\> >0
	\end{cases}\,,\quad &&m\text{ odd}\,,
\end{alignat}
\end{subequations}
and the sequence
\begin{align}\label{eq:POS_sign-flip-def-amp-Y}
	\{\<Y12\cdots (m-1)m\>,\ldots,\<Y12\cdots (m-1)n\>\}\,,
\end{align}
has exactly $K$ sign flips.

For the physically relevant case when $m=4$ we can also give a sign-flip definition of the loop amplituhedron by additionally imposing
\begin{align}
	&\< Y \Lcal^{(l)} i \,i+1\>>0\,,\quad \<Y\Lcal^{(l_1)}\Lcal^{(l_2)}\>>0\\
	& \{\<Y \Lcal^{(l)}12\>,\ldots, \<Y \Lcal^{(l)} 1n\> \}\quad \text{has $K+2$ sign-flips}\,.
\end{align}

\subsection{Description in Momentum Twistor Space}

Next, we focus on a description of the amplituhedron which lives directly in the space of `momentum twistor variables' (they only reduce to actual momentum twistor variables for the case $m=4$), without referencing the auxiliary Grassmannian space $G(K,K+m)$. From the physical point of view, this is more natural, as the scattering amplitudes in \nf are naturally phrased as questions in momentum twistor space. To do this, we first notice that $Y^\perp$ denotes an element in $G(m,K+m)$, which defines an $m$-plane in $n$ dimensions via 
\begin{align}
	z=Y^\perp\cdot Z^T= (C\cdot Z^T)^\perp\cdot Z\in G(m,n)\,.
\end{align}
This defines a map from $G(K,m+K)$ to momentum twistor space $G(m,n)$\footnote{Due to the little group action rescaling the columns of the $m\times n$ matrix $z$, momentum twistor space is more properly understood as the configuration space of $n$ points $\Pbb^{m-1}$, rather than $G(m,n)$.}, which we shall denote $\Xi$.
It is clear that in this case
\begin{align}
	\<i_1\cdots i_m\>_z = \<Y i_1\cdots i_m\>\,.
\end{align}
We can now map the amplituhedron into momentum twistor space by composing the maps $\Phi_Z$ and $\Xi$.

Comparing to formula \eqref{eq:binary-matrices}, we see that if we identify $Z$ with the element of $G_+(K+m,n)$ defined by its row span, then we can interpret $z$ to be 
\begin{align}
	z &= Z\cap C^\perp=Z\setminus C=C^\perp\setminus Z^\perp\\
	&= (C\cdot Z^T)^\perp\cdot Z = (Z^\perp\cdot (C^\perp)^T)^\perp\cdot C^\perp\,.
\end{align}
Clearly the new ambient space $G(m,n)$ is not $Km$ dimensional, and hence the amplituhedron is not top-dimensional in this space. To define the amplituhedron in a proper dimensional way, we proceed analogous to the `projection through $Y$', which we encountered for dual polytopes in section \ref{sec:GRASS_projective-volume}, except that we are now projecting through the $K$-plane $Y$, rather than through a line. Explicitly, let us fix the $GL(K)$ redundancy to write
\begin{align}
	Y_\alpha^A = \begin{pmatrix}
		-y_\alpha^a & \unit_{K\times K}
	\end{pmatrix}\,\implies\, (Y^\perp)_A^a =\begin{pmatrix}
	\unit_{m\times m} & y_\alpha^a
\end{pmatrix}\,,
\end{align}
where $y_\alpha^a$ is a $K\times m$ matrix, $\alpha=1,\ldots k$, $a=1,\ldots, m$. If we decompose the matrix $Z$ as
\begin{align}
	Z_i^A = \begin{pmatrix}
		z_i^{*a}\\ \Delta_i^\alpha\,,
	\end{pmatrix}
\end{align}
where $z^*$ and $\Delta$ are fixed elements of $G(K,n)$ and $G(m,n)$, respectively, then
\begin{align}
	z_i^a = (Y^\perp)_\alpha^A Z_i^A = z_i^{*a}+y_\alpha^a \Delta_i^\alpha\,.
\end{align}
Therefore, the amplituhedron lives in the $Km$ dimensional subspace of momentum twistor space defined by
\begin{align}
	\Vcal_{n,K,m}[Z]\coloneqq \{z_i^a\colon z_i^a =z_i^{*a}+ Y_\alpha^a\Delta_i^\alpha \}\,.
\end{align}
Morally, our external data $Z$ defines an $m$-plane $z^*$, and we are considering the affine subspaces obtained by translating $z^*$ in some direction lying in the fixed $k$-plane $\Delta$.

We can further define the `winding space' $\Wcal_{n,K,m}$ by taking the sign-flip conditions of equations \eqref{eq:POS_boundary-def-amp-Y} and \eqref{eq:POS_sign-flip-def-amp-Y} and projecting it though $Y$ by sending $\<Y a_1\cdots a_m\>\to \<a_1\cdots a_m\>$. Explicitly, we define $\Wcal_{n,K,m}$ as the subspace of momentum twistor space where
\begin{subequations}
\begin{alignat}{2}
	&\<i_1\, i_1+1\,\cdots i_{\frac{m}{2}} \,i_{\frac{m}{2}}+1\>>0,\quad &&m\text{ even}\,,\\
	&\begin{cases}
		(-1)^k\<1\,i_1\, i_1+1\,\cdots i_{\frac{m-1}{2}} \,i_{\frac{m-1}{2}}+1\> >0\\
		\<i_1\, i_1+1\,\cdots i_{\frac{m-1}{2}} \,i_{\frac{m-1}{2}}+1\, 1\,n\> >0
	\end{cases}\,,\quad &&m\text{ odd}\,,
\end{alignat}
\end{subequations}
and the sequence
\begin{align}
\{\<12\cdots (m-1)m\>,\ldots,\<12\cdots (m-1)n\>\}\,,
\end{align}
has exactly $K$ sign flips. Then we can define the amplituhedron directly in momentum twistor space as 
\begin{align}
	\Acal_{n,K,m}^z=\Vcal_{n,K,m}[Z]\cap \Wcal_{n,K,m}\,.
\end{align}
We use the superscript $z$ on the amplituhedron to identify that this is the version which lives directly in momentum twistor space, rather than an auxiliary Grassmannian space. However, $\Acal_{n,K,m}$ and $\Acal_{n,K,m}^z$ are homeomorphic, and we often abuse notation by dropping the superscript. We often use the two descriptions interchangeably, and it should be clear from context which ambient space of the amplituhedron we are referring to.

As an added bonus, the way we retrieve the superamplitude is even more transparent in momentum twistor space: we simply interpret the $\dd z_i^a$ as the Grassmann variables $\eta_i^a$
\begin{align}
	\hat{A}_{n,K} = \left.\Omega(\Acal_{n,K})\right|_{\dd z_i^a\to \eta_i^a}\,.
\end{align}
The anticommutivity of the Grassmann numbers can now be given the more familiar interpretation as the anticommutivity of the wedge product.

We can also give the loop amplituhedron an interpretation directly in momentum twistor space. We define
\begin{align}
	(AB)_l \coloneqq \Lcal^{(l)} \cdot (Y^\perp)^T = D^{(l)}\cdot z^T\in G(2,4)\,.
\end{align}
We see that we can interpret the loop part of the amplituhedron as a collection of $L$ 2-planes (or, projectively, lines) living inside momentum twistors space $z\in G(4,n)$: $(AB)_l\in G(2,z)$. The notation $(AB)$ is used to emphasise that a line in momentum twistor space can be parametrised by two momentum twistors $z_A,z_B$. The fact that linear combinations of $z_A$ and $z_B$ still parametrise the same line leads to the standard notion of $GL(2)$ invariance of an element of the Grassmannian. The additional sign-flip conditions on the loop amplituhedron in momentum twistor space are
\begin{align}
	&\< (AB)_l i \,i+1\>>0\,,\quad \<(AB)_i (AB)_j\>>0\\
	& \{\<(AB)_l 12\>,\ldots, \<(AB)_l 1n\> \}\quad \text{has $K+2$ sign-flips}\,.
\end{align}

\section{The Momentum Amplituhedron}\label{sec:POS_mom-amp}

As we saw in the previous section, the amplituhedron describes scattering amplitudes in \nf in momentum twistor space normalised by a tree level MHV amplitude. In some sense, the amplituhedron should thus be understood as calculating Wilson loops, which are known to be $T$-dual to scattering amplitudes in \nf. Furthermore, the use of momentum twistors manifests that we are considering massless, planar, four dimensional scattering amplitudes, as loosening any of these constraints invalidates the definition of momentum twistors. These things combined make it difficult to imagine a generalisation of the amplituhedron to different theories or to different spacetime dimensions. 

This motivates the importance of the \emph{momentum amplituhedron} \cite{Damgaard:2019ztj, Ferro:2020lgp, Damgaard:2020eox, Damgaard:2021qbi, Ferro:2022abq}, which is a positive geometry that encodes \emph{amplitudes} (rather than Wilson loops) of \nf in spinor-helicity space. For this, we need to write our amplitudes in non-chiral superspace, where the $n$ point $\text{N}^{k-2}\text{MHV}$ superamplitude has Grassmann degree $2(n-k)$ in $\eta$, and $2k$ in $\tilde\eta$, with an overall Grassmann degree of $2n-4$. The definition is in many ways similar to the amplituhedron, and we shall discuss both a definition of the momentum amplituhedron in an auxiliary Grassmannian space, as well as a sign-flip definition directly in spinor-helicity space.

Given a positive matrix $\tilde\Lambda \in M_+(k+2,n)$, and a \emph{twisted positive} matrix $\Lambda\in M_+^\perp(n-k+2,n)$ (meaning that its orthogonal complement, $\Lambda^\perp$, is a positive $(k-2)\times n$ matrix), then we define the momentum amplituhedron $\Mcal_{n,k}$ as the image of the map
\begin{align}\label{eq:POS_mom-amp-def}
	\Phi_{\Lambda,\tilde\Lambda}\colon G_+(k,n)&\to G(k,k+2)\times G(n-k,n-k+2)\\
	C&\mapsto (\tilde{Y},Y)=(C\cdot \tilde\Lambda^T,C^\perp\cdot\Lambda^T)\,.
\end{align}
We can map this directly into spinor-helicity space by defining
\begin{subequations}
\begin{align}
	\lambda &=Y^\perp\cdot \Lambda \in G(2,n)\,,\\
	\tilde\lambda &= \tilde{Y}^\perp\cdot\tilde\Lambda \in G(2,n)\,.
\end{align}
\end{subequations}
Using the results from section \ref{sec:GRASS_binary}, we interpret this as
\begin{subequations}
\begin{alignat}{3}
	\lambda &=\Lambda \cap C &&= C\setminus \Lambda^\perp&&=\Lambda\setminus C^\perp\,,\\
	\tilde\lambda &= \tilde\Lambda\cap C^\perp &&= C^\perp\setminus \tilde\Lambda &&= \tilde{\Lambda}\setminus C\,.
\end{alignat}
\end{subequations}
In terms of matrices we can write this as
\begin{subequations}\label{eq:POS_lambda-def-mom-amp}
\begin{alignat}{2}
	\lambda &= (C^\perp\cdot \Lambda^T)^\perp\cdot \Lambda &&= (\Lambda^\perp\cdot C^T)^\perp\cdot C\,,\\
	\tilde\lambda &= (C\cdot \tilde\Lambda^T)^\perp\cdot\tilde\Lambda &&= (\tilde\Lambda^\perp\cdot(C^\perp)^T)^\perp\cdot C^\perp\,,
\end{alignat}
\end{subequations}
such that
\begin{align}
	\<Yij\>= \<ij\>\,,\quad [\tilde{Y}ij]=[ij]\,.
\end{align}
Since $\lambda \subseteq C,\,\tilde\lambda\subseteq C^\perp$ we see that $\lambda$ and $\tilde\lambda$ must be orthogonal to each other:
\begin{align}
	\lambda\cdot\tilde\lambda^T=\nul_{2\times 2}\,,
\end{align}
which has the interpretation of momentum conservation, as explained in section \ref{sec:KIN_spin-hel}. A consequence of this is that the image of the map \eqref{eq:POS_mom-amp-def} lives on a codimension-4 `momentum conserving' hypersurface. The dimension of the momentum amplituhedron is thus
\begin{align}
	\dim(\Mcal_{n,k}) = \dim(G(k,k+2))+\dim(G(n-k,n-k+2)) -4 = 2n-4\,.
\end{align}
We can give a definition of the momentum amplituhedron in kinematic space by defining the winding space $\Wcal_{n,k}$ as the region of spinor-helicity space which satisfies 
\begin{alignat}{2}
	&\<ii+1\>>0\,,\quad &&[ii+1]>0\,,\\
	&\{\<12\>,\<13\>,\ldots,\<1n\>\}\quad &&\text{ has $k-2$ sign flips,}\\
	&\{[12],[13],\ldots,[1n]\}\quad &&\text{ has $k$ sign flips,}
\end{alignat}
and the additional requirement that planar Mandelstam variables are positive
\begin{align}\label{eq:POS_mom-amp-planar-mand}
	s_{i i+1\ldots j}=\sum_{\{a,b\}\in\binom{\{i, i+1,\ldots, j\}}{2}} \<ab\>[ab] > 0\,.
\end{align}
Unlike the positivity of $\<ii+1\>$ and $[ii+1]$, the positivity of the planar Mandelstam variables does not follow directly from the definition of the momentum amplituhedron as the image of $\Phi_{\Lambda,\tilde\Lambda}$. Instead, this requirement imposes some non-trivial constraints on the matrices $\Lambda$, $\tilde\Lambda$, which we shall return to in the next subsection. 

Similarly to the amplituhedron, we decompose
\begin{align}
	\Lambda_i^A = \begin{pmatrix}
		\lambda^{*\alpha}_i\\ \Delta_i^a
	\end{pmatrix}\,, \quad \tilde\Lambda_i^{\dot{A}} = \begin{pmatrix}
	\tilde\lambda_i^{*\alphadot}\\\tilde\Delta_i^{\dot{a}}
\end{pmatrix}\,,
\end{align}
and define the $2n-4$ dimensional subspace of spinor helicity space
\begin{align}
	\Vcal_{n,k}[\Lambda,\tilde\Lambda] = \{(\lambda,\tilde\lambda)\colon \lambda\cdot\tilde\lambda^T=\nul\,, \lambda_i^\alpha=\lambda^{*\alpha}_i+y_a^\alpha\Delta_i^a\,,\tilde\lambda_i^{*\alphadot}+\tilde{y}_{\dot{a}}^\alphadot\tilde\Delta_i^{\dot{a}} \}\,.
\end{align}
The spinor-helicity space momentum amplituhedron is then given by
\begin{align}
	\Mcal_{n,k}=\Vcal_{n,k}[\Lambda,\tilde\Lambda]\cap\Wcal_{n,k}\,.
\end{align}
To extract the superamplitude from the canonical form of the momentum amplituhedron living in the auxiliary Grassmannian space, we first need to uplift the form to a top-form on $G(k,k+2)\times G(n-k,n-k+2)$. We do this by noting that
\begin{align}
	\delta^{4}(P)\dd^4 P = 1\,,\quad P^{\alpha\alphadot} = \sum_{i=1}^n \left(Y^\perp\cdot\Lambda^T\right)_i^\alpha\left(\tilde{Y}^\perp\cdot\tilde\Lambda^T\right)_i^\alphadot\,,
\end{align}
and wedging this with $\Omega(\Mcal_{n,k})$. Then we extract the canonical function $\omega_{n,k}$ as
\begin{align}
	\Omega(\Mcal_{n,k})\wedge\dd^4P \delta^4(P) = \prod_{\alpha=1}^{n-k}\<Y_1\cdots Y_{n-k}\dd^2Y_\alpha\>\prod_{\alphadot=1}^k[\tilde{Y}_1\cdots\tilde{Y}_k\dd^2\tilde{Y}_\alphadot]\delta^4(P)\omega_{n,k}\,.
\end{align}
The tree-level amplitude in \nf is obtained by integrating out the auxiliary Grassmann variables
\begin{align}
	A_{n,k}=\delta^4(p)\int \dd\phi_a^1\cdots\dd\phi_a^{n-k}\int\dd\tilde{\phi}_{\dot{a}}^1\cdots\dd\tilde{\phi}_{\dot{a}}^k \omega_{n,k}(Y_0,\tilde{Y}_0;\Lambda,\tilde\Lambda)\,,
\end{align}
where we localise on 
\begin{align}\label{eq:POS_Ystar-mom-amp}
	Y_0 = \left(\begin{array}{c}	\nul_{2\times (n-k)} \\ \hdashline[2pt/2pt]	\unit_{(n-k)\times (n-k)} \\ \end{array}\right)\,,\quad \tilde{Y}_0 = \left(\begin{array}{c}	\nul_{2\times k} \\ \hdashline[2pt/2pt]	\unit_{k\times k} \\ \end{array}\right)\,.
\end{align}
Alternatively, when considering the momentum amplituhedron in spinor-helicity space, we recover the superamplitude simply by replacing
\begin{align}
	A_{n,k} = \left.\Omega(\Mcal_{n,k})\right|_{{\dd\lambda_i^\alpha\to\eta_i^\alpha,\dd\tilde\lambda_i^\alphadot\to\tilde{\eta}_i^\alphadot}}\,.
\end{align}

\subsection{Connection to the Amplituhedron}\label{sec:POS_mom-amp-amp-connection}

Given $(\lambda,\tilde\lambda)\in\Mcal_{n,k}$, we can map this into a point in momentum twistor space by simply defining $\mu_i = x_{1i}\lambda_i$. We claim that this maps a point from the momentum amplituhedron $\Mcal_{n,k}$ into the amplituhedron $\Acal_{n,k-2}$, as long as the matrix $Z=\big( \tilde\Lambda^\perp\cdot Q \big)^\perp$ is a positive $(k-2)\times n$ matrix.
We remind ourselves that in section \ref{sec:AMP_nf} we saw that the matrix $\hat{C}=(C\setminus\lambda)\cdot Q$ is an element of $G_+(k-2,n)$.\footnote{ For $\lambda$ constructed via the momentum amplituhedron, the matrix $\hat{C}$ can be written as $\hat{C}=\Lambda^\perp\cdot C^T\cdot C\cdot Q$.}
It then follows that $Z\cap\hat{C}^\perp$ is an element of the amplituhedron $\Acal_{n,k-2}$. We further argue that this $z$ is also exactly the four-plane of momentum twistors we constructed from $\lambda$ and $\tilde\lambda$. To show this, it is sufficient to show that
\begin{align}
	\lambda\subseteq z\,,
\end{align}
and
\begin{align}
	\tilde\lambda=(z\setminus\lambda)\cdot Q\,.
\end{align}
The first statement is trivial: since $Z^\perp$ and $\hat{C}$ are both inside $\lambda^\perp$ (they are both constructed by multiplying something by $Q$), then both $Z$ and $\hat{C}^\perp$ must contain $\lambda$, hence $\lambda\subseteq Z\cap\hat{C}^\perp$. The second statement is less trivial, and it follows from 
\begin{align}\label{eq:POS_C-Ccheck-conj}
	C^\perp = \big(\hat{C}^\perp\setminus\lambda\big)\cdot Q\,.
\end{align}
Then, from \eqref{eq:POS_lambda-def-mom-amp}, we have
\begin{subequations}
\begin{align}
	\tilde\lambda &= \big(\tilde\Lambda^\perp\cdot (C^\perp)^T\big)\cdot C^\perp\\
	&= \big(\tilde\Lambda^\perp\cdot ((\hat{C}^\perp\setminus\lambda)\cdot Q)^T\big)\cdot \big(\hat{C}^\perp\setminus\lambda\big)\cdot Q\\
	&= \big( Z^\perp\cdot (\hat{C}^\perp\setminus\lambda)^T \big)^\perp \cdot (\hat{C}^\perp\setminus\lambda)\cdot Q\\
	&= \big( Z^\perp\cap (\hat{C}^\perp\setminus\lambda) \big)\cdot Q\\
	&= \big( (\hat{C}\setminus\lambda)\setminus Z^\perp \big)\cdot Q\\
	&= (z\setminus\lambda)\cdot Q\,.
\end{align}
\end{subequations}
We use the result $(A^\perp\cdot B^T)^\perp\cdot B= A\cap B$ from section \ref{sec:GRASS_binary}, and $z\setminus\lambda = (\hat{C}^\perp\setminus Z^\perp)\setminus \lambda = (\hat{C}^\perp\setminus\lambda)\setminus Z^\perp$.

This correspondence holds some non-trivial information about the momentum amplituhedron. We recall that any point in the amplituhedron satisfies $\<ii+1jj+1\>>0$. Furthermore, since $\<ii+1jj+1\>=\<ii+1\>\<jj+1\>X_{ij}$, and $\<ii+1\>>0$ in the momentum amplituhedron, this then implies that $X_{ij}>0$. Thus, our assumption that $\big( \tilde\Lambda^\perp\cdot Q \big)^\perp$ is a positive matrix is sufficient to ensure that all planar Mandelstam variables are positive, which was a lingering requirement for the momentum amplituhedron. From this point on, we will therefore assume that we have chosen matrices $\Lambda$ and $\tilde\Lambda$ such that $\big( \tilde\Lambda^\perp\cdot Q \big)^\perp$ is positive. 

In addition, for the amplituhedron $\Acal_{n,k-2}$ we know that the sequence 
\begin{align}
	\{\<i-1\,i\,i+1\,i+1\>,\<i-1\,i\,i+1\,i+2\>,\ldots,\<i-1\,i\,i+1\,i-2\>\,,
\end{align}
has $k-2$ sign-flips. We can write these invariants in momentum space as
\begin{align}
	\<i-1\,i\,i+1\,j\> = \<i-1\,i\>\<i+1\,j\>(x_i-\ell^\star_{i+1\,j})^2=\<i-1\,i\>\<i\,i+1\>\<j|x_{ij}|i]\,,
\end{align}
Since $\<i\,i+1\>>0$, we find that points in the momentum amplituhedron satisfy
\begin{align}
	\{\<i+1|x_{i\,i+1}|i], \<i+2|x_{i\,i+2}|i],\ldots,\<i-2|x_{i\,i-2}|i]\}\quad \text{has $k-2$ sign flips.}
\end{align}
We note that the above derivations rely on equation \eqref{eq:POS_C-Ccheck-conj}, which is a conjectural way we can retrieve the $C$ matrix from the $\hat{C}$ matrix and $\lambda$, and can thus be seen as an inverse to the relation $\hat{C}=(C\setminus\lambda)\cdot Q$. This conjecture has been verified for many numeric cases. It is worth noting that a similar formula also allows us to invert the relation $\tilde\lambda=(z\setminus\lambda)\cdot Q$:
\begin{align}
	z^\perp = (\tilde\lambda^\perp\setminus\lambda)\cdot Q\,.
\end{align}
We remind the reader that the ``$\setminus\lambda$'' appearing in these relations is only to make the dimensions of the matrices work out: since $\lambda\cdot Q=\nul_{2\times n}$ and $\lambda\subset C$, $C\cdot Q$ will be a $k\times n$ matrix with rank $k-2$. The row-span of $C\cdot Q$ defines $\hat{C}$. To obtain an explicit $(k-2)\times n$ matrix realisation of $\hat{C}$ we can first `remove' $\lambda$ from $C$ using equation \eqref{eq:binary-matrices} to obtain a $(k-2)\times n$ matrix $C\setminus\lambda$ that spans the $(k-2)$ subplane of $C$ that is orthogonal to $\lambda$. Now that we have removed the kernel of $Q$ from the domain, when we then apply $Q$ from the right we get a full-rank $(k-2)\times n$ matrix: $\hat{C}=(C\setminus\lambda)\cdot Q$. The same story holds when replacing $C\to z$ and $\hat{C}\to\tilde\lambda$. Thus, if we are comfortable interpreting matrices that are not full-rank as the element of the Grassmannian defined by their row-span, then we can simply write
\begin{alignat}{4}
	&\hat{C} &&= C\cdot Q\,,\qquad && C^\perp &&= \hat{C}^\perp\cdot Q\,,\\
	&\tilde{\lambda} &&= z\cdot Q\,, && z^\perp &&= \tilde\lambda^\perp\cdot Q\,.
\end{alignat}

\subsection{The Loop Momentum Amplituhedron}\label{sec:POS_loop-mom-amp}

The loop extension of the momentum amplituhedron was proposed in \cite{Ferro:2022abq}. The idea is that, now that we have a definition of the tree-level amplitudes in spinor-helicity space, we can simply reuse the definition of loop momenta from the amplituhedron by recasting the definition into the appropriate variables. We will discuss a different construction for loop integrands for the momentum amplituhedron in section \ref{sec:DUAL_nf}, which will be set in the space of dual momenta.

The loop amplituhedron contains $L$ lines $(AB)_l=D^{(l)}\cdot z^T$, where $z$ are momentum twistors and $D^{(l)}$ are elements of the Grassmannian $G(2,n)$ which satisfy the loop positivity conditions explained around equation \eqref{eq:POS_loop-positivity}. We further have the sign-flip definition:
\begin{align}
	&\<(AB)_l ii+1\>>0\,,\quad \<(AB)_i(AB)_j\>>0\\
	&\{\<(AB)_l 12\>,\ldots,\<(AB)_l1n\>\}\quad\text{ has $K+2=k$ sign-flips}\,.
\end{align}
Although the tree-level forms and geometries are fundamentally different between the amplituhedron and the momentum amplituhedron, the $4L$-form part of the canonical form can be translated between the two directly. 

Writing $z_i = \begin{pmatrix} \lambda_i & \mu_i \end{pmatrix}^T$, we have
\begin{align}
	(AB)=D\cdot z^T = \begin{pmatrix}
		\lambda_A & \lambda_B\\ \mu_A &\mu_B
	\end{pmatrix}\,,
\end{align}
where $\lambda_\gamma = D_\gamma\cdot \lambda^T = \sum_{i=1}^n D_{\gamma i}\lambda_i^\alpha\,, \mu_\gamma =D_\gamma\cdot \mu^T=\sum_{i=1}^nD_{\gamma i}\mu_i^\alphadot,\, \gamma=A,B$. We know from section \ref{sec:KIN_mom-twistor} that we can translate the line $(AB)$ in momentum twistor space into a point in dual space as
\begin{align}
	(AB)\mapsto \frac{\lambda_B\mu_A-\lambda_A\mu_B}{\<AB\>}\,.
\end{align}
Expanding the numerator, we find
\begin{align}
	\lambda_B\mu_A-\lambda_A\mu_B= \sum_{i,j=1}^n 
	(D_{Ai}D_{Bj}- D_{Bi}D_{Aj})\lambda_j\mu_i = \sum_{i,j=1}^n (ij)_D \lambda_j\mu_i=\sum_{i<j} (ij)_D (\lambda_j\mu_i-\lambda_i\mu_j)\,,
\end{align}
where $(ij)_D$ is the \Pluck variable $p_{ij}(D)$. Comparing with \eqref{eq:KIN_ls-def}, we see that we can equivalently write 
\begin{align}\label{eq:POS_ell-mom-amp-def}
	(AB)\mapsto y = \frac{\sum_{i<j} \<ij\> (ij)_D \ls_{ij}}{\<AB\>}=\frac{\sum_{i<j} \<ij\> (ij)_D \ls_{ij}}{\sum_{i<j}\<ij\>(ij)_D}\,.
\end{align}
To go from dual space to spinor-helicity space we need to break translation invariance by picking an origin in dual space. Our convention is to set $x_1=0$, in which case we can expand $\smash{\ls_{ij}}$ as in equation \eqref{eq:KIN_ls-def}. All the positivity conditions roll over from the standard loop amplituhedron, however we now have to be careful to use the $\hat{C}$ matrix rather than the $C$ matrix.

The full definition of the loop momentum amplituhedron is then 
the image of the map
\begin{align}
	\Phi_{\Lambda,\tilde\Lambda}^{(L)}\colon G_+(k,n) \dot\times G(2,n)^L &\to G(2,n)\times G(2,n)\times (M_{2,2})^L\\
	(C,D^{(1)},\ldots,D^{(L)})&\mapsto (\lambda,\tilde\lambda, \ell^{(1)},\ldots,\ell^{(L)})\,,
\end{align}
where $\lambda,\tilde\lambda$ are defined in \eqref{eq:POS_lambda-def-mom-amp}, $\ell^{(l)}$ is defined in \eqref{eq:POS_ell-mom-amp-def}, and $M_{2,2}$ denotes the space of real $2\times 2$ matrices. The product `$\dot\times$' in the domain is used to indicate that
\begin{align}\label{eq:POS_loop-positivity-mom amp}
	\left(
	\begin{array}{c}
		\hat{C}
	\end{array}
	\right)\quad \left(
	\begin{array}{c}
		D^{(l_1)}\\ \hdashline[2pt/2pt]\rule{0pt}{2.6ex}
		\hat{C}
	\end{array}
	\right)\quad \left(
	\begin{array}{c}
		D^{(l_1)}\\ \hdashline[2pt/2pt]\rule{0pt}{2.6ex}
		D^{(l_2)}\\ \hdashline[2pt/2pt]\rule{0pt}{2.6ex}
		\hat{C}
	\end{array}
	\right)\quad\cdots\quad \left(
	\begin{array}{c}
		D^{(l_1)}\\ \hdashline[2pt/2pt]\rule{0pt}{2.6ex}
		\vdots\\ \hdashline[2pt/2pt]\rule{0pt}{2.6ex}
		D^{(l_L)}\\ \hdashline[2pt/2pt]\rule{0pt}{2.6ex}
		\hat{C}
	\end{array}
	\right)\,,
\end{align}
are all positive matrices.

The sign-flips can again be translated directly from the amplituhedron. Using our `momentum twistor space -- dual space' dictionary, we find
\begin{align}
	\<ABii+1\> &= \<AB\>\<ii+1\> (y-x_i)^2\,,\\
	\<ABCD\> &= \<AB\>\<CD\>(y_1-y_2)^2\,,\\
	\<AB1i\> &= \<AB\>\<1i\>(y-\ell^\star_{1i})^2\,.
\end{align} 
Since $\<ii+1\>>0$ in the momentum amplituhedron, and we can fix $\<AB\>>0$, we have the following sign-flips:
\begin{align}
	&(y_l-x_i)^2>0\,, (y_{l_1}-y_{l_2})^2>0\,,\\
	&\{\<12\>(y-\ell^\star_{12})^2,\ldots,\<1n\>(y-\ell^\star_{1n})^2\}\quad\text{ has $k$ sign-flips}\,.
\end{align}

\subsection{Boundary Structure}\label{sec:POS_mom-amp-boundaries}

There is a one-to-one correspondence between boundaries of the momentum amplituhedron and singularities of the tree-level scattering amplitude. This statement follows directly from the assumption that the momentum amplituhedron is a positive geometry whose canonical form is the amplitude. Thus, by studying the boundary stratification of the momentum amplituhedron, we learn about the set of all singularities of the scattering amplitude and how they are related.

We have seen that the momentum amplituhedron is the image of the positive Grassmannian $G_+(k,n)$, and its canonical form gives the full scattering amplitude, which is the on-shell function associated to the top-cell. This correspondence goes one step further, and for any positroid cell $S_\sigma$, the canonical form of the image $\Phi_{\Lambda,\tilde\Lambda}(S_\sigma)$ gives the on-shell function associated to this positroid cell:
\begin{align}\label{eq:POS_mom-amp-on-shell-function}
	\Omega(\Phi_{\Lambda,\tilde\Lambda}(S_\sigma))=f_\sigma\,.
\end{align}
We recall from section \ref{sec:AMP_nf} that certain on-shell function (like factorisation channels) are only nonvanishing on the support of certain constraints on the external kinematics. Since equation \eqref{eq:POS_mom-amp-on-shell-function} is non-zero for any positroid cell, this means that any $(\lambda,\tilde\lambda)\in\Phi_{\Lambda,\tilde\Lambda}(S_\sigma)$ must satisfy these kinematic constraints. 

That this is the case is easy to see in certain situations. The momentum amplituhedron map generates $\lambda,\tilde\lambda$ such that $\lambda\subseteq C$ and $\tilde\lambda\subseteq C^\perp$. We note that a black lollipop on $i$ means that $C$ has only zeroes in its $i$\textsuperscript{th} column, and a white vertex connecting $i$ and $i+1$ means that the columns $\bm{c}_i$ and $\bm{c}_{i+1}$ are proportional to each other. Writing $C$ as in equation \eqref{eq:AMP_C-with-lambda}, it is then clear that $\lambda$ will also satisfy these same properties. Other kinematic constraints, like the vanishing of a planar Mandelstam variables in a factorisation channel, are more difficult to see from first principles.

Thus, on-shell diagrams are perfectly suited to label singularities of scattering amplitudes and boundaries of the momentum amplituhedron: the additional delta-function constraints of the on-shell function tell us which residues need to be taken to arrive at this boundary, and the associated positroid cell tells us which element of the positive Grassmannian maps to the corresponding boundary of the momentum amplituhedron. 

A systematic study of which positroid cells map to boundaries of the momentum amplituhedron (and hence represent singularities of the scattering amplitude) has been performed in \cite{Ferro:2020lgp}, based on a method introduced in \cite{Lukowski:2019kqi}. In the remainder of this section, we will briefly recall their methodology and main results. 

First, we introduce some basic notation. For each positroid cell $S_\sigma$, we label its dimension in the positive Grassmannian as $\dim_C\sigma$, and the dimension of its image in the momentum amplituhedron as $\dim_M \sigma$. It is generally the case that
\begin{align}
	\dim_C\sigma \geq \dim_M\sigma\,.
\end{align}
There is a simple way to find the dimension of the image of a positroid cell. Starting from a positive parametrisation $C_\sigma(\bm{\alpha})$, we can find $\dim_M\sigma$ by finding the rank of the Jacobian matrix of the map $\Phi_{\Lambda,\tilde\Lambda}(C_\sigma(\bm{\alpha}))$.
We further let $\partial_C \sigma$ and $\partial^{-1}_C\sigma$ denote the positroid stratification and inverse stratification of $\sigma$. That is, $\partial_C \sigma$ is the set of all positroid cells which are boundaries of $S_\sigma$, and $\partial^{-1}_C\sigma$ is the set of all positroid cells which have $S_\sigma$ in their positroid stratification.

To find all boundaries of the momentum amplituhedron, we proceed as follows. We assume that we have found all boundaries with a momentum amplituhedron dimension larger than $d$. We then find the set of positroid cells for which $\dim_M\sigma=d$, and separate them into two categories:
\begin{enumerate}
	\item All inverse boundaries of $\sigma$ have a higher momentum amplituhedron dimension than $\sigma$:$\displaystyle \forall \sigma' \in \partial^{-1}_C\sigma\colon \dim_M\sigma' > d $.
	\item There exist an inverse boundary of $\sigma$ which has the same momentum amplituhedron dimension as $\sigma$: $\displaystyle \exists \sigma\in\partial^{-1}_C\sigma \colon \dim_M\sigma' =d$ (assuming $\sigma \neq\sigma'$).
\end{enumerate}
We are only interested in the highest dimensional positroid cells which map to any given boundary, and we thus discard all cells in the second category. The first category contains a unique representative for all $d$-dimensional boundaries of the momentum amplituhedron, however it might also include additional spurious boundaries. To remove the spurious boundaries, we observe that any `real' $d$-dimensional boundaries belong to two or more $(d+1)$-dimensional boundaries, whereas spurious boundaries always belong only to a single $(d+1)$-dimensional boundary. After removing all spurious boundaries, we are left with exactly the set of $d$-dimensional boundaries of the momentum amplituhedron. Repeating this process recursively, we find the full boundary stratification of the momentum amplituhedron. This algorithm has been implemented in the \texttt{Mathematica} package \texttt{amplituhedronBoundaries} \cite{Lukowski:2020bya}.

All results found using this algorithm are consistent with the following, which we will conjecture to be true for all $n$ and $k$ and have been verified explicitly for all $k$, $n\leq 9$. Firstly, the set of positroid cells which represent boundaries of the momentum amplituhedron are precisely the positroid cells whose on-shell diagram can be represented as a \emph{Grassmannian forest}, which are Grassmannian graphs (see section \ref{sec:GRAS_positroid-combinatorics}) which do not have any internal loops. Furthermore, for any Grassmannian forest $F$ we can directly read of what the dimension of its image, $\dim_M (F)$, is. To do this, we first take any vertex $v$ in the forest and define its `dimension' as
\begin{align}
	\dim_M v \coloneqq \begin{cases}
		2 \deg(v)-4\,,\quad &\text{ if } 2\leq h(v)\leq \deg(v)-2\\
		\deg(v)-1\,, & \text{ if } h(v)=1,\deg(v)-1
	\end{cases}\,,
\end{align}
where $h(v)$ denotes the helicity of this vertex. Next, we define a \emph{Grassmannian tree} as a connected Grassmannian forest, and define its dimension as
\begin{align}
	\dim_M T \coloneqq \begin{cases}
		1+\sum_{v\in\Vcal_{\text{int}}(T)} (\dim_M(v)-1)\,,\quad &\text{ for }n>3\\
		n-1\,, & \text{ for } n\leq 2
	\end{cases}\,,
\end{align} 
where $\Vcal_{\text{int}}(T)$ denotes the set of internal vertices of the Grassmannian tree $T$.
Then, the momentum amplituhedron dimension of a Grassmannian forest $F$ can be found by summing over all trees which make up the forest:
\begin{align}
	\dim_M(F) \coloneqq \sum_{T\in\text{Tree}(F)}\dim_M(T)\,.
\end{align}
Let us fix $n,k$, and we let $f_d$ be the number of Grassmannian forests with $\dim_M(F)=d$, which we summarise in the \emph{$f$-vector} $\bm{f}=(f_0,f_1,\ldots,f_{2n-4})$. The Euler characteristic of the momentum amplituhedron $\Mcal_{n,k}$ is then given by $\chi=\sum_{i=0}^{2n-4} (-1)^i f_i$ (assuming that the boundaries are indeed labelled by Grassmannian graphs). At has been proven in \cite{Moerman:2021cjg} that $\chi=1$ for any $n,k$, which gives strong evidence that the momentum amplituhedron is topologically a closed ball.

\begin{figure}
	\centering
	\includegraphics[height=30mm]{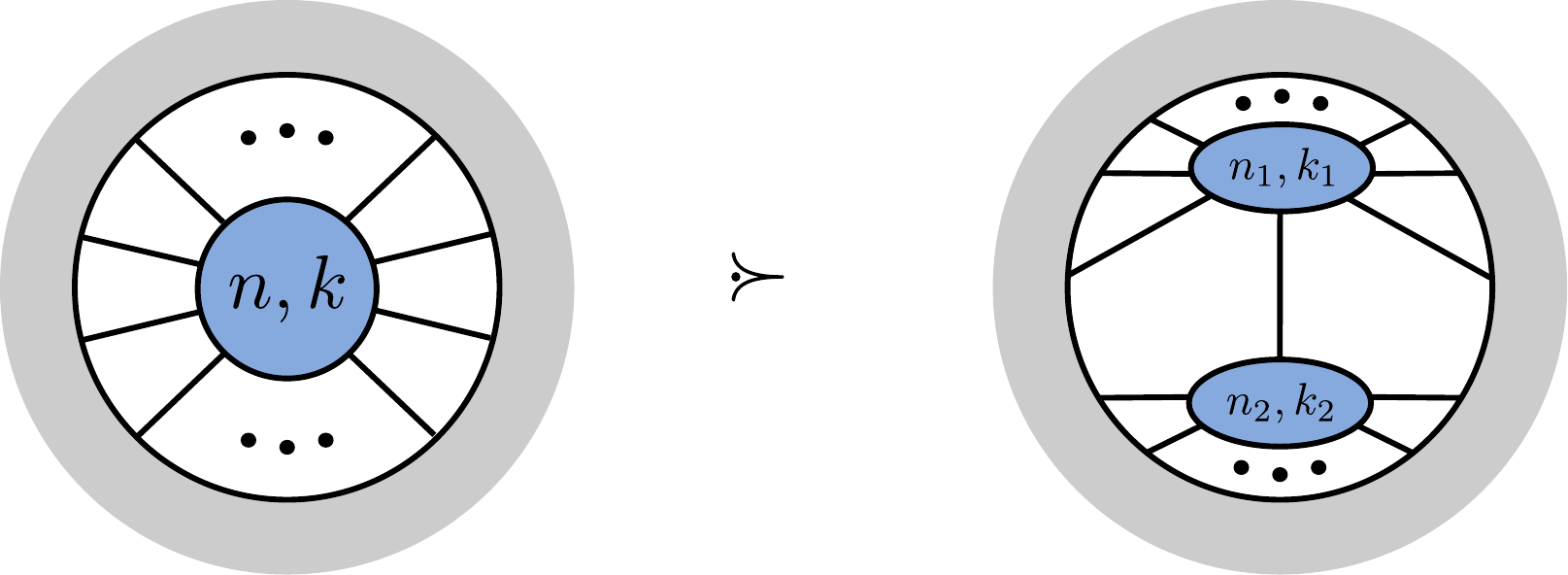}
	\caption{Factorisation channels are codimension-1 boundaries of generic vertices. The numbers in the vertices represent their degree and helicity, respectively. For these factorisation channels we require $n_1+n_2=n+2$, $k_1+k_2=k+1$, and we assume the vertices are neither white nor black: $2\leq k \leq n-2$, $2\leq k_1\leq n_1-2$, $2\leq k_2\leq n_2-2$.}
	\label{fig:Grass-forest-factorisation}
\end{figure}

\begin{figure}
	\centering
	\includegraphics[height=30mm]{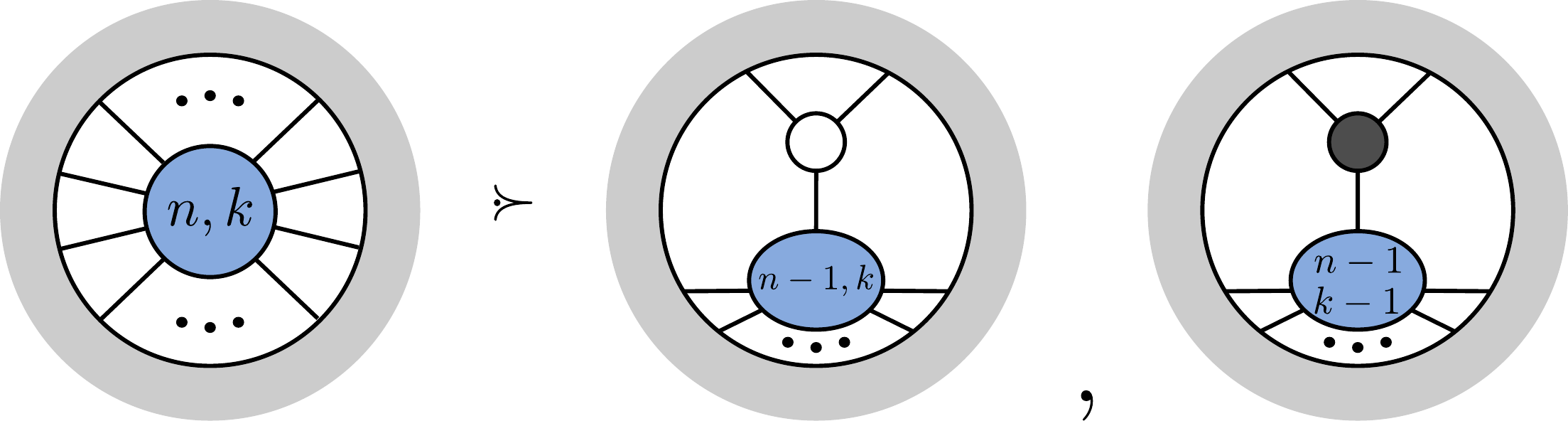}
	\caption{Collinear limits are codimension-1 boundaries of generic vertices.}
	\label{fig:Grass-forest-collinear}
\end{figure}

\begin{figure}
	\centering
	\includegraphics[width=0.9\textwidth]{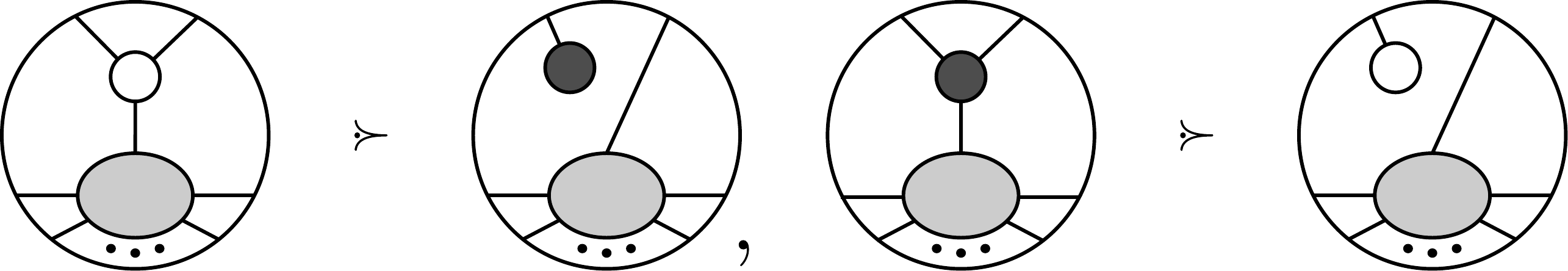}
	\caption{Soft limits are codimension-1 boundaries of black or white vertices. Note that, unlike the other codimension-1 boundaries, these diagrams are not embedded in a grey disk. This is to signify that these soft limits only occur for external particles. The analogue of soft limits for internal edges are depicted in figure \ref{fig:Grass-forest-p-zero}. The grey blobs indicate arbitrary subgraphs.}
	\label{fig:Grass-forest-soft}
\end{figure}

\begin{figure}
	\centering
	\includegraphics[width=0.9\textwidth]{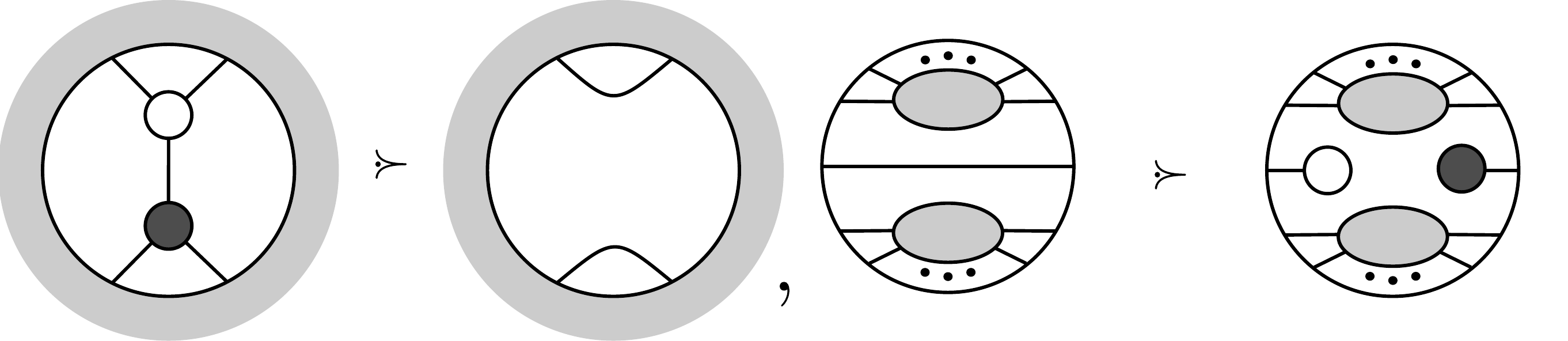}
	\caption{On the left it is depicted that removing an edge between a black and a white vertex is a codimension-1 boundaries of a Grassmannian forest. On the right we see that an edge connecting only to the boundary can be dissolved into a white and a black lollipop.}
	\label{fig:Grass-forest-p-zero}
\end{figure}

We can further describe the exact \emph{covering relations} of the momentum amplituhedron. That is, for any given Grassmannian forest we know exactly which Grassmannian forests represent a codimension-1 boundary. They accurately capture the singularity structure expected from physical considerations. These covering relations can be described as follows. Any generic vertex $v$ (\emph{i.e.} not a white or a black vertex) with helicity $h$ can be replaced by a factorisation channel: a vertex $v_1$ with helicity $h_1=2,\ldots,\deg(v_1)-2$ connected through an internal edge to vertex $v_2$ with helicity $h_2=2,\ldots,\deg(v_2)-2$ such that $\deg(v_1)+\deg(v_2)=\deg(v)+2$ and $h_1+h_2=h+1$. This is depicted in figure \ref{fig:Grass-forest-factorisation}.
Generic vertices also cover \emph{collinear limits}, which are similar in nature to the factorisation channels defined above, except we have one (or both) of the daughter vertices with degree three and helicity $1$ or $2$. This is represented in figure \ref{fig:Grass-forest-collinear}.
Next, black or white vertices can no longer factorise any further, and instead admit \emph{soft-limits}, which allow you to take away a white or black lollipop from a black or white vertex, respectively. This corresponds to the kinematic limit where $\lambda_i^\alpha$ or $\tilde\lambda_i^\alphadot$ go to zero, respectively, and is represented in figure \ref{fig:Grass-forest-soft}. 
Further, if a Grassmannian forest contains a white and a black vertex connected through an internal edge, then this covers the Grassmannian forest with this internal edge removed. This corresponds to the kinematic limit where the on-shell momentum corresponding to this edge goes to zero, and is shown in the left of figure \ref{fig:Grass-forest-p-zero}. This can be though of as an analogue of the soft limit for internal edges. Lastly, a Grassmannian graph with an edge which connects only to the boundary covers the same diagram with the edge split into a white and a black lollipop, which is depicted on the right of figure \ref{fig:Grass-forest-p-zero}.

These covering relations transitively close to a partial order on Grassmannian forests. The set of all Grassmannian forests of type $(n,k)$ equipped with this partial order gives the full boundary stratification of $\Mcal_{n,k}$.

\section{The ABJM Momentum Amplituhedron}\label{sec:POS_ABJM-mom-amp}

The idea of the momentum amplituhedron has been generalised to supersymmetry reduced ABJM theory in \cite{He:2021llb, Huang:2021jlh}. The definition is directly analogous to the \nf momentum amplituhedron, except that we restrict our domain to the positive orthogonal Grassmannian defined in section \ref{sec:GRASS_orth}. Alternatively, the ABJM momentum amplituhedron can be interpreted as the slice of the \nf momentum amplituhedron where $\tilde\lambda=\lambda\cdot\eta$, where $\eta=\diag(1,-1,1,\ldots,-1)$. 

Explicitly, we fix $\Lambda\in M_+(2k,k+2)$, and define the ABJM momentum amplituhedron\footnote{The ABJM momentum amplituhedron is referred to as the `orthogonal momentum amplituhedron' in \cite{Huang:2021jlh} and \cite{Lukowski:2021fkf}, due to its connection to the orthogonal Grassmannian.} $\Ocal_k$ as the image of the map
\begin{align}
	\Phi_\Lambda\colon OG_+(k)&\to G(k,k+2)\\
	C&\mapsto Y= C\cdot \Lambda^T\,.
\end{align}
Like the momentum amplituhedron, the image of $\Phi_\Lambda$ is not top-dimensional in $G(k,k+2)$, but instead lives on some `momentum conserving hyperplane'. We define
\begin{align}\label{eq:POS_ABJM-lambda-def}
	\lambda = Y^\perp\cdot \Lambda\,,
\end{align}
which satisfies
\begin{align}
	\lambda\cdot\eta\cdot\lambda^T=\nul_{2\times 2}\,,
\end{align}
such that $\lambda \in OG(2,2k)$. This imposes three additional constraints on the image of $\Phi_\Lambda$, and hence
\begin{align}
	\dim \Ocal_k= \dim G(k,k+2) - 3 = 2k-3\,.
\end{align}
We define the winding space $\Wcal_k$ in three-dimensional spinor-helicity space as the region where
\begin{align}
	&\<ii+1\>>0\,,\quad s_{ii+1\cdots j} >0\,,\\
	&\{\<12\>,\<13\>,\ldots,\<1n\>\}\quad\text{ has $k$ sign-flips,}
\end{align}
where we define the planar Mandelstam variables as
\begin{align}
	s_{ii+1\cdots j}=\sum_{i\leq a<b\leq j} (-1)^{a+b+1}\<ab\>^2\,.
\end{align}
As was the case for the \nf momentum amplituhedron, the positivity of planar Mandelstam variables is not manifest, and this should be interpreted as an additional condition on the $\Lambda$ matrix. The nature of this additional constraint is currently not well understood. We further decompose
\begin{align}
	\Lambda_i^A = \begin{pmatrix}
		\lambda^{*\alpha}_i\\\Delta_i^a
	\end{pmatrix}\,,
\end{align}
and define an $(n-3)$-dimensional subspace of spinor-helicity space
\begin{align}
	\Vcal_k[\Lambda]\coloneqq \{\lambda\colon \lambda\cdot\eta\cdot\lambda^T=\nul\,, \lambda_i^\alpha=\lambda_i^{*\alpha}+y_a^\alpha\Delta_i^a \}\,.
\end{align}
We can then define the ABJM momentum amplituhedron in momentum space as
\begin{align}
	\Ocal_k = \Vcal_k[\Lambda]\cap\Wcal_k\,.
\end{align}
To extract the $n=2k$ scattering amplitude in supersymmetry reduced ABJM theory we can again proceed in two ways. First, if we interpret $\Lambda$ as a matrix with bosonised kinematic variables:
\begin{align}
	\Lambda = \begin{pmatrix}
		\lambda\\\phi_I\cdot\eta^I
	\end{pmatrix}\,,
\end{align}
and we extract the canonical function $\omega_k$ as
\begin{align}
	\Omega(\Ocal_k)\wedge\dd^3P \delta^3(P) = \prod_{\alpha=1}^k\<Y_1\cdots Y_{k-2} \dd^2 Y_\alpha\> \delta^3(P)\omega_k\,,
\end{align}
then the $2k$-point amplitude is obtained by localising on $Y_0$ (as defined in \eqref{eq:POS_Ystar-mom-amp} with $n=2k$) and integrating out the auxiliary Grassmann variables:
\begin{align}
	A_{2k}=\delta^3(p)\int \dd \phi_a^1\cdots\dd\phi_a^k \omega_k(Y_0;\Lambda)\,.
\end{align}
Alternatively, we can take the canonical form in spinor-helicity space and make the substitution
\begin{align}
	A_{2k}=\left.\Omega(\Ocal_k) \right|_{\dd\lambda_i^\alpha\to \eta_i^\alpha}\,.
\end{align}
We note that $\Ocal_k$ can equivalently be defined as the momentum amplituhedron $\Mcal_{2k,k}$ intersected with the `orthogonal slice' where $\lambda=\tilde\lambda\cdot\eta$. To this extent the ABJM momentum amplituhedron is nothing but a dimensional reduction of the \nf momentum amplituhedron. This statement is true on the level of the geometry, but it does not hold on the level of the canonical form or scattering amplitude: a dimensional reduction of the \nf amplitude \emph{does not} yield an ABJM amplitude. The fact that their geometries are related in this way is a highly non-trivial statement. We note that, due to the parity symmetry, our choice to keep $\tilde\lambda$ and impose constraints on $\lambda$ is completely arbitrary. We could have equivalently kept $\lambda$ instead. This redefinition essentially swaps $\lambda\to\lambda\cdot\eta$ in the formulae above, which changes the sign-flip patterns and the (twisted) positivity constraints on $\Lambda$ slightly. To see that the two definitions are trivially identical, we only need to observe that if $\Lambda \in M_+(2k,k+2)$, then $\Lambda\cdot\eta\in M_+^\perp(2k,k+2)$, and vice versa.

\subsection{The Loop ABJM Momentum Amplituhedron}\label{sec:POS_loop-ABJM-mom-amp}

The ABJM momentum amplituhedron also has an extension to include loop integrands by similarly reducing the loop momenta from \eqref{eq:POS_ell-mom-amp-def} to three dimensions. The loop momentum is defined essentially identically as
\begin{align}\label{eq:POS_loop-mom-def-ABJM}
	\ell = \frac{\sum_{i<j} \<ij\>(ij) \ell^\star_{ij}}{\sum_{i<j}\<ij\>(ij)}\,,
\end{align}
where the only difference is that we now require this $2\times 2$ matrix $\ell$ to be \emph{symmetric}, and that $\ell^\star_{ij}$ is defined as in \eqref{eq:KIN_ls-def}, but with $\tilde\lambda=\lambda\cdot\eta$ \cite{Lukowski:2023nnf}. This imposes additional constraint on the $D$-matrices. Beyond $n=4$ this constraint is currently not fully understood. Additionally, the $2\times 2$ matrices $\ell^\star_{ij}$ are typically not symmetric, and can therefore not be given the interpretation of three-dimensional momentum vectors or triple intersections of lightcones. For these reasons the definition for $n>4$ is not the preferred way to define the loop ABJM momentum amplituhedron. In chapter \ref{sec:DUAL_ABJM}, we will give an alternative definition of the loop ABJM momentum amplituhedron in dual space which sidesteps all these issues. 

For now, let us consider the only well-understood case of the loop ABJM momentum amplituhedron, which is the case when $n=2k=4$. We note that we have
\begin{align}
	\<12\>=\<34\>\,,\quad \<13\>=\<24\>\,,\quad \<14\>=\<23\>\,,
\end{align}
and
\begin{alignat}{2}
	&\ell^\star_{12}= p_1=\lambda_1\lambda_1\,,\quad &&\ell^\star_{23}=p_1-p_2=\lambda_1\lambda_1-\lambda_2\lambda_2\,,\\
	&\ell^\star_{34}=p_1-p_2+p_3=p_4=\lambda_4\lambda_4\,,\quad &&\ell^\star_{14}=p_1-p_2+p_3-p_4=0\,,\\
	&\ell^\star_{13}=-\frac{\<12\>}{\<13\>}\lambda_1\lambda_4\,,\quad &&\ell^\star_{24}=-\frac{\<12\>}{\<24\>}\lambda_4\lambda_1\,.
\end{alignat}
Obviously, the matrices $\ell^\star_{ii+1}$ are symmetric, and only $\ell^\star_{13}$ and $\ell^\star_{24}$ are not symmetric. When expanding \eqref{eq:POS_loop-mom-def-ABJM}, we see that $\ell$ is symmetric if
\begin{align}
	(13)\<13\>\ell^\star_{13}+(24)\<24\>\ell^\star_{24} = -\<12\>\big((13)\lambda_1\lambda_4+(24)\lambda_4\lambda_1\big)\,,
\end{align}
is symmetric. This happens when $(13)=(24)$, which puts and extra constraint on the $D$ matrix. Since $D\in G_+(2,4)$, the \Pluck relations imply
\begin{align}
	(13)=\sqrt{(12)(34)+(14)(23)}\,.
\end{align}
We thus only have four independent \Pluck variables, which we can take to be $(12)$, $(23)$, $(34)$, and $(14)$. We cannot set all of these \Pluck variables to zero, since this would decrease the rank of the $D$ matrix, and, as always, the \Pluck variables are interpreted up to an overall scale. This shows that the subspace of $G_+(2,4)$ where $(13)=(24)$ is isomorphic to $\Pbb^3$, where we have the natural set of homogenous coordinates given by $\big((12),(23),(34),(14)\big)$.

To summarise, the $L$-loop four-point ABJM momentum amplituhedron $\Ocal_2^{(L)}$ is defined as the image of the map
\begin{align}
	\Phi^{(L)}\colon OG_+(2)\dot\times (\Pbb^3)^L &\to OG(2,4)\times (Sym_{2\times 2})^L\\
	(C,D^{(1)},\ldots,D^{(L)})&\mapsto (\lambda,\ell_1,\ldots,\ell_L)\,,
\end{align}
where $\lambda$ is defined in \eqref{eq:POS_ABJM-lambda-def}, $\ell_l$ is defined as
\begin{align}
	\ell_l\coloneqq \frac{(12)_l x_{12} + (23)_l x_{13} + (34)_l x_{14} - \<12\> (13)_l (\lambda_1\lambda_4+\lambda_4\lambda_1)}{\sum_{i<j}(ij)_l\<ij\>}\,,\quad x_{1i}=\sum_{j=1}^{i-1} (-1)^j \lambda_j\lambda_j\,,
\end{align}
$(ij)_l=(ij)_{D^{(l)}}$, and the symbol $\dot\times$ is used to indicate that $d_{ij}>0$, where
\begin{align}
	d_{ij}\coloneqq \left|
	\begin{array}{c}
		D^{(i)}\\ \hdashline[2pt/2pt]
		D^{(j)}
	\end{array}
	\right|= &(12)_i(34)_j+(23)_i(14)_j+(34)_i(12)_j+(14)_i(23)_j\\\nonumber&-2\sqrt{\big((12)_i(34)_i+(14)_i(23)_i\big)\big((12)_j(34)_j+(14)_j(23)_j\big)}\,,
\end{align}
and $Sym_{2\times 2}$ is the space of symmetric $2\times 2$ matrices.

A straight-forward calculation show that
\begin{subequations}
\begin{align}
	\ell_i^2 &=\frac{\<23\>\<12\>^2 (23)_i}{\<AB\>_i}\,,\\
	(\ell_i-p_1)^2 &= \frac{\<12\>\<23\>^2(34)_i}{\<AB\>_i}\,,\\
	(\ell_i-p_1+p_2)^2 &= -\frac{\<23\>\<12\>^2(14)_i}{\<AB\>_i}\,,\\
	(\ell_i+p_4)^2 &= \frac{\<12\>\<23\>^2(12)_i}{\<AB\>_i}\,,\\
	(\ell_i-\ell_j)^2 &= -\frac{\<12\>^2\<23\>^2}{\<AB\>_i\<AB\>_j}d_{ij}\,,
\end{align}
\end{subequations}
where $\<AB\>_i=\sum_{a<b} (ab)_i\<ab\>$. Since
\begin{align}
	\Omega(\Ocal_2)=\dd\log\frac{\<12\>}{\<23\>}\,,
\end{align}
and the canonical form of $\Pbb^3$ is given by
\begin{align}
	\Omega(\Pbb^3)=\dd\log\frac{(12)}{(14)}\wedge\dd\log\frac{(23)}{(14)}\wedge\dd\log\frac{(34)}{(14)}\,,
\end{align}
it is easy to see that the canonical form of the one-loop four-point ABJM momentum amplituhedron is given by
\begin{align}
	\Omega(\Ocal_2^{(1)}) = \dd\log\frac{\<12\>}{\<23\>}\wedge\dd\log\frac{\ell^2}{(\ell_i+p_4)^2}\wedge\dd\log\frac{	(\ell_i-p_1)^2}{(\ell_i+p_4)^2}\wedge\dd\log\frac{	(\ell_i-p_1+p_2)^2}{(\ell_i+p_4)^2}\,.
\end{align}
This agrees with the known result for the one-loop integrand given in \cite{Chen:2011vv}.

\subsection{Boundary Structure}\label{sec:POS_ABJM-mom-amp-boundaries}

In this section we will give a complete description of the boundary structure of $\Ocal_k$. The method and arguments closely resemble those encountered in section \ref{sec:POS_mom-amp-boundaries} when discussing the boundary structure of the momentum amplituhedron.

We recall from section \ref{sec:GRASS_orthitroid} that we can label orthitroid cells by permutations which consists of $k$ transpositions, or by equivalence classes of orthogonal Grassmannian (OG) graphs. Of particular interest will be \emph{OG forests} (loopless OG graphs), and \emph{OG trees} (connected OG forests). We know that the top-cell of $OG_+(k)$ has an OG diagram which consists of a single vertex of degree $2k$, whose image under the map $\Phi_\Lambda$ is $2k-3$ dimensional, with the exception of $k=1$, where the degree-2 vertex has a zero-dimensional image. We will associate to each OG forest a dimension $\dim_\Ocal$ which simply adds up the dimension of all the vertices in the diagram. Explicitly, for an OG tree $T$, we define
\begin{align}
	\dim_\Ocal T = \begin{cases}
		0\quad &\text{ if } |\Vcal_{\text{ext}}(T)|=2\\
		\sum_{v\in\Vcal_{\text{int}}(T)}\deg(v)-3 &\text{ if } |\Vcal_{\text{ext}}(T)|>2
\end{cases}\,.
\end{align} 
Furthermore, if $F$ is an OG forest, then we define
\begin{align}\label{eq:POS_ABJM-mom-amp-forest-dim}
	\dim_\Ocal F = \sum_{T\in \text{Trees}(F)}\dim_\Ocal T\,.
\end{align}
One can use the algorithm described in \cite{Kim:2014hva} to generate a positive parametrisation of any orthitroid cell $O$. By determining the rank of the Jacobian matrix of $\tilde\Phi_\Lambda(O)$ we can find the dimension of the image of this cell. Using the \texttt{Mathematica} package \texttt{orthitroids} \cite{Lukowski:2021fkf}, which automates these processes, we were able to verify up to $n\leq 14$ that the dimension of the image of any OG forest $F$ is equal to $\dim_\Ocal F$, and we conjecture this to hold for higher $n$ as well. 

We can study the boundaries of $\Ocal_k$ by seeing how different boundaries (orthitroid cells) of $OG_+(k)$ map under $\Phi_\Lambda$. The method is identical to what was described in section \ref{sec:POS_mom-amp-boundaries} to find the boundary structure of the momentum amplituhedron. To summarise, given some orthitroid cell $O$ such that $\Phi_\Lambda(O)$ is on the boundary of $\Ocal_k$, and assuming that any $O'$ in the inverse OG stratification of $O$ has $\dim\Phi_\Lambda(O')>\dim\Phi_\Lambda(O)$, then $\Phi_\Lambda(O)$ is a full boundary of $\Ocal_k$. By identifying which OG graphs in the OG stratification of $OG_+(k)$ satisfy these conditions, we then know the full boundary stratification of $\Ocal_k$. 

\begin{figure}[t]
	\centering
	\includegraphics[width=\textwidth]{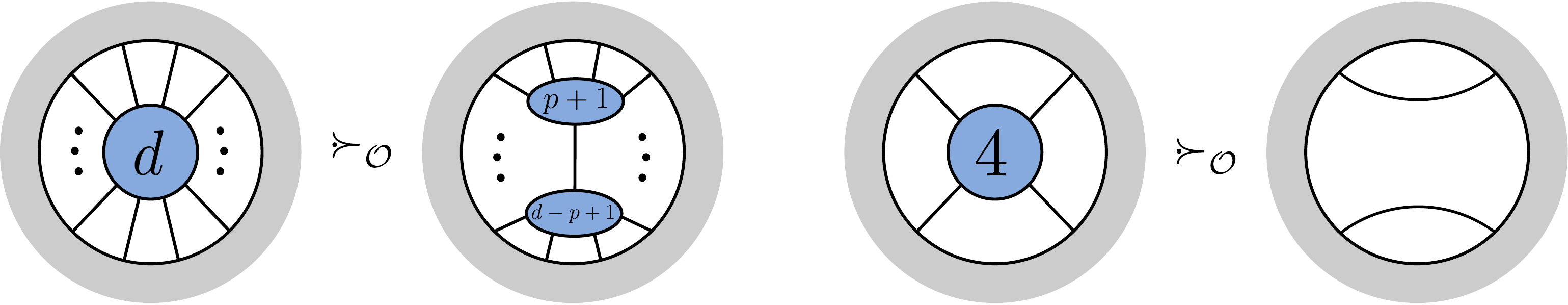}
	\caption{A graphical depiction of the covering relations $\succdot_\Ocal$.}
	\label{fig:ABJM-covering-relations}
\end{figure}

\begin{figure}[t]
	\begin{center}
		\includegraphics[scale=0.3]{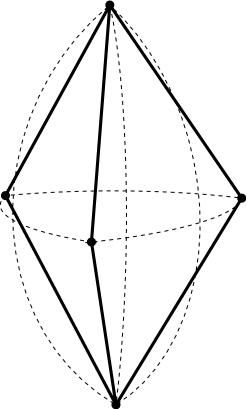}\qquad\qquad\includegraphics[scale=0.3]{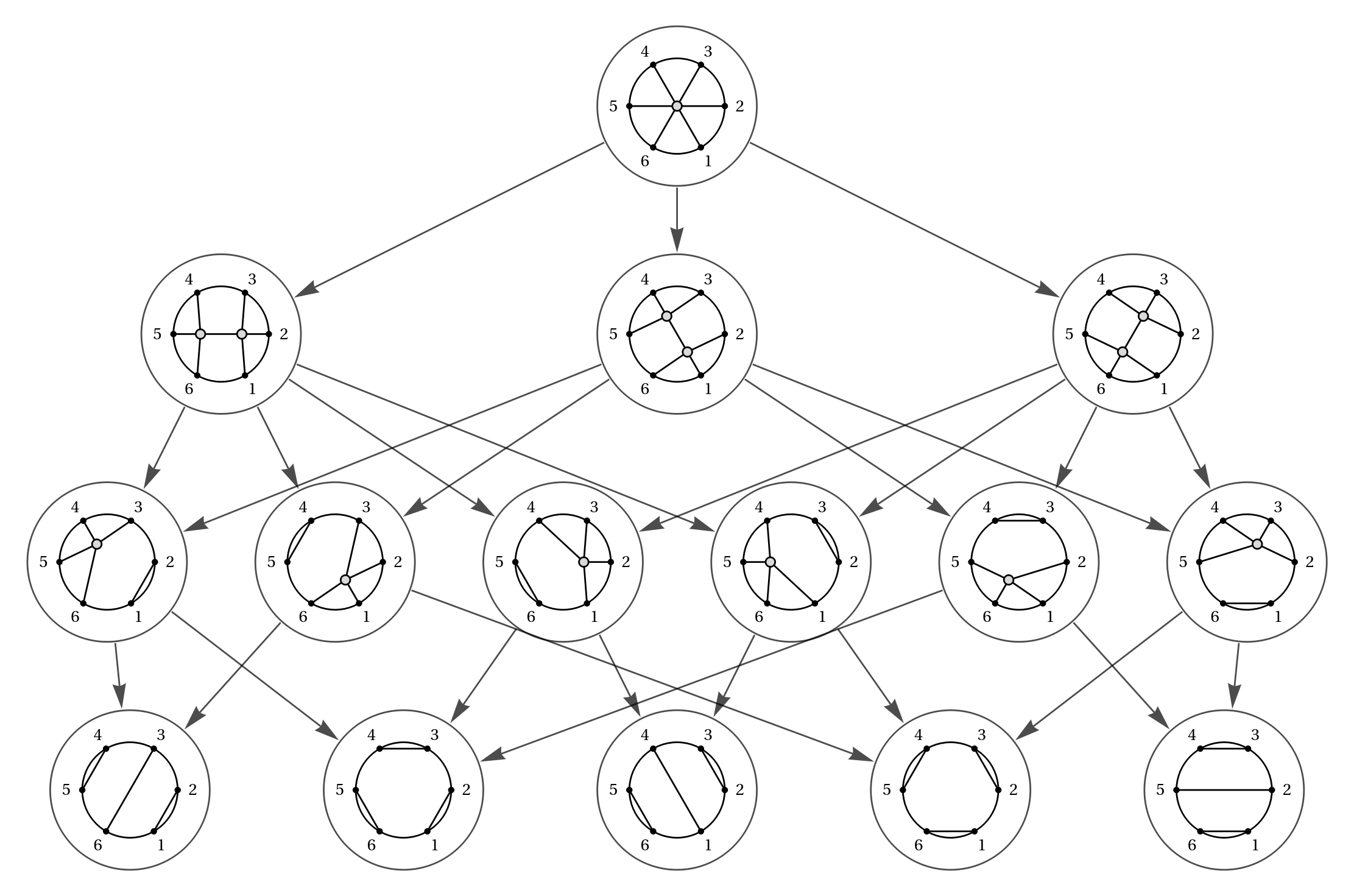}
	\end{center}
	\caption{The ABJM momentum amplituhedron $\Ocal_3$ and a Hasse diagram of its boundary poset.}
	\label{fig:k3}
\end{figure}

Using this algorithm, also implemented in \texttt{orthitroids}, we explicitly studied the boundary stratification of $\Ocal_k$ for $k\leq 7$. Our results are consistent with the following discussion, which we conjecture to hold for all $k$. First, the set of all OG graphs whose images are boundaries of $\Ocal_k$ is \emph{exactly} the set of OG forests. Furthermore, the boundary stratification of $\Ocal_k$ is generated by the following \emph{covering relations} (boundary operator): if an OG forest $F$ contains some vertex of degree $d>4$, then $F \succdot_\Ocal F'$, where $F'$ is any OG forest that is identical to $F$, except that we broke apart the $d$-vertex into two vertices of degree $p+1$ and $d-p+1$ connected by an internal edge for some odd $p=3,\ldots,d-3$. In the case where $d=4$, the covering relations break apart the 4-vertex into two non-intersecting edges. These covering relations are summarised in figure \ref{fig:ABJM-covering-relations}. The covering relations $\succdot_\Ocal$ extend transitively to a partial order $\succeq_\Ocal$ on the set of all OG forests of type $k$, and this poset is identical to the boundary poset of $\Ocal_k$. We depict $\Ocal_3$ and its boundary poset in figure \ref{fig:k3}.

\subsubsection{Enumerating All Boundaries}
To summarise, we know that the boundaries of $\Ocal_k$ are labelled by OG forests of type $k$, and we know that their dimension is given by \eqref{eq:POS_ABJM-mom-amp-forest-dim}. We can now use the tools developed in \cite{Moerman:2021cjg} to find the Euler characteristic $\chi$ of $\Ocal_k$. The idea is as follows: if we consider planar tree graphs with some independent internal structure on the vertices encoded by some function $f$ (which only depends on the degree), then we can find the generating function of the entire class of objects from the generating function $F$ of $f$. In our case, $f$ maps a vertex of degree $d\in 2\Zbb_{\geq 2}$ to $q^{d-3}$, which reflects that this vertex contributes $d-3$ to $\dim_\Ocal T$. The generating function $F$ which enumerates the dimension contribution of a vertex is therefore
\begin{align}
	F(x,q)=\sum_{d=4,6,\ldots} f(d) x^d = q^{-3}\sum_{k=2}^\infty (xq)^{2k}=\frac{x^4 q}{1-(xq)^2}\,,
\end{align}
such that the coefficient of $x^d$ in $F(x,q)$ is equal to $f(d)=q^{d-3}$. We denote the extraction of the coefficient of $x^d$ as
\begin{align}
	[x^d]F(x)=q^{d-3}\,.
\end{align}
Using the results from \cite{Moerman:2021cjg}, the generating function for OG trees of type $k$ is then given by
\begin{align}\label{eq:POS_Gcal-tree}
	\Gcal_\text{tree}^\Ocal (x,q) \coloneqq x \left( x-\frac{1}{x}F(x,q) \right)^{\<-1\>}_x\,,
\end{align}
where the notation $(\cdots)^{\<-1\>}_x$ denotes the compositional inverse with respect to $x$. We can then extract the number of OG trees of type $k$ with $\dim_\Ocal T=r$ as the coefficient $[x^{2k}q^r]\Gcal_\text{tree}^\Ocal (x,q)$. The compositional inverse in \eqref{eq:POS_Gcal-tree} can be calculated explicitly using the \emph{Lagrange inversion formula}, from which we find
\begin{align}\label{eq:POS_Gcal-tree-2}
	\Gcal_{\text{tree}}^\Ocal(x,q) = x^2 \left( 1+\sum_{k=1}^\infty \sum_{l=1}^\infty \frac{1}{k}\binom{k}{l} \binom{2k+l}{2k+1} x^{2k} q^{2k-l} \right)\,.
\end{align} 
Now that we know the generating function for OG trees, we can use \emph{Speicher's analogue of the exponential formula} \cite{Speicher1994MultiplicativeFO} to find the generating function for OG forests. We find
\begin{align}\label{eq:POS_Gcal-forest}
	\Gcal_{\text{forest}}^\Ocal (x,q) \coloneqq \frac{1}{x}\left( \frac{x}{1+\Gcal_{\text{tree}}^\Ocal} \right)^{\<-1\>}_x\,,
\end{align}
which gives
\begin{align}
	[x^n]\Gcal_{\text{forest}}^\Ocal (x) = \frac{1}{n+1} [x^n]\left(1+\Gcal_{\text{tree}}^\Ocal (x,q)\right)^{n+1}\,.
\end{align}
Using this formula, we can find the $f$-vector which encodes its $i$\textsuperscript{th} entry the number of OG forests (and hence, the number of boundaries of $\Ocal_k$) with $\dim_\Ocal F = 2k-i+1$. We record the $f$-vectors for the first few values of $k$ in table \ref{tab:f-vector}.
\begin{table}
	\centering
	\begin{tabular}{|c|l|c|}
		\hline
		$k$ & $f$-vector & $\chi$ \\\thickhline
		2&$(1,2)$&1\\\hline
		3&$(1,3,6,5)$&1
		\\\hline
		4&$(1,8,20,28,28,14)$&1
		\\\hline
		5&$(1,15,65,145,195,180,120,42)$&1
		\\\hline
		6&$(1,24,168,562,1131,1518,1430,990,495,132)$&1
		\\\hline
		7&$(1,35,364,1764,5019,9436,12558,12285,9009,5005,2002,429)$&1\\
		\hline
	\end{tabular}
	\caption{The $f$-vector of $\Ocal_k$ for $k\le 7$. We see that $\chi=1$ in all cases.}
	\label{tab:f-vector}
\end{table}

The Euler characteristic $\chi$ is defined by the alternating sum $\chi = f_0-f_1+f_2-f_3+\ldots $, where $f_d$ is the number of boundaries of dimension $d$. We can extract the Euler characteristic of $\Ocal_k$ from our generating function simply as
\begin{align}
	\chi = [x^{2k}]\Gcal_{\text{forest}}^\Ocal (x,-1)\,.
\end{align}
From \eqref{eq:POS_Gcal-tree-2}, we find
\begin{align}
	\Gcal_{\text{tree}}^\Ocal (x,-1) = -\frac{1}{2}(1+\sqrt{1+4x^2})\,,
\end{align} 
and using \eqref{eq:POS_Gcal-forest}
\begin{align}
	\Gcal_{\text{forest}}^\Ocal (x,-1) = \frac{1}{x}\left(\frac{x}{1+\Gcal_{\text{tree}}^\Ocal (x,-1)}\right)^{\<-1\>}_x = \frac{1}{1-x^2}= \sum_{k=0}^\infty x^{2k}\,.
\end{align}
Hence, we find
\begin{align}
	\chi = [x^{2k}]\Gcal_{\text{forest}}^\Ocal (x,-1) = 1\,,
\end{align}
which shows that the Euler characteristic of $\Ocal_k$ is equal to one for any $k$. This is a strong indication that $\Ocal_k$ are topologically closed balls.

\subsubsection{\texorpdfstring{Physical Singularities and Relation to $\tr{\phi^3}$}{Physical Singularities and Relation to Tr(phi\^3)}}

The boundaries of the ABJM momentum amplituhedron are in a one-to-one correspondence to the singularities of tree-level ABJM scattering amplitudes. Our classification thus completely characterises all singularities of ABJM amplitudes and how they are related. The covering relations we found (see figure \ref{fig:ABJM-covering-relations}) have the natural physical interpretations of factorisation, and the two-particle poles special to four-point amplitudes. A special property of ABJM is that we only have factorisations into even particle amplitudes, such that the codimension-1 boundaries correspond to the vanishing of an odd-particle planar Mandelstam variables (\textit{i.e.} a planar Mandelstam $X_{ij}$ with $|i-j|$ odd, corresponding to an odd sum of momentum vectors being squared), the poles corresponding to the vanishing of even-particle Mandelstam variables only appear at a higher codimension. These statements can be understood from a surprising relation to the ABHY associahedron. We will see a more direct relation between the ABJM momentum amplituhedron and the ABHY associahedron in section \ref{sec:POS_pf-through-scattering-eq}, for now we make some observations regarding the relation between their boundaries. 

\begin{figure}
	\centering
	\includegraphics[width=0.6\textwidth]{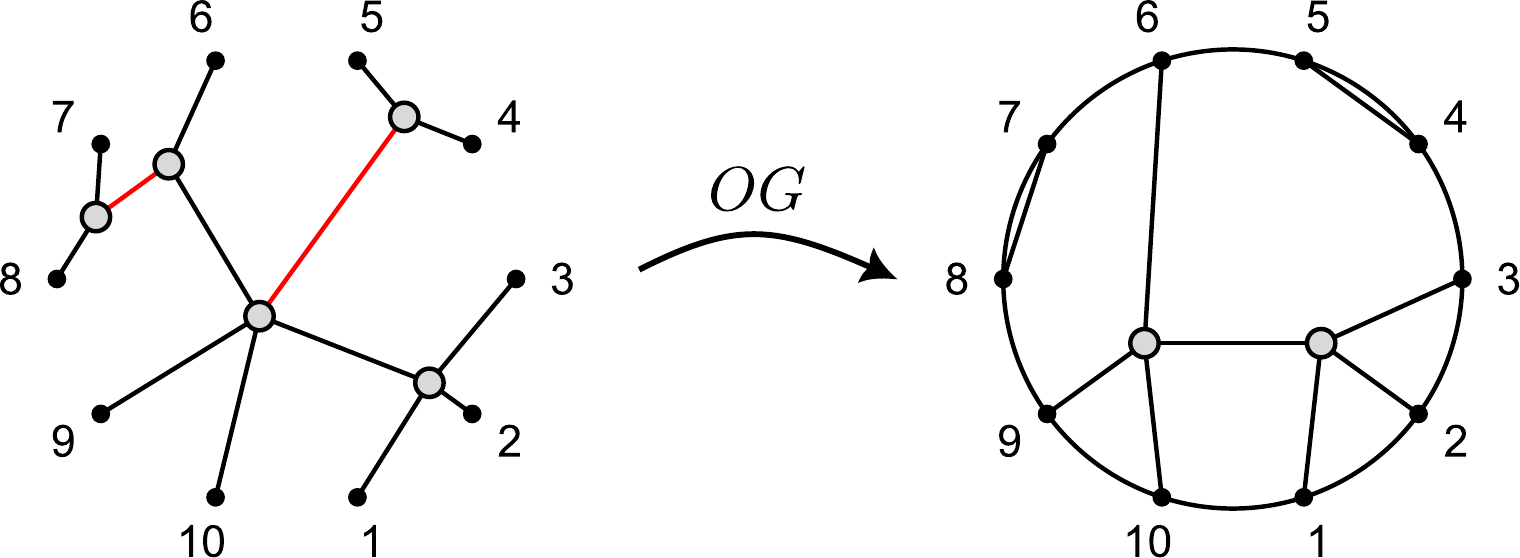}
	\caption{An example of how the map $OG$ acts on a factorisation graph. The internal edges corresponding to even-particle Mandelstam variables are depicted in red.}
	\label{fig:OG-map}
\end{figure}

Recall that the boundaries of the ABHY associahedron $\Ascr_n$ are labelled by `factorisation diagrams', and we have given the covering relations $\succdot_A$ in section \ref{sec:POS_ABHY}. We now consider the following diagrammatic map from factorisation diagrams to OG forests: for any factorisation diagram $\Gamma$, we define $OG(\Gamma)$ to be the OG forest obtained by the following steps:
\begin{enumerate}
	\item embed $\Gamma$ in a disk,
	\item remove all internal edges corresponding to even-particle Mandelstam variables,
	\item replace all vertices of degree 2 with a straight edge,
	\item remove all subgraphs that are not connected to a boundary component.
\end{enumerate}
We show an example of the map $OG$ in figure \ref{fig:OG-map}.

Consider some internal degree-$p$ vertex $v$ of $\Gamma$ which has incident edges labelled by the planar Mandelstam variables $s_{A_1},\ldots,s_{A_p}$, where the index sets $A_i$ are cyclic sets on $[n]=[2k]$. Note that the fact that we have an even number of boundary components implies that $|A_1|+\ldots+|A_p|$ has to be even. When considering $OG(\Gamma)$, we remove all edges $A_i$ where $|A_i|$ is odd. There are necessarily an even number of them, and hence the corresponding vertex in $OG(\Gamma)$ will be of even degree. This shows that $OG(\Gamma)$ is indeed an OG forest. Furthermore, it is clear that \emph{any} OG forest can be obtained from at least one factorisation diagram. This can be seen by drawing the OG forest of interest, and adding edges between disconnected components (taking care to preserve planarity), and interpreting the resulting graph as a factorisation diagram. Any edge added in this way will correspond to an even-particle Mandelstam invariant, and hence when we apply the map $OG$ we will end up with the desired OG forest. Thus, we have shown that $OG$ is an injective map from factorisation graphs to OG forests.

\begin{figure}
	\centering
	\includegraphics[width=0.8\textwidth]{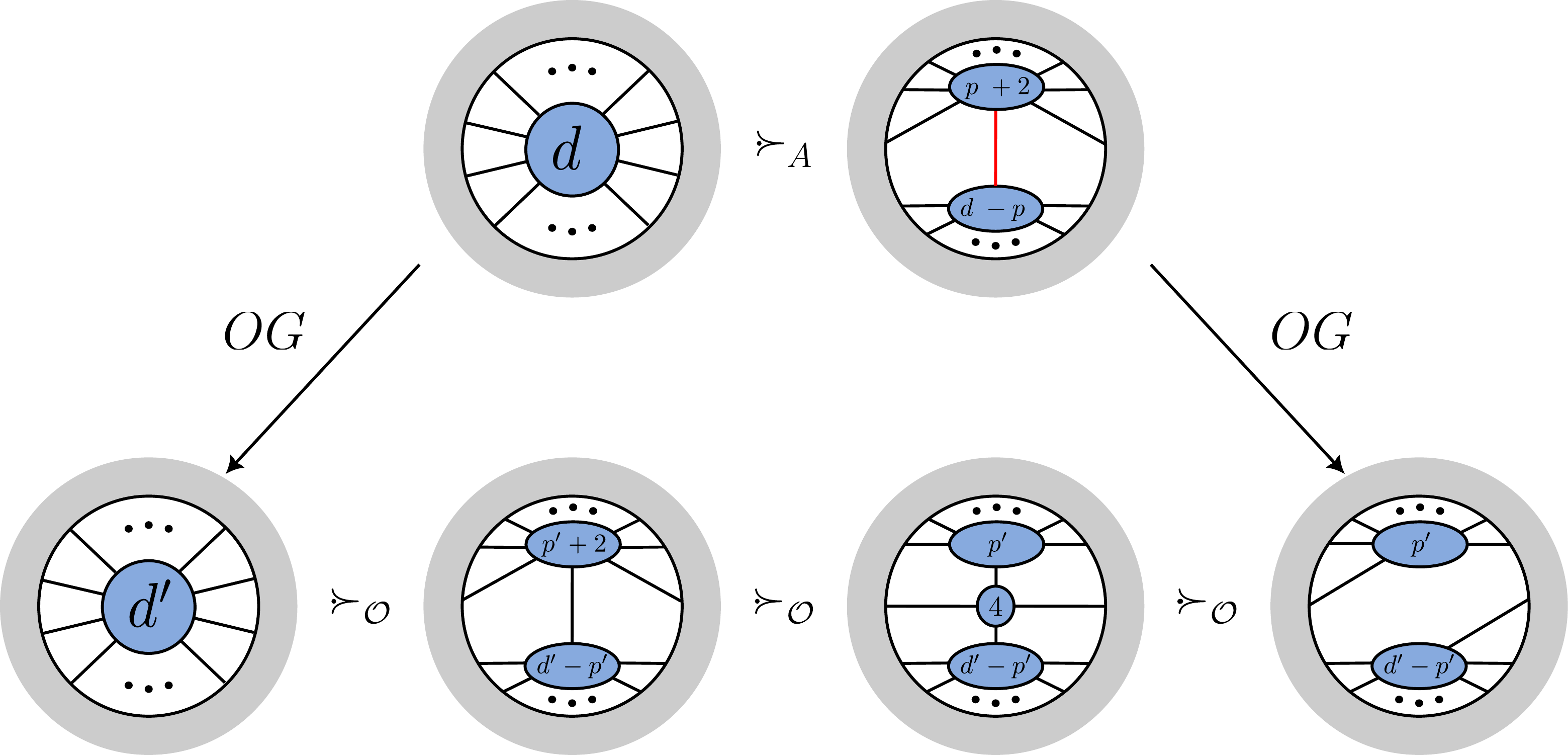}
	\caption{A sequence of covering relations which establish that $OG(\Gamma_2)\succeq_\Ocal OG(\Gamma_1)$ for $\Gamma_2 \succdot_A \Gamma_1$. We denote internal edges corresponding to even-particle Mandelstam invariants in red.}
	\label{fig:POS_OG-map-preserves-order}
\end{figure}

The partial order $\preceq_A$ on factorisation diagrams induces a partial ordering on OG forests. We will now argue that this is \emph{exactly} the partial order $\preceq_\Ocal$ we introduced above. To start, we will show that for any two factorisation diagrams such that $\Gamma_1 \preceq_A \Gamma_2$, then $OG(\Gamma_1)\preceq_\Ocal OG(\Gamma_2)$. It is sufficient to show the above statement in the case when $\Gamma_1 \precdot_A \Gamma_2$, \textit{i.e.} when $\Gamma_1$ is a codimension-1 boundary of $\Gamma_2$. In this case, $\Gamma_1$ has one additional edge with respect to $\Gamma_2$. If this extra edge is an odd-particle Mandelstam variable, then the map $OG$ will not affect it, and $OG(\Gamma_1)\precdot OG(\Gamma_2)$ from the definition of $\precdot_\Ocal$. If, however, the extra edge is an even particle Mandelstam, then it will be removed in $OG(\Gamma_1)$. The partial order between $OG(\Gamma_1)$ and $OG(\Gamma_2)$ is established by a sequence of covering relations depicted in figure \ref{fig:POS_OG-map-preserves-order}.
This proves that the $OG$-induced partial order on the set of OG forests is contained in the partial order $\preceq_\Ocal$.

To prove that the induced partial order on OG forests is exactly $\preceq_\Ocal$ we need to additionally show that all links of $\preceq_\Ocal$ are also present in the induced poset. That is, for all $F_1\preceq_{\Ocal} F_2$, we must show that there exist $\Gamma_1 \in OG^{-1}(F_1), \Gamma_2\in OG^{-1}(F_2)$ such that $\Gamma_1\preceq_A \Gamma_2$. It is again sufficient to restrict ourselves to the case where $F_1\precdot_{\Ocal} F_2$. From the covering relations for OG forests depicted in figure \ref{fig:ABJM-covering-relations} it is easy to see that it is always possible to find such $\Gamma_1$ and $\Gamma_2$. Assume that we have found some $\Gamma_2\in OG^{-1}(F_2)$ which can be obtained by `filling in the grey disk' around the vertex labelled $d$ or $4$ in figure \ref{fig:ABJM-covering-relations} (this is always possible by inserting new vertices on existing edges and connecting all the disconnected pieces by new edges). If $F_1$ differs from $F_2$ by an extra factorisation channel (depicted on the left in figure \ref{fig:ABJM-covering-relations}), then we simply fill in the grey disk in the same manner, and we are done. It is clear that the resulting factorisation diagrams satisfy $\Gamma_1\preceq_A \Gamma_2$, since they are equivalent except that $\Gamma_1$ has one extra factorisation channel. For the second case, where $F_1$ is obtained by dissolving a four-points vertex of $F_2$ (depicted on the right in figure \ref{fig:ABJM-covering-relations}), we proceed similarly, except we first reconnect the two newly disconnected pieces of $\Gamma_1$ by adding two new nodes in the edges in the rightmost diagram in figure \ref{fig:ABJM-covering-relations} and connecting them by an edge. Again, it is easy to see that the resulting $\Gamma_1,\Gamma_2$ satisfy $\Gamma_1\preceq_A \Gamma_2$.

\begin{figure}
	\centering
	\begin{equation*}
		\vcenter{\hbox{\includegraphics[width=75mm]{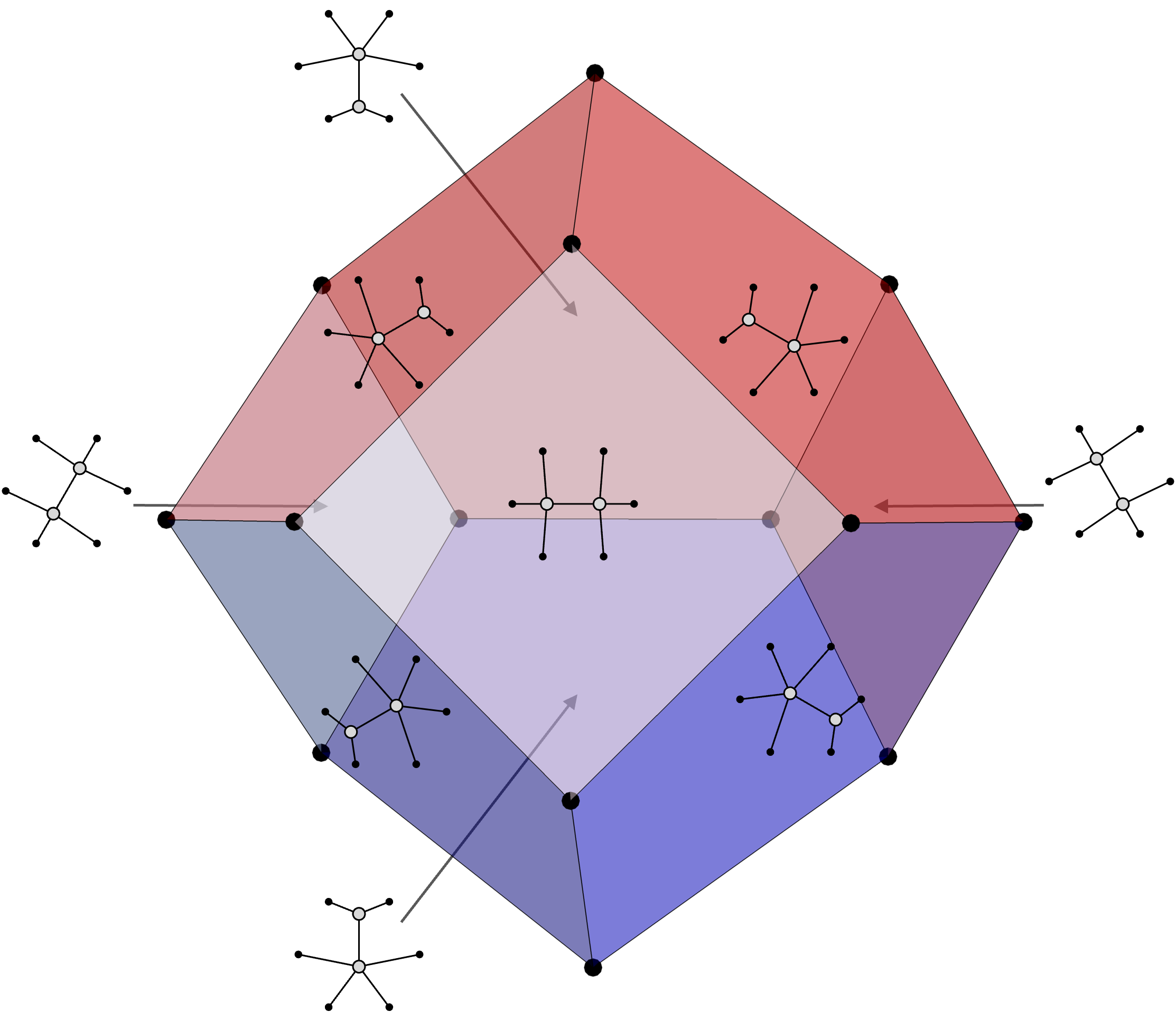}}}\quad \text{\Huge$\to$} \vcenter{\hbox{\includegraphics[width=55mm]{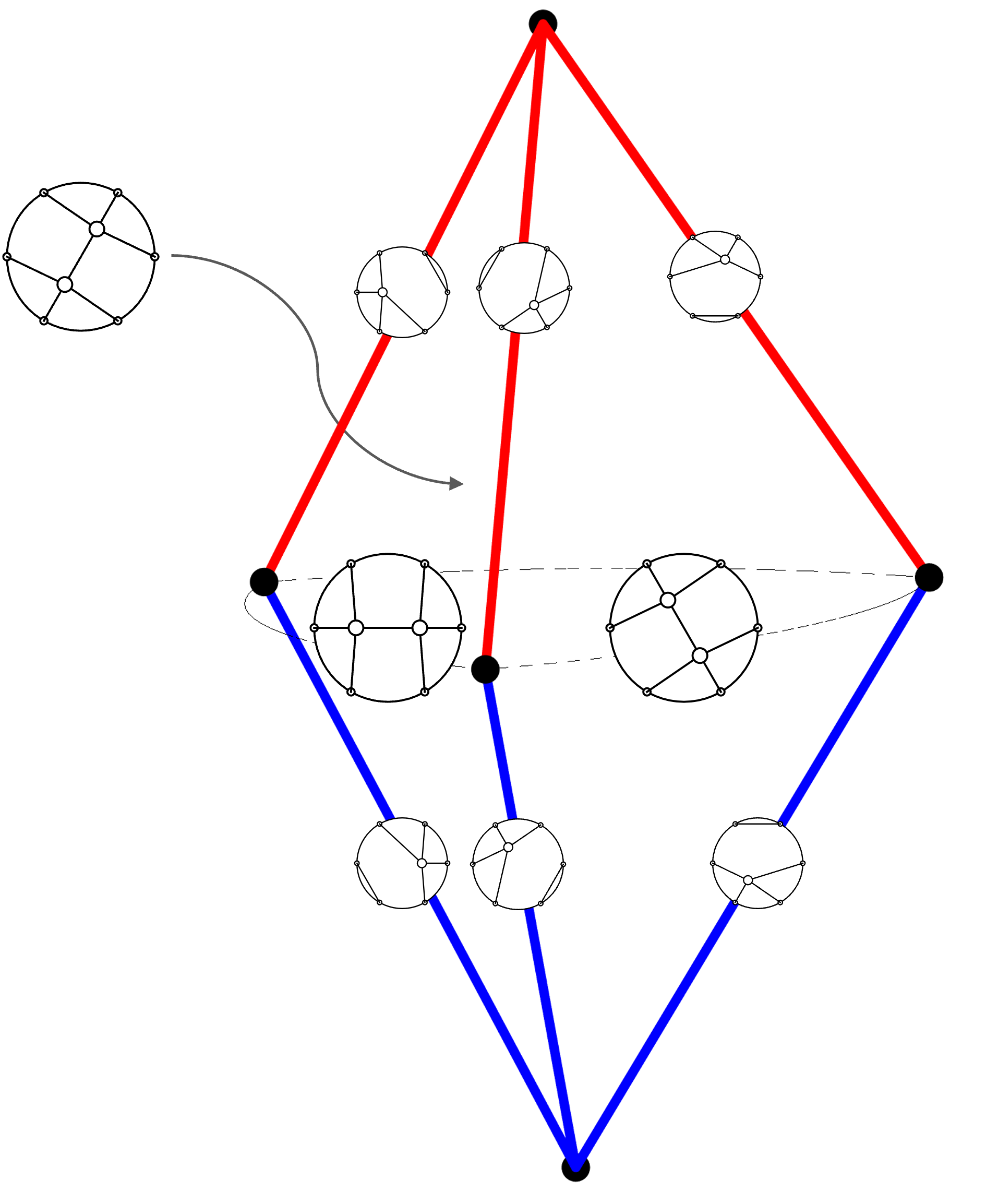}}}
	\end{equation*}
	\caption{We can imagine collapsing all the boundaries of $\Ascr_6$ corresponding to even-particle Mandelstam variables. We end up with an object which is topologically $\Ocal_3$.}
	\label{fig:ABHY-ABJM-collapse}
\end{figure}

This argument shows that any partial order between OG forests $F_1\preceq_\Ocal F_2$ can be induced by the map $OG$ from some partial order $\Gamma_1\preceq_A\Gamma_2$. Together with the previous statement that all partial orders induced by the map $OG$ are compatible with $\preceq_\Ocal$, this is sufficient to show the map $OG$ is a surjective map from the set of all $2k$-point factorisation diagrams to OG forests such that the partial order $\succeq_A$ induces the partial order $\succeq_\Ocal$. This shows that the singularities of tree-level ABJM amplitudes are a sub-poset of the singularities of $\tr{\phi^3}$. This can be understood at the level of geometry: starting from the ABHY associahedron $\Ascr_{2k}$, we `collapse' any boundary which has an even-particle Mandelstam, and we end up with a geometry which is homeomorphic to $\Ocal_k$. This is illustrated in figure \ref{fig:ABHY-ABJM-collapse} for the case $k=3$. We see that the codimension-one boundaries associated to odd-particle Mandelstam variables are mapped to codimension-one boundaries, whereas codimension-one boundaries associated to even-particle Mandelstam variables are mapped to lower dimensional boundaries, as one would expect from physical considerations.

\section[head={Push Forwards \& Scattering Equations},tocentry={Push Forwards Through the Scattering Equations}]{Push Forwards Through the Scattering Equations}\label{sec:POS_pf-through-scattering-eq}

We now turn to a striking framework which captures the ABHY associahedron, the momentum amplituhedron, and the ABJM momentum amplituhedron. The idea is that these positive geometries can be obtained as the image of a \emph{push forward map} from a \emph{world-sheet moduli space} induced by the \emph{scattering equations}. This was first observed for the ABHY associahedron in \cite{Arkani-Hamed:2017mur}, and later conjectured to extend to the momentum amplituhedron and ABJM momentum amplituhedron in \cite{He:2018okq} and \cite{He:2021llb}. This is a story which is deeply intertwined with twistor string formulations of scattering amplitudes.

We recall that the scattering equations relate the moduli space $\Mfrak_{0,n}$ to the kinematic space $\Kbb_n$. Furthermore, on $\Mfrak_{0,n}$ there is a natural positive geometry called the \emph{positive moduli space} 
\begin{align}
	\Mfrak_{0,n}^+ = \{(z_1,\ldots,z_n) \in \Mfrak_{0,n}(\Rbb)\colon z_1<z_2<\ldots<z_n \}/SL(2)\,.
\end{align}
Its canonical form is given by the \emph{world-sheet Parke-Taylor form}
\begin{align}
	\omega^{\text{WS}}_n \coloneqq \frac{1}{\vol[SL(2)]}\frac{\dd z_1\wedge\cdots\wedge\dd z_n}{(z_1-z_2)(z_2-z_3)\cdots(z_n-z_1)}\,.
\end{align}
In \cite{Arkani-Hamed:2017mur}, a map from $\Mfrak_{0,n}^+$ to the ABHY associahedron $\Ascr_n$ was given by solving the scattering equations for the planar Mandelstam variables:
\begin{align}\label{eq:POS_Xab-SE-map}
	X_{ab}=\sum_{\substack{1 \leq i<a \\ a<j<b}} z_{a, j} \frac{c_{i j}}{z_{i, j}}+\sum_{\substack{a \leq i<b \\ b \leq j<n}} z_{i, b-1} \frac{c_{i j}}{z_{i, j}}+\sum_{\substack{1 \leq i<a \\ b \leq j<n}} z_{a, b-1} \frac{c_{i j}}{z_{i, j}}\,,
\end{align}
where we use $z_{i,j}=z_i-z_j$. We have already seen this map in action in the example surrounding figure \ref{fig:ABHY-map-pentagon}, where we considered the push forward from $\Mfrak^+_{0,5}$ to $\Ascr_{5}$ with $c_{ij}=1$, and with a relabelling of the variables.

To see an analogous push forward map relating the moduli space to the (ABJM) momentum amplituhedron, we first describe how to embed $\Mfrak_{0,n}^+$ in $G_+(k,n)$ and $OG_+(k)$. We note that we have the isomorphism $G_+(2,n)\simeq \Mfrak^+_{0,n}\times T^+_n$, where the positive torus is defined as
\begin{align}
	T^+_n\coloneqq \{(t_1,t_2,\ldots,t_n)\in (\Rbb^+)^n\}/GL(1)\simeq G_+(1,n-1)\,,
\end{align}
such that we can write
\begin{align}\label{eq:POS_G2n-form-mod}
	\Omega(G_+(2,n))=\frac{1}{\vol[GL(2)]}\frac{\dd z_1\wedge\cdots\wedge\dd z_n}{(z_1-z_2)(z_2-z_3)\cdots(z_n-z_1)}\wedge\dd\log t_1\wedge\cdots\wedge\dd\log t_n\,.
\end{align}
In a matrix form this parametrization of $G_+(2,n)$ looks like
\begin{align}
	\begin{pmatrix}
		t_1 & t_2 & \cdots & t_n\\ t_1 z_1 & t_2 z_2 & \cdots & t_n z_n
	\end{pmatrix}\,,
\end{align}
which has positive minors when $t_i>0, z_{i+1}-z_i>0$. We can embed $G_+(2,n)$ into $G_+(k,n)$ using the \emph{Veronese map}
\begin{align}
	\begin{pmatrix}
		t_1 & t_2 & \cdots & t_n\\ t_1 z_1 & t_2 z_2 & \cdots & t_n z_n
	\end{pmatrix} \mapsto C(\bm{z},\bm{t})= \begin{pmatrix}
	t_1 & t_2 & \cdots & t_n\\
	t_1 z_1 & t_2 z_2 & \cdots & t_n z_n\\
	\vdots & \vdots & \ddots & \vdots \\
	t_1 z_1^{k-1} & t_2 z_2^{k-1}& \cdots & t_n z_n^{k-1}
\end{pmatrix}\,.
\end{align}
The maximal minors of $C(\bm{z},\bm{t})$ are Vandermonde determinants
\begin{align}\label{eq:POS_C-mat-veronese}
	(i_1 i_2\cdots i_k)_{C(\bm{z},\bm{t})} =t_{i_1} t_{i_2}\cdots t_{i_k} (z_{i_1}-z_{i_2})(z_{i_1}-z_{i_3})\cdots (z_{i_{k-1}}-z_{i_k})\,.
\end{align}
Therefore, these minors are also positive when $t_i>0, z_{i+1}-z_i>0$. Its orthogonal complement is given by
\begin{align}\label{eq:POS_Cperp-veronese}
	C(\bm{z},\bm{t})^\perp = \begin{pmatrix}
		\tilde{t}_1 & \tilde{t}_2 & \cdots & \tilde{t}_n\\
		\tilde{t}_1 z_1 & \tilde{t}_2 z_2 & \cdots & \tilde{t}_n z_n\\
		\vdots & \vdots & \ddots & \vdots \\
		\tilde{t}_1 z_1^{n-k-1} & \tilde{t}_2 z_2^{n-k-1}& \cdots & \tilde{t}_n z_n^{n-k-1}\end{pmatrix}\,,
\end{align}
where 
\begin{align}
	\tilde{t}_i = \frac{1}{t_i \prod_{j\neq i} (z_j-z_i)}\,.
\end{align}
When we restrict to $n=2k$, we can further constrain this matrix to be part of the positive orthogonal Grassmannian $OG_+(k)$. We will use our $GL(1)$ redundancy to fix $t_n\to 1$, and we will divide our $C^\perp$ matrix by $\tilde{t}_n$. Orthogonality is then achieved by imposing $\tilde{t}_i/\tilde{t}_n=(-1)^i t_i$, since this would turn the matrix \eqref{eq:POS_Cperp-veronese} into $C\cdot\eta$. This imposes the following constraint on the $t$ variables:
\begin{align}\label{eq:POS_t-solve-orth}
	t_i^2 = (-1)^i \frac{\prod_{j\neq n} (z_n-z_j)}{\prod_{j\neq i} (z_i-z_j)}\,.
\end{align}
We note that the numerator of \eqref{eq:POS_t-solve-orth} is manifestly positive, while the product in the numerator has $2k-i$ negative elements, thus $t_i^2$ is always positive. Furthermore, to ensure that the matrix $C(\bm{z})$ has positive maximal minors, we require $t_i$ to be positive and we have to restrict to
\begin{align}
	t_i = \sqrt{(-1)^i \frac{\prod_{j\neq n} (z_n-z_j)}{\prod_{j\neq i} (z_i-z_j)}}\,.
\end{align} 
Therefore, requiring orthogonality and positivity of the matrix \eqref{eq:POS_C-mat-veronese} fully fixes the $\bm{t}$ variables, while leaving $\bm{z}$ a generic point in $\Mfrak_{0,n}^+$. The map $C(\bm{z})$ then provides an embedding of $\Mfrak_{0,n}^+$ into $OG_+(k)$. As noted in section \ref{sec:AMP_CHY} and appendix \ref{sec:scatt-eq-4D}, the four dimensional scattering equations can be written as
\begin{align}
	C(\bm{z},\bm{t})^\perp\cdot \lambda^T = 0\,,\quad C(\bm{z},\bm{t})\cdot \tilde\lambda^T=0\,,
\end{align}
and they appear in the twistor-string formula for scattering amplitudes in \nf we encountered in equation \eqref{eq:AMP_twistor-string-Grassmannian}. The three-dimensional equation
\begin{align}
	C(\bm{z})\cdot \lambda^T=0\,,
\end{align}
appear in the twistor-string formula \eqref{eq:AMP_ABJM-reduced-twistor-string} for scattering amplitudes in supersymmetry reduced ABJM theory. We shall refer to these equations as the \emph{three-dimensional scattering equations}. The number of solutions to the three-dimensional scattering equations is counted by the \emph{tangent number} (also known as the \emph{Euler zag numbers}) \cite{Cachazo:2013iaa}. 

\begin{figure}
	\centering
	\includegraphics[width=0.7\textwidth]{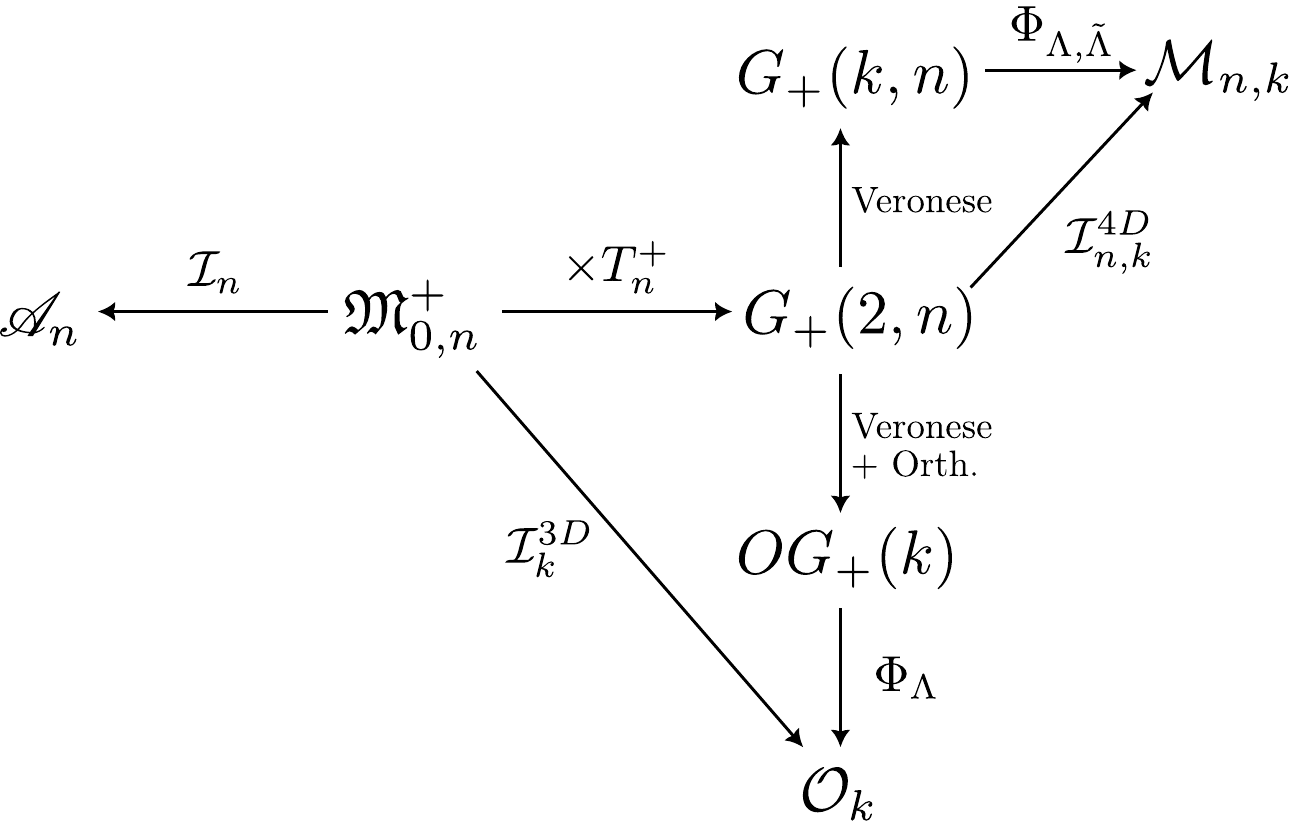}
	\caption{A web which summarises the various maps between the moduli space, Grassmannians, and kinematic geometries.}
	\label{fig:mod-grass-mom-web}
\end{figure}

Now that we have an embedding of the positive moduli spaces $\Mfrak_{0,n}^+$ and $\Mfrak_{0,n}^+\times T_n^+$ into $OG_+(k)$ and $G_+(k,n)$, respectively, we can consider the maps $\Phi_\Lambda(C(\bm{z}))$ into $\Ocal_k$ and $\Phi_{\Lambda,\tilde\Lambda}(C(\bm{z},\bm{t}))$ into $\Mcal_{n,k}$, whereas equation \eqref{eq:POS_Xab-SE-map} provides a map from $\Mfrak_{0,n}^+$ into the ABHY associahedron $\Ascr_n$. We summarise these maps in the diagram \ref{fig:mod-grass-mom-web}. The conjectures from \cite{Arkani-Hamed:2017mur} and \cite{He:2021llb} are that these three maps are morphisms (see section \ref{sec:POS_def}) between the positive moduli spaces and these kinematic geometries. 

These morphisms imply that we can find the canonical form of $\Ascr_n$, $\Mcal_{n,k}$, and $\Ocal_k$ by calculating the push forward. Furthermore, as explained in section \ref{sec:POS_push-forward}, we don't need to know the exact map to calculate a pushforward, we can instead do an implicit pushforward through the ideal generated by the scattering equations. It is thus accurate to say that the canonical forms are related by doing a \emph{push forward through the scattering equations}. That is, we let $\Ical_n$, $\Ical^{4D}_{n,k}$, and $\Ical_k^{3D}$ are the ideals generated by the scattering equations, the four-dimensional scattering equations, and the three-dimensional scattering equations, respectively. Then, the push forward equation gives us
\begin{subequations}\label{eq:POS_push-forward-SE}
\begin{alignat}{2}
	\Omega(\Ascr_n) &= \Ical_{n*} \omega_n^{\text{WS}} &&= \sum_{\bm{z}\in \Vcal(\Ical_n)} \omega_n^{\text{WS}}\,,\\
	\Omega(\Mcal_{n,k}) &= \Ical^{4D}_{n,k*} \omega_n^{\text{WS}}\wedge(\dd\log t)^n &&= \sum_{(\bm{z},\bm{t})\in \Vcal(\Ical_{n,k}^{4D})} \omega_n^{\text{WS}}\wedge(\dd\log t)^n\,,\\
	\Omega(\Ocal_k) &= \Ical^{3D}_{k*} \omega_n^{\text{WS}} &&= \sum_{\bm{z}\in \Vcal(\Ical^{3D}_{k})} \omega_n^{\text{WS}}\,,
\end{alignat}
\end{subequations}
where $\Vcal(\Ical)$ denotes the variety of the ideal $\Ical$, which is the set of all solutions to the respective scattering equations. If we want to use these formulae to find the canonical form of the positive geometries in kinematic space, we are now faced with an immediate problem: to calculate the push forward we (naively) need to sum over all solutions to the scattering equations! There are generally no closed form solutions to the scattering equations, which poses a serious problem in calculating these push forwards. In \cite{Lukowski:2022fwz}, we presented three methods which allows us to sidestep this issue altogether, instead allowing us to calculate push forward without needing to solve the scattering equations. In the remainder of this section we will explain the main ideas and results from this paper.

\subsection{Recovering Scattering Amplitudes}

Before moving on to the intricacies of calculating pushforwards, let us first see how \eqref{eq:POS_push-forward-SE} recovers the correct scattering amplitudes. For \nf and ABJM, the delta function expression of the push forward reproduces the twistor string formulas \eqref{eq:AMP_twistor-string-Grassmannian} and \eqref{eq:AMP_ABJM-reduced-twistor-string}, respectively, and as such give the correct tree-level scattering amplitudes in the respective theories. To see that we also correctly recover the $\tr{\phi^3}$ amplitudes, we need to make use of the CHY formula \eqref{eq:AMP_CHY-bas}. Some effort is needed to recast the push forward formula into the CHY formula, and we will spend the remainder of this subsection on the derivation. This will be a rather technical discussion and can serve as a useful study case to see how to manipulate expressions in the push forward. A shorter, yet less illuminating, derivation was given in \cite{Arkani-Hamed:2017mur}. 

To start, we note that the polynomial scattering equations \eqref{eq:AMP_scatt-eq-poly} relate the moduli space $\Mfrak_{0,n}^+$ to the space of Mandelstam variables $s_A$, which we first need to recast in terms of planar Mandelstam variables $X_{ij}$. After we take the push forward of $\smash{\omega_n^{\text{WS}}}$ through these scattering equations, we end up with a form in terms of all $X_{ij}$ variables which lives in the kinematic space $\Kbb_n$, which was called the \emph{planar scattering form} in \cite{Arkani-Hamed:2017mur}, which we will denote $\omega_n^{\text{ABHY}}$. As noted in section \ref{sec:POS_ABHY}, we can retrieve the canonical form of $\Ascr_n$ and the $n$-point scattering amplitude in $\tr{\phi^3}$ by pulling back $\omega_n^{\text{ABHY}}$ onto the hypersurface $H_n$. 

We choose $\{X_{13},\ldots,X_{1n-1}\}$ as our basis of $H_n$. To decrease the risk of confusion and to compact some formulae, we will use the variables $\{a_i\}_{i=1}^{n(n-3)/2}$ on $\Kbb_n$, and $\{b_i\}_{i=1}^{n-3}$ on $H_n$:
\begin{align}
	X_{ij}&=a_{(i-1)n+j-\binom{i+2}{2}+\delta_{i,1}}\,,\\
	X_{1i}&=b_{i-2}\,.
\end{align}
The ideal $\Ical_n$ allows us to push forward from $\Mfrak_{0,n}^+$ to $\Kbb_n$ is generated by the polynomial scattering equations
\begin{align}
	\Ical_{n}&\coloneqq \<f_1,\ldots,f_{n-3}\>\subseteq \Qbb(\bm{a})[\bm{z}]\,,\\
	f_i &\coloneqq \lim_{z_n\to\infty} \sum_{A\in\binom{[n]}{i+1}}\left.s_A z_A/z_n \right|_{z_1\to0,z_{n-1}\to1}\,,
\end{align}
where we have used the $SL(2)$ freedom to fix $z_1\to0,z_{n-1}\to1,z_n\to\infty$, and we use $\bm{z}=(z_2,\ldots,z_{n-2})$. The ideal $\Jcal_n$ subsequently allows us to pull back onto $H_n$:
\begin{align}
	&\Jcal_n \coloneqq\<X_{i,j}-g_{i,j}(\bm{b})\>_{1\leq i<j\leq n}\,,\\
	&g_{1,j}(\bm{b}) = b_{j-2}\,,\quad g_{i>1,j}(\bm{b})=b_{j-2}-b_{i-1}+C_{i,j}\,.
\end{align}
Comparing to equations \eqref{eq:POS_pushforward-ideal} and \eqref{eq:POS_pushforward-special-fn}, we have
\begin{align}
	\omega^{\text{WS}}_n = \PT(\bm{z})\dd^{n-3}\bm{z}\,,\quad \PT(\bm{z}) \coloneqq \frac{1}{z_{12}\cdots z_{n1}}/SL(2)\,,
\end{align}
and 
\begin{align}
	\overline\omega_J(\bm{z},\bm{a})=(-1)^{n-3} \PT(\bm{z}) \left| \frac{\partial \bm{f}}{\partial \bm{a}} \right|_J \left| \frac{\partial \bm{f}}{\partial \bm{z}} \right|^{-1}\,.
\end{align}
We can then write
\begin{align}
	\omega_n^{\text{ABHY}} = \Ical_{n*}\omega_n^{\text{WS}} = \sum_{J\in\binom{[n(n-3)/2]}{n-3}}\Ical_{n*}\overline\omega_J\dd \bm{a}^J\,,
\end{align}
and we find the canonical form of $\Ascr_n$ (and hence the amplitudes $m_n$) by pulling back through $\Jcal_n$:
\begin{align}
	\Omega(\Ascr_n)=m_n \dd^{n-3}\bm{b} = \Jcal_n^* \omega_n^{\text{ABHY}}=\Jcal_n^*\big(\Ical_{n*}\omega_n^{\text{WS}}\big)\,.
\end{align}
We note that 
\begin{align}
	\Jcal_n^* \dd \bm{a}^J = \left|\frac{\partial \bm{g}}{\partial \bm{b}}\right|^J\dd^{n-3}\bm{b}\,,
\end{align}
and the pull back of a function $\Jcal_n^*\big(\Ical_{n*}\overline\omega_J\big)$ simply substitutes $X_{ij}\to g_{ij}(\bm{b})$. We can equivalently do this substitution in the ideal $\Ical_n$:
\begin{align}
	\Jcal_n^*\big(\Ical_{n*}\overline\omega_J\big) = \big(\Jcal_n^*\Ical_n\big)_* \overline\omega_J\,.
\end{align}
Here we use the notation $\Jcal_n^*\Ical_n$ to denote the ideal generated by the functions $\smash{\Jcal_n^*f_i}$, which are the functions $f_i$ pulled back onto $H_n$, \emph{i.e.} the function $f_i$ with $X_{ij}$ replaced by $g_{ij}(\bm{b})$.

Furthermore, since
\begin{align}
	\sum_{J\in\binom{[n(n-3)/2]}{n-3}}\left| \frac{\partial(\Jcal_n^*\bm{f})}{\partial\bm{a}} \right|_J \left| \frac{\partial\bm{g}}{\partial\bm{b}} \right|^J = \left| \frac{\partial (\Jcal_n^*\bm{f})}{\partial \bm{b}} \right|\,,
\end{align}
we can write
\begin{align}
	m_n \dd^{n-3}\bm{b} = \big(\Jcal_n^*\Ical_n\big)_* \overline\omega_J\ = (\Jcal_n^*\Ical_{n})_* F_n \dd^{n-3}\bm{b}\,,
\end{align}
where
\begin{align}
	F_n(\bm{z};\bm{b})\coloneqq \sum_{J\in\binom{[n(n-3)/2]}{n-3}}\Jcal_n^*\overline\omega_J(\bm{z})\left| \frac{\partial\bm{g}}{\partial\bm{b}} \right|^J=\PT(\bm{z})\left| \frac{\partial(\Jcal_n^*\bm{f})}{\partial\bm{z}} \right|^{-1}\left| \frac{\partial (\Jcal_n^*\bm{f})}{\partial\bm{b}} \right|\,.
\end{align}
We can explicitly calculate the Jacobian of pulling back onto $H_n$ to be
\begin{align}
	\left| \frac{\partial (\Jcal_n^*\bm{f})}{\partial\bm{b}} \right| = (-1)^{n+1}\prod_{i=1}^{n-3}\prod_{j=i+2}^{n-1}(z_i-z_j)\,.
\end{align}
We see that the push forward and the pull back are `associative' in the sense that $\Jcal^*(\Ical_*\omega) = (\Jcal^*\Ical)_*\omega$. We now arrive at the result
\begin{align}
	\Omega(\Ascr_n) = m_n \dd^{n-3}\bm{b}=\sum_{\bm{\xi}\in\Vcal(\Jcal_n^*\Ical)} F_n(\bm{\xi})\dd^{n-3}\bm{b}\,,
\end{align}
which gives us the $n$-point scattering amplitude as a rational function of $\bm{b}$ and $C_{i,j}$

These are the appropriate variables to use on the subspace $H_n$, however it is more natural to write the scattering amplitude in terms of planar Mandelstam variables. This can be achieved by re-substituting $b_i\to X_{1,i+2}$, and $C_{i,j}\to X_{i,j}+X_{1,i+1}-X_{1,j}$. Or, equivalently, we can replace $|\partial (\Jcal_n^*\bm{f})/\partial \bm{z}|$ by $|\partial\bm{f}/\partial \bm{z}|$ and summing over $\Vcal(\Ical_n)$ instead:
\begin{align}
	m_n=\sum_{\bm{\xi}\in\Vcal(\Ical_n)}F'_n(\bm{\xi})\,,
\end{align}
where
\begin{align}
	F_n'(\bm{z}) = \PT(\bm{z})(-1)^{n+1}\prod_{i=1}^{n-3}\prod_{j=i+2}^{n-1}(z_i-z_j)\left|\frac{\partial\bm{f}}{\partial\bm{z}}\right|^{-1}\,.
\end{align}
Although similar, this is not quite the usual CHY summand, while it will sum to give the correct $\tr{\phi^3}$ scattering amplitude. We note that there is a distinct flavour of the \emph{polynomial} scattering equations appearing in this summand, not only in the expected Jacobian $|\partial\bm{f}/\partial\bm{z}|$, but also in a Jacobian $|\partial(\Jcal_n^*\bm{f})/\partial\bm{b}|$, which we picked up from the pull back onto $H_n$. Indeed, different equivalent forms of the scattering equations will, through the process outlined above, yield different CHY-like formulas for $\tr{\phi^3}$. Specifically, for any set of $n-3$ functions $h_1,\ldots,h_{n-3}$ which have the same solutions as the scattering equations, then 
\begin{align}\label{eq:POS_CHY-summand-general}
	\PT(\bm{z}) \left|\frac{\partial\bm{h}}{\partial\bm{z}}\right|^{-1}\left|\frac{\partial(\Jcal_n^*\bm{h})}{\partial\bm{b}}\right|\,, 
\end{align}
gives a valid CHY summand. 

Let us consider the standard scattering equations $E_1,\ldots,E_n$ defined in \eqref{eq:AMP_Scatt-Eq-CHY}. Since only $n-3$ of these equations are independent, we can freely remove $E_q, E_r$ and $E_s$. We find that the Jacobian picked up from the pull back to $H_n$ is given by
\begin{align}
	\left|\frac{\partial(\Jcal^*_n \bm{E})}{\partial\bm{b}}\right|^{[n]\setminus\{q,r,s\}} = (-1)^{n+q+r+s+1} \frac{(z_q-z_r)(z_r-z_s)(z_s-z_q)}{(z_1-z_2)(z_2-z_3)\cdots(z_n-z_1)} \,.
\end{align}
In this case equation \eqref{eq:POS_CHY-summand-general} gives the well-known standard CHY summand for $\tr{\phi^3}$ amplitudes. We see that, from the current perspective, the fundamental part of a CHY summand is just a \emph{single} Parke-Taylor form coming from $\omega^{\text{WS}}_n$. The second Parke-Taylor factor is picked up from the pull back onto $H_n$, and a different specific form of the scattering equations might yield a different factor. This further raises an interesting question regarding possible extensions of the ABHY formalism to different theories. If we can find a different subspace of $\Kbb_n$ to which we can pull back with the ideal $\Kcal_n$ such that $|\partial(\Kcal_n^*\bm{f})/\partial\bm{b}|$ yields some different desired half-integrand, then the calculations above show that the pull back on this subspace of $\omega_n^{\text{ABHY}}$ will give a differential form whose canonical function is the scattering amplitude for a new theory. 

\subsection{How to do Push Forwards?}\label{sec:POS_pf}

In the preceding sections we have argued that we can find the canonical form of positive geometries in kinematic space by calculating the push forward through the scattering equations. However, as we already noted above, this naively runs in to the problem of having to solve the scattering equations explicitly. We will now summarise the three methods we introduced in \cite{Lukowski:2022fwz} to calculate push forwards which circumvent this obstacle. We will proceed in some generality, and we will discuss push forwards of arbitrary differential forms through arbitrary zero-dimensional ideals. These methods are heavily reliant on tools from computational algebraic geometry and \Grob bases. To maintain a streamlined discussion, many technical aspects have been delegated to appendix \ref{sec:APP_alg-geom}. For the ease of reading, we will briefly summarise some of the basic notions which are introduced in the appendix, and refer to \cite{cattani2005introduction, cox2013ideals} for more detailed definitions. 

We let the functions $f_i\in\Cbb(\bm{a})[\bm{z}]$ be polynomials in $\bm{z}=(z_1,\ldots,z_n)$ whose coefficients are rational functions of $\bm{a}=(a_1,\ldots,a_m)$ which generate the zero-dimensional ideal
\begin{align}
	\Ical\coloneqq \<f_1,\ldots,f_n\> \subseteq \Cbb(\bm{a})[\bm{z}]\,.
\end{align}
That the ideal is zero-dimensional means that the variety $\Vcal(\Ical)$ consists of a finite number of points:
\begin{align}
	\Vcal(\Ical)=\{\bm{\xi}_i\}_{i=1}^d\,,
\end{align}
where we can interpret the solutions $\bm{\xi}_i$ to be functions of the $a$-variables. We will further assume that the $a$-variables are \emph{generic}, which means that the ideal $\Ical$ is a radical ideal. 

Given some monomial ordering $\prec$ on $\Cbb[\bm{z}]$, we let $\Gcal\equiv\Gcal_\prec(\Ical)$ be the \emph{\Grob basis} of $\Ical$ with respect to $\prec$. For any zero-dimensional ideal $\Ical$, the \emph{quotient ring} $Q= \Cbb(\bm{a})[\bm{z}]/\Ical$ is a $(d=|\Vcal(\Ical)|)$-dimensional vector space over $\Cbb(\bm{a})$, which admits a \emph{standard basis} $\Bcal\equiv \Bcal_\prec(\Ical)= \{e_\alpha\}_{\alpha=1}^d$.

A key property of \Grob bases is that if we write some arbitrary polynomial $F$ as $F= c_1 g_1+\ldots+ c_t g_t +r$ for some polynomials $c_i,r\in\Cbb(\bm{a})[\bm{z}]$, then the remainder $r$ is always uniquely defined. We denote this remainder by $\overbar{F}^{\Gcal}$ and we can decompose it in the standard basis
\begin{align}
	\overbar{F}^{\Gcal}=\sum_{\alpha=1}^d F_\alpha e_\alpha\,,
\end{align}
where $F_\alpha\in\Cbb(\bm{a})$. We note that, since $g_i\in\Ical$,
\begin{align}
	F(\bm{\xi})=\overbar{F}^{\Gcal}(\bm{\xi})=\sum_{\alpha=1}^d F_\alpha e_\alpha(\bm{\xi})\,,\quad\forall F\in \Cbb(\bm{a})[\bm{z}], \bm{\xi}\in\Vcal(\Ical)\,,
\end{align}
which motivates our interest in remainders and the quotient ring.

We further recall a few formulae from section \ref{sec:POS_push-forward}. The pushforward of a differential form
\begin{align}
	\omega = \sum_{I\in\binom{[n]}{p}}\omega_I(\bm{z})\dd \bm{z}^I\,,
\end{align}
can be found by calculating the push forwards of the rational functions
\begin{align}\label{eq:POS_pf-function}
	\overline\omega_J(\bm{z};\bm{a})=(-1)^p\sum_{I\in\binom{[n]}{p}}\omega_I(\bm{z})\left|\left[\frac{\partial\bm{f}}{\partial\bm{z}}\right]^{-1}\frac{\partial\bm{f}}{\partial\bm{a}}\right|^I_J\,,
\end{align}
as
\begin{align}
	\Ical_*\omega = \sum_{J\in\binom{[m]}{p}}\big( \Ical_*\overline\omega_J \big)\dd \bm{a}^J\,.
\end{align}
Before moving on, we make some remarks about \Grob bases, which are ubiquitous in this story and they are used in some form in all techniques presented below. There are many algorithms for finding \Grob bases, such as the very efficient \emph{Faug\`ere's F4/F5 algorithms} \cite{faugere1999new, faugere2002new}. However, regardless of these algorithms, the calculation of \Grob bases forms the main computational bottleneck. Most \Grob basis techniques are optimised over finite fields, rather than $\Cbb(\bm{a})$. We can use this to our advantage, by evaluating the $\bm{a}$-variables over a finite field, after which the push forward of a rational functions yields a numeric value, which is the function we want to find evaluated on this choice of $\bm{a}$-variables. By doing this repeatedly for different values for $\bm{a}$, it is then possible to reconstruct the final answer using \emph{finite field reconstruction}, as has been implemented in \texttt{FiniteFlow} \cite{Peraro:2019svx} and \texttt{FireFly} \cite{Klappert:2019emp}.

\subsubsection{Push Forwards via Companion Matrices}
The idea of companion matrices is motivated by the observation that multiplication of all $f\in Q$ by $z_i$ can be viewed as an endomorphism on $Q$
\begin{align}
	\times z_i\colon Q\to Q\,,\quad f\mapsto z_i f\,.
\end{align} 
Since $Q$ is a finite dimensional vector space, this can be represented by a matrix $T_i\in M_{d\times d}(\Cbb(\bm{a}))$ in the standard basis $\Bcal$. $T_i$ is the \emph{$i$\textsuperscript{th} companion matrix}, and its components can be found as
\begin{align}
	(T_i)_{\alpha\beta} e_\beta = \overbar{z_i e_\alpha}^\Gcal\,.
\end{align}
Since multiplication is commutative, this means that all companion matrices also mutually commute. The companion matrices thus define an isomorphism between $Q$ and a commutative subalgebra of $M_{d\times d}(\Cbb(\bm{a}))$:
\begin{align}
	\Cbb(\bm{a})[T_1,\ldots,T_n]\simeq Q\,,\quad T_i\mapsto z_i\,.
\end{align}
Our interest in companion matrices is mainly due to a result known as \nameref{thm:stickelberger}, which states that 
\begin{center}
	\textit{the variety $\Vcal(\Ical)$ is precisely the set of simultaneous eigenvalues of the companion matrices.}
\end{center}

We have assumed that the ideal $\Ical$ is radical. A consequence of this assumption is that the companion matrices are simultaneously diagonalisable: $T_i = SD_iS^{-1}$, where $D_i$ are diagonal matrices. Any rational function evaluated on the companion matrices is similar to that function evaluated on the diagonalised companion matrices: $r(\bm{z})|_{z_i\to T_i}\equiv r(\bm{T})= Sr(\bm{D})S^{-1}$. From Stickelberger's theorem, we can then sum the rational function $r$ over all elements of the variety by simply taking the trace:
\begin{align}\label{eq:POS_com-mat-trace}
	\Ical_*r = \sum_{\bm{\xi}\in\Vcal(\Ical)}r(\bm{\xi})=\tr{r(\bm{D})}=\tr{r(\bm{T})}\,.
\end{align} 
Applying this to the rational function \eqref{eq:POS_pf-function} then yields the desired result. Applications of companion matrices to scattering amplitudes have been studied in \cite{Huang:2015yka} (see also \cite{Cardona:2015eba, Cardona:2015ouc}).

\subsubsection{Push Forward via Derivatives of Companion Matrices}

We have shown how we can calculate push forwards of differential forms by summing the rational function \eqref{eq:POS_pf-function} over the variety $\Vcal(\Ical)$. In the previous section we showed how to do this by evaluating \eqref{eq:POS_pf-function} on the companion matrices. Since the companion matrices are $d\times d$ matrices, and $d$ can generally be very large ($(n-3)!$ for the scattering equations), and given the complicated structure of \eqref{eq:POS_pf-function} arising from the $p\times p$ minor, this is easier said than done and this can be a computationally intensive task. In this section we explain a method to calculate push forwards which does not resort to summing \eqref{eq:POS_pf-function}, and hence does not run in to this issue. This method requires knowledge of the \emph{derivatives of companion matrices}. 

We recall from equation \eqref{eq:POS_push-forward-dxi} that we can calculate the push forward as
\begin{align}
	\Ical_*\omega=\sum_{J\in\binom{[m]}{p}}\sum_{I\in\binom{[n]}{p}}\left[ \sum_{\bm{\xi}\in\Vcal(\Ical)}\omega_I(\bm{\xi})\left| \frac{\partial \bm{\xi}}{\partial\bm{a}} \right|^I_J \right]\dd\bm{a}^J\,.
\end{align}
Remarkably, we can use companion matrices in this equation directly by simply replacing $\bm{\xi}\to\bm{T}$:
\begin{align}\label{eq:POS_comp-mat-der}
	\sum_{\bm{\xi}\in\Vcal(\Ical)}\omega_I(\bm{\xi})\left| \frac{\partial \bm{\xi}}{\partial\bm{a}} \right|^I_J = \tr{\omega_I(\bm{T})\sum_{\sigma\in S_p}\sgn(\sigma)\frac{\partial T_{i_{\sigma(1)}}}{\partial a_{j_1}}\cdots \frac{\partial T_{i_{\sigma(p)}}}{\partial a_{j_p}}}\,.
\end{align}
This statement is non-trivial, since, unlike the companion matrices, the derivatives $\partial T_i/\partial a_j$ generally don't commute, and hence the matrix in the trace is not similar to the same expression with $T_i\to D_i$. A proof of equation \eqref{eq:POS_comp-mat-der} is given in appendix \ref{sec:APP_compt-mat-der}.

Once the companion matrices $T_i(\bm{a})$ and their derivatives $\partial T_i/\partial a_j$ are calculated, it is often more efficient to evaluate \eqref{eq:POS_comp-mat-der} than $\tr{\overline{\omega}_J(\bm{T})}$, which gives some computation merit to this method over the previous one. However, as argued above, it can be computationally beneficial to use \Grob basis techniques when the $\bm{a}$-variables are evaluated on some finite field. It is not clear how to find $\partial T/\partial a$ without knowing the explicit $\bm{a}$-dependent $T$ matrices first. In appendix \ref{sec:APP_comp-mat-der-numeric} we include an algorithm which sidesteps this issue and allows to find $\partial T/\partial a$ evaluated on numerical values of $\bm{a}$ without the need to know $T(\bm{a})$.

\subsubsection{Push Forward via Global Residues}
One of the downsides of using companion matrices is that they quickly grow in size. For example, for the scattering equations they are $(n-3)!\times (n-3)!$ matrices. We now discuss a way to calculate push forwards using global residues and a special \emph{dual basis} for the quotient ring $Q=\Cbb(\bm{a})[\bm{z}]/\Ical$. These calculations are not dependent on companion matrices. Instead, as explained in appendix \ref{sec:APP_decompose-unity}, the method we present below relies on a \emph{Bezoutian matrix}, which is only $(n-3)\times(n-3)$ in size.

\paragraph{Global Residues.}
We start by considering some arbitrary rational function $r$ which we want to sum over all $\bm{\xi}\in\Vcal(\Ical)$. We uplift $r$ to a rational differential form with poles on $\Vcal(\Ical)$ as:
\begin{align}
	\Omega[r]\coloneqq r(\bm{z})\frac{\dd z_1\wedge\cdots\wedge\dd z_n}{f_1(\bm{z})\cdots f_n(\bm{z})}\,.
\end{align}
The \emph{local residue} of $\Omega[r]$ at some $\bm\xi\in\Vcal(\Ical)$ is defined through a multi-dimensional contour integral
\begin{align}
	\Res_{\bm{\xi}}\Omega[r]=\frac{1}{(2\pi i)^n}\oint_{\Gamma_{\bm{\xi}}}\Omega[r]\,,
\end{align}
where $\Gamma_{\bm{\xi}}$ is a some contour surrounding $\bm{\xi}$. The precise definition of $\Gamma_{\bm{\xi}}$ is not important for the discussion at hand, an explicit definition can be found in \cite{Sogaard:2015dba}. Using the multi-dimensional Cauchy theorem we find that
\begin{align}
	\Res_{\bm{\xi}}\Omega[r] = r(\bm{\xi})\left|\frac{\partial\bm{f}}{\partial\bm{z}}(\bm{\xi})\right|^{-1}\,.
\end{align}
Next, we define the \emph{global residue} of $r$ as
\begin{align}\label{eq:POS_global-residue-def}
	\Res(r)\coloneqq \sum_{\bm{\xi}\in\Vcal(\Ical)} \Res_{\bm{\xi}}\Omega[r]\,,
\end{align}
such that the push forward of $r$ can be written as the global residue
\begin{align}
	\Ical_*r=\Res\left( \left|\frac{\partial\bm{f}}{\partial\bm{z}}\right| r \right)\,.
\end{align}
We can calculate this global residue by making use of the \nameref{thm:global-duality}. We will first restrict to a polynomial function $p$, and we later explain how to deal with rational functions.

\paragraph{The Dual Basis.}
We define the following symmetric inner product on $Q$:
\begin{align}\label{eq:POS_Q-Q-inner-product}
	\<\bullet,\bullet\>\colon Q\times Q&\to \Cbb(\bm{a})\\\nonumber
	(p_1,p_2)&\mapsto \<p_1,p_2\> = \Res(p_1 p_2)\,.
\end{align}
The \nameref{thm:global-duality} tells us that that this is a non-degenerate inner product, which implies that there must exist a \emph{dual basis} $\Bcal^\vee=\{\Delta_\alpha\}_{\alpha=1}^d$ which satisfies
\begin{align}
	\<e_\alpha,\Delta_\beta\>=\delta_{\alpha\beta}\,.
\end{align}
Since $1\in Q$, we can \emph{decompose unity in the dual basis}:
\begin{align}
	1=\sum_{\alpha=1}^d \mu_\alpha \Delta_\alpha\,,
\end{align}
where $\mu_\alpha\in \Cbb(\bm{a})$ are the \emph{components of unity} in the dual basis $\Bcal^\vee$. We further decompose the remainder of the polynomial $|\partial\bm{f}/\partial\bm{z}|p$ in the standard basis as
\begin{align}
	\overbar{\left(\left|\frac{\partial\bm{f}}{\partial\bm{z}}\right|p\right)}^\Gcal = \sum_{\alpha=1}^d \left(\left|\frac{\partial\bm{f}}{\partial\bm{z}}\right|p\right)_\alpha e_\alpha\,,
\end{align}
which allows us to calculate the global residue as
\begin{align}
	\Ical_* p &= \Res(\left|\frac{\partial\bm{f}}{\partial\bm{z}}\right|p) = \left\<\left|\frac{\partial\bm{f}}{\partial\bm{z}}\right|p,1\right\>=\left\<\overbar{\left(\left|\frac{\partial\bm{f}}{\partial\bm{z}}\right|p\right)}^\Gcal,1\right\>\\\nonumber
	&=\left\<\sum_{\alpha=1}^d \left(\left|\frac{\partial\bm{f}}{\partial\bm{z}}\right|p\right)_\alpha e_\alpha,\sum_{\beta=1}^d \mu_\beta \Delta_\beta\right\> = \sum_{\alpha=1}^d \left(\left|\frac{\partial\bm{f}}{\partial\bm{z}}\right|p\right)_\alpha \mu_\alpha\,.
\end{align}
We have thus reduced the problem of calculating the push forward of $p$ to finding the universal coefficients of unity in the dual basis, and finding the coefficients of $|\partial\bm{f}/\partial\bm{z}|p$ in the standard basis. The latter follows directly from the division algorithm with respect to $\Gcal$, whereas an algorithm for the former is presented in appendix \ref{sec:APP_decompose-unity}. 

Naively, the arguments presented above only work for polynomial functions. However, it can be applied more generally to \emph{rational} functions $r(\bm{z})=p(\bm{z})/q(\bm{z})$, $p,q\in\Cbb(\bm{a})[\bm{z}]$ by finding a \emph{polynomial inverse} $q_{\text{inv}}\in\Cbb(\bm{a})[\bm{z}]$ of $q$. This polynomial inverse satisfies
\begin{align}
	\frac{1}{q(\bm\xi)} = q_{\text{inv}}(\bm\xi)\quad\forall \bm\xi\in\Vcal(\Ical)\,.
\end{align}
An algorithm to find polynomial inverses is presented in appendix \ref{sec:APP_poly-inverse}. Once we know the polynomial inverse $q_{\text{inv}}$, the push forward of $r$ can be computed as
\begin{align}\label{eq:POS_push-forward-GR}
	\Ical_*r = \Res\left(\left|\frac{\partial\bm{f}}{\partial\bm{z}}\right|p q_{\text{inv}}\right) = \sum_{\alpha=1}^d \left(\left|\frac{\partial\bm{f}}{\partial\bm{z}}\right|p q_{\text{inv}}\right)_\alpha \mu_\alpha\,.
\end{align}
Lastly, we point out that some simplifications occur when considering top-forms, which is relevant in the cases related to positive geometries. The rational function $\overline{\omega}_J(\bm{z})$ then simplify to
\begin{align}
	\overline\omega_J(\bm{z}) = (-1)^n \underline{\omega}(\bm{z})\left|\frac{\partial\bm{f}}{\partial\bm{z}}\right|\left|\frac{\partial \bm{f}}{\partial \bm{a}}\right|_J\,,
\end{align}
and we notice that the Jacobian factor $|\partial\bm{f}/\partial\bm{z}|$ cancels with the one in \eqref{eq:POS_push-forward-GR}. This means that the only denominator factors come from the canonical function $\underline\omega$, and when considering a polynomial top-form no polynomial inverses need to be calculated at all!

\section{Summary}

The field of positive geometries is an important recent contribution to the study of scattering amplitudes and QFT as a whole. The entirety of this thesis is motivated by ideas based on positive geometry. In this chapter we have defined what positive geometries are, and we have studied some of their important properties, including \emph{triangulations} and \emph{push forwards}. Due to the importance of these ideas for the entirety of this thesis, numerous examples have been included which have hopefully been helpful to build some intuition for these concepts. We have given a short introduction to the positive geometry of simple polytopes, which will nonetheless play an important role in the following chapter, where it serves as a guiding example which leads to the discovery of new formulae for loop integrands. 

We have given a description of the \emph{ABHY associahedron}, which captures the structure of tree-level scattering amplitudes in $\tr{\phi^3}$. Next, we defined the \emph{amplituhedron}, which describes Wilson loops (T-dual to scattering amplitudes) in \nf. We have given a description of the amplituhedron as the image of a positive linear map from $G_+(K,n)$ to some auxiliary Grassmannian space $G(K,K+m)$, and we have seen how to extract scattering amplitudes from its canonical form. Additionally, we have given a topological description of the amplituhedron directly in the space of momentum twistors. We further saw that we can extend the amplituhedron to include loop integrands if we generalise the positive Grassmannian to the positive loop Grassmannian. Very similar ideas have been applied also for the \emph{momentum amplituhedron}, which encodes tree-level scattering amplitudes in \nf in spinor-helicity space, and admits both an `image through a linear map from the positive Grassmannian' and a topological definition. We have further given an explicit map from the momentum amplituhedron $\Mcal_{n,k}$ into the amplituhedron $\Acal_{n,K}$. Additionally, we have seen that the loop extension of the amplituhedron can be used to define a loop version of the momentum amplituhedron. The last positive geometry which we introduced in this chapter is the \emph{ABJM momentum amplituhedron}, which describes tree-level amplitudes in supersymmetry reduced ABJM theory. The construction is similar to that of the momentum amplituhedron, except we restrict to three-dimensional spinor-helicity variables and the positive orthogonal Grassmannian instead. We further made some comments on the extension of the ABJM momentum amplituhedron to loop level, although in the next chapter we will discuss a new formalism which is more well-suited for this purpose.

The boundaries of the ABHY associahedron, the momentum amplituhedron, and the ABJM momentum amplituhedron, capture the singularity structure of tree-level scattering amplitudes in $\tr{\phi^3}$, \nf, and ABJM theory, respectively. We have given a detailed account of these boundaries, including a full classification of the \emph{covering relations} (the codimension-1 boundaries). In particular, we saw that the boundaries of the momentum amplituhedron are labelled by Grassmannian forests, whereas the boundaries of the ABJM momentum amplituhedron are labelled by orthogonal Grassmannian forests. We have provided a proof that the ABJM momentum amplituhedron has Euler character $\chi=1$. This statement has been proven for the momentum amplituhedron in \cite{Moerman:2021cjg}. The classification of the boundaries of these positive geometries is equivalent to a full characterisation of the singularities of the respective scattering amplitudes. We further uncovered a surprising connection between the boundaries of the ABJM momentum amplituhedron and the boundaries of the ABHY associahedron, which shows that the singularities of ABJM amplitudes form a sub-poset of the singularities of $\tr{\phi^3}$ amplitudes.

Next, we have given a detailed account of the idea that positive geometries can be obtained from the positive moduli space by taking push forwards through the scattering equations. We have seen how this correctly reproduces the CHY formalism or twistor string formulae, which ensures that we obtain the correct scattering amplitudes. Furthermore, we have given three easily implementable algorithms to calculate push forwards of general differential forms using tools from computational algebraic geometry. This provides valuable tools to calculate the push forward through the scattering equations, as this would otherwise necessitate one to find algebraic solutions to the scattering equations, which do not exist in general.

%% file: chapters/dualspace.tex
\chapter{Positive Geometries in Dual Space}\label{sec:DUAL}

In the previous chapter, we have encountered positive geometries which describe scattering amplitudes in various kinematic spaces. The ABHY associahedron $\Ascr_n$ lives in the space of planar Mandelstam variables, the amplituhedron $\Acal_{n,K}$ lives in the space of momentum twistors, and the (ABJM) momentum amplituhedron $\Mcal_{n,k}$ ($\Ocal_k$) lives in the space of (three-dimensional) spinor-helicity variables. Apart from the ABHY associahedron, these positive geometries also have a suitable extension to include loop integrands. The definition of these loop (momentum) amplituhedra rely on a non-trivial notion of positivity between a matrix $C$ in the positive Grassmannian and some subset of matrices $D$ in $G(2,n)$, as was explained in sections \ref{sec:POS_loop-amp}, \ref{sec:POS_loop-mom-amp}, and \ref{sec:POS_loop-ABJM-mom-amp}. Alternatively, to avoid the need to explicitly use these positive loop Grassmannians, one can construct the loop geometries directly in kinematic space from their sign-flip definitions. These positive loop Grassmannians and the loop (momentum) amplituhedra are not nearly as well studied as their non-loop counterparts. This is partly because their structure is a lot more complicated, and although their definitions are very straight forward, it is far from trivial to use them to calculate loop integrands. Furthermore, the kinematic spaces where these positive geometries live make potential generalisations to different theories opaque. This is especially true for momentum twistor space, which is only well-defined for massless, planar theories in four dimensions. Some of these restrictions are lifted by going to spinor helicity space, although the number of spacetime dimensions (and to some degree the masslessness) are still restricted.

In this chapter we will encounter a novel class of positive geometries which describe loop integrands in planar \nf and ABJM. The ambient space of these positive geometries is \emph{dual space}, which we introduced in section \ref{sec:KIN_dual}. The definition of these positive geometries is remarkably simple, and relies only on the notion of \emph{lightcones} (or, more generally, \emph{null-cones}) in dual space. As such, there is no need to consider a map from some positive loop Grassmannian, nor any sign-flip conditions\footnote{In practice, this construction is equivalent to the loop (momentum) amplituhedron, and the sign-flip conditions are therefore still satisfied. Although they are not necessary for the definition, we still find it useful to resort to sign-flips for certain calculations.}. Furthermore, we recall that to introduce momentum twistors in section \ref{sec:KIN_mom-twistor}, we started from spinor-helicity variables and needed to go through dual space. This shows that dual space is situated `in between' spinor-helicity space and momentum twistor spaces. This makes it particularly easy to translate any expression we obtain from our dual space geometries to either the amplituhedron or the momentum amplituhedron. Although these null-cone geometries in dual space are perfectly capable of describing higher loop integrands, in this chapter we will mainly focus on one-loop integrands. The extension to higher loops is currently a work in progress.

We note that this framework still relies on the definition of a tree-level (momentum) amplituhedron to serve as a `seed' for the loop-level geometry. We will see in section \ref{sec:DUAL_chambers} that we can triangulate the tree-level geometry in terms of \emph{chambers}, and we associate a loop-level geometry to each of these chambers. Apart from this, the framework is remarkably general, and generalises with minimal footnotes between ABJM theory and \nf. Dual space is not restricted to any number of spacetime dimensions or masslessness, and we expect that this construction will give a natural loop-level generalisation for any potential tree-level (momentum) amplituhedra which may be discovered in the future. The definition of dual momenta is inherently reliant on a notion of planarity, however recent progress in defining loop integrands for non-planar theories \cite{Arkani-Hamed:2023lbd} might offer a way forward. 

\section{Chambers}\label{sec:DUAL_chambers}

The idea of chambers was first introduced in \cite{He:2023rou}. The chambers provide a specific triangulation of tree-level amplituhedra or momentum amplituhedra (for both \nf and ABJM) based on the geometry of the \emph{loop fibre}. To illustrate the notion of a chamber triangulation, we will restrict our attention to the amplituhedron for now. The main ideas and definitions seamlessly generalise to the other (momentum) amplituhedra.

\subsubsection{Chambers From Loop Geometry}
We recall from section \ref{sec:POS_amplituhedron} that a point in the $L$-loop amplituhedron is specified by a $z\in G(4,n)$ in the tree-level amplituhedron together with $L$ two-planes $(AB)_i$ inside $z$. We define a projection map $\pi$ which projects out the tree-level part:
\begin{align}
	\pi \colon \Acal_{n,K}^{(L)} \to \Acal_{n,K}, \quad(z,(AB)_1,\ldots, (AB)_L)\mapsto z\,.
\end{align} 
Said differently, we interpret $\Acal_{n,K}^{(L)}$ as the total space of a fibre bundle with base space $\Acal_{n,K}$ and bundle projection map $\pi$. The \emph{fibre} $\pi^{-1} z$ of a point $z\in\Acal_{n,K}$ is the space of all two-planes $(AB)_1,\ldots, (AB)_L$ which satisfy 
\begin{align}
	&\<(AB)_l i i+1\>>0\,,\quad \<(AB)_i(AB)_j\>>0\,,\\
	&\{\<AB12\>,\ldots,\<AB1n\>\}\quad \text{ has $K+2$ sign flips.}
\end{align}
It is clear that these fibres encode the loop-level structure of the amplituhedron, and this thus provides an interpretation of the loop geometry as a \emph{fibration} over the tree geometry.

A \emph{chamber} $\cfrak$ of the tree-amplituhedron is defined as a top-dimensional region of the amplituhedron where the fibres are \emph{combinatorially equivalent}:
\begin{align}
	\cfrak\subseteq \Acal_{n,K}\colon \forall z_1, z_2 \in \cfrak,\; \pi^{-1}(z_1) \simeq \pi^{-1}(z_2)\,.
\end{align}
We will not give a mathematical definition of combinatorial equivalence, however the essential property is that combinatorially equivalent positive geometries have the same canonical form:
\begin{align}
	\Omega(\pi^{-1}(z_1))=\Omega(\pi^{-1}(z_2))\equiv \Omega^{L-\text{loop}}(\cfrak)\,.
\end{align}
Note that, in a literal sense, the preceding formula is not correct. The canonical form of a fibre will depend on the specific point $z$ in the base space, and when evaluating the canonical forms on numeric values for $z_1$ and $z_2$ their canonical forms will not agree. However, the \emph{functional dependence} of the canonical form on $z$ is identical, so when considering the $z$ dependence of $\Omega(\pi^{-1}(z))$ in a formal sense, then the canonical forms are equivalent.

This suggests a natural triangulation of the tree-amplituhedron into a (finite) set of chambers:
\begin{align}\label{eq:DUAL_chambers-tree-amp}
	\Acal_{n,K}=\bigcup_{\cfrak\in \Cfrak(\Acal_{n,K})}\cfrak\implies \Omega(\Acal_{n,K})=\sum_{\cfrak\in\Cfrak(\Acal_{n,K})}\Omega(\cfrak)\,,
\end{align}
where $\Cfrak(\Acal_{n,K})$ denotes the set of all chambers of $\Acal_{n,K}$. For extra clarity, we will generally refer to $\Omega(\cfrak)$, the canonical form of the chamber, as $\Omega^{\text{tree}}(\cfrak)$.

The full $L$-loop amplituhedron is triangulated by sets $(\cfrak,\pi^{-1}(\cfrak))$. Since, by definition, the canonical form of all fibres in a chamber are the same, we have
\begin{align}
	\Omega\big((\cfrak,\pi^{-1}(\cfrak))\big) = \Omega^{\text{tree}}(\cfrak)\wedge\Omega^{L-\text{loop}}(\cfrak)\,.
\end{align}
From this follows the main result of this discussion, which is that we can write the canonical form of the $L$-loop amplituhedron as
\begin{align}\label{eq:DUAL_chamber-expansion-loop-amp}
	\Omega(\Acal_{n,K}^{(L)})=\sum_{\cfrak\in\Cfrak(\Acal_{n,K})} \Omega^{\text{tree}}(\cfrak)\wedge\Omega^{L-\text{loop}}(\cfrak)\,.
\end{align}

\subsubsection{Chambers From Intersections of Tiles}
\begin{figure}
	\centering
	\includegraphics[width=\textwidth]{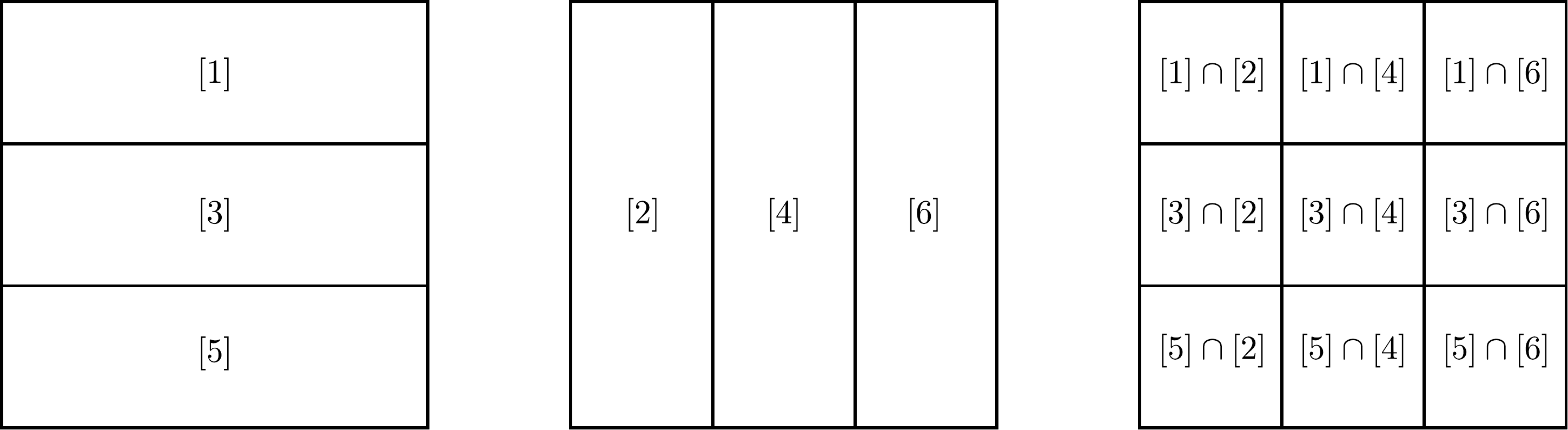}
	\caption{A schematic depiction which shows that $\Acal_{6,1}$ can be triangulated by three BCFW tiles in two different ways, or by the nine chambers.}
	\label{fig:amp_6-1-triangulation}
\end{figure}

There is another sense in which the notion of `chambers' of the amplituhedron makes an appearance in the literature. We recall that the amplituhedron can be triangulated using \emph{BCFW tiles}, which are images of $4K$-dimensional positroid cells associated to the (T-dual of) the BCFW cells considered in section \ref{sec:AMP_nf}. There are generally many different BCFW tilings of $\Acal_{n,K}$, which means that some set of tiles must intersect. This then leads to a new triangulation of the amplituhedron, where each triangle is the \emph{maximal intersection} of BCFW tiles. As an example, we recall that the amplituhedron $\Acal_{6,1}$ has a triangulation in terms of BCFW tiles as
\begin{align}
	\Acal_{6,1}=[1]\cup[3]\cup[5]=[2]\cup[4]\cup[6]\,,
\end{align}
where $[i]$ denotes the tile that is the image of the positroid cell $C\in G_+(1,6)$ which has a zero in the $i$\textsuperscript{th} column. The maximal intersections of these tiles indicate a triangulation of the form
\begin{alignat}{4}
	\Acal_{6,1}&=&&[1]\cap[2]\quad&&\cup[1]\cap[4]\quad&&\cup[1]\cap[6]\\\nonumber
	&\cup&&[3]\cap[2]\quad&&\cup[3]\cap[4]\quad&&\cup[3]\cap[6]\\\nonumber
	&\cup&&[5]\cap[2]\quad&&\cup[5]\cap[4]\quad&&\cup[5]\cap[6]\,,
\end{alignat}
which is illustrated schematically in figure \ref{fig:amp_6-1-triangulation}. 
These maximal intersection of BCFW tiles $[i]\cap[j]$ are also called `chambers' of the amplituhedron $\Acal_{6,1}$. This might seem like a confusing abuse of terminology, but it turns out that the two notions of chambers are related. We will see that in section \ref{sec:DUAL_nf-chambers-integrand} that the one-loop chambers of the amplituhedron are given by maximal intersection of cells corresponding to leading singularities, rather than BCFW cells. This interpretation of chambers is not exclusive to the $m=4$ amplituhedron (whereas the `loop fibre' definition of chambers is), and has been studied in detail for the case of $m=2$ in \cite{Parisi:2021oql}. As another example, we return to the case $\Acal_{4,1,2}$ which we encountered in section \ref{sec:POS_amplituhedron}. We saw that $\Acal_{4,1,2}$ is triangulated by the tiles $[1],[3]$, or $[2],[4]$. As can clearly be seen from figure \ref{fig:m2-amplituhedron-example}, some of these tiles intersect, giving us a triangulation in terms of the chambers $[1]\cap [2], [1]\cap [4], [3]\cap[2]$, and $[3]\cap[4]$.

Lastly, we note that the tilings of the amplituhedron are \emph{T-dual} to tilings of the momentum amplituhedron. There is a way in which we can interpret the set of triangulations as being complementary to the set of all chambers, a statement which is made more precise in appendix \ref{sec:APP_chambers}. A consequence of this statement is that also the \emph{chambers} of the amplituhedron and the momentum amplituhedron are T-dual. Hence, the classification of chambers of $\Acal_{n,1}$ is equivalent to the chambers of $\Mcal_{n,3}$. This statement is non-trivial beyond NHMV, however we expect it to hold for any helicity sector. In addition, since the loop part of the momentum amplituhedron introduced in section \ref{sec:POS_loop-mom-amp} is equivalent to the loop part of the amplituhedron, the geometry and canonical forms of the loop fibres are the same between the amplituhedron and momentum amplituhedron, they are just written in different variables! This means that the chamber expansion for $\Omega(\Mcal_{n,K+2}^{(L)})$ is identical to \eqref{eq:DUAL_chamber-expansion-loop-amp}, with the only difference being that the $\Omega^{\text{tree}}(\cfrak)$ takes a different form, the chambers in the sum and $\Omega^{L-\text{loop}}(\cfrak)$ being equivalent.

\section[Null-Cone Geometries]{Null-Cone Geometries in Dual Space}\label{sec:DUAL_nullcone-geometry}

We now turn to the study of loop fibres and their geometries. We will do this in \emph{dual space} (see section \ref{sec:KIN_dual}), which is situated in-between momentum twistor space and spinor helicity space. We will take the momentum amplituhedron as our starting point, because, at least conceptually, it is easier to generalise to a different number of spacetime dimensions. Although certain calculations might be easier from the amplituhedron point of view, due to the duality between chambers and loop fibres of the amplituhedron and momentum amplituhedron this does not end up being an issue, and we have the freedom to translate back and forth between the two languages as we see fit. We further note that we currently don't restrict ourselves to the \nf momentum amplituhedron, as the story presented below is equally applicable to ABJM theory. For this reason, we will kick off this discussion for a general number of spacetime dimensions and in a theory agnostic setting. We will restrict our attention to ABJM and \nf in sections \ref{sec:DUAL_ABJM} and \ref{sec:DUAL_nf}, respectively. 

Assume we have some momentum amplituhedron which encodes massless\footnote{The restriction to masslessness can be loosened, as we will see for a toy example in section \ref{sec:DUAL_toy}. However, for ABJM and \nf we will mainly be interested in massless theories.} tree-level amplitudes of some theory in some $d$-dimensional spacetime $\Dbb^d$ (for $d=3$ we take $\Dbb^3=\Rbb^{1,2}$, and for $d=4$ we take $\Dbb^4=\Rbb^{2,2}$). A point in the momentum amplituhedron specifies for us an ordered set of $n$ momentum vectors $p_1^\mu,\ldots,p_n^\mu$, which we can translate into dual space:
\begin{align}
	x_i^\mu = \sum_{j=1}^{i-1}p_i^\mu\,,
\end{align}
such that $p_i^\mu=x_{i+1}^\mu-x_i^\mu$. The masslessness condition $p_i^2=0$ then implies that $X_{ii+1}\coloneqq (x_i-x_{i+1})^2=0$. The points $x_1^\mu,\ldots,x_n^\mu$ thus define a polygon in dual space whose edges are all segments of null-rays, \textit{i.e.} a \emph{null-polygon}. Momentum amplituhedra typically generate momenta with positive planar Mandelstam variables, which in our language means that $X_{ij}\coloneqq (x_i-x_j)^2>0$. We define the null-cone of a point $x$ as
\begin{align}
	\Ncal_x\coloneqq\{y\in\Dbb^d\colon (y-x)^2=0\}\,.
\end{align}
The null-cone divides $\Dbb^d$ into a positively-separated and a negatively-separated region (analogous to the familiar space-like and time-like separation in Minkowski signature)
\begin{align}
	\Ncal^+_x &\coloneqq \{y\in\Dbb^d\colon (y-x)^2>0\}\,,\\
	\Ncal^-_x &\coloneqq \{y\in\Dbb^d\colon (y-x)^2<0\}\,,
\end{align}
and we similarly define $\Ncal^{\geq0}_x$ and $\Ncal^{\leq0}_x$. 

We now ask the following question \emph{what is the region of $\Dbb^d$ where all points are positively separated from all $x_i$?} We denote this region $\Kcal(\bm{x})$\footnote{We use the bold-face $\bm{x}$ to emphasise that we are talking about the collection of all $x$'s: $\bm{x}=(x_1,\ldots,x_n)$. }:
\begin{align}
	\Kcal(\bm{x})\coloneqq \Ncal^{\geq0}_{x_1}\cap\cdots\cap\Ncal^{\geq0}_{x_n}\,,
\end{align}
which is a top-dimensional region of $\Dbb^d$. Surprisingly, in all cases we studied (\textit{i.e.} for \nf and ABJM), $\Kcal(\bm{x})$ naturally splits up into a \emph{compact} region $\Delta(\bm{x})$ `inside' the null-polygon, and a \emph{non-compact} region $\overline{\Delta}(\bm{x})$ `outside' the null-polygon:
\begin{align}
	\Kcal(\bm{x})=\Delta(\bm{x})\cup\overline{\Delta}(\bm{x})\,.
\end{align}
We claim that the compact part $\Delta(\bm{x})$ is exactly the one-loop fibre geometry of the point in the momentum amplituhedron we started with:
\begin{align}
	\Omega^{1-\text{loop}}(\cfrak) = \Omega(\Delta(\bm{x}))\,.
\end{align}
This is a non-trivial statement, and we will motivate it with many examples for ABJM and \nf in the sections to follow.

Our claim also extends to higher loops. We consider the configuration of $L$ mutually positively separated points inside $\Delta(\bm{x})$:
\begin{align}
	\Delta^{(L)}(\bm{x})\coloneqq \{(y_1,\ldots,y_L)\in \big(\Delta(\bm{x})\big)^L\colon (y_i-y_j)^2\geq 0\,,\quad\forall i,j=1,\ldots,L\}\,.
\end{align}
Then, $\Delta^{(L)}(\bm{x})$ is precisely the $L$-loop fibre geometry for this point:
\begin{align}
	\Omega^{L-\text{loop}}(\cfrak) = \Omega(\Delta^{(L)}(\bm{x}))\,.
\end{align}

\subsubsection{General Properties of $\Delta(\bm{x})$}

Let us remark some general properties of the geometry $\Delta(\bm{x})$, stemming directly from the geometry of null-cones and from basic physical considerations. Firstly, we note that translational invariance in dual space implies that the canonical forms can only be dependent on Lorentz invariants made up from the \emph{differences} of points in dual space. This means that we will encounter expression of the form $(a-b)^2$, but never just $a^2$ or $a\cdot b$. For the moment, we will no longer assume $X_{ii+1}=0$, and instead consider a collection of generic points in $\Dbb^d$. In addition to the squared distance between points $(x_i-x_j)^2$, there are some other Lorentz invariants which make a recurring appearance. We define the three `epsilon invariants' in $\Dbb^d$:
\begin{align}
	\epsilon(1,2,\ldots,d)&\coloneqq \begin{vmatrix}
		x_1 & x_2 & \cdots & x_d
	\end{vmatrix}\,,\\
	\epsilon(1,2,\ldots,d,d+1)&\coloneqq \begin{vmatrix}
		1 & 1 & \cdots & 1 & 1 \\
		x_1 & x_2 & \cdots & x_d & x_{d+1}
	\end{vmatrix}\,,\\
	\epsilon(1,2,\ldots,d,d+1,d+2)&\coloneqq \begin{vmatrix}
		1 & 1 & \cdots & 1 & 1 \\
		-x_1^2 & -x_2^2 & \cdots & -x_{d+1}^2 & -x_{d+2}^2\\
		x_1 & x_2 & \cdots & x_{d+1} & x_{d+2}
	\end{vmatrix}\,.
\end{align}
The first of these is the standard epsilon Lorentz invariant in $d$ spacetime dimensions, which is not invariant under translations and will therefore not explicitly appear in our expressions. As we saw in section \ref{sec:GRASS_proj}, the second epsilon invariant can be interpreted as the volume of a simplex with vertices $x_1,\ldots,x_{d+1}$, which is obviously a translation invariant statement. We can expand the determinant with respect to the first row to find
\begin{align}
	\epsilon(1,2,\ldots,d,d+1) = \epsilon(2,3,\ldots,d+1)-\epsilon(1,3,\ldots,d,d+1)+\ldots+ (-1)^{d+1} \epsilon(1,2,\ldots,d)\,.
\end{align}
The third and last epsilon invariant is the determinant of points in \emph{embedding space} (see section \ref{sec:KIN_embedding}), and plays an important role in theories with dual conformal invariance.

Any $d$ distinct points $x_1,\ldots,x_d$ define a hyperplane on which they lie, which we denote  as
\begin{align}
	H^0(x_1,\ldots,x_d)\coloneqq \{y\in\Dbb^d \colon \epsilon (1,\ldots,d,y)=0\}\,.
\end{align}
This hyperplane divides $\Dbb^d$ into two regions, distinguished by the sign of $\epsilon (1,\ldots,d,y)$, which we shall denote $H^\pm(x_1,\ldots,x_d)$. Furthermore, the null-cones of these $d$ points generally intersect in 2 points:
\begin{align}
	\{q^+_{12\ldots d},q^-_{12\ldots d}\} = \Ncal_{x_1}\cap\cdots\cap\Ncal_{x_d}\,,
\end{align}
defined such that 
\begin{align}
	q^\pm_{12\cdots d} \in H^\pm(x_1,\ldots,x_d)\,.
\end{align}
We give an explicit formula for these maximal intersections of null-cones in appendix \ref{sec:APP_schubert}.

We can interpret $\Delta(\bm{x})$ as being `cut out' by the null-cones $\Ncal_{x_i}$, and the codimension-1 boundaries of $\Delta$ are thus given by $(y-x_i)^2=0$. The vertices of $\Delta$ consists of the all $x_i$, together with some set of $q^\pm$. The local geometry close to a vertex $q^\pm_{a_1\cdots a_d}$ is as follows: we have $d$ facets meeting at this point, given by $(y-x_{a_1})^2=0,\ldots,(y-x_{a_d})^2=0$, and there are also $d$ edges meeting at this vertex, defined by the intersection of $d-1$ of these null-cones. This reminds us of the structure of a \emph{simple polytope}, which we discussed in section \ref{sec:POS_simple-polytopes}. It is therefore tempting to write the canonical form $\Omega(\Delta(\bm{x}))$ as a `sum over vertices'. The curved nature of the facets, and the non-simple structure of the vertices $x_i$ makes this less trivial than in the polytopal case. However, as we will see below, this line of thinking actually proves to be quite fruitful.

\subsection{\texorpdfstring{A Toy Example in $\Rbb^{1,1}$}{A Toy Example in R1,1}}\label{sec:DUAL_toy}

\begin{figure}
	\centering
	\includegraphics[scale=1.2]{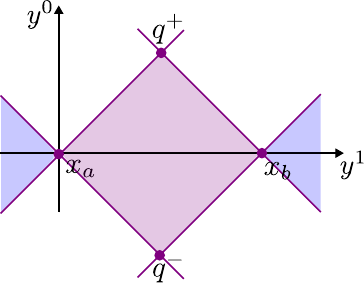}
	\caption{The shaded region represents $\Kcal(x_a,x_b)$. We see that is splits up into a compact part $\Delta$ shaded in purple, and a non-compact part $\overbar\Delta$ shaded in blue.}
	\label{fig:2d-toy-model}
\end{figure}

As a warm-up, let us consider a two-point toy model in $\Rbb^{1,1}$. We assume that we are given two points $x_a,x_b$\footnote{Typically, they would be labelled $x_1,x_2$, however in this example we want to reserve the numeric subscript for loop variables.}, which have the interpretation as the base point in some tree-level momentum amplituhedron over which we want to study the loop fibre. To make this simple toy model slightly less trivial, we will forgo the masslessness condition and instead assume that $X_{ab}=(x_a-x_b)^2>0$. The region $\Kcal(x_a,x_b)$ is defined as the region which is space-like (positively) separated from both $x_a,x_b$, which we have depicted in figure \ref{fig:2d-toy-model}. By inspection, it is clear that $\Kcal(x_a,x_b)$ naturally splits into a `compact' part $\Delta$ and a `non-compact' part $\overbar\Delta$.

It is natural to work in the light-cone coordinates 
\begin{align}
	y_i^\mu=\begin{pmatrix} y_i^0\\y_i^1 \end{pmatrix} \to \begin{pmatrix} \lambda_i \\ \tilde\lambda_i \end{pmatrix} = \begin{pmatrix}	y_i^1+y_i^0\\y_i^1-y_i^0 \end{pmatrix}\,,
\end{align}
and we define
\begin{align}
	\<ij\>=\lambda_i-\lambda_j\,,\quad [ij]=\tilde\lambda_i-\tilde\lambda_j\,.
\end{align}
In these coordinates, then 
\begin{align}\label{eq:DUAL_2d-Xij-factor}
	(y_i-y_j)^2 = \<ij\>[ij]\,.
\end{align}
These brackets are anti-symmetric and can be given the interpretation of a determinant,
\begin{align}
	\<ij\>=\begin{vmatrix}
		1 & 1 \\\lambda_j & \lambda_i
	\end{vmatrix}\,,
\end{align}
which means that they also satisfy the Schouten identity
\begin{align}
	\<ij\>\<kl\>-\<ik\>\<jl\>+\<il\>\<jk\>=0\,.
\end{align}
However, the analogy to spinor-helicity variables can also be misleading. For example, since the polynomials $\<ij\>$ are not homogeneous in particle index, there is no $GL(1)$ `little group' that leaves $(y_1-y_2)^2$ invariant, and there is no `momentum conservation' $\sum_i\lambda_i\tilde\lambda_i\neq0$. Some similarity to spinor-helicity is not at all surprising, as 4D spinor-helicity can be interpreted as a version of `double lightcone coordinates', where two $\Rbb^{1,1}$ subspaces of $\Rbb^{2,2}$ are both rotated into lightcone coordinates (for $\Rbb^{1,3}$ we need to Wick rotate one of the space dimensions into a time dimension, hence the need for complex spinor-helicity variables in this signature).

\subsubsection{One Loop}
To further the analogy to the loop (momentum) amplituhedron, which we will encounter in more detail later in this chapter, we note that we can isolate the compact part $\Delta(x_a,x_b)$ in terms of sign-flip conditions as the set of all $y\in\Kcal(x_a,x_b)$ satisfying
\begin{subequations}
\begin{align}
	&\<ya\>>0\,,\\
	&\{\<ya\>,\<yb\>\}\qquad\text{has $1$ sign-flip.}
\end{align}
\end{subequations}
In terms of light-cone coordinates we can equivalently write
\begin{align}
	\Delta = \{y\in \mathds{R}^{1,1}\colon \lambda_a\leq\lambda\leq\lambda_b, \tilde\lambda_a\leq\tilde\lambda\leq\tilde\lambda_b\}\,.
\end{align}
This makes it explicit that $\Delta$ is just the product of two line segments (\emph{i.e.} a rectangle). The line segments are given by $\lambda_a\leq\lambda\leq\lambda_b$ and $\tilde\lambda_a\leq\tilde\lambda\leq\tilde\lambda_b$, and hence the canonical form is given by
\begin{align}
	\Omega(\Delta) = \dd\log\frac{\lambda-\lambda_a}{\lambda-\lambda_b}\wedge\dd\log\frac{\tilde\lambda-\tilde\lambda_a}{\tilde\lambda-\tilde\lambda_b}=\frac{X_{ab}\dd^2y}{(y-x_a)^2(y-x_b)^2}\,.
\end{align}
We noted above that we are tempted to write the canonical form as a sum over vertices. In this case it is trivially possible, as $\Delta$ is a simple polytope, but to highlight certain aspects we will encounter later, let us proceed under the guise of ignorance. In general dimensions, the $x$ vertices are not simple, whereas the $q^\pm$ vertices are. In this case, the vertex $q^+$ is incident to the two facets $(y-x_a)^2=0$ and $(y-x_b)^2=0$, and hence, based on the discussion on simple polytopes in section \ref{sec:POS_simple-polytopes}, we are tempted to associate it to the form
\begin{align}
	\omega_{ab}=\dd\log(y-x_a)^2\wedge\dd\log (y-x_b)^2=\frac{-2\epsilon(a,b,y)\dd^2 y}{(y-x_a)^2(y-x_b)^2}\,.
\end{align}
However, this form also has a non-zero residue at $q^-$, which goes contrary to the `simple polytope' idea of summing over vertices. This can be remedied by introducing
\begin{align}
	\omega^{\text{bubble}}_{ab} = \pm\dd\log\frac{(y-x_a)^2}{(y-q^\pm)^2}\wedge\dd\log\frac{(y-x_b)^2}{(y-q^\pm)^2}=\frac{X_{ab}\dd^2y}{(y-x_a)^2(y-x_b)^2}\,,
\end{align}
such that the combination
\begin{align}
	\omega^\pm_{ab} &= \frac{\omega^{\text{bubble}}_{ab} \pm \omega_{ab}}{2}\,,
\end{align}
satisfies
\begin{align}
	\Res_{y=q^\pm}\omega^\pm=1\,,\quad \Res_{y=q^\pm}\omega^\mp =0\,.
\end{align}
At this stage, we don't know what to assign to the vertices $x_a$, as it is only incident to a single facet: $(y-x_a)^2=0$ (the fact that this vertex is actually simple in this toy example only comes after the observation that $(y-x_a)^2$ factorises as in equation \eqref{eq:DUAL_2d-Xij-factor}, which doesn't generalise for higher dimensions). Interestingly, we can actually ignore the contribution from $x_a$ and $x_b$ entirely, as summing over the vertices $q^+$ and $q^-$ already gives the correct result:
\begin{align}\label{eq:DUAL_2d-toy-1-loop-sum}
	\Omega(\Delta) = \omega_{ab}^++\omega_{ab}^-=\omega^{\text{bubble}}_{ab}\,.
\end{align}
The fact that this canonical form has a correct residues at $x_a$ and $x_b$ follow from a `composite' residue of $\omega^{\text{bubble}}_{ab}$. Specifically, when taking a residue at $(y-x_a)^2=0$, we pick up a Jacobian factor in the denominator, such that $\omega^\pm_{ab}$ has a residue of $1/2$ at $x_a$. In the current example this can easily be seen if we write the form in light-cone coordinates. Although the sum \eqref{eq:DUAL_2d-toy-1-loop-sum} is not equivalent to the `sum over vertices' we would get from a simple polytope, it is interesting to note that we are led to discover a similar formula where we only sum over the `maximal intersections of light-cones' $q^\pm$.

\subsubsection{All Loops}
Next, we define the $L$-loop geometry as
\begin{align}
	\Delta^{(L)}(x_a,x_b)\coloneqq\{(y_1,\ldots,y_L)\in\Delta(x_a,x_b)^L\colon (y_i-y_j)^2\geq 0 \quad\forall i,j = 1,\ldots,L\}\,.
\end{align}
Clearly, there is a distinct notion of ordering between $L$ points inside $\Delta$, based on their $y^1$ component. This suggests a triangulation of $\Delta^{(L)}$ as
\begin{align}\label{eq:L-loop-triangulation}
	\Delta^{(L)} = \bigcup_{\sigma\in S_L} \Delta^{(L)}_{\sigma(1)\cdots\sigma(L)}\equiv \bigcup_{\sigma\in S_L} \Delta^{(L)}_\sigma\,,
\end{align}
where
\begin{align}
	\Delta^{(L)}_{i_1\cdots i_L}\coloneqq \{(y_1,\ldots,y_L)\in \Delta^{(L)}\colon y_{i_1}^1\leq\ldots\leq y_{i_L}^1\}\,
\end{align}
In terms of light-cone coordinates, this becomes
\begin{align}
	\Delta^{(L)}_{i_1\cdots i_L} = \{(y_1,\ldots,y_L)\in\Delta^L\colon & \lambda_a\leq \lambda_{i_1}\leq \lambda_{i_2}\leq\ldots\leq \lambda_{i_L}\leq \lambda_b, \notag \\ 
	& \tilde\lambda_a\leq \tilde\lambda_{i_1}\leq \tilde\lambda_{i_2}\leq\ldots\leq \tilde\lambda_{i_L}\leq \tilde\lambda_b\}\,.
\end{align}
We see that $\Delta^{(L)}_{i_1\cdots i_L}$ is, much like the one-loop case, the product of two independent geometries in $\lambda$ and $\tilde\lambda$ variables separately. We note that the region $\lambda_a\leq \lambda_{i_1}\leq \lambda_{i_2}\leq\ldots\leq \lambda_{i_L}\leq \lambda_b$ is isomorphic to the positive moduli space $\mathcal{M}^+_{0,L+3}$, after we use $SL(2)$ fix $z_1\to \lambda_a, z_{L+2}\to \lambda_b, z_{L+3}\to\infty$. The canonical form of $\mathcal{M}^+_{0,L+3}$ is simply the world-sheet Parke-Taylor form, and hence we find
\begin{align}
	\Omega[\Delta^{(L)}_{i_1\cdots i_L}] &= \frac{\<ab\> \dd \lambda_{i_1}\wedge\cdots\wedge\dd \lambda_{i_L}}{\<a i_1\>\<i_2 i_3\>\cdots\<i_L b\>}\wedge \frac{[ab] \dd \tilde\lambda_{i_1}\wedge\cdots\wedge\dd \tilde\lambda_{i_L}}{[a i_1][i_1 i_2]\cdots[i_l b]}\\
	&=\frac{X_{ab} \dd^2y_1\wedge\cdots\wedge\dd^2y_L}{(x_a-y_{i_1})^2(y_{i_1}-y_{i_2})^2\cdots(y_{i_L}-x_b)^2}\,.
\end{align}
The full $L$-loop form is then given by
\begin{align}\label{eq:massless-form}
	\Omega[\Delta^{(L)}] = \sum_{\sigma\in S_L} \Omega[\Delta^{(L)}_{\sigma}]\,.
\end{align}

\subsubsection{Boundaries}
Before moving on, let us briefly look at some interesting aspects on the boundary structure of $\Delta^{(L)}$. We have seen that the full $L$-loop geometry is naturally split into $L!$ `loop ordered' regions $\Delta^{(L)}_\sigma$. Each of these loop ordered regions is geometrically an $(L-\text{simplex})^2$. For example, the one loop case consists of one region which is geometrically a 1-simplex (line segment) times a 1-simplex, \emph{i.e.} a rectangle. 

In general, the loop ordering labels topologically distinct regions. If we consider $\Delta_\sigma,\Delta_\sigma'$ for $\sigma,\sigma'\in S_L, \sigma \neq \sigma'$, then their interiors do not intersect, and they only share some boundary of codimension 2 or higher. In particular, all $L!$ loop ordered geometries share a 2-dimensional boundary given by $\lambda_1=\lambda_2=\ldots=\lambda_L,\,\tilde\lambda_1=\tilde\lambda_2=\ldots=\tilde\lambda_L$, ensuring that $\<ij\>=[ij]=0\,\forall i,j$. This two-dimensional boundary is geometrically a one-loop geometry. We note that 
\begin{align}
	\mathop{\Res}_{\lambda_1=\cdots=\lambda_L}\mathop{\Res}_{\tilde\lambda_1=\cdots=\tilde\lambda_L} \Omega(\Delta^{(L)}_\sigma)=\Omega(\Delta)\,,
\end{align}
for all permutations $\sigma$. This means that the entire $L$-loop form satisfies
\begin{align}
	\mathop{\Res}_{\lambda_1=\cdots=\lambda_L}\mathop{\Res}_{\tilde\lambda_1=\cdots=\tilde\lambda_L} \Omega(\Delta^{(L)})=L!\Omega(\Delta)\,.
\end{align}
If we continue taking residues, localising all loop momenta on the same vertex $x_a$ of $\Delta$, then we find a residue of $L!$. This means that this canonical form does not satisfy that all vertices have a residue of $\pm1$, which is a defining property of positive geometries (see section \ref{sec:POS_def})! This is a property which is also present for the amplituhedron at two loops and higher. For this reason, \emph{weighted positive geometries} were introduced in \cite{Dian:2022tpf}. Weighted positive geometries do not have the requirement that the maximal residue needs to be $\pm1$, and hence these loop geometries should be understood in this generalised sense instead. 

We take this moment to point out another well-known property of canonical forms, which is that the order in which we take residues has an influence on the resulting differential form. We note that we can first localise $y_1$ on $x_a$ by taking the residue at $\<a1\>=0$ and $[a1]=0$. If we then subsequently take the residue as $y_2\to x_a,\ldots,y_L\to x_a$, then we end up at the same vertex of $\Delta^{(L)}$ as we considered above. However, the residue at this vertex is now $\pm1$, since only $\Delta_{12\cdots L}^{(L)}$ contributes to this residue. This also shows that the residue of $L!$ we found before is not just a consequence of a bad normalisation.

\section{ABJM}\label{sec:DUAL_ABJM}

We recall from section \ref{sec:POS_ABJM-mom-amp} that a point in the ABJM momentum amplituhedron $\Ocal_k$ is given by a $\lambda\in OG(2,2k)$ which satisfies that
\begin{align}
	\<ii+1\> >0\,,\quad s_{ii+1\cdots j}>0\,,\quad \{\<1i\>\}_{i=2}^{2k}\text{ has $k$ sign-flips,}
\end{align}
where orthogonality is defined with respect to $\eta=\diag(1,-1,1,\ldots,-1)$. From a point $\lambda\in\Ocal_k$, we translate into dual space by defining the matrix
\begin{align}
	x_a^{\alpha\beta}\coloneqq\sum_{b=1}^{a-1}(-1)^b \lambda_b^\alpha\lambda_b^\beta\,,
\end{align}
which we can further interpret as a point in $\Rbb^{1,2}$ as
\begin{align}
	x^{\alpha\beta}=\begin{pmatrix}
		-x^0+x^2 & x^1 \\ x^1 & -x^0-x^2
	\end{pmatrix}\Leftrightarrow x^\mu = \begin{pmatrix}
		-(x^{11}+x^{22})/2 \\ x^{12} \\ (x^{11}-x^{22})/2
	\end{pmatrix}\,.
\end{align}
The points $x_1,\ldots,x_n$ form a null-polygon with $X_{ij}>0$ and the even-indexed points $x_i$ in the future of their odd-indexed neighbours.

We recall from section \ref{sec:POS_loop-ABJM-mom-amp} that we can define a loop extension of the ABJM momentum amplituhedron by reducing the loop momentum amplituhedron from section \ref{sec:POS_loop-mom-amp} to three dimensions. To summarise, we introduce the map
\begin{align}
	\Phi_\lambda\colon G(2,2k)^L&\to M_{2,2}^L\\
	(D_1,\ldots,D_L)&\mapsto (\ell_1,\ldots,\ell_L)\,,
\end{align}
where
\begin{align}
	\ell_i = \frac{\sum_{a<b}(ab)_i \<ab\>\ls_{ab}}{\sum_{a<b}(ab)_i\<ab\>}\,,
\end{align}
$(ab)_i$ denotes the \Pluck variables $p_{ab}(D_i)$, and
\begin{align}
	\ls_{ab}=\frac{1}{\<ab\>}\left( \sum_{c=b+1}^n (-1)^c \lambda_a\<bc\>\lambda_c - \sum_{c=a+1}^n (-1)^c\lambda_b\<ac\>\lambda_c \right)\,.
\end{align}
These $\ls$ are not necessarily symmetric matrices, and we need to restrict the $D$ matrices such that the matrices in \eqref{eq:POS_loop-positivity-mom amp} are all positive, \emph{and} such that the matrices $\ell_i$ are symmetric. Then $(\lambda,\Phi_\lambda(D_1,\ldots,D_L))$ described the loop ABJM momentum amplituhedron. In the language introduced in section \ref{sec:DUAL_chambers}, the image $ \Phi_\lambda(D_1,\ldots,D_L)$ is the loop fibre with base point $\lambda$. 

\subsection{Loops from Lightcones}

Following our discussion in section \ref{sec:DUAL_nullcone-geometry}, we define the region $\Kcal(\bm{x})\subseteq \Rbb^{1,2}$ to be the region of Minkowski space which is space-like separated from all points $x_1,\ldots,x_{2k}$:
\begin{align}
	\Kcal(\bm{x})\coloneqq \{y\in\Rbb^{1,2}\colon (y-x_i)^2\geq 0\,, i=1,\ldots,2k\}\,.
\end{align}
This region is naturally decomposed into a compact part $\Delta(\bm{x})$ and a non-compact part $\overbar{\Delta}(\bm{x})$. We claim that $\Delta(\bm{x})$ is exactly the one-loop fibre geometry at base point $\lambda$, and coincides with the image $\Phi_\lambda(D)$ when translated into dual space. For additional clarity, since we don't often write out the list $\bm{x}$ explicitly, we sometimes indicate the number of particles as a subscript as $\Delta_n(\bm{x})$.

\begin{figure}
	\begin{center}
		\includegraphics[height=55mm]{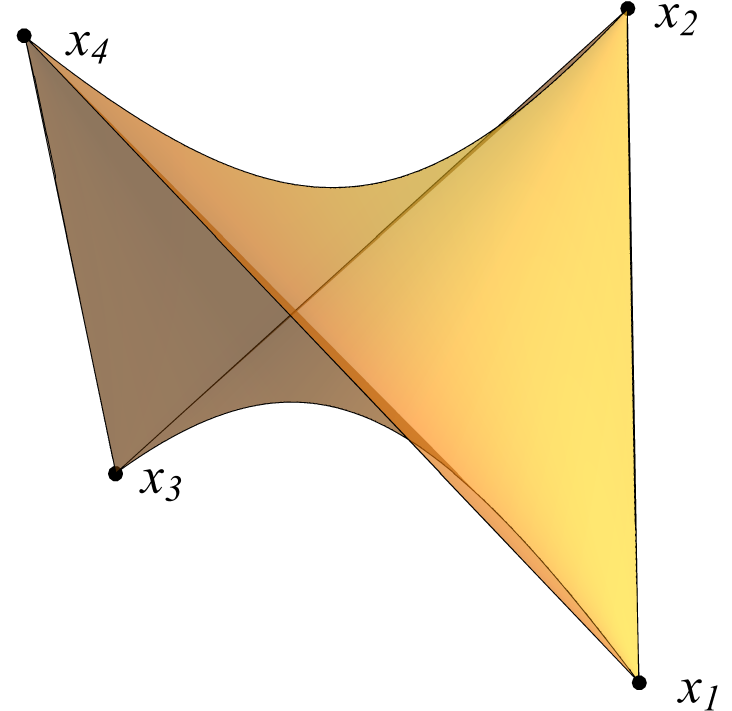}\quad
		\includegraphics[height=55mm]{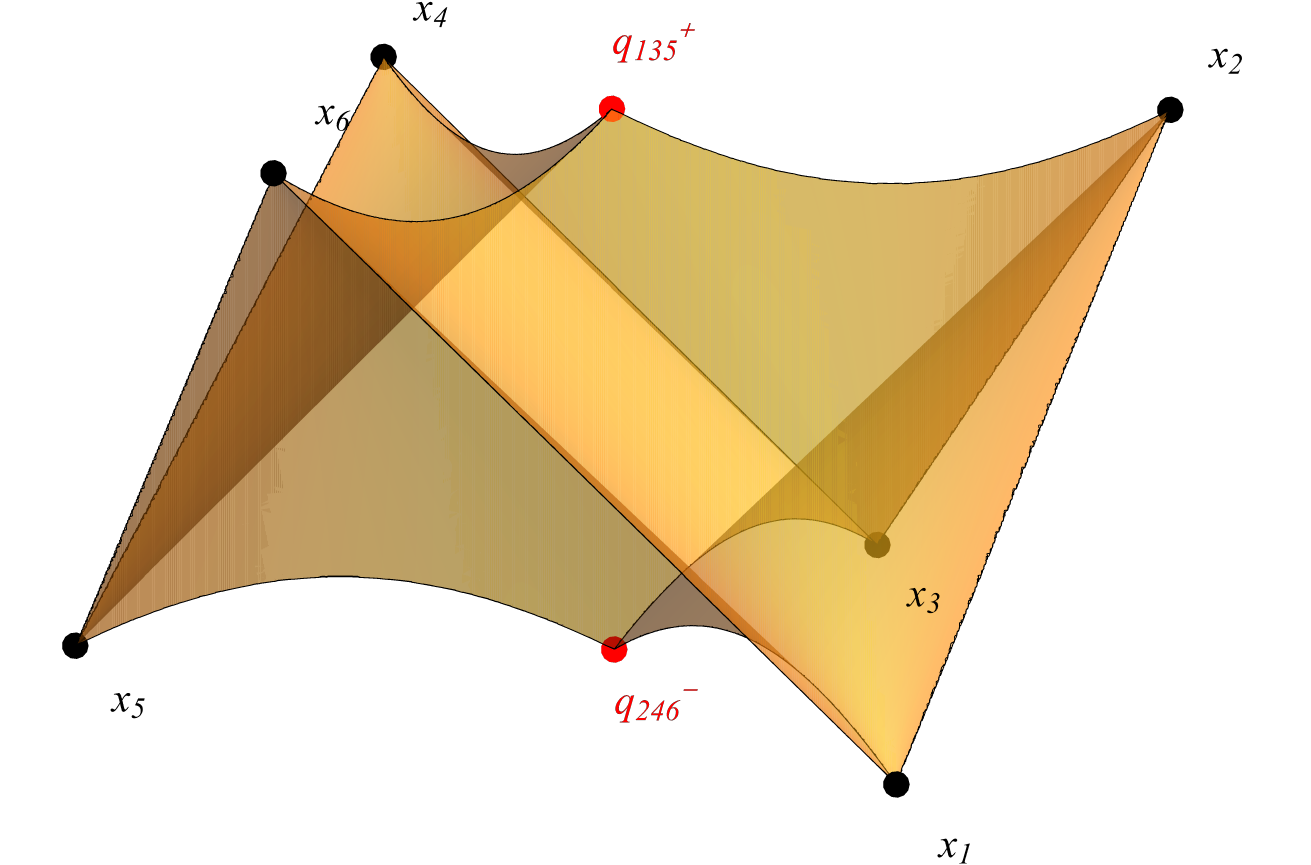}
	\end{center}
	\caption{The region $\Delta_4(\bm{x})$ (left) and $\Delta_6(\bm{x})$ (right).}
	\label{fig:ABJM-Delta-4-6}
\end{figure}

We have depicted an example of $\Delta(\bm{x})$ for the cases $n=4$ and $n=6$ in figure \ref{fig:ABJM-Delta-4-6}. We note that they are `curvy' versions of a tetrahedron and cube, respectively. These examples showcase some general structures. First, the facets are `curvy' degree-two curves corresponding to the lightcones $(y-x_i)^2=0$. Furthermore, the vertices of $\Delta(\bm{x})$ are separated into two types: the $n$ points $x_i$, and the triple intersections $q^\pm_{abc}$. It follows from their definition that the vertices $q^\pm_{abc}$ are incident to the three facets $\Ncal_{x_a}$, $\Ncal_{x_b}$ and $\Ncal_{x_c}$, and they are incident to three edges $\Ncal_{x_a}\cap\Ncal_{x_b}$, $\Ncal_{x_a}\cap\Ncal_{x_c}$, $\Ncal_{x_b}\cap\Ncal_{x_c}$. It is slightly less trivial to see that this counting also works for the vertices $x_i$. Since $(x_i-x_{i-1})^2=(x_i-x_i)^2=(x_i-x_{i+1})^2=0$, we see that $x_i$ is incident to the facets $\Ncal_{x_{i-1}}$, $\Ncal_{x_i}$, and $\Ncal_{x_{i+1}}$. The edges incident to a vertex $x_i$ are given by the light-rays connecting it to $x_{i-1}$ and $x_{i+1}$ (which are precisely the intersections $\Ncal_{x_{i-1}}\cap\Ncal_{x_i}$ and $\Ncal_{x_i}\cap\Ncal_{x_{i+1}}$), and the `curvy' edge $\Ncal_{x_{i-1}}\cap\Ncal_{x_{i+1}}$. 

This counting of the local boundaries around the vertices shows that $\Delta(\bm{x})$ can be interpreted as a `curvy' version of a three-dimensional simple polytope, as is indeed the case for the curvy tetrahedron and the curvy cube in figure \ref{fig:ABJM-Delta-4-6}. We recall from section \ref{sec:POS_simple-polytopes} that we can find the canonical form of a simple polytope as a sum over vertices, where each term consists of a wedge product of $\dd\log$s of the adjacent facets. By analogy, the differential form we would attribute to the vertices of $\Delta(\bm{x})$ are given by
\begin{align}
	x_a & \to \omega_{a-1aa+1}\coloneqq\dd\log(y-x_{a-1})^2\wedge\dd\log(y-x_{a})^2\wedge\dd\log(y-x_{a+1})^2\,,\\
	q^\pm_{abc} & \to \omega_{abc}\coloneqq\dd\log(y-x_{a})^2\wedge\dd\log(y-x_{b})^2\wedge\dd\log(y-x_{c})^2\,.
\end{align}
The naive canonical form of $\Delta$ would then be given by
\begin{align}\label{eq:DUAL_ABJM-Omega-naive}
	\Omega^{\text{naive}}(\Delta(\bm{x})) = \sum_{a=1}^n \sgn(x_a)\omega_{a-1aa+1}+\sum_{q^\pm_{abc}\in\Vcal(\Delta(\bm{x}))}\sgn(q^\pm_{abc})\omega_{abc}\,,
\end{align}
where $\Vcal(\Delta(\bm{x}))$ denotes the set of vertices of $\Delta(\bm{x})$, and the signs $\sgn(v)$ can be found by requiring that the form is projective, similar to section \ref{sec:POS_ABHY}. That is, the signs are fixed such that $\Omega^{\text{naive}}(\Delta(\bm{x}))$ is invariant under the local $GL(1)$ transformation $(y-v)^2\to \Lambda(y)(y-v)^2$ for any vertex $v$. This is essentially equivalent to the statement that $\Omega^{\text{naive}}(\Delta(\bm{x}))$ only depends on ratios $\smash{(y-v_1)^2/(y-v_2)^2}$, and has as a consequence that the form is free of poles at infinity.

There is, however, an obvious problem with this naive proposal. Although the vertices $x_a$ are uniquely determined by the intersection of the facets $\Ncal_{x_{a-1}}\cap\Ncal_{x_a}\cap\Ncal_{x_{a+1}}$, the same is not true for $q^\pm_{abc}$. That is, the form $\omega_{a-1aa+1}$ only has a maximal residue of $1$ at the point $x_a$, whereas the form $\omega_{abc}$ has a maximal residue of $1$ at \emph{both} $q^+_{abc}$ and $q^-_{abc}$. This is not a desirable property in the case where only one of $q^\pm_{abc}$ is a vertex of the geometry. To remedy this, we make the following observation: the form 
\begin{align}
	\frac{\dd^3y}{(y-x_a)^2(y-x_b)^2(y-x_c)^2}\,,
\end{align}
also has residues exactly on $q^\pm_{abc}$. However, from the global residue theorem, the sum of these residues has to add up to zero, and hence the residue of this form at $q^\pm_{abc}$ has to be plus or minus one over the Jacobian
\begin{align}
	\begin{vmatrix}
		\frac{\partial (y-x_{a})^2}{\partial y^\mu} & \frac{\partial (y-x_{b})^2}{\partial y^\mu} & \frac{\partial (y-x_{c})^2}{\partial y^\mu}
	\end{vmatrix}_{y=q^\pm_{abc}}&=8 \epsilon(a,b,c,q^\pm_{abc}) = \pm8 \sqrt{-\det\Xcal_{abc}} \notag\\
	&= \pm 4\sqrt{X_{ab}X_{bc}X_{ca}}\,,
\end{align}
where the matrix $\Xcal_{abc}$ is introduced in appendix \ref{sec:APP_schubert}. This means that 
\begin{align}
	\omega^\triangle_{abc}\coloneqq \frac{4 \sqrt{X_{ab}X_{bc}X_{ca}}}{(y-x_a)^2(y-x_b)^2(y-x_c)^2}\,,
\end{align}
has a residue of $\pm1$ at the vertices $q_{abc}^\pm$. We note that this differential form can be written as a $\dd\log$ form as
\begin{align}\label{eq:DUAL_ABJM-omega-tri-dlog}
	\omega^\triangle_{abc} =\pm \dd\log\frac{(y-x_a)^2}{(y-q^\pm_{abc})^2}\wedge\dd\log\frac{(y-x_b)^2}{(y-q^\pm_{abc})^2}\wedge\dd\log\frac{(y-x_c)^2}{(y-q^\pm_{abc})^2}\,.
\end{align}
The combination 
\begin{align}\label{eq:DUAL_omega-pm-ABJM}
	\omega^\pm_{abc}=\omega^\triangle_{abc}\pm \omega_{abc}\,,
\end{align}
thus only has a maximal residue exactly at $q^\pm_{abc}$ and not at $q^\mp_{abc}$. We then propose that the canonical form of $\Delta(\bm{x})$ can be obtained from equation \eqref{eq:DUAL_ABJM-Omega-naive} by replacing $\omega_{abc}\to\omega^\pm_{abc}$. After using projective invariance to fix the relative signs, we find that the canonical form of $\Delta(\bm{x})$ is given by
\begin{align}\label{eq:DUAL-omega-delta-ABJM}
	\Omega(\Delta(\bm{x}))=\sum_{a=1}^n (-1)^a \omega_{a-1aa+1}+\sum_{q^\pm_{abc}\in\Vcal(\Delta(\bm{x}))} \omega^\pm_{abc}\,.
\end{align}

\subsection{Chambers of the ABJM Momentum Amplituhedron} 
We remind ourselves why we are interested in the canonical form of $\Delta(\bm{x})$, following the discussion of section \ref{sec:DUAL_chambers}. The ABJM momentum amplituhedron is subdivided into chambers $\cfrak\in\Cfrak(\Ocal_k)$, where $\Cfrak(\Ocal_k)$ is the set of all chambers of $\Ocal_k$. In each chamber, the loop fibres have the same canonical form, which allows us to write the canonical form of the one-loop ABJM momentum amplituhedron as
\begin{align}
	\Omega(\Ocal_k^{(1)})=\sum_{\cfrak\in\Cfrak(\Ocal_k)} \Omega^{\text{tree}}(\cfrak)\wedge \Omega^{1-\text{loop}}(\cfrak)\,.
\end{align}
We have seen that the one-loop fibres are described by $\Delta(\bm{x})$, and its canonical form is uniquely determined by the vertex set $\Vcal(\Delta(\bm{x}))$. At one loop, we can therefore classify chambers of $\Ocal_k$ by the set of vertices of $\Delta(\bm{x})$. Varying a base point $\lambda$ in $\Ocal_k$ corresponds to changing the positions of the $x_a$ in dual space, the one-loop chambers of $\Ocal_k$ are then specified by the intersection patterns of the lightcones of these points $x_a$.

Using the Mathematica package \texttt{orthitroids}, we can generate many $\lambda\in\Ocal_k$. Using these points, we have analysed many instances of $\Delta(\bm{x})$ and we came to the following conclusions. First of all, we find that the vertices $q^\pm$ come in two different types: $q^+_{abc}$ where $a,b,c$ are odd, and $q^-_{abc}$ where $a,b,c$ are even. Secondly, there is a completely predictable pattern which dictates which vertices can appear together in $\Delta(\bm{x})$. To illustrate this, let us have a look at the complete list of one-loop chambers for $\Ocal_{4}$, which we label by the vertices of $\Delta(\bm{x})$:
\begin{subequations}
\begin{align}
	\Vcal_1&=\{x_a,q^+_{135},q^+_{157},q^-_{246},q^-_{268}\}\,,\\
	\Vcal_2&=\{x_a,q^+_{137},q^+_{357},q^-_{246},q^-_{268}\}\,,\\
	\Vcal_3&=\{x_a,q^+_{135},q^+_{157},q^-_{248},q^-_{468}\}\,,\\
	\Vcal_4&=\{x_a,q^+_{137},q^+_{357},q^-_{248},q^-_{468}\}\,.
\end{align}
\end{subequations}
These chambers can be categorised by choosing between two sets of odd-$q$s: $\{q^+_{135},q^+_{157}\}$ or $\{q^+_{137},q^+_{357}\}$, and independently choosing between two sets of even-$q$s: $\{q^-_{246},q^-_{268}\}$ or $\{q^-_{248},q^-_{468}\}$. Furthermore, these sets correspond precisely to the different triangulations of a square with corners labelled $1,3,5,7$ or $2,4,6,8$:
\begin{alignat}{2}
	&\{q^+_{135},q^+_{157}\} \leftrightarrow \begin{gathered} \includegraphics[scale=0.5]{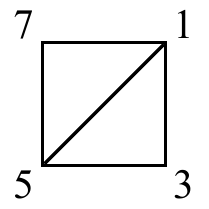} \end{gathered}\,,\qquad && \{q^+_{137},q^+_{357}\} \leftrightarrow \begin{gathered} \includegraphics[scale=0.5]{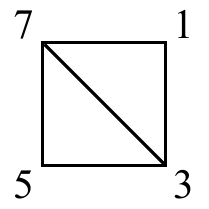} \end{gathered}\,,\nonumber\\
	&\{q^-_{246},q^-_{268}\} \leftrightarrow \begin{gathered} \includegraphics[scale=0.5]{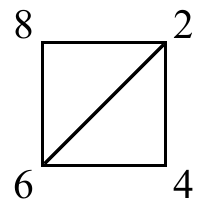} \end{gathered}\,, &&\{q^-_{248},q^-_{468}\} \leftrightarrow \begin{gathered} \includegraphics[scale=0.5]{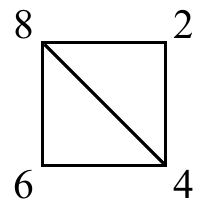} \end{gathered}\,.
\end{alignat}
We thus see that the vertices which define a chamber are labelled by two triangulations of a square, and, furthermore, all combinations of triangulations are realised in some chamber. We have verified up to $n=10$ that this pattern continues: the one-loop chambers of $\Ocal_k$ are labelled by the triangulations of two $k$-gons, one with odd corners and one with even corners. This is in agreement with the results from \cite{He:2023rou}, and we conjecture that it will continue to hold for any $k$. The number of chambers of $\Ocal_k$ is then $C_{k-2}^2$, where $C_p$ is the $p$\textsuperscript{th} Catalan number.

If we take some chamber $\cfrak$ of $\Ocal_k$ specified by the triangulations $T_1$ of an odd $k$-gon and $T_2$ of an even $k$-gon, then the canonical form of this chamber is given by
\begin{align}
	\Omega^{1-\text{loop}}(\cfrak) \equiv \Omega^{1-\text{loop}}_{T_1,T_2} = \sum_{a=1}^n (-1)^a \omega_{a-1aa+1} + \sum_{(a,b,c)\in T_1} \omega^+_{abc}+\sum_{(a,b,c,)\in T_2} \omega^-_{abc}\,.
\end{align}
Then, if we let $\Tcal_k^o$ and $\Tcal_k^e$ denote the set of all triangulations of a $k$-gon with odd and even labelled corners, respectively, then the full one-loop integrand can be written as
\begin{align}\label{eq:DUAL_ABJM-1-loop-chambers}
	A_{2k}^{(1)} = \sum_{\substack{T_1\in \Tcal_k^o\\ T_2\in \Tcal_k^e}}\Omega^{\text{tree}}_{T_1,T_2}\wedge \Omega^{1-\text{loop}}_{T_1,T_2}\,,
\end{align}
where $\Omega^{\text{tree}}_{T_1,T_2}$ is the canonical form of the respective chamber of $\Ocal_k$.

We note that $\omega^\pm_{abc}$, as defined in equation \eqref{eq:DUAL_omega-pm-ABJM}, has a residue of 2 at $q^\pm_{abc}$. At the same time, the sum in \eqref{eq:DUAL-omega-delta-ABJM} only has a residue of 1 at the vertices $x_a$. We could remedy this by normalising $\omega^\pm$ by a factor of $1/2$, in which case all vertices would have a residue of 1. However, this would not give the correct result, and, in fact, this difference in residues appears to be necessary to ensure that the canonical form is projectively invariant. We already noted before that it is necessary to move away the traditional requirement that positive geometries have unit leading residues, as was pointed out for the two-loop amplituhedron in \cite{Dian:2022tpf}. In this case, however, this discrepancy seems to have a slightly different origin, which is still not fully understood. 

It is possible that these strange leading residues are related to the fact that the full ABJM integrand is only recovered once we have summed over $2^{k-2}$ different \emph{branches}, as was argued in \cite{He:2023rou}. Our current construction only captures the so-called \emph{positive branch} of the ABJM integrands. The integrands which correspond to the other branches are very similar to \eqref{eq:DUAL_ABJM-1-loop-chambers}, except with some different signs for a subsets of $\omega^\triangle_{abc}$. In our geometric construction, this can be understood as interchanging certain $q^+_{abc}\leftrightarrow q^-_{abc}$, although it is not clear if all of these geometries can actually be realised. In practice, if we want to find a canonical form of one of the other branches, we can start from equation \eqref{eq:DUAL_ABJM-1-loop-chambers} and apply one of the \emph{parity operations} which were considered in \cite{He:2023rou}. To recover the full integrand, we simply sum \eqref{eq:DUAL_ABJM-1-loop-chambers} over all $2^{k-2}$ parity operations.

\subsection{Higher Loops}
Following the discussion of section \ref{sec:DUAL_chambers}, we can easily generalise to higher loops. The $L$-loop fibre in dual space can be defined as
\begin{align}
	\Delta^{(L)}(\bm{x})\coloneqq\{(y_1,\ldots,y_L)\in\Delta(\bm{x})^L\colon (y_i-y_j)^2\geq 0\,,\quad \forall i,j=1,\ldots,n\}\,,
\end{align}
from which we find the $L$-loop integrand
\begin{align}
	\Omega(\Ocal_k^{(L)})=\sum_{\cfrak\in\Cfrak(\Ocal_k)} \Omega^{\text{tree}}(\cfrak)\wedge \Omega^{L-\text{loop}}(\cfrak)\,,
\end{align}
where $\Omega^{L-\text{loop}}(\cfrak)$ is the canonical form of any $\Delta^{(L)}(\bm{x})$ for $\bm{x}$ coming from $\lambda\in\cfrak$.

\begin{figure}
	\centering
	\includegraphics[width=0.45\linewidth]{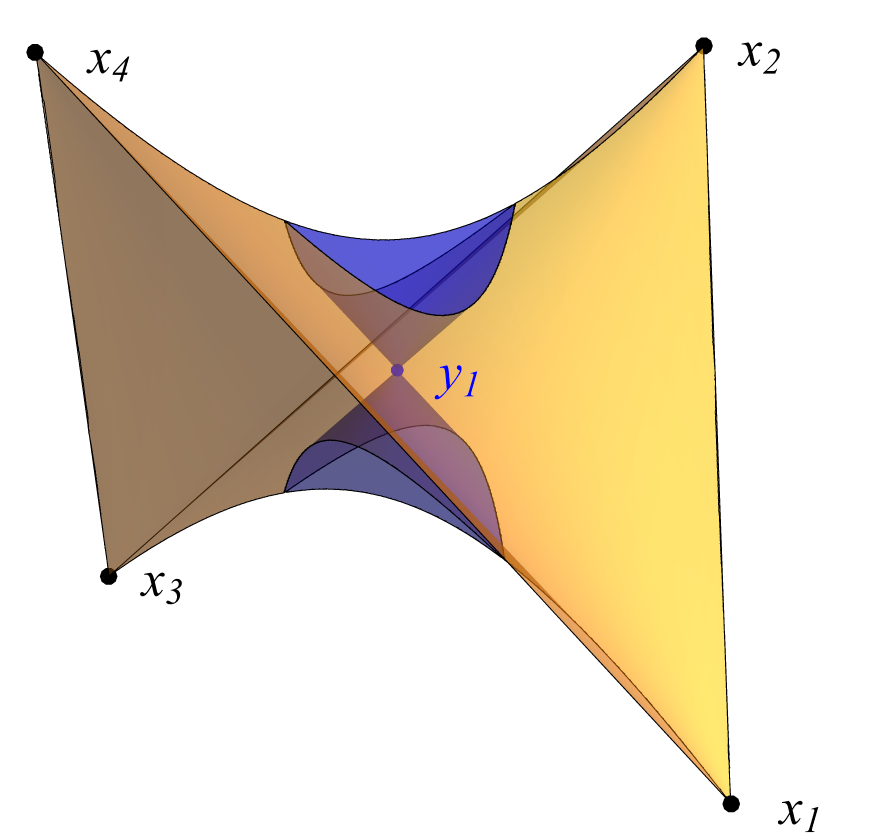}
	\caption{An illustration of the two-loop positive geometry (orange) and negative geometry (blue) for $\Delta^{(2)}(x_1,x_2,x_3,x_4)$ with a fixed position of $y_1$.}
	\label{fig:ABJM-posneg-n4}
\end{figure}

\begin{figure}
	\begin{center}
		\includegraphics[height=55mm]{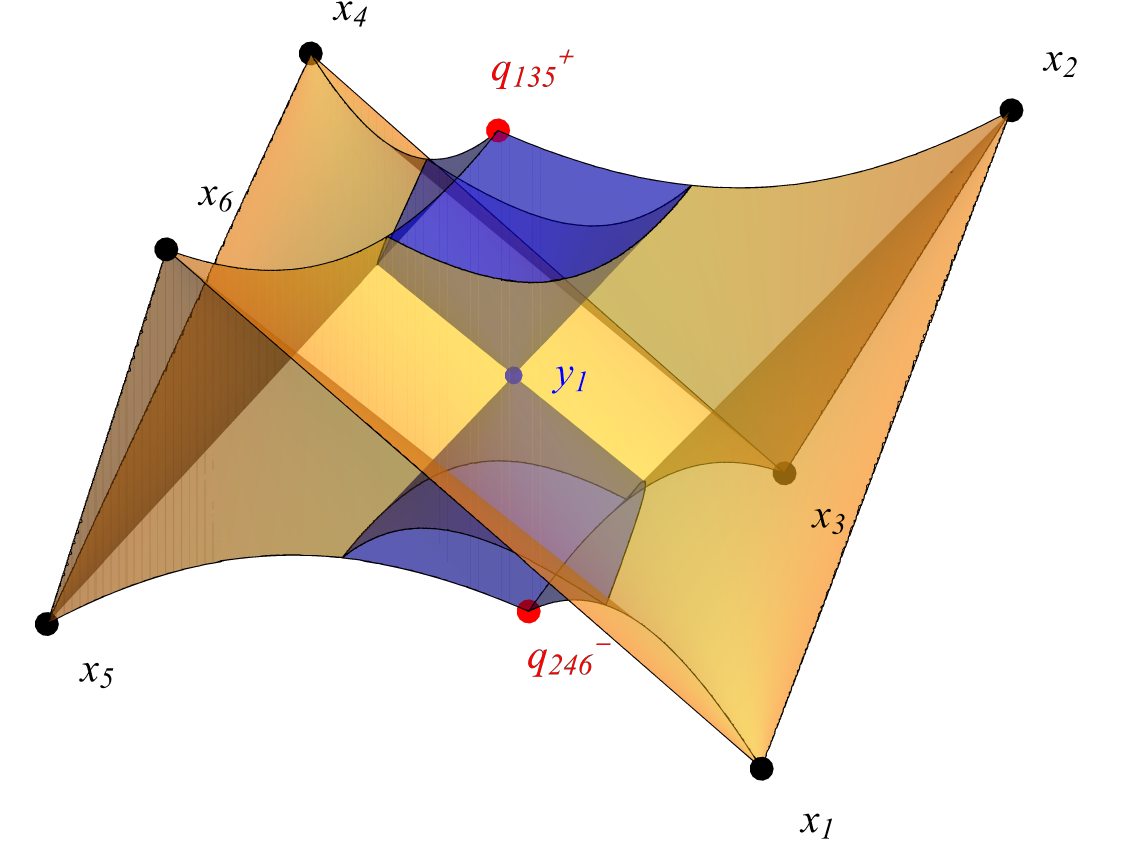}\quad
		\includegraphics[height=55mm]{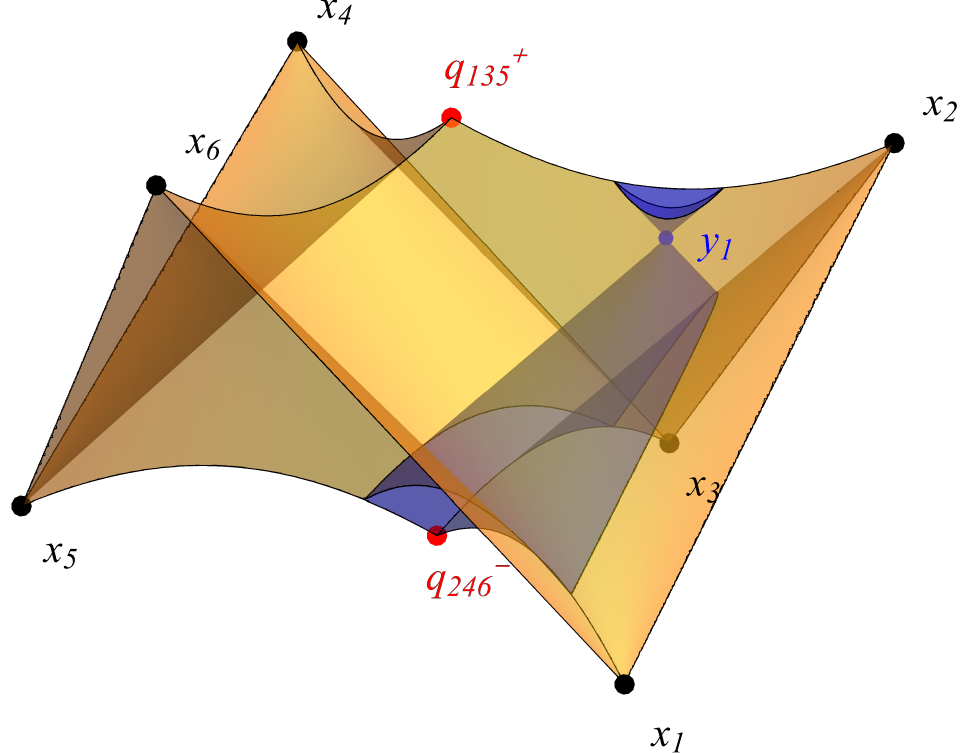}
	\end{center}
	\caption{Two combinatorially distinct regions for $y_2$ when fixing $y_1\in\Delta_6(\bm{x})$, labelled by $\{1,3,5,2,4,6\}$ (left), and $\{1,3,2,4,6\}$ (right).}
	\label{fig:ABJM-6-posneg-chambers}
\end{figure}

At this stage, we note that the notion of \emph{negative geometries} also has a natural interpretation in our current framework. The only difference between positive and negative geometries is that we require that for negative geometries some subset of the $y$s are \emph{time-like} separated, instead of space-like, and the remaining $(y_i-y_j)^2$ are unconstrained. By summing over the canonical form of all these negative geometries, we obtain the \emph{logarithm} of the integrand, rather than the standard integrand \cite{Arkani-Hamed:2021iya}. The study of negative geometries has lead to many new insights regarding higher loop integrands and geometric interpretations of the cusp anomalous dimension at 4-points for both \nf \cite{Arkani-Hamed:2021iya, Brown:2023mqi} and ABJM \cite{He:2022cup, Henn:2023pkc, He:2023exb, Li:2024lbw, Lagares:2024epo}. At two loops there is only one negative geometry, which is defined as
\begin{align}
	\Delta^{(2)}_{\text{neg}}(\bm{x})\coloneqq\{(y_1,y_2)\in\Delta(\bm{x})^2\colon (y_1-y_2)^2\leq 0\}\,.
\end{align}
The difference between the two-loop positive and negative geometries is illustrated in figure \ref{fig:ABJM-posneg-n4} for the case $n=4$ for a fixed value of $y_1$.
For two loops, it is clear that adding together the positive and the negative geometry results in no constraints between $y_1$ and $y_2$ at all. In terms of canonical forms, this means that
\begin{align}
	\Omega(\Delta^{(2)})+\Omega(\Delta^{(2)}_{\text{neg}}) = \big(\Omega(\Delta)\big)^2\,.
\end{align}
From figure \ref{fig:ABJM-posneg-n4} it is further clear that the negative geometry is split into two distinct regions: the region $y_1\prec y_2$ is defined to have the time component of $y_1$ smaller than the time component of $y_2$. That is,
\begin{align}
	\Omega(\Delta^{(2)}_{\text{neg}})= \Omega(\Delta^{(2)}_{\text{neg},y_1\prec y_2})+ \Omega(\Delta^{(2)}_{\text{neg},y_2\prec y_1})\,.
\end{align}
For $n=4$, the region $\Delta^{(2)}_{\text{neg},y_1\prec y_2}$ has facets at $(y_1-x_2)^2=0$, $(y_1-x_4)^2=0$, $(y_2-x_1)^2=0$, $(y_2-x_3)^2=0$, and $(y_1-y_2)^2=0$. Following \cite{He:2022cup}, there is a natural differential form to associate to this region:
\begin{align}
	\Omega(\Delta^{(2)}_{\text{neg},y_1\prec y_2})=\frac{X_{13}X_{24}\dd^3y_1\wedge\dd^3y_2}{(y_1-x_2)^2(y_1-x_4)^2(y_1-y_2)^2(y_2-x_1)^2(y_2-x_3)^2}\,.
\end{align}
The form associated to $\Delta^{(2)}_{\text{neg},y_2\prec y_1}$ is identical but with $y_1\leftrightarrow y_2$. Using these forms we can arrive at the correct two-loop integrand as
\begin{align}
	\Omega(\Delta^{(2)}) = \big(\Omega(\Delta)\big)^2 - \Omega(\Delta^{(2)}_{\text{neg},y_1\prec y_2}) - \Omega(\Delta^{(2)}_{\text{neg},y_2\prec y_1})\,.
\end{align}
The positive or negative geometries in figure \ref{fig:ABJM-posneg-n4} for fixed $y_1$ are combinatorially equivalent for any position of $y_1\in\Delta_4(\bm{x})$. This is not the case for a larger number of particles. Already for $n=6$, we find that there are 13 combinatorially distinct regions. We can classify these region based on the facets of the negative geometry. Excluding the facet $(y_1-y_2)^2=0$, the 13 regions are given by
\begin{subequations}\label{eq:DUAL_ABJM-n6-2loop-chambers}
\begin{alignat}{3}
	&\{1,3,5,2,4,6\}\,,\\
	&\{1,3,2,4,6\}\,,\quad&&\{1,5,2,4,6\}\,,\quad&&\{3,5,2,4,6\}\,,\\
	&\{1,3,5,2,4\}\,,\quad&&\{1,3,5,2,6\}\,,\quad&&\{1,3,5,4,6\}\,,\\
	&\{1,3,2,4\}\,,\quad&&\{1,5,2,6\}\,,\quad&&\{3,5,4,6\}\,,\\
	&\{1,3,2,6\}\,,\quad&&\{1,5,4,6\}\,,\quad&&\{3,5,2,4\}\,,
\end{alignat}
\end{subequations}
where we use the notation $\{1,3,5,2,4,6\}$ to denote that $\Ncal_{x_1},\Ncal_{x_3},\Ncal_{x_5},\Ncal_{x_2},\Ncal_{x_4},\Ncal_{x_6}$ are facets of the negative geometry. In particular, we note the absence of $\{1,3,4,6\}$ and its three cyclic permutations, which correspond to `opposite' edges of the curvy cube. We have depicted two of these regions in figure \ref{fig:ABJM-6-posneg-chambers}.
This suggests that the two-loop canonical form can be written as a sum over these 13 combinatorially distinct regions. To make this statement more precise, we introduce the notion of `fibrations over fibrations'. The idea is remarkably similar to the idea of chambers of the tree-level momentum amplituhedron as introduced in section \ref{sec:DUAL_chambers}. We define a projection map
\begin{align}
	\pi_{(1)}\colon \Delta^{(2)}(\bm{x})\to \Delta(\bm{x})\qquad (y_1,y_2)\mapsto y_1\,.
\end{align} 
We then divide $\Delta(\bm{x})$ into `chambers' $\cfrak_{(1)}\in\Cfrak(\Delta(\bm{x}))$ where the fibres $\pi_{(1)}^{-1}(y_1)$ are combinatorially equivalent for all $y_1$ in a chamber. The canonical form of $\Delta^{(2)}(\bm{x})$ can then be found as
\begin{align}\label{eq:DUAL_ABJM-2loop-chambers}
	\Omega(\Delta^{(2)}(\bm{x}))=\sum_{\cfrak_{(1)}\in\Cfrak(\Delta(\bm{x}))} \Omega^{1-\text{loop}}(\cfrak_{(1)})\wedge\Omega^{2-\text{loop}}(\cfrak_{(1)})\,,
\end{align} 
where $\Omega^{1-\text{loop}}(\cfrak_{(1)})$ denotes the canonical form of the chamber $\cfrak_{(1)}$ of $\Delta(\bm{x})$, and $\Omega^{2-\text{loop}}(\cfrak_{(1)})$ denotes the canonical form of the second loop fibre $\pi_{(1)}^{-1}(y_1)$ for some $y_1\in\cfrak_{(1)}$. The idea of obtaining higher-loop integrands from these `fibrations over fibrations' is currently still an active work in progress in collaboration with L. Ferro, R. Glew, and T. \L{}ukowski (see \cite{Ferro:2024vwn} for recent work on `fibrations of fibrations' for \nf).

In this language, the regions \eqref{eq:DUAL_ABJM-n6-2loop-chambers} label the different chambers of $\Delta(\bm{x})$, and equation \eqref{eq:DUAL_ABJM-2loop-chambers} means that the canonical form of $\Delta_6^{(2)}$ can be written as a sum over these 13 chambers. This is reminiscent of the sum for the $n=6$ 2-loop integrand given in \cite{He:2023rou} (see the discussion around equation 3.61). However, the sum presented there also includes contributions which appear to correspond to the $\{1,3,4,6\}$-type chambers. As discussed, these chambers are empty, and our formalism of fibrations over fibrations thus suggests that their answer can be rewritten in a way which only includes the 13 realised chambers.

\subsection{The Dual Amplituhedron}

In this section we make some observations regarding the \emph{dual} of the lightcone geometry $\Delta_{2k}(\bm{x})$. This would be an important step in the formulation of a dual amplituhedron for ABJM theory. The dual of a polytope is a well-defined notion, and has been useful in describing scattering amplitudes in $\tr{\phi^3}$ (see \cite{Arkani-Hamed:2017mur}). Given some polytope, its dual is a new polytope whose boundary poset is obtained by turning the poset of the original polytope 'upside down', as was explained in section \ref{sec:GRASS_proj}. This means that the facets of the dual polytope correspond to vertices, edges (1-dimensional boundaries) to codimension-2 boundaries, etc. As we have seen repeatedly, positive geometries encode the scattering amplitude in their canonical form , whereas it would be encoded in the \emph{volume} of the dual geometry. It is not obvious if the dual of the amplituhedron is a well-defined notion, but some evidence for this has been given in \cite{Arkani-Hamed:2014dca}, where it is further argued that the description of the dual amplituhedron would make a connection to weakly coupled string theory at strong 't Hooft coupling. Some progress towards the construction of the dual amplituhedron has been made in \cite{Herrmann:2020qlt} for the one-loop MHV amplituhedron.

\begin{figure}
	\centering
	\includegraphics[width=0.9\textwidth]{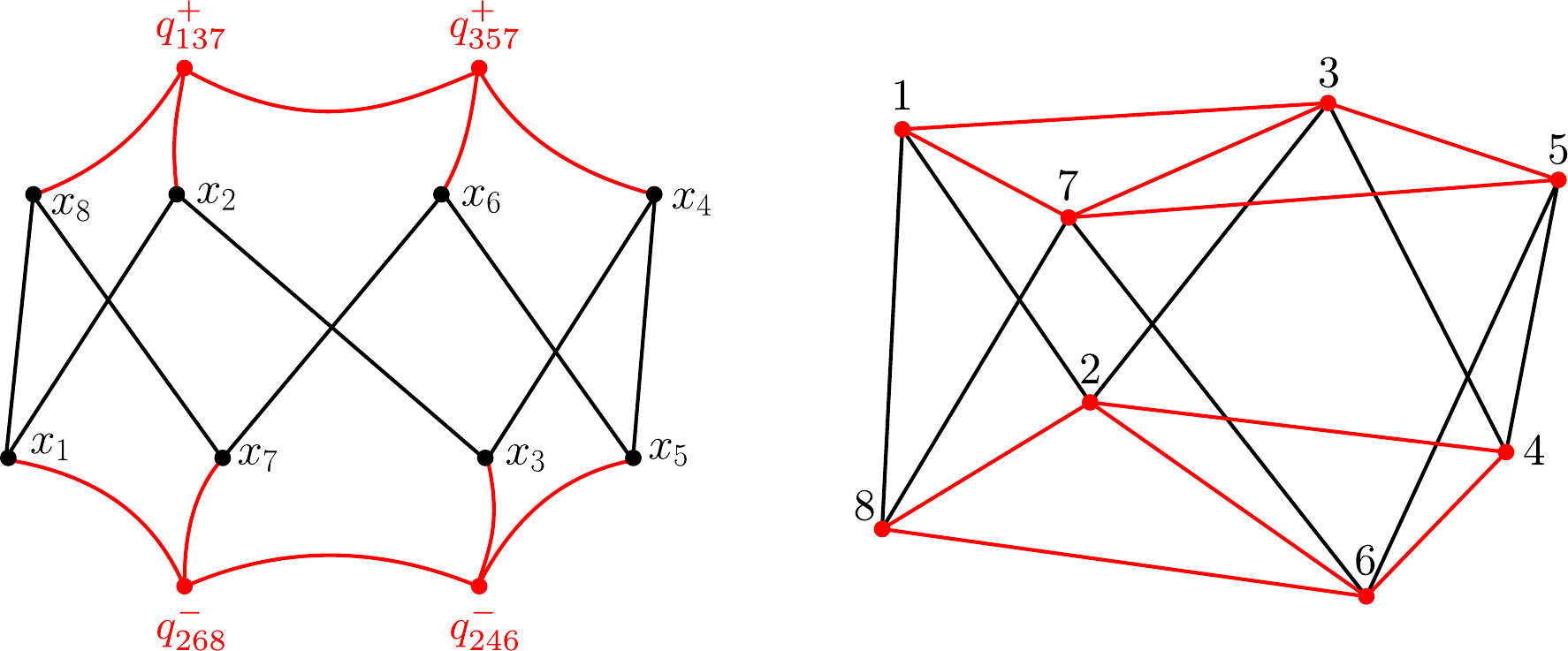}
	\caption{A skeleton of $\Delta_8(\bm{x})$ in one of the four chambers (left) and a skeleton of its dual geometry (right). Note that the $q^\pm_{abc}$ vertices of $\Delta_8(\bm{x})$ correspond to triangles in the triangulations of an odd-labelled and an even-labelled square in the dual.}
	\label{fig:Delta-8-ABJM}
\end{figure}

We first note that our classification of chambers in terms of triangulations of an odd-labelled $k$-gon and an even-labelled $k$-gon has a natural interpretation in terms of the dual geometry. Consider embedding an even-labelled $k$-gon and an odd-labelled $k$-gon in three-dimensional space, and connecting the corners $i$ and $i+1$, forming a $k$-antiprism. A chamber of $\Ocal_k$ can be specified by triangulating the two $k$-gons. The resulting shape \emph{is the skeleton of the dual geometry of $\Delta_{2k}(\bm{x})$}. The corner $i$ of one of the $k$-gons correspond to a facet $\Ncal_{x_i}$ of $\Delta_{2k}(\bm{x})$, the triangles $(i-1,i,i+1)$ in the antiprism correspond to vertices $x_i$, the edges $(i-1,i+1)$ of one of the $k$-gons correspond to $\Ncal_{x_{i-1}}\cap\Ncal_{x_{i+1}}$, which we argued above are edges of $\Delta_{2k}(\bm{x})$, triangles $(a,b,c)$ part of one of the triangulations of a $k$-gon correspond to vertices $q^\pm_{abc}$, etc. The statement that all the facets of the dual geometry are triangles is equivalent to the statement that $\Delta_{2k}(\bm{x})$ is {simple}, as it shows that every vertex of $\Delta_{2k}(\bm{x})$ is incident to exactly three edges and tree facets. We have depicted $\Delta_{8}(\bm{x})$ and its dual schematically in figure \ref{fig:Delta-8-ABJM} for one of the four chambers. Our classification of the different chambers of $\Ocal_k$ is thus more naturally understood as a classification of all possible dual geometries of $\Delta_{2k}(\bm{x})$. 

We further want to make a remark about the canonical form $\Omega(\Delta(\bm{x}))$ and the volume of the dual geometry. We first define the box integrand\footnote{Not to be confused with the box integrand in 4D, which we will encounter in the next section.}
\begin{subequations}\label{eq:DUAL_ABJM-omega-box-dlog}
	\begin{align}
		\omega^{\square}_{abcd} &= \dd\log\frac{(y-x_a)^2}{(y-x_d)^2}\wedge\dd\log\frac{(y-x_b)^2}{(y-x_d)^2}\wedge\dd\log\frac{(y-x_c)^2}{(y-x_d)^2}\\
		&= \frac{-16\epsilon(a,b,c,d,y)\dd^3y}{(y-x_a)^2(y-x_b)^2(y-x_c)^2(y-x_d)^2}\\
		&= \omega_{abc}-\omega_{abd}+\omega_{acd}-\omega_{bcd}\,.\label{eq:DUAL_ABJM-omega-box-dlog-expand}
	\end{align}
\end{subequations}
We note that, since this $\dd\log$ form only contains physical propagators, it will integrate to zero \cite{Herrmann:2019upk}. From its definition it is clear that $\omega^\square$ is projective invariant. Furthermore, from equation \eqref{eq:DUAL_ABJM-omega-tri-dlog} we see that also $\omega^\triangle$ is projectively invariant. If we manage to re-sum all $\omega_{abc}$ in $\Omega(\Delta(\bm{x}))$ into box integrands, then the projective invariance of the canonical form is manifest. This turns out to always be possible, and not in a unique way. For $n=4$ we have
\begin{align}
	\Omega(\Delta_4) &= \omega_{123}-\omega_{124}+\omega_{134}-\omega_{234} = \omega^\square_{1234}\\
	&=\omega^\square_{\star 123}-\omega^\square_{\star 124}+\omega^\square_{\star 134} - \omega^\square_{\star 234}\,.
\end{align}
For $n=6$ we have multiple options:
\begin{subequations}\label{eq:DUAL_ABJM-n6-box-all}
\begin{align}
	\Omega(\Delta_6) &= \omega_{123}-\omega_{234}+\omega_{345}-\omega_{456}+\omega_{561}-\omega_{612}+\omega_{135}-\omega_{246}+\omega^\triangle_{135}+\omega^\triangle_{246}\\
	&= \omega^\square_{1234}+\omega^\square_{1456}+\omega^\square_{1246}-\omega^\square_{1345}+\omega^\triangle_{135}+\omega^\triangle_{246}\label{eq:DUAL_ABJM-n6-box-expand}\\
	&=\omega^\square_{1235}+\omega^\square_{1256}+\omega^\square_{2456}-\omega^\square_{2345}+\omega^\triangle_{135}+\omega^\triangle_{246}\\
	&=\omega^\square_{1236}+\omega^\square_{1356}+\omega^\square_{3456}-\omega^\square_{2346}+\omega^\triangle_{135}+\omega^\triangle_{246}\,.
\end{align}		
\end{subequations}
In general, we can write $\Omega(\Delta(\bm{x}))$ in terms of $\omega^\square$ by simply replacing $\omega_{abc}\to\omega^\square_{\star abc}$, for some arbitrary reference point $x_\star$, which, for simplicity, we can set to be any $x_i$.

We note that $\Delta_4(\bm{x})$ is a curvy tetrahedron with canonical form $\omega^\square_{1234}$. Furthermore, equation \eqref{eq:DUAL_ABJM-omega-box-dlog} looks very reminiscent to the canonical form of a simplex (see section \ref{sec:POS_def}). It is therefore natural to ask whether we can interpret $\omega^\square_{abcd}$ as the canonical form of a curvy simplex with facets $\Ncal_{x_a},\Ncal_{x_b},\Ncal_{x_c},\Ncal_{x_d}$, to which the answer is affirmative in the $n=4$ case. In general this is not the case, as can be seen by counting residues. Expanding $\omega^\square$ as in \eqref{eq:DUAL_ABJM-omega-box-dlog-expand} and recalling our discussion on the residues of $\omega_{ijk}$, it is clear that $\omega^\square_{abcd}$ will have maximal residues at eight points: $q^\pm_{abc},q^\pm_{abd},q^\pm_{acd},q^\pm_{bcd}$, which is not the correct number of vertices for a tetrahedron. However, we notice that in $\Omega(\Delta(\bm{x}))$ there are additional $\omega^\triangle$'s which will cancel some of the residues of these vertices. So, perhaps a combination of $\omega^\square$ and $\omega^\triangle$ will give the appropriate residues.

Let's take a closer look at $n=6$, specifically the expression \eqref{eq:DUAL_ABJM-n6-box-expand}. We note that $\omega^\square_{1234}$ has appropriate residues for a simplex: its vertices are $x_2,x_3$ and $q_{124},q_{134}$\footnote{We remember that $\Ncal_{x_i}\cap\Ncal_{x_{i+1}}$ is a light-ray, and a light-ray generically only intersects a light-cone in a single point, rather than in two points. Hence $\Ncal_{x_i}\cap\Ncal_{x_{i+1}}\cap\Ncal_{x_j}$ only consists of one point: $q_{ii+1j}$. This can equivalently be seen by noting that the square root in \eqref{eq:SCH_q-3D-gen} vanishes when any two particles are light-like separated.}, and similarly for $\omega^\square_{1456}$. This is \emph{not} the case for $\omega^\square_{1246}$, which has support on $x_1,q_{124},q_{146},q^+_{246},q^-_{246}$. However, we notice that the presence of $\omega^\triangle_{246}$ will kill the residue at $q^-_{246}$, and hence the combination of $\omega^\square_{1246}+\omega^\triangle_{246}$ has the appropriate residue structure for a simplex, and similarly for $\omega^\square_{1345}+\omega^\triangle_{135}$. 

\begin{figure}
	\centering
	\includegraphics[width=0.5\textwidth]{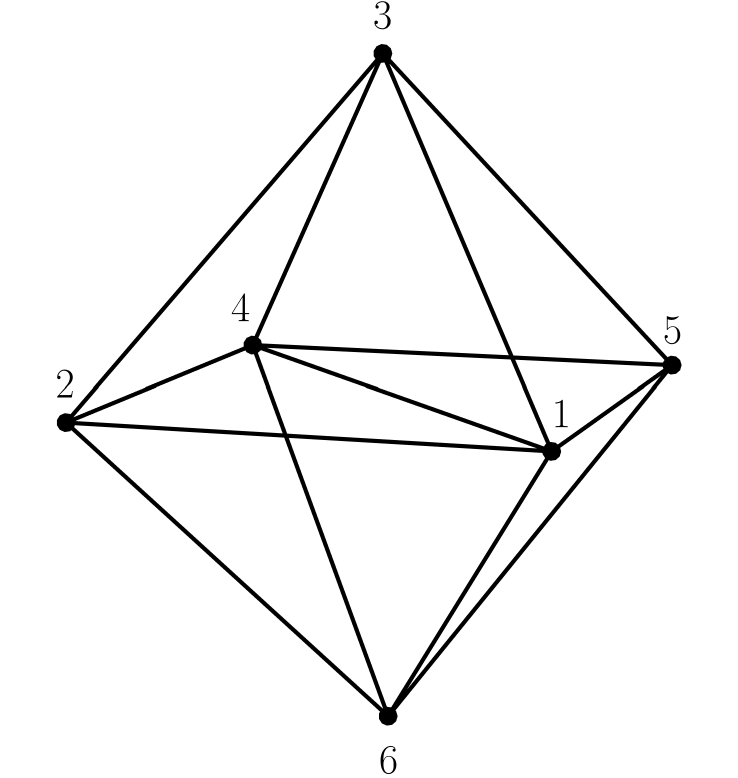}
	\caption{The dual of $\Delta_6(\bm{x})$ is combinatorially an octahedron which has a natural internal triangulation in terms of simplices with vertices $(1,2,3,4)$, $(1,2,4,6)$, $(1,3,4,5)$, and $(1,4,5,6)$.}
	\label{fig:dual-Delta6-triangles}
\end{figure}

Let us see schematically what these simplices look like in the dual geometry of $\Delta_6(\bm{x})$. The dual of a cube is an octahedron (the skeleton of the dual geometry can be equivalently obtained from the discussion above as a 3-antiprism) with vertices labelled 1 through 6, which we depict in figure \ref{fig:dual-Delta6-triangles}. The `simplex' corresponding to $\omega^\square_{1234}$ corresponds to a `simplex' in the dual with vertices $1,2,3,4$, and $\omega^\square_{1246}+\omega^\triangle_{246}$ describes a simplex in the dual with vertices $1,2,4,6$, etc. We see that the octahedron admits an internal triangulation in terms of these four tetrahedra by drawing a spurious edge between vertices 1 and 4. The other expressions for $\Omega(\Delta_6)$ in \eqref{eq:DUAL_ABJM-n6-box-all} simply correspond to different internal triangulations of the dual, with spurious edges 2,5 and 3,6 respectively. This general pattern continues to hold for all chambers for a higher number of particles as well. Our expression for the canonical form can be written in terms of $\omega^\square$'s in multiple different ways, each of them corresponding to a different internal triangulation of the dual geometry. 

The observations in this section are suggestive of the existence of a dual geometry for $\Delta(\bm{x})$. we have explained how our classification of chambers is more naturally interpreted as a classification of dual geometries, and how the canonical form $\Omega(\Delta(\bm{x}))$ has a natural formulation which would correspond to an internal triangulation of the dual geometry in terms of simplices. An explicit formulation of the dual geometry whose volume encapsulates the canonical form is currently not known, and this remains an important open question.

\section{\texorpdfstring{\nf}{N=4 SYM}}\label{sec:DUAL_nf}

We now turn to the loop momentum amplituhedron in \nf. We recall from section \ref{sec:POS_mom-amp} that the momentum amplituhedron $\Mcal_{n,k}$ consists of $(\lambda,\tilde\lambda)\in G(2,n)\times G(2,n)$ which satisfy $\lambda\cdot\tilde\lambda^T=\nul$, and 
\begin{alignat}{2}
	&\<ii+1\>>0\,,\quad[ii+1]>0\,,\quad &&s_{ii+1\cdots j} >0\,,\\
	&\{\<12\>,\<13\>,\ldots,\<1n\>\}\quad&&\text{has $k-2$ sign flips}\,,\\
	&\{[12],[13],\ldots,[1n]\}\quad&&\text{has $k$ sign flips}\,.
\end{alignat}
We translate $(\lambda,\tilde\lambda)$ into dual space by defining
\begin{align}
	x_i^{\alpha\alphadot}=\sum_{j=1}^{i-1}\lambda_j^\alpha\tilde\lambda_j^\alphadot\,,
\end{align}
and we interpret these $2\times2$ matrices as points in $\Rbb^{2,2}$ via
\begin{align}
	x^{\alpha\alphadot} = \begin{pmatrix} x^0+x^2 & x^1+x^3\\ -x^1+x^3 & x^0-x^2 \end{pmatrix} \leftrightarrow x^\mu = \frac{1}{2}\begin{pmatrix} x^{1\dot{1}} + x^{2\dot{2}}\\ x^{1\dot{2}}-x^{2\dot{1}}\\ x^{1\dot{2}}+x^{2\dot{1}}\\ x^{1\dot{1}}-x^{2\dot{2}} \end{pmatrix}\,.
\end{align}
These points in dual space satisfy $(x_i-x_{i+1})^2=X_{ii+1}=0$, and $(x_i-x_j)^2=X_{ij}>0$. Since $\Rbb^{2,2}$ is not a Minkowski space, we will talk about \emph{positive}, \emph{negative}, and \emph{null} separation, rather than \emph{space-like}, \emph{time-like}, and \emph{light-like} separation.

Following the discussion in section \ref{sec:POS_loop-mom-amp}, we can extend the momentum amplituhedron to loop level in dual space by defining the map
\begin{align}
	\Phi_{\lambda,\tilde\lambda}\colon G(2,n)^L\to (\Rbb^{2,2})^L,\qquad (D_1,\ldots,D_L)\mapsto (y_1,\ldots, y_L)\,,
\end{align}
where
\begin{align}\label{eq:DUAL_mom-amp-y-def}
	y_i^\mu = \frac{\sum_{a<b}(ab)_i\<ab\>{\ls_{ab}}^\mu}{\sum_{a<b}(ab)_i\<ab\>}\,,
\end{align}
where $(ab)_i=p_{ab}(D_i)$, and ${\ls_{ab}}^\mu$ is the the four-vector form of the $\ls_{ab}$ introduced in equation \eqref{eq:KIN_ls-def}. If we restrict the $D$ matrices to satisfy the positivity of \eqref{eq:POS_loop-positivity-mom amp}, then the combination of $(\lambda,\tilde\lambda,\Phi_{\lambda,\tilde\lambda}(D_1,\ldots,D_L))$ specifies a point in the loop momentum amplituhedron. From the sign-flip definition of the loop momentum amplituhedron we know that these points satisfy 
\begin{align}\label{eq:DUAL_mom-amp-sign-flip-loop}
	&(y-x_i)^2>0\,, \quad(y_i-y_j)^2>0\,,\\
	&\{\<ii+1\> (y-\ls_{ii+1})^2\,, \<ii+2\> (y-\ls_{ii+2})^2\,,\ldots \<ii+n-1\> (y-\ls_{ii+n-1})^2 \}\quad\text{ has $k$ sign-flips}\,.
\end{align} 
We note that we pick up a factor of $(-1)^{k-1}$ for $\<i a\>$, for $a>n$ due to \emph{twisted cyclic symmetry}.

We divide the momentum amplituhedron $\Mcal_{n,k}$ into chambers $\cfrak\in\Cfrak(\Mcal_{n,k})$ such that the loop fibre (which is the image of $\Phi_{\lambda\tilde\lambda}$ over the domain of $D$ matrices satisfying the correct positivity) is combinatorially equivalent for all $\lambda,\tilde\lambda$ in a chamber. Similar to the cases considered earlier in this chapter, we claim that the region
\begin{align}
	\Kcal(\bm{x})\coloneqq\{y\in\Rbb^{2,2}\colon (y-x_i)^2\geq0\,, i=1,\ldots,n\}\,,
\end{align}
splits into a compact part $\Delta(\bm{x})$ and a non-compact part $\overbar{\Delta}(\bm{x})$:
\begin{align}
	\Kcal(\bm{x})=\Delta(\bm{x})\cup\overbar\Delta(\bm{x})\,,
\end{align}
and that the one-loop fibre is given precisely by $\Delta(\bm{x})$. To keep track of the values of $n$ and $k$ of the tree-amplituhedron which seeds $\Delta(\bm{x})$, we sometimes explicitly write $\Delta_{n,k}(\bm{x})$. To make the relation to the loop momentum amplituhedron a bit more explicit, we note that the constraints $(y-x_i)^2>0$ in equation \eqref{eq:DUAL_mom-amp-sign-flip-loop} imply that the loop fibre must be inside $\Kcal(\bm{x})$. We claim that the sign-flip conditions exactly isolate the compact part $\Delta(\bm{x})$. Furthermore, the $L$-loop fibre consists of $L$ mutually positively separated points inside $\Delta(\bm{x})$:
\begin{align}
	\Delta^{(L)}(\bm{x})\coloneqq\{(y_1,\ldots,y_L)\in\Delta(\bm{x})^L\colon (y_i-y_j)^2\geq 0\}\,.
\end{align}
For the remainder of this section we will focus on the geometry $\Delta(\bm{x})$ and its associated canonical form. The study of the higher-loop geometries is an ongoing project. But, before we move on, let us remark some of the differences inherent in the definitions of the amplituhedron, the momentum amplituhedron, their loop extensions, and the new null-cone geometries.

One of the essential differences between the amplituhedron and the momentum amplituhedron is that these positive geometries live in different kinematic spaces: the amplituhedron lives in momentum twistor space, and the momentum amplituhedron lives in spinor-helicity space. Momentum twistors are the variables in which most symmetries of \nf are manifest, giving arguably the simplest and least constrained description. The exception to this rule is \emph{parity symmetry}, which relates $A_{n,k}$ and $A_{n,n-k}$ amplitudes, a property which is far from obvious in the amplituhedron description: $\Acal_{n,K}$ and $\Acal_{n,n-K-4}$ are geometrically distinct, and don't even have the same dimension ($4K$ and $4(n-K-4)$, respectively)! This is not an issue for the momentum amplituhedron. The use of spinor-helicity variables and non-chiral superspace makes parity symmetry trivial: we simply interchange $\lambda\leftrightarrow \tilde\lambda$, which relates $\Mcal_{n,k}$$\leftrightarrow\Mcal_{n,n-k}$ (for superamplitudes we also swap $\eta\leftrightarrow\tilde\eta$).

We have seen that the definition of the loop momentum amplituhedron essentially borrows the definition of the loop amplituhedron, and reformulates it into spinor-helicity space. In particular, this means that the definition of the loop momentum amplituhedron is no longer parity symmetric, unlike the tree-level momentum amplituhedron! This can be seen from the definition of the loop momenta in equation \ref{eq:DUAL_mom-amp-y-def}, which singles out $\lambda$ over $\tilde\lambda$. The fact that the parity-conjugate
\begin{align}
	\ell' = \frac{\sum_{i<j} (ij)[ij]\lstil_{ij}}{\sum_{i<j}(ij)[ij]}\,,
\end{align}
gives rise to the \emph{same} integrands is a non-trivial statement\footnote{There is some evidence that a `mixed' version of the loop momenta of the form $\sum_{i<j}(ij)\<ij\>\lstil_{ij}/\sum_{i<j}(ij)\<ij\>$, or $\sum_{i<j}(ij)[ij]\ls_{ij}/\sum_{i<j}(ij)[ij]$, also gives the correct integrands. This is an even less trivial statement, which goes beyond standard parity symmetry.}.

The null-cone geometries which we consider in this chapter offer a different perspective for loop level geometries. They provide a unified framework in the space of dual momenta which describes loop geometries for the amplituhedron and the momentum amplituhedron \emph{at the same time}. This new way of thinking about the loop momentum amplituhedron makes it manifestly parity symmetric: since $\Mcal_{n,k}$ and $\Mcal_{n,n-k}$ are parity dual, so are their chambers and the associated loop geometries. On the level of $\Delta(\bm{x})$, this parity conjugation simply swaps $q^\pm_{abcd}\leftrightarrow q^\mp_{abcd}$.

\subsection{\texorpdfstring{The Geometry of Null-Cones and the Structure of $\Delta_{n,k}$}{The Geometry of Null-Cones and the Structure of Delta(n,k)}}\label{sec:DUAL_nullcone-geometry-4D}

We now continue to study some generic properties of null-cones in $\Rbb^{2,2}$ and what they teach us about the geometry of $\Delta_{n,k}(\bm{x})$. 

\subsubsection{The Local Geometry of Vertices}
We define the null-cone
\begin{align}
	\Ncal_x\coloneqq \{y\in\Rbb^{2,2}\colon (x-y)^2=0\}\,.
\end{align}
As before, the vertices of $\Delta(\bm{x})$ come in two different types, the vertices $x_i$, and the quadruple intersections $\{q^+_{abcd},q^-_{abcd}\}=\Ncal_{x_{a}}\cap\Ncal_{x_b}\cap\Ncal_{x_c}\cap\Ncal_{x_d}$. A specific formula for these quadruple intersections can be found in equation \eqref{eq:SCH_q-4D-gen}. The local geometry around a quadruple intersection is trivial. Incident to $q^\pm_{abcd}$ are four facets $\Ncal_{x_a},\Ncal_{x_b},\Ncal_{x_c},\Ncal_{x_d}$, and four edges
\begin{align}
	\{e^+_{ijk},e^-_{ijk}\}=\Ncal_{x_i}\cap\Ncal_{x_j}\cap\Ncal_{x_k}\,,
\end{align}
coming from intersections of three of these null-cones. The local geometry of the vertices $x_i$ is more interesting, and it naturally leads to the notion of \emph{white} and \emph{black planes}.

If we have $(x_i-x_{i-1})^2=0$, then $\Ncal_{x_{i-1}}\cap\Ncal_{x_i}$ is the union of two affine planes, which we call `white' and `black' planes. These planes intersect along the null-ray connecting $x_{i-1}$ and $x_{i}$. This is a familiar statement from spinor-helicity space, and is equivalent to the `three particle special kinematics' which distinguish the MHV and \MHVbar 3-point amplitudes in four spacetime dimensions. To see this, let us consider a point $y\in\Ncal_{x_{i-1}}\cap\,\Ncal_{x_{i}}$, such that $(y-x_i)^2=(y-x_{i-1})^2=(x_i-x_{i-1})^2=0$. Using spinor-helicity variables we can then write $(x_i-x_{i-1})^{\alpha\alphadot}=\lambda_{i}^\alpha\tilde\lambda_i^\alphadot$, $(y-x_{i})^{\alpha\alphadot}=\kappa_i^\alpha\tilde\kappa_i^\alphadot$, $(y-x_{i-1})^{\alpha\alphadot}=\kappa_{i-1}^\alpha\tilde\kappa_{i-1}^\alphadot$. There are two solutions to these equations, given by
\begin{alignat}{3}
	\<\lambda_i \kappa_{i-1}\> &= \<\lambda_i \kappa_{i}\> = \<\kappa_{i-1}\kappa_{i}\>=0\quad &&\implies \quad &&\lambda_i \propto \kappa_{i-1} \propto \kappa_{i}\,,\\
	[\tilde\lambda_i \tilde\kappa_{i-1}] &= [\tilde\lambda_i\tilde\kappa_{i}] = [\tilde\kappa_{i-1}\tilde\kappa_{i}]=0\quad &&\implies\quad &&\tilde\lambda_i\propto \tilde\kappa_{i-1}\propto\tilde\kappa_{i}\,.
\end{alignat}
These are linear constraints and thus define two affine planes in $\Rbb^{2,2}$ which are the white and black planes, respectively:
\begin{align}
	W_{i-1 i}&\coloneqq \{y\in \Ncal_{x_{i-1}}\cap\Ncal_{x_{i}}\colon \lambda_i \propto \kappa_{i-1} \propto \kappa_{i}\}\,,\\
	B_{i-1 i}&\coloneqq \{y\in \Ncal_{x_{i-1}}\cap\Ncal_{x_{i}}\colon\tilde\lambda_i\propto \tilde\kappa_{i-1}\propto\tilde\kappa_{i}\}\,.
\end{align}
From this definition it is further clear that any two points $y_1,y_2\in W_{i-1 i}$ satisfy $(y_1-y_2)^2=0$, and similarly for $B_{i-1 i}$. We continue the discussion of these white and black planes in appendix \ref{sec:APP_white-black}.

If we additionally consider a point $x_{i+1}$ such that $(x_i-x_{i+1})^2=0$, then it is clear that $W_{i-1i}$ and $W_{ii+1}$ generically only intersect at the point $x_i$, as anything more would imply that $\lambda_i\propto \lambda_{i+1}$ which would imply that $(x_{i-1}-x_{i+1})^2=0$, and this would impose additional restrictions on the points $x_{i-1}, x_i, x_{i+1}$. We instead find that $\Ncal_{x_{i-1}}\cap \Ncal_{x_i}\cap \Ncal_{x_{i+1}}$ consists of two null-rays given by $W_{i-1i}\cap B_{ii+1}$ and $B_{i-1 i}\cap W_{i i+1}$. Since these are the intersections of affine planes, they are necessarily straight lines in $\Rbb^{2,2}$, and since any two points in a white or black plane are null separated, this means the line is a null-ray. We label these rays
\begin{align}
	e_{i-1ii+1}^+&=W_{i-1i}\cap B_{ii+1}\,,\\
	e_{i-1ii+1}^-&=B_{i-1i}\cap W_{ii+1}\,,
\end{align}
which satisfy
\begin{align}
	e_{i-1ii+1}^+\cap \Ncal_j &= q^+_{i-1ii+1 j}\,,\\
	e_{i-1ii+1}^-\cap \Ncal_j &= q^-_{i-1ii+1 j}\,,
\end{align}
for any $x_j$ such that $X_{i-1j}>0, X_{ij}>0, X_{i+1j}>0$.

The local geometry is thus as follows: the vertex $x_i$ has four incident null-edges given by $e_{i-1 i i +1}^\pm$ and the two null-rays connecting $x_i$ to $x_{i-1}$ and $x_{i+1}$, four incident planes $W_{i-1 i}$, $B_{i-1 i}$, $W_{i i+1}$, $B_{i i+1}$, and three incident lightcones $\Ncal_{i-1}$, $\Ncal_{i}$, $\Ncal_{i+1}$.

The fact that all vertices $x_i,q^\pm_{abcd}$ have four incident edges is again reminiscent of a \emph{curvy simple geometry}. However, in contrast to simple polytopes, not all vertices have four incident facets, as the vertices $x_a$ only have three incident facets.

\subsubsection{Finding the Vertices of $\Delta(\bm{x})$}
Next, we turn to the question of determining which quadruple intersections are vertices of the geometry $\Delta(\bm{x})$, starting from some point $(\lambda,\tilde\lambda)\in\Mcal_{n,k}$. We can check whether a point $y$ is \emph{inside} $\Delta$ by verifying if the sign-flips of \eqref{eq:DUAL_mom-amp-sign-flip-loop} hold. However, when we consider a point $y$ on the boundary of $\Delta$, then some of the invariants will be zero, and we can't simply check the sign-flips. To remedy this, we consider a slight deviation of the point, which allows us to replace the zeroes in the sequence by small positive or negative numbers. If there is a way to uplift the invariants in such a way that \emph{all} the sequences are simultaneously satisfied, then the point $y$ is on the boundary of $\Delta(\bm{x})$. 

As an example, we take an arbitrary point $(\lambda,\tilde\lambda)\in\Mcal_{4,2}$, and we want to find out if $\ls_{13}=q^+_{1234}$ is a vertex of $\Delta_{4,2}(\bm{x})$. We define $\sgn_{ab}\coloneqq \sgn(\<ab\>(\ls_{13}-\ls_{ab})^2$). For any point in $\Mcal_{4,2}$ we have $\<ab\>>0$, and furthermore $(\ls_{13}-\ls_{24})^2<0$\footnote{This can easily be seen by going to momentum twistor variables and back: $(\ls_{13}-\ls_{24})^2=\<1324\>/(\<13\>\<24\>)=-\<1234\>/(\<13\>\<24\>)=-X_{13}\<12\>\<34\>/(\<13\>\<24\>)$. Since $X_{13}>0$, $\<ab\>>0$, we find that this expression is negative.}. We record all the sequences of \eqref{eq:DUAL_mom-amp-sign-flip-loop} in a matrix:
\begin{align}
	\begin{pmatrix}
		\sgn_{12} & \sgn_{13} & \sgn_{14}\\
		\sgn_{23} & \sgn_{24} & - \sgn_{21}\\
		\sgn_{34} & -\sgn_{31} & - \sgn_{32}\\
		-\sgn_{41} & -\sgn_{42} & -\sgn_{43}
	\end{pmatrix} = 
	\begin{pmatrix}
		0 & 0 & 0\\
		0 & -1 & 0\\
		0 & 0 & 0\\
		0 & -1 & 0
	\end{pmatrix}\,,
\end{align}
where the minus signs on the left hand side come from twisted cyclic symmetry. If $\ls_{13}$ is a vertex of the geometry, then there must be a way to resolve the zero elements of this matrix to make every row in this matrix have exactly two sign-flips, where we have to keep in mind that $\sgn_{ab}=-\sgn_{ba}$. In this case, we find that the resolution
\begin{align}
	\sgn_{12}\to 1\,, \sgn_{13}\to -1\,,\sgn_{14}\to 1\,,\sgn_{23}\to 1\,,\sgn_{24}\to -1\,, \sgn_{34}\to 1\,,
\end{align}
satisfies these conditions. Hence, we conclude that $\ls_{13}$ is indeed a vertex of $\Delta_{4,2}(\bm{x})$. The fact that the values of $\sgn_{ab}$ are not dependent on the choice of base point $(\lambda,\tilde\lambda)\in\Mcal_{4,2}$ is a consequence of the fact that there is only a single chamber for $\Mcal_{4,2}$: the combinatorial structure of $\Delta_{4,2}(\bm{x})$ is independent of the choice of base point, and both $\ls_{13}=q^+_{1234}$ and $\lstil_{13}=\ls_{24}=q^-_{1234}$ are vertices.

\subsubsection{From Vertices to Canonical Form}
Based on the `simple' structure of $\Delta_{n,k}(\bm{x})$ and the success for ABJM theory, we are encouraged to write the canonical form as a sum over vertices. For the vertices $q^\pm_{abcd}$ we proceed in complete analogy to the three-dimensional case studied in section \ref{sec:DUAL_ABJM}. We introduce
\begin{align}
	\omega_{abcd}=\dd\log(y-x_a)^2\wedge\dd\log(y-x_b)^2\wedge\dd\log(y-x_c)^2\wedge\dd\log(y-x_d)^2\,,
\end{align}
and to kill the contribution of $q^\mp_{abcd}$ we introduce
\begin{align}\label{eq:DUAL_omega-box-def-4D}
	\omega^\square_{abcd}&=\frac{16\sqrt{\det\Xcal_{abcd}}\dd^4y}{(y-x_a)^2(y-x_b)^2(y-x_c)^2(y-x_d)^2}\\
	&=\pm\dd\log\frac{(y-x_a)^2}{(y-q^\pm_{abcd})^2}\wedge\dd\log\frac{(y-x_b)^2}{(y-q^\pm_{abcd})^2}\wedge\dd\log\frac{(y-x_c)^2}{(y-q^\pm_{abcd})^2}\wedge\dd\log\frac{(y-x_d)^2}{(y-q^\pm_{abcd})^2}\,,
\end{align}
where $16\det\Xcal_{abcd}=2\big[ (X_{ab}X_{cd})^2+(X_{ac}X_{bd})^2+(X_{ad}X_{bc})^2 \big]-(X_{ab}X_{cd}+X_{ac}X_{bd}+X_{ad}X_{bc})^2$ is defined in appendix \ref{sec:APP_schubert}. We combine these forms into
\begin{align}
	\omega^\pm_{abcd}=\frac{\omega^\square_{abcd}\pm\omega_{abcd}}{2}\,,
\end{align}
which is the form we associate to the vertex $q^\pm_{abcd}$.

It is less clear what form we should associate to a vertex $x_i$, as it has only three incident facets. We now make the remarkable claim that we don't need to add contribution from these vertices \emph{at all}\footnote{This is analogous to the toy example studied in section \ref{sec:DUAL_toy}, as was observed around equation \eqref{eq:DUAL_2d-toy-1-loop-sum}. Similar to that example, the expansion in equation \eqref{eq:DUAL_nf-Delta-form} no longer has the interpretation as the sum over all vertices, but rather as a sum over all maximal intersections of null-cones.}. That is, we claim that the canonical form of $\Delta_{n,k}(\bm{x})$ can be written as
\begin{align}\label{eq:DUAL_nf-Delta-form}
	\Omega(\Delta_{n,k}(\bm{x})) = \sum_{q^\pm_{abcd}\in \Vcal(\Delta_{n,k}(\bm{x}))} \omega^\pm_{abcd}\,,
\end{align}
where $\Vcal(\Delta_{n,k}(\bm{x}))$ denotes the set of vertices of $\Delta_{n,k}(\bm{x})$. By construction, this form will have a residue of $\pm1$ at all vertices $q^\pm_{abcd}$ of the geometry. To see how this sum also gives rise to a residue of $\pm1$ at the vertices $x_i$, we make the following observations. First, we note that $\omega^\square_{i-1ii+1j}$ has a \emph{composite singularity} at $x_i$ (see appendix \ref{sec:APP_max-intersection-massive}). If we take a residue at $(y-x_{i-1})^2=(y-x_i)^2=(y-x_{i+1})^2=0$, then we are geometrically localising on either edge $e^\pm_{i-1ii+1}$, and we can subsequently localise on $x_i$.

Thus, we find that
\begin{align}
	\Res_{y=x_i}\omega^\square_{i-1ii+1j}=1\quad\implies\quad \Res_{y=x_i}\omega^\pm_{i-1ii+1j}=\frac{1}{2}\,.
\end{align}
Furthermore, we note that $\Delta_{n,k}$ always has exactly one vertex of the form $q^+_{i-1ii+1j}$, and one of the form $q^-_{i-1ii+1 k}$. We already know that the rays $e^\pm_{i-1ii+1}$ are part of the geometry. Whichever null-cone these rays intersect first will contribute a vertex to $\Delta(\bm{x})$. The forms $\omega^+_{i-1ii+1j}$ and $\omega^-_{i-1ii+1k}$ are the only forms in the sum \eqref{eq:DUAL_nf-Delta-form} which contribute a residue to $x_i$, and hence we conclude that 
\begin{align}
	\Res_{y=x_i}\Omega(\Delta(\bm{x}))=1\,.
\end{align}
We will give further evidence for the correctness of equation \eqref{eq:DUAL_nf-Delta-form} in section \ref{sec:DUAL_nf-chambers-integrand}, where we show that it reproduces the correct one-loop integrand in planar \nf.

The relative signs in \eqref{eq:DUAL_nf-Delta-form} have again been fixed by requiring projective invariance. In analogy to the ABJM case, we can once again re-sum all $\omega_{abcd}$ into manifestly projectively invariant objects. We define
\begin{align}
	\omega^{\pentagon}_{abcde}&=\dd\log\frac{(y-x_a)^2}{(y-x_e)^2}\wedge\dd\log\frac{(y-x_b)^2}{(y-x_e)^2}\wedge\dd\log\frac{(y-x_c)^2}{(y-x_e)^2}\wedge\dd\log\frac{(y-x_d)^2}{(y-x_e)^2}\\
	&=\frac{32\epsilon(a,b,c,d,e,y)\dd^4 y}{(y-x_a)^2(y-x_b)^2(y-x_c)^2(y-x_d)^2(y-x_e)^2}\\
	&=\omega_{abcd}-\omega_{abce}+\omega_{abde}-\omega_{acde}+\omega_{bcde}\,.
\end{align}
Making the substitution $\omega_{abcd}\to\omega^{\pentagon}_{1abcd}$ leave the canonical form $\Omega(\Delta_{n,k}(\bm{x})) $ invariant, and it renders it manifestly projectively invariant. Furthermore, since all propagators in the $\dd\log$ form $\omega^{\pentagon}$ are local poles, it must integrate to zero \cite{Herrmann:2019upk}. Hence, only the $\omega^\square$ have a non-zero contribution to the integrated answer, the $\omega$ and $\omega^{\pentagon}$ are only present to ensure that the integrand has the appropriate residues.

\subsection{Boundary Diagrams and Positroid Cells}

In this section we develop a graphical representation for boundaries of $\Delta_{n,k}(\bm{x})$. These diagrams are, in some sense, two-dimensional projections of points and null-rays of $\Delta_{n,k}(\bm{x})$. The starting point of these diagrams is an $n$-gon with corners labelled clockwise 1 through $n$, this represents a projection of the null-polygon $x_1,\ldots,x_n$. A marked point inside this $n$-gon represents a point $y\in\Delta_{n,k}(\bm{x})$. We further add edges between the marked point and a corner $i$ to indicate that $(y-x_i)^2=0$, which we shall refer to as `cuts'. Since $\Delta_{n,k}(\bm{x})$ is a four dimensional geometry, we cannot connect more than four edges to the marked point, since five or more null-cones generically don't intersect. When two or more cuts are present, the $n$-gon is subdivided into smaller polygons $P_a$. Next, we decorate each (sub) $m$-gon $P$ with an integer $h$ which can take values $h=2,3,\ldots,m-2$, which we call the \emph{helicity}. At this stage there is no physics motivation for this, as it is purely a statement about topologically distinct `types' of null-polygons in $\Rbb^{2,2}$. The exception to the helicity assignment is for $m=3$: triangles in the boundary diagram can have helicity equal to one or two, corresponding to white or black planes, respectively. From our discussion in the previous section, it is clear that it is not possible to make cuts which isolate two adjacent triangles with the same helicity, as this would impose additional constraint on the $x_i$'s. If the boundary diagram is divided into polygons $P_1,\ldots,P_q$ with helicities $h_1,\ldots,h_q$, and $r$ cuts (either $r=q$ or $r=0$), then we define the helicity of the total diagram to be $h_1+\ldots+ h_q-r$. A boundary diagram with $n$ corners and helicity $k$ is said to be of \emph{type $(n,k)$}. Since each cut imposes one constraint on $y$, the dimension of a boundary diagram is equal to $4-r$. We further introduce two special types of boundary diagrams. The one-dimensional configuration where $y$ is on the null-polygon between $x_i$ and $x_{i+1}$ is denoted by a marking the edge between corners $i$ and $i+1$, and the zero-dimensional case where $y$ is localised on $x_i$ is represented by a boundary diagram with a marked point on the corner $i$. 

We conjecture that every boundary of $\Delta_{n,k}(\bm{x})$ can be labelled by a boundary diagram of type $(n,k)$. As an example, we list all the boundaries of $\Delta_{4,2}(\bm{x})$ in table \ref{tab:n4-boundaries}. We see that the boundaries are in a bijection to boundaries of $G_+(2,4)$, as is expected from the definition of $\Mcal_{4,2}^{(1)}$.

\begin{table}[t]
	\begin{center}	\begin{tabular}{|c| c c c|} 
			\hline
			$d=4$ & $\begin{gathered}
				\includegraphics[ height=2.2cm]{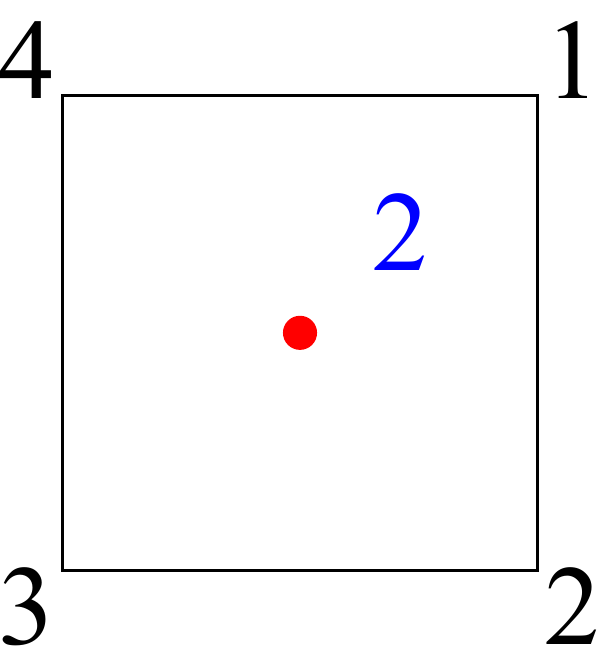} \end{gathered}\times 1$ & & \\ 
			\hline
			$d=3$ & $\begin{gathered}
				\includegraphics[ height=2.2cm]{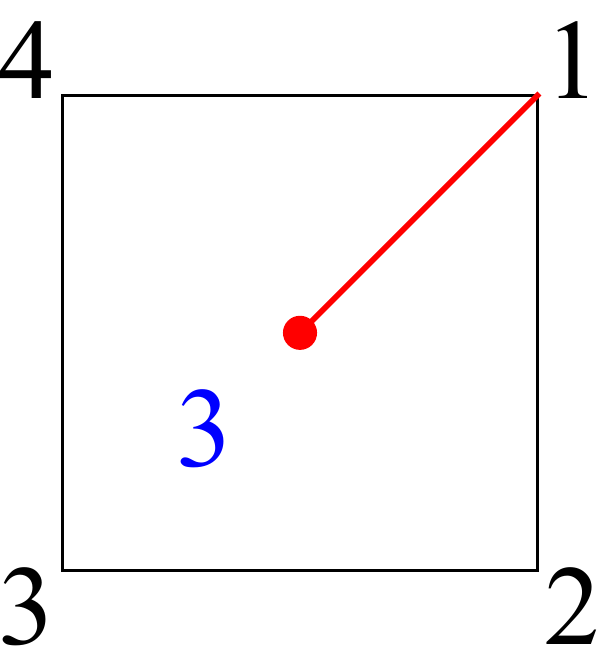} \end{gathered}\times 4$ & & \\
			\hline
			$d=2$ & $\begin{gathered}
				\includegraphics[ height=2.2cm]{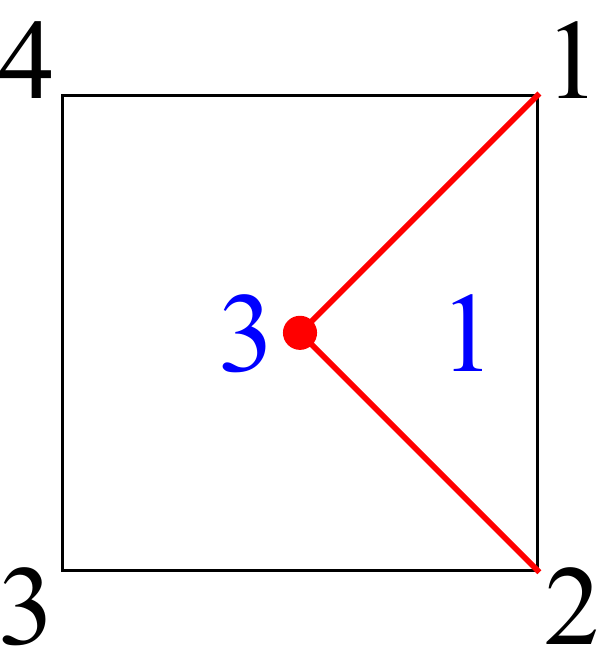} \end{gathered}\times 4$ & $\begin{gathered}
				\includegraphics[ height=2.2cm]{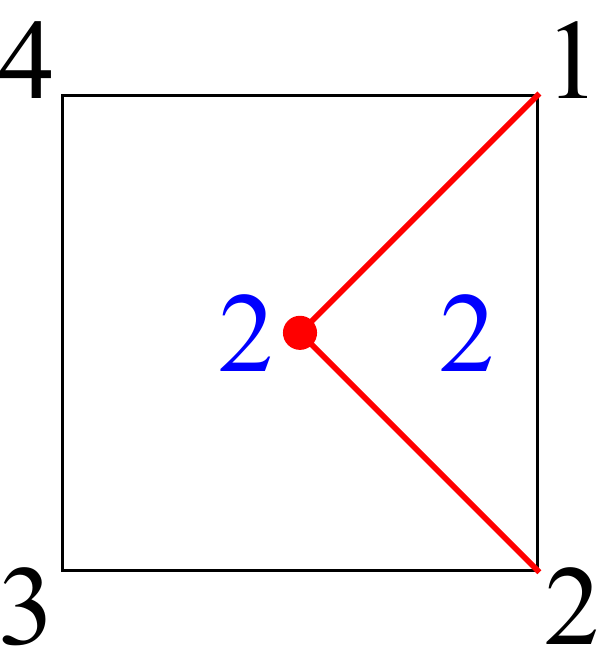} \end{gathered}\times 4$ & $\begin{gathered}
				\includegraphics[ height=2.2cm]{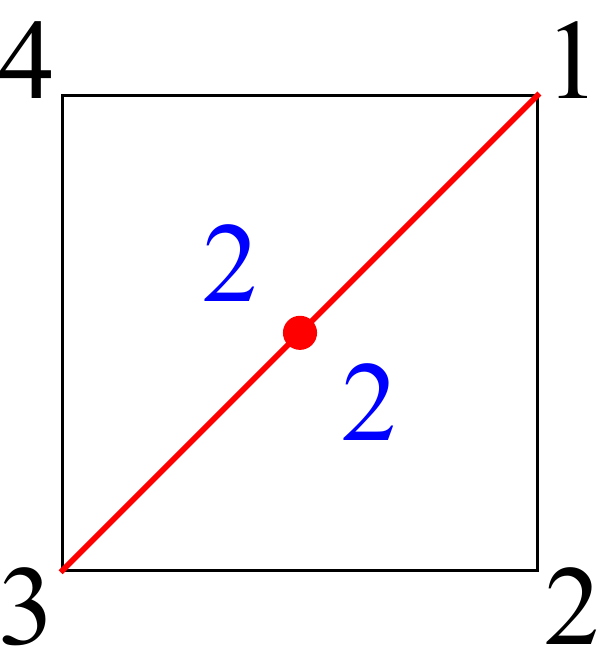} \end{gathered}\times 2$ \\
			\hline
			$d=1$ & $\begin{gathered}
				\includegraphics[ height=2.2cm]{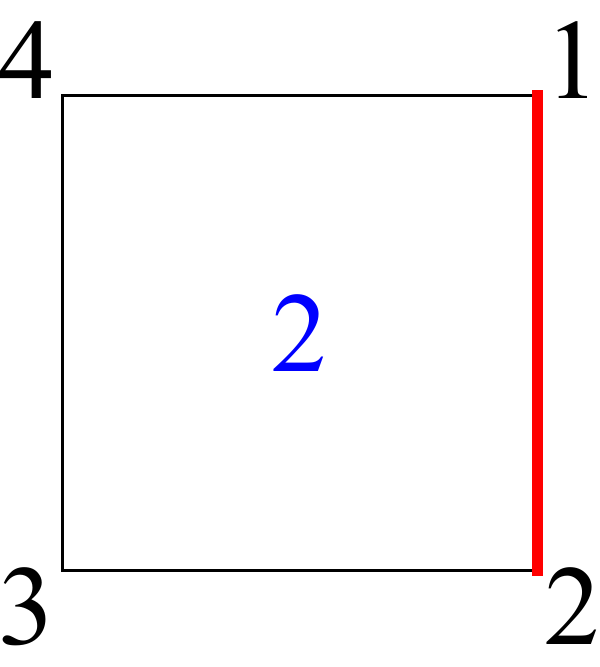} \end{gathered}\times 4$ & $\begin{gathered}
				\includegraphics[ height=2.2cm]{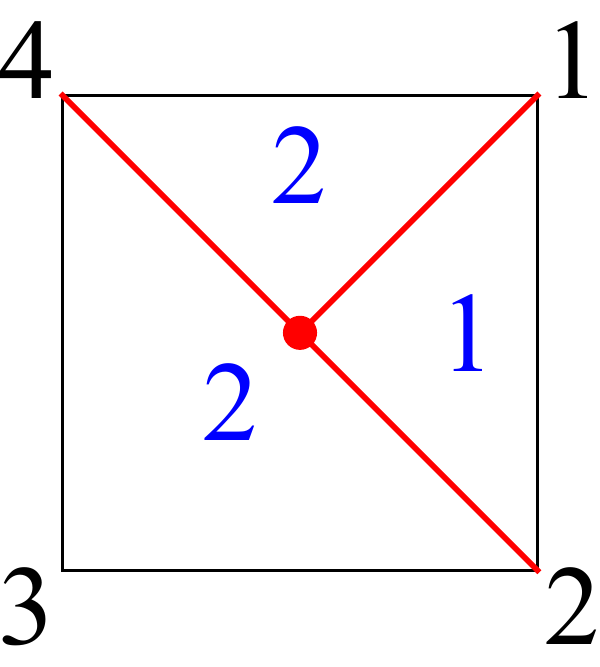} \end{gathered}\times 8$ & \\
			\hline
			$d=0$ &$\begin{gathered}
				\includegraphics[ height=2.2cm]{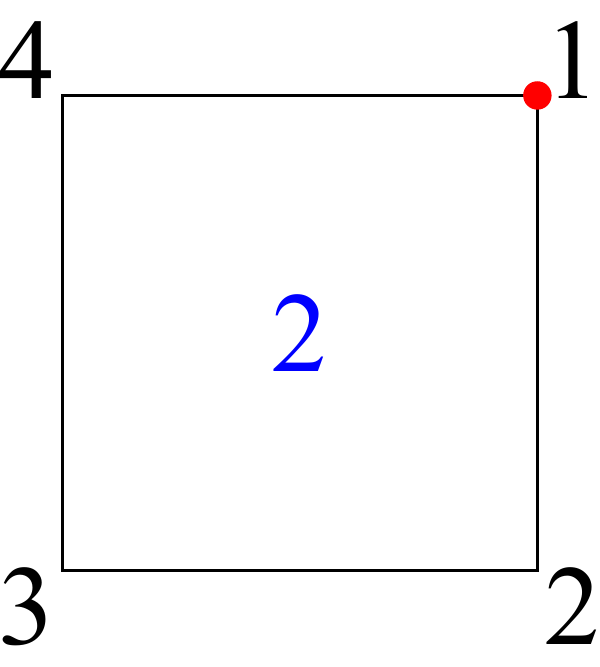} \end{gathered}\times 4$ & $\begin{gathered}		\includegraphics[ height=2.2cm]{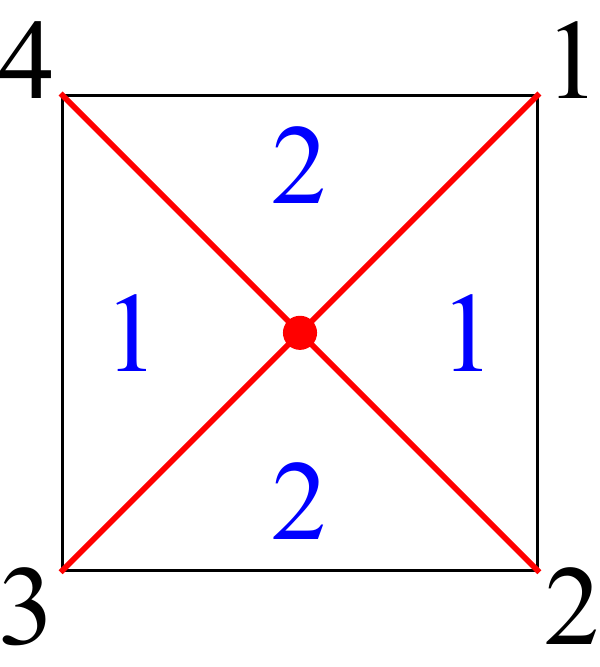} \end{gathered}\times 2$ & \\ [1ex] 
			\hline
		\end{tabular}
	\end{center}
	\caption{Boundary diagrams associated to all boundaries of $\Delta_{4,2}$. We only show dihedral representatives of each boundary, the `$\times n$' indicates the number of boundaries of this type.}
	\label{tab:n4-boundaries}
\end{table}

\subsubsection{Quadruple Cut Diagrams}
\begin{figure}[t]
	\centering
	\includegraphics[width=0.8\textwidth]{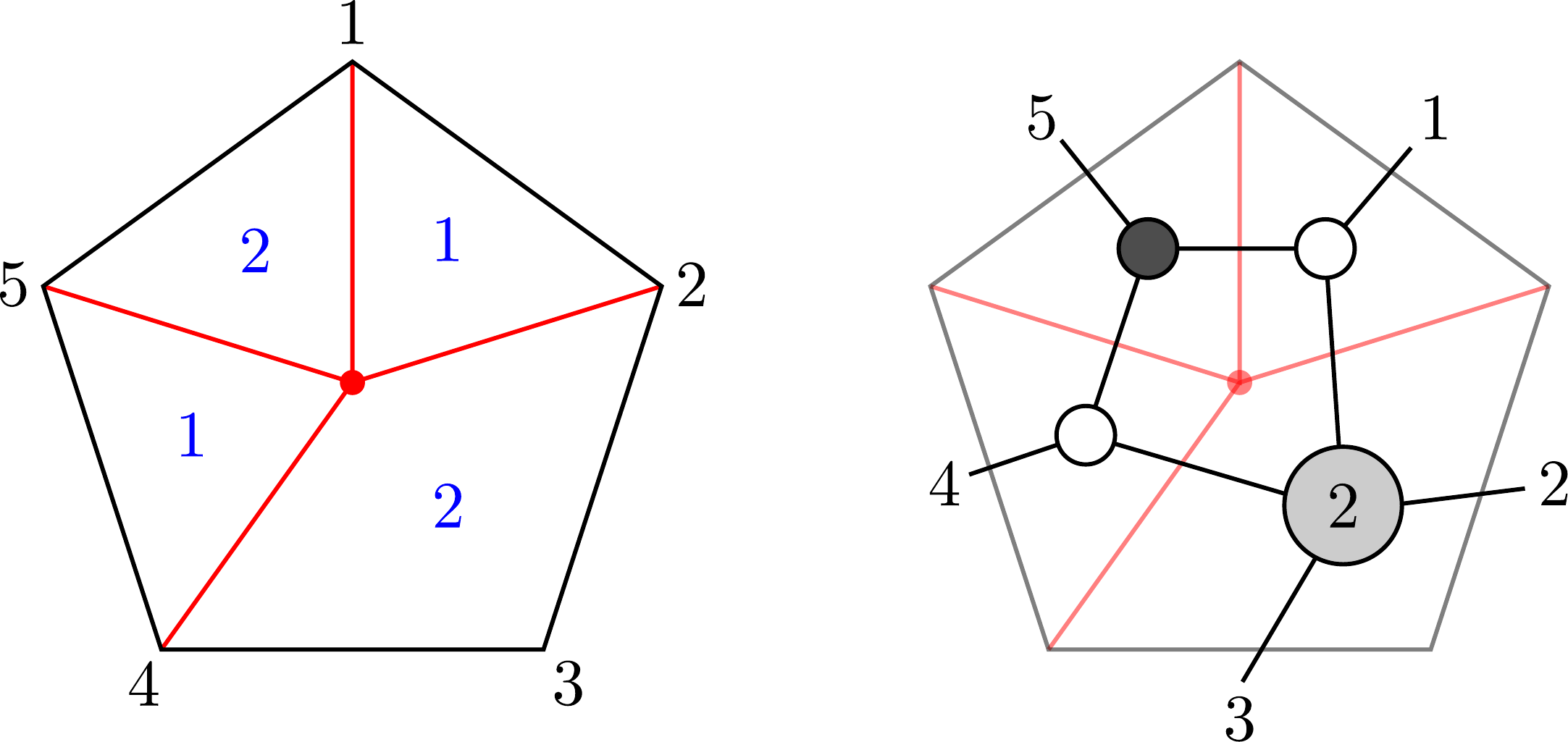}
	\caption{An illustration of the duality between boundary diagrams corresponding to quadruple cuts, and on-shell diagrams for leading singularities.}
	\label{fig:DUAL_boundary-diagram-leading-singularity}
\end{figure}

The boundary diagrams corresponding to quadruple intersections of null-cones are of a particular interest. They label the non-trivial vertices of $\Delta_{n,k}(\bm{x})$, and, furthermore, they correspond to \emph{leading singularities} of the one-loop integrand, as the vertices $q^\pm_{abcd}$ correspond precisely to the kinematic configuration where the loop variable $y$ has been cut four times. We briefly encountered leading singularities in section \ref{sec:AMP_nf}, where we argued that leading singularities of the one-loop integrand are interpreted as on-shell diagrams consisting of four tree-level amplitudes. This is also clear from our boundary diagrams, as the quadruple cuts divide the $n$-gon into four smaller null-polygons. These boundary diagrams are precisely \emph{dual} to the on-shell diagram corresponding to the leading singularity, this is illustrated in figure \ref{fig:DUAL_boundary-diagram-leading-singularity}.

\begin{figure}[t]
	\centering
	\includegraphics[height=5cm]{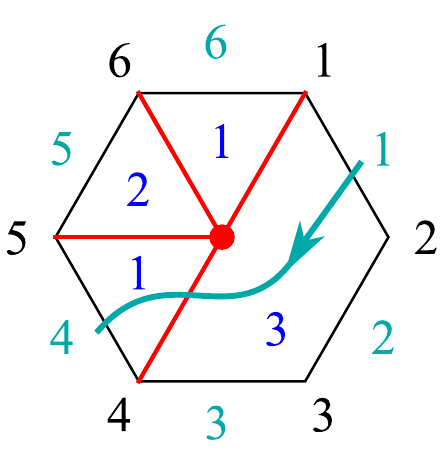}
	\caption{An illustration of the `rules of the road' which shows that this quadruple cut boundary diagram has an associated permutation with $\sigma(1)=4$. The full permutation associated to this diagram is $\{4, 5, 1, 6, 2, 3\}$, or, decorated, $\{4,5,7,6,8,9\}$.}
	\label{fig:diag-path}
\end{figure}

We further recall that each on-shell diagram has a corresponding positroid cell. The permutation labelling the positroid cell corresponding to a quadruple cut boundary diagram can be read off by following the `rules of the road': we find $\sigma(i)$ by following a path starting at the edge between corners $i$ and $i+1$. If we enter a (sub-)polygon with helicity $h$, then we take the $h$\textsuperscript{th} clockwise exit. We repeat this until we exit the boundary diagram between edges $j$ and $j+1$, in which case $\sigma(i)=j$. This is illustrated in figure \ref{fig:diag-path}.

Let us assume that we have two quadruple cut diagrams corresponding to vertices $q^\pm_{abcd}$ and $q^\pm_{ijkl}$ which have the same associated positroid cell. Any two maximal cuts with the same permutation have the same leading singularity. This means that the canonical form of the one-loop momentum amplituhedron will have the same residue at these vertices:
\begin{align}
	\Res_{y=q^\pm_{abcd}} \Omega(\Mcal_{n,k}^{(1)}) = \Res_{y=q^\pm_{ijkl}} \Omega(\Mcal_{n,k}^{(1)})\,.
\end{align}

\subsection{\texorpdfstring{Chambers of $\Mcal_{n,k}$ and the Full One-Loop Integrand}{Chambers of M(n,k) and the Full One-Loop Integrand}}\label{sec:DUAL_nf-chambers-integrand}

We recall that the tree-level momentum amplituhedron $\Mcal_{n,k}$ is the image of $G_+(k,n)$ under the map \eqref{eq:POS_mom-amp-def}. We now make a non-trivial statement regarding the image of cells $C\in G_+(k,n)$ corresponding to the leading singularities/quadruple cuts discussed in the previous section. In all cases we studied, we found that
\begin{align}\label{eq:DUAL_positroid-map-vertices}
	q^\pm_{abcd} \sim C \quad \Leftrightarrow \quad q^\pm_{abcd} \in \Vcal\big( \Delta(\Phi(C)) \big)\,,
\end{align} 
where $q^\pm_{abcd} \sim C$ denotes that $C$ associates to $q^\pm_{abcd}$, and $\Phi(C)$ should be understood as the momentum amplituhedron map into spinor-helicity space composed with the map to dual space. In different words, if $C$ is in a positroid cell associated to the quad cut $q^\pm_{abcd}$, then $\smash{\Phi_{\Lambda,\tilde\Lambda}(C)}$ gives points $\bm{x}$ in dual space such that $\Delta(\bm{x})$ has the vertex $q^\pm_{abcd}$. 

Furthermore, we found in section \ref{sec:DUAL_nullcone-geometry-4D} that, similar to the ABJM case, the canonical form of $\Delta(\bm{x})$ is completely determined by the set of vertices $\Vcal(\Delta(\bm{x}))$. In particular, this means that the one-loop chambers $\cfrak\in\Cfrak(\Mcal_{n,k})$ are also uniquely determined by the vertices of $\Delta(\bm{x})$. Combined with the observation \eqref{eq:DUAL_positroid-map-vertices}, this implies that the chambers of $\Mcal_{n,k}$ are the maximal intersections of the image of positroid cells corresponding to quadruple cuts. This is a refinement of the statement given in section \ref{sec:DUAL_chambers}, where we encountered a definition of chambers as the maximal intersection of BCFW tiles. 

Since the images of positroid cells are determined by the sign of certain `functionaries', we can study whether the image of quad cut positroid cells intersect by checking if their inequalities are compatible or not. We have used this method to find all chambers of $\Mcal_{n,3}$ for $n\leq 10$. This method is summarised in more detail in appendix \ref{sec:APP_chambers}.

From \eqref{eq:POS_lambda-def-mom-amp} it is clear that any $C\in G_+(k,n)$ maps to $(\lambda,\tilde\lambda)$ which satisfy $\lambda\subseteq C\,,\tilde\lambda\subseteq C^\perp$. This means that we can check whether a point $(\lambda,\tilde\lambda)$ is in the image of some positroid cell $S_\sigma$ by checking if
\begin{align}\label{eq:DUAL_lambda-in-C}
	C^\perp_\sigma(\bm{\alpha})\cdot \lambda = \nul\,,\quad C_\sigma(\bm{\alpha})\cdot\tilde\lambda=\nul\,,
\end{align}
has any solutions with all $\alpha\geq0$. Here, $C_\sigma(\bm{\alpha})$ denotes a positive parametrisation of matrices in the positroid cell $S_\sigma$, which can be generated efficiently using the \texttt{Mathematica} package \texttt{positroids}\cite{Bourjaily:2012gy}. This gives us a way of finding which chamber a given point $(\lambda,\tilde\lambda)$ is in: we simply find all positroid cells which admit a positive solution to \eqref{eq:DUAL_lambda-in-C}. The list of positroid cells which satisfy this condition specify the chamber. This additionally gives us a new way of finding the vertices (and hence the canonical form) of $\Delta(\bm{x})$. Once the set of positroid cells defining the chamber is known, we can list all quadruple cuts corresponding to these positroid cells. The union of all these vertices (combined with the vertices $x_i$) gives the vertex set $\Vcal(\Delta(\bm{x}))$. We have checked for MHV (all $n$), NMHV (up to $n=9$) and N\textsuperscript{2}MHV (a large number of randomly generated points for $n=8$) that this way of finding vertices is equivalent to the method discussed in section \ref{sec:DUAL_nullcone-geometry-4D}.

\begin{figure}
	\centering
	\includegraphics[width=0.7\textwidth]{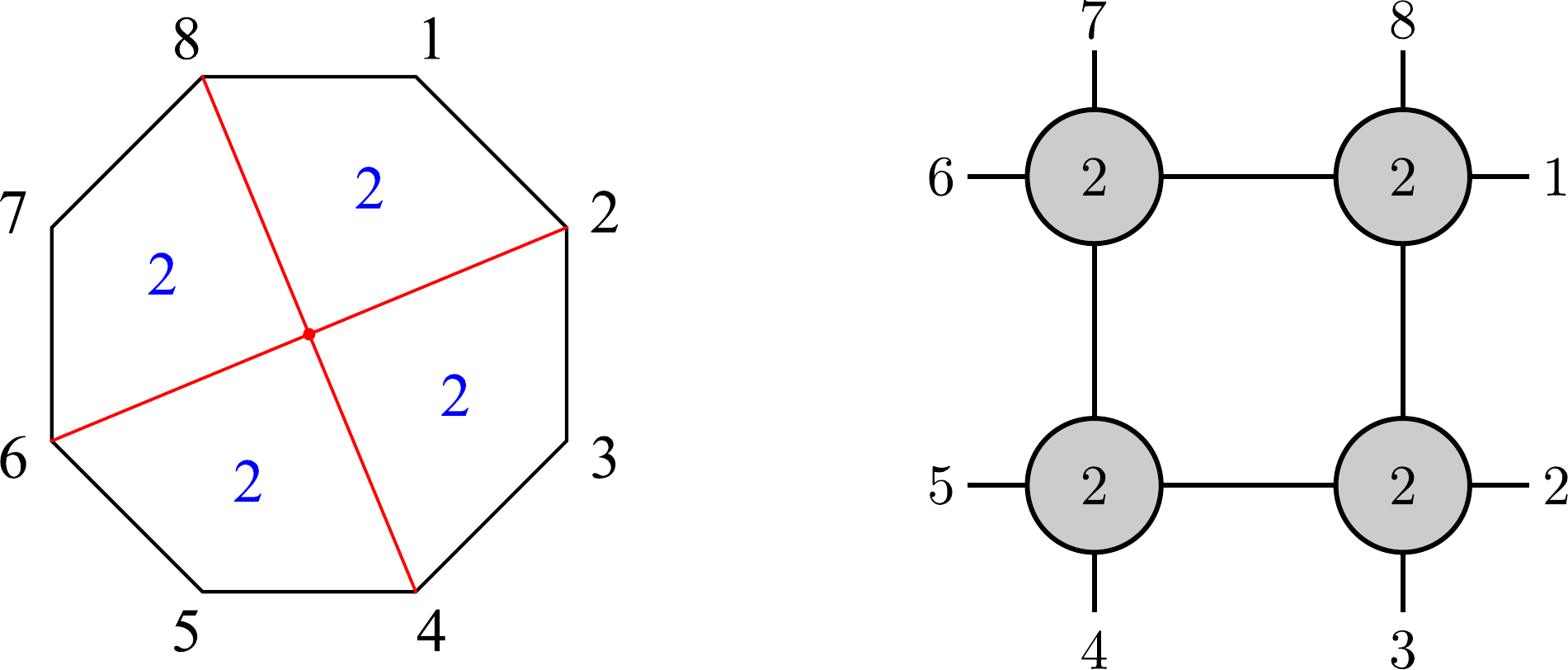}
	\caption{The boundary diagram (left) and Grassmannian graph (right) corresponding to the four-mass-box positroid cell with permutation $\{6, 5, 8, 7, 10, 9, 12, 11\}$. Its rotation, with permutation $\{4, 7, 6, 9, 8, 11, 10, 13\}$, labels the other four-mass-box of $G_+(4,8)$.}
	\label{fig:DUAL_four-mass-box}
\end{figure}

There is one notable complication to the above discussion, which regards the famous \emph{four-mass-box positroid cells}. These are a positroid cells in $G_+(4,8)$ which are associated to a specific type of quadruple cut. We depict one of the four-mass-box quad cut diagrams and its corresponding Grassmannian graph in figure \ref{fig:DUAL_four-mass-box}.
These positroid cells are the first example of $2n-4$ dimensional cells which do not map injectively into the momentum amplituhedron, and is instead said to have `intersection number' two. That is, although both the positroid cell and its image have the same dimension, there are families of two points in the positroid cell that map to the same point in the momentum amplituhedron. Furthermore, due to the absence of white/black planes in \ref{fig:DUAL_four-mass-box}, we do not know a priori whether to associate this diagram to $q^+_{2468}$ or $q^-_{2468}$. That is, for $\bm{x}=\Phi(C)$ for some $C\in S_{\{6, 5, 8, 7, 10, 9, 12, 11\}}$, it is not clear if $q^+_{2468}\in\Vcal(\Delta(\bm{x}))$, or $q^-_{2468}\in\Vcal(\Delta(\bm{x}))$. Using the method to determine vertices explained in section \ref{sec:DUAL_nullcone-geometry-4D}, we have done extensive numerical tests to find that there are three possibilities: either $q^+_{2468}$, or $q^-_{2468}$, or both $q^+_{2468}$ and $q^-_{2468}$ are vertices of $\Delta(\bm{x})$. This suggests that we should `split up' the four-mass-box positroid cell into two smaller cells, one of which has an image in the momentum amplituhedron with $q^+_{2468}\in\Vcal(\Delta(\bm{x}))$, and the other with $q^-_{2468}\in\Vcal(\Delta(\bm{x}))$, in such a way that the images of these two sub-cells partially overlap. How to properly define this division into sub-cells is still an open question. It is worth mentioning that there are also positroid cells with intersection numbers larger than two. We suspect that these do not contribute any vertices to the geometry. In particular, we have checked that the cells $\{10,8,12,7,11,9,16,14,18,13,17,15\}$ and $\{11,5,16,10,15,9,20,14,19,13,24,18,23,17,28,22\}$ with intersection number four do not have any corresponding quadruple cuts.

\subsubsection{The Full One-Loop Integrand}
Let $\Vcal(\sigma)$ denote the set of all quadruple cuts associated to a positroid cell $S_\sigma$, and $\Sigma(q^\pm_{abcd})$ be the positroid cell associated to this quadruple cut. In the previous section we have argued that all chambers of $\Mcal_{n,k}$ can be obtained as the maximal intersection $\cfrak =\Phi(\sigma_1)\cap\cdots\cap\Phi(\sigma_m)\in \Cfrak(\Mcal_{n,k})$ for some set of positroid cells $S_{\sigma_1},\ldots,S_{\sigma_m}$ with $\Vcal(\sigma_i)\neq\emptyset$. We will denote the vertex set of $\Delta(\bm{x})$ corresponding to points $(\lambda,\tilde\lambda)$ in this chamber as $\Vcal(\sigma_1\cap\cdots\cap\sigma_m)$, which is then found as
\begin{align}
	\Vcal(\sigma_1\cap\cdots\cap\sigma_m)=\Vcal(\sigma_1)\cup\cdots\cup\Vcal(\sigma_m)\,.
\end{align}
The canonical form of the $\Delta(\bm{x}$) corresponding to this chamber is given as
\begin{align}
	\Omega^{1-\text{loop}}(\cfrak) = \sum_{q^\pm_{abcd}\in \Vcal(\sigma_1)\cup\cdots\cup\Vcal(\sigma_m)} \omega^\pm_{abcd}\,,
\end{align}
which we can subsequently substitute into the expression for the full one-loop integrand 
\begin{align}
	\Omega(\Mcal_{n,k}^{(1)})=\sum_{\cfrak\in\Cfrak(\Mcal_{n,k})} \Omega^{\text{tree}}(\cfrak)\wedge\Omega^{1-\text{loop}}(\cfrak)\,.
\end{align}
The coefficient of any particular $\omega^\pm_{abcd}$ is the sum of all $\Omega^{\text{tree}}(\cfrak)$ where $q^\pm_{abcd}\in\Vcal(\cfrak)$. Thus, their union is precisely $\Phi(\Sigma(q^\pm_{abcd}))$, and the coefficient of $\omega^\pm_{abcd}$ is therefore exactly the leading singularity associated to this quadruple cut. This leading singularity can be found explicitly by calculating the on-shell function of the associated Grassmannian graph, or equivalently by finding the canonical form $\Omega\big( \Phi(\Sigma(q^\pm_{abcd})) \big)$. 

Thus, if we let $\Qcal_{n,k}$ denote the set of all positroid cells with associated quadruple cut diagrams of type $(n,k)$, and we recall that $\Omega(\Phi(\sigma))$ is the leading singularity associated to the positroid cell $\sigma$, then the above expression can be re-summed as
\begin{align}\label{eq:DUAL_nf-full-one-loop}
	\Omega(\Mcal_{n,k}^{(1)})=\sum_{\sigma\in\Qcal_{n,k}} \sum_{q^\pm_{abcd}\in\Vcal(\sigma)} \Omega(\Phi(\sigma))\wedge \omega^\pm_{abcd}\,. 
\end{align} 
This shows that we can derive an explicit formula for the one-loop integrand without needing to know all the chambers explicitly! Furthermore, since the only part of $\omega^\pm$ which doesn't integrate to zero is $\omega^\square$, this formula is equivalent to the well-known statement that the one-loop integrand of \nf can be written as a sum of a leading singularity times a box integrand over all possible cuts, which we encountered in equation \eqref{eq:AMP_1-loop-nf}.

\subsection{Examples}\label{sec:DUAL_examples}
In this section we study some explicit examples of the geometry $\Delta_{n,k}(\bm{x})$ and their canonical forms.

\subsubsection{MHV}
\begin{figure}
	\centering
	\includegraphics[height=4.5cm]{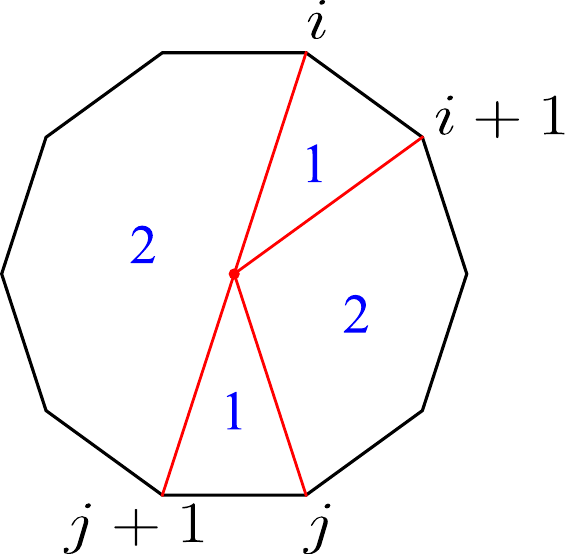}
	\caption{All quadruple cut diagrams of type $(n,2)$. They correspond to vertices $q^+_{i\,i+1\,j\,j+1}=\ls_{ij}$, and the associated positroid cell is the top cell $G_+(2,n)$.}
	\label{fig:diag-MHV}
\end{figure}

\begin{figure}
	\centering
	\includegraphics[width=\textwidth]{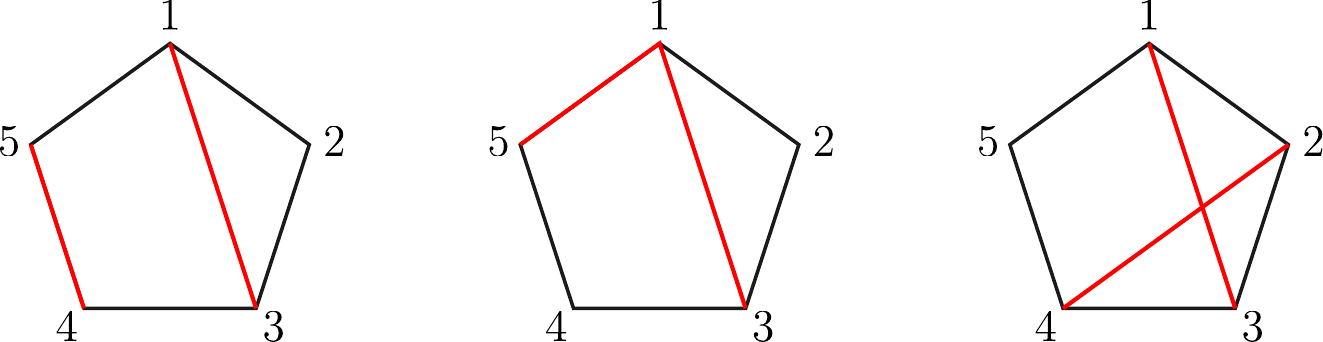}
	\caption{These pentagons show that the vertices of $\Delta_{5,2}(\bm{x})$ satisfy $(x_5-\ls_{13})^2>0$ (left), $(x_1-\ls_{13})^2=0$ (middle), $(\ls_{13}-\ls_{24})^2<0$ (right).}
	\label{fig:lstar-ngons}
\end{figure}

The discussion for $\Mcal_{n,2}$ is remarkably simple. This is because there is only a \emph{single} chamber for the MHV momentum amplituhedron. This can be seen from the fact that $\dim(\Mcal_{n,2})=\dim(G_+(2,n))=2n-4$. Thus, there are no multiple positroid cells whose image can intersect in the momentum amplituhedron. Furthermore, the only quadruple cut diagrams of type $(n,2)$ are those corresponding to $q^+_{ii+1jj+1}$, and the associated positroid cell of these diagrams is always the top cell $G_+(2,n)$. We have depicted these quad cut diagrams in figure \ref{fig:diag-MHV}.
The fact that there is only a single chamber is also reflected by the fact that the signs in the sequences \eqref{eq:DUAL_mom-amp-sign-flip-loop} are completely determined. In the MHV case with $a<b<c<d$, the brackets $\<ab\>$ are positive, and, as argued in section \ref{sec:POS_mom-amp-amp-connection}, also $\<abcd\>$ are positive. Since $(\ls_{ab}-\ls_{cd})^2=\<abcd\>/(\<ab\>\<cd\>)$ the signs of differences of $\ls$s are determined by the following simple rule: if the chords $(ab)$ and $(cd)$ of an $n$-gon intersect/don't intersect/share a vertex, then $(\ls_{ab}-\ls_{cd})^2$ is negative/positive/zero. This is summarised in figure \ref{fig:lstar-ngons}.

We further remark that the definition of $\Delta_{n,2}(\bm{x})$ as the image of the map
\begin{align}
	y = \frac{\sum_{i<j}(ij)\<ij\>\ls_{ij}}{\sum_{i<j}(ij)\<ij\>}\,,
\end{align}
has the `weights' $(ij)\<ij\>$ manifestly positive. This formula looks remarkably similar to the definition of the convex hull of these points, which we encountered in section \ref{sec:GRASS_proj_poly}. The difference being that the weights satisfy some non-trivial relations coming from the Schouten identity. It is precisely these identities which make the facets of this `convex hull' curvy, but other than that we find that $\Delta_{n,2}$ is geometrically very similar to the convex hull of the vertices $\ls_{ij}$.

We find that the canonical form of $\Omega_{n,2}(\bm{x})$ is given by
\begin{align}
	\Omega(\Delta_{n,2}(\bm{x})) = \sum_{i<j}\omega^+_{ii+1jj+1} = \frac{1}{2} \sum_{i<j}\omega^\square_{ii+1jj+1}+ \frac{1}{2}\sum_{1<i<j} \omega^{\pentagon}_{1ii+1jj+1}\,,
\end{align}
and the full one-loop integrand is
\begin{align}
	\Omega(\Mcal^{(1)}_{n,2})=\Omega(\Mcal_{n,2})\wedge\Omega(\Delta_{n,2}(\bm{x}))\,.
\end{align}
The only exception is for $n=3$, where the only vertices are $x_1,x_2,x_3$ which all lie on a black plane. The geometry $\Delta_{3,2}(\bm{x})$ is just a triangle living on this plane, and has canonical form
\begin{align}
	\Omega(\Delta_{3,2}(\bm{x}))=\dd\log\frac{\<\ell 1\>}{\<\ell 3\>}\wedge\dd\log\frac{\<\ell 2\>}{\<\ell 3\>}\,.
\end{align}
Since this is only a two-form, rather than a four-form, it will integrate to zero over any four-dimensional contour. This reflects the fact that the three-point amplitude is perturbatively exact at tree level, and does not have any loop corrections.

To give some explicit examples, here are the first few $\Delta_{n,2}$ worked out in detail. For $n=4$, we have two quad cuts: $\ls_{13}=q^+_{1234}$ and $\lstil_{13}=\ls_{24}=q^-_{1234}$, with a canonical form
\begin{align}
	\Omega(\Delta_{4,2})=\omega_{1234}^++\omega_{1234}^-=\omega^\square_{1234}\,.
\end{align}
For $n=5$ we have vertices 
\begin{align}
	\Vcal(\Delta_{5,2})=\{x_i,q^+_{1234},q^+_{1245},q^+_{2345},q^+_{2351}=q^-_{1235}, q^+_{3451}=q^-_{1345}\}\,,
\end{align}
and the canonical form
\begin{align}
	\Omega(\Delta_{5,2}) &= \omega^+_{1234}+\omega^+_{1245}+\omega^+_{2345}+\omega^-_{1235}+\omega^-_{1345}\\
	&=\frac{1}{2}\left( \omega^\square_{1234}+\omega^\square_{1245}+\omega^\square_{2345}+\omega^\square_{1235}+\omega^\square_{1345} + \omega^{\pentagon}_{12345}\right) \,.
\end{align}
For $n=6$ we find the vertex set
\begin{align}
	\Vcal(\Delta_{6,2})=\{x_i,q^+_{1234},q^+_{1245},q^+_{1256},q^+_{2345}, q^+_{2356},q^-_{1236}, q^+_{3456}, q^-_{1346},q^-_{1456}\}\,,
\end{align}
which gives the canonical form
\begin{align}
	\Omega(\Delta_{6,2})=\sum_{i<j}\omega^+_{ii+1jj+1}= \frac{1}{2}\sum_{i<j}\omega^\square_{ii+1jj+1}+\frac{1}{2}\left( \omega^{\pentagon}_{12346}+\omega^{\pentagon}_{12456}+\omega^{\pentagon}_{23456} \right)\,.
\end{align}

\paragraph{\MHVbar.}
Results for \MHVbar ($k=n-2$) can be found from the MHV results by a simple `parity conjugation' which swaps $q^+_{abcd}\leftrightarrow q^-_{abcd}$. The vertices of $\Delta_{n,n-2}$ are the set of $\lstil_{ij}=q^-_{ii+1jj+1}$, and the canonical form is given by 
\begin{align}
	\Omega(\Delta_{n,n-2}(\bm{x})) = \sum_{i<j}\omega^-_{ii+1jj+1} = \frac{1}{2} \sum_{i<j}\omega^\square_{ii+1jj+1}- \frac{1}{2}\sum_{1<i<j} \omega^{\pentagon}_{1ii+1jj+1}\,.
\end{align}

\subsubsection{NMHV}

\begin{figure}
	\centering
	\includegraphics[width=\textwidth]{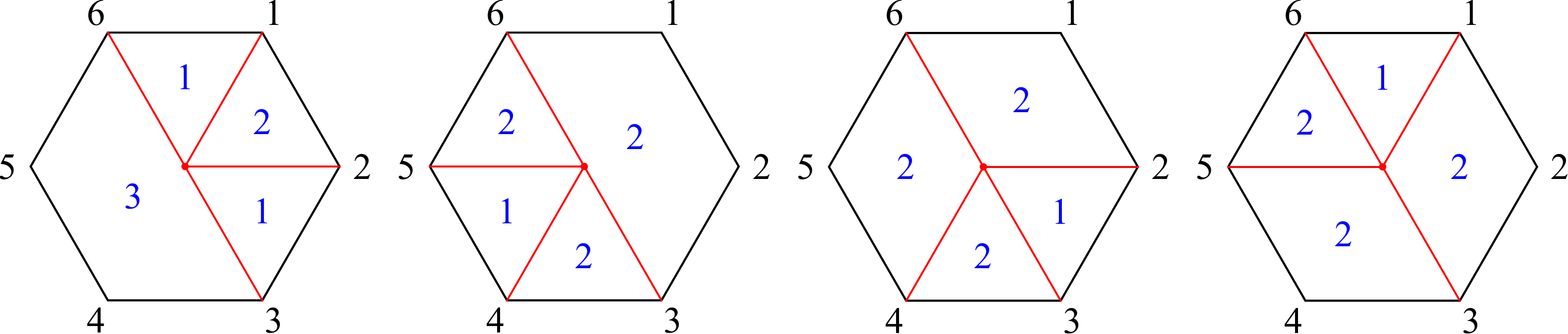}
	\caption{The four quadruple cuts associated to the positroid cell $(1)$. From left to right, the corresponding vertices are $q^-_{1236}$, $q^-_{3456}$, $q^+_{2346}$, and $q^-_{1356}$.}
	\label{fig:DUAL_NMHV-6-quad-cuts-1}
\end{figure}

\paragraph{NMHV\textsubscript{6}.}
The first non-trivial case is $n=6$, $k=3$. We briefly touched on this case in section \ref{sec:DUAL_chambers}. We will repeat the main points here. There are six different positroid cells of dimension $8$ whose image in the momentum amplituhedron is also 8-dimensional. These are precisely the six BCFW cells. Their T-dual are 4-dimensional positroid cells of $G_+(1,6)$, and their representative matrices have one zero element. For convenience, we choose to label the BCFW cells by the zero element $(i)$ of their T-dual, and we denote the associated tile as $[i]=\Phi_{\Lambda,\tilde\Lambda}\big((i)\big)$. The leading singularity of this positroid cell is then equal to $\Omega([i])$. To be completely explicit,
\begin{subequations}
\begin{align}
	(1) &= \{4,5,6,7,9,8\}\,,\\
	(2) &= \{3,5,6,7,8,10\}\,,\\
	(3) &= \{5,4,6,7,8,9\}\,,\\
	(4) &= \{4,6,5,7,8,9\}\,,\\
	(5) &= \{4,5,7,6,8,9\}\,,\\
	(6) &= \{4,5,6,8,7,9\}\,.
\end{align}
\end{subequations}
In the notation of the previous section, we have
\begin{align}
	\Qcal_{6,3}=\{(1),(2),(3),(4),(5),(6)\}\,.
\end{align}
We further recall that $\Mcal_{6,3}$ is triangulated by $[1] \cup [3] \cup [5] = [2] \cup [4] \cup [6]$, and hence
\begin{align}\label{eq:DUAL_NHMV6-homological-id}
	\Omega(\Mcal_{6,3})&=\Omega([1])+\Omega([3])+\Omega([5])\\
	&=\Omega([2])+\Omega([4])+\Omega([6])\,.
\end{align}
As derived in detail in appendix \ref{sec:APP_chambers}, there are nine chambers given by
\begin{alignat}{3}
	\Cfrak(\Mcal_{6,3})=\{ &[1]\cap[2]\,,\quad&& [1]\cap[4]\,,\quad&&[1]\cap[6]\,,\notag\\\notag
	&[3]\cap[2]\,, && [3]\cap[4]\,, && [3]\cap [6]\,,\\
	& [5]\cap[2]\,, &&[5]\cap[4]\,, &&[5]\cap [6] \}\,,
\end{alignat} 
which we already encountered in section \ref{sec:DUAL_chambers} (see also figure \ref{fig:amp_6-1-triangulation}). The quadruple cut diagrams which contribute to the positroid cell $(1)$ are depicted in figure \ref{fig:DUAL_NMHV-6-quad-cuts-1}, corresponding to the vertices
\begin{align}
	\Vcal\big((1)\big)=\{q^-_{1236},q^-_{3456},q^+_{2346},q^-_{1356}\}\,.
\end{align}
The vertices corresponding to the other cells follow cyclically. In particular, if $(\lambda,\tilde\lambda)\in[i]\cap[j]$, then the vertices in $\Delta(\bm{x})$ are given by
\begin{alignat}{4}
	\Vcal\big((i)\cap(j)\big) = \{&q^+_{i-1ii+1i+2} ,&&q^-_{i+2 i+3 i+4 i+5} , &&q^+_{i+1i+2i+3i+5} ,&&q^-_{i-1i-2ii+2} ,\notag\\
	&q^+_{j-1jj+1j+2} ,&&q^-_{j+2 j+3 j+4 j+5} ,&&q^+_{j+1j+2j+3j+5} ,&&q^-_{j-1j-2jj+2} \}\,.
\end{alignat}
If we define
\begin{align}\label{eq:DUAL_NMHV6-positroid-form}
	\Omega^{[i]}_{6,3}=\sum_{q^\pm_{abcd}\in\Vcal([i])}\omega^\pm_{abcd} = \omega^+_{i-1ii+1i+2}+\omega^-_{i+2i+3i+4i+5}+\omega^+_{i+1+2i+3i+5}+\omega^-_{i-2-1ii+1}\,,
\end{align}
then we find the canonical form of $\Delta(\bm{x})$ for the chamber $[i]\cap[j]$ to be
\begin{align}
	\Omega^{1-\text{loop}}([i]\cap[j])=\Omega^{[i]}_{6,3}+\Omega^{[j]}_{6,3}\,.
\end{align}
There are two different types of chamber up to cyclic permutations: $[1]\cap[2]$ and $[1]\cap[4]$. Their canonical forms are explicitly given by
\begin{subequations}\label{eq:DUAL_NMHV6-12-form}
\begin{alignat}{2}
	\Omega^{1-\text{loop}}([1]\cap[2]) &= &&\omega^-_{6123}+\omega^+_{3456}+\omega^-_{2346}+\omega^-_{1234}+\omega^+_{4561}+\omega^-_{3451}+\omega^+_{2461}\\
	&=\frac{1}{2}\Big( &&\omega^\square_{1236}+\omega^\square_{3456}+\omega^\square_{2346}+\omega^\square_{1234}+\omega^\square_{1456}+\omega^\square_{1345}+\omega^\square_{1246}\notag\\
	& &&+ \omega^{\pentagon}_{13456}-\omega^{\pentagon}_{12346} \Big)\,,
\end{alignat}
\end{subequations}
and
\begin{subequations}\label{eq:DUAL_NMHV6-14-form}
\begin{align}
	\Omega^{1-\text{loop}}([1]\cap[4])&=\omega^+_{1236}+\omega^+_{1356}+\omega^-_{2346}+\omega^+_{3456}+\omega^-_{1236}+\omega^-_{1356}+\omega^+_{2346}+\omega^-_{3456}\\
	&=\omega^\square_{1236}+\omega^\square_{1356}+\omega^\square_{2346}+\omega^\square_{3456}\,.
\end{align}	
\end{subequations}
The full one-loop integrand is found to be
\begin{align}
	\Omega(\Mcal_{6,3}^{(1)})&=\sum_{\substack{i\in\{1,3,5\}\\j\in\{2,4,6\}}}\Omega^\text{tree}([i]\cap[j])\wedge\Omega^{1-\text{loop}}([i]\cap[j])\\
	&=\sum_{\substack{i\in\{1,3,5\}\\j\in\{2,4,6\}}}\Omega^\text{tree}([i]\cap[j])\wedge\big(\Omega^{[i]}_{6,3}+\Omega^{[j]}_{6,3}\big)\\
	&= \sum_{i=1}^6 \Omega([i])\wedge\Omega^{[i]}_{6,3}\,,\label{eq:DUAL_NMHV6-full-integrand}
\end{align}
where in the last line we use that
\begin{align}
	\Omega([1])=\Omega^\text{tree}([1]\cap[2])+\Omega^\text{tree}([1]\cap[4])+\Omega^\text{tree}([1]\cap[6])\,.
\end{align}
We see from \eqref{eq:DUAL_NMHV6-12-form} and \eqref{eq:DUAL_NMHV6-14-form} that, as promised, the canonical form of the loop fibres are projectively invariant. As a consequence, the full one-loop integrand will also be projective invariant. However, the form associated to a positroid cell in equation \eqref{eq:DUAL_NMHV6-positroid-form} is \emph{not} projective invariant. The result in \eqref{eq:DUAL_NMHV6-full-integrand} is therefore not manifestly projective invariant. We can use the `homological' identity in \eqref{eq:DUAL_NHMV6-homological-id} to restore this invariance
\begin{align}
	\Omega(\Mcal_{6,3}^{(1)})&=\frac{1}{2}\left( \Omega([1])+\Omega([4]) \right)\wedge\left(\omega^{\square}_{1236}+\omega^{\square}_{3456}+\omega^{\square}_{2346}+\omega^{\square}_{1356}\right)\notag\\
	&+\frac{1}{2}\left(\Omega([2])+\Omega([5])\right)\wedge\left(\omega^{\square}_{1234}+\omega^{\square}_{1456}+\omega^{\square}_{1345}+\omega^{\square}_{1246}\right)\notag\\
	&+\frac{1}{2}\left(\Omega([3])+\Omega([6])\right)\wedge\left(\omega^{\square}_{1256}+\omega^{\square}_{2345}+\omega^{\square}_{1235}+\omega^{\square}_{2456}\right)\notag\\
	&+\frac{1}{2}\left(\Omega([1])-\Omega([4])\right)\wedge\left(\omega^{\pentagon}_{12356}-\omega^{\pentagon}_{23456}\right)\notag\\
	&+\frac{1}{2}\left(\Omega([2])-\Omega([5])\right)\wedge\left(\omega^{\pentagon}_{12345}-\omega^{\pentagon}_{23456}\right)\,.
\end{align}

\paragraph{NMHV\textsubscript{7}.}
In appendix \ref{sec:APP_chambers} we show that $\Mcal_{7,3}$ has 71 chambers, which come in 11 cyclic classes. We again label positroid cells by the position of the zeroes of their T-dual. Unlike the NMHV\textsubscript{6} case, we notice something interesting regarding the types of positroid cells which contribute quad cut vertices: they are not all of the same dimension! Indeed, we find
\begin{align}
	\Qcal_{7,3}=\{(1),(1,2),(1,3),(1,4),+\text{ cyclic}\}\,,
\end{align}
where $\dim((i))= 11$, and $\dim((i,j))=10$ (for reference, $\dim(\Mcal_{7,3})=10$). The vertices associated to these positroid cells are
\begin{subequations}
	\begin{align}
		\Vcal\big((1)\big) &= \{q^-_{1237}\}\,,\\
		\Vcal\big((1,2)\big) &= \{ q^-_{1347}\,,q^-_{1467}\,,q^-_{4567}\,,q^+_{3457} \}\,,\\
		\Vcal\big((1,3)\big) &= \{q^+_{2357}\}\,,\\
		\Vcal\big((1,4)\big) &= \{q^-_{1367}\,,q^-_{3567}\}\,.
	\end{align}
\end{subequations}
We note that, since $(1,2)$ is a positroid boundary of $(1)$, $[1,2]$ is completely contained in $[1]$. This means that if $(\lambda,\tilde\lambda)$ is in a chamber containing $[1,2]$, it will automatically also be inside $[1]$. Thus, without loss of generality, we can denote our chamber by the maximal intersections of the images of the lowest dimensional positroid cells in $\Qcal_{7,3}$, keeping in mind that are automatically also in the image of any positroid cell in their \emph{inverse positroid stratification}. For example, we can define a chamber as $[1,2]\cap [1,3]\cap [2,3]$, but when listing the vertices of this chamber we have
\begin{align}
	\Vcal\big((1,2)\cap(1,3)\cap(2,3)\big)= \Vcal\big((1,2)\big)\cup\Vcal\big((1,3)\big)\cup\Vcal\big((2,3)\big)\cup\Vcal\big((1)\big)\cup\Vcal\big((2)\big)\cup\Vcal\big((3)\big)\,.
\end{align}
We again define a differential form associated to each positroid cell as
\begin{align}
	\Omega^\sigma_{7,3} = \sum_{q^\pm_{abcd}\in \Vcal(\sigma)} \omega^\pm_{abcd}\,,\qquad \sigma\in \Qcal_{7,3}\,.
\end{align}
When summing over all chambers we once again recover \eqref{eq:DUAL_nf-full-one-loop}:
\begin{align}
	\Omega(\Mcal_{7,3}^{(1)})=\sum_{i<j} \Omega([i,j])\wedge \Omega^{[i,j]}_{7,3}+\sum_{i=1}^7\Omega([i])\wedge\Omega^{[i]}_{7,3}\,,
\end{align}
where we used the fact that
\begin{align}
	 \Omega([1])= \Omega([1,2])+ \Omega([1,4])+ \Omega([1,6]) = \Omega([1,3])+ \Omega([1,5])+ \Omega([1,7])\,,
\end{align}
to re-sum some of the leading singularities.

\begin{figure}
	\centering
	\begin{subfigure}{.5\textwidth}
		\centering
		\includegraphics[width=0.6\textwidth]{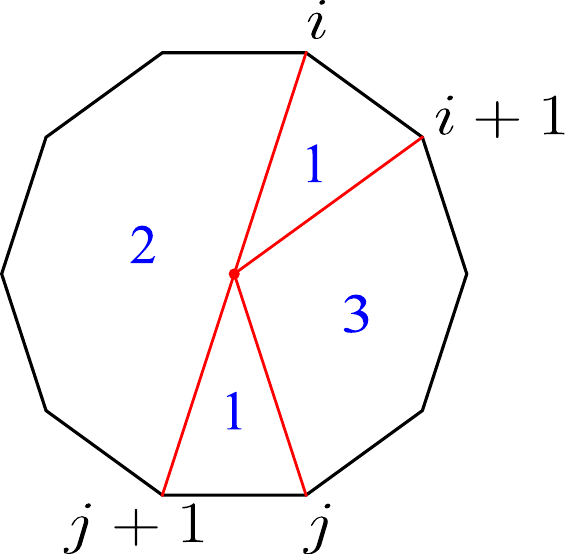}
	\end{subfigure}%
	\begin{subfigure}{.5\textwidth}
		\centering
		\includegraphics[width=0.6\textwidth]{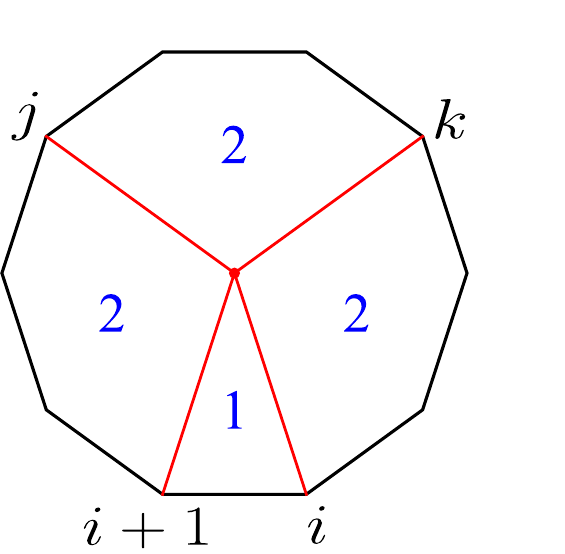}
	\end{subfigure}
	\caption{The two families of quad cut diagrams of type $(n,3)$. They correspond to the vertices $q^+_{i\,i+1\,j\,j+1}$ (left), and $q^+_{i\,i+1\,j\,k}$ (right).}
	\label{fig:DUAL_diag-NMHV}
\end{figure}

\paragraph{General NMHV.}
The types of vertices which contribute to NMHV integrands fall into two different families, which we depict in figure \ref{fig:DUAL_diag-NMHV}.
To be completely explicit, we let $\{i,j,\ldots\}$ denote the positroid tile corresponding to the cell in $G_+(3,n)$ whose T-dual has non-zero entries on locations $i,j,\ldots$ (this is the opposite of the notation $[i,j,\ldots]$ we use above). We can then write the general 1-loop NMHV integrand as
\begin{align}
	\Omega(\Mcal_{n,3}^{(1)}) =&\sum_{a=5}^{n-1}\Omega(\{1,2,\ldots,a\})\wedge\omega^+_{12aa+1}+ \sum_{a=4}^{n-2}\Omega(\{1,2,3,a,a+1\})\wedge\omega^+_{123a+1}\notag\\
	+&\sum_{a=3}^{ \left \lfloor{ \frac{n}{2} }\right \rfloor } \sum_{b=2a-1}^{n-1} \Omega(\{1,2,a,a+1,b\})\wedge\omega^+_{2a+1bb+1}	+ \Omega(\{1,2,3,4,5\})\wedge\omega_{2345}^- \notag\\ +& \sum_{a=3}^{ \left \lfloor{ \frac{n-1}{2} }\right \rfloor }\sum_{b=a+2}^{n-a+1}\Omega(\{1,2,a,b,b+1\})\wedge\omega^-_{2aa+1b+1}
+\text{cyc.}
\end{align}

\section{Summary}

In this chapter we discussed a novel class of positive geometries set in the space of dual momenta. We triangulated (momentum) amplituhedra in terms of chambers such that the geometry factorises in a tree part and a loop part. We have seen that the loop part of this geometry has a natural interpretation in dual space in terms of compact regions which are positively (space-like) separated from the points $\{x_i\}_{i=1}^n$, which form the corners of a null-polygon. We have argued that the one-loop geometry $\Delta(\bm{x})$ can be completely specified by the vertices which emerge as the maximal intersections of the null-cones (lightcones) of these points. Furthermore, we have seen that there is a simple prescriptive way to find the canonical form of $\Delta(\bm{x})$ over these vertices.

This framework is completely general, and the above summary is applicable to the amplituhedron, momentum amplituhedron, and ABJM momentum amplituhedron at the same time. This has allowed us to derive simple expression for the general one-loop integrands for both \nf and ABJM. For the ABJM momentum amplituhedron we have additionally given a full classification of the one-loop chambers. This classification has the natural interpretation as characterising the skeleton of a \emph{dual geometry}. We have further argued that our expression for the canonical form of $\Delta(\bm{x})$ can be interpreted as an internal triangulation of this putative dual geometry. We have also given a short introduction to how this formalism can be extended to higher loops. In particular, we have briefly encountered the idea that higher loop geometries can be described as `fibrations over fibrations', a topic which is currently still under investigation (see \cite{Ferro:2024vwn} for some recent results). For the amplituhedron and the momentum amplituhedron we have not managed to give a full classification of the one-loop chambers, however we noted that they must consist of the maximal intersections of images of positroid cells corresponding to leading singularities (up to complications coming from four-mass-box type cells, which require us to subdivide the positroid cells into two parts). This further allowed us to partially re-sum the full sum over chambers into contributions which only depend only on the leading singularities. This leads to a general formula for the one-loop integrand which does not depend on an explicit knowledge of the chamber structure. 

%% file: chapters/conclusion.tex
\chapter{Conclusions and Outlook}\label{sec:CONC}

The main topic of study in this thesis are positive geometric descriptions of scattering amplitudes in planar \nf and ABJM. We have seen that we can define positive geometries for \nf on both sides of the scattering amplitude/Wilson loop duality: the amplituhedron on the Wilson loop side, and the momentum amplituhedron on the scattering amplitude side. For ABJM theory we have encountered the ABJM momentum amplituhedron, which describes tree-level amplitudes in supersymmetry reduced ABJM theory. In addition, we have discussed the ABHY associahedron for tree-level scattering amplitudes in bi-adjoint $\phi^3$ theory. 

Let us remark some of the similarities and differences of the geometries which we have encountered. To start, it is natural to group the ABHY associahedron with the momentum amplituhedron and the ABJM momentum amplituhedron, leaving the amplituhedron as the odd one out. One reason for this is because these three geometries describe scattering amplitudes, rather than Wilson loops. This means that their boundaries are in a one-to-one correspondence with the singularities of the respective amplitudes. We have given a detailed account of the boundaries of these geometries, providing a complete diagrammatic description. In particular, it has been shown in \cite{Moerman:2021cjg} (based on the results of \cite{Ferro:2020lgp, Lukowski:2020bya}) that the Euler characteristic of $\Mcal_{n,k}$ is $\chi=1$. In section \ref{sec:POS_ABJM-mom-amp-boundaries} we provide a proof that also $\Ocal_k$ has an Euler characteristic $\chi=1$ \cite{Lukowski:2021fkf}. Thus, it is likely that the tree-level geometries for these three theories are all homeomorphic to a closed ball. On the mathematics side, it would be interesting to prove this statement, for which it would be sufficient to show that their boundary posets are \emph{thin} and \emph{shellable} \cite{BJORNER19847}. 

In addition, the ABHY associahedron, momentum amplituhedron, and ABJM momentum amplituhedron all admit a description based on push forwards through the scattering equations. We have seen that we can find the canonical form of these positive geometries by summing the world-sheet Parke-Taylor form over the solutions to the scattering equations. In section \ref{sec:POS_pf} we developed three methods to calculate these push forwards without needing to find the specific solutions to the scattering equations. These methods allow a further investigation and a possible extension of the web of connections between these various positive geometries.

On the other hand, it is also natural to group the three (momentum) amplituhedra, as they all share a similar definition. They can be defined as the image of a positive linear map from the positive (orthogonal) Grassmannian into some other Grassmannian space. Alternatively, they admit a definition directly in the kinematic space based on the sign-flip patterns of the appropriate Lorentz invariant brackets. Additionally, we have seen that they all have extensions to loop integrands by extending the domain of the positive linear map to the appropriate positive loop Grassmannian. These properties do not hold for the ABHY associahedron, which is only defined for tree-level and does not have a Grassmannian or sign-flip definition. 

In chapter \ref{sec:DUAL} we have developed a novel definition of these geometries for loop integrands in \nf and ABJM, this time set in the space of dual momenta. It is a unifying framework which captures the loop-level geometry of the amplituhedron, momentum amplituhedron, and the ABJM momentum amplituhedron at the same time (it also extends to the recently introduced ABJM amplituhedron \cite{He:2023rou}). The formalism outlined in chapter \ref{sec:DUAL} is remarkably simple and effective. After we triangulated our tree-level geometry in terms of chambers, the loop-level construction relies solely on the intersections of lightcones or null-cones in dual space, and it allowed us to derive general statements for the one-loop integrand in \nf and ABJM with relative ease.

There are many future directions of research which emerge naturally from the topics discussed in this thesis. First and foremost, it is natural to ask if our dual space formalism will yield similarly simple results beyond one loop. One interesting new approach to these higher-loop geometries is to proceed via the `fibrations of fibrations' method, which we briefly touched upon in section \ref{sec:DUAL_ABJM}, which was also explored in the recent paper \cite{Ferro:2024vwn}. Furthermore, we gave some evidence for the supposed existence of a dual geometry for $\Delta(\bm{x})$. We gave a classification of its skeleton, and we argued that the canonical form which we find for $\Delta(\bm{x})$ is naturally written in a way which should correspond to an internal triangulation of the dual geometry in terms of simplices. A proper understanding of these dual geometries would be an important step towards a definition of non-polytopal dual amplituhedra. 

On the topic of classifying boundaries of momentum amplituhedra, it would be interesting to extend the study to loop level. We have checked up to $k=7$ that $\smash{\Ocal_k^{(1)}}$ \emph{also} has $\chi=1$, and we expect that this is a general property of $\smash{\Ocal_k^{(1)}}$ and $\smash{\Mcal_{n,k}^{(1)}}$. Beyond one loop, however, this is no longer expected to be the case. We have already seen in the toy example in section \ref{sec:DUAL_toy} that, starting from two loops, the geometry is no longer a ball. We note, however, that the procedure outlined in section \ref{sec:POS_mom-amp-boundaries} is unlikely to generalise to higher loops. This is in part due to the lacking understanding of the positive loop Grassmannian, but also because of the presence of a denominator in the map \eqref{eq:POS_ell-mom-amp-def}, which means that we first have to provide a non-trivial blow-up of the domain to study the boundaries of the image. The classification of the boundary stratification of the (momentum) amplituhedron at two loops and higher is currently an interesting open problem, the study of which could reveal many interesting properties of these geometries (it has already lead to the discover of internal boundaries of the amplituhedron in \cite{Dian:2022tpf}, which showed that it is necessary to move away from the traditional definition of positive geometries). 

Another natural question is whether we can extend our list of positive geometries by calculating push forwards through the scattering equations. There are many twistor string/CHY formulae for various theories, and using the methods developed in section \ref{sec:POS_pf} we can start calculating push forwards. A natural starting point would be the search for a momentum amplituhedron for six-dimensional theories \cite{He:2021llb}, which admit Grassmannian, CHY and twistor string formulas \cite{Cachazo:2018hqa, Schwarz:2019aat, Geyer:2018xgb}. This would add to a web of connections between positive geometries, part of which we already explained in the diagram \ref{fig:mod-grass-mom-web}. In addition, we have encountered a surprising connection between the ABJM momentum amplituhedron and the ABHY associahedron. We argued that we can (at least combinatorially) obtain $\Ocal_k$ by `collapsing' certain boundaries of $\Ascr_{2k}$. Additionally, it has been shown in \cite{Damgaard:2020eox} that the ABHY associahedron is hidden inside the `little group invariant part' of the momentum amplituhedron. A further investigation of these geometric connections can expose interesting relations between the scattering amplitudes of their theories. The study of these connections will undoubtedly require the calculations of push forwards. Although we have provided easily implementable algorithms which allow the calculation of these push forwards, we also remarked that these methods necessarily use \Grob bases, which forms the main computational bottleneck. It would be interesting to investigate if these push forwards can be done in a way which does not rely on these \Grob bases (perhaps using techniques similar to the \emph{Macaulay matrix}, which has been used to bypass \Grob bases for Feynman integral calculations in \cite{Chestnov:2022alh}), which would aid in this endeavour.

These are but a small subset of the possible directions of continuations which emerge from the topics of this thesis. Undoubtedly, new connections will be found and new ideas will synthesise as these topics are further explored, and it will be fascinating to see what the future holds for this lively and exciting field of research.

%% file: appendices/alggeom.tex
\chapter{Algebraic Geometry}\label{sec:APP_alg-geom}

\section[head={Basics},tocentry={Basics of Algebraic Geometry}]{Basics of Algebraic Geometry}\label{sec:APP_alg-geom-gen}

In this appendix we will review some basic notions from algebraic geometry which we use throughout the thesis. In particular section \ref{sec:POS_pf} relies heavily on the topics discussed here. There are a plethora of introductory texts on algebraic geometry, but for a more computational focus we mention \cite{cox2013ideals} as an excellent introductory resource which encompasses all topics discussed here.

Given some field $K$, we define $K[z_1,\ldots,z_n]$ to be the ring of polynomials in $z_1,\ldots,z_n$ with coefficients in $K$. Furthermore, we let $K(a_1,\ldots,a_m)$ be the field of rational functions in variables $a_1,\ldots,a_m$ with coefficients in $K$. Hence, $K(a_1,\ldots,a_m)[z_1,\ldots, z_n]$ is the ring of polynomials in $z_1,\ldots,z_n$ whose coefficients are rational functions in $a_1,\ldots, a_m$ with coefficients in $K$. We often use the shorthand $\bm{z}=(z_1,\ldots,z_n)$, $\bm{a}=(a_1,\ldots, a_m)$, and $K(\bm{a})[\bm{z}]=K(a_1,\ldots,a_m)[z_1,\ldots, z_n]$. For convenience, we will assume that $K$ is algebraically closed, and in particular we will typically take $K=\Cbb$.

An \emph{ideal} $\Ical$ is a subset of $ \Cbb(\bm{a})[\bm{z}]$ which contains the zero polynomial, is closed under addition, and satisfies that $hf\in \Ical$ for all $h\in\Cbb(\bm{a})[\bm{z}]$, $f\in\Ical$. Given some set of polynomials $f_1,\ldots,f_s$ in $\Cbb(\bm{a})[\bm{z}]$, we define the \emph{ideal generated by $f_1,\ldots, f_s$} as
\begin{align}
	\<f_1,\ldots,f_s\>\coloneqq \left\{ \sum_{i=1}^s h_if_i \colon h_1,\ldots,h_s\in \Cbb(\bm{a})[\bm{z}] \right\}\,.
\end{align}
The \emph{variety} $\Vcal(\Ical)$ is the subset of $\Cbb(\bm{a})^n$ where all elements of the ideal vanish. An ideal is called \emph{radical} if for all elements of $\Ical$ of the form $f^m$ for some $f\in\Cbb(\bm{a})[\bm{z}]$, then also $f\in\Ical$. 

Given some $\bm{\alpha}=(\alpha_1,\alpha_2,\ldots,\alpha_n)\in \Zbb_{\geq 0}^n$, we will often abbreviate monomials as
\begin{align}
	\bm{z}^{\bm{\alpha}} \equiv z_1^{\alpha_1}z_2^{\alpha_2}\cdots z_n^{\alpha_n}\,.
\end{align}
A \emph{monomial ordering} $\succ$ on $\Cbb(\bm{a})[\bm{z}]$ is an ordering on the set of all monomials in $\Cbb(\bm{a})[\bm{z}]$ which allows one to compare any two monomials such that
\begin{itemize}
	\item $ \displaystyle\bm{z}^{\bm\alpha}\succ\bm{z}^{\bm\beta}, \bm{z}^{\bm\beta}\succ\bm{z}^{\bm\gamma}\implies \bm{z}^{\bm\alpha}\succ\bm{z}^{\bm\gamma}$,
	\item $ \displaystyle\bm{z}^{\bm\alpha}\succ\bm{z}^{\bm\beta} \implies \bm{z}^{\bm\alpha+\bm\gamma}\succ\bm{z}^{\bm\beta+\bm\gamma}\quad\forall \gamma \in \Zbb_{\ge0}^n$. 
\end{itemize}
Given some polynomial $p\in\Cbb(\bm{a})[\bm{z}]$, its \emph{leading term} is the largest monomial in $p$ with respect to $\succ$, denoted $\text{LT}(p)$. 

We will typically define $z_1\succ z_2\succ\ldots\succ z_n$. A few particularly useful monomial orders are
\begin{itemize}
	\item \emph{Lexicographic ordering (lex)} is defined such that $\bm{z}^{\bm{\alpha}}\succ \bm{z}^{\bm\beta}$ if the leftmost non-zero entry of $\bm\alpha-\bm\beta$ is positive. 
	\item \emph{Reverse lexicographic ordering (rlex)} is defined such that $\bm{z}^{\bm{\alpha}}\succ \bm{z}^{\bm\beta}$ if the rightmost non-zero entry of $\bm\alpha-\bm\beta$ is positive. 
	\item \emph{Graded lexicographic ordering (grlex)} defines $\bm{z}^{\bm{\alpha}}\succ \bm{z}^{\bm\beta}$ if $|\bm\alpha|>|\bm\beta|$ (where $|\bm\alpha|\coloneqq\sum_{i=1}^n \alpha_i$). In the case $|\bm\alpha|=|\bm\beta|$ ties are broken using lexicographic ordering.
	\item \emph{Graded reverse lexicographic ordering (grevlex)} defines $\bm{z}^{\bm{\alpha}}\succ \bm{z}^{\bm\beta}$ if $|\bm\alpha|>|\bm\beta|$. In the case $|\bm\alpha|=|\bm\beta|$ ties are broken using reverse lexicographic ordering.
\end{itemize}
In the case where a choice of monomial ordering is not specified, the reader may assume that the subsequent results are independent of this choice.

A \emph{\Grob basis} $\Gcal=\{g_1,\ldots,g_t\}\neq\{0\}$ of an ideal $\Ical\subseteq\Cbb(\bm{a})[\bm{z}]$ is a finite subset of $\Ical$ such that
\begin{align}
	\<\text{LT}(g_1),\ldots,\text{LT}(g_t)\> = \<\text{LT}(I)\>\,,
\end{align}
with respect to some fixed monomial ordering. Equivalently, we can define $\Gcal$ such that every (non-zero) $f\in\Ical$ has a leading term with respect to $\prec$ which is divisible by some element $g_i$ of $\Gcal$, and $\Ical=\<g_1,\ldots,g_t\>$. 

In general, given any set $p_1,\ldots,p_s$ of polynomials in $\Cbb(\bm{a})[\bm{z}]$, we can write any $f\in \Cbb(\bm{a})[\bm{z}]$ as $f=q_1p_1+\ldots+q_sp_s+r$ with $q_i,r\in\Cbb(\bm{a})[\bm{z}]$, and either $r=0$, or none of the terms in $r$ are divisible by the leading terms $\text{LT}(p_1),\ldots,\text{LT}(p_n)$. $r$ is called the \emph{remainder} of $f$ on dividing by $p_1,\ldots,p_n$, and polynomial long division provides an algorithm to find it. This expansion, and in particular also the remainder $r$, are generally not unique. An important property of \Grob bases $\Gcal=\{g_1,\ldots,g_t\}$ is that the remainder of any polynomial $f$ on division by $g_1,\ldots,g_t$ is unique, and we will denote it as $\overbar{f}^\Gcal$.

Given some ideal $\Ical$, we define the \emph{quotient ring} as
\begin{align}
	Q= \Cbb(\bm{a})[\bm{z}]/\Ical=\{[f]_\sim\colon f \in \Cbb(\bm{a})[\bm{z}]\}\,,\quad \text{ where }f\sim g \Leftrightarrow f-g\in\Ical\,.
\end{align} 
If $\Ical$ is a zero-dimensional ideal, then the quotient ring is a vector space with dimension $\dim(Q)=|\Vcal(\Ical)|=d$. Given some \Grob basis $\Gcal=\{g_1,\ldots,g_t\}$ of $\Ical$, we can find a \emph{standard monomial basis} $\Bcal=\{e_i\}_{i=1}^d$ of $Q$ as the set of all monomials $e_i=\bm{z}^{\bm\alpha}$ which do not divide any $\text{LT}(g_1),\ldots,\text{LT}(g_t)$. The remainder of some $f\in\Cbb(\bm{a})[\bm{z}]$ under division by $\Gcal$ can be written $\overbar{f}^\Gcal = \sum_{i=1}^d f_i e_i$, for some coefficients $f_i\in \Cbb(\bm{a})$.

\subsection{Selected Theorems}

We will now state several well-known results from algebraic geometry which we use in this thesis. 
\begin{theorem}[Stickelberger's Theorem \cite{sturmfels2002solving}]\label{thm:stickelberger}
	Given a zero-dimensional ideal $\Ical \subseteq \Cbb[\bm{z}]$ with $|\Vcal(\Ical)|=d$, then the $d$ complex zeroes are the vectors of simultaneous eigenvalues $\bm{\lambda}=(\lambda_1,\ldots,\lambda_n)$ of the companion matrices $T_1,\ldots,T_n$ of $\Ical$. That is, if ${\bm{v}\in\Cbb^d\setminus\{\bm{0}\}}$ is an eigenvector of $T_i$, then
	\begin{align}
		\Vcal(\Ical)=\left\{\bm{\lambda}\in\Cbb^n\,\colon T_i\cdot\bm{v} = \lambda_i\,\bm{v}\,,\quad \forall {i\in[n]}\right\}\,,
	\end{align}
	where $\bm{\lambda}=(\lambda_1,\ldots,\lambda_n)$.
\end{theorem}

\begin{theorem}[Global Duality Theorem \cite{cattani2005introduction}]\label{thm:global-duality}
	Let $\Ical=\langle f_1,\ldots,f_n\rangle\subseteq\Cbb[z_1,\ldots,z_n]$ be a zero-dimensional ideal with corresponding quotient ring $Q=\Cbb[z_1,\ldots,z_n]/\Ical$. The symmetric inner product
	\begin{align}
		\<\bullet,\bullet\>:Q\times Q\to\Cbb,\; (p_1,p_2)\mapsto\Res(p_1\,p_2)\,,
	\end{align}
	is non-degenerate, where $\Res$ is the global residue with respect to $f_1,\ldots,f_n$, defined in equation \eqref{eq:POS_global-residue-def}.
\end{theorem}

\begin{theorem}[Hilbert's Weak Nullstellensatz \cite{cox2013ideals}]\label{thm:nullstellensatz}
	Any ideal $\Ical\subseteq \Cbb[z_1,\ldots,z_n]$ satisfying $\Vcal(\Ical)=\emptyset$ must include every polynomial in $\bm{z}$ with complex coefficients. That is, $\Ical=\Cbb[z_1,\ldots,z_n]$.
\end{theorem}

\begin{theorem}[Elimination Theorem \cite{cox2013ideals}]\label{thm:elimination}
	Let $\succ$ be the lexicographic monomial order with $z_1\succ\ldots\succ z_n$. Let $\Gcal$ be a \Grob basis of some given ideal $\Ical\subseteq \Cbb[z_1,\ldots,z_n]$ with respect to $\prec$. Then for every $\ell\in\Zbb$ with $0\le\ell\le n$ the intersection $\Gcal\cap\Cbb[z_{\ell+1},\ldots,z_n]$ is a \Grob basis of the $\ell$\textsuperscript{th} elimination ideal $\Ical\cap\Cbb[z_{\ell+1},\ldots,z_n]$.
\end{theorem}

\section[head={Additional Information},tocentry={Additional Information on Various Statements}]{Additional Information on Various Statements}

\subsection{\texorpdfstring{Proof of Equation \eqref{eq:POS_comp-mat-der}}{Proof of Equation (6.263)}}\label{sec:APP_compt-mat-der}
In this section we will prove equation \eqref{eq:POS_comp-mat-der}, which states that
\begin{align}
	\sum_{\bm{\xi}\in\Vcal(\Ical)}\omega_I(\bm{\xi})\left| \frac{\partial \bm{\xi}}{\partial\bm{a}} \right|^I_J = \tr{\omega_I(\bm{T})\sum_{\sigma\in S_p}\sgn(\sigma)\frac{\partial T_{i_{\sigma(1)}}}{\partial a_{j_1}}\cdots \frac{\partial T_{i_{\sigma(p)}}}{\partial a_{j_p}}}\,.
\end{align}
The idea is similar to equation \eqref{eq:POS_com-mat-trace}. We recall that we can simultaneously diagonalise the companion matrices $T_i = S D_i S^{-1}$, and that, by \nameref{thm:stickelberger}, the diagonal entries of the $D$ matrices are exactly the elements of our variety. It then follows directly that
\begin{align}
	\sum_{\bm{\xi}\in\Vcal(\Ical)}\omega_I(\bm{\xi})\left| \frac{\partial \bm{\xi}}{\partial\bm{a}} \right|^I_J = \tr{\omega_I(\bm{D})\sum_{\sigma\in S_p}\sgn(\sigma)\frac{\partial D_{i_{\sigma(1)}}}{\partial a_{j_1}}\cdots \frac{\partial D_{i_{\sigma(p)}}}{\partial a_{j_p}}}\,.
\end{align}
It is therefore sufficient to prove that
\begin{align}\label{eq:APP_tr-T-tr-D}
	\tr{\omega_I(\bm{T})\sum_{\sigma\in S_p}\sgn(\sigma)\frac{\partial T_{i_{\sigma(1)}}}{\partial a_{j_1}}\cdots \frac{\partial T_{i_{\sigma(p)}}}{\partial a_{j_p}}} = \tr{\omega_I(\bm{D})\sum_{\sigma\in S_p}\sgn(\sigma)\frac{\partial D_{i_{\sigma(1)}}}{\partial a_{j_1}}\cdots \frac{\partial D_{i_{\sigma(p)}}}{\partial a_{j_p}}}\,.
\end{align}
The problem with respect to equation \eqref{eq:POS_com-mat-trace} is that, unlike the companion matrices $T_i$, their partial derivatives $\partial T_i/\partial a_j$ generally don't commute. This means that the matrices inside the trace of equation \eqref{eq:APP_tr-T-tr-D} are not similar to each other. 

To prove equation \eqref{eq:APP_tr-T-tr-D} we first note that from $T_i= S D_i S^{-1}$ we get
\begin{align}
	\frac{\partial T_i}{\partial a_{j}} = S\left(\frac{\partial D_i}{\partial a_j}+[\Gamma_i,D_j]\right)S^{-1}\,,\quad\text{ where } \Gamma_i \coloneqq S^{-1}\frac{\partial S}{\partial a_i}\,.
\end{align}
Substituting this into the left-hand-side of equation \eqref{eq:APP_tr-T-tr-D}, and using the fact that $\omega_I(\bm{T}) = S  \omega_I(\bm{D}) S^{-1}$ yields
\begin{align}\label{eq:APP_tr-sum-Tr-IJ}
		\tr{\omega_I(\bm{T})\sum_{\sigma\in S_p}\sgn(\sigma)\frac{\partial T_{i_{\sigma(1)}}}{\partial a_{j_1}}\cdots \frac{\partial T_{i_{\sigma(p)}}}{\partial a_{j_p}}} = \sum_{r=0}^p \sum_{K\in\binom{[p]}{r}}\sum_{\sigma\in S_p} \sgn(\sigma) \text{Tr}^I_J (K;\sigma)\,,
\end{align}
where we defined
\begin{align}
	\text{Tr}^I_J (K;\sigma)\coloneqq \tr{\omega_I(\bm{D})\prod_{k=1}^p 
	\begin{rcases}
		\begin{dcases}
			\partial D_{i_{\sigma(k)}}/\partial a_{j_k}\,\quad &\text{ if }k\not\in K\\ 
			[\Gamma_{j_k},D_{i_{\sigma(k)}}] &\text{ if }k\in K
		\end{dcases}
	\end{rcases} }\,.
\end{align}
In the case when $r=0$, the multiindex $K=\emptyset$ and there are only partial derivatives of $D$ in $\text{Tr}^I_J(\emptyset;\sigma)$. Hence, in this case the right hand side of equation \eqref{eq:APP_tr-sum-Tr-IJ} exactly yields the desired result. It remains to be shown that all contributions on the right hand side of \eqref{eq:APP_tr-sum-Tr-IJ} vanish for $r>0$. To see this, we will fix $K=\{k_1,\ldots,k_r\}\in\binom{[p]}{r}$ and proceed in matrix components. Since $D_i$ is a diagonal matrix, we can define its matrix components as $[D_i]_{\alpha\beta} = \lambda_i^{(\alpha)}\delta_{\alpha\beta}$. The matrix components of $[\Gamma_{i},D_j]$ are given by $([\Gamma_{i},D_j])_{\alpha\beta}=(\Gamma_i)_{\alpha\beta} (\lambda^{(\alpha)}_j-\lambda^{(\beta)}_j)$, and $(\partial D_i/\partial a_j)_{\alpha\beta} = \delta_{\alpha\beta}\partial \lambda^{(\alpha)}_i/\partial a_j$. We then find
\begin{align}
	&\text{Tr}^I_J(K;\sigma) = \sum_{\alpha_1,\ldots,\alpha_r=1}^d \omega_I(\bm{\lambda}^{(\alpha_1)})\times\\ \notag
	& \left[\prod_{s=1}^r \left(\prod_{k=k_{s-1}+1}^{k_s-1} \frac{\partial \lambda_{i_{\sigma(k)}}^{(\alpha_s)}}{\partial a_{j_k}} \right)(\Gamma_{j_{k_s}})_{\alpha_s\alpha_{s+1}} \left( \lambda_{i_{\sigma(k_s)}}^{(\alpha_{s+1})}- \lambda_{i_{\sigma(k_s)}}^{(\alpha_{s})} \right)\right]\left[\prod_{k=k_r+1}^p \frac{\partial \lambda_{i_{\sigma(k)}}^{(\alpha_1)}}{\partial a_{j_k}} \right]\,,
\end{align}
where $k_0=0$ and $\alpha_{r+1}=\alpha_1$. At this stage, we split up into two separate cases, given by $r<p$ and $r=p$. 
\begin{itemize}
	\item In the case where $r<p$, we define $K'=[p]\setminus K = \{k_1',\ldots,k_{p-r}'\}$. We will split up our sum over all permutations in $S_p$ as
	\begin{align}
		\sum_{\sigma\in S_p} = \sum_{L'\in\binom{[p]}{p-r}}\sum_{\pi\in S_{p-r}}\sum_{\sigma\in S_p(K',L';\pi)}\,,
	\end{align}
	where the last sum is defined to be the sum over the $r!$ permutations in $S_p$ mapping $K'$ to $L'$ while keeping $\pi$ fixed:
	\begin{align}
		S_p(K',L';\pi)\coloneqq \{\sigma\in S_p\colon \sigma(k_\gamma')=l_{\pi(\gamma)}'\,,\quad \forall \gamma\in [p-r]\}\,.
	\end{align}
	The reason we split the sum over permutations in this way is because it allows us to isolate the sum
	\begin{align}
		\sum_{\sigma\in S_p(K',L';\pi)}\sgn(\sigma)\prod_{s=1}^r \left( \lambda_{i_{\sigma(k_s)}}^{(\alpha_{s+1})} - \lambda_{i_{\sigma(k_s)}}^{(\alpha_{s})}\right)\,.
	\end{align}
	It is easy to show that this vanishes for any choice of $\alpha_1,\ldots,\alpha_r$ and for all $L'\in\binom{[p]}{p-k}$. As a result, we find that
	\begin{align}
		\sum_{\sigma \in S_p} \text{Tr}^I_J(K;\sigma) =0\,.
	\end{align}
	\item In the case where $r=p$, we follow a similar line of reasoning. There are no derivative terms left in the expansion, and we are left with a product containing
	\begin{align}
		\sum_{\sigma\in S_p}\sgn(\sigma)\prod_{s=1}^p\left( \lambda_{i_{\sigma(k_s)}}^{(\alpha_{s+1})} - \lambda_{i_{\sigma(k_s)}}^{(\alpha_{s})}\right)=0\,.
	\end{align}
\end{itemize}
We have just shown that
\begin{align}
	\sum_{\sigma\in S_p}\sgn(\sigma)\text{Tr}^I_J(K;\sigma) = \begin{cases}
		0\qquad&\text{ if } K \neq \emptyset\\
		\tr{\omega_I(\bm{D})\sum_{\sigma\in S_p}\sgn(\sigma)\frac{\partial D_{i_{\sigma(1)}}}{\partial a_{j_1}}\cdots \frac{\partial D_{i_{\sigma(p)}}}{\partial a_{j_p}}}&\text{ if } K = \emptyset
	\end{cases}\,.
\end{align}
Hence, equation \eqref{eq:APP_tr-sum-Tr-IJ} implies equation \eqref{eq:APP_tr-T-tr-D}, which concludes our proof of equation \eqref{eq:POS_comp-mat-der}.
 
\subsection{\texorpdfstring{An Algorithm to Find $\partial T/\partial a$ Numerically}{An Algorithm to Find dT/da Numerically}}\label{sec:APP_comp-mat-der-numeric}

In section \ref{sec:POS_pf}, combined with the results from appendix \ref{sec:APP_compt-mat-der}, we have shown that we can calculate the push forwards of differential forms by evaluating 
\begin{align}\label{eq:APP_Tr-dT-da}
	\tr{\omega_I(\bm{T})\sum_{\sigma\in S_p}\sgn(\sigma)\frac{\partial T_{i_{\sigma(1)}}}{\partial a_{j_1}}\cdots \frac{\partial T_{i_{\sigma(p)}}}{\partial a_{j_p}}}\,.
\end{align}
To construct the companion matrices $T_i$, we need to resort to \Grob basis techniques which can quickly become computationally intensive. This provides the main bottleneck for actually calculating these push forwards. To partially overcome this obstacle, we note that \Grob basis techniques are optimised over finite fields. We can use this to our advantage by substituting numeric values for our $a$-variables, and evaluate the rational functions over this finite field, after which the final answer can be obtained through \emph{rational reconstruction} (see, for example, \cite{Peraro:2019svx, Klappert:2019emp}). This is an efficient workaround for the `push forward via companion matrices' and `push forward via global residues' methods outlined in section \ref{sec:POS_pf}. However, it is less obvious if these finite field methods can be used for the `push forward via derivative of companion matrix' method, as equation \eqref{eq:APP_Tr-dT-da} naively requires an explicit $a$-dependent form of the companion matrices, which would then allow us to find the matrices $\partial T_i/\partial a_j$.

In this appendix, we will provide an algorithm to find $\partial T_i/\partial a_j$ numerically, allowing us to make use of the more optimised finite field methods for \Grob bases. We start by defining the polynomial rings
\begin{align}
	S & \coloneqq \Cbb[z_1,\ldots,z_n]\,,\\
	S_j & \coloneqq \Cbb[\frac{\partial z_1}{\partial a_j},\ldots,\frac{\partial z_n}{\partial a_j},z_1,\ldots,z_m]\,,
\end{align}
where $\partial z_i/\partial a_j$ should be regarded as formal variables. We recall that the functions $f(\bm{z})$ which generate the ideal $\Ical = \<f_1,\ldots,f_n\>\subseteq S$ are implicitly (rational) functions of the $a$-variables. Hence, we regard the ideal as being dependent on the $a$-variables, and we write $\Ical=\Ical(\bm{a})$. We further define the ideals
\begin{align}
	\frac{\dd \Ical}{\dd a_j}(\bm{a})\coloneqq\left\< \frac{\dd f_1}{\dd a_j},\ldots,\frac{\dd f_n}{\dd a_j} \right\> = \left\< \frac{\partial f_1}{\partial a_j}+\frac{\partial f_1}{\partial z_i}\frac{\partial z_i}{\partial a_j},\ldots,\frac{\partial f_n}{\partial a_j}+\frac{\partial f_n}{\partial z_i}\frac{\partial z_i}{\partial a_j} \right\>\subseteq S_j\,.
\end{align}
Next, we define the ideal $\Ical_j$ to be the \emph{sum} of the ideals $\Ical$ and $\dd \Ical/\dd a_j$:
\begin{align}
	\Ical_j(\bm{a})\coloneqq \Ical(\bm{a})+\frac{\dd\Ical}{\dd a_j}(\bm{a})\subseteq S_j\,,
\end{align}
which is generated by the union of the generators of $\Ical$ and $\dd \Ical/\dd a_j$.

We fix a lexicographic monomial ordering $\succ$ with 
\begin{align}
	\frac{\partial z_1}{\partial a_j}\succ \cdots \succ \frac{\partial z_n}{\partial a_j}\succ z_1\succ \cdots\succ z_n\,,
\end{align}
and define $\Gcal(\bm{a})$ and $\Gcal_j(\bm{a})$ to be the \Grob bases for the ideals $\Ical(\bm{a})$ and $\Ical_j(\bm{a})$, respectively. We note that the standard bases for the quotient rings $S /\Ical$ and $S_j / \Ical_j$ are equivalent for all cases relevant for this thesis (we do not know a general statement for when this is the case), which we shall denote $\Bcal(\bm{a})=\{e_\alpha\}_{\alpha=1}^d$.

The companion matrices and their derivatives can now be found using the division algorithm with respect to the \Grob basis $\Gcal_j(\bm{a})$ as
\begin{align}
	\overbar{z_i e_\alpha}^{\Gcal_j(\bm{a})} &= T_i(\bm{a})_{\alpha\beta} e_\beta\,, \label{eq:APP_Ti-remainder}\\
	\overbar{\frac{\partial(z_ie_\alpha)}{\partial a_j}-T_i(\bm{a})_{\alpha\beta}\frac{\partial e_\beta}{\partial a_j}}^{\Gcal_j(\bm{a})} &= \frac{\partial T_i(\bm{a})_{\alpha\beta}}{\partial a_j}e_\beta\,.\label{eq:APP_dT-remainder}
\end{align}
The first of these equations follows from the fact that $\Ical(\bm{a})$ is the \emph{$n$\textsuperscript{th} elimination ideal} of $\Ical_j(\bm{a})$, which follows from the definition of of $\Ical_j(\bm{a})$ and our choice of monomial ordering. Since we have chosen lexicographic ordering, the \nameref{thm:elimination} implies that $\Gcal_j(\bm{a})\cap S = \Gcal(\bm{a})$. Hence, the remainders on division by $\Gcal_j(\bm{a})$ and $\Gcal(\bm{a})$ are the same.

The second equation, \eqref{eq:APP_dT-remainder}, follows from the following arguments. First, in the quotient ring $S_j / \Ical_j$ we have the equality
\begin{align}
	z_i e_\alpha = T_i(\bm{a})_{\alpha\beta}e_\beta\,.
\end{align}
Differentiating this with respect to $a_j$ yields
\begin{align}
	\frac{\partial(z_i e_\alpha)}{\partial a_j} = \frac{\partial T_i(\bm{a}_{\alpha\beta})}{\partial a_j}e_\beta + T_i(\bm{a})_{\alpha\beta}\frac{\partial e_\beta}{\partial a_j}\,.
\end{align}
Rearranging this equation and taking the remainder with respect to the \Grob basis $\Gcal_j(\bm{a})$ yields equation \eqref{eq:APP_dT-remainder}.

To summarise, we can fix some generic numeric values $\bm{a}^\star$ for the $\bm{a}$-variables. We can then find the ideals $\Ical(\bm{a}^\star)$ and $\dd \Ical/\dd a_j (\bm{a}^\star)$ by evaluating the generating functions on $\bm{a}=\bm{a}^\star$. This then allows us to find the \Grob basis $\Gcal_j(\bm{a}^\star)$ via the steps described above. Then, we find both the companion matrices $T_i(\bm{a}^\star)$ and their derivatives $\partial T_i/\partial a_j (\bm{a}^\star)$ \emph{evaluated on $\bm{a}^\star$}, by finding the remainder with respect to this \Grob basis as in equations \eqref{eq:APP_Ti-remainder} and \eqref{eq:APP_dT-remainder}. Now that the numeric values for the companion matrices and their derivatives are known, we can obtain a numeric value for \eqref{eq:APP_Tr-dT-da}. Doing this repeatedly for different numeric values of $\bm{a}$ then allows us to use rational reconstruction methods to find the final answer, which will be a rational function in $\bm{a}$.

\subsection{Decomposition of Unity in the Dual Basis}\label{sec:APP_decompose-unity}

We recall that the quotient ring $Q=\Cbb(\bm{a})[\bm{z}]$ is a $d$-dimensional vector space with a standard basis $\Bcal = \{e_\alpha\}_{\alpha=1}^d$. In equation \eqref{eq:POS_Q-Q-inner-product} we defined a non-degenerate inner product
\begin{align}\
	\<\bullet,\bullet\>\colon Q\times Q&\to \Cbb(\bm{a})\\\nonumber
	(p_1,p_2)&\mapsto \<p_1,p_2\> = \Res(p_1 p_2)\,,
\end{align}
which implies the existence of a \emph{dual basis} $\Bcal^\vee=\{\Delta_\alpha\}_{\alpha=1}^d$ which satisfies
\begin{align}
	\<e_\alpha,\Delta_\beta\>=\delta_{\alpha\beta}\,.
\end{align}
In section \ref{sec:POS_pf} we have shown that one can calculate pushforwards of any polynomial function (using the results from appendix \ref{sec:APP_poly-inverse} this can be extended to rational functions) by decomposing unity in the dual basis:
\begin{align}
	1=\sum_{\alpha=1}^d \mu_\alpha \Delta_\alpha\,.
\end{align}
In this appendix we will explain how to find the dual basis, which will allow us to find the \emph{components of unity} $\mu_\alpha\in \Cbb(\bm{a})$.

To start, we calculate the \Grob basis $\Gcal$ of our ideal $\Ical$ with graded lexicographic or graded reverse lexicographic ordering. We define the \emph{Bezoutian matrix} $B$ with components
\begin{align}
	B_{ij}\coloneqq \frac{f_i(y_1,\ldots,y_{j-1},z_j,\ldots,z_n)-f_i(y_1,\ldots,y_j,z_{j+1},\ldots,z_n)}{z_j-y_j}\,,
\end{align}
where $y_1,\ldots,y_n$ are auxiliary variables. Next, we will find the remainder of $\det B$ on division by $\Gcal\cup\tilde\Gcal$, where $\tilde\Gcal\coloneqq\Gcal|_{\bm{z}\to\bm{y}}$, and decompose it in the standard basis:
\begin{align}
	\overbar{\det B}^{\Gcal\cup\tilde\Gcal} \eqqcolon \sum_{\alpha=1}^d (\det B)_\alpha e_\alpha\,,
\end{align}
where the components of the Bezoutian determinant, $(\det B)_\alpha$, are now polynomials in $\bm{y}$. The dual basis is then found by evaluating these functions on $\bm{y}=\bm{z}$ \cite{cattani2005introduction}
\begin{align}
	\Delta_\alpha = (\det B)_\alpha (\bm{z})\,.
\end{align} 

\subsection{Finding Polynomial Inverses}\label{sec:APP_poly-inverse}

In section \ref{sec:POS_pf} (using results from appendix \ref{sec:APP_decompose-unity}) we showed how to calculate the push forward of polynomial function $p\in\Cbb(\bm{a})[\bm{z}]$ using the global duality of residues. In this appendix we will show how this result can be generalised to rational functions $p/q$, $q\in \Cbb(\bm{a})[\bm{z}]$ by finding the \emph{polynomial inverse} of $q$, where we assume that $q$ does not have any common zeroes with $f_1,\ldots,f_n$. 

Since $f_1,\ldots,f_n,q$ do not have any common zeroes, we have $\Vcal(\<f_1,\ldots,f_n,q)\>=\emptyset$. Then, since $1\in\Cbb(\bm{a})[\bm{z}]$, \nameref{thm:nullstellensatz} implies the existence of the polynomials $\tilde{f}_1,\ldots,\tilde{f}_n,q_{\text{inv}}$ such that
\begin{align}
	\tilde{f}_1 f_1+\cdots + \tilde{f}_n f_n +q q_{\text{inv}} =1 \implies \overbar{q q_{\text{inv}}}^\Gcal = 1\,.
\end{align}
We will introduce an auxiliary variable $y$ and define the ideal
\begin{align}
	\Jcal = \<f_1,\ldots,f_n,yq-1\>\subseteq \Cbb(\bm{a})[y,z_1,\ldots,z_n]\,.
\end{align}
Next, we define the monomial `block' order $\succ$ such that $z_1\succ \cdots\succ z_n$, and
\begin{align}
	\begin{cases}
		y^{\alpha_1} \bm{z}^{\bm{\beta}_1}\succ y^{\alpha_2} \bm{z}^{\bm{\beta}_2}\quad &\text{ if } \alpha_1 > \alpha_2\\
		y^\alpha \bm{z}^{\bm{\beta}_1}\succ y^\alpha \bm{z}^{\bm{\beta}_2}\quad &\text{ if } \bm{z}^{\bm{\beta}_1} \succ_{\text{grlex/grevlex}} \bm{z}^{\bm{\beta}_2}
	\end{cases}\,,
\end{align}
where $\succ_{\text{grlex/grevlex}}$ can be either graded lexicographic or graded reverse lexicographic ordering. 

Inside the \Grob basis $\Gcal_\succ(\Jcal)$ there will be a single polynomial linear in $y$ of the form
\begin{align}
	y-q_{\text{inv}}(\bm{z})\in\Gcal_{\succ}(\Jcal)\,,
\end{align}
where $q_{\text{inv}}(\bm{z}) \in \Cbb(\bm{a})[\bm{z}]$ is the polynomial inverse of $q$.

%% file: appendices/schubert.tex
\chapter{Intersections of Lightcones}\label{sec:APP_schubert}

In this appendix we explore some of the mathematics of lightcones (or null-cones), and how they intersect. Throughout this thesis we frequently encounter intersections of null-cones, notably they appear as boundaries of the positive geometry $\Delta(\bm{x})$, which plays a leading role in chapter \ref{sec:DUAL}. In section \ref{sec:APP_max-intersection} we first derive a completely general formula for the maximal intersection of $D$ lightcones in $D$-dimensional Minkowski space, which appear as the vertices of $\Delta(\bm{x})$. In section \ref{sec:APP_max-intersection-3D4D} we specialise to three and four dimensions. We note that in four dimensions we can phrase everything in terms of twistor variables, in which case the maximal intersection of lightcones corresponds to the intersection of four lines in twistor space, which is sometimes referred to as the \emph{Schubert problem}. We further show that the results we found in section \ref{sec:APP_max-intersection} can easily be generalised to the massive case in section \ref{sec:APP_max-intersection-massive}, where we give a general formula for the maximal intersection of mass-shells in $D$-dimensional Minkowski space with general masses. This further allows us to give an explicit formula for non-maximal intersections of lightcones. We note that our results in Minkowski space can easily be generalised to an arbitrary signature $\Rbb^{d_1,d_2}$ at the cost of a few minus signs.

\section[head={Maximal},tocentry={Maximal Intersection of Lightcones}]{Maximal Intersection of Lightcones}\label{sec:APP_max-intersection}

We consider $D$ generic points in $\Rbb^{1,D-1}$ with metric $\eta=\diag(-1,1,1,\ldots)$. We define
\begin{align}
	X_{ab} = (x_a-x_b)^2 = \eta_{\mu\nu}(x_a-x_b)^\mu(x_a-x_b)^\nu,\quad (a \cdot b) = \eta_{\mu\nu}x_a^\mu x_b^\nu\,.
\end{align}
The \emph{lightcone} of $x_a^\mu$ is defined as
\begin{align}
	\Ncal_a\coloneqq \{y^\mu\in\Rbb^{1,D-1}\colon (y-x_a)^2=0\}.
\end{align}
The maximal intersection of the lightcones of points $x_1,\ldots,x_D$ generically consists of two points $\{q^+_{},q^-_{}\}=\Ncal_1\cap \Ncal_2\cap\cdots\cap\Ncal_D$. Our aim is to find an explicit formula for $q^\pm$.

We define 
\begin{subequations}
\begin{align}
	H^0&\coloneqq \{ y\in\Rbb^{1,D-1}\colon \epsilon(1,2,\ldots,D,y) = 0\}\,,\\
	H^+&\coloneqq \{ y\in\Rbb^{1,D-1}\colon \epsilon(1,2,\ldots,D,y) > 0\}\,,\\
	H^-&\coloneqq \{ y\in\Rbb^{1,D-1}\colon \epsilon(1,2,\ldots,D,y) < 0\}\,,
\end{align}
\end{subequations}
where
\begin{align} 
	\epsilon(1,2,\ldots,D,y) =\begin{vmatrix} 1&1&\cdots&1&1\\x_1^\mu&x_2^\mu&\cdots&x_D^\mu&y^\mu \end{vmatrix}.
\end{align}
$H^0 $ is a $(D-1)-$dimensional hyperplane that passes through all points $x_1,\ldots,x_D$, and $q^\pm_{}\in H^\pm $. We define $S^\mu$ to be the symmetric combination of $q^+$ and $q^-$:
\begin{align}
	S^\mu \coloneqq \frac{ q^+_{} + q^-_{}}{2}\,,
\end{align}
which is fully symmetric with respect to permutations of $x_1,x_2,\ldots, x_D$. We further define
\begin{align}\label{eq:equi-set-def}
	\Ecal \coloneqq\{y\in\Rbb^{1,D-1}\colon (y-x_1)^2=(y-x_2)^2=\ldots=(y-x_D)^2\}\,,
\end{align}
which is the set of the points that are equidistant from all points $x_1,x_2\ldots,x_D$.

\begin{figure}
	\centering
	\includegraphics{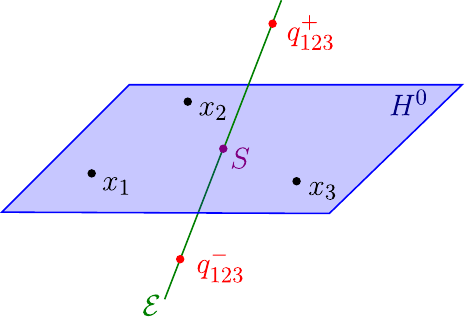}
	\caption{An illustration depicting the set-up for the case when $D=3$.}
	\label{fig:D3-ex}
\end{figure}

Since $(y-x_a)^2-(y-x_b)^2=x_a^2-x_b^2-2y\cdot(x_a-x_b)=0,\,\forall y \in \Ecal $, we see that \eqref{eq:equi-set-def} can be written as a $\binom{D}{2}$ linear relation ($D-1$ of which independent) on $y$. Thus, $\Ecal $ is a straight line in $\Rbb^{1,D-1}$ that passes through $ q^+_{}, \, q^-_{}$, and $S $. In fact, we find that 
\begin{align}
	S^\mu = H^0 \cap \Ecal \,.
\end{align}
These definitions are illustrated for three dimensions in figure \ref{fig:D3-ex}.

Assuming $\epsilon(x_1,\ldots,x_D)\neq0$ (assuming the points are generic we can always do a translation to ensure this), we can expand
\begin{align}
	S^\mu = \sum_{i=1}^D f_i x_i^\mu\,.
\end{align}
Since $S\in\Ecal$, we have $(S-x_a)^2-(S-x_b)^2=0$, which now becomes
\begin{align}\label{eq:fi-rel-1}
	\sum_{i=1}^D f_i\, [(i\cdot a)-(i\cdot b)] = \frac{x_a^2-x_b^2}{2}\,,
\end{align}
which imposes $D-1$ constraints on the $f_i$. Furthermore, since $S\in H^0$, we have the constraint
\begin{align}
	0=\epsilon(1,2,\ldots,D,S) = (f_1+f_2+\ldots+f_D-1)
	\eps_{\mu_1 \, \mu_2\ldots\mu_D}x_1^{\mu_1}x_2^{\mu_2}\cdots x_D^{\mu_D}\,,
\end{align}
so we get the final constraint on $f_i$: $\sum_{i=1}^D f_i=1$. Using the fact that $X_{ab}=x_a^2+x_b^2-2(a\cdot b)$, we can rewrite equation \eqref{eq:fi-rel-1} as
\begin{align}
	&\sum_{i=1}^D f_i \frac{x_i^2+x_a^2-X_{ia}-x_i^2-x_b^2+X_{ib}}{2} = \frac{x_a^2-x_b^2}{2}\implies \sum_{i=1}^Df_i\frac{X_{ib}-X_{ia}}{2}=0\\
	&\implies \sum_{i=1}^D f_i \frac{X_{ia}}{2}= \lambda\quad\forall a\,,
\end{align}
for some constant $\lambda$ (we keep the factor of 1/2 for future simplifications). This condition, together with $\sum_{i} f_i =1$ allows us to solve for the $f_i$. We introduce\footnote{As is the case for $q^\pm$, we drop the subscript which labels the points $x_1,\ldots,x_D$ in this appendix, although we often write them explicitly in the main body of this thesis.}
\begin{align}\label{eq:APP_Xmat-def}
	\Xcal_{12\cdots D}=\frac{1}{2}\begin{pmatrix}
		0 & X_{12} & X_{13} & \cdots & X_{1D}\\
		X_{12} & 0 & X_{23} & \cdots & X_{2D}\\
		X_{13} & X_{23} & 0 & \cdots & X_{3D}\\
		\vdots & \vdots & \vdots & \ddots & \vdots\\
		X_{1D} & X_{2D} & X_{3D} & \cdots & 0\\
	\end{pmatrix}\,,
\end{align}
and we use
\begin{align}
	\bm{f}=\begin{pmatrix}
		f_1 \\ f_2 \\ \vdots \\ f_n
	\end{pmatrix}\,,\quad \bm{1}=\begin{pmatrix}
		1 \\ 1 \\ \vdots \\ 1
	\end{pmatrix}\,.
\end{align}
Our constraints on $\bm{f}$ can then be written in matrix form as
\begin{align}
	\bm{1}^T\cdot\bm{f}&=1\,,\\
	\Xcal\cdot\bm{f}&=\lambda\bm{1}\,,
\end{align}
which has the solution
\begin{align}
	\bm{f}&=\lambda\,\Xcal^{-1}\cdot\bm{1}\,,\\
	\lambda^{-1}&=\bm{1}^T\cdot\Xcal^{-1}\cdot\bm{1}=\text{su}(\Xcal^{-1})\,,
\end{align}
where $\text{su}(\Xcal^{-1})$ denotes the \emph{grand sum} of $\Xcal^{-1}$, \emph{i.e.} the sum of all elements. We find an explicit expression for $S^\mu$:
\begin{align}\label{eq:S-sol}
	S^\mu = \frac{\bm{1}^T\cdot \Xcal^{-1}\cdot\bm{x}^\mu}{\text{su}(\Xcal^{-1})}\,,
\end{align}
where the dot product in the numerator should be interpreted as
\begin{align}
	\bm{1}^T\cdot \Xcal^{-1}\cdot\bm{x}^\mu = \sum_{i,j=1}^D [\Xcal^{-1}]_{ij}\,x_i^\mu\,.
\end{align}
Now that we have solved for the symmetric combination of $q^+_{}$ and $q^-_{}$, we only have to translate into the $\Ecal$ direction to find $q^\pm_{}$. We note that any point in $\Ecal $ can be decomposed as $S^\mu + \alpha W^\mu$, where $W^\mu$ is any vector that satisfies
\begin{align}
	W\cdot x_a = W\cdot x_b\quad\forall a,b\,.
\end{align}
We can see that this is true by noting that
\begin{align}
	\big((S + \alpha W)-x_a\big)^2=\alpha^2 W^2+(S-x_a)^2 +2 \alpha W\cdot S -2 \alpha W\cdot x_a\,,
\end{align}
and since $(S-x_a)^2$ has the same value for all $a$, we find that this is indeed independent of particle label as long as $W\cdot x_a$ takes the same value for all $a$. It follows directly from \eqref{eq:S-sol} that $W\cdot S = W\cdot x_a$. A natural choice for $W^\mu$ is
\begin{align}
	W^\mu&=\sum_{i=1}^D (-1)^{i+1} \eps^\mu_{ \;\nu_1 \nu_2\cdots \nu_{D-1}} x_1^{\nu_1} x_2^{\nu_2}\cdots \hat{x}_i\cdots x_D^{\nu_{D-1}}\\
	&=\eta^{\mu\nu}\frac{\partial}{\partial y^\nu} \epsilon(1,2,\ldots,D,y)\,,
\end{align}
where the hat denotes omission. This choice of $W^\mu$ satisfies $W\cdot x_i = W\cdot S = \epsilon(1,2,\ldots,D)$.
We note that $q^\pm_{}$ are defined as the points on $\Ecal$ such that
\begin{align}
	\big((S + \alpha W)-x_a\big)^2=\alpha^2 W^2 + (S-x_a)^2=0\,.
\end{align}
Solving this for $\alpha$ gives us the following expression for the maximal intersection of $D$ lightcones in $\Rbb^{1,D-1}$:
\begin{align}
	q^\pm_{} = S^\mu \pm \sqrt{-\frac{(S-x_i)^2}{W^2}}W^\mu\,.
\end{align}
We note in passing that there is an equivalent way to write $W^\mu$ that looks more similar to our expression for $S^\mu$. If we write $W^\mu=\sum_i g_i x_i^\mu$, then $W\cdot x_a=\epsilon(1,2,\ldots,D)$ implies
\begin{align}
	W^\mu =\epsilon(1,2,\ldots,D) \bm{1}^T\cdot \mathcal{G}^{-1}\cdot\bm{x}^\mu\,,
\end{align}
where $\mathcal{G}$ is the \emph{Gram matrix} with elements $[\mathcal{G}]_{ij}=(i \cdot j)$.

For the sake of reference we will now record several identities that relate the quantities defined in this appendix. This will allow us to simplify our expression for $q^\pm_{}$.
\begin{subequations}
	\begin{align}
		&W^2=\det\Xcal \;\text{su}(\Xcal^{-1})= \det\mathcal{G}\; \text{su}(\mathcal{G}^{-1})\,,\\
		&\epsilon(1,2,\ldots,D,q^\pm_{}) = 
		\pm2^{-D}\left|\frac{\partial (y-x_1)^2\cdots \partial (y-x_D)^2}{\partial y^\mu}\right|_{y^\mu=q^\pm_{}} = \pm\sqrt{-\det\Xcal}\,,\\
		&\lambda=(S-x_a)^2=1/\text{su}(\Xcal^{-1})=-(q^+_{} - q^-_{})^2/4\,,\\
		&\det\mathcal{G}=\epsilon(1,2,\ldots,D)^2=(W\cdot x_a)^2=(W\cdot S)^2\,,\\
		&W^\mu = \frac{2\sqrt{-\det\Xcal}}{(q^+_{}- q^-_{})^2} (q^+_{}-q^-_{})\,,\\
		&\epsilon(1,2,\ldots,D,y)=\frac{\sqrt{-\det\Xcal}}{(q^+_{}- q^-_{})^2} \big((y-q^-_{})^2-(y-q^+_{})^2\big)\,.
	\end{align}
\end{subequations}
Using these identities, we can rewrite our expression for $q^\pm_{}$ as 
\begin{align}\label{eq:APP_qpm-gen}
	q^\pm_{} = \frac{\det\Xcal\;(\bm{1}^T\cdot\Xcal^{-1}\cdot \bm{x}^\mu)\pm\sqrt{-\det\Xcal}W^\mu}{W^2}\,.
\end{align}

\subsection{Maximal Intersections in Three and Four Dimensions}\label{sec:APP_max-intersection-3D4D}

Explicitly, for $D=3$ we find
\begin{align}\label{eq:SCH_q-3D-gen}
	q^\pm_{abc}=\frac{1}{4 W^2}\Big[&x_a^\mu X_{bc}\big(X_{ab}+X_{ac}-X_{bc}\big)+x_b^\mu X_{ac}\big(X_{ab}+X_{bc}-X_{ac}\big)+x_c^\mu X_{ab}\big(X_{ac}+X_{bc}-X_{ab}\big)\nonumber\\& \pm 2\sqrt{-X_{ab}X_{bc}X_{ca}}W^\mu
	\Big]\,,
\end{align}
where
\begin{align}
	W^\mu &= \eps^\mu_{\;\nu\rho}(x_a^\nu x_b^\rho - x_a^\nu x_c^\rho + x_b^\nu x_c^\rho),\\
	W^2 &= -\frac{1}{4}[X_{ab}^2+X_{ac}^2+X_{bc}^2-2(X_{ab}X_{ac}+X_{ab}X_{bc}+X_{ac}X_{bc})]\,.
\end{align}
For $D=4$ we pick up a minus sign in front of each $\det\Xcal$ due to the change in signature from $\Rbb^{1,3}$ to $\Rbb^{2,2}$:
\begin{alignat}{3}\label{eq:SCH_q-4D-gen}
	q^\pm_{abcd}=\frac{1}{8W^2}\Big[ &x_a^\mu\big( &&X_{ab}X_{cd}(X_{bc}+X_{bd}-X_{cd}) &&+X_{ac}X_{bd}(X_{bc}+X_{cd}-X_{bd}) \notag\\ 
	&+ &&X_{ad}X_{bc}(X_{bd}+X_{cd}-X_{bc}) &&-2 X_{bc}X_{bd}X_{cd}\big)+ \notag\\
	&x_b^\mu\big( &&X_{ab}X_{cd}(X_{ac}+X_{ad}-X_{cd}) &&+X_{ac}X_{bd}(X_{ad}+X_{cd}-X_{ac}) \notag\\ 
	&+ &&X_{ad}X_{bc}(X_{ac}+X_{ad}-X_{cd}) &&-2 X_{ac}X_{ad}X_{cd}\big)+ \notag\\ 
	&x_c^\mu\big( &&X_{ab}X_{cd}(X_{ad}+X_{bd}-X_{ab}) &&+X_{ac}X_{bd}(X_{ab}+X_{ad}-X_{bd}) \notag\\ 
	&+ &&X_{ad}X_{bc}(X_{ab}+X_{bd}-X_{ad}) &&-2 X_{ab}X_{ad}X_{bd}\big)+ \notag\\
	&x_d^\mu\big( &&X_{ab}X_{cd}(X_{ac}+X_{bc}-X_{ab}) &&+X_{ac}X_{bd}(X_{ab}+X_{bc}-X_{ac}) \notag\\ 
	&+ &&X_{ad}X_{bc}(X_{ab}+X_{ac}-X_{bc}) &&-2 X_{ab}X_{bc}X_{ab}\big) \notag\\
	&\pm&& 2 \sqrt{\Delta} W^\mu\Big]\,,
\end{alignat}
where
\begin{align}
	W^\mu&=\eps^\mu_{\;\nu\rho\sigma} (x_a^\nu x_b^\rho x_c^\sigma - x_a^\nu x_b^\rho x_d^\sigma +x_a^\nu x_c^\rho x_d^\sigma -x_b^\nu x_c^\rho x_d^\sigma )\,,\\
	W^2&=\frac{1}{4}\big[X_{ab}X_{ac}X_{bd}-X_{ab}X_{ac}X_{bc}-X_{ab}^2X_{cd}+\substack{\text{perms.}\\\text{(no duplicates)}}\big]\,,\\
	\Delta&=2\big[ (X_{ab}X_{cd})^2+(X_{ac}X_{bd})^2+(X_{ad}X_{bc})^2 \big]-(X_{ab}X_{cd}+X_{ac}X_{bd}+X_{ad}X_{bc})^2 \notag\\
	&=32\det\Xcal_{abcd}\,.
\end{align}
When $X_{ab}=0$, we see that $\Delta=(X_{ad}X_{bc}-X_{ac}X_{bd})^2$, such that the square root disappears from $q^\pm$ when any two of the points $x_a,x_b,x_c,x_d$ are null-separated.

\section[head={Non-Maximal},tocentry={Maximal Intersections of Mass-Shells, and Non-Maximal Intersections of Lightcones}]{Maximal Intersections of Mass-Shells, and Non-Maximal Intersections of Lightcones}\label{sec:APP_max-intersection-massive}

If we consider the maximal intersection of $D$ mass-shells, instead of $D$ lightcones, we get a remarkably similar formula to \eqref{eq:APP_qpm-gen}. Explicitly, if we want to solve
\begin{align}
	(y-x_1)^2=m_1^2,\,(y-x_2)^2=m_2^2,\ldots,(y-x_D)^2=m_D^2\,,
\end{align}
for some arbitrary masses $m_1,\ldots,m_D$, then the solution has the same general form:
\begin{align}\label{eq:APP_mass-shell-intersect}
	Q^\pm_{} = \frac{\det\overline{\Xcal}\;\big(\bm{1}^T\cdot\overline{\Xcal}^{-1}\cdot\bm{x}^\mu\big)\pm\sqrt{-\det\overline{\Xcal}}W^\mu}{W^2}\,,
\end{align}
where the only difference is in the $\Xcal$ matrix, which is now defined as
\begin{align}\label{eq:APP_Xmat-massive-def}
	\overline{\Xcal}\coloneqq\frac{1}{2}\begin{pmatrix}
		-2m_1^2 & X_{12}-m_1^2-m_2^2 & X_{13}-m_1^2-m_3^2 & \cdots & X_{1D}-m_1^2-m_D^2\\
		X_{12} -m_1^2-m_2^2& -2m_2^2 & X_{23}-m_2^2-m_3^2 & \cdots & X_{2D}-m_2^2-m_D^2\\
		X_{13}-m_1^2-m_3^2 & X_{23} -m_2^2-m_3^2& -2m_3^2 & \cdots & X_{3D}-m_3^2-m_D^2\\
		\vdots & \vdots & \vdots & \ddots & \vdots\\
		X_{1D} -m_1^2-m_D^2& X_{2D} -m_2^2-m_D^2& X_{3D}-m_3^2-m_D^2 & \cdots & -2m_D^2\\
	\end{pmatrix}\,.
\end{align}
We note that we can change our signature from $\Rbb^{1,D-1}$ to $\Rbb^D$, in which case equation \eqref{eq:APP_mass-shell-intersect} gives an explicit formula for the intersection of $D$ spheres in $D$ dimensions with radii $m_1,\ldots,m_D$.

This formula also allows us to write down a general solution to non-maximal intersections of lightcones. Consider the case where we want to find the solutions in $\Rbb^{1,D-1}$ to
\begin{align}\label{eq:non-max-int}
	(y-x_1)^2=(y-x_2)^2=\ldots=(y-x_k)^2=0\,.
\end{align}
We note that we can rewrite
\begin{align}
	&(y-x_i)^2=(y^0-x_i^0)^2-(y^1-x_i^1)^2-\ldots-(y^{D-1}-x_i^{D-1})^2=0\implies\\
	&(y^0-x_i^0)^2-(y^1-x_i^1)^2-\ldots-(y^{k-1}-x_{i}^{k-1})^2=\sum_{a=k}^{D-1} (y^a-x_i^a)^2\,,
\end{align}
where the second line now looks like the the mass-shell equation in $\Rbb^{1,k-1}$. Hence, solving \eqref{eq:non-max-int} is equivalent to finding the intersection of $k$ mass-shells in $\Rbb^{1,k-1}$:
\begin{align}
	(y-x_1)^2=m_1^2,\,(y-x_2)^2=m_2^2,\ldots,(y-x_k)^2=m_k^2\,,
\end{align}
where $m_i^2=\sum_{a=k}^{D-1} (y^a-x_i^a)^2$.

As an example, we can use this to prove the existence of \emph{composite singularities} in four dimensions. To stay in line with the main text, we will do this in $\Rbb^{2,2}$. Composite singularities emerge when considering the form
\begin{align}
	\frac{\dd^4 y}{(y-x_{i-1})^2(y-x_i)^2(y-x_{i+1})^2}\,,
\end{align}
and taking the residue at
\begin{align}
	(y-x_{i-1})^2=(y-x_i)^2=(y-x_{i+1})^2=0\,,
\end{align}
where we assume $(x_i-x_{i-1})^2=(x_{i+1}-x_i)^2=0$. Using the formula above, we can find an explicit expression for this non-maximal intersection of null-cones (we will give an explicit formula in terms of bi-spinors in appendix \ref{sec:APP_white-black}, they correspond to $y$ being on the null-ray $e^\pm_{i-1ii+1}$). However, this explicit solution is not what we are after at the moment. The Jacobian we pick up from such a residue is precisely given by $\smash{\sqrt{-\det\overline{\Xcal}_{i-1ii+1}}}$. Using $m_a= y^3-x_a^3$ in \eqref{eq:APP_Xmat-massive-def}, we find the explicit answer
\begin{align}
	\det\overline{\Xcal}_{i-1ii+1}=-\frac{1}{4}(y^3-x_i^3)^2 X_{i-1\,i+1}^2\,,
\end{align}
and hence
\begin{align}
	\mathop{\Res}_{y=e^\pm_{i-1ii+1}} \frac{\dd^4 y}{(y-x_{i-1})^2(y-x_i)^2(y-x_{i+1})^2} = \pm \frac{\dd y^3}{4 X_{i-1\,i+1}(y^3-x_i^3)}\,.
\end{align}
We see that we have developed a new pole coming from the Jacobian in the denominator. Localising on this pole restricts $y=x_i$.
Thus, we find that the form
\begin{align}
	\frac{4 X_{i-1\,i+1}\dd^4 y}{(y-x_{i-1})^2(y-x_i)^2(y-x_{i+1})^2}\,,
\end{align}
has a residue of $1$ at $y=x_i$. A similar calculation yields that $\omega^\square_{i-1ii+1j}$, as defined in equation \eqref{eq:DUAL_omega-box-def-4D}, also has a residue of $1$ at $y=x_i$.

%% file: appendices/black-white.tex
\chapter{White and Black Planes}\label{sec:APP_white-black}

In this appendix we continue the study of white and black planes in $\Rbb^{2,2}$ initiated in section \ref{sec:DUAL_nullcone-geometry-4D}. We recall that when $(x_i-x_{i-1})^2=0$ the intersection of two null-cones $\Ncal_{x_{i-1}}\cap\Ncal_{x_i}$ decomposes into two affine planes given by
\begin{align}
	W_{i-1 i}&\coloneqq \{y\in \Ncal_{x_{i-1}}\cap\Ncal_{x_{i}}\colon \lambda_i \propto \kappa_{i-1} \propto \kappa_{i}\}\,,\\
	B_{i-1 i}&\coloneqq \{y\in \Ncal_{x_{i-1}}\cap\Ncal_{x_{i}}\colon\tilde\lambda_i\propto \tilde\kappa_{i-1}\propto\tilde\kappa_{i}\}\,,
\end{align}
where we use spinor-helicity variables to write $(x_i-x_{i-1})^{\alpha\alphadot}=\lambda_{i}^\alpha\tilde\lambda_i^\alphadot$, $(y-x_{i})^{\alpha\alphadot}=\kappa_i^\alpha\tilde\kappa_i^\alphadot$, $(y-x_{i-1})^{\alpha\alphadot}=\kappa_{i-1}^\alpha\tilde\kappa_{i-1}^\alphadot$.

As an example, we consider $W_{12}$. A first interesting observation is that intersecting this plane with \emph{any} null-cone will be a straight line (a null-ray). To see this, we note that
\begin{align}
	y\in W_{12} \implies y^{\alpha\alphadot} = \lambda_1^\alpha \tilde\kappa^\alphadot + x_1^{\alpha\alphadot}\,,
\end{align}
where the position of this point is completely determined by the 2-vector $\tilde\kappa$. If we further require that this $y$ is also on $\Ncal_i$ for some $x_i$, then
\begin{align}
	\det(\lambda_1\tilde\kappa+x_1-x_i)=0\implies X_{1i}+\<\lambda_1|x_{1i}|\tilde\kappa]=0\,,
\end{align}
which is a linear constraint on $\tilde\kappa$, and thus defines a straight line in $W_{12}$ (we use $x_{1i}=x_i-x_1$). Explicitly solving $\tilde\kappa^{\dot{2}}$ in terms of $\tilde\kappa^{\dot{1}}$ gives
\begin{align}
	\tilde\kappa^{\dot{2}} = \frac{-X_{1i}+ \<\lambda_1|x_{1i}^{\dot{2}}\tilde\kappa^{\dot{1}}}{\<\lambda_1|x_{1i}^{\dot{1}}}\,,
\end{align}
where $\<\lambda_1|x_{1i}^{\dot{\alpha}}=\epsilon_{\alpha\beta}\lambda_1^\alpha x_{1i}^{\beta\alphadot}$. 

If we further intersect with $\Ncal_{j}$, we find a unique point
\begin{align}
	y=x_1+\lambda_1\tilde\kappa \in W_{12}\cap \Ncal_i\cap \Ncal_j \implies \tilde\kappa^\alphadot= \frac{X_{1i}\<\lambda_1|x_{1j}^\alphadot-X_{1j}\<\lambda_1|x_{1i}^\alphadot}{\<\lambda_1|x_{1i}\cdot x_{1j} | \lambda_1\>}\,,
\end{align}
where $\<\lambda_1|x_{1i}\cdot x_{1j} | \lambda_1\> = \epsilon_{\alpha\beta}\epsilon_{\gamma\delta}\epsilon_{\alphadot\betadot} \lambda_1^\alpha x_{1i}^{\beta\alphadot}x_{1j}^{\gamma\betadot}\lambda_1^\delta$. It is clear that this point is null-separated from $x_1,x_2,x_i,x_j$, and this is thus a bi-spinor formula for $q^\pm_{12ij}$, the opposite $q^\mp_{12ij}$ being given by its parity conjugate\footnote{In particular, if $X_{1i}X_{2j}-X_{1j}X_{2i}>0$, then $q^+_{12ij}\in W_{12}\cap \Ncal_i\cap \Ncal_j$, otherwise it is $q^-_{12ij}$.}.

Some significant simplifications occur when $j=i+1$, where we assume $(x_{i+1}-x_i)^2=0$. In this case
\begin{align}
	W_{12}\cap\Ncal_i\cap\Ncal_{i+1}=W_{12}\cap W_{ii+1}\,,
\end{align}
which means that 
\begin{align}
	y=x_1+\lambda_1\tilde\kappa = x_i+\lambda_i\tilde\kappa'\,.
\end{align}
Contracting this equation with $\lambda_i$ allows us to solve for $\tilde\kappa$, which gives
\begin{align}
	\tilde\kappa^\alphadot = -\frac{\<\lambda_i|x_{1i}^\alphadot}{\<1i\>}\,.
\end{align}
Using $x_{1i}=\sum_{l=1}^{i-1}\lambda_l\tilde\lambda_l$ we find
\begin{align}
	y-x_1 = \frac{1}{\<1i\>}\sum_{l=1}^{i-1}\lambda_1\<il\>\tilde\lambda_l\,,
\end{align}
in agreement with the definition of $\ls_{1i}$ in equation \eqref{eq:KIN_ls-def}. 

From a similar line of reasoning we can find a simple bi-spinor parametrisation of the null-rays $e^\pm_{i-1ii+1}$:
\begin{align}
	y \in e^+_{i-1ii+1}=W_{i-1i}\cap\Ncal_{i+1} \implies \<\lambda_{i-1}|x_{i i+1}|\tilde{\kappa}]=\<\lambda_{i-1}\lambda_i\>[\tilde\lambda_i\tilde{\kappa}]=0\,,
\end{align}
which has the simple solution $\tilde\kappa \propto \tilde\lambda_i$. So
\begin{alignat}{3}
	y &\in e^+_{i-1ii+1}=W_{i-1i}&&\cap\Ncal_{i+1} &&\implies y = x_i+\alpha \lambda_{i-1}\tilde\lambda_i\,,\\
	y &\in e^-_{i-1ii+1}=B_{i-1i}&&\cap\Ncal_{i+1} &&\implies y = x_i+\alpha \lambda_{i}\tilde\lambda_{i-1}\,.\label{eq:APP_emin}
\end{alignat}
Intersecting this with $\Ncal_{j}$ allows us to solve for $\alpha$, giving us a simple bi-spinor formula for $q^\pm_{i-1ii+1j}$:
\begin{align}
	q^+_{i-1ii+1j}&=x_i-\frac{X_{ij}}{\<i-1|x_{ij}|i]}\lambda_{i-1}\tilde\lambda_{i}\,,\\
	q^-_{i-1ii+1j}&=x_i-\frac{X_{ij}}{\<i|x_{ij}|i-1]}\lambda_i\tilde\lambda_{i-1}\,.
\end{align}

\section[head={Relation to BCFW},tocentry={Relation to BCFW Shifts}]{Relation to BCFW Shifts}\label{sec:APP_WB-BCFW}

\begin{figure}[t]
	\centering
	\includegraphics[width=0.6\textwidth]{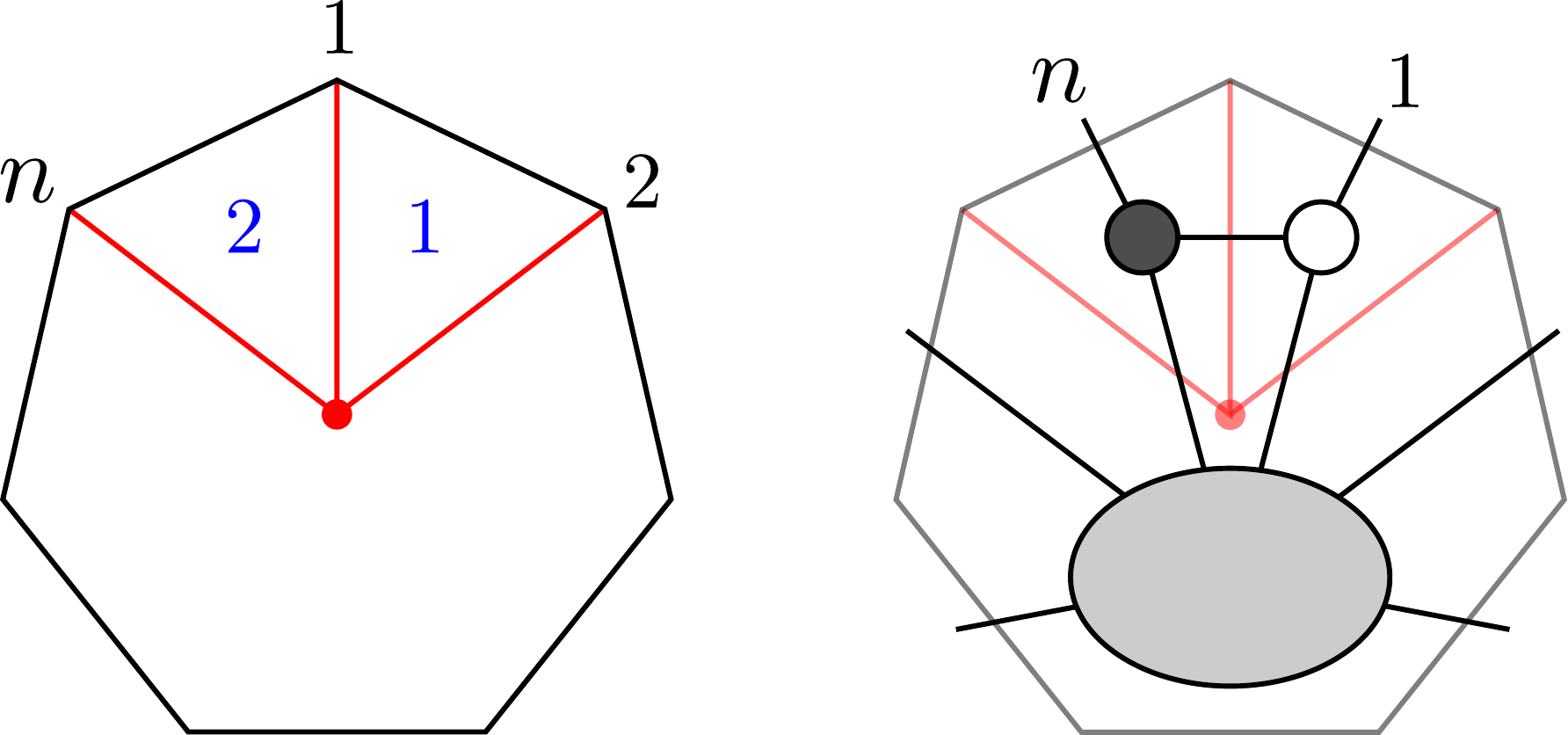}
	\caption{The boundary diagram corresponding the the null-ray $e^-_{n12}$ (left) is dual to an on-shell diagram with a BCFW bridge (right)}
	\label{fig:edge-bcfw}
\end{figure}

We now make an interesting observation regarding the edge $e^-_{n12}$ from equation \eqref{eq:APP_emin} and the BCFW shift discussed in section \ref{sec:AMP_BCFW}.

Assume that we have a null-polygon in dual space with corners $\{x_i\}_{i=1}^n$ which represent the massless momenta $p_i=x_{i+1}-x_i$. Let us consider shifting $x_1$ along the null-ray $e^-_{n12}$. That is,
\begin{align}
	x_1\to \hat{x}_1=x_1+\alpha \lambda_1\tilde\lambda_n\,.
\end{align}
The effect of this shift on the momentum vectors is
\begin{align}
	p_1\to\;&\hat{p}_1(\alpha)=p_1-\alpha \lambda_1\tilde\lambda_n\,,\\
	p_n\to\;&\hat{p}_n(\alpha)=p_n+\alpha \lambda_1\tilde\lambda_n\,.
\end{align}
We note that this shift is \emph{precisely} the BCFW shift we introduced in equation \eqref{eq:AMP_BCFW-shift-mom-vec}! We thus see that a BCFW shift has a very natural geometric interpretation in dual space: it simply shifts the point $x_1$ along $e^-_{n12}$. Shifting $x_1$ along the null-ray $e^+_{n12}$ instead would correspond to the parity-conjugate BCFW shift. When $\hat{x}_1(\alpha)$ approaches another null-cone $\Ncal_{x_i}$, then $\hat{X}_{1i}= (\hat{x}_1-x_i)^2\to 0$, which corresponds to a pole of the shifted amplitude. Hence, all the terms in the BCFW expansion correspond to sending $\hat{x}_1$ to some $q^-_{n12 i}$. We further note that the boundary diagram corresponding to the edge $e^-_{n12}$ is precisely dual to a BCFW bridge, as can be seen in figure \ref{fig:edge-bcfw}.

%% file: appendices/scatt-eq.tex
\chapter{Scattering Equations in Four Dimensions}\label{sec:scatt-eq-4D}

In this thesis we have encountered various equivalent formulations for the scattering equations. In this appendix we will derive several versions of the four-dimensional scattering equations which we encountered in section \ref{sec:AMP_scatt-eq-4D}.

We have seen that the scattering equations are equivalent to the condition that the vector
\begin{align}
	P^\mu(z)=\sum_{a=1}^nk_a^\mu \prod_{b\neq a}(z-z_b),
\end{align}
is a null-vector (\emph{i.e.} $P^2(z)=0$ for all $z$). In four dimensions, this condition can be solved by using spinor-helicity variables: 
\begin{align}\label{eq:scatt-eq-4D-constraint}
	P^{\alpha\dot\alpha}(z)=\pi^\alpha(z)\tilde\pi^{\dot\alpha}(z)=\sum_{a=1}^n \lambda_a^\alpha \tilde\lambda_a^{\dot\alpha}\prod_{b\neq a}(z-z_b).
\end{align}
We see that $P^{\alpha\dot\alpha}(z)$ is a polynomial in $z$ of degree $n-2$ (the term proportional to $z^{n-1}$ drops out on the support of momentum conservation) which factorizes into two polynomials $\pi^\alpha(z)$ and $\tilde\pi^{\dot\alpha}(z)$ of degree $d$ and $n-d-2$ respectively. For generic kinematics, there are no solutions for $d=0,n-2$, and hence we see that the scattering equations split up into $n-3$ sectors for $d=1,\ldots,n-3$. From \eqref{eq:scatt-eq-4D-constraint} we see that 
\begin{align}
	\pi^\alpha(z_a)\tilde\pi^{\dot\alpha}(z_a)=\lambda_a^\alpha\tilde\lambda_a^{\dot\alpha}\prod_{b\neq a}(z_a-z_b)\implies \lambda_a^\alpha\tilde\lambda_a^{\dot\alpha}=\frac{\pi^\alpha(z_a)\tilde\pi^{\dot\alpha}(z_a)}{\prod_{b\neq a}(z_a-z_b)}.
\end{align}
We make a general ansatz for $\pi$ and $\tilde\pi$ as degree $d$ and degree $n-d-2$ polynomials as
\begin{align}
	\pi^\alpha(z)=\sum_{m=0}^d \rho^\alpha_m z^m,\quad \tilde\pi^{\dot\alpha}(z)=\sum_{\tilde{m}=0}^{n-d-2}\tilde{\rho}_{\tilde{m}}^{\dot\alpha}z^{\tilde{m}},
\end{align}
which gives
\begin{align}
	\lambda_a^\alpha\tilde\lambda_a^{\dot\alpha}=\frac{\big(\sum_{m=0}^d \rho^\alpha_m z_a^m\big)\big(\sum_{\tilde{m}=0}^{n-d-2}\tilde{\rho}_{\tilde{m}}^{\dot\alpha}z_a^{\tilde{m}}\big)}{\prod_{b\neq a}(z_a-z_b)}.
\end{align}
This is solved by 
\begin{subequations}\label{eq:scatt-eq-WRSV-Cachazo}
	\begin{align}
		\lambda_a^\alpha&=t_a\sum_{m=0}^d \rho_m^\alpha z_a^m, \\
		\tilde\lambda_a^{\dot\alpha}&=\tilde{t}_a\sum_{\tilde{m}=0}^{n-d-2} \tilde{\rho}_{\tilde{m}}^{\dot\alpha}z_a^{\tilde{m}},\label{eq:RSV-Cachaczo-ltilde}	\\
		t_a\tilde{t}_a&=\prod_{b\neq a}\frac{1}{z_a-z_b}.\label{eq:t-constraint}
	\end{align}
\end{subequations}
This gives us $5n$ equations for $5n$ variables ($2(d+1)$ $\rho$'s, $2(n-d-1)$ $\tilde\rho$'s, $n$ $z$'s, $n$ $t$'s, and $n$ $\tilde{t}$'s). These equations are equivalent to the scattering equations in the following sense: when solving \eqref{eq:scatt-eq-WRSV-Cachazo} for the $5n$ variables (after the appropriate gauge fixing) in terms of the kinematic variables $\lambda,\tilde\lambda$, the solutions for $z$ are equivalent to the solutions for $z$ of the standard scattering equations (when restricted to 4D kinematics). On such a solution for $z$, the equation \eqref{eq:scatt-eq-WRSV-Cachazo} then completely fixes solutions for the $\rho,\tilde\rho,t,\tilde{t}$ variables. This form of the scattering equations was first introduced in \cite{Cachazo:2013zc} as a parity invariant form of the Witten-RSV equations.

We define $k=d+1$ and introduce the $k\times n$ matrix $C$ with components
\begin{align}
	C_{ma}(\bm{z},\bm{t})=t_az_a^{m-1},\quad a=1,\ldots,n,\quad m=1,\ldots,k\,,
\end{align}
and the $(n-k)\times n$ matrix $\tilde{C}$ with components
\begin{align}
	\tilde{C}_{\tilde{m}a}(\bm{z},\tilde{\bm{t}})=\tilde{t}_az_a^{\tilde{m}-1},\quad a=1,\ldots,n,\quad \tilde{m}=1,\ldots,n-k\,.
\end{align}
A simple calculation shows that \eqref{eq:t-constraint} implies that
\begin{align}
	\sum_{a=1}^n C_{ma}\tilde{C}_{\tilde{m}a}=0\,,
\end{align}
and hence \eqref{eq:scatt-eq-WRSV-Cachazo} is equivalent to 
\begin{subequations}\label{eq:scatt-eq-WRSV-Cachazo-Grass}
\begin{align}
	&\lambda=\rho\cdot C(\bm{z},\bm{t}),\label{eq:APP_lambda-rho-C-Cachazo}\\
	&\tilde\lambda = \tilde\rho\cdot \tilde{C}(\bm{z},\tilde{\bm{t}})\,,\label{eq:APP_tilde-lambda-rho-C-Cachazo}\\
	&C(\bm{z},\bm{t})\cdot \tilde{C}(\bm{z},\tilde{\bm{t}})^T=\nul_{k\times (n-k)}\,,\label{eq:APP_C-Cperp-Cachazo}
\end{align}
\end{subequations}
for some $2\times k$ matrix $\rho$ and $2\times (n-k)$ matrix $\tilde\rho$. If we interpret $C$ as an element of $G(k,n)$, then \eqref{eq:APP_lambda-rho-C-Cachazo} implies that $\lambda\subseteq C$, with the matrix $\rho$ indicating which specific linear combination of column vectors of $C$ yields $\lambda$. The constraint \eqref{eq:APP_C-Cperp-Cachazo} then means that $C$ and $\tilde{C}$ are orthogonal, \emph{i.e.} $\tilde{C}=C^\perp$. Hence, the constraint \eqref{eq:APP_tilde-lambda-rho-C-Cachazo} has the interpretation that $\tilde\lambda\subseteq C^\perp$, which we can equivalently write as $C\cdot\tilde\lambda^T=\nul$. Hence, the four dimensional scattering equations are equivalent to 
\begin{subequations}
\begin{align}
	\lambda &= \rho\cdot C(\bm{z},\bm{t})\,,\\
	\nul &= C(\bm{z},\bm{t})\cdot\tilde\lambda^T\,,
\end{align}
\end{subequations}
or, in components,
\begin{subequations}\label{eq:Witten-RSV-eq}
	\begin{align}
		\lambda_a^\alpha&=t_a\sum_{m=0}^d \rho_m^\alpha z_a^m\,, \\
		0&=\sum_{a=1}^nt_a\tilde\lambda_a^{\dot\alpha}z_a^m\,.
	\end{align}
\end{subequations}
These are known as the \emph{Witten-RSV equations} \cite{Witten:2003nn, Roiban:2004yf}. Their equivalence to the scattering equations was first shown in \cite{Cachazo:2013iaa}, and their relation to the Grassmannian was first pointed out in \cite{Arkani-Hamed:2009kmp}. A benefit of the Witten-RSV equations is that they are purely polynomial, which makes them well-suited for the \Grob basis techniques we study in section \ref{sec:POS_pf}.

We can further interpret the constraint that $\lambda\subseteq C$ as $\lambda$ being orthogonal to $C^\perp$. That is, in terms of delta functions
\begin{align}
	\int \dd^{k\times 2}\rho \;\delta^{2\times n}\big(\lambda-\rho\cdot C\big)=\delta^{(n-k)\times 2}(C^\perp \cdot \lambda^T)\,.
\end{align}
This means that after integrating out the $\rho$'s, \eqref{eq:Witten-RSV-eq} turns into the \emph{Grassmannian scattering equations}
\begin{subequations}\label{eq:scatt-eq-Grass}
\begin{align}
	C^\perp(\bm{z},\bm{t})\cdot \lambda^T &= \nul_{(n-k)\times 2}\,,\\ C(\bm{z},\bm{t})\cdot\tilde\lambda^T&=\nul_{k\times 2}\,.
\end{align}
\end{subequations}
At this point we make a few remarks about the gauge redundancies in this description of the scattering equations. Clearly, the $SL(2,\Cbb)$ redundancy in the $z_a$ variables is still in place. In addition, when writing $P^{\alpha\dot\alpha}=\pi^\alpha\tilde\pi^{\dot\alpha}$ we introduced a $GL(1,\Cbb)$ `little group' redundancy as $\pi\to s \pi,\;\tilde\pi\to s^{-1}\tilde\pi$. These combine into a $GL(2,\Cbb)$ redundancy which allows us to fix the location of three $z$'s, as well as one $t$. Our convention is to fix $z_1\to0,z_2\to1,z_n\to \infty,t_n\to 1$. Furthermore, equation \eqref{eq:scatt-eq-Grass} is invariant under a $GL(k)$ transformation of $C$, and a $GL(n-k)$ transformation of $C^\perp$. 

Following \cite{He:2016vfi}, we can fix the $GL(k)$ redundancy of $C$ by setting the submatrix of of the first $k$ columns to the identity matrix. The non-trivial information is then encoded in the $k\times (n-k)$-dimensional submatrix $c$, whose elements are given by
\begin{align}\label{eq:link-param}
	c_{ij}=\frac{t_j\prod_{a\neq i}^k (z_j-z_a)}{t_i\prod_{a\neq i}^k (z_i-z_a)},\quad i=1,\ldots,k,\;j=k+1,\ldots,n,
\end{align}
The corresponding orthogonal complement $C^\perp$ will have a the submatrix defined by columns $\{k+1,\ldots,n\}$ set to the identity matrix, and the non-trivial $(n-k)\times k$ submatrix has elements $-c^T$,
with $c$ as in \eqref{eq:link-param}. Schematically,
\begin{align}
	C =\left[
	\begin{array}{c;{2pt/2pt}c}
		\unit_{k\times k}& c_{k\times(n-k)} 
	\end{array} \right],\quad C^\perp=\left[\begin{array}{c;{2pt/2pt}c} (-c^T)_{(n-k)\times k} & \unit_{(n-k)\times (n-k)} \end{array}\right].
\end{align}
In this choice of gauge fixing, the scattering equations \eqref{eq:scatt-eq-Grass} become
\begin{subequations}
	\begin{alignat}{3}
		0&=\lambda_j^\alpha - \sum_{i=1}^k &&c_{ij}\lambda_j^\alpha,\qquad &&j=k+1,\ldots,n,\\
		0&=\tilde\lambda_i^{\dot\alpha}+\sum_{j=k+1}^n &&c_{ij}\tilde\lambda_j^{\dot\alpha},&& i=1,\ldots,k.
	\end{alignat}
\end{subequations}
We make the following substitution:
\begin{align}
	\begin{cases}
		s_i=\big[t_i\prod_{a \neq i}^k (z_i-z_a)\big]^{-1},\quad & i=1,\ldots,k,\\
		s_j=t_j\prod_{a=1}^k (z_j-z_a),\quad&j= k+1,\ldots,n,
	\end{cases}
\end{align}
such that \eqref{eq:link-param} becomes $c_{ij}=s_i s_j/(z_i-z_j)$. The Grassmannian scattering equations now take the form
\begin{subequations}\label{eq:scatt-eq-4D-ambi}
	\begin{alignat}{3}
		\lambda_j^\alpha &= \sum_{i=1}^k &&\frac{\lambda_i^\alpha}{(ij)},\qquad &&j=k+1,\ldots,n,\label{eq:scatt-eq-lambda}\\
		\tilde\lambda_i^{\dot\alpha}&=\sum_{j=k+1}^n &&\frac{\tilde\lambda_j^{\dot\alpha}}{(ji)},&& i=1,\ldots,k.
	\end{alignat}
\end{subequations}
where $(ab)=z_{ab}/s_a s_b$ is the minor $p_{ab}$ of the $2\times n$ matrix
\begin{align}
	\begin{pmatrix}
		1/s_1 & 1/s_2 & \cdots & 1/s_n\\ z_1/s_1 &z_2/s_2&\cdots&z_n/s_n
	\end{pmatrix}.
\end{align}
The $GL(2,\Cbb)$ redundancy acting on this matrix means that we can interpret it as an element of $G(2,n)$. We refer to \eqref{eq:scatt-eq-4D-ambi} as the \emph{ambitwistor scattering equations}, and it was first derived in the context of ambitwistor strings in \cite{Geyer:2014fka}. The relation to the Witten-RSV equations was explored in \cite{He:2016vfi}.

There are a few things to note about these ambitwistor scattering equations. First of all, the choice of which submatrix of $C(\bm{z},\bm{t})$ to set to identity is of course arbitrary, and different equivalent forms of the ambitwistor scattering equations can be derived by a different gauge fixing. Second, of these $2n$ equations, only $2n-4$ are independent. The equations \eqref{eq:scatt-eq-4D-ambi} imply momentum conservation, since
\begin{align}
	\sum_{a=1}^n\lambda_a^\alpha\tilde\lambda_a^{\dot\alpha}=\sum_{i=1}^k\left( \lambda_i^\alpha\sum_{j=k+1}^n\frac{\tilde\lambda_j^{\dot\alpha}}{(ji)}\right) + \sum_{j=k+1}^n\left( \tilde\lambda_i^{\dot\alpha} \sum_{i=1}^k \frac{\lambda_i^\alpha}{(ij)}\right)=\sum_{i=1}^k\sum_{j=k+1}^n\left(\frac{\lambda_i^\alpha\tilde\lambda_j^{\dot\alpha}}{(ij)}+\frac{\lambda_i^\alpha\tilde\lambda_j^{\dot\alpha}}{(ji)}\right),
\end{align}
which vanishes since $(ab)=-(ba)$.

\subsubsection{MHV Scattering Equations}

In four dimensions, the scattering equations are exactly solvable in the $k=2$ (MHV) and $k=n-2$ ($\overline{\mbox{MHV}}$) regimes. We start from the ambitwistor scattering equations and fix $z_1\to0,z_2\to1,z_n\to\infty,s_n\to 1$. The brackets become 
\begin{align}
	(ij)=\begin{cases}
		(z_i-z_j)/s_is_j,\quad &j\neq n,\\ 1/s_i, & j=n,
	\end{cases}\qquad (\text{for }i< j),
\end{align}
When considering $k=2$ we can restrict out attention to the $2n-4$ equations given by \eqref{eq:scatt-eq-lambda}. Contracting with $\lambda_{1\alpha}$ and $\lambda_{2\alpha}$, the MHV scattering equations read
\begin{align}
	\<{1i}\> = \frac{\<{12}\>}{(2i)},\quad \<{2i}\> = \frac{\<{21}\>}{(i1)}.
\end{align}
The solutions to the MHV scattering equations are given by
\begin{align}
	z_i=\frac{\<{1i}\>\<{2n}\>}{\<{12}\>\<{in}\>},\; s_1=\frac{\<{2n}\>}{\<{12}\>},\;s_2=-\frac{\<{1n}\>}{\<{12}\>},\;s_i=\frac{\<{1i}\>\<{2i}\>}{\<{12}\>\<{in}\>}.
\end{align}

%% file: appendices/chambers.tex
\chapter{Chambers of the NMHV (Momentum) Amplituhedron}\label{sec:APP_chambers}

In this appendix we expand upon the ideas put forward in chapter \ref{sec:DUAL}, and explain an algorithm which can be used to find all chambers of the NMHV (momentum) amplituhedron. The chambers of the amplituhedron $\Acal_{n,1}$ and the momentum amplituhedron $\Mcal_{n,3}$ are T-dual to each other, and for this reason we can restrict our attention to the easier object to study, which is the amplituhedron. We expect that the T-duality between chambers of the amplituhedron and the momentum amplituhedron holds for higher $k$ as well.

We argued in section \ref{sec:DUAL_nf-chambers-integrand} that chambers of the amplituhedron are given by the maximal intersections of images of positroid cells, which we call \emph{tiles} (up to complications coming from cells with intersection number greater than one, like the four-mass-box starting at N\textsuperscript{2}MHV, see the discussion in section \ref{sec:DUAL_nf-chambers-integrand}). We propose a method of determining whether two tiles intersects based on the compatibility of the signs of their \emph{functionaries}. Functionaries are expressions of the kinematic variables, such that all spurious boundaries of tiles are on the zero set of some functionary. By `spurious boundaries' we mean boundaries of tiles that are not boundaries of the amplituhedron. The tiles can be completely determined by the list of signs of the functionaries \cite{Even-Zohar:2023del}. By checking the compatibility of these lists for two tiles, we can then determine whether or not they intersect.

Since $\Acal_{n,1}$ is four dimensional, we consider only the intersection of tiles of four dimensional positroid cells. As argued at the NMHV\textsubscript{7} example in section \ref{sec:DUAL_examples}, the higher dimensional positroid cells can have an influence on the canonical form of $\Delta(\bm{x})$, but they are not necessary to determine the chambers. All four dimensional positroid cells in $G_+(1,n)$ will have exactly five non-zero entries. We label these cells $(a_1a_2a_3a_4a_5)$ by the position of these non-zero entries (we assume $a_1<a_2<a_3<a_4<a_5$). We note that not all of these positroid cells correspond to leading singularities. That is, the vertex set $\Vcal(\sigma)$ might be empty for some of these cells. The discussion below gives a \emph{refinement} of the actual one-loop chambers of the amplituhedron, which means that some of the chambers we find might have combinatorially equivalent one-loop fibres. However, when we move to higher loops it is expected that these refinements will classify the higher-loop chamber structure of $\Acal_{n,1}$. It is expected that this will not continue indefinitely, and starting from 10-point N\textsuperscript{3}MHV at two loops we will start encountering chambers which can not be defined as the maximal intersection of tiles in the amplituhedron. This is because of the presence of elliptic curves \cite{Caron-Huot:2012awx}.

A point in the amplituhedron $\Acal_{n,1}$ is defined by
\begin{align}
	Y^I = \sum_{a=1}^n C_a Z_a^I\,.
\end{align}
If we take the matrix $C$ to be in the positroid cell $(a_1a_2a_3a_4a_5)$, then
\begin{align}
	Y^I= C_{a_1}Z_{a_1}^I+C_{a_2}Z_{a_2}^I+C_{a_3}Z_{a_3}^I+C_{a_4}Z_{a_4}^I+C_{a_5}Z_{a_5}^I\,,
\end{align}
and hence $\<Ya_1a_2a_3a_4\>= C_{a_5}\<a_5 a_1 a_2 a_3 a_4\>$, where the five-bracket $\<a_5 a_1 a_2 a_3 a_4\>$ denotes a minor of the matrix $Z\in M_+(5,n)$. Since both $C_{a_5}$ and $\<a_1 a_2 a_3 a_4 a_5\>$ are positive, we see that any point in this tile will satisfy $\<Ya_1a_2a_3a_4\>>0$. Similar calculations show that any $Y$ in the tile $\Phi_Z((a_1a_2a_3a_4a_5))$ satisfies
\begin{align}
	\<Ya_1a_2a_3a_4\>>0\,, \<Ya_1a_2a_3a_5\><0\,,\<Ya_1a_2a_4a_5\>>0\,,\<Ya_1a_3a_4a_5\><0\,,\<Ya_2a_3a_4a_5\>>0\,.
\end{align}
The boundaries of this tile are given by the zeroes of these quantities. The boundaries of $\Acal_{n,1}$ are given by $\<Yii+1jj+1\>=0$, and the functionaries for the NHMV amplituhedron are therefore the set of all $\<Yabcd\>$, minus the set of  $\<Yii+1jj+1\>$. 

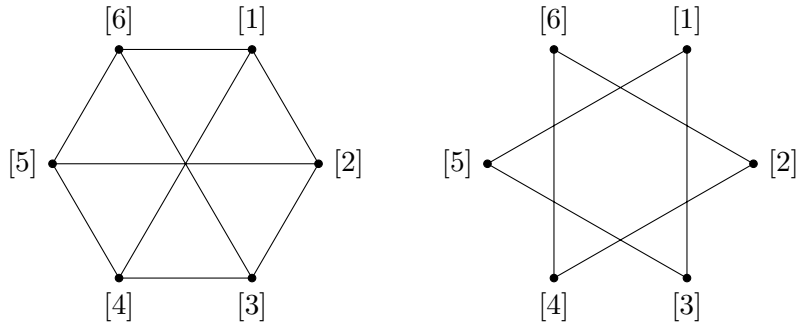
\begin{figure}[h]
	\center
	\begin{tikzpicture}[scale=0.7]
		\newdimen\R
		\R=2.5cm
		\draw (0:\R) \foreach \x in {60,120,...,360} {  -- (\x:\R) };
		\draw (0:\R) \foreach \x in {60,120,180} { (\x-180:\R) -- (\x:\R) };
		\foreach \x/\l/\p in
		{ 60/{[1]}/above,
			120/{[6]}/above,
			180/{[5]}/left,
			240/{[4]}/below,
			300/{[3]}/below,
			360/{[2]}/right
		}
		\node[inner sep=1pt,circle,draw,fill,label={\p:\l}] at (\x:\R) {};
	\end{tikzpicture}\qquad
	\begin{tikzpicture}[scale=0.7]
		\newdimen\R
		\R=2.5cm
		%\draw (0:\R) \foreach \x in {60,120,...,360} {  -- (\x:\R) };
		\draw (0:\R) \foreach \x in {60,120,...,360} { (\x+120:\R) -- (\x:\R) };
		\foreach \x/\l/\p in
		{ 60/{[1]}/above,
			120/{[6]}/above,
			180/{[5]}/left,
			240/{[4]}/below,
			300/{[3]}/below,
			360/{[2]}/right
		}
		\node[inner sep=1pt,circle,draw,fill,label={\p:\l}] at (\x:\R) {};
	\end{tikzpicture}
	\caption{The compatibility graph (left) and incompatibility graph (right) of tiles in $\Acal_{6,1}$.}
	\label{fig:adjacency_NMHV_6}
\end{figure}

As an explicit example, let us look at the chambers of $\Acal_{6,1}$. The functionaries are 
\begin{align}
	\<Y1235\>\,,\<Y1246\>\,,\<Y1345\>\,,\<Y1356\>\,,\<Y2346\>\,,\<Y2456\>\,.
\end{align}
We summarise the signs of these functionaries for all six tiles in table \ref{tab:NHMV6-functionaries}.
\begin{table}
	\centering
	\begin{tabular}{|c||c|c|c|c|c|c|}\hline
		& $\<Y1235\>$ & $\<Y1246\>$ & $\<Y1356\>$ & $\<Y2456\>$ & $\<Y1345\>$ & $\<Y2346\>$ \\ \hline\hline
		$\Phi_Z((12345))=[6]$  & $-$ &  &  &  & $-$ &  \\\hline
		$\Phi_Z((12346))=[5]$  &  & $+$ &  &  &  & $+$ \\\hline
		$\Phi_Z((12356))=[4]$  & $+$ &  & $-$ &  &  &  \\\hline
		$\Phi_Z((12456))=[3]$  &  & $-$ &  & $+$ &  &  \\\hline
		$\Phi_Z((13456))=[2]$  &  &  & $+$ &  & $+$ &  \\\hline
		$\Phi_Z((23456))=[1]$  &  &  &  & $-$ &  & $-$ \\\hline
	\end{tabular}
	\caption{All tiles in $\Acal_{6,1}$ and the signs of the functionaries.}
	\label{tab:NHMV6-functionaries}
\end{table}
We see that, for example, the signs for tiles $[1]$ and $[2]$ are \emph{compatible}, meaning that they intersect, whereas $[1]$ doesn't intersect $[3]$, because they are on opposite sides of the spurious boundary $\<Y2456\>=0$. We can summarise these results in a \emph{compatibility graph}, and its complement, the \emph{incompatibility graph}, which are graphs with nodes representing tiles, and we connect nodes with edges if they are compatible/incompatible. We show both the compatibility and the incompatibility graph for $\Acal_{6,1}$ in figure \ref{fig:adjacency_NMHV_6}.
Chambers are given by maximal collections of intersecting tiles, whereas triangulations are given by maximal collections of non-intersecting tiles. Chambers and triangulations can therefore be found as \emph{maximal cliques} (\textit{i.e.} maximal complete subgraphs) of the compatibility and incompatibility graphs, respectively. In the case of $\Acal_{6,1}$, we find the chambers
\begin{alignat}{3}
	\Cfrak(\Acal_{6,1})=\{ &[1]\cap[2]\,,\quad&& [1]\cap[4]\,,\quad&&[1]\cap[6]\,,\notag\\\notag
	&[3]\cap[2]\,, && [3]\cap[4]\,, && [3]\cap [6]\,,\\
	& [5]\cap[2]\,, &&[5]\cap[4]\,, &&[5]\cap [6]  \}\,,
\end{alignat} 
and the triangulations
\begin{align}
	\Acal_{6,1} = [1]\cup [3]\cup [5] = [2]\cup [4]\cup [6]\,.
\end{align}
In practice, \texttt{IGraphM} \cite{Horvat2023} can be used to find all maximal cliques very efficiently. We have used this algorithm to find all chambers and triangulations for $\Acal_{n,1}$ for $n\leq 10$. We summarise the number of chambers and triangulations in table \ref{tab:number_of_chambers}.
\begin{table}
	\begin{center}
		\begin{tabular}{|c||c|c|c|c|c|c|c|}
			\hline
			$n$&5&6&7&8&9&10\\
			\hline\hline
			$\#$ tiles &1&6&21&56&126&252\\
			\hline
			$\#$ chambers&1&9&71&728&15979&1144061\\
			\hline
			$\#$ triangulations&1&2&7&40&357&4824\\
			\hline
		\end{tabular}
	\end{center}
	\caption{The number of chambers and triangulations of the NMHV (momentum) amplituhedron for $n\leq 10$.}
	\label{tab:number_of_chambers}
\end{table}
For example, for $n=7$ we find 71 chambers coming in 11 cyclic classes. They are explicitly given by
\begin{align}
	\Cfrak(\Acal_{7,1})= \{& [1,2]\cap[1,3]\cap[2,4],\notag\\\notag
	&[1,2]\cap[1,5]\cap[2,5],\\\notag
	&[1,2]\cap[1,3]\cap[2,5]\cap[3,5],\\\notag
	&[1,2]\cap[1,3]\cap[2,7]\cap[3,7],\\\notag
	&[1,2]\cap[1,5]\cap[2,7]\cap[5,7],\\\notag
	&[1,3]\cap[1,4]\cap[3,7]\cap[4,7],\\\notag
	&[1,2]\cap[1,5]\cap[2,7]\cap[3,5]\cap[5,7],\\\notag
	&[1,3]\cap[1,4]\cap[2,4]\cap[2,5]\cap[3,5],\\\notag
	&[1,3]\cap[1,4]\cap[3,5]\cap[4,7]\cap[5,7],\\\notag
	&[1,3]\cap[1,4]\cap[2,4]\cap[2,7]\cap[3,5]\cap[5,7],\\
	&[1,3]\cap[1,6]\cap[2,4]\cap[2,7]\cap[3,5]\cap[4,6]\cap[5,7],	+\text{ cyclic}\}\,,
\end{align}
where we use the notation introduced in section \ref{sec:DUAL_examples}, \emph{i.e.} $[1,2]=\Phi_Z((34567))$, etc.